\documentclass[PHD]{macro/neu_msthesis}

\title{Design and Evaluation of Next-Generation Cellular Networks through Digital and Physical Open and Programmable Platforms}

\author{Davide Villa}

\nuid{001560652}

\dept{Electrical and Computer Engineering}

\degreename{Electrical and Computer Engineering}

\field{Electrical and Computer Engineering}

\submitdate{December 2025}

\numberofmembers{3}
\principaladviser{Prof. Tommaso Melodia}
\firstreader{Prof. Stefano Basagni}
\secondreader{Prof. Josep M. Jornet}
\thirdreader{Dr. Third Member}
\fourthreader{Dr. Fourth Member}
\fifthreader{Dr. Fifth Member}
\chairman{Dr. Miriam Leeser}
\dean{Dr. Thomas C. Sheahan}

\usepackage{amsfonts,amssymb,amsmath}			
\usepackage{times}		
\usepackage{bm}

\newcommand{\ifno}[1]{}

\usepackage{multirow}
\makeindex
\usepackage{makeidx}

\usepackage{url}
\ifnum\pdfoutput>0
\usepackage[pdftex]{graphicx}
\usepackage{epstopdf}
\else
\usepackage{graphicx}
\fi

\ifnum\pdfoutput>0
\usepackage[pdftex,
bookmarks=true,
bookmarksnumbered=true,
hypertexnames=false,
breaklinks=true           
]{hyperref}
\else
\usepackage[hypertex,
bookmarks=true,
bookmarksnumbered=true,
hypertexnames=false,
breaklinks=true           
]{hyperref}
\fi

\hypersetup{				
pdfauthor = {\authorRef},
pdftitle = {\titleRef},
pdfsubject = {\expandafter{\degreeRef} thesis submitted to Northeastern University},
pdfkeywords = {add keywords here}
pdfcreator = {LaTeX with hyperref package},
pdfproducer = {dvips + ps2pdf}}

\usepackage{microtype}

\usepackage[acronyms,nonumberlist,nopostdot,nomain,nogroupskip,acronymlists={hidden}]{glossaries}
\newglossary[algh]{hidden}{acrh}{acnh}{Hidden Acronyms}

\newacronym{3gpp}{3GPP}{3rd Generation Partnership Project}
\newacronym{1g}{1G}{First Generation}
\newacronym{2g}{2G}{Second Generation}
\newacronym{3g}{3G}{Third Generation}
\newacronym{4g}{4G}{Fourth Generation}
\newacronym{5g}{5G}{Fifth Generation}
\newacronym{5gc}{5GC}{5G Core}
\newacronym{6g}{6G}{Sixth Generation}
\newacronym{abr}{ABR}{Adaptive Bitrate Streaming}
\newacronym{adc}{ADC}{Analog to Digital Converter}
\newacronym{adl}{ADL}{Aerial Data Lake}
\newacronym{aerpaw}{AERPAW}{Aerial Experimentation and Research Platform for Advanced Wireless}
\newacronym{ai}{AI}{Artificial Intelligence}
\newacronym{aimd}{AIMD}{Additive Increase Multiplicative Decrease}
\newacronym{am}{AM}{Acknowledged Mode}
\newacronym{amc}{AMC}{Adaptive Modulation and Coding}
\newacronym{amf}{AMF}{Access and Mobility Management Function}
\newacronym{aoa}{AoA}{Angle of Arrival}
\newacronym{aops}{AOPS}{Adaptive Order Prediction Scheduling}
\newacronym{api}{API}{Application Programming Interface}
\newacronym{apn}{APN}{Access Point Name}
\newacronym{aqm}{AQM}{Active Queue Management}
\newacronym{arc}{ARC}{Aerial RAN CoLab}
\newacronym{arc-ota}{ARC-OTA}{Aerial RAN CoLab Over-the-Air}
\newacronym{asic}{ASIC}{Application-Specific Integrated Circuit}
\newacronym{ausf}{AUSF}{Authentication Server Function}
\newacronym{avc}{AVC}{Advanced Video Coding}
\newacronym{awgn}{AWGN}{Additive White Gaussian Noise}
\newacronym{balia}{BALIA}{Balanced Link Adaptation Algorithm}
\newacronym{bbu}{BBU}{Base Band Unit}
\newacronym{bdp}{BDP}{Bandwidth-Delay Product}
\newacronym{ber}{BER}{Bit Error Rate}
\newacronym{bf}{BF}{Beamforming}
\newacronym{bler}{BLER}{Block Error Rate}
\newacronym{bmca}{BMCA}{Best Master Clock Algorithm}
\newacronym{bom}{BoM}{Bill of Materials}
\newacronym{bpsk}{BPSK}{Binary Phase-shift keying}
\newacronym{brr}{BRR}{Bayesian Ridge Regressor}
\newacronym{bs}{BS}{Base Station}
\newacronym{bsr}{BSR}{Buffer Status Report}
\newacronym{bss}{BSS}{Business Support System}
\newacronym{ca}{CA}{Carrier Aggregation}
\newacronym{caas}{CaaS}{Connectivity-as-a-Service}
\newacronym{cast}{\textit{CaST}}{Channel emulation generator and Sounder Toolchain}
\newacronym{cb}{CB}{Code Block}
\newacronym{cbrs}{CBRS}{Citizen Broadband Radio Service}
\newacronym{cc}{CC}{Congestion Control}
\newacronym{ccid}{CCID}{Congestion Control ID}
\newacronym{cco}{CC}{Carrier Component}
\newacronym{cd}{CD}{Continuous Delivery}
\newacronym{cdd}{CDD}{Cyclic Delay Diversity}
\newacronym{cdf}{CDF}{Cumulative Distribution Function}
\newacronym{cdma}{CDMA}{Code-Division Multiple Access}
\newacronym{cdn}{CDN}{Content Distribution Network}
\newacronym{cfo}{CFO}{Carrier Frequency Offset}
\newacronym{cfr}{CFR}{Code of Federal Regulations}
\newacronym{cicd}{CI/CD}{Continuous Integration/Continuous Delivery}
\newacronym{ci}{CI}{Continuous Integration}
\newacronym{cir}{CIR}{Channel Impulse Response}
\newacronym{cn}{CN}{Core Network}
\newacronym{cnn}{CNN}{Convolutional Neural Network}
\newacronym{codel}{CoDel}{Controlled Delay Management}
\newacronym{comac}{COMAC}{Converged Multi-Access and Core}
\newacronym{cord}{CORD}{Central Office Re-architected as a Datacenter}
\newacronym{cornet}{CORNET}{COgnitive Radio NETwork}
\newacronym{cosmos}{COSMOS}{Cloud Enhanced Open Software Defined Mobile Wireless Testbed for City-Scale Deployment}
\newacronym{cots}{COTS}{Commercial Off-the-Shelf}
\newacronym{cp}{CP}{Control Plane}
\newacronym{cpri}{CPRI}{Common Public Radio Interface}
\newacronym{cpu}{CPU}{Central Processing Unit}
\newacronym{cqi}{CQI}{Channel Quality Information}
\newacronym{cr}{CR}{Cognitive Radio}
\newacronym{cran}{CRAN}{Cloud \gls{ran}}
\newacronym{crc}{CRC}{Cyclic Redundancy Check}
\newacronym{crs}{CRS}{Cell Reference Signal}
\newacronym{csi}{CSI}{Channel State Information}
\newacronym{csirs}{CSI-RS}{Channel State Information - Reference Signal}
\newacronym{cu}{CU}{Central Unit}
\newacronym{cubb}{cuBB}{CUDA Baseband}
\newacronym{cublas}{cuBLAS}{CUDA Basic Linear Algebra Subroutines}
\newacronym{cuda}{CUDA}{Compute Unified Device Architecture}
\newacronym{cudnn}{cuDNN}{CUDA Deep Neural Network library}
\newacronym{cuphy}{cuPHY}{CUDA Physical layer}
\newacronym{d2tcp}{D$^2$TCP}{Deadline-aware Data center TCP}
\newacronym{d3}{D$^3$}{Deadline-Driven Delivery}
\newacronym{dac}{DAC}{Digital to Analog Converter}
\newacronym{dag}{DAG}{Directed Acyclic Graph}
\newacronym{dapp}{dApp}{distributed Application}
\newacronym{darpa}{DARPA}{Defense Advanced Research Projects Agency}
\newacronym{das}{DAS}{Distributed Antenna System}
\newacronym{dash}{DASH}{Dynamic Adaptive Streaming over HTTP}
\newacronym{dc}{DC}{Direct Current}
\newacronym{dccp}{DCCP}{Datagram Congestion Control Protocol}
\newacronym{dce}{DCE}{Direct Code Execution}
\newacronym{dci}{DCI}{Downlink Control Information}
\newacronym{dcl}{DCL}{Dear Colleague Letter}
\newacronym{dctcp}{DCTCP}{Data Center TCP}
\newacronym{dl}{DL}{Downlink}
\newacronym{dmr}{DMR}{Deadline Miss Ratio}
\newacronym{dmrs}{DMRS}{DeModulation Reference Signal}
\newacronym{dnn}{DNN}{Deep Neural Network}
\newacronym{dpu}{DPU}{Data Processing Unit}
\newacronym{drl}{DRL}{Deep Reinforcement Learning}
\newacronym{drlcc}{DRL-CC}{Deep Reinforcement Learning Congestion Control}
\newacronym{drs}{DRS}{Discovery Reference Signal}
\newacronym{dsp}{DSP}{Digital Signal Processing}
\newacronym{dt}{DT}{Digital Twin}
\newacronym{dtmn}{DTMN}{Digital Twins for Mobile Networks}
\newacronym{dtn}{DTN}{Digital Twin Network}
\newacronym{dtran}{DT-RAN}{Digital Twin for Radio Access Network}
\newacronym{dtwn}{DTWN}{Digital Twin Wireless Network}
\newacronym{du}{DU}{Distributed Unit}
\newacronym{e2e}{E2E}{end-to-end}
\newacronym{e2ap}{E2AP}{E2 Application Protocol}
\newacronym{e2sm}{E2SM}{E2 Service Model}
\newacronym{e3ap}{E3AP}{E3 Application Protocol}
\newacronym{e3sm}{E3SM}{E3 Service Model}
\newacronym{ecaas}{ECaaS}{Edge-Cloud-as-a-Service}
\newacronym{ecdf}{eCDF}{Empirical Cumulative Distribution Function}
\newacronym{ecn}{ECN}{Explicit Congestion Notification}
\newacronym{ecpri}{eCPRI}{enhanced CPRI}
\newacronym{edf}{EDF}{Earliest Deadline First}
\newacronym{eirp}{EIRP}{Effective Isotropic Radiated Power}
\newacronym{em}{EM}{Electro-Magnetic}
\newacronym{embb}{eMBB}{Enhanced Mobile Broadband}
\newacronym{empower}{EMPOWER}{EMpowering transatlantic PlatfOrms for advanced WirEless Research}
\newacronym{enb}{eNB}{evolved Node Base}
\newacronym{endc}{EN-DC}{E-UTRAN-\gls{nr} \gls{dc}}
\newacronym{epc}{EPC}{Evolved Packet Core}
\newacronym{eps}{EPS}{Evolved Packet System}
\newacronym{es}{ES}{Edge Server}
\newacronym[firstplural=Estimated Times of Arrival (ETAs)]{eta}{ETA}{Estimated Time of Arrival}
\newacronym{etsi}{ETSI}{European Telecommunications Standards Institute}
\newacronym{eutran}{E-UTRAN}{Evolved Universal Terrestrial Access Network}
\newacronym{faas}{FaaS}{Function-as-a-Service}
\newacronym{fapi}{FAPI}{Functional Application Platform Interface}
\newacronym{fcc}{FCC}{Federal Communications Commission}
\newacronym{fdd}{FDD}{Frequency Division Duplexing}
\newacronym{fdm}{FDM}{Frequency Division Multiplexing}
\newacronym{fdma}{FDMA}{Frequency Division Multiple Access}
\newacronym{fed4fire}{FED4FIRE+}{Federation 4 Future Internet Research and Experimentation Plus}
\newacronym{fh}{FH}{Fronthaul}
\newacronym{fid}{FID}{Fréchet Inception Distance}
\newacronym{fir}{FIR}{Finite Impulse Response}
\newacronym{fit}{FIT}{Future \acrlong{iot}}
\newacronym{fl}{FL}{Federated Learning}
\newacronym{fn}{FN}{False Negative}
\newacronym{fp}{FP}{False Positive}
\newacronym{fp16}{FP16}{Float16}
\newacronym{fp32}{FP32}{Float32}
\newacronym{fpga}{FPGA}{Field Programmable Gate Array}
\newacronym{fps}{FPS}{Frames per second}
\newacronym{fr1}{FR1}{Frequency Range 1}
\newacronym{fr2}{FR2}{Frequency Range 2}
\newacronym{frand}{FRAND}{Fair, Reasonable, And Non-Discriminatory}
\newacronym{fs}{FS}{Fast Switching}
\newacronym{fscc}{FSCC}{Flow Sharing Congestion Control}
\newacronym{ftp}{FTP}{File Transfer Protocol}
\newacronym{fw}{FW}{Flow Window}
\newacronym{ga128}{Ga}{Golay Sequence type A}
\newacronym{gan}{GAN}{Generative Adversarial Network}
\newacronym{ge}{GE}{Gaussian Elimination}
\newacronym{genai}{GenAI}{Generative Artificial Intelligence}
\newacronym{gentwin}{Gen-TWIN}{GenerativeAI-enabled Digital Twin}
\newacronym{gh}{GH}{Grace Hopper}
\newacronym{glfsr}{GLFSR}{Galois Linear Feedback Shift Register}
\newacronym{gnb}{gNB}{Next Generation Node Base}
\newacronym{gold}{Gold}{Gold}
\newacronym{gop}{GOP}{Group of Pictures}
\newacronym{gpr}{GPR}{Gaussian Process Regressor}
\newacronym{gpu}{GPU}{Graphics Processing Unit}
\newacronym{grpc}{gRPC}{gRPC Remote Procedure Calls}
\newacronym{gtp}{GTP}{GPRS Tunneling Protocol}
\newacronym{gtpc}{GTP-C}{GPRS Tunnelling Protocol Control Plane}
\newacronym{gtpu}{GTP-U}{GPRS Tunnelling Protocol User Plane}
\newacronym{gtpv2c}{GTPv2-C}{\gls{gtp} v2 - Control}
\newacronym{gui}{GUI}{Graphical User Interface}
\newacronym{gw}{GW}{Gateway}
\newacronym{harq}{HARQ}{Hybrid Automatic Repeat Request}
\newacronym{hdf5}{HDF5}{Hierarchical Data Format version 5}
\newacronym{hdr}{HDR}{High Dynamic Range}
\newacronym{hest}{$\hat{\mathbf{H}}$}{Channel Estimates}
\newacronym{hetnet}{HetNet}{Heterogeneous Network}
\newacronym{hh}{HH}{Hard Handover}
\newacronym{hol}{HOL}{Head-of-Line}
\newacronym{hqf}{HQF}{Highest-quality-first}
\newacronym{hss}{HSS}{Home Subscription Server}
\newacronym{http}{HTTP}{HyperText Transfer Protocol}
\newacronym{ia}{IA}{Initial Access}
\newacronym{iab}{IAB}{Integrated Access and Backhaul}
\newacronym{ic}{IC}{Incident Command}
\newacronym{ietf}{IETF}{Internet Engineering Task Force}
\newacronym{ifw}{IFW}{Interference Free Window}
\newacronym{imsi}{IMSI}{International Mobile Subscriber Identity}
\newacronym{imt}{IMT}{International Mobile Telecommunication}
\newacronym{int32}{INT32}{32-bit Integer}
\newacronym{io}{I/O}{Input/Output}
\newacronym{iot}{IoT}{Internet of Things}
\newacronym{ip}{IP}{Internet Protocol}
\newacronym{ipc}{IPC}{Inter-Process Communication}
\newacronym{iq}{I/Q}{In-phase and Quadrature}
\newacronym{irc}{IRC}{Interference Rejection Combining}
\newacronym{isac}{ISAC}{Integrated Sensing and Communication}
\newacronym{itu}{ITU}{International Telecommunication Union}
\newacronym{iw}{IW}{Interference Whitening}
\newacronym{kl}{KL}{Kullback–Leibler}
\newacronym{kpi}{KPI}{Key Performance Indicator}
\newacronym{kpm}{KPM}{Key Performance Measurement}
\newacronym{kvm}{KVM}{Kernel-based Virtual Machine}
\newacronym{ldpc}{LDPC}{Low Density Parity-Check}
\newacronym{leo}{LEO}{Low Earth Orbit}
\newacronym{lfsr}{LFSR}{Linear Feedback Shift Register}
\newacronym{los}{LOS}{Line-of-Sight}
\newacronym{ls}{LS}{Loosely Synchronised}
\newacronym{lsm}{LSM}{Link-to-System Mapping}
\newacronym{lstm}{LSTM}{Long Short Term Memory}
\newacronym{lte}{LTE}{Long Term Evolution}
\newacronym{lxc}{LXC}{Linux Container}
\newacronym{m2m}{M2M}{Machine to Machine}
\newacronym{mac}{MAC}{Medium Access Control}
\newacronym{manet}{MANET}{Mobile Ad Hoc Network}
\newacronym{mano}{MANO}{Management and Orchestration}
\newacronym{mc}{MC}{Multi-Connectivity}
\newacronym{mcc}{MCC}{Mobile Cloud Computing}
\newacronym{mchem}{MCHEM}{Massive Channel Emulator}
\newacronym{mcs}{MCS}{Modulation and Coding Scheme}
\newacronym{mec}{MEC}{Multi-access Edge Computing}
\newacronym{mec2}{MEC}{Mobile Edge Cloud}
\newacronym{mfc}{MFC}{Mobile Fog Computing}
\newacronym{mgen}{MGEN}{Multi-Generator}
\newacronym{mi}{MI}{Mutual Information}
\newacronym{mib}{MIB}{Master Information Block}
\newacronym{miesm}{MIESM}{Mutual Information Based Effective SINR}
\newacronym{mig}{MIG}{Multi-Instance GPU}
\newacronym{mimo}{MIMO}{Multiple Input, Multiple Output}
\newacronym{ml}{ML}{Machine Learning}
\newacronym{mlp}{MLP}{Multilayer Perceptron}
\newacronym{mlr}{MLR}{Maximum-local-rate}
\newacronym[plural=\gls{mme}s,firstplural=Mobility Management Entities (MMEs)]{mme}{MME}{Mobility Management Entity}
\newacronym{mmse}{MMSE}{Minimum Mean Square Error}
\newacronym{mmtc}{mMTC}{Massive Machine-Type Communications}
\newacronym{mmwave}{mmWave}{millimeter wave}
\newacronym{mpdccp}{MP-DCCP}{Multipath Datagram Congestion Control Protocol}
\newacronym{mps}{MPS}{Multi-Process Service}
\newacronym{mptcp}{MPTCP}{Multipath TCP}
\newacronym{mr}{MR}{Maximum Rate}
\newacronym{mrdc}{MR-DC}{Multi \gls{rat} \gls{dc}}
\newacronym{mse}{MSE}{Mean Square Error}
\newacronym{mss}{MSS}{Maximum Segment Size}
\newacronym{mt}{MT}{Mobile Termination}
\newacronym{mtc}{MTC}{Machine-type Communications}
\newacronym{mtd}{MTD}{Machine-Type Device}
\newacronym{mtu}{MTU}{Maximum Transmission Unit}
\newacronym{mumimo}{MU-MIMO}{Multi-user \gls{mimo}}
\newacronym{mvno}{MVNO}{Mobile Virtual Network Operator}
\newacronym{nalu}{NALU}{Network Abstraction Layer Unit}
\newacronym{nas}{NAS}{Network Attached Storage}
\newacronym{nbiot}{NB-IoT}{Narrow Band IoT}
\newacronym{near-rt}{near-RT}{near-Real-Time}
\newacronym{near-rt-ric}{near-RT-RIC}{near-RT-\gls{ric}}
\newacronym{nextg}{NextG}{Next Generation}
\newacronym{nf}{NF}{Network Function}
\newacronym{nfv}{NFV}{Network Function Virtualization}
\newacronym{nfvi}{NFVI}{Network Function Virtualization Infrastructure}
\newacronym{nic}{NIC}{Network Interface Card}
\newacronym{nlb}{NLB}{Non-local Block}
\newacronym{nlos}{NLOS}{Non-Line-of-Sight}
\newacronym{nn}{NN}{Neural Network}
\newacronym{non-rt}{non-RT}{non-Real-Time}
\newacronym{now}{NOW}{Non Overlapping Window}
\newacronym[type=hidden]{nr}{NR}{New Radio}
\newacronym{nrdz}{NRDZ}{National Radio Dynamic Zone}
\newacronym{nrf}{NRF}{Network Repository Function}
\newacronym{nrmse}{NRMSE}{Normalized Root Mean Squared Error}
\newacronym{nsa}{NSA}{Non Stand Alone}
\newacronym{nse}{NSE}{Network Slicing Engine}
\newacronym{nsf}{NSF}{National Science Foundation}
\newacronym{nsm}{NSM}{Network Service Mesh}
\newacronym{nssf}{NSSF}{Network Slice Selection Function}
\newacronym{ntp}{NTP}{Network Time Protocol}
\newacronym{nvipc}{NVIPC}{NVIDIA Inter-Process Communication}
\newacronym{o2i}{O2I}{Outdoor to Indoor}
\newacronym{oai}{OAI}{OpenAirInterface}
\newacronym{oaic}{OAIC}{Open AI Cellular}
\newacronym{oaicn}{OAI-CN}{\gls{oai} \acrlong{cn}}
\newacronym{oairan}{OAI-RAN}{\acrlong{oai} \acrlong{ran}}
\newacronym{oam}{OAM}{Operations, Administration and Maintenance}
\newacronym[plural=\gls{obu}s,firstplural=Onboard Units (OBUs)]{obu}{OBU}{Onboard Unit}
\newacronym{odc}{ODC}{ORAN Development Company}
\newacronym{ofdm}{OFDM}{Orthogonal Frequency Division Multiplexing}
\newacronym{olia}{OLIA}{Opportunistic Linked Increase Algorithm}
\newacronym{omec}{OMEC}{Open Mobile Evolved Core}
\newacronym{onap}{ONAP}{Open Network Automation Platform}
\newacronym{onf}{ONF}{Open Networking Foundation}
\newacronym{onnx}{ONNX}{Open Neural Network Exchange}
\newacronym{onos}{ONOS}{Open Networking Operating System}
\newacronym{oom}{OOM}{\gls{onap} Operations Manager}
\newacronym{opnfv}{OPNFV}{Open Platform for \gls{nfv}}
\newacronym{oran}{Open RAN}{Open \gls{ran}}
\newacronym{orbit}{ORBIT}{Open-Access Research Testbed for Next-Generation Wireless Networks}
\newacronym{ort}{ORT}{ONNX Runtime}
\newacronym{os}{OS}{Operating System}
\newacronym{osc}{OSC}{O-RAN Software Community}
\newacronym{osm}{OSM}{OpenStreetMap}
\newacronym{oss}{OSS}{Operations Support System}
\newacronym{ota}{OTA}{Over-the-Air}
\newacronym{otc}{OTC}{Over-the-Cable}
\newacronym{p5g}{P5G}{Private 5G}
\newacronym{pa}{PA}{Position-aware}
\newacronym{pase}{PASE}{Prioritization, Arbitration, and Self-adjusting Endpoints}
\newacronym{pawr}{PAWR}{Platforms for Advanced Wireless Research}
\newacronym{pbch}{PBCH}{Physical Broadcast Channel}
\newacronym{pca}{PCA}{Principal Component Analysis}
\newacronym{pcef}{PCEF}{Policy and Charging Enforcement Function}
\newacronym{pcfich}{PCFICH}{Physical Control Format Indicator Channel}
\newacronym{pci}{PCI}{Peripheral Component Interconnect}
\newacronym{pcrf}{PCRF}{Policy and Charging Rules Function}
\newacronym{pdcch}{PDCCH}{Physical Downlink Control Channel}
\newacronym{pdcp}{PDCP}{Packet Data Convergence Protocol}
\newacronym{pdp}{PDP}{Power Delay Profile}
\newacronym{pdsch}{PDSCH}{Physical Downlink Shared Channel}
\newacronym{pdu}{PDU}{Packet Data Unit}
\newacronym{pf}{PF}{Proportional Fair}
\newacronym{pgw}{PGW}{Packet Gateway}
\newacronym{ph}{PH}{Power Headroom}
\newacronym{phich}{PHICH}{Physical Hybrid ARQ Indicator Channel}
\newacronym{phy}{PHY}{Physical}
\newacronym{pl}{PL}{Path Loss}
\newacronym{pmch}{PMCH}{Physical Multicast Channel}
\newacronym{pmi}{PMI}{Precoding Matrix Indicators}
\newacronym{pnf}{PNF}{Physical Network Function}
\newacronym{powder}{POWDER}{Platform for Open Wireless Data-driven Experimental Research}
\newacronym{ppo}{PPO}{Proximal Policy Optimization}
\newacronym{ppp}{PPP}{Poisson Point Process}
\newacronym{prach}{PRACH}{Physical Random Access Channel}
\newacronym{prb}{PRB}{Physical Resource Block}
\newacronym{prtc}{PRTC}{Primary Reference Time Clocks}
\newacronym{psd}{PSD}{Power Spectral Density}
\newacronym{psnr}{PSNR}{Peak Signal to Noise Ratio}
\newacronym{pss}{PSS}{Primary Synchronization Signal}
\newacronym{ptp}{PTP}{Precision Timing Protocol}
\newacronym{pucch}{PUCCH}{Physical Uplink Control Channel}
\newacronym{pusch}{PUSCH}{Physical Uplink Shared Channel}
\newacronym{qam}{QAM}{Quadrature Amplitude Modulation}
\newacronym{qci}{QCI}{\gls{qos} Class Identifier}
\newacronym{qoe}{QoE}{Quality of Experience}
\newacronym{qos}{QoS}{Quality of Service}
\newacronym{qsfp28}{QSFP28}{Quad Small Form-factor Pluggable 28}
\newacronym{qtgui}{QT-GUI}{QT Graphical User Interface}
\newacronym{quic}{QUIC}{Quick UDP Internet Connections}
\newacronym{rach}{RACH}{Random Access Channel}
\newacronym{ran}{RAN}{Radio Access Network}
\newacronym[firstplural=Radio Access Technologies (RATs)]{rat}{RAT}{Radio Access Technology}
\newacronym{rb}{RB}{Resource Block}
\newacronym{rcn}{RCN}{Research Coordination Network}
\newacronym{rdma}{RDMA}{Remote Direct Memory Access}
\newacronym{re}{RE}{Resource Element}
\newacronym{rec}{REC}{Radio Edge Cloud}
\newacronym{red}{RED}{Random Early Detection}
\newacronym{renew}{RENEW}{Reconfigurable Eco-system for Next-generation End-to-end Wireless}
\newacronym{rf}{RF}{Radio Frequency}
\newacronym{rfc}{RFC}{Request for Comments}
\newacronym{rfr}{RFR}{Random Forest Regressor}
\newacronym{ric}{RIC}{RAN Intelligent Controller}
\newacronym{rlc}{RLC}{Radio Link Control}
\newacronym{rlf}{RLF}{Radio Link Failure}
\newacronym{rlnc}{RLNC}{Random Linear Network Coding}
\newacronym{rmr}{RMR}{RIC Message Router}
\newacronym{rmse}{RMSE}{Root Mean Squared Error}
\newacronym{rnis}{RNIS}{Radio Network Information Service}
\newacronym{rnti}{RNTI}{Radio Network Temporary Identifier}
\newacronym{rr}{RR}{Round Robin}
\newacronym{rrc}{RRC}{Radio Resource Control}
\newacronym{rrm}{RRM}{Radio Resource Management}
\newacronym{rru}{RRU}{Remote Radio Unit}
\newacronym{rs}{RS}{Remote Server}
\newacronym{rsrp}{RSRP}{Reference Signal Received Power}
\newacronym{rsrq}{RSRQ}{Reference Signal Received Quality}
\newacronym{rss}{RSS}{Received Signal Strength}
\newacronym{rssi}{RSSI}{Received Signal Strength Indicator}
\newacronym{rsu}{RSU}{Road-Side Unit}
\newacronym{rt}{RT}{Real-Time}
\newacronym{rtt}{RTT}{Round Trip Time}
\newacronym{ru}{RU}{Radio Unit}
\newacronym{rw}{RW}{Receive Window}
\newacronym{rx}{RX}{Receiver}
\newacronym{s1ap}{S1AP}{S1 Application Protocol}
\newacronym{sa}{SA}{standalone}
\newacronym{sack}{SACK}{Selective Acknowledgment}
\newacronym{sap}{SAP}{Service Access Point}
\newacronym{sas}{SAS}{Spectrum Access System}
\newacronym{sc2}{SC2}{Spectrum Collaboration Challenge}
\newacronym{scef}{SCEF}{Service Capability Exposure Function}
\newacronym{scf}{SCF}{Small Cell Forum}
\newacronym{sch}{SCH}{Secondary Cell Handover}
\newacronym{scoot}{SCOOT}{Split Cycle Offset Optimization Technique}
\newacronym{sctp}{SCTP}{Stream Control Transmission Protocol}
\newacronym{sd}{SD}{Standard Deviation}
\newacronym{sdap}{SDAP}{Service Data Adaptation Protocol}
\newacronym{sdk}{SDK}{Software Development Kit}
\newacronym{sdm}{SDM}{Space Division Multiplexing}
\newacronym{sdma}{SDMA}{Spatial Division Multiple Access}
\newacronym{sdn}{SDN}{Software-defined Networking}
\newacronym{sdr}{SDR}{Software-defined Radio}
\newacronym{seba}{SEBA}{SDN-Enabled Broadband Access}
\newacronym{sfp+}{SFP+}{Small Form-factor Pluggable Plus}
\newacronym{sgsn}{SGSN}{Serving GPRS Support Node}
\newacronym{sgw}{SGW}{Service Gateway}
\newacronym{shm}{SHM}{Shared Memory}
\newacronym{si}{SI}{Study Item}
\newacronym{sib}{SIB}{Secondary Information Block}
\newacronym{sinr}{SINR}{Signal to Interference plus Noise Ratio}
\newacronym{sip}{SIP}{Session Initiation Protocol}
\newacronym{sir}{SIR}{Signal to Interference Ratio}
\newacronym{siso}{SISO}{Single Input, Single Output}
\newacronym{sla}{SLA}{Service Level Agreement}
\newacronym{sm}{SM}{Service Model}
\newacronym{smf}{SMF}{Session Management Function}
\newacronym{smo}{SMO}{Service Management and Orchestration}
\newacronym{sms}{SMS}{Short Message Service}
\newacronym{smsgmsc}{SMS-GMSC}{\gls{sms}-Gateway}
\newacronym{snssai}{S-NSSAI}{Single Network Slice Selection Assistance Information}
\newacronym{snr}{SNR}{Signal-to-Noise-Ratio}
\newacronym{soc}{SoC}{System-on-Chip}
\newacronym{son}{SON}{Self-Organizing Network}
\newacronym{sp}{S-plane}{Synchronization Plane}
\newacronym{sptcp}{SPTCP}{Single Path TCP}
\newacronym{srb}{SRB}{Service Radio Bearer}
\newacronym{srn}{SRN}{Standard Radio Node}
\newacronym{srs}{SRS}{Sounding Reference Signal}
\newacronym{ss}{SS}{Synchronization Signal}
\newacronym{sss}{SSS}{Secondary Synchronization Signal}
\newacronym{sst}{SST}{Slice/Service Type}
\newacronym{st}{ST}{Spanning Tree}
\newacronym{stft}{STFT}{Short-Time Fourier Transform}
\newacronym{svc}{SVC}{Scalable Video Coding}
\newacronym{synce}{SyncE}{Synchronous Ethernet}
\newacronym{ta}{TA}{Timing Advance}
\newacronym{tai}{TAI}{International Atomic Time}
\newacronym{tb}{TB}{Transport Block}
\newacronym{tcp}{TCP}{Transmission Control Protocol}
\newacronym{tdd}{TDD}{Time Division Duplexing}
\newacronym{tdl}{TDL}{Tapped Delay Line}
\newacronym{tdm}{TDM}{Time Division Multiplexing}
\newacronym{tdma}{TDMA}{Time Division Multiple Access}
\newacronym{tensorrt}{TensorRT}{Tensor RealTime}
\newacronym{tf}{TF}{Tensorflow}
\newacronym{tfl}{TfL}{Transport for London}
\newacronym{tfrc}{TFRC}{TCP-Friendly Rate Control}
\newacronym{tft}{TFT}{Traffic Flow Template}
\newacronym{tgen}{TGEN}{Traffic Generator}
\newacronym{tip}{TIP}{Telecom Infra Project}
\newacronym{tl}{TL}{Transfer Learning}
\newacronym{tm}{TM}{Transparent Mode}
\newacronym{to}{TO}{Telco Operator}
\newacronym{toa}{ToA}{Time of Arrival}
\newacronym{tr}{TR}{Technical Report}
\newacronym{trp}{TRP}{Transmitter Receiver Pair}
\newacronym{trt}{TRT}{Tensor RealTime}
\newacronym{ts}{TS}{Technical Specification}
\newacronym{tti}{TTI}{Transmission Time Interval}
\newacronym{ttt}{TTT}{Time-to-Trigger}
\newacronym{tx}{TX}{Transmitter}
\newacronym{uas}{UAS}{Unmanned Aerial System}
\newacronym{uav}{UAV}{Unmanned Aerial Vehicle}
\newacronym{uci}{UCI}{Uplink Control Indication}
\newacronym{udm}{UDM}{Unified Data Management}
\newacronym{udp}{UDP}{User Datagram Protocol}
\newacronym{udr}{UDR}{Unified Data Repository}
\newacronym{ue}{UE}{User Equipment}
\newacronym{uhd}{UHD}{\gls{usrp} Hardware Driver}
\newacronym{ul}{UL}{Uplink}
\newacronym{um}{UM}{Unacknowledged Mode}
\newacronym{uml}{UML}{Unified Modeling Language}
\newacronym{up}{UP}{User Plane}
\newacronym{upa}{UPA}{Uniform Planar Array}
\newacronym{upf}{UPF}{User Plane Function}
\newacronym{urllc}{URLLC}{Ultra Reliable and Low Latency Communication}
\newacronym{usa}{U.S.}{United States}
\newacronym{usim}{USIM}{Universal Subscriber Identity Module}
\newacronym{usrp}{USRP}{Universal Software Radio Peripheral}
\newacronym{utc}{UTC}{Coordinated Universal Time}
\newacronym{v2x}{V2X}{Vehicle-to-everything}
\newacronym{vim}{VIM}{Virtualization Infrastructure Manager}
\newacronym{vlan}{VLAN}{Virtual Local Area Network}
\newacronym{vm}{VM}{Virtual Machine}
\newacronym{vnf}{VNF}{Virtual Network Function}
\newacronym{volte}{VoLTE}{Voice over \gls{lte}}
\newacronym{voltha}{VOLTHA}{Virtual OLT HArdware Abstraction}
\newacronym{vr}{VR}{Virtual Reality}
\newacronym{vran}{vRAN}{Virtualized \gls{ran}}
\newacronym{vss}{VSS}{Video Streaming Server}
\newacronym{wbf}{WBF}{Wired Bias Function}
\newacronym{wf}{WF}{Wired-first}
\newacronym{wi}{WI}{Wireless InSite}
\newacronym{wlan}{WLAN}{Wireless Local Area Network}
\newacronym{wsn}{WSN}{Wireless Sensor Network}
\newacronym{xapp}{xApp}{RAN Application}
\newacronym{zmq}{ZMQ}{ZeroMQ}
\makeglossaries
\glsdisablehyper

\newcommand{\blue}[1]{{#1}}

\usepackage{booktabs}
\usepackage{subcaption}
\usepackage{ragged2e}

\usepackage[dvipsnames]{xcolor}
\usepackage{soul}

\usepackage[]{algorithmic}
\usepackage{algorithm}

\usepackage{tikz}
\usepackage{pgfplots}

\pgfplotsset{compat=newest}
\pgfplotsset{plot coordinates/math parser=false}
\newlength\fheight
\newlength\fwidth
\usetikzlibrary{plotmarks,patterns,decorations.pathreplacing,backgrounds,calc,arrows,arrows.meta,spy,matrix,scopes,fit,positioning}
\usepgfplotslibrary{patchplots,groupplots}
\usepackage{tikzscale}

\newif\ifexttikz
\exttikzfalse

\ifexttikz
  \usetikzlibrary{external}
\fi

\definecolor{inputcolor}{HTML}{4A90D9}
\definecolor{convcolor}{HTML}{5CB85C}
\definecolor{bncolor}{HTML}{F0AD4E}
\definecolor{relucolor}{HTML}{D9534F}
\definecolor{poolcolor}{HTML}{9B59B6}
\definecolor{fccolor}{HTML}{3498DB}
\definecolor{dropcolor}{HTML}{95A5A6}
\definecolor{softmaxcolor}{HTML}{E74C3C}
\definecolor{outputcolor}{HTML}{1ABC9C}
\definecolor{resblockcolor}{HTML}{ECF0F1}
\definecolor{arrowcolor}{HTML}{2C3E50}

\definecolor{lightgray204}{RGB}{204,204,204}
\definecolor{rawcolor}{RGB}{160,160,160}
\definecolor{avgcolor}{RGB}{230,159,0}
\definecolor{kalmancolor}{RGB}{86,180,233}
\definecolor{gtcolor}{RGB}{0,158,115}

\begin{document}

\pdfbookmark[1]{Cover}{cover}

\titlepage

\begin{frontmatter}


\begin{dedication}
To my family.

\vspace{70pt}

\begin{flushleft}
"There are only two ways to live your life. One is as though nothing is a miracle.\\The other is as though everything is a miracle."

\vspace{5pt}
\noindent --- Attributed to Albert Einstein
\end{flushleft}
\end{dedication}


\pdfbookmark[1]{Table of Contents}{contents}
\tableofcontents
\listoffigures
\newpage\ssp
\listoftables



\chapter*{List of Acronyms}
\addcontentsline{toc}{chapter}{List of Acronyms}
\label{chap:list_acronyms}

\renewcommand{\glossarysection}[2][]{}
\renewcommand{\arraystretch}{1.1}
\footnotesize
\setlength{\glsdescwidth}{0.75\columnwidth}
\printglossary[style=index,type=\acronymtype]
\normalsize


\begin{abstract}

The evolution of the Radio Access Network (RAN) in 5G and 6G technologies marks a shift toward open, programmable, and softwarized architectures, driven by the Open RAN paradigm.
This approach emphasizes open interfaces for RAN telemetry sharing, intelligent data-driven control loops for network optimization and control, and the virtualization and disaggregation of multi-vendor RAN components.
While promising, this transition introduces significant challenges, including the need to design interoperable solutions, acquire datasets to train and test Artificial Intelligence (AI)/Machine Learning (ML) algorithms for network inference, forecasting, and control, and develop comprehensive testbeds to benchmark and validate these solutions.
To this end, experimental wireless platforms and private 5G deployments play a key role, providing architectures comparable to real-world systems and enabling the design, prototyping, and testing of a wide variety of solutions in realistic environments.

This Ph.D. dissertation focuses on the development and evaluation of complementary experimental platforms: Colosseum---the world's largest Open RAN digital twin---and X5G---an open, programmable, multi-vendor, end-to-end private 5G O-RAN testbed with GPU acceleration.
It further discusses how these platforms can be leveraged to advance use cases in spectrum sharing, AI-driven modeling, security, network slicing, and Integrated Sensing and Communication (ISAC) for next-generation cellular research.
The main contributions of this dissertation include:
(i) the development of CaST (Channel emulation Scenario generator and Sounder Toolchain), enabling the automated creation and validation of digital twin wireless scenarios for Colosseum through 3D modeling, ray-tracing, and channel sounding;
(ii) the validation of Colosseum digital twins at scale, demonstrating that emulated environments can closely reproduce real-world setups and providing RAN emulation blueprints for the research community;
(iii) the design and deployment of X5G, an open, programmable, multi-vendor private 5G O-RAN testbed that integrates NVIDIA Aerial GPU-accelerated PHY layer processing with OpenAirInterface higher layers;
(iv) the integration of a GPU-accelerated dApp framework for real-time inference and control within the RAN, exposing PHY/MAC telemetry and enabling sub-millisecond control loops for AI-native applications, including ISAC; and
(v) the realization of intelligent RAN applications and pipelines spanning spectrum sharing, interference detection, network slicing, security assessment, and CSI-based sensing.

Overall, this dissertation provides an end-to-end methodology and set of platforms that bridge digital and physical experimentation, offering a practical blueprint for designing, validating, and operating next-generation open and programmable cellular networks.

\end{abstract}

\end{frontmatter}

\pagestyle{headings}


\chapter{Introduction}
\label{chap:intro}

The ever-growing wireless industry is driving an ongoing demand for systems that deliver higher key performance metrics, including increased throughput, reduced latency, and the ability to connect a greater number of \glspl{ue} simultaneously.
Over the past few decades, each cellular generation has reshaped how people, devices, and services connect. This evolution is not only bringing higher data rates and lower latency, but also increasing the scope of what wireless networks can support, from traditional \gls{embb}, \gls{mmtc}, and \gls{urllc} to sensing, localization, and \gls{ai}-native applications~\cite{ituIMT2030Vision,6gsurvey}.
At the same time, the way in which these networks are built is changing. Earlier generations were mainly developed on vertically integrated hardware platforms in a closed-box approach, while modern \glspl{ran} are becoming more open, programmable, and softwarized. Open interfaces, virtualized network functions, and \gls{ai}-enabled control loops are turning the \gls{ran} into a platform on which new services and intelligent applications can be developed and deployed~\cite{bonati2020open}. This transition, while promising, also makes the network substantially more complex to design, test, and operate.

This dissertation sits at this intersection and focuses on how experimental platforms can be leveraged to ease the design, deployment, and evaluation of next-generation, open, programmable, and \gls{ai}-native cellular networks.
Throughout the thesis, we embark on an experimental journey that starts with large-scale wireless digital emulators and continues through \gls{ota} testing on a physical private \gls{5g} network, including the development and testing of various intelligent applications for monitoring, control, sensing, and security.

\section{Motivation and Context}
\label{sec:intro-motivation}

\subsection{The Evolution of Cellular Networks}

\begin{figure}[htb]
    \centering
    \includegraphics[width=0.85\columnwidth]{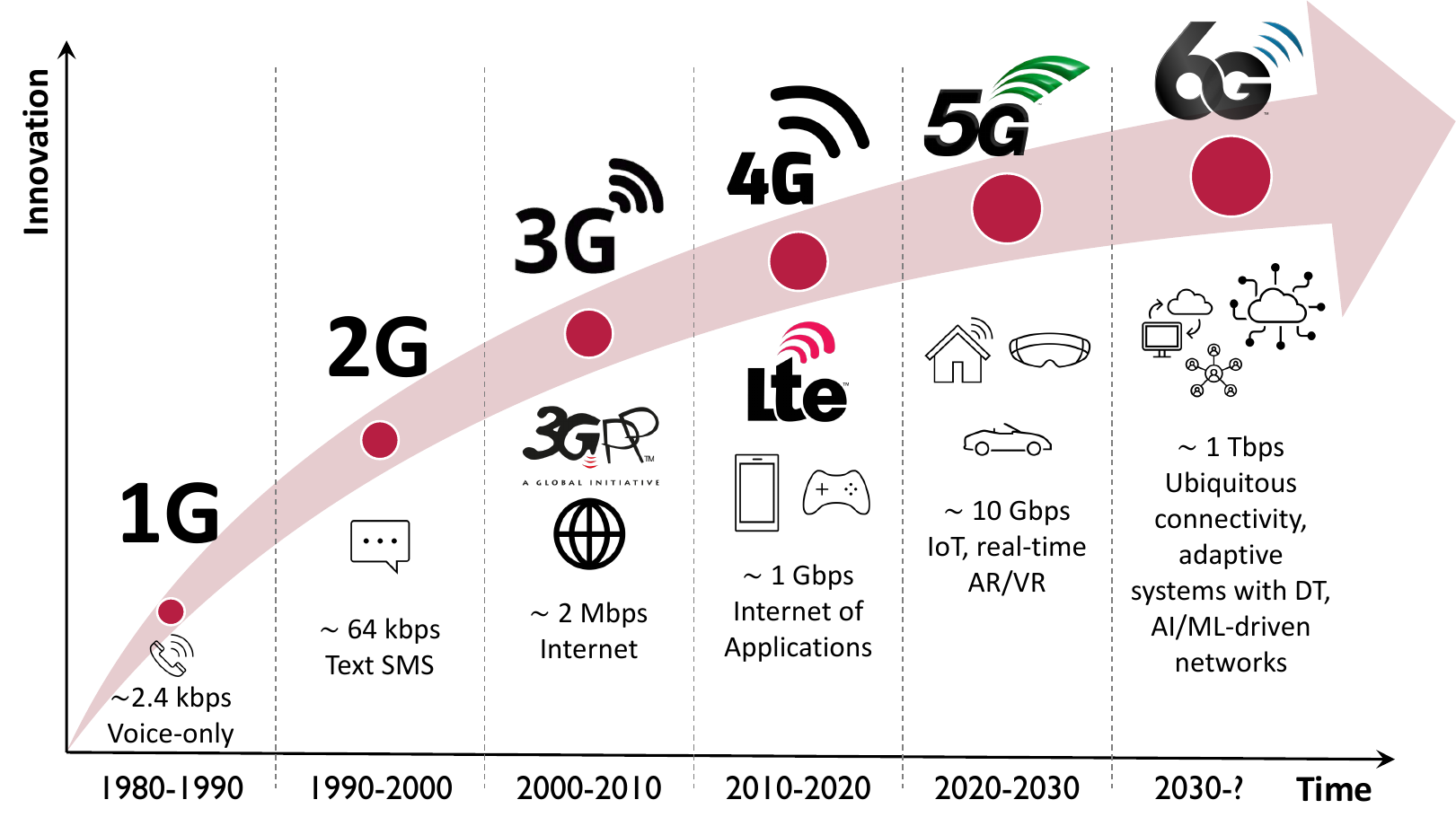}
    \caption{Evolution of cellular networks from \acrshort{1g} to \acrshort{6g}, showing the progression in data rates, capabilities, and the increasing demands on next-generation systems.}
    \label{chap1-fig:cellular-evolution}
\end{figure}

Cellular networks have historically experienced a generational update cycle of approximately ten years, with each generation bringing advances in capabilities and services. As shown in Figure~\ref{chap1-fig:cellular-evolution}, this progression started with \gls{1g} in the 1980s, offering voice-only services at approximately $2.4$~kbps. The 1990s introduced \gls{2g} with text messaging capabilities at $64$~kbps. The new millennium brought \gls{3g} and the advent of mobile Internet access at $2$~Mbps, fundamentally changing how people retrieved and accessed information. The 2010s delivered \gls{4g}/\gls{lte} and the Internet of applications at $1$~Gbps, coinciding with the smartphone revolution that put powerful computing devices in the pockets of billions worldwide~\cite{atzori2010iot,villa2021iot}.
The current decade is characterized by \gls{5g}, with data rates approaching $10$~Gbps, enabling the \gls{iot}, real-time augmented and virtual reality, smart devices, and connected vehicles, and leveraging new frequency bands and advanced spectrum-sharing techniques. Looking ahead, \gls{6g} development is already underway and is expected to push toward 1~Tbps, with ubiquitous connectivity, adaptive systems powered by \glspl{dt}, and \gls{ai}/\gls{ml}-driven network intelligence~\cite{giordani2020toward,zhang20196g}.
This progression reflects an ever-growing demand for higher data rates, more connected devices, and richer services, all with lower latency, energy consumption, and operational costs. This raises the challenge of how to drive this innovation and generational evolution efficiently.

\subsection{Open RAN and AI-RAN: Opening the Network}

\begin{figure}[htb]
    \centering
    \includegraphics[width=0.95\columnwidth]{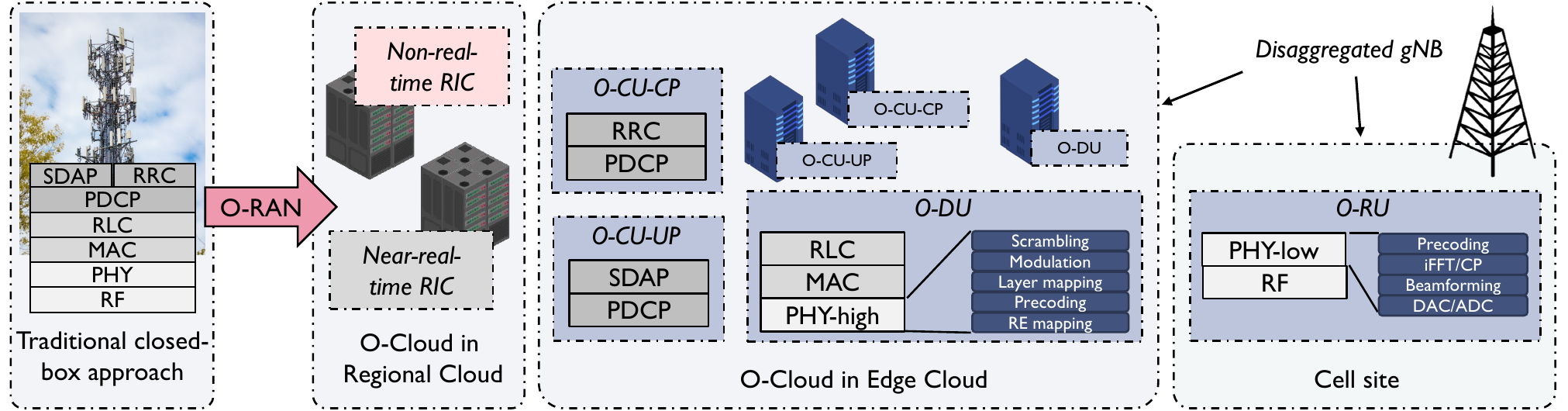}
    \caption{The Open RAN architecture, illustrating the transition from traditional closed-box base stations to disaggregated, software-defined, and intelligent network components connected through open interfaces~\cite{polese2023understanding}.}
    \label{chap1-fig:openran}
\end{figure}

In recent years, the Open RAN paradigm has emerged as a key answer to this question, reshaping how next-generation networks are designed, deployed, and operated. As shown in Figure~\ref{chap1-fig:openran}, Open RAN is built on a few key pillars~\cite{polese2023understanding}. \emph{Disaggregation} splits the traditional monolithic base station into atomic functional units, \gls{cu}, \gls{du}, and \gls{ru}, each dedicated to running a specific part of the \gls{ran}. \emph{Open interfaces} enable these units to communicate via standardized protocols, enabling a multi-vendor ecosystem in which equipment from different manufacturers can interoperate seamlessly. \emph{Softwarization} delivers network functions as software running on general-purpose or accelerated hardware, enabling high scalability, rapid updates, and virtualized deployments. Finally, \emph{intelligence and virtualization} introduce data-driven optimization by deploying intelligent applications in the \glspl{ric} (rApps, xApps) or colocating them with the \gls{gnb} (dApps) at various closed-loop control timescales~\cite{bonati2021intelligence}. This introduces the notion of \emph{AI-RAN}, in which a programmable \gls{ran} exposes telemetry, and \gls{ai}/\gls{ml} models can be used to enhance network efficiency and enable new use cases, such as \gls{isac}.
However, this shift toward open and programmable networks also introduces greater complexity in properly designing, building, and operating a wireless system.

\subsection{Complexity and the Need for Testbeds}

\begin{figure}[htb]
    \centering
    \includegraphics[width=0.8\columnwidth]{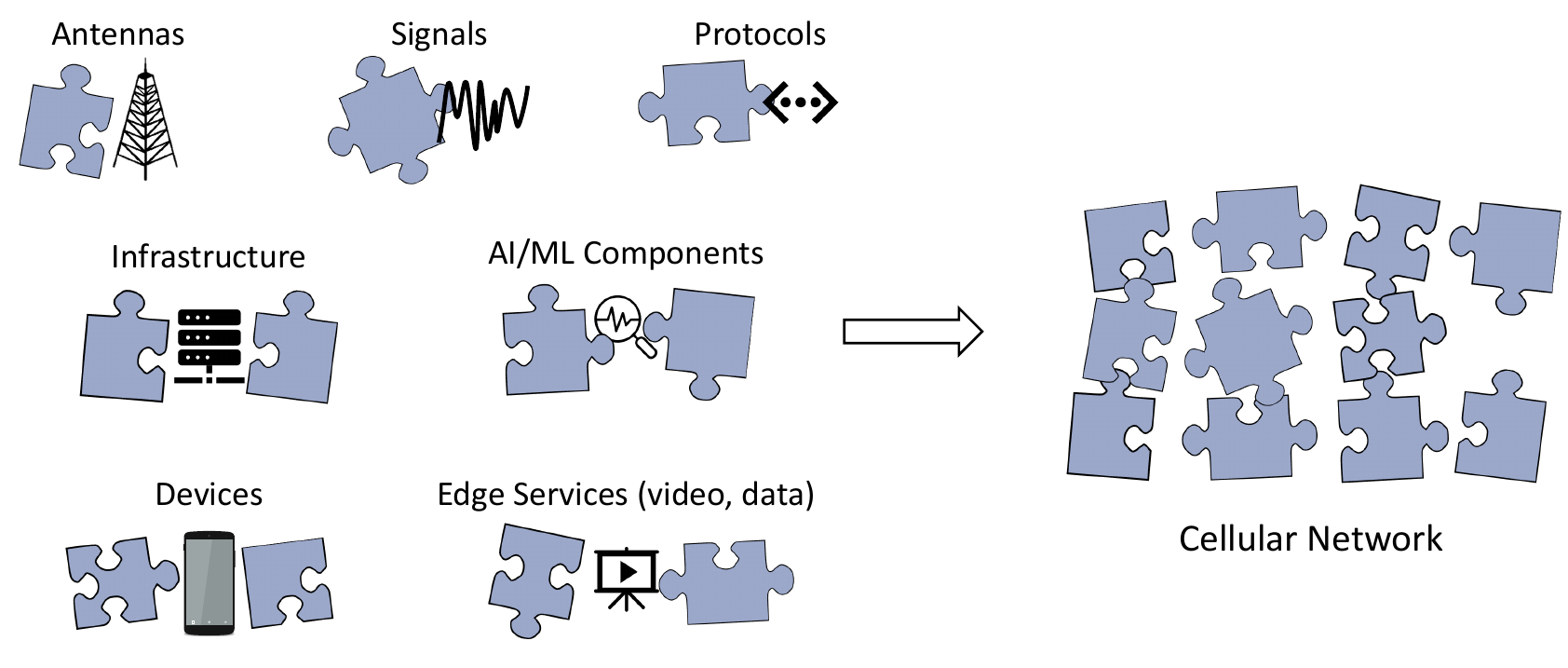}
    \caption{Continuous growth in complexity, where modern cellular networks can be seen as a puzzle of interoperable hardware, software, and AI components.}
    \label{chap1-fig:puzzle}
\end{figure}

A useful way to visualize the increasing complexity of current networks is through the puzzle metaphor shown in Figure~\ref{chap1-fig:puzzle}. Modern cellular systems require many components to work seamlessly together to provide connectivity to end users: antennas and radio hardware, signal processing chains, protocol stacks, computing infrastructure, \gls{ai}/\gls{ml} components, edge services, and diverse user devices. Each of these elements must integrate correctly with the others, much like pieces of a puzzle that must fit together precisely to form a complete picture. Completing such a puzzle takes time and resources, and the increasing number of pieces, particularly with the addition of intelligent components and multi-vendor interoperability requirements, and the possibility of swapping existing ones, opens the floor to new challenges in how to connect all these pieces effectively while allowing the system to still function correctly.  

To tackle these problems, this dissertation proposes the answer of \emph{testbeds}. Testbeds are experimental platforms that can support wireless research in numerous areas, from \gls{iot} and \gls{wsn}~\cite{sheshashayee2024experimental} to satellite communications, and address these challenges in three fundamental ways. First, they provide an open and programmable environment where researchers can design and develop new solutions with complete control over each component of the network stack. Second, they can serve as data factories capable of generating large volumes of realistic \gls{rf} data under diverse conditions, data that is essential for training and testing \gls{ai}/\gls{ml} algorithms. Third, they offer a controlled playground where solutions can be evaluated and validated before deployment, allowing researchers to test bold ideas, break things, and recover without risking commercial infrastructure or end-user services.

\section{Research Questions and Challenges}
\label{sec:intro-challenges}

While testbeds offer potential for advancing wireless research, fundamental questions must be addressed to use these systems effectively. Throughout this dissertation, four key research dimensions have guided the work.

\textbf{Deployment.} Given the complexity of modern cellular systems, how can we design and deploy experimental platforms that are realistic, reproducible, and scalable? This includes questions on the architecture, orchestration, and automation of the platforms, as well as challenges related to hardware and software integration.

\textbf{Realism.} Do testbeds accurately reflect real-world conditions? Even the most sophisticated emulator is ultimately an approximation of reality, and physical testbeds are deployed in specific locations with their own constraints. The value of experimental results depends critically on how well the testbed environment matches the characteristics of production deployments, such as channel propagation, protocol behavior, and device interactions.

\textbf{Usability.} How can we ensure interoperability and accessibility for the broader research community? The Open RAN ecosystem presents challenges related to the \gls{e2e} integration of components from multiple vendors. Beyond building capable testbeds, it is essential to provide tools, documentation, and interfaces that lower the entry barrier for researchers.

\textbf{Use Cases.} How can experimental platforms help address key challenges and advance research? Ultimately, testbeds should demonstrate their value by enabling concrete research contributions that would be difficult or impossible to achieve solely through simulation, and by supporting the transfer of these contributions toward commercialization.

\section{The Journey of an Experimental Idea}
\label{sec:intro-journey}

\begin{figure}[htb]
    \centering
    \includegraphics[width=0.95\columnwidth]{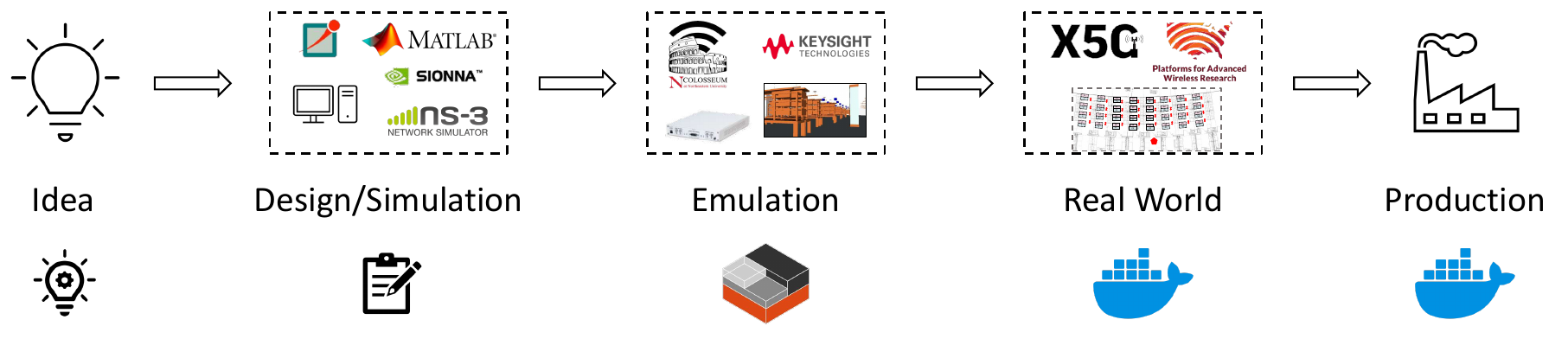}
    \caption{The journey of an experimental idea: from initial concept through design and simulation, to emulation at scale, to real-world \acrshort{ota} testing, and finally to production deployment.}
    \label{chap1-fig:journey}
\end{figure}

To address these challenges, this dissertation develops complementary platforms that together support the lifecycle of wireless research, as illustrated in Figure~\ref{chap1-fig:journey}. A researcher can follow the journey of an experimental idea through progressive stages of increasing fidelity and realism. An initial concept moves from design and simulation using tools such as MATLAB, Sionna, or ns-3, to large-scale emulation and \glspl{dt} on platforms like Colosseum, where hardware-in-the-loop experiments can be conducted at scale, and finally to real-world \gls{ota} validation on physical testbeds like X5G before potential production deployment. Each stage offers different trade-offs between control, scale, realism, and risk, and the ability to move ideas smoothly through this pipeline accelerates the path from concept to deployment.

\section{Contributions and Dissertation Structure}
\label{sec:intro-contributions}

The main contributions of this dissertation align directly with the experimental journey outlined above and address research questions related to both emulation and real-world validation. 
In particular, this work:
\begin{itemize}
    \item Develops CaST, a toolchain to enable the automated creation and validation of \gls{dt} wireless scenarios for large-scale wireless emulators through 3D modeling, ray-tracing, and channel sounding.

    \item Validates \glspl{dt} on emulation platforms like Colosseum at scale, demonstrating that emulated environments can closely reproduce real-world setups and providing \gls{ran} emulation blueprints for the research community.

    \item Designs and deploys X5G, a \gls{gpu}-accelerated private \gls{5g} O-RAN platform for \gls{ota} experimentation with commercial devices and multi-vendor components.

    \item Integrates a \gls{gpu}-based dApp framework for real-time inference and control within the \gls{ran}, enabling sub-millisecond control loops and \gls{ai}-native applications.

    \item Realizes intelligent \gls{ran} applications and pipelines for monitoring and control, spectrum sensing, \gls{isac}, network slicing, and security assessment, showcasing the capabilities of the proposed platforms to advance the development of next-generation wireless networks.
\end{itemize}

The remainder of this dissertation is organized as follows.
Chapter~\ref{chap:2} (based on~\cite{bonati2021colosseum,villa2022cast,villa2024dt,johari2025patent}) introduces \glspl{dt} for wireless networking and presents the CaST toolchain and the Colosseum testbed as core systems for large-scale emulation.
Chapter~\ref{chap:3} (based on~\cite{villa2023wintech,saeizadeh2024camad,saeizadeh2025airmap,basaran2025gentwin}) explores use cases enabled by Colosseum, including spectrum sharing with incumbent radar systems, \gls{ai}-driven radio map generation, and generative models for synthetic \gls{rf} data augmentation.
Chapter~\ref{chap:4} (based on~\cite{villa2024x5g,villa2025tmc,villa2025cusense}) describes the design, deployment, and evaluation of X5G, an open, programmable, and multi-vendor private \gls{5g} testbed, as well as the development of a \gls{gpu}-accelerated dApp framework for real-time applications.
Chapter~\ref{chap:5} (based on~\cite{villa2025cusense,neasamoni2025interforan,Cheng2024oranslice,groen2024timesafe}) presents intelligent \gls{ran} applications realized on X5G, including cuSense for \gls{csi}-based \gls{isac}, InterfO-RAN for real-time interference detection, ORANSlice for dynamic resource management, and TIMESAFE for fronthaul security assessment.
Finally, Chapter~\ref{chap:conclusion} summarizes the contributions, revisits the research questions, and discusses the broader implications of this work for next-generation wireless research.


\chapter{Large-Scale Wireless Network Emulation with Digital Twins}
\label{chap:2}

\section{Introduction}

Powerful experimental wireless platforms have recently been developed to provide an ecosystem for advanced wireless research through repeatable, reproducible experimentation and the creation of large datasets. These platforms are becoming the nexus of \gls{ai}-enabled wireless research, where researchers can design, develop, train, and test new solutions for \gls{nextg} wireless systems.
Examples include the US~NSF \gls{pawr} program with its four at-scale, outdoor programmable platforms~\cite{pawr}, and indoor testbeds including the Drexel Grid~\cite{dandekar2019grid}, ORBIT~\cite{kohli2021openaccess}, and Arena~\cite{bertizzolo2020comnet}. 
While these testbeds expose realistic indoor and outdoor wireless propagation environments, their physical scope is inherently tied to a specific deployment, and scaling them to capture the dynamics of real-world topologies, traffic conditions, and radio environments can be costly, time-consuming, and sometimes infeasible.

To complement these \gls{ota} platforms, large-scale wireless emulation systems have been proposed to support site-independent experimentation. By emulating a virtually unlimited variety of scenarios, these instruments are becoming a key resource for designing, developing, and validating networking solutions in quasi-realistic environments, at scale, and with a diverse set of fully customizable \gls{rf} channel conditions, traffic scenarios, and network topologies~\cite{bonati2021colosseum,sichitiu2020aerpaw,dandekar2019grid}.
Solution development and testing for \gls{nextg} networks are, in fact, evolving toward the integration of actual networked systems with a digital model that provides a replica of the physical network for continuous prototyping, testing, and self-optimization of the living network. These \emph{\glspl{dt}}~\cite{BarricelliCF19} are at the forefront of wireless research testing and prototyping~\cite{Saracco22}.
However, the reliability of the solutions developed and tested on \gls{dt} platforms depends critically on the precision of the emulation process and the underlying channel models, as well as on the ability to validate the emulated environment against real measurements.
The realization of high-fidelity emulation-based \glspl{dt} for wireless systems as a whole, namely a \gls{dtmn}~\cite{ericssondtmn, spirent5G}, remains a challenge that has not been fully addressed.

To this end, this chapter centers on \gls{cast}~\cite{villa2022cast}, a complete toolchain for creating and validating wireless \glspl{dt} on large-scale emulation testbeds, such as Colosseum---the world's largest wireless network emulator with hardware-in-the-loop and Open RAN Digital Twin~\cite{bonati2021colosseum,polese2024colosseum}. We leverage \gls{cast} together with Colosseum to realize \glspl{dtmn} of representative \gls{ota} platforms and environments, and we run experiments on both systems to assess how faithfully the \gls{dt} reproduces the behavior of a physical system.
This chapter primarily targets three of the four key challenges introduced in Chapter~\ref{chap:intro}: the \emph{deployment} of a scalable \gls{dt} platform, the \emph{realism} of its emulated environments, and the \emph{usability} of a reusable toolchain. The fourth challenge, namely the \emph{use cases}, is the focus of Chapter~\ref{chap:3}.
The main contributions of this chapter are the following:
\begin{itemize}
    \item We present \gls{cast}, a streamlined framework for creating and validating realistic wireless scenarios with mobility support, based on precise ray-tracing methods for emulation platforms such as Colosseum~\cite{bonati2021colosseum}.
    \item We leverage the \gls{cast} toolchain to generate a variety of \gls{dt} emulation scenarios, both indoor and outdoor, including a \gls{dt} of the \gls{ota} physical platform Arena~\cite{bertizzolo2020comnet}. 
    \item We develop a \gls{cicd} pipeline for real-time twinning of selected protocol stacks, e.g., cellular and Wi-Fi, enabling researchers to automatically test new releases of open-source stacks.
    \item We validate the fidelity of the twinning process by comparing representative \glspl{kpm}, such as throughput and \gls{sinr}, across experiments on digital and physical platforms.
\end{itemize}
Results show that the twinning process can faithfully reproduce real-world \gls{rf} environments. The remainder of this chapter is organized as follows. Section~\ref{chap2-sec:dt} provides a brief primer on \glspl{dt} in the context of wireless network emulators. Section~\ref{chap2-sec:dtplatforms} presents the platforms used to implement our \glspl{dtmn}. Section~\ref{chap2-sec:digitizing} describes the \gls{cast} workflow, while Section~\ref{chap2-sec:results} presents our experimental setup and validation results. Section~\ref{chap2-sec:relatedwork} surveys related work, and Section~\ref{chap2-sec:conclusion} summarizes the main findings.

\section{Digital Twins}
\label{chap2-sec:dt}



\begin{figure}[th]
    \centering
    \includegraphics[width=0.6\columnwidth]{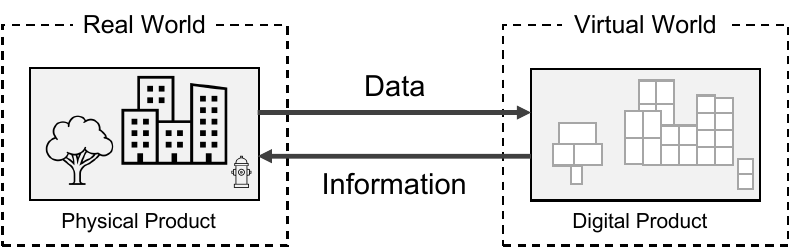}
    \caption{High-level representation of digital twin components.}
    \label{chap2-fig:dt-architecture}
\end{figure}

The \gls{dt} concept is finding increasing momentum
as a means of enhancing the performance of physical systems by using their virtual counterparts~\cite{Jones2020characteristing}.
The origin of this name is universally credited to Grieves and Vickers~\cite{grieves2015dt}, who 
define a \gls{dt} as a system consisting of three primary elements (Figure~\ref{chap2-fig:dt-architecture}): (i)~a physical product in the real world; (ii)~a virtual representation of the product in the virtual world; and (iii)~a connection of data and information tying the two.

Over the years, the \gls{dt} idea has extended into new domains, starting from its description and adding different flavors to this concept.
For example, some works consider \gls{dt} as an enabler for Industry~4.0 applications, as detailed in~\cite{rolle2020architecture}, while others suggest its use in areas such as product design, assembly, or production planning~\cite{Tao2018dt}.
Moreover, the continuous evolution of \gls{dt}s and their applications ushered the concept of \glspl{dtn}, as systems interconnect multiple \glspl{dt}~\cite{wu2021dtn}.
%
Finally, \glspl{dt} have been adopted in the context of the wireless communications ecosystem and cellular networks.

\begin{figure*}[b]
    \centering
    \includegraphics[width=\textwidth]{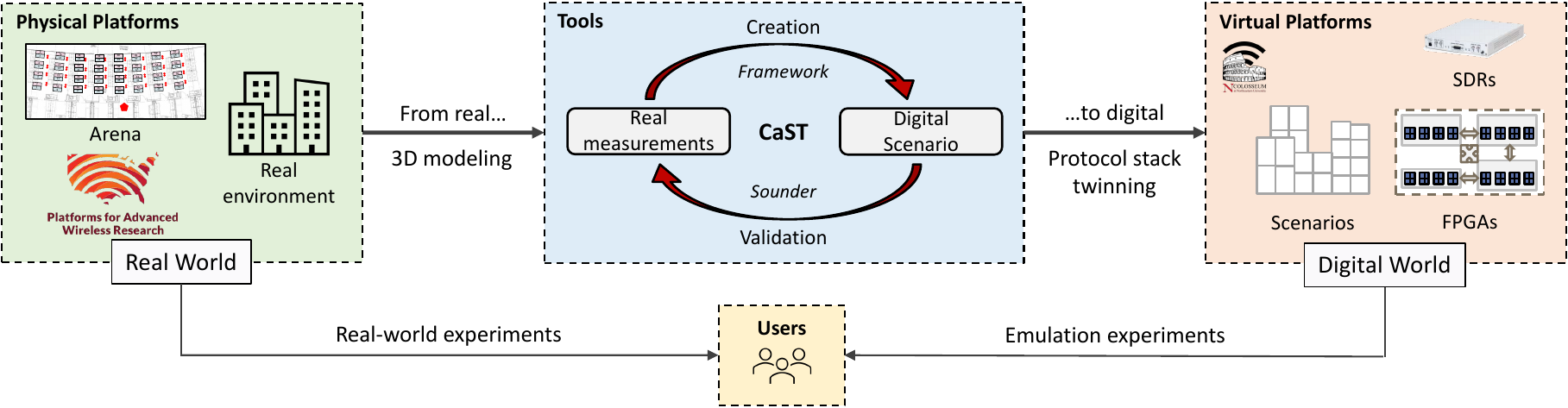}
    \caption{Main components of our high-level representation of a \acrshort{dt}.}
    \label{chap2-fig:dt-colosseum}
\end{figure*}

In this work, we apply the concept of \gls{dt} to experimental wireless research, and, to the best of our knowledge, in what is the \textit{first example of \gls{dtmn} for real-world applications.}
Figure~\ref{chap2-fig:dt-colosseum} shows a high-level diagram of all main components of our \gls{dt} representation.
Specifically, we develop a set of tools to create and validate a comprehensive digital representation of a particular real-world system inside a virtual environment. 
%
%
%
This would enable researchers to run wireless experiments inside a \gls{dt} of virtually any type of physical environment; develop and test new algorithms; and derive results as accurately and as close as possible to the behavior that they would obtain in the real-world environment.

To this aim, we propose Colosseum, the world's largest wireless network emulator~\cite{bonati2021colosseum}, as a \gls{dt} for real-world wireless experimental testbeds and environments.
Thanks to its large-scale emulation capabilities, Colosseum twins both the real and digital worlds by capturing conditions of real environments and reproducing them in a high-fidelity emulation. 
This is done through so-called \gls{rf} scenarios that model the characteristics of the physical world (e.g., channel effects, propagation environment, mobility, etc.) and convert them into digital emulation terrains to be used for wireless experimentation. 
%
%
Colosseum can also twin the protocol stack itself, i.e., it allows the deployment of the same generic software-defined stacks that can replicate the functionalities of real-world wireless networks, e.g., O-RAN managed cellular protocol stacks~\cite{bonati2021intelligence,polese2022coloran}, orchestrators~\cite{doro2022orchestran}, and different \gls{tcp} congestion control algorithms~\cite{pinto2023wintech}.

Colosseum is not limited to \gls{sdr} devices, but it also supports the integration of \gls{cots} devices---\blue{as long as they expose \gls{rf} connectors compatible with those of the \gls{mchem} \glspl{usrp}, e.g., SMA connectors or equivalent ones}---as demonstrated in~\cite{baldesi2022charm}, where \gls{soc} boards running OpenWiFi are used.
%
Additionally, custom equipment and waveforms can be integrated within Colosseum,
as demonstrated in prior work where we integrated a proprietary jammer~\cite{robinson2023esword}, and a radar waveform~\cite{villa2023wintech}, within the system.

Through the utilization of these scenarios and the twinning of protocol stacks and generic waveforms, users can collect data and test solutions in many different environments representative of real-world deployments, and fine-tune their solutions before deploying them in production networks to ensure they perform as expected.
An example of this is provided by~\cite{bonati2022openrangym-pawr}, where data-driven solutions for cellular networking were prototyped and tested on Colosseum before moving them to other platforms.
%
Overall, this allows users to retain full control over the digitized virtual world, to reproduce all---and solely---the desired channel effects, and to repeat and reproduce experiments at scale.
This is particularly important for \gls{ai}/\gls{ml} applications~\cite{bonati2022openrangym-pawr}, where: (i)~access to a large amount of data is key to designing solutions as general as possible; and (ii)~\gls{ai} agents need to be thoroughly tested and validated in different conditions to be sure they do not cause harm to the commercial infrastructure.

To enable RF twinning between physical and digital worlds in Colosseum, we utilize our recently developed tool \gls{cast}, an end-to-end toolchain to create and characterize realistic wireless network scenarios with a high degree of fidelity and accuracy~\cite{villa2022cast}.
\gls{cast} is composed of two main parts: (i)~a streamlined framework to create realistic mobile wireless scenarios from real-world environments (thus digitizing them); and (ii)~a \gls{sdr}-based channel sounder to characterize emulated \gls{rf} channels.
The protocol stack twinning is enabled by a \gls{ci} and \gls{cd} platform that can deploy in the Colosseum system the latest, or a specifically desired, version of a wireless protocol stack. We support any software-defined stack that has been designed for real-world experiments and have implemented a specific version of a \gls{cicd} framework for the OpenAirInterface 5G cellular implementation~\cite{KALTENBERGER2020107284}.


As proof of concept, we use \gls{cast} to create the \gls{dt} of a publicly available over-the-air indoor testbed for sub-6\:GHz research, namely Arena~\cite{bertizzolo2020comnet}.
This allows us to showcase the capabilities of Colosseum as a \gls{dt} platform, as well as the level of fidelity that can be achieved by the twinning process and operations.

%

\section{Digital Twin Platforms}
\label{chap2-sec:dtplatforms}

In this section, we describe the two platforms that are part of our \gls{dt} ecosystem: (i)~Colosseum, for large-scale emulation/digitization of physical environments, is described in Section~\ref{chap2-sec:colosseum}; and (ii)~Arena, for over-the-air real-world experimentation, in Section~\ref{chap2-sec:arena}.

\subsection{Large-scale Emulation: Colosseum}
\label{chap2-sec:colosseum}

\begin{figure*}[t]
    \centering
    \includegraphics[width=\textwidth]{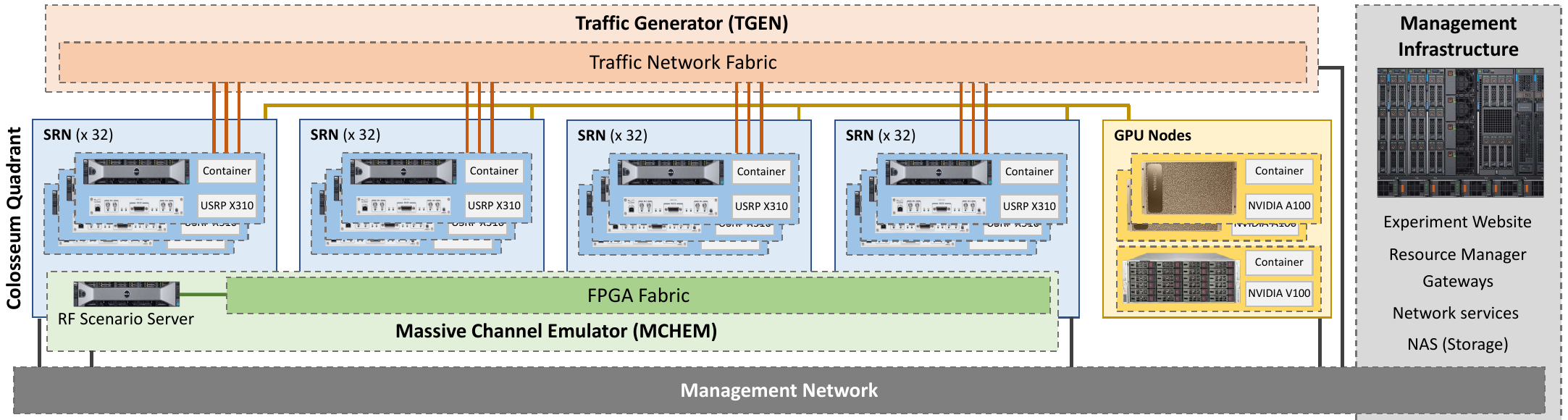}
    \caption{Colosseum architecture, adapted from~\cite{bonati2021colosseum}.}
    \label{chap2-fig:colosseum-architecture}
\end{figure*}

Colosseum is the world's largest publicly available wireless network emulator with hardware-in-the-loop.
At a high level, Colosseum consists of five main components, depicted in Figure~\ref{chap2-fig:colosseum-architecture}~\cite{bonati2021colosseum}: (i)~128 \acrfullpl{srn}; (ii)~the \acrfull{mchem}; (iii)~the \acrfull{tgen}; (iv)~the GPU nodes; and (v)~the management infrastructure.

The \glspl{srn}, which are divided into four quadrants, comprise 128~high-performance Dell PowerEdge R730 compute servers, each driving a dedicated USRP X310 \gls{sdr}---able to operate in the $[10\:\mathrm{MHz}, 6\:\mathrm{GHz}]$ frequency range---through a 10~Gbps fiber cable.
These servers are equipped with Intel Xeon E5-2650 CPUs with 48~cores, as well as NVIDIA Tesla K40m GPUs, to support heavy computational loads (e.g., \gls{ai}/\gls{ml} applications) and be able to properly drive their dedicated \gls{sdr}.
Users of the testbed can reserve \glspl{srn} for their experiments through a web-based \gls{gui}, as well as specify the date/time, and amount of time they need these resources for.
At the specified reservation time, Colosseum exclusively allocates the requested resources to the users and instantiates on them a softwarized protocol stack---also specified by the user when reserving resources---in the form of a \gls{lxc}.
After these operations have been carried out, users of the testbed can access via SSH to the allocated \glspl{srn}, and use the softwarized protocol stack instantiated on them (e.g., cellular, Wi-Fi, etc.) to drive the \glspl{sdr} and test solutions for wireless networking in a set of diverse environments emulated by Colosseum.

\begin{figure}[ht]
    \centering
    \includegraphics[width=0.7\columnwidth]{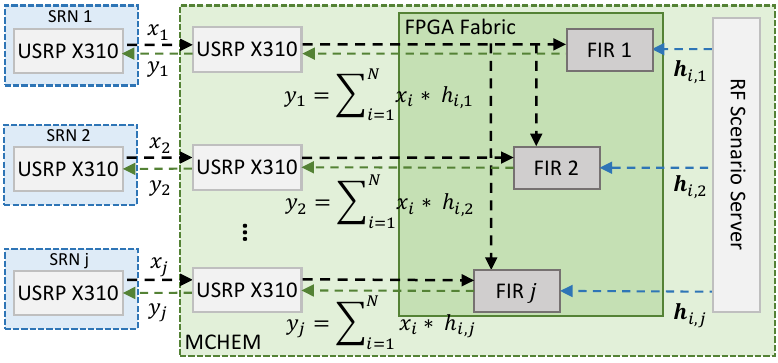}
    \caption{FPGA-based RF scenario emulation in Colosseum, from~\cite{bonati2021colosseum}.}
    \label{chap2-fig:mchem_diagram}
\end{figure}

These environments---called \gls{rf} scenarios in the Colosseum jargon---are emulated by Colosseum \gls{mchem}.
\gls{mchem} is formed of 16~NI ATCA 3671~\gls{fpga} distributed across the four quadrants of Colosseum.
Each ATCA module includes 4~Virtex-7 690T \glspl{fpga} that process through \gls{fir} filters the signals from/to an array of \glspl{usrp} X310 (32~USRPs per \gls{mchem} quadrant, for a total of 128~USRPs across the four quadrants of Colosseum) connected one-to-one, through SMA cables, to the \glspl{usrp} driven by the \glspl{srn} controlled by the users (see Figure~\ref{chap2-fig:mchem_diagram}).

Instead of being transmitted over the air, signals generated by the \gls{srn} USRPs are sent to the corresponding USRP on the \gls{mchem} side.
From there, they are converted in baseband and to the digital domain, and processed by the \gls{fir} filters of the \gls{mchem} \glspl{fpga} that apply the \gls{cir} corresponding to the \gls{rf} scenario chosen by the user of the testbed (see Figure~\ref{chap2-fig:mchem_diagram}).

Specifically, these \gls{fir} filters comprise 512~complex-valued taps that are set to reproduce the conditions and characteristics of wireless channels in real-world environments, i.e., the \gls{cir} among each pair of \gls{srn}.
As an example, and as depicted in Figure~\ref{chap2-fig:mchem_diagram}, signal $x_i$ generated by one of the \glspl{srn} is received by the USRP of \gls{mchem} and transmitted to its \glspl{fpga}.
Here, the \gls{fir} filters load the vector ${h}_{i,j}$ corresponding to the 512-tap \gls{cir} between nodes $i$ and $j$ (with $i, j \in \{1,...,N\}$ set of \glspl{srn} active in the user experiment) from the \gls{rf} scenario server, which contains a catalog of the scenario available on Colosseum.
Then, they apply these taps to $x_i$ through a convolution operation.
The signal $y_{j} = \sum_{i=1}^{N} x_i \ast h_{i,j}$ resulting from this operation, i.e., the originally transmitted $x_i$ signal with the \gls{cir} of the emulated channel, is finally sent to \gls{srn} $j$.
Analogous operations also allow Colosseum to perform superimposition of signals from different transmitters, and to consider interfering signals (besides the intended ones), as it would happen in a real-world wireless environment~\cite{ashish2018scalable}.
In this way, thus, Colosseum can emulate effects typical of real and diverse wireless environments, including fading, multi-path, and path loss, in terrains up to $1\:\mathrm{km^2}$ of emulated area, and with up to $80$\:MHz bandwidth, and can support the simultaneous emulation of different scenarios from multiple users.
Furthermore, Colosseum is capable of emulating node mobility discretely.
Every millisecond, the RF Scenario Server loads different pre-defined channel taps into the Colosseum FPGAs, effectively mimicking changes in channel conditions resulting from node position changes.

Similarly to the emulation of \gls{rf} environments, the \gls{tgen} allows users of the testbed to emulate different IP traffic flows among the reserved nodes.
This tool, which is based on the U.S.\ Naval Research Laboratory's \gls{mgen}~\cite{mgen}, enables the creation of flows with specific packet arrival distributions (e.g., Poisson, uniform, etc.), packet size, and rate.
These traffic flows, namely \textit{traffic scenarios}, are sent to the \glspl{srn} of the user experiment that, then, handles them through the specific protocol stack instantiated on the \glspl{srn} (e.g., Wi-Fi, cellular, etc.).

Recently, Colosseum added various GPU nodes to the pool of resources that can be reserved by users.
These include two NVIDIA DGX servers, state-of-the-art computing solutions with 8~NVIDIA A100 GPUs each and interconnected through a Tbps internal NVlink switching interface, and one large memory node (Supermicro SuperServer 8049U-E1CR4T) with 6~NVIDIA V100 GPUs, 128-core Intel Xeon Gold 6242 CPUs, and 3\:TB of RAM.
These resources, which can be reserved from the same web-based \gls{gui} used for the \glspl{srn}, can stream data in real-time from/to the \glspl{srn} through high-speed links and have the capability of powering computational-intensive workloads, such as those typical of \gls{ai}/\gls{ml} applications.

Finally, Colosseum includes a management infrastructure---not accessible by the users---that is used to maintain the rest of the system operational (see Figure~\ref{chap2-fig:colosseum-architecture}).
Some of the services offered by this include: (i)~servers that run the website used to reserve resources on the testbed; (ii)~resource managers to schedule and assign \glspl{srn} and GPU nodes to users; (iii)~multiple \gls{nas} systems to store experiment data and container images; (iv)~gateways and firewalls to enable user access and isolation throughout experiments; and (v)~precise timing servers and components to synchronize the \glspl{srn}, the GPU nodes, and the \glspl{sdr}.

\subsection{Over-the-Air Experimentation: Arena}
\label{chap2-sec:arena}

\begin{figure*}[t]
    \centering
    \includegraphics[width=\textwidth]{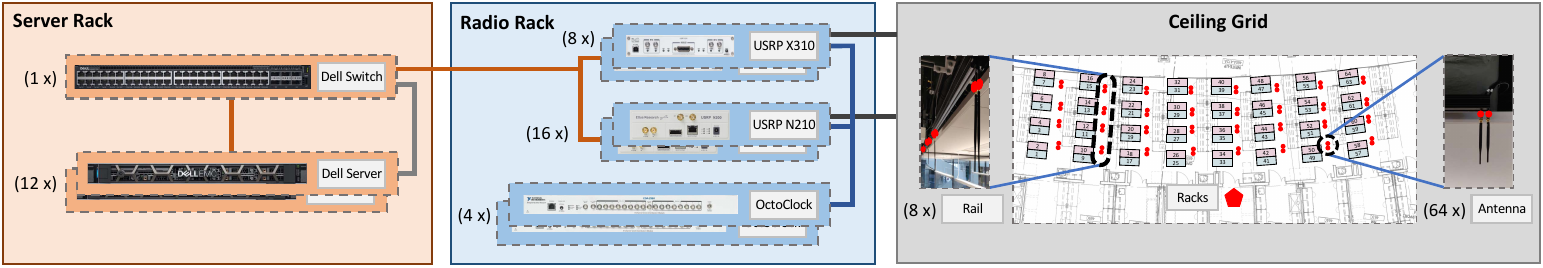}
    \caption{Arena architecture.}
    \label{chap2-fig:arena-architecture}
\end{figure*}

Arena is an over-the-air wireless testbed deployed on the ceiling of an indoor laboratory space~\cite{bertizzolo2020comnet}.
The architecture of Arena is depicted at a high level in Figure~\ref{chap2-fig:arena-architecture}.
Its main building blocks are: (i)~the ceiling grid; (ii)~the radio rack; and (iii)~the server rack.

The ceiling grid concerns~64 VERT2450 omnidirectional antennas hung off a $2450\:\mathrm{ft^2}$ indoor office space.
These are deployed on sliding rails and arranged in an $8 \times 8$ configuration to support \gls{mimo} applications.
The antennas of the ceiling grid are cabled through $100$\:ft low-attenuation coaxial cables to the radio rack.
This is composed of~24 USRP \glspl{sdr} (16~USRP N210 and 8~USRP X310) synchronized in phase and frequency through four OctoClock clock distributors.
Similarly to the USRPs on Colosseum, these \glspl{sdr} can be controlled through softwarized protocol stacks (e.g., cellular, Wi-Fi, etc.) deployed on the compute nodes of the server rack, to which they are connected through a Dell S4048T-ON \gls{sdn} programmable switch.
The server rack includes 12~Dell PowerEdge R340 compute nodes that are powerful enough to drive the \glspl{sdr} of the radio rack and use them for wireless networking experimentation in a real wireless propagation environment.
%

\smallskip
Because of the similarities offered by these two testbeds, software containers can be seamlessly transferred between the Colosseum and Arena testbeds with minimal modifications (e.g., specifying the network interface used to communicate with the \glspl{sdr}), as discussed in Section~\ref{chap2-sec:protocol-stack-twinning}.)
As we will show in Section~\ref{chap2-sec:results}, this allows users to design and prototype solutions in the controlled environment provided by the Colosseum \textit{digital twin}, to transfer them on Arena, and to validate these solutions in a real and dynamic wireless ecosystem.

\section{Digitizing Real-world Environments}
\label{chap2-sec:digitizing}

The process of digitizing real-world environments into their \gls{dt} representation is composed of different steps: (i)~\gls{rf} scenario twinning, in which the physical environment is represented into a virtual scenario and validated thereafter; and (ii)~protocol stack twinning, in which softwarized protocol stacks are swiftly transferred from the real world to the \gls{dt}, thus allowing users to evaluate their performance in the designed virtual scenarios.
We will describe these steps in the remainder of this section.

\subsection{RF Scenario Twinning}
\label{chap2-sec:scenario-twinning}

The \gls{rf} scenario twinning operations are performed by our \acrfull{cast}~\cite{villa2022cast}, which we made publicly available to the research community.\footnote{\url{https://github.com/wineslab/cast}}
This tool allows users to characterize a physical real-world \gls{rf} environment and to convert it into its digital representation, to be used in a digital twin, such as the Colosseum wireless network emulator.
\gls{cast} is based on an open \gls{sdr}-based implementation that enables: (i)~the creation of virtual scenarios from physical terrains; and (ii)~their validation through channel sounding operations to ensure that the characteristics of the designed \gls{rf} scenarios closely mirror the behavior of the real-world wireless environment.

\subsubsection{Scenario Creation}

\begin{figure}[ht]
    \centering
    \includegraphics[width=0.8\columnwidth]{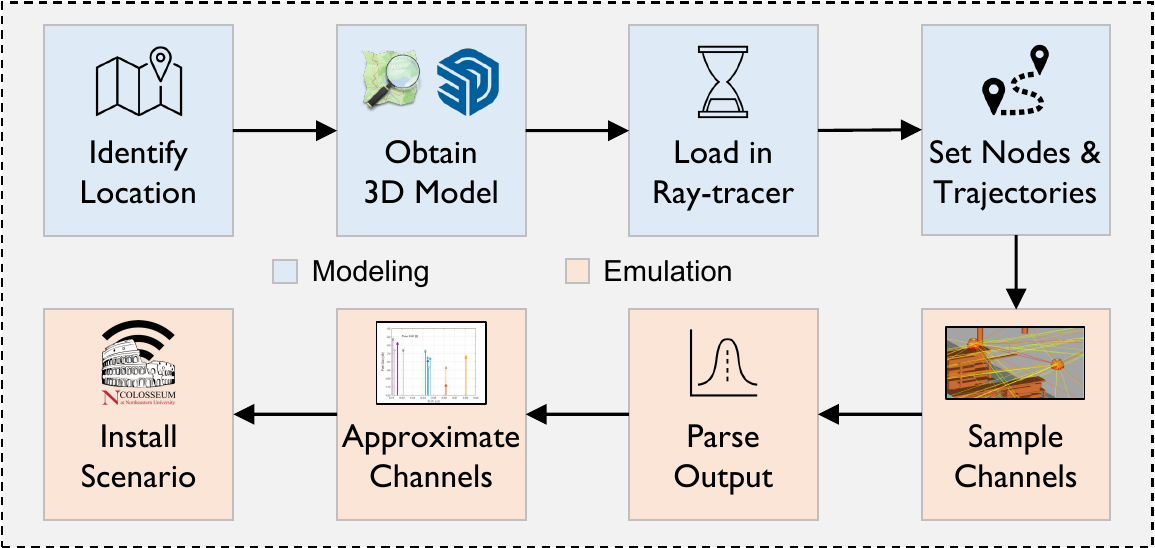}
    \caption{CaST scenario creation workflow.}
    \label{chap2-fig:scenariocreation}
\end{figure}

The scenario creation framework consists of several steps that capture the characteristics of a real-world propagation environment and model it into an \gls{rf} emulation scenario to install on Colosseum.
These steps, which are shown in Figure~\ref{chap2-fig:scenariocreation}, concern: (i)~identifying the wireless environment to emulate; (ii)~obtaining a 3D model of the environment; (iii)~loading the 3D model in a ray-tracing software; (iv)~modeling nodes and defining their trajectories; (v)~sampling the channels between each pair of nodes; (vi)~parsing the ray-tracing output of the channel samples; (vii)~approximating the obtained channels in a format suitable for the emulation platform (e.g., Colosseum \gls{mchem} \glspl{fpga}); and, finally, (viii)~installing the scenario on Colosseum.

\textbf{Identify the Wireless Environment.}
The first step consists of identifying the wireless environment, i.e., the physical location to twin in the channel emulator.
The area to model can be of different sizes, and representative of different environments, e.g., indoor (see Section~\ref{chap2-sec:arenadtscen}), outdoor (as shown in~\cite{villa2022cast}), urban, or rural.

\textbf{Obtain the 3D Model.}
The second step concerns obtaining the 3D model of the area to digitize.
This can be obtained from various databases, e.g., \gls{osm}, which is publicly available for outdoor environments or it needs to be designed using 3D modeling software, e.g., SketchUp.

\textbf{Load the Model in the Ray-tracer and Assign Material Properties.}
The 3D model obtained in the previous step needs to be converted into a file format (e.g., STL) suitable to be loaded into a ray-tracing software, e.g., the MATLAB ray-tracer or \gls{wi}, a commercial suite of ray-tracing models and high-fidelity \gls{em} solvers developed by Remcom~\cite{WI}.
Each object in the 3D model imported by the ray-tracing software consists of surfaces, and the material properties of these surfaces should be set to have reasonable ray-tracing results. The level of granularity in this step may depend on the ray-tracer platform, e.g., in the \gls{wi}, the material properties can be assigned to each surface. In the current version of MATLAB ray-tracer, this assignment is limited to the terrain and the buildings. The flexibility in assigning materials with a high level of detail leads to complex structures in the environment objects and accurate ray-tracing results.

\textbf{Model Nodes and Define Trajectories.}
Once the 3D model of the environment has been loaded in the ray-tracing software and the material properties are assigned, the radio nodes need to be modeled, which includes setting the nodes' radio parameters, modeling the antenna pattern, and defining locations of the nodes in the physical environment.
These nodes can be either static or mobile, in which case their trajectories and movement speeds need to also be defined.
The radio parameters of the nodes, e.g., carrier frequency, bandwidth, transmit power, receiver noise figure, ambient noise density, and antenna characteristics, need to be set as well.

\textbf{Sample the Channels.}
At this point, the channel is sampled through the ray-tracing software with a predefined sampling time interval $T_s$, which allows for capturing the mobility of the nodes in a discrete way.
To this aim, the node trajectories are spatially sampled with a spacing $D_i = V_i \cdot T_s$, where $V_i$ is the speed of node $i$.
Since spatial consistency plays a key role in providing a consisting correlated scattering environment in the presence of mobile nodes, we follow the \gls{3gpp} recommendations and consider a coherence distance of $15$\:m to guarantee an apt spatial consistency~\cite{3gppModel}.

\textbf{Parse the Output.}
The next step consists of parsing the ray-tracer output to extract a synchronized channel between each pair of nodes in the scenario for each discrete time instant $t$ spaced at least $1\:\mathrm{ms}$.
The temporal characteristic of the wireless channels is considered as a \gls{fir} filter, where the \gls{cir} is time-variant and expressed by:

\begin{equation}
\label{chap2-eq:time_variant_cir}
    h(t,\tau) = \sum_{i = 1}^{N_t} \Tilde{c_i}(t) \cdot \delta(t-\tau_i(t)),
\end{equation}

\noindent
where $N_t$ is the number of paths at time $t$, and $\tau_i$ and $c_i$ are the \gls{toa} and the path gain coefficient of the $i$-th path, respectively.
The latter is a complex number with magnitude $a_i$ and phase $\varphi_i$

\begin{equation}
\label{chap2-eq:cir_coefficient}
    \Tilde{c_i}(t) = a_i(t) \cdot e^{j\varphi_{i}(t)}
\end{equation}

\textbf{Approximate the Channels.}
The \gls{cir} characterized in the previous steps needs to be converted in a format suitable for \gls{mchem} \glspl{fpga}, e.g., 512~channel taps, 4~of which assume non-zero values, spaced with steps of $10\:\mathrm{ns}$ and with a maximum excess delay of $5.12\:\mathrm{\mu s}$.
To do this, we leverage a \gls{ml}-based clustering technique to reduce the taps found by the ray-tracing software, align the tap delays, and finalize their dynamic range, whilst ensuring the accuracy of the emulated scenario~\cite{tehrani2021creating}.

\textbf{Install the Scenario.}
Finally, the channel taps resulting from the previous steps are fed to Colosseum scenario generation toolchain, which converts them in \gls{fpga}-friendly format and installs the resulting \gls{rf} scenario on the \gls{dt}, ready to be loaded on-demand by the RF Scenario Server.

\subsubsection{Scenario Validation}

Now that the scenario has been created and installed in the \gls{dt}, we validate its correct functioning through the channel sounder embedded in \gls{cast}~\cite{villa2022cast}.
In doing this, we also ensure that the scenario installed in the \gls{dt} closely follows the behavior experienced in the real-world environment.
%

\begin{figure}[htbp]
    \centering
    \includegraphics[width=0.8\columnwidth]{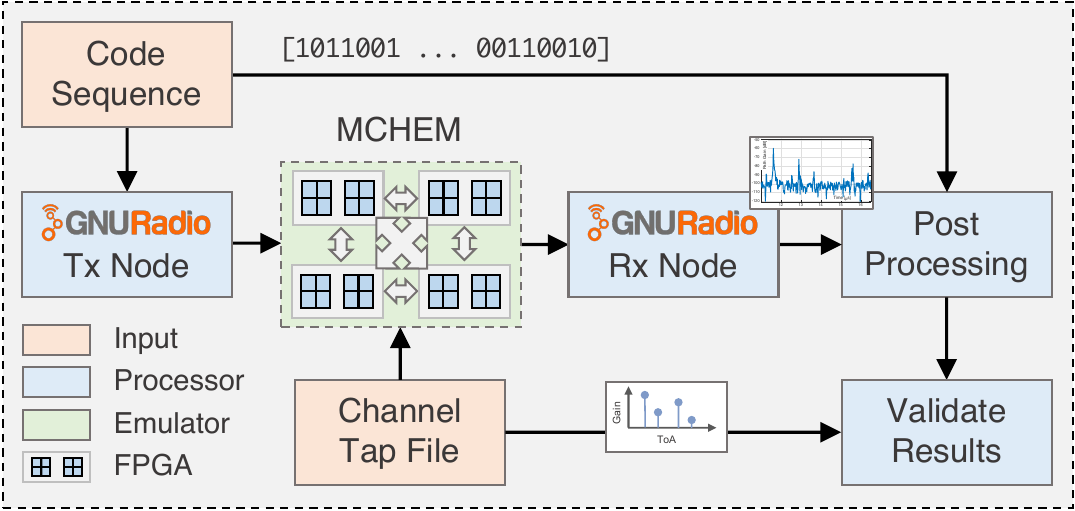}
    \caption{CaST channel sounding workflow.}
    \label{chap2-fig:soundingbd}
\end{figure}

The main steps of \gls{cast} channel sounder, shown in blue shades in Figure~\ref{chap2-fig:soundingbd}, are: (i)~the transmission of a known code sequence used as a reference for the channel sounding operations; (ii)~the reception of the transmitted code sequence, processed by \gls{mchem} through the channel taps of the emulated \gls{rf} scenario; (iii)~the post-processing of the received data and its correlation with the originally transmitted code sequence; and (iv)~the validation of the results with the modeled channel taps.

The \gls{cast} sounder uses a transmitter and a receiver node implemented through the GNU Radio open-source \gls{sdr} development toolkit~\cite{gnuradio}.
This software toolkit allows implementing and programming \glspl{sdr} through provided signal processing blocks that can be interconnected to one another.

In our sounding application, the transmitter takes as input a known code sequence---how to derive the specific code sequence is described in Section~\ref{chap2-sec:cast-tuning}---and transmits it to the receiver node through the wireless channel emulated by the Colosseum \gls{dt} through the \gls{rf} scenario to evaluate.
The transmitted signal is composed of sequential repetitions of the code sequence encoded through a \gls{bpsk} modulation.
While other modulation types are not restricted, we leverage \gls{bpsk} because it offers sufficient channel information for the sounding in Colosseum.
Additionally, it allows for simple data computations that are less susceptible to errors and approximations, resulting in a cleaner and less disrupted signal.
%
Data is streamed to the USRP controlled by the \gls{srn} that transmits it to the receiving node through \gls{mchem}.
For increased flexibility of the channel sounder, \gls{cast} allows users to set \blue{various USRP parameters, such as clock source, sample rate, and frequency.}

At the receiver side, the \gls{srn} USRP samples the signal sent by \gls{mchem}, i.e., the transmitted signal processed with the channel taps of the emulated scenario.
%
%
This signal is cross-correlated with the originally transmitted known code sequence to extract the \gls{cir} $h(t)$ of the emulated scenario, and the \gls{pl} $p(t)$.
The \gls{cir} is then used to obtain the \gls{toa} of each multi-tap component of the transmitted signal, which allows measuring the distance between taps, while the \gls{pl} allows measuring the intensity and attenuation of such components as a function of the time delay.
To perform the above post-processing operations, let $c(t)$ be the $N$-bit known code sequence, and $s^{IQ}(t)$ and $r^{IQ}(t)$ the \gls{iq} components of the transmitted ($s(t)$) and received ($r(t)$) signals, respectively.
The \gls{iq} components of the \gls{cir} are computed by separately correlating $r^{I}(t)$ and $r^{Q}(t)$ (i.e., the $I$ and $Q$ components of $r^{IQ}(t)$) with the $I$ and $Q$ components of $s(t)$ divided by the inner product of the transmitted known sequence with its transpose:

\begin{align}
    h^{I}(t) &= \frac{r^{I}(t) \otimes s^{I}(t)}{s^{I^T}(t) \times s^{I}(t)},\label{chap2-eq:hicorr} \\[1em]
    h^{Q}(t) &= \frac{r^{Q}(t) \otimes s^{Q}(t)}{s^{Q^T}(t) \times s^{Q}(t)},\label{chap2-eq:hqcorr}
\end{align}

\noindent
where $\otimes$ is the cross-correlation operation~\cite{buck2002computer} between two discrete-time sequences $x$ and $y$, which measures the similarity between $x$ and shifted (i.e., lagged) repeated copies of $y$ as a function of the lag $k$ \blue{following:}
\begin{equation}\label{chap2-eq:xcorrcast}
    \blue{\chi(k) = \sum_{n=1}^{N} x(n) \cdot y(n+k),}
\end{equation}
\blue{with $\chi$ denoting the cross-correlation and $N$ the length of the $x$ sequence.}
It is worth noticing that if the considered modulation is a \gls{bpsk}, the denominator is equal to the length $N$ of $c(t)$.
The amplitude of the \gls{cir} can be computed as:

\begin{equation}
    \label{chap2-eq:habs}
    |h(t)| = \sqrt{(h^{I}(t))^2 + (h^{Q}(t))^2}
\end{equation}

\noindent
and the path gains as:

\begin{equation}
    \label{chap2-eq:pgdb}
    G_p(t) [dB] = 20log_{10}(|h(t)|) - P_t - G_t - G_r,
\end{equation}

\noindent
where $P_t$ is the power of the transmitted signal, and $G_t$ and $G_r$ are the transmitter and receiver antenna gains expressed in $dB$.

\subsection{Protocol Stack Twinning}
\label{chap2-sec:protocol-stack-twinning}

The twinning of protocol stacks from real to virtual environments (and back) is key in the \gls{dt} ecosystem, as it allows users to swiftly transfer and evaluate real-world solutions in a controlled setup through automated tools.
Twinning at the protocol stack level, combined with the RF scenario twinning discussed in Section~\ref{chap2-sec:scenario-twinning}, makes it possible to seamlessly prototype, test, and transition end-to-end, full-stack solutions for wireless networks to and from digital and physical worlds.
After validation in the controlled environment of the \gls{dt}---to make sure whatever is tested works as expected---the protocol stack solutions can be transitioned back to real-world deployments where they are ultimately used on a production network. As an example, in our prior works~\cite{bonati2022openrangym-pawr,polese2022coloran}, we have shown how \gls{ai} solutions for 5G cellular networks trained and tested on the digital twin environment---Colosseum---can be effective and also work on real-world environments---Arena and the \gls{pawr} platforms~\cite{pawr}. 

At a high level, the twinning of the protocol stack involves: (i) tracking one or multiple remote, centralized version control systems that host the code of the protocol stack; and (ii) providing pipelines that can automatically replicate the same software build in the digital and physical domains. In addition, it is possible to embed automated steps for the performance validation (i.e., profiling of relevant performance metrics), similar to the scenario validation step of \gls{cast}.

\begin{figure}[htbp]
    \centering
    \includegraphics[width=0.8\columnwidth]{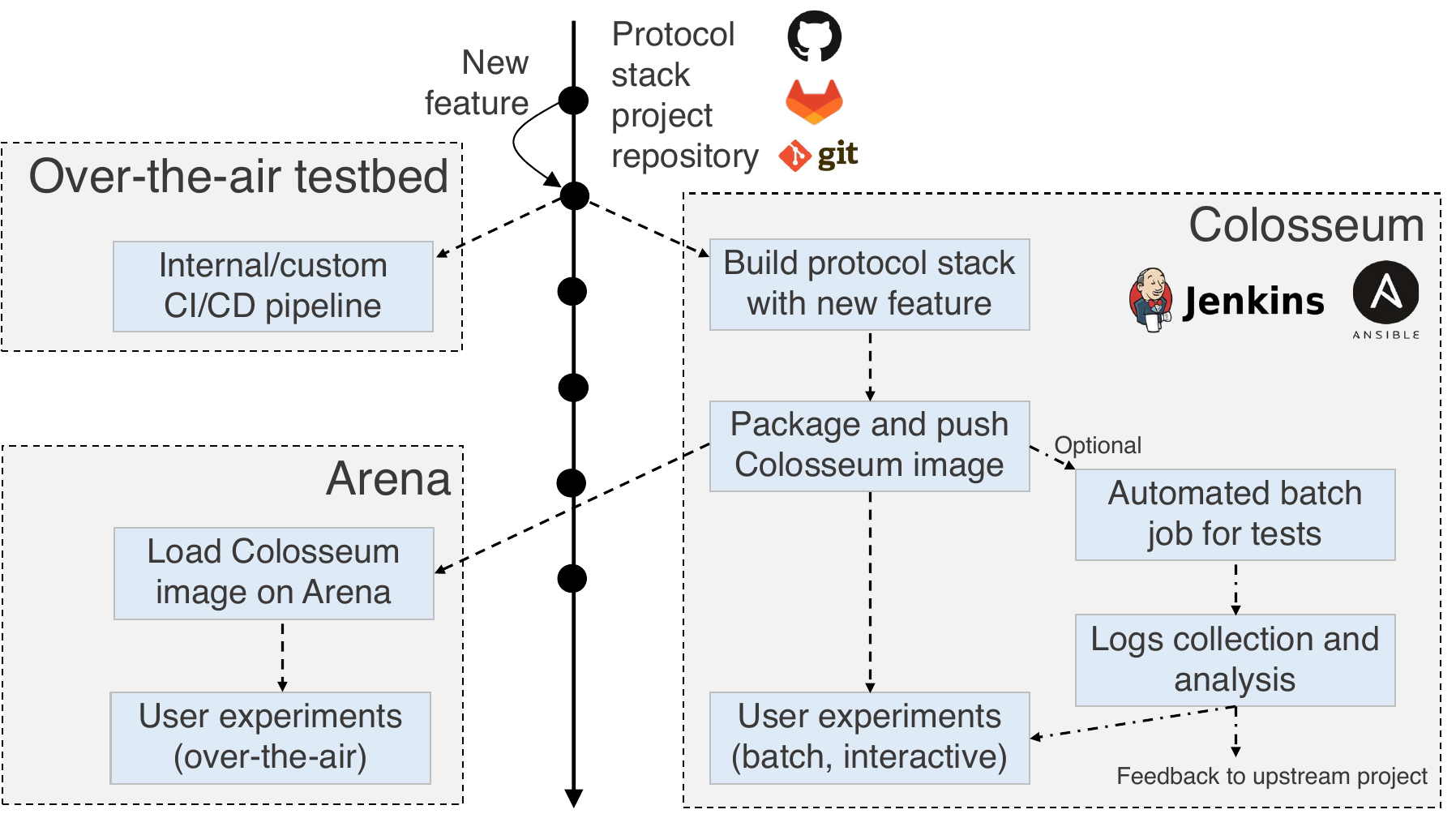}
    \caption{Protocol stack twinning workflow across a digital environment, i.e., Colosseum, and two physical environments, i.e., Arena and a generic over-the-air testbed.}
    \label{chap2-fig:protocol-twinning}
\end{figure}

Figure~\ref{chap2-fig:protocol-twinning} illustrates how the protocol stack twinning is implemented in Colosseum (right), with extensions to a generic over-the-air testbed (top left) and the specific integration with Arena (bottom left). The figure refers to a project repository for a sample protocol stack, hosted on a versioning platform that supports the \texttt{git} version control system (e.g., GitHub, GitLab, among others). In Colosseum, we implemented this pipeline for the OpenAirInterface reference stack for 5G base stations and \glspl{ue}, and are working toward the integration of additional components for O-RAN testing~\cite{bonati2022openrangym-pawr}. 

Whenever a new feature (i.e., a commit on selected branches, or a pull request) is pushed to the target project repository, 
a \gls{ci}/\gls{cd} framework implemented with Jenkins and Ansible triggers the automated process in the digital Colosseum domain. Specifically, Jenkins monitors the remote repository and orchestrates the kickoff of the build job, and Ansible applies the relevant configuration parameters to the machine that actually executes the build job (e.g., a Colosseum \gls{srn} or a dedicated virtual machine on AWS). Once the build is successful, the Jenkins job packages the output of the process into an LXC image which is stored on the Colosseum \gls{nas}.  

Once this is done, Colosseum can be further used to perform automated testing, e.g., to automatically test solutions and algorithms on the \gls{dt} and collect relevant metrics from such experiments. These can be shared with the relevant stakeholders, e.g., the developers of the protocol stack framework being deployed and tested. Moreover, since no over-the-air transmissions happen in Colosseum, as the channels are emulated through \gls{mchem} (see Section~\ref{chap2-sec:colosseum}), this \gls{dt} environment enables users to test networking solutions over frequencies and bandwidths that would normally require compliance with the \gls{fcc} regulations.

Finally, the image with the relevant components can be used by experimenters in Colosseum or moved to the physical domain, e.g., Arena, for validation on a real-world infrastructure. In addition, the protocol stack can be twinned in other over-the-air testbeds following their internal and custom \gls{cicd} pipelines, as long as the centralized repository that the different testbeds track provides shared specifications for the build environment (e.g., operating system, compiler versions, packages, etc.). As an example, the Colosseum protocol stack twinning process already replicates internal \gls{cicd} pipelines used in the Eurecom/OpenAirInterface facilities~\cite{eurecomCi}.
%
%
%
%
%
%

%

\section{Experimental Evaluation}
\label{chap2-sec:results}

This section discusses the capabilities of our \gls{dt} system through experimental evaluations, and it is organized as follows: (i)~we showcase \gls{cast} tuning process (Section~\ref{chap2-sec:cast-tuning}); (ii)~we leverage \gls{cast} to validate Colosseum scenarios, both with single and multiple taps (Section~\ref{chap2-sec:cast-colosseum-validation}); (iii)~we present a ray-tracing V2X use-case scenario in Tampa, FL and validate its emulated channels on Colosseum (Section~\ref{chap2-sec:tampa}); (iv)~we describe the Arena scenario designed as part of this work (Section~\ref{chap2-sec:arenadtscen}); (v)~we compare some experimental use cases (e.g., for cellular networking and Wi-Fi applications) both in the Arena testbed and in its \gls{dt} representation (Section~\ref{chap2-sec:usecases}).

%

%
%
%

\subsection{CaST Tuning}
\label{chap2-sec:cast-tuning}

\begin{figure}[ht]
\centering
    \centering
    \includegraphics[width=0.75\linewidth]{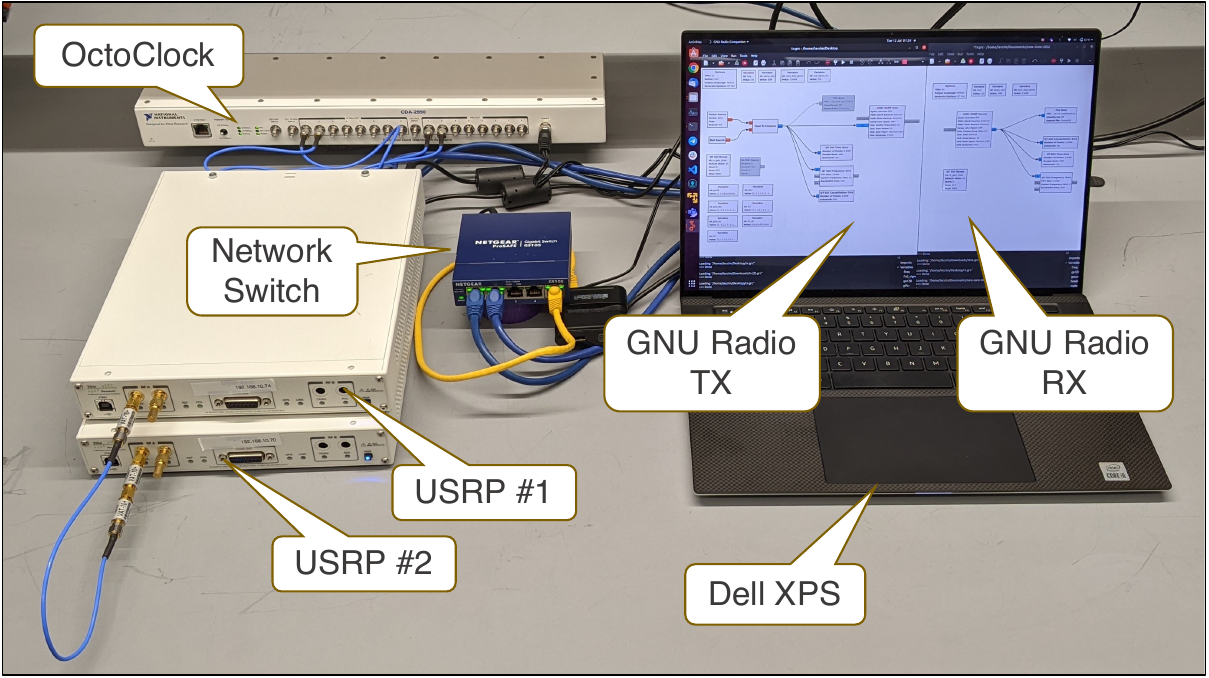}
    \caption{Controlled laboratory environment used for the \acrshort{cast} tuning process.}
    \label{chap2-fig:localtestbed}
\end{figure}

As a first step, we tune \gls{cast} parameters and configurations (see Section~\ref{chap2-sec:scenario-twinning}) outside the Colosseum channel emulator to identify a code sequence with high auto-correlation and low cross-correlation between transmitted code sequence and received signal which functions well within our combination of software and hardware.
It is worth noting that the tuning of CaST is a one-time operation, and it would need to be repeated only upon changes in the Colosseum hardware (e.g., radio devices and channel emulation system).
This step, which is key for \gls{cast} to be able to derive taps from arbitrary \glspl{cir}, is performed in the controlled environment shown in Figure~\ref{chap2-fig:localtestbed}.

This consists of two \gls{usrp} X310 \glspl{sdr} equipped with a UBX-160 daughterboard and synchronized in phase and frequency through an OctoClock clock distributor to mirror the same deployment used in Colosseum.
Differently from the Colosseum deployment, however, the two \glspl{usrp} are connected through a $12$\:inches SMA cable, and $30$\:dB attenuators (to shield the circuitry of the daughterboard from direct power inputs, as indicated in their datasheet).
This is done to derive the above-mentioned code sequences in a baseline and controlled setup without additional effects introduced by over-the-air wireless channels, or channel emulators.
The USRPs are connected through a network switch to a Dell XPS laptop, used to drive them.
The sounding parameters used in this setup are summarized in Table~\ref{chap2-table:localconfig}.
We consider different values for the gains of the \glspl{usrp} (i.e., in $[0, 15]$\:dB) to evaluate their effect on the sounding results.
The receiving period time and data acquisition are set to $3$\:s.
\begin{table}[htbp]
    \centering
    \caption{Configuration parameters used in the controlled laboratory setup.}
    \label{chap2-table:localconfig}
    \scriptsize
    \setlength{\tabcolsep}{4pt}
    \begin{tabular}{lc}
        \toprule
        \textbf{Parameter} & \textbf{Value} \\
        \midrule
        Center frequency & $1$\:GHz \\
        Sample rate & $[1, 50]$\:MS/s \\
        USRP transmit gain & $[0, 15]$\:dB \\
        USRP receive gain & $[0, 15]$\:dB \\
        \bottomrule
    \end{tabular}
\end{table}

\begin{figure}[ht]
    \centering
    \subfloat[Gold sequence]{\label{chap2-fig:cirlocalgold}\includegraphics[width=0.48\columnwidth]{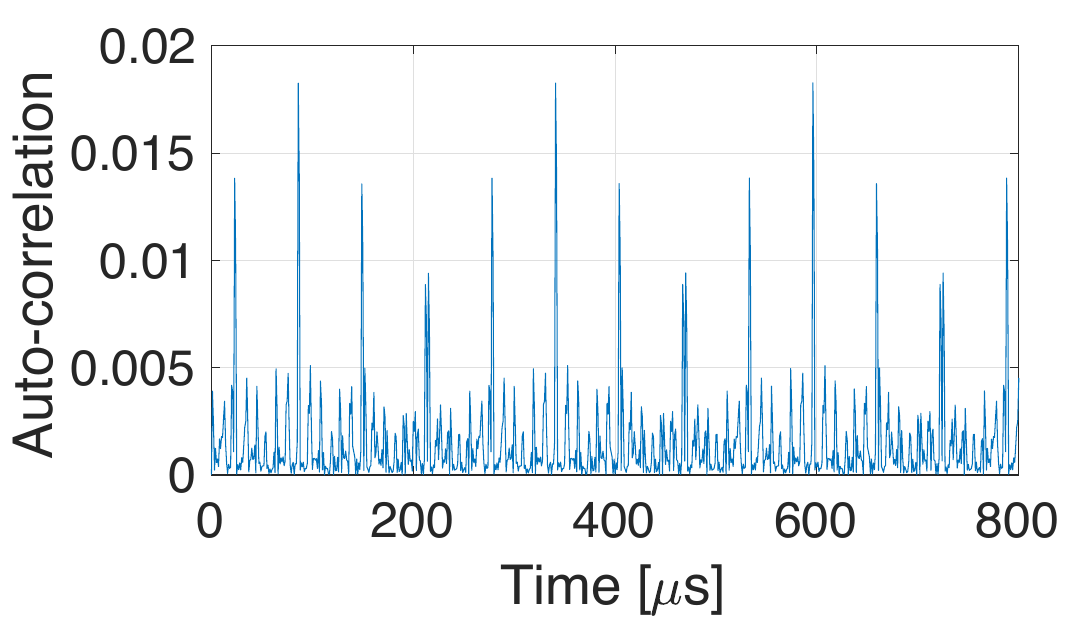}}%
    \hfill
    \subfloat[Ga$_{128}$ sequence]{\label{chap2-fig:cirlocalga}\includegraphics[width=0.48\columnwidth]{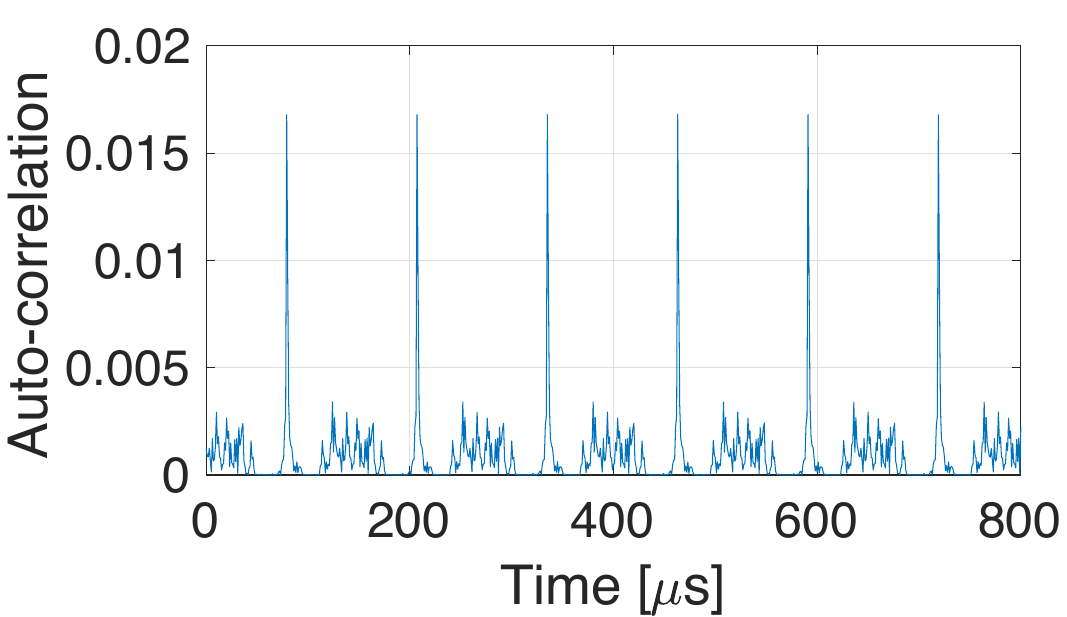}}
    \hfill
    \subfloat[LS sequence]{\label{chap2-fig:cirlocalls1}\includegraphics[width=0.48\columnwidth]{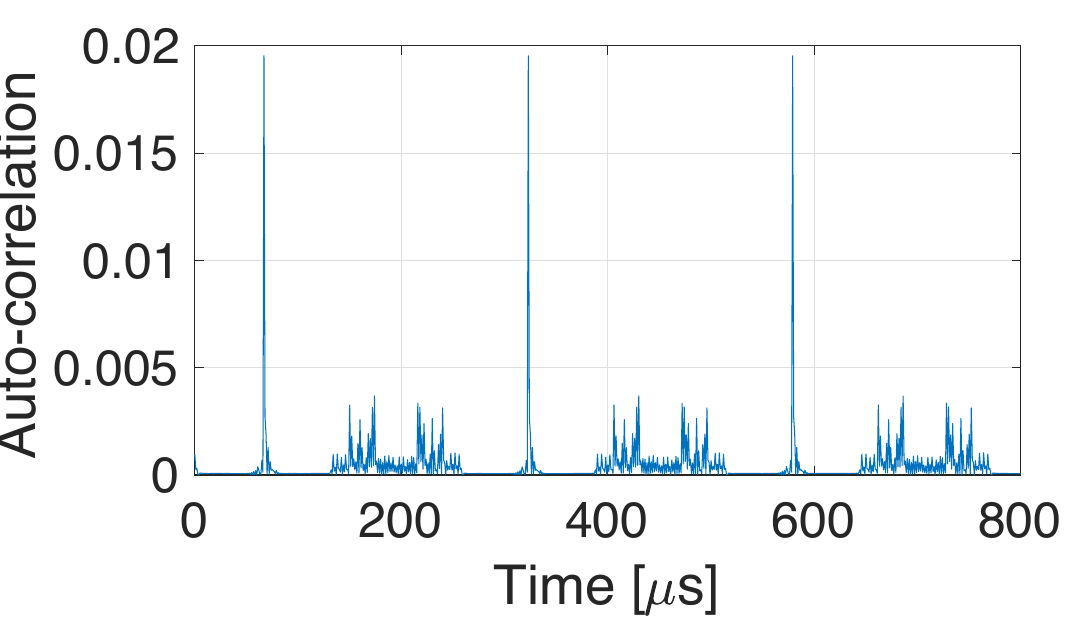}}
    \hfill
    \subfloat[GLFSR sequence]{\label{chap2-fig:cirlocalglfsr}\includegraphics[width=0.48\columnwidth]{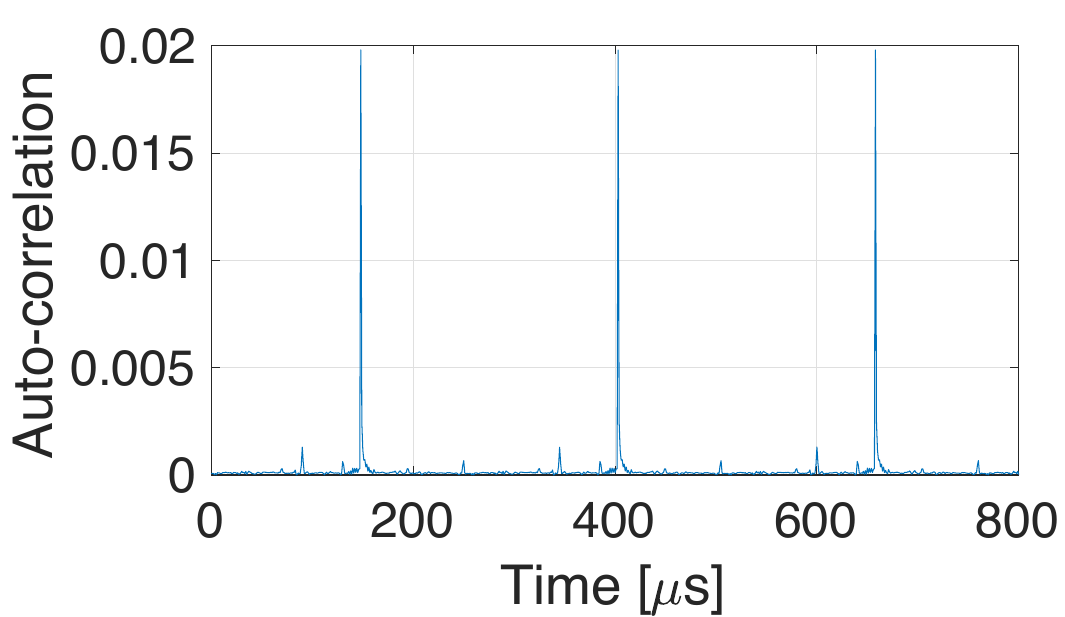}}
    \caption{Correlation of different code sequences in the controlled laboratory environment.}
    \label{chap2-fig:cirlocal}
\end{figure}

\textbf{Finding the Code Sequence.}
Code sequences have been widely investigated in the literature because of their role in many different fields~\cite{velazquez2016sequence,stanczak2001are}.
Good code sequences achieve a high auto-correlation (i.e., the correlation between two copies of the same sequence), and a low cross-correlation (i.e., the correlation between two different sequences).
For our channel-sounding characterization, we consider and test four different code sequences by leveraging the laboratory environment shown in Figure~\ref{chap2-fig:localtestbed}:
\begin{itemize}
    \item \textit{Gold sequence}. These sequences
    are created by leveraging the XOR operator in various creation phases applied to a pair of codes, $u$, and $v$, which are called a preferred pair. This pair of sequences must satisfy specific requirements to qualify as suitable for a gold sequence, as detailed in~\cite{zhang2011analysis}.
    Gold sequences have small cross-correlation within a set, making them useful when more nodes are transmitting in the same frequency range. They are mainly used in telecommunication (e.g., in \gls{cdma}) and in satellite navigation systems (e.g., in GPS).
    In this work, we use a Gold sequence of $255\:\mathrm{bits}$ generated with the MATLAB Gold sequence generator system object with its default first and second polynomials, namely $z^6+z+1$ and $z^6+z^5+z^2+z+1$, for the generation of the preferred pair sequences.
    
    \item \textit{Golay complementary sequence}. Being complementary, these sequences have the property that the sum of their out-of-phase aperiodic auto-correlation coefficients is equal to $0$~\cite{golay1961seq}. Their applications range from multi-slit spectrometry and acoustic measurements to Wi-Fi networking and \gls{ofdm} systems. In our tests, we use a $128$-bit type~A Golay Sequence (Ga$_{128}$) as defined in the IEEE 802.11ad-2012 Standard~\cite{ieee-802_11ad-standard}.
    
    \item \textit{\gls{ls} sequence}.
    These sequences exhibit the property of reaching very low auto-correlation and cross-correlation values in a certain portion of time, based on the maximum delay dispersion of the channel, called \gls{ifw}. This allows the mitigation of the interference if the maximum transmission delay is smaller than the \gls{ifw} length.
    In our experiments, we use an \gls{ls} sequence generated following the directions in~\cite{garcia2010generation}, and only leveraging the first codeset of $\{-1, 1\}$ without including the \gls{ifw}.
    
    \item \textit{\gls{glfsr} sequence}. These sequences add time offsets to \gls{lfsr} codes by leveraging extra XOR gates at the output of the \gls{lfsr}.
    %
    This allows to achieve a higher degree of randomness if compared to the classic \gls{lfsr}, making them more efficient and fast in detecting potential faults with increased auto-correlation results~\cite{pradhan1999glfsr}.
    %
    %
    In this work, we leverage GNU Radio
    to generate a $255$-bits sequence with the following parameters: shift register degree 8, bit mask 0, and seed 1.
\end{itemize}
Each of these sequences has been separately used by the transmitter node to construct the sending signal and to send it to the receiver node with a sample rate of $1$\:MHz.
After that, the receiver node performs the post-processing operations.
Results of $800\:\mathrm{\mu s}$ \gls{cir} for each code sequence are shown in Figure~\ref{chap2-fig:cirlocal}.
We can notice that all code sequences are able to correctly identify the starting position of the transmitted signal, as shown by the peak values.
The distance $D_{peak}$ of each peak can be written as a function of the code length $N$ and the sampling rate $SR$.

\begin{equation}
\label{chap2-eq:dpeak}
    D_{peak} = \frac{N}{SR}
\end{equation}

\noindent
Therefore, $D_{peak}$ is equal to $255\:\mathrm{\mu s}$ for the Gold, \gls{ls}, and \gls{glfsr} codes, each showing 3~transmitted sequences in Figure~\ref{chap2-fig:cirlocal}, and to $128\:\mathrm{\mu s}$ for the Ga$_{128}$ code, which displays 6~sequences instead.
We notice that \gls{glfsr} shows the highest auto-correlation and lowest cross-correlation among the four considered code sequences.
This results in an overall cleaner \gls{cir}.
For these reasons, we adopt the \gls{glfsr} code sequence in our experimental evaluation through \gls{cast}.

\begin{figure}[ht]
    \centering
    \includegraphics[width=0.7\columnwidth]{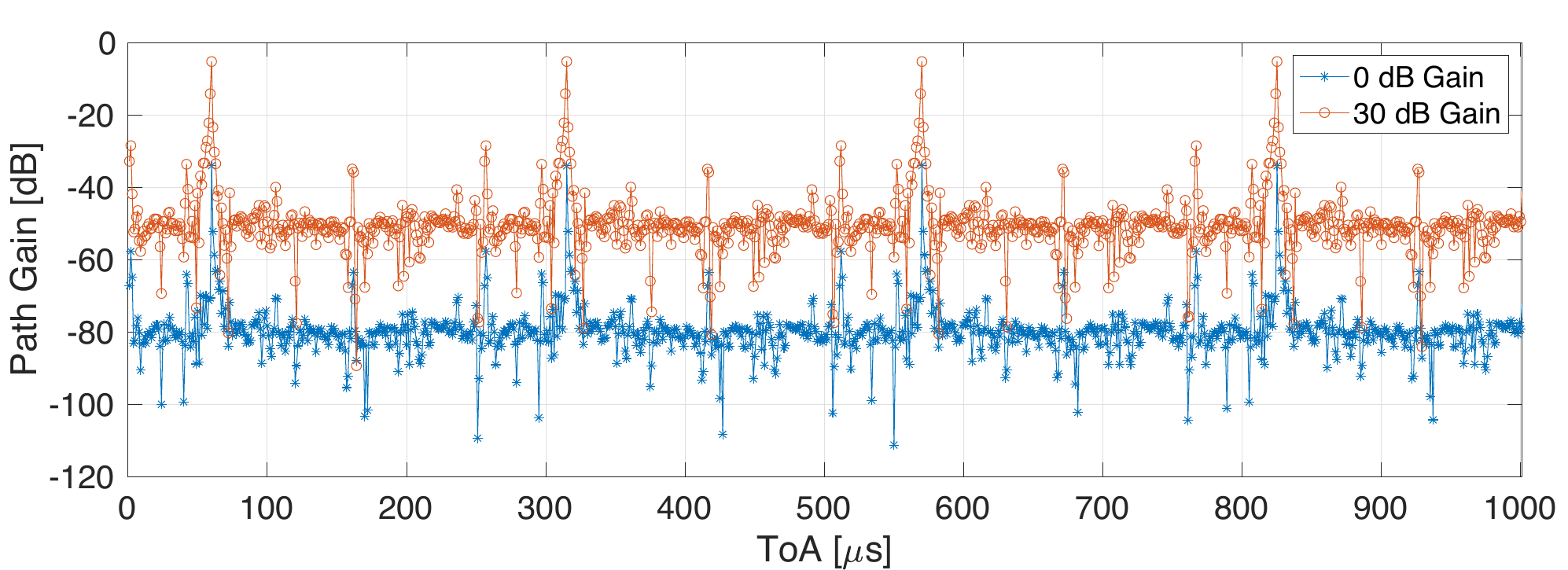}
    \vspace{-5pt}
    \caption{Received path gains in the controlled laboratory environment with $0$ and $30$\:dB total transmit and receive gains use cases ($15$\:dB at both transmitter and receiver sides).}
    \label{chap2-fig:pglocal}
\end{figure}

\textbf{\gls{cast} Validation in a Laboratory Environment.}
After identifying the code sequence for our application, we evaluate \gls{cast} in the laboratory setup shown in Figure~\ref{chap2-fig:localtestbed}.
To this aim, we test our sounder with a \gls{glfsr} code sequence and various configuration parameters, e.g., sample rate, center frequency, and antenna gains, to study its behavior and gather reference information to be leveraged in the Colosseum experiments.
Figure~\ref{chap2-fig:pglocal} shows a time frame of the received path gains for the case with $0$\:dB (blue line in the figure), and $30$\:dB total transmit and receive gains ($15$\:dB at both transmitter and receiver sides, orange line).

The figure shows signals that repeat based on the length of the transmitted code sequence, i.e., every 255 sample points (or equivalently every $255\:\mathrm{\mu s}$, since one point equals to $1/\mathrm{sample\_rate} = 1\:\mathrm{\mu s}$).
The peaks represent the path loss of the single tap of this experiment, which are equal to $34.06$\:dB for the $0$\:dB case, and $5.24$\:dB for the $30$\:dB case.
Since we have $30$\:dB attenuation in this validation setup, these results are in line with our expectations (with some extra loss due to the physical components of the setup, e.g., cable attenuation and noise).
We also notice that in the $30$\:dB case, the measured loss is slightly more severe due to imperfections in the power amplifiers of the \glspl{usrp}.
We use these results as a reference for our channel-sounding operations.

\subsection{Validation of Colosseum Scenarios through CaST}
\label{chap2-sec:cast-colosseum-validation}

After the tuning and validation in the controller laboratory environment, we can leverage \gls{cast} to validate the behavior of Colosseum \gls{mchem}.
We deploy the \gls{cast} sounder on the Colosseum wireless network emulator by creating an \gls{lxc} container from the open-source \gls{cast} source code.
This container, which has been made publicly available on Colosseum, contains all the required libraries and software to perform channel-sounding operations, as well as for the post-processing of the obtained results.
This enables the re-usability of the sounder with different \glspl{srn} and scenarios, as well as portability to different testbeds (e.g., to the Arena testbed described in Section~\ref{chap2-sec:arena}).
It also allows the automation of the channel sounding operations through automatic runs supported by Colosseum, namely \textit{batch jobs}.

To achieve our goal of characterizing \gls{mchem}, we test a set of synthetic \gls{rf} scenarios (i.e., single- and multi-tap \gls{rf} scenarios) on Colosseum, i.e., scenarios created specifically for the purpose of channel sounding.
These scenarios have been manually generated with specific channel characteristics to validate the behavior of \gls{mchem}, and have been made publicly available for all Colosseum users.
The parameters used in this evaluation are the same as the ones in Table~\ref{chap2-table:localconfig} with the only exception of the sample rate that is set at $50$\:MS/s to have a $20$\:ns resolution (thus being able to properly retrieve tap delays and gains), and the \gls{glfsr} code sequence found above.

\begin{figure}[ht]
    \centering
    \includegraphics[width=0.8\columnwidth]{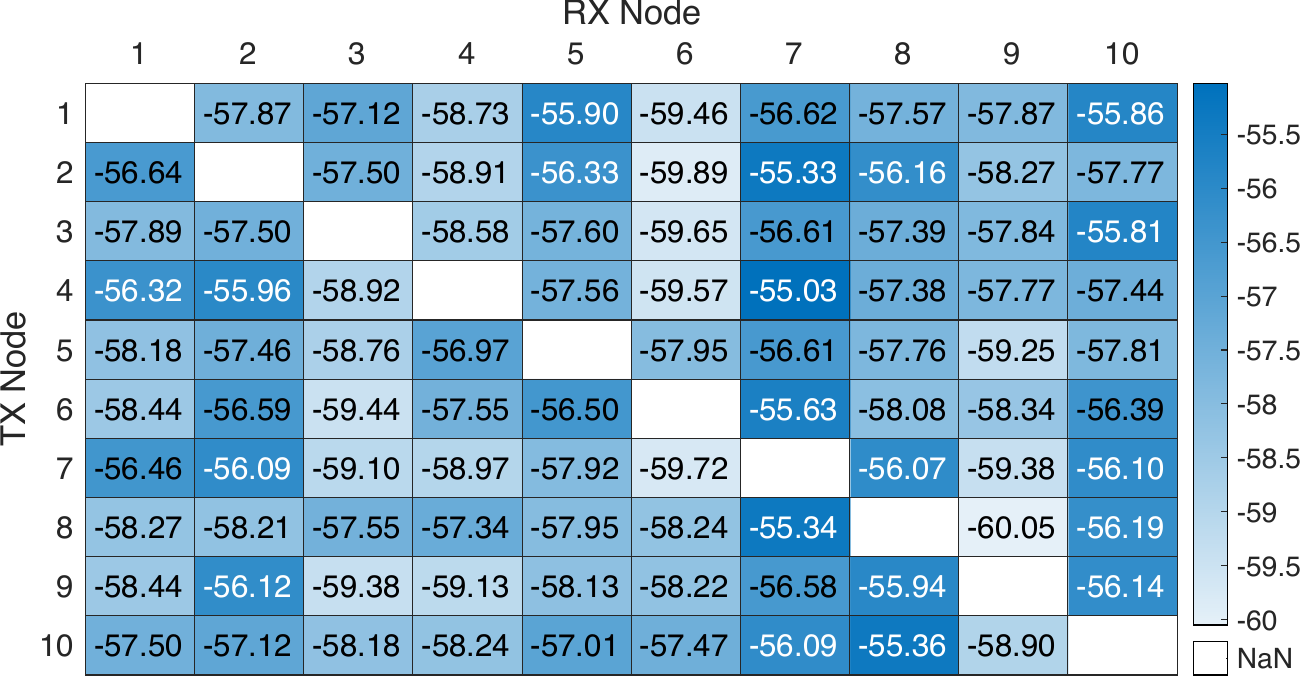}
    \caption{Path loss heatmap as measured by \acrshort{cast} in a $0$\:dB Colosseum \acrshort{rf} scenario with~10 SRNs.}
    \label{chap2-fig:heatmap0db}
\end{figure}

\textbf{Single-tap Scenario.}
The first synthetic \gls{rf} scenario that we consider is a single-tap scenario with nominal $0$\:dB path loss (i.e., $0$\:dB of path loss added to the inherent loss of the hardware components of the testbed).
To find the base loss of \gls{mchem}, i.e., the loss due to Colosseum hardware-in-the-loop, we instantiate \gls{cast} on 10~\glspl{srn}, and sound the channels among them, measuring the path loss of each link, shown in Figure~\ref{chap2-fig:heatmap0db}.
%

Each cell in the figure represents the average path loss for $2$\:s of reception time between transmitter (row) and receiver (column) nodes.
%
Results show an average Colosseum base loss of $57.55$\:dB with a \gls{sd} of $1.23$\:dB.
We also observe that the current dynamic range of Colosseum is approximately $43$\:dB, i.e., between the $57.55$\:dB base loss at $1$\:GHz and
the noise floor of $-100$\:dB.
%

\begin{figure}[htbp]
\centering
    \centering
    \includegraphics[width=0.7\columnwidth]{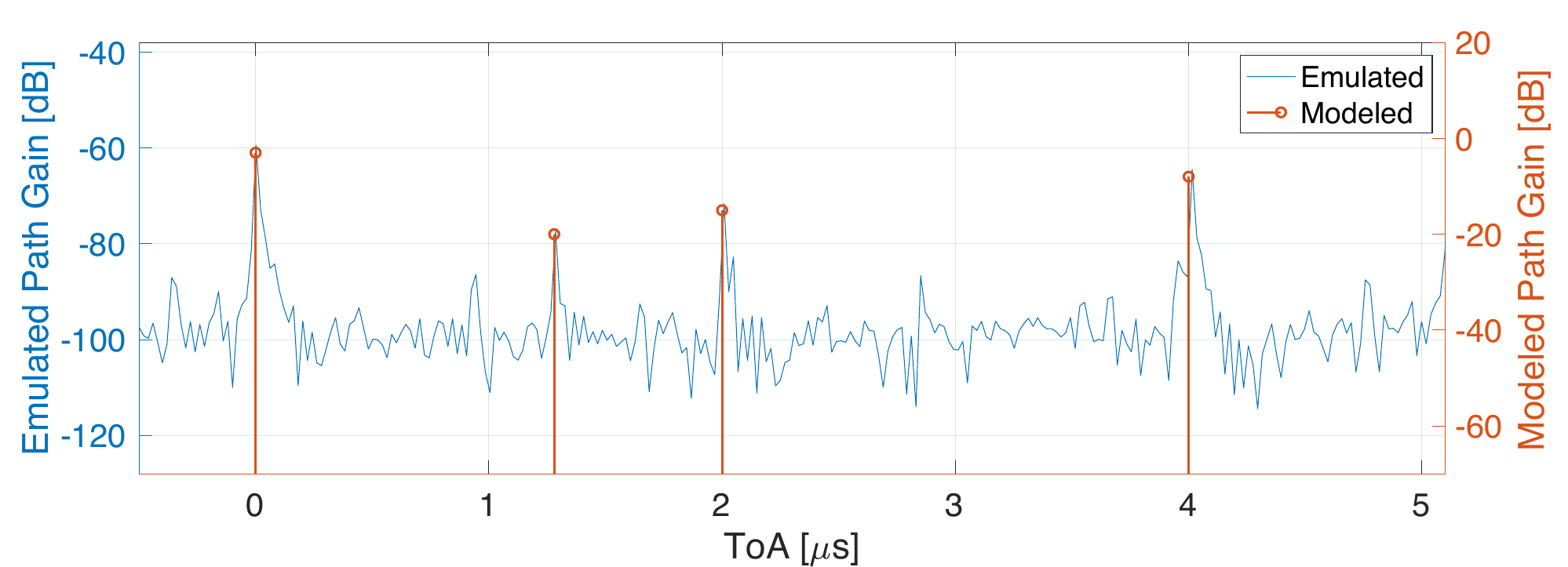}
    \caption{Comparison between emulated and modeled path gains in Colosseum for a single time frame.}
    \label{chap2-fig:cirpgcomparison}
\end{figure}

\textbf{Multi-tap Scenario.}
The second synthetic \gls{rf} scenario that we consider is a four-tap scenario in which taps have different delays and path gains.
%
We characterize such a scenario on Colosseum through \gls{cast} channel sounding operations.
Results for the emulated and modeled path gains for a single time frame are shown in Figure~\ref{chap2-fig:cirpgcomparison} in blue and orange, respectively.

We notice that the \glspl{toa} match between the modeled \gls{cir} and the taps emulated by the Colosseum \gls{rf} scenario, namely they occur at $0$, $1.28$, $2$, and $4\:\mathrm{\mu s}$.
We also notice that the received powers are in line with our expectations.
Indeed, by adding the Colosseum base loss computed in the previous step to the power measured by \gls{cast} (in blue in the figure), we obtain the modeled taps (corresponding to $-3$, $-20$, $-15$, and $-8$\:dB, shown in orange in the figure).

We now analyze the accuracy of the measurements performed with \gls{cast} by computing the relative difference between the emulated taps over time.
We do so by considering $1,500$ time frames.
Results show that the average difference between the strongest tap of each time frame is in the order of $10^{-6}$\:dB, with a \gls{sd} of $0.03$\:dB.
Analogous results occur for the second tap---which is the weakest tap in our modeled \gls{cir}---with a \gls{sd} of $0.17$\:dB, and for the third and fourth taps.
Finally, differences between the first and second taps of each time frame (i.e., between strongest and weakest taps in our modeled \gls{cir}) amount to $0.52$\:dB with a \gls{sd} of $0.18$\:dB.
These results are a direct consequence of the channel noise, which impacts weaker taps more severely.

Overall, results demonstrate \gls{mchem} accuracy in emulating wireless \gls{rf} scenarios in terms of received signal, tap delays, and gains.
This also shows \gls{cast} effectiveness in achieving a $20$\:ns resolution, thus sustaining a $50$\:MS/s sample rate, and a tap gain accuracy of $0.5$\:dB, which allows \gls{cast} to capture even small differences between the modeled and emulated \gls{cir}.

\subsection{A V2X Use-Case Scenario in Tampa, FL}
\label{chap2-sec:tampa}

\subsubsection{V2X Scenario Description}
\begin{figure}[htb]
\centering
    \includegraphics[width=0.7\columnwidth]{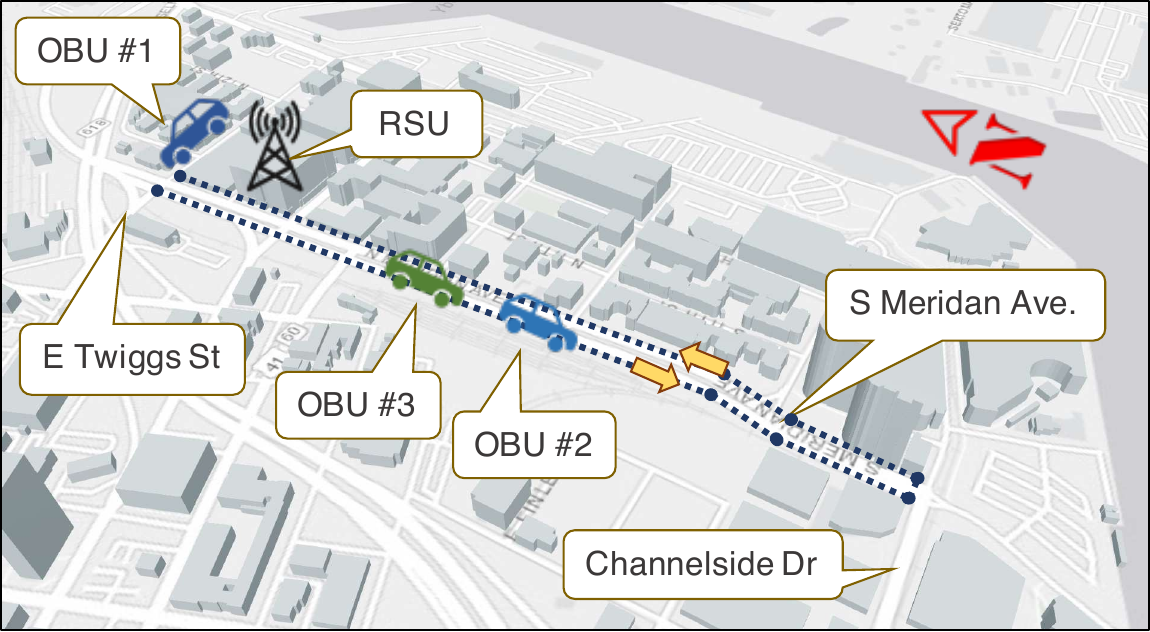}
    \caption{Tampa, FL, V2X scenario simulation environment in WI.}
    \label{chap2-fig:Tampa_scenario}
\end{figure}

We consider a scenario around the Tampa Hillsborough Expressway in Tampa, FL.
In order to simulate the scenario in \gls{wi}, we need a 3D model of the wireless environment. 
We obtained such a model from \gls{osm} in XML format and converted it to an STL file, which is supported by \gls{wi}.
The resulting scenario is depicted in Figure~\ref{chap2-fig:Tampa_scenario}.

In this scenario, we consider a \gls{v2x} model with four nodes.
One is a \gls{rsu} mounted on the traffic light at the intersection of E.\ Twiggs St.\ and N.\ Meridan Ave.
The other three nodes are \glspl{obu} installed on three vehicles: one is stationary and parked in the parking lot of the Tampa Expressway Authority; the other two vehicles are following each other at a constant speed of $25\:\mathrm{Mph}$ on Meridan Avenue, from E.\ Twiggs St.\ to Channelside Dr., and then back to E.\ Twiggs St. (Figure~\ref{chap2-fig:Tampa_scenario}). 
The radio parameters of the nodes are listed in Table~\ref{chap2-table:Tampa_wirelessParam}.

\begin{table}[hbp]
\centering
\caption{Wireless parameters for the Tampa simulation scenario.}
\begin{tabular}{@{}ll@{}}
\toprule
\textbf{Parameters} & \textbf{Values for \gls{v2x}} \\
\midrule
Carrier frequency & 5.915 [GHz] \\
Signal bandwidth & 20 [MHz] \\
Transmit power & 20 [dBm] \\
Antenna pattern & Omnidirectional \\
Antenna gain & 5 [dBi] \\
Antenna Height & \gls{rsu}: 16 [ft], \glspl{obu}: 5 [ft] \\
Ambient noise density & -172.8 [dBm/Hz] \\ \bottomrule
\end{tabular}
\label{chap2-table:Tampa_wirelessParam}
\end{table}

\subsubsection{V2X Mobile Scenario Validation}

\begin{figure}[htb]
\centering
    \centering
    \includegraphics[width=0.8\linewidth]{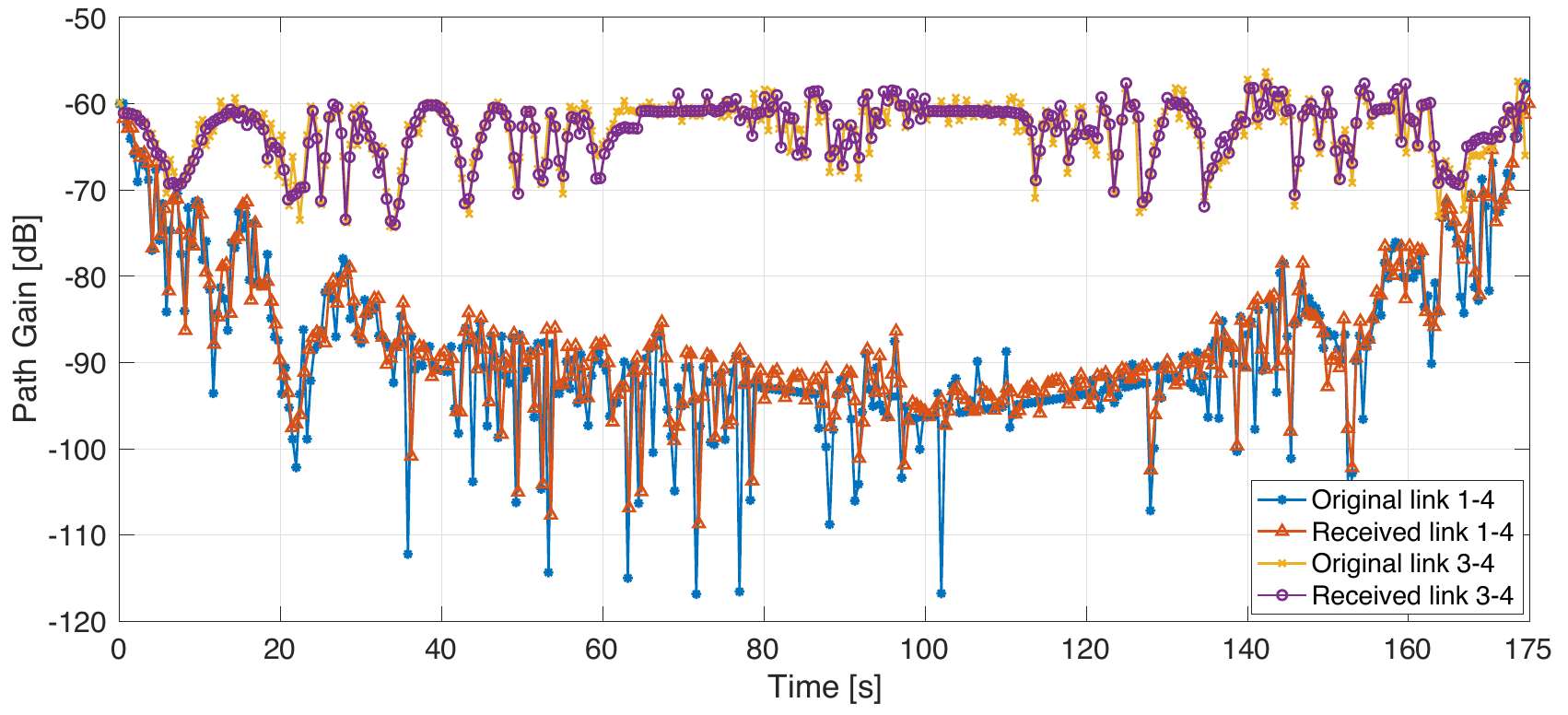}
    \caption{Comparison between original and received path gains for \acrshort{obu}\#3 (node 4) against \acrshort{rsu} (node 1) and \acrshort{obu}\#2 (node 3).}
    \label{chap2-fig:recorigpathloss}
\end{figure}

\begin{figure}[htb]
    \centering
    \begin{subfigure}{0.49\linewidth}
        \includegraphics[width=\linewidth]{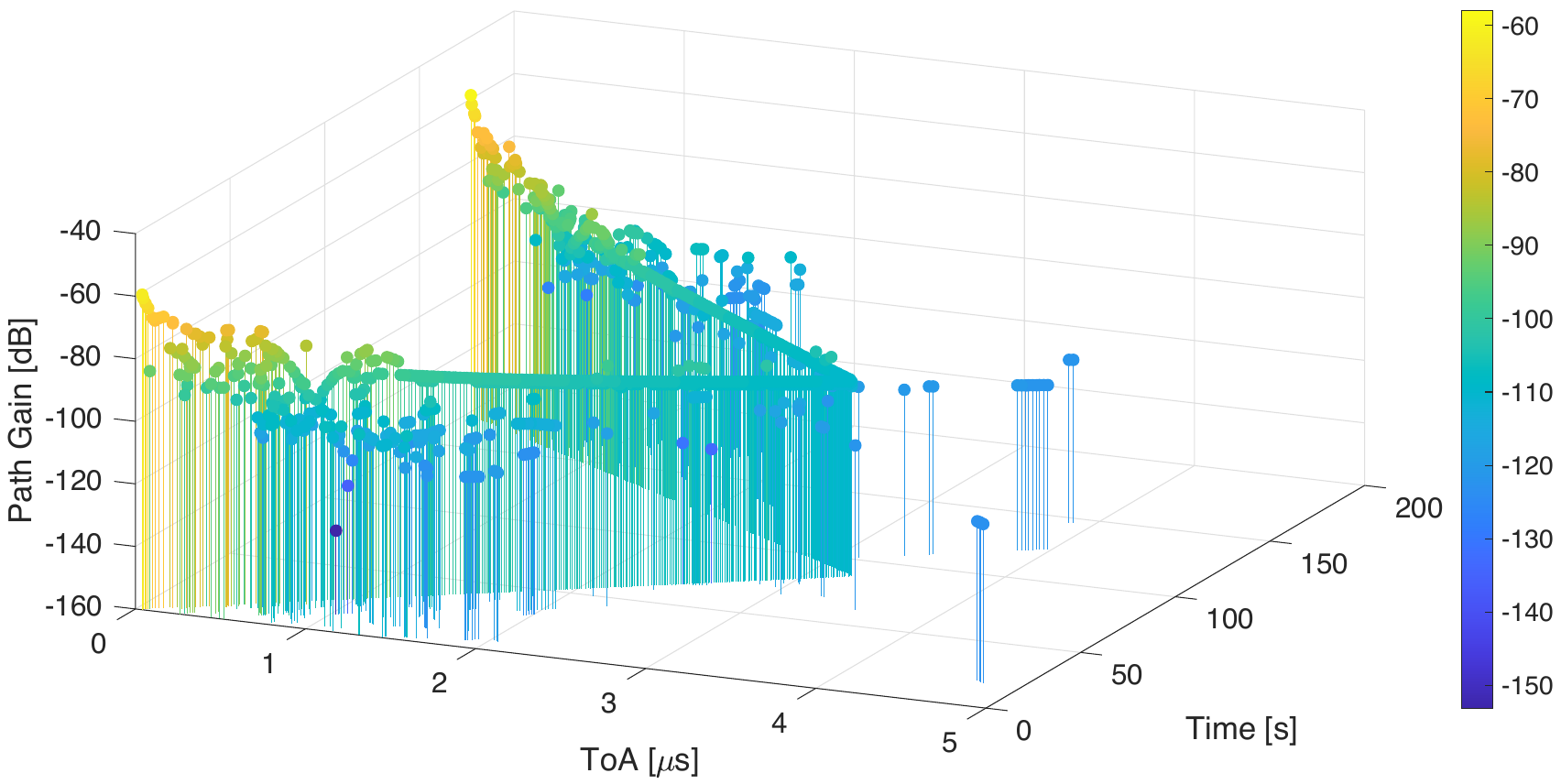}
        \caption{Original taps representation.}
        \label{chap2-fig:mobileorigpaths}
    \end{subfigure}
    \begin{subfigure}{0.49\linewidth}
        \includegraphics[width=\linewidth]{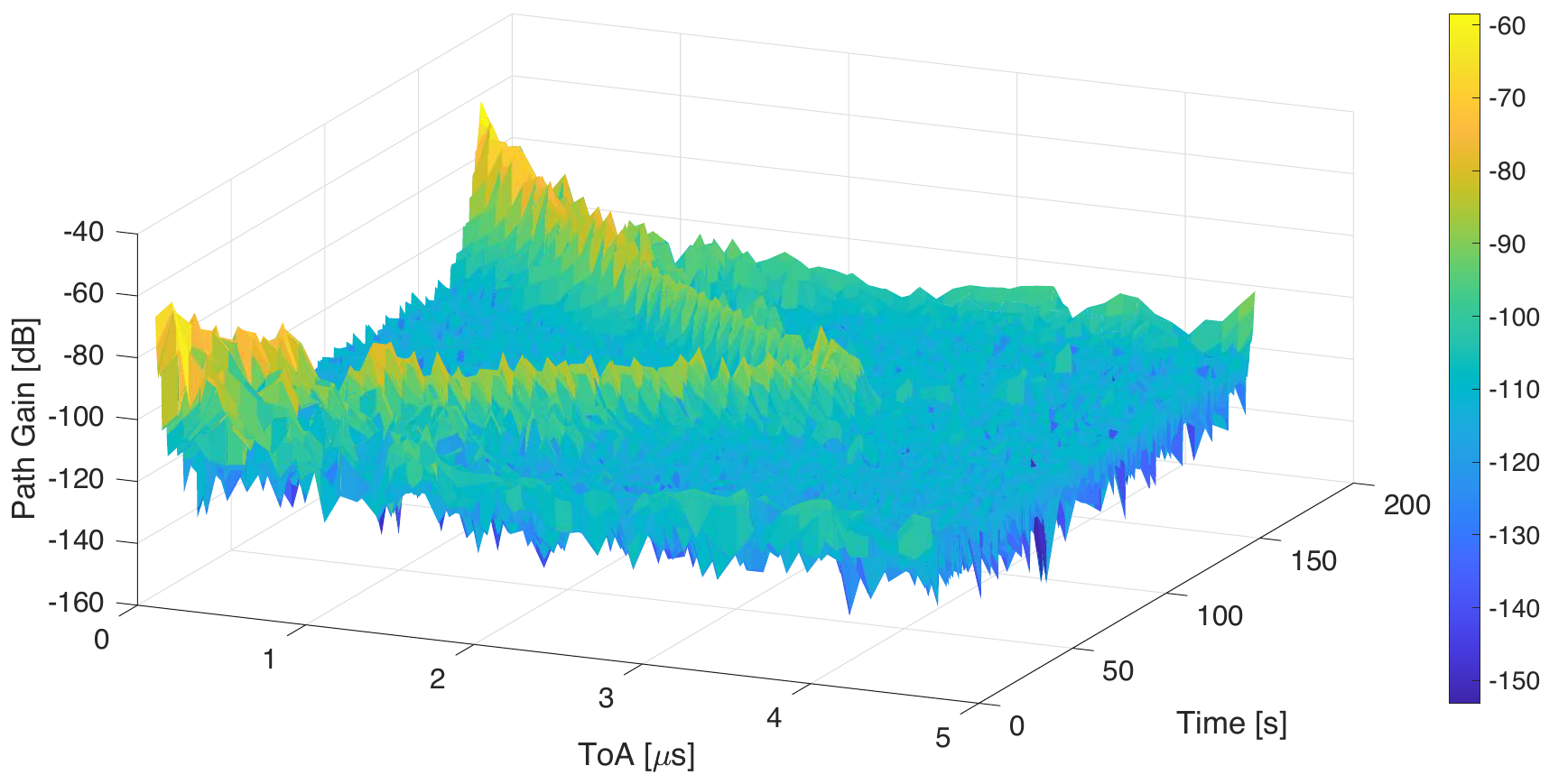}
        \caption{Received path gains.}
        \label{chap2-fig:mobileallpg}
    \end{subfigure}
    \caption{Results comparison for the mobile use case scenario.}
    \label{chap2-fig:3dorigmobile}
\end{figure}

In this set of experiments, we test the \gls{v2x} mobile use-case \gls{dt} in Tampa, FL. 
This scenario has been installed in Colosseum, with an increase of $60\:\mathrm{dB}$ across all taps, to fall within the Colosseum dynamic range.
The parameters for the sounding process are the same as those listed in Table~\ref{chap2-table:localconfig} with $15\:\mathrm{dB}$ gains and $10\:\mathrm{MS/s}$ sample rate. 
Since the total scenario time is $175\:\mathrm{s}$ and processing all the data together would require extreme memory, the rx time is divided into three chunks of around $60\:\mathrm{s}$ each. In this way, each chunk is around $5\:\mathrm{GB}$ in size, and it takes about $30\:\mathrm{minutes}$ to be processed. The results from each chunk are cleaned and merged to produce the final outcome. The received path gains have been adjusted by removing the Colosseum base loss and adding the original $60\:\mathrm{dB}$ increase.
Figure~\ref{chap2-fig:mobileallpg} shows how the received path gains (xy-axis) vary over the scenario time (z-axis). We can notice that the strongest tap resembles the same `V-shape' behavior seen in the original tap representation (Figure~\ref{chap2-fig:mobileorigpaths}) as a direct consequence of the movement to and from the static \gls{rsu} node.
Moreover, Figure~\ref{chap2-fig:recorigpathloss} shows the link path loss of node 1 (\gls{rsu}) and mobile node 3 (\gls{obu}\#2) against the mobile node 4 (\gls{obu}\#3). 

The original and received results are fluctuating due to multi-path fading, but they align almost perfectly. In this case, for links 1-4, a very similar received-power trend with a convex `U-shape` is noticeable. The mobile node first moves away from the \gls{rsu}, heading south, decreasing the gains and increasing the \gls{toa}. Then the pattern reverses as the vehicle turns around and travels back to the intersection and the \gls{rsu}. On the other hand, links 3-4 exhibit a more stable trend because the two vehicle nodes travel together.
These results confirm that Colosseum emulates the channel correctly even in a mobile scenario and validate the twinning capabilities of \gls{cast} of creating \glspl{dt} even when the metrics change over time.

\subsection{Arena Digital Twin Scenario}
\label{chap2-sec:arenadtscen}

\begin{figure}[ht]
    \centering
    \subfloat[Real-world location]{\label{chap2-fig:arena-real-loc}\includegraphics[width=0.48\columnwidth]{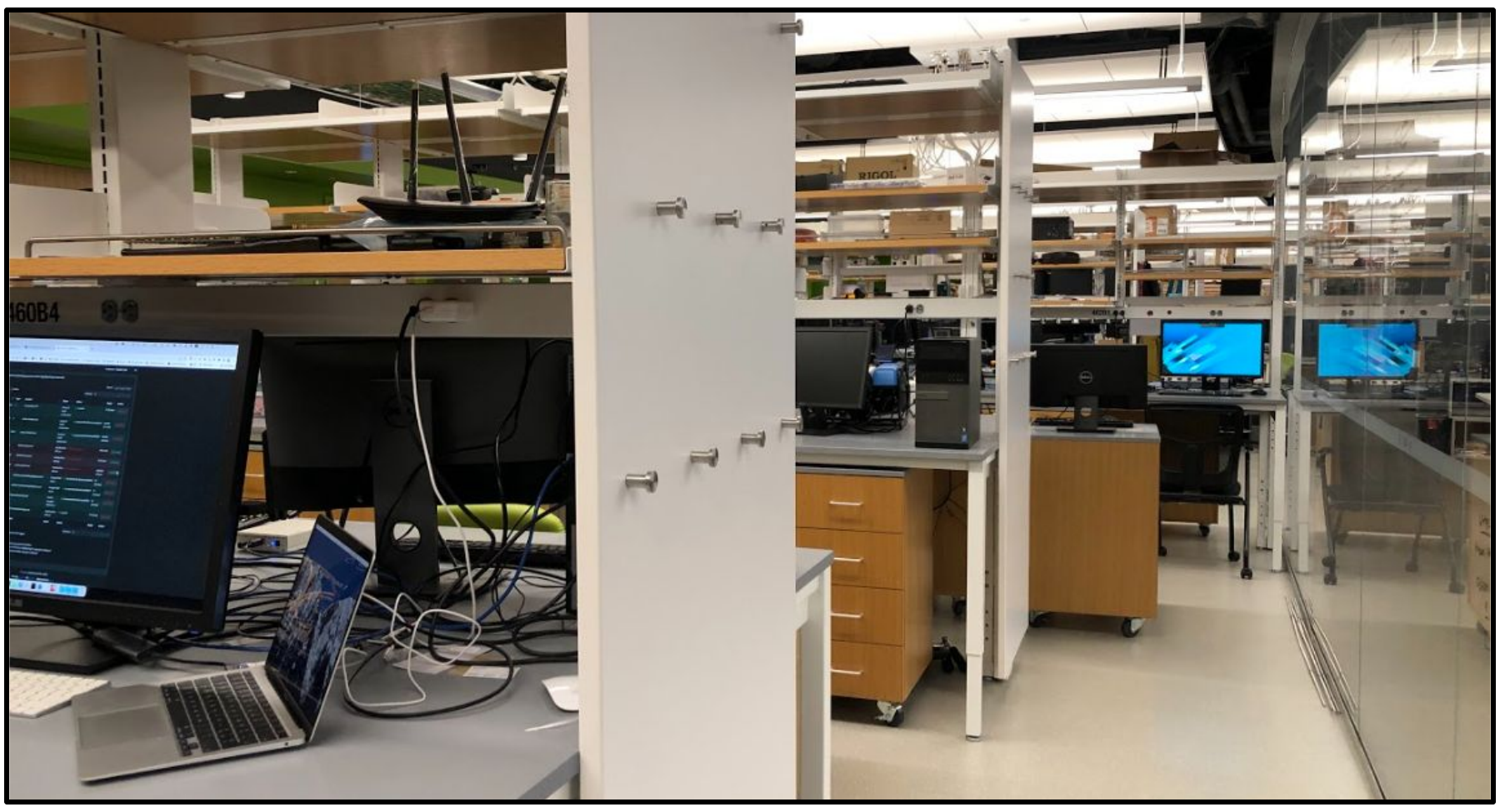}}%
    \hfill
    \subfloat[Digital-twin scenario]{\label{chap2-fig:arena-twin-loc}\includegraphics[width=0.48\columnwidth]{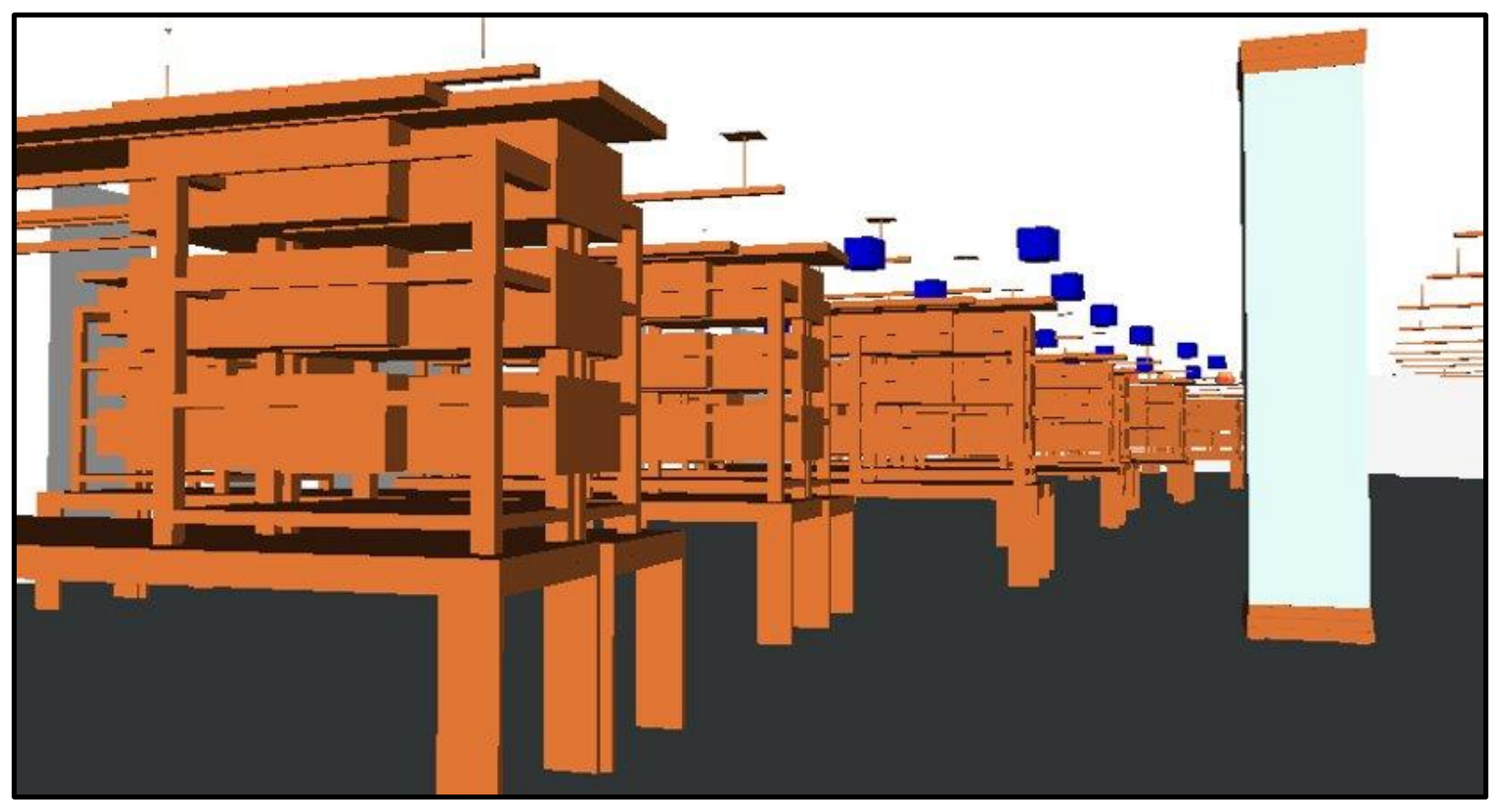}}
    \hfill
    \caption{The \blue{conversion} from a real-world location, into a digital medium scenario used to create the digital twin representation.}
    \label{chap2-fig:arena-real-to-dig}
\end{figure}

We use Sketchup~\cite{sketchup} software to create a 3D representation of the Arena testbed.
This software allows users to model a broad range of environments starting from an architectural layout (e.g., of the Arena testbed, a picture of which is shown in Figure~\ref{chap2-fig:arena-real-loc}), and with different surface renderings, e.g., glass walls and windows, wooden walls, carpeted floors~\cite{sketchup}.
%
%
%
The resulting 3D model (shown in Figure~\ref{chap2-fig:arena-twin-loc}) is then fed to the ray-tracing software, Wireless inSite~\cite{WI} in this case, to create a \gls{dt} scenario on Colosseum following the steps described in Section~\ref{chap2-sec:scenario-twinning}.
\begin{figure}[htb]
    \centering
    \includegraphics[width=\columnwidth]{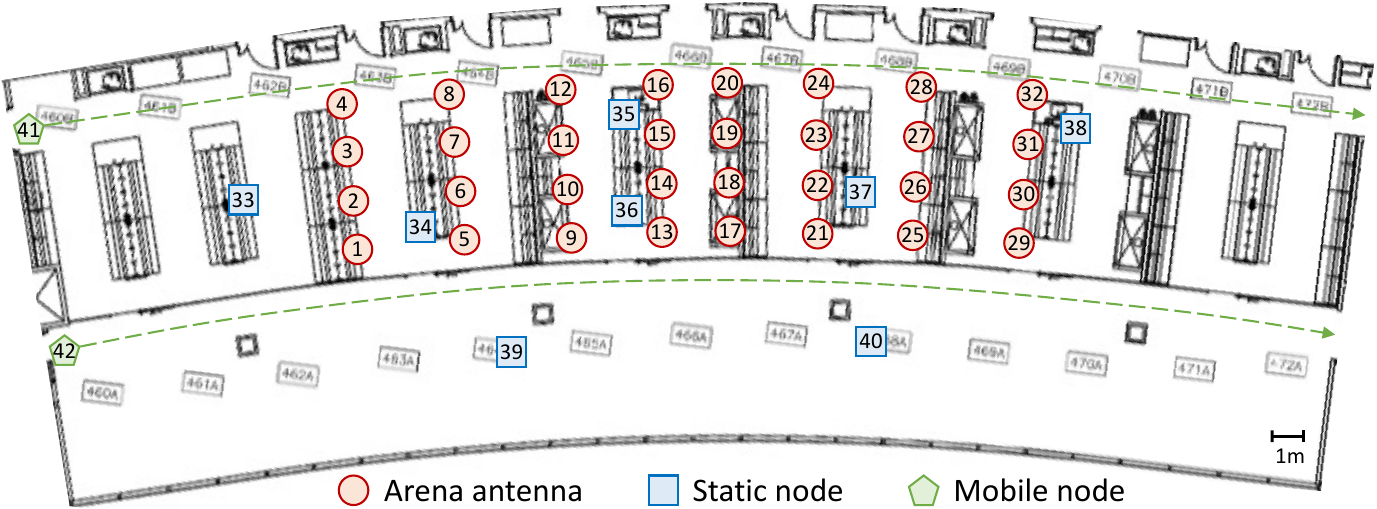}
    \caption{Location of the nodes in an Arena \acrshort{dt} scenario.}
    \label{chap2-fig:dt-arena-nodes}
\end{figure}
\begin{figure}[htb]
    \vspace{-10pt}
    \centering
    \includegraphics[width=\columnwidth]{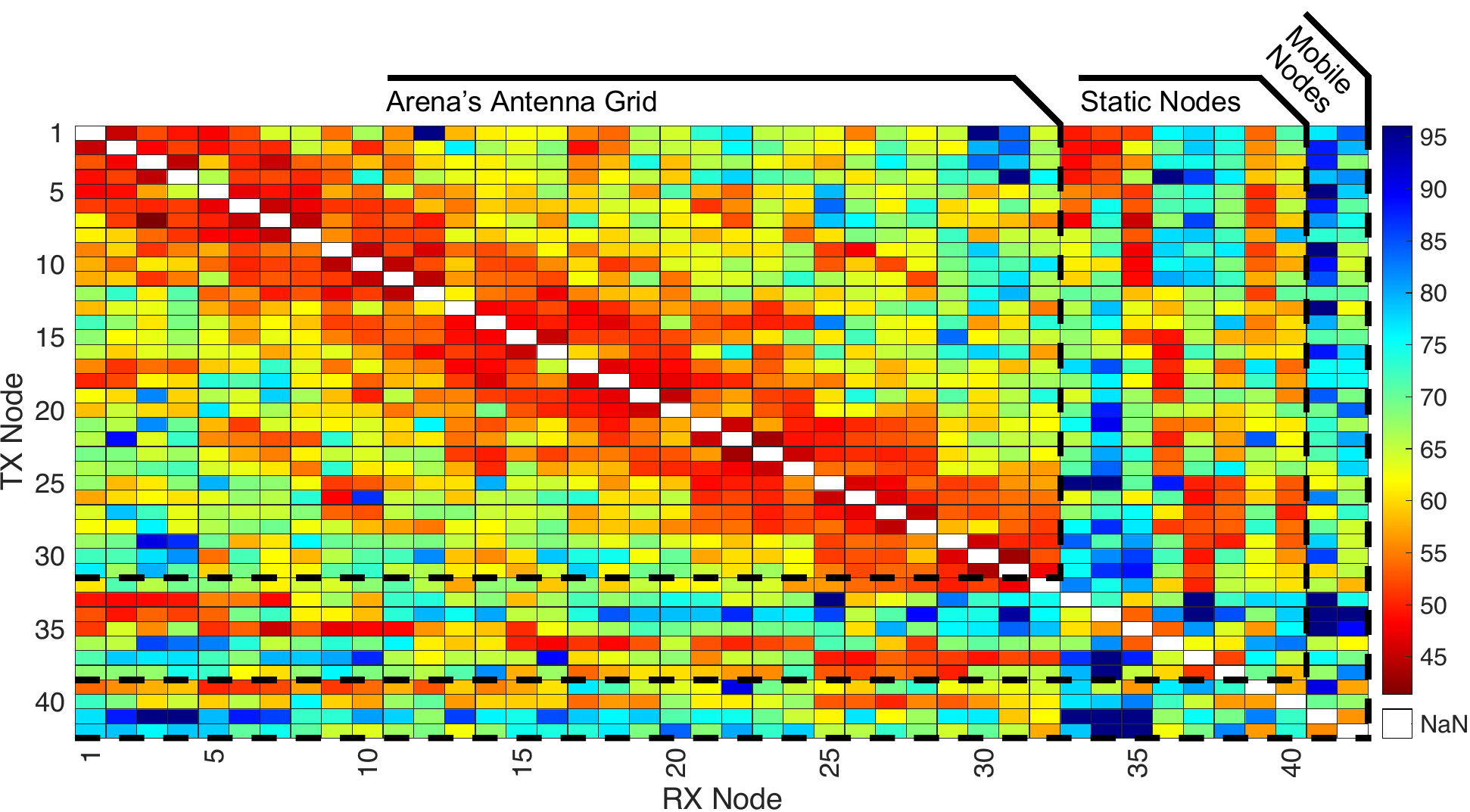}
    \caption{Heat map of the path loss among the nodes of Figure~\ref{chap2-fig:dt-arena-nodes}, with a line separator between antenna, static, and mobile. The mobile nodes are considered in the starting position on the left.}
    \label{chap2-fig:dt-arena-heatmap}
\end{figure}



For the developed Arena scenario, we model the antenna points of the Arena testbed in 32 locations (one for each antenna pair), as well as 8~static nodes distributed in their surroundings, and 2~mobile nodes traversing the laboratory space at a constant speed of $1.2$\:m/s. The height of the nodes (both static and mobile) is set to $1$\:m, e.g., to emulate handheld devices, or devices lying on table surfaces.
The modeled locations and nodes are shown in Figure~\ref{chap2-fig:dt-arena-nodes}, where the red circles represent the antenna pairs of Arena, while the blue squares and green pentagons identify the static and mobile nodes, respectively.
The dashed green arrows denote the movement direction of the mobile nodes.
By using a Dell T630 machine with 2 Xeon E52660 $14$\:cores CPU, $128$\:GB RAM, and Tesla K40 GPU, the ray-tracing operations of this scenario took $14$\:h and $21$\:m, while the channel approximation process $2$\:h and $33$\:m.
Additionally, the installation process in Colosseum required around $19$\:h and $30$\:m by leveraging a virtual machine hosted on a Dell PowerEdge M630 Server with 24 CPU cores and $96$\:GB of RAM.
It is worth noting that these are one-time operations, and the scenario is played on demand right away afterward.

Figure~\ref{chap2-fig:dt-arena-heatmap} shows the heat map of the path loss among the transmit-receive node pairs (the mobile nodes are considered in the starting position on the left).
As expected, closer nodes experience a lower path loss, which increases with the distance between the nodes.
A similar trend is also visible for the static nodes, even though this is less noticeable due to their scattered locations.
On the other hand, due to their remote starting locations on the side of the room, the mobile nodes exhibit a very high path loss against all nodes, as depicted in Figure~\ref{chap2-fig:dt-arena-heatmap}. These path losses decrease as they get closer to each node on their path, and increase again while reaching their end locations on the other side of the room.
%

\subsection{Arena Digital Twin Experimental Use Cases}
\label{chap2-sec:usecases}

In this section, we show outcomes of relevant experimental use cases run on both the Arena testbed, as well as on its \gls{dt} representation.
The first use case involves the deployment of a cellular networking system using the srsRAN software suite, while the second one encompasses a Wi-Fi adversarial jamming use case built on GNU Radio.

\subsubsection{\blue{Cellular Networking}}
\begin{figure}[ht]
    \centering
    \includegraphics[width=0.7\columnwidth]{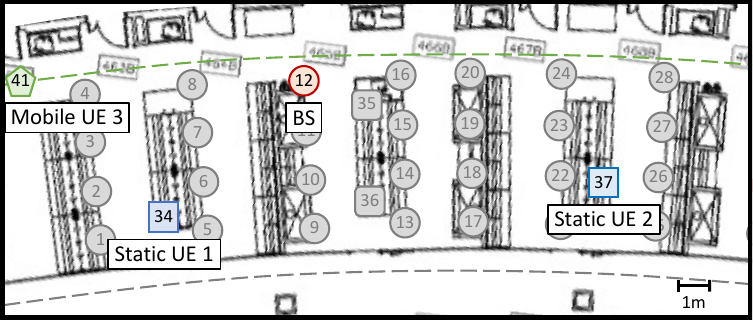}
    \caption{Location of the nodes in the cellular experiment.}
    \label{chap2-fig:cellular-nodes-singlebs}
\end{figure}

%
\blue{\textbf{Single BS.}} In the first cellular networking use case, we leverage SCOPE~\cite{bonati2021scope}---an open-source framework based on srsRAN~\cite{gomez2016srslte} for experimentation of cellular networking technologies---to deploy a twinned \gls{ran} protocol stack with one \gls{bs} and three \glspl{ue} in the Arena over-the-air testbed and in the Colosseum emulation system.
The same node positions, shown in Figure~\ref{chap2-fig:cellular-nodes-singlebs}, are used in the two platforms: the \gls{bs}, which transmits over a $10$\:MHz spectrum, is located on node~12, two static \glspl{ue} on nodes~34 and~37, and one mobile \gls{ue} on node~41.
In Arena, \glspl{ue} are implemented through commercial smartphones (Xiaomi Redmi Go), while on Colosseum, they are deployed on the \glspl{sdr} of the testbed.

We conduct two experiments on each system: the first one involves a downlink \gls{udp} traffic sent at a $5$\:Mbps rate; the second one \gls{tcp} downlink traffic.
The traffic generation for each experiment is achieved using iPerf, a benchmarking tool designed for assessing the performance of IP networks~\cite{iperf}.
The following results show the average of at least 5 separate experiment realizations.
%

%
%

\begin{figure}[htbp]
    \centering
    \subfloat[Arena]{\label{chap2-fig:cellular-arena-results-udp}\includegraphics[width=0.49\columnwidth]{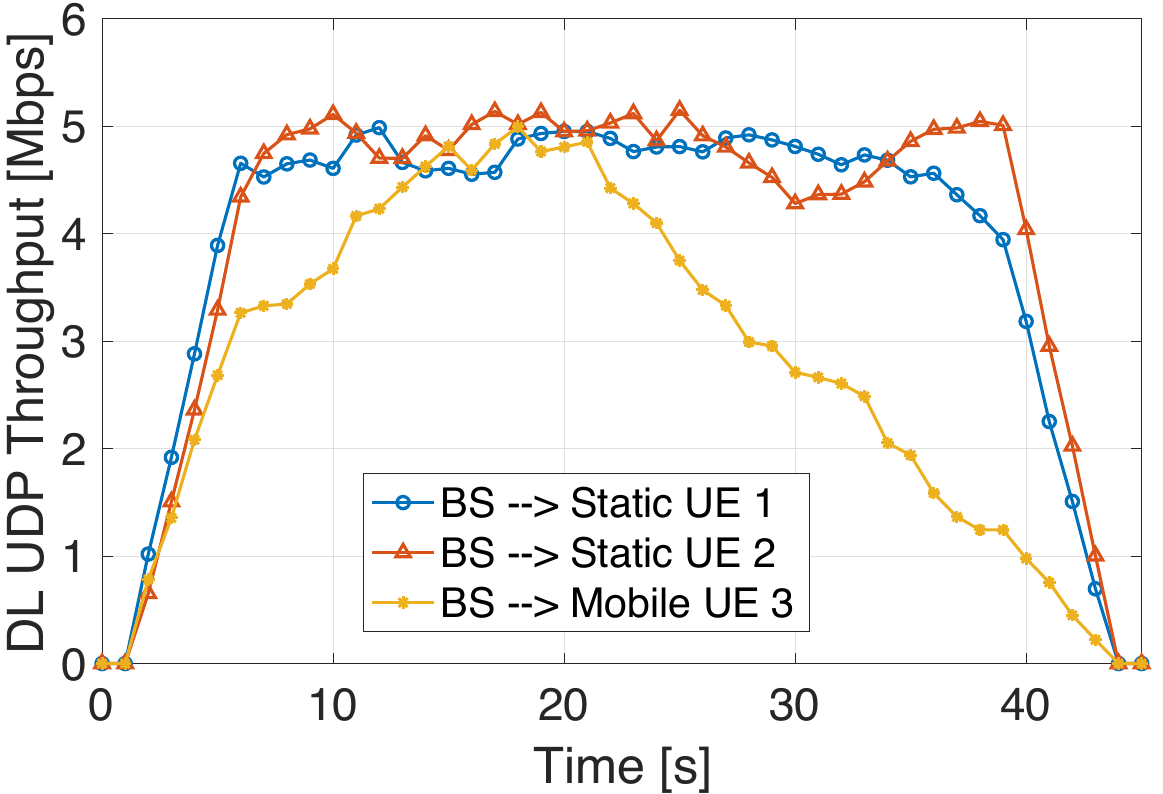}}
    \hfill
    \subfloat[Colosseum]{\label{chap2-fig:cellular-colosseum-results-udp}\includegraphics[width=0.49\columnwidth]{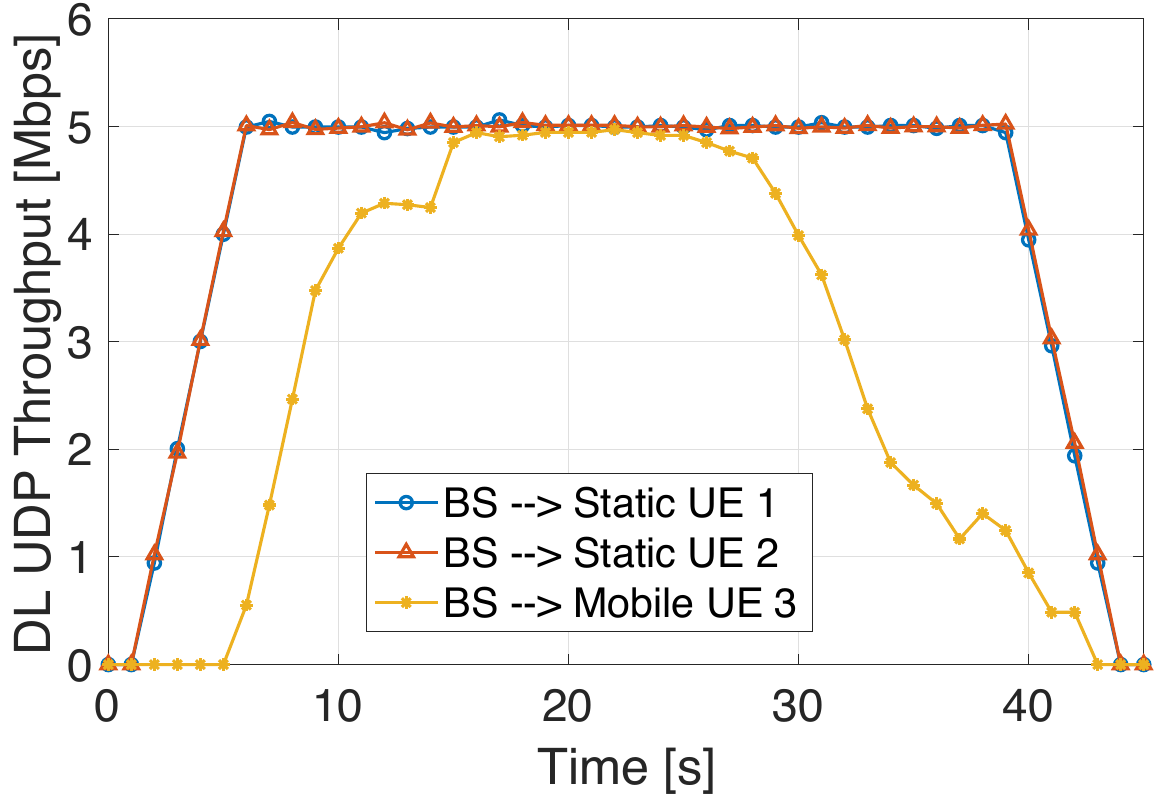}}
    \hfill
    \caption{\acrshort{udp} downlink throughput of the cellular use case on the Arena and Colosseum testbeds.}
    \label{chap2-fig:cellular-results-udp}
\end{figure}

Figure~\ref{chap2-fig:cellular-results-udp} shows the \gls{udp} downlink throughput for static (blue and orange lines), and mobile (yellow line) nodes on the Arena (Figure~\ref{chap2-fig:cellular-arena-results-udp}) and Colosseum (Figure~\ref{chap2-fig:cellular-colosseum-results-udp}) testbeds. 
We can notice similar trends and patterns exhibited on both testbeds. 
Specifically, the throughput of the static nodes remains stable around $5$\:Mbps in both Colosseum and Arena, where we notice a less stable behavior due to the use of over-the-air communications, and potential external interference.
%
%
As expected, the throughput of the mobile node---that starts from the top-left location shown in Figure~\ref{chap2-fig:cellular-nodes-singlebs} and travels to the right along the trajectory depicted with the green line in the figure---increases as the node gets closer to the \gls{bs} (where it reaches a $5$\:Mbps peak), and then decreases as the node gets farther away.

\begin{figure}[htbp]
    \centering
    \subfloat[Arena]{\label{chap2-fig:cellular-arena-results-tcp}\includegraphics[width=0.49\columnwidth]{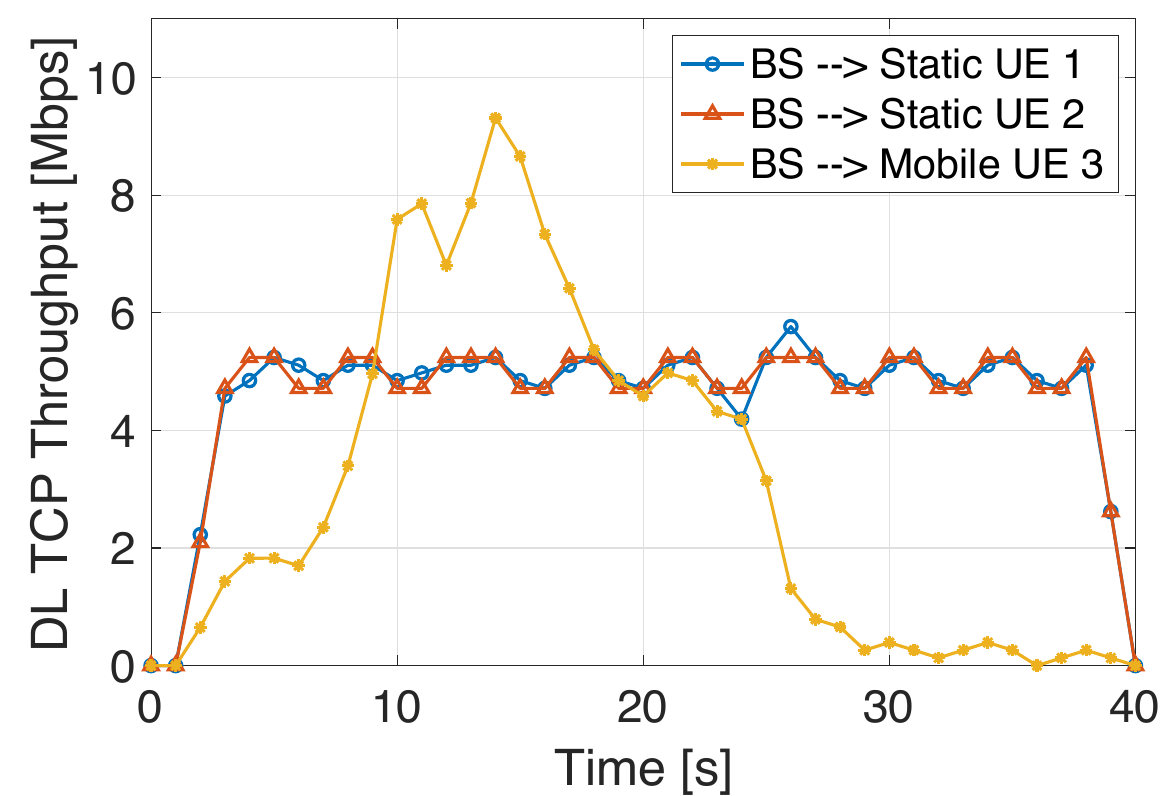}}
    \hfill
    \subfloat[Colosseum]{\label{chap2-fig:cellular-colosseum-results-tcp}\includegraphics[width=0.49\columnwidth]{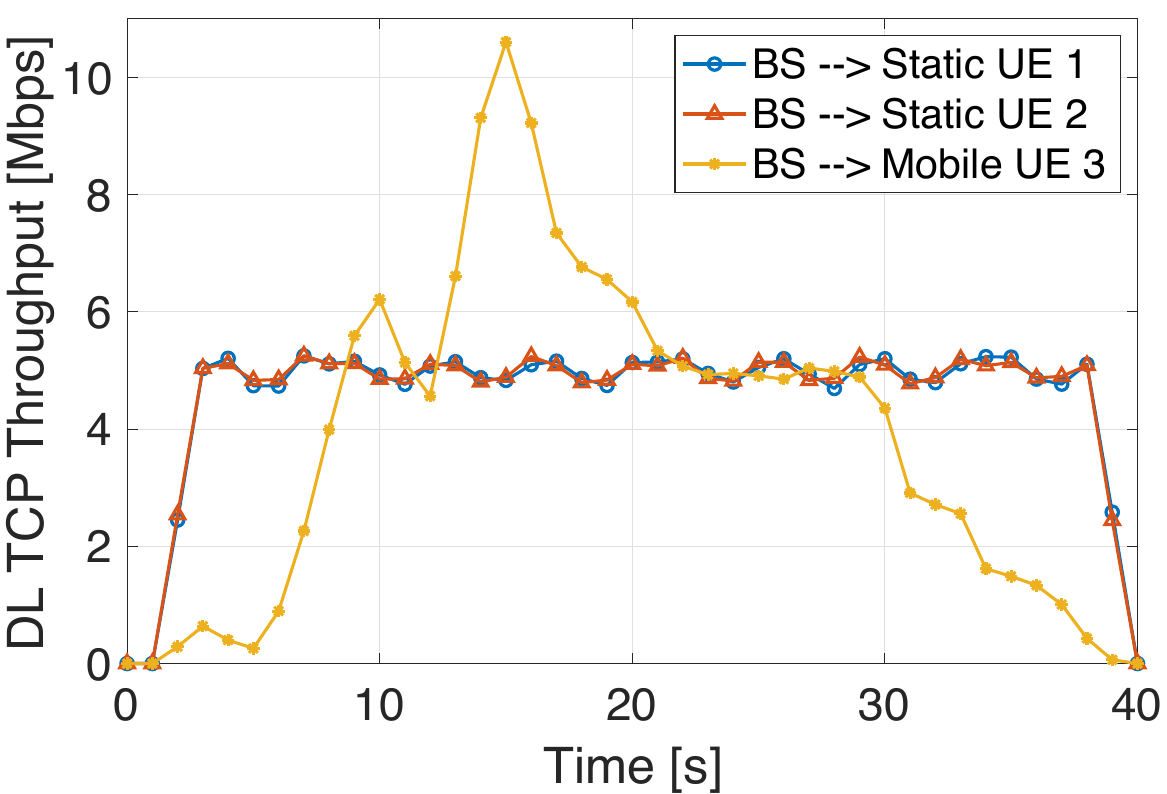}}
    \hfill
    \caption{\acrshort{tcp} downlink throughput of the cellular use case on the Arena and Colosseum testbeds.}
    \label{chap2-fig:cellular-results-tcp}
\end{figure}

\begin{figure}[htbp]
    \centering
    \subfloat[Arena]{\label{chap2-fig:sinr-cellular-arena-results-tcp}\includegraphics[width=0.49\columnwidth]{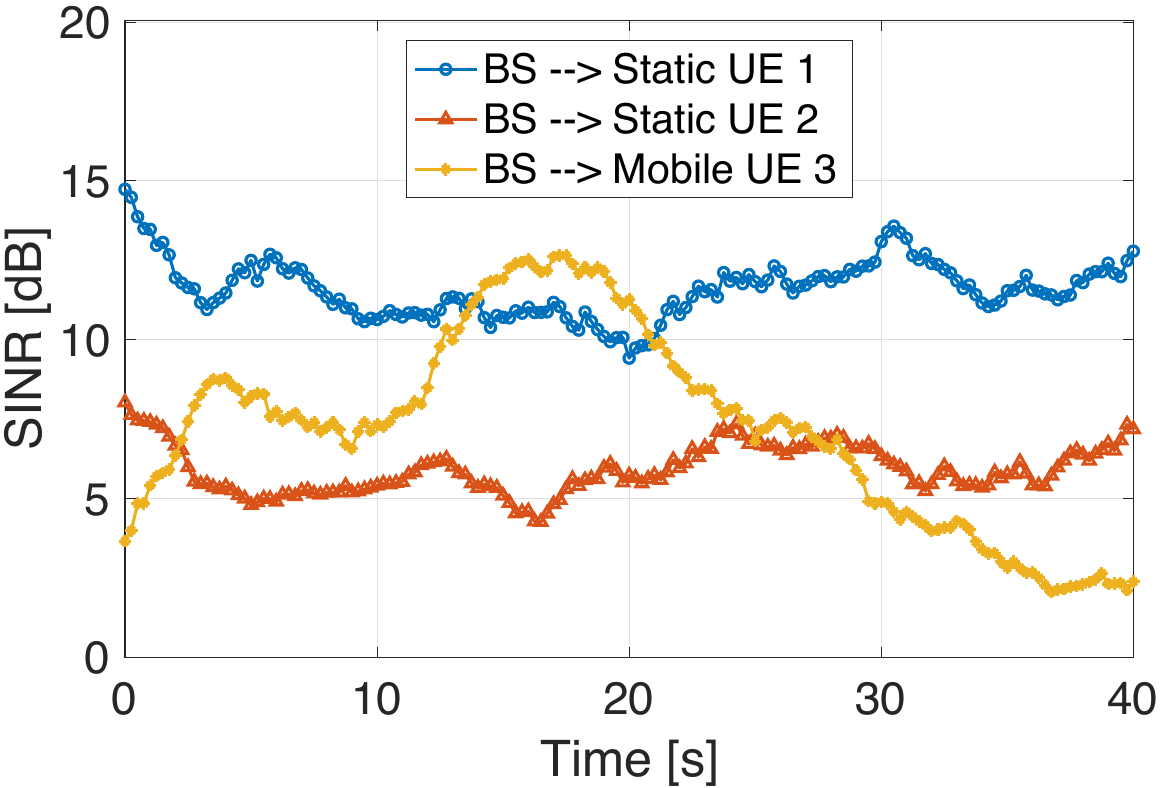}}
    \hfill
    \subfloat[Colosseum]{\label{chap2-fig:sinr-cellular-colosseum-results-tcp}\includegraphics[width=0.49\columnwidth]{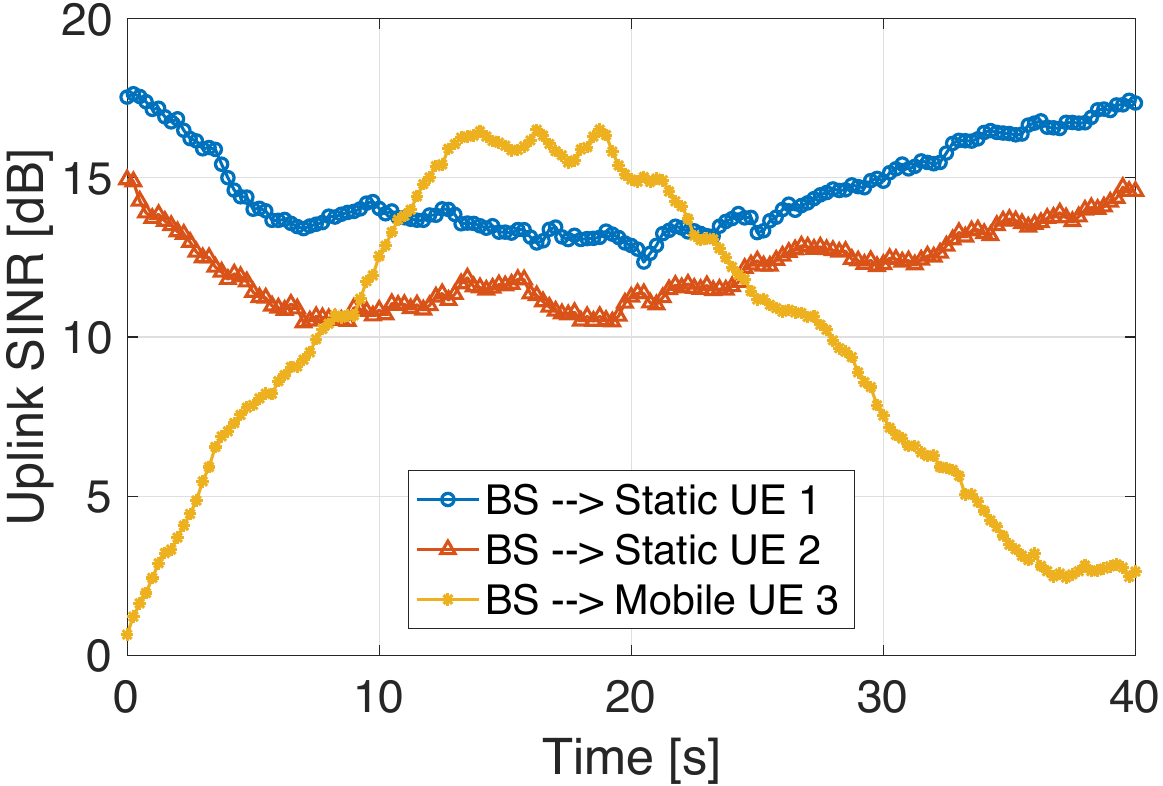}}
    \hfill
    \caption{\acrshort{tcp} \acrshort{sinr} of the cellular use case on the Arena and Colosseum testbeds.}
    \label{chap2-fig:sinr-cellular-results-tcp}
\end{figure}

Figure~\ref{chap2-fig:cellular-results-tcp} and Figure~\ref{chap2-fig:sinr-cellular-results-tcp} plot the \gls{tcp} downlink throughput and \gls{sinr} results of the second experiment for static (blue and orange) and mobile (yellow) nodes on Arena (Figure~\ref{chap2-fig:cellular-arena-results-tcp} and Figure~\ref{chap2-fig:sinr-cellular-arena-results-tcp}) and Colosseum (Figure~\ref{chap2-fig:cellular-colosseum-results-tcp} and Figure~\ref{chap2-fig:sinr-cellular-colosseum-results-tcp}).
%
Also in this use case, a similar pattern is clearly noticeable.
In particular, the two static nodes maintain a relatively stable throughput of the nominal $5$\:Mbps \gls{tcp} traffic on both testbeds with no apparent impact during the passage of the mobile \gls{ue} close to the static \glspl{ue}, which in the Arena case is moved manually.
This behavior is visible in the \gls{sinr} results, which show a decrease due to the created interference when the mobile node approaches each of the static \glspl{ue}.
The mobile node exhibits a similar trend with two high peaks in throughput
on both systems.
These peaks can be attributed to the \gls{tcp} protocol, which retransmits data to ensure delivery in case of packet failures as soon as the signal improves, resulting in higher application throughput values compared to the nominal $5$\:Mbps.
Specifically, each peak corresponds to the time right after the mobile device transitions close to static node 1 (around time $10$\:s), and then to static node 2 (around time $15$\:s).
Due to the increased interference, the mobile node loses more packets, which will eventually be retransmitted.
%

\begin{figure}[ht]
    \centering
    \includegraphics[width=0.7\columnwidth]{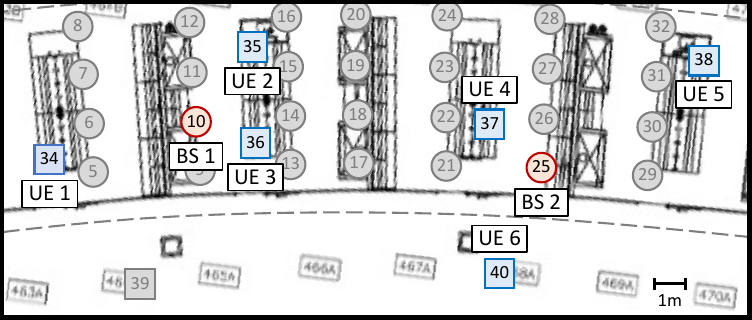}
    \caption{\blue{Location of the nodes in the multiple \acrshortpl{bs} cellular use-case, consisting of 2 \acrshortpl{bs} (BS 1, BS 2), and 6 static \acrshortpl{ue}, equally assigned to each \acrshort{bs}: UE 1, UE 2, and UE 3 for BS 1; UE 4, UE 5, and UE 6 for BS 2.}}
    \label{chap2-fig:cellular-nodes-multiplebs}
\end{figure}

\blue{\textbf{Multiple \glspl{bs}.} In the second cellular networking use case, we use a similar setup as for the first experiment by leveraging SCOPE and Xiaomi Redmi Go phones to deploy a twinned srsRAN protocol stack with two \glspl{bs} located in positions 10 (BS 1) and 25 (BS 2), as shown in Figure~\ref{chap2-fig:cellular-nodes-multiplebs}.
We assign three \glspl{ue} to each \gls{bs}, for a total of 6~\glspl{ue} (i.e., the number of smartphones currently at our disposal).
%
Specifically, UE 1, UE 2, and UE 3 are assigned to BS 1, while UE 4, UE 5, and UE 6 are assigned to BS 2.}

\begin{figure}[ht]
    \centering
    \includegraphics[width=.7\columnwidth]{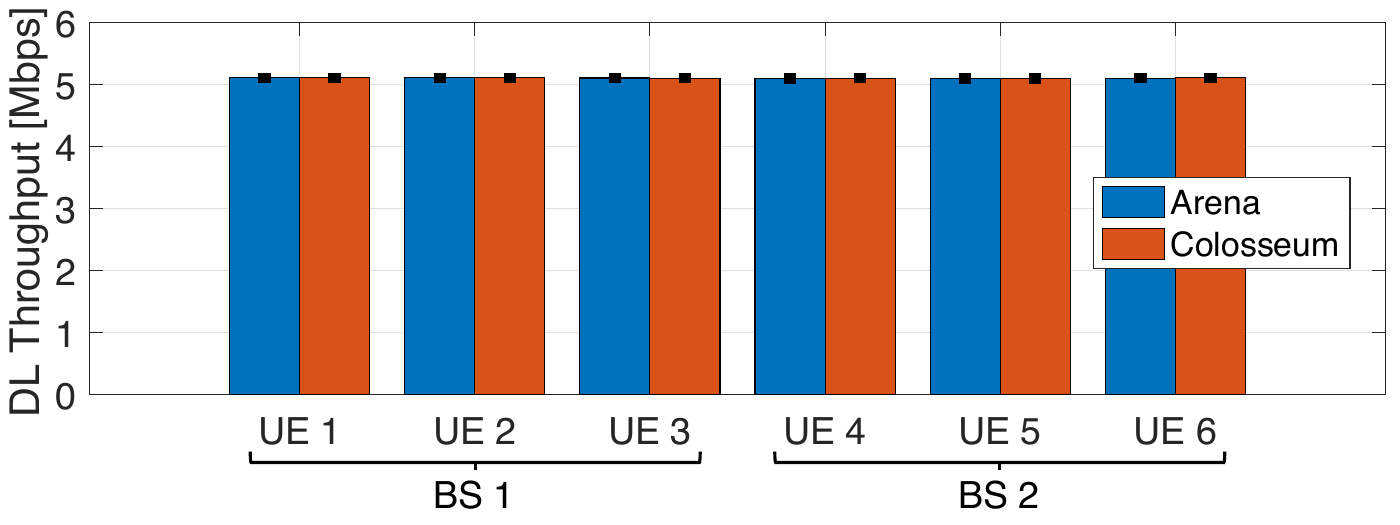}
    \caption{\blue{Downlink throughput average results and 95\% confidence interval of the second cellular networking use-case with 2 \acrshortpl{bs} and 6 \acrshortpl{ue}.}}
    \label{chap2-fig:th-cellular-multiplebs}
\end{figure}

\blue{Once each \gls{ue} has completed the attachment procedures, we initiate a continuous $5$\:Mbps UDP downlink traffic stream from each smartphone using iPerf towards its corresponding \gls{bs}. 
Figure~\ref{chap2-fig:th-cellular-multiplebs} shows the average downlink throughput results over a period of more than $30$\:minutes of data collection, along with 95\% confidence intervals.
These metrics are collected at the data-link layer at the \gls{bs} level through the SCOPE framework.
We observe that all \glspl{ue} are able to consistently ensure the $5$\:Mbps throughput with a very low 95\% confidence interval showing an average margin of error of $0.02$\:Mbps.
This demonstrates the accuracy of the digital twin representation, even in the presence of multiple \glspl{bs} and \glspl{ue}, showcasing the scalability of our \gls{dtmn}.
}

\textbf{\blue{Comparison.}} By analyzing the results after the \glspl{ue} have completed the attachment procedures, \blue{we can assess the similarity between the two testbeds by following}
\begin{equation}
    \label{chap2-eq:xcorr}
    \blue{
    \rho(k) = \frac{\sum_{n=1}^{N} (x(n) - \bar{x})(y(n+k) - \bar{y})}{\sqrt{\sum_{n=1}^{N} (x(n) - \bar{x})^2 \sum_{n=1}^{N} (y(n) - \bar{y})^2}},
    }
\end{equation}
\blue{which measures the normalized cross-correlation $\rho(k)$ between the Arena data sequence, denoted as $x(n)$, and shifted lagged copies of the Colosseum data sequence, denoted as $y(n)$, as a function of the lag $k$.
In this context, $\bar{x}$ and $\bar{y}$ represent the means of the input sequences and $N$ is their length.
If $x(n)$ and $y(n)$ have different lengths, zeros are appended to the end of the shorter vector to ensure both sequences have the same length $N$.}

\blue{Table~\ref{chap2-table:cellular-corr-results} presents the normalized cross-correlation results and their averages for throughput and \gls{sinr} of each \gls{ue}, considering the maximum values between 10~lags, of the \gls{udp} and \gls{tcp} \blue{single \gls{bs}} use-case experiments.
The multiple \glspl{bs} experiments would yield similar comparison results.}
\begin{table}[htbp]
    \centering
    \caption{Normalized cross-correlation results, and their averages, considering the maximum values between 10 lags for the single \acrshort{bs} cellular experiment.}
    \label{chap2-table:cellular-corr-results}
    \scriptsize
    \setlength{\tabcolsep}{4pt}
    \begin{tabular}{lcccc}
        \toprule
        \textbf{Metric} & \textbf{Static UE 1} & \textbf{Static UE 2} & \textbf{Mobile UE 3} & \textbf{Average} \\
        \midrule
        UDP Throughput & 0.986 & 0.998 & 0.999 & 0.994 \\
        TCP Throughput & 0.937 & 0.998 & 0.998 & 0.978 \\
        TCP SINR & 0.997 & 0.994 & 0.982 & 0.991 \\
        \bottomrule
    \end{tabular}
\end{table}

\noindent
We can observe a very high similarity between the two testbeds in all use-case experiments, with individual \gls{ue} values consistently above 0.93 and an average exceeding 0.97 for each use-case metric.
It is worth noting, as observed in Figure~\ref{chap2-fig:sinr-cellular-results-tcp}, that the \gls{sinr} results in Arena are quite similar from \gls{ue} 1 and \gls{ue} 3, and they present a fixed difference of about $5$\:dB compared to Colosseum for \gls{ue} 2.
%
%
This difference can mainly be attributed to the additional and uncontrolled interference and impairments of a real-world \gls{rf} environment, as well as the different power levels between a simulated \gls{ue} and a real smartphone.
However, these variances can be compensated for in the \gls{dt} by adjusting factors such as the node gains at the transmitter and receiver, as well as by adding stochastically representative interference models to the channel that represents the real-world behavior more closely.
%
%
These findings confirm the capabilities of the \gls{dt} to perform emulated cellular experiments that closely replicate the behavior of real-world setups and environments, even in the presence of mobile nodes.

\subsubsection{\blue{Wi-Fi Jamming}}
Adversarial jamming has continuously plagued the wireless spectrum over the years with the ability to disrupt, or fully halt, communications between parties.
%
%
While there are potential solutions to specific types of jamming, due to the open nature of wireless communication, this kind of attack continues to find ways to be effective.
However, the development of new techniques to counter this attack is not always straightforward, as even experimenting with possible solutions requires complying with strict \gls{fcc} regulations~\cite{noauthor_jammer_2011}.
%
%
Even though some environments allow for jamming research, e.g., anechoic chambers or Faraday cages, these setups can hardly capture the characteristics and scale of real-world network deployments.
To bridge this gap, a \gls{dt} environment---such as the Colosseum wireless network emulator---could be fundamental in further developing techniques for jamming mitigation research 
as shown in our previous work in~\cite{robinson2023esword} where we implement jamming software within Colosseum to test the impact that jamming signals have within a cellular scenario as well as compare real-world and \gls{dt} throughput results.

\begin{figure}[ht]
    \centering
    \includegraphics[width=0.7\columnwidth]{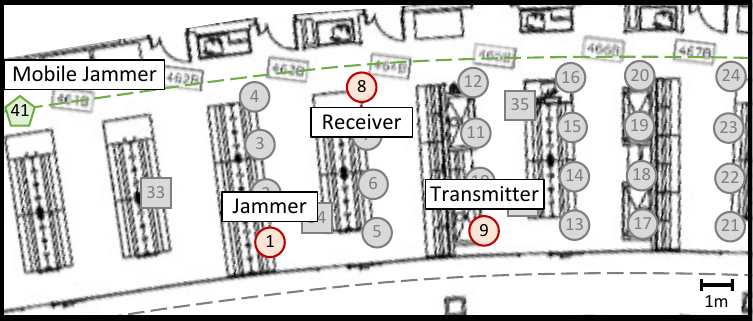}
    \caption{Location of the nodes in the jamming experiment, consisting of three static (1, 8, 9) and one mobile (41).}
    \label{chap2-fig:jamming-map}
\end{figure}

\begin{figure}[t]
    \centering
    \subfloat[Narrowband]{\label{chap2-fig:jamming-narrowband-static-results}\includegraphics[width=0.49\columnwidth]{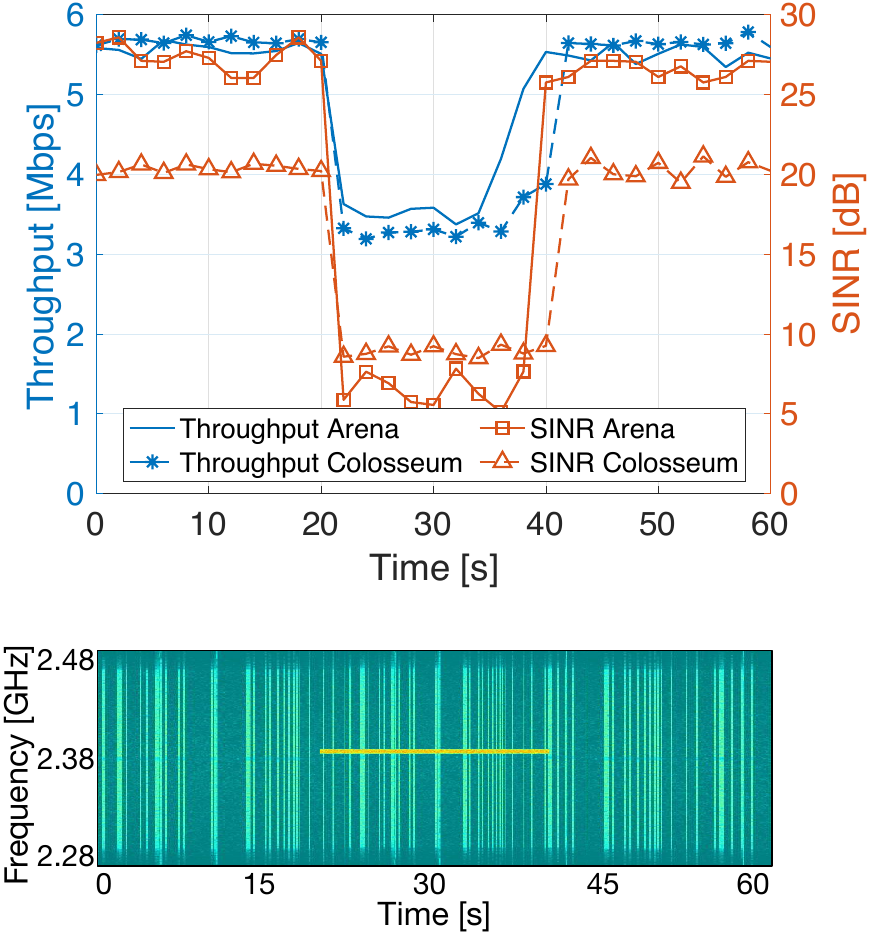}}
    \hfill
    \subfloat[Wideband]{\label{chap2-fig:jamming-wideband-static-results}\includegraphics[width=0.49\columnwidth]{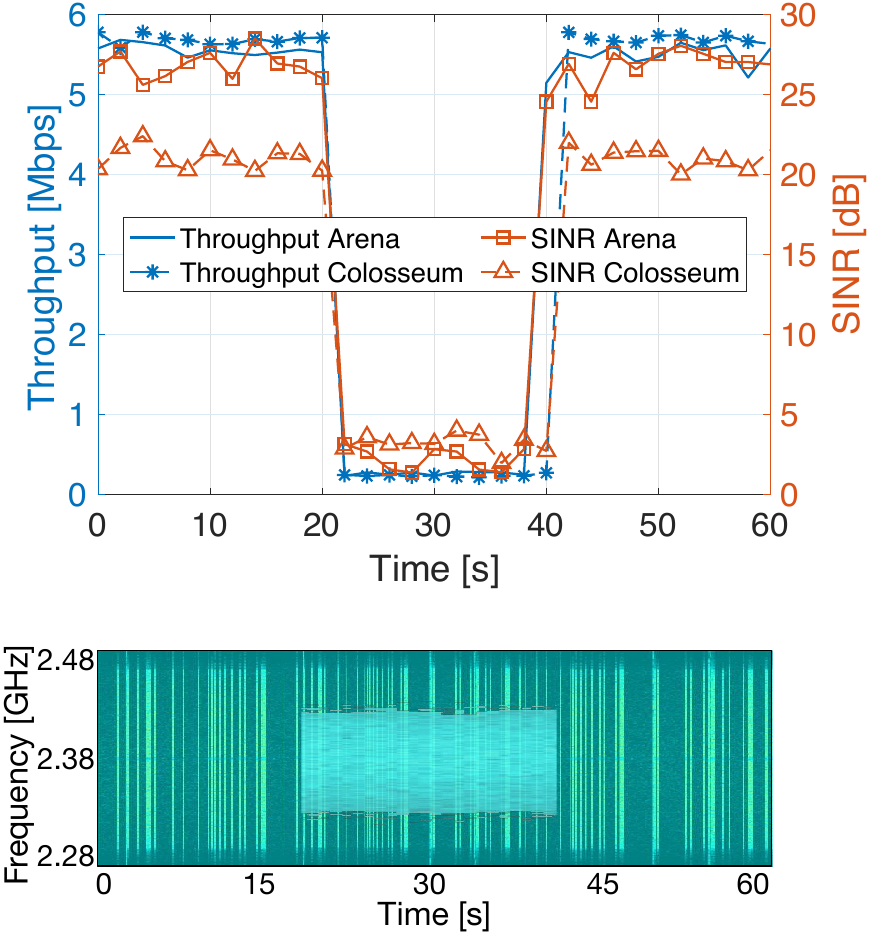}}
    \hfill
    \caption{Throughput and \acrshort{sinr} results on the Arena and Colosseum testbeds of the jamming experiments for the narrowband and wideband use cases. The spectrogram is shown for both forms of jamming, showing the wideband and narrowband signals over a channel.} 
    \label{chap2-fig:jamming-static-results}
    \vspace{-10pt}
\end{figure}

Here, we leverage the GNU Radio-based IEEE 802.11 implementation~\cite{bloessl_ieee_2022} to deploy two Wi-Fi nodes (\gls{tx} and \gls{rx}) communicating over a $20$\:MHz spectrum on the Arena testbed~\cite{bertizzolo2020comnet}.
Additionally, we leverage GNU Radio to deploy a jammer (both stationary and mobile) that transmits Gaussian noise signals to hamper the correct functioning of our Wi-Fi network.
Our setup can be seen in Figure~\ref{chap2-fig:jamming-map}.
For the sake of fairness in the transmitted signals, in the stationary case, we deployed our nodes so that the Wi-Fi transmitter and jammer are at the same distance from the Wi-Fi receiver.
We consider two common forms of static jamming: (i)~jamming through narrowband signals (shown in Figure~\ref{chap2-fig:jamming-narrowband-static-results}); and (ii)~jamming through wideband signals (Figure~\ref{chap2-fig:jamming-wideband-static-results}).

The first type of jamming only occupies a small portion of the Wi-Fi bandwidth (i.e., $\sim$$156$\:kHz), resulting in a minimal displacement of the Wi-Fi signals.
On the contrary, the latter covers half of the spectrum used by the Wi-Fi nodes (i.e., $10$\:MHz), causing larger disruptions in the network.

\blue{\textbf{Static Jamming.}} Figure~\ref{chap2-fig:jamming-static-results} evaluates how narrowband and wideband stationary jammers impact the throughput and \gls{sinr} of a Wi-Fi network in the real and \gls{dt}-based scenarios.
%
%
In this experiment, the Wi-Fi nodes communicate for $60$ seconds, and the jammer starts transmitting at second $20$ for a duration of $20$ seconds.
Specifically, Figure~\ref{chap2-fig:jamming-narrowband-static-results} shows Wi-Fi throughput and \gls{sinr} for the narrowband jamming experiment in both the real-world and \gls{dt}, while the wideband jamming experiment throughput and \gls{sinr} results as perceived by the Wi-Fi nodes are shown in Figure~\ref{chap2-fig:jamming-wideband-static-results}.

By looking at the narrowband jamming case, we notice that in the real-world experiment, the Wi-Fi throughput achieves between $5$ and $6$\:Mbit/s when there is no jammer (Figure~\ref{chap2-fig:jamming-narrowband-static-results}).
Once the jammer starts (at second 20), we notice a rapid decrease in the throughput (i.e., between 37\% and 43\% decrease).
%
The wideband jammer (Figure~\ref{chap2-fig:jamming-wideband-static-results}), instead, has a more severe impact on the Wi-Fi throughput, causing a performance drop between 94\% and 96\% (with the throughput achieving values between $220$ and $290$\:kbit/s).
In both narrowband and wideband cases, we notice that the behavior obtained in the \gls{dt} is consistent with that of the real-world scenario.
%
%
%
%
Analogous trends can be seen for the \gls{sinr} of both signal types, where the narrowband jammer causes an \gls{sinr} decrease of approximately $20$\:dB (i.e., $\sim$77\% decrease), while the wideband jammer of approximately $25$\:dB (i.e., $\sim$92\% decrease) in the real-world scenario.
Similarly to the previous case, results are consistent with those of the \gls{dt}.

\begin{figure}[htbp]
    \centering
\begin{subfigure}{0.49\linewidth}
    \centering
    \includegraphics[width=\linewidth]{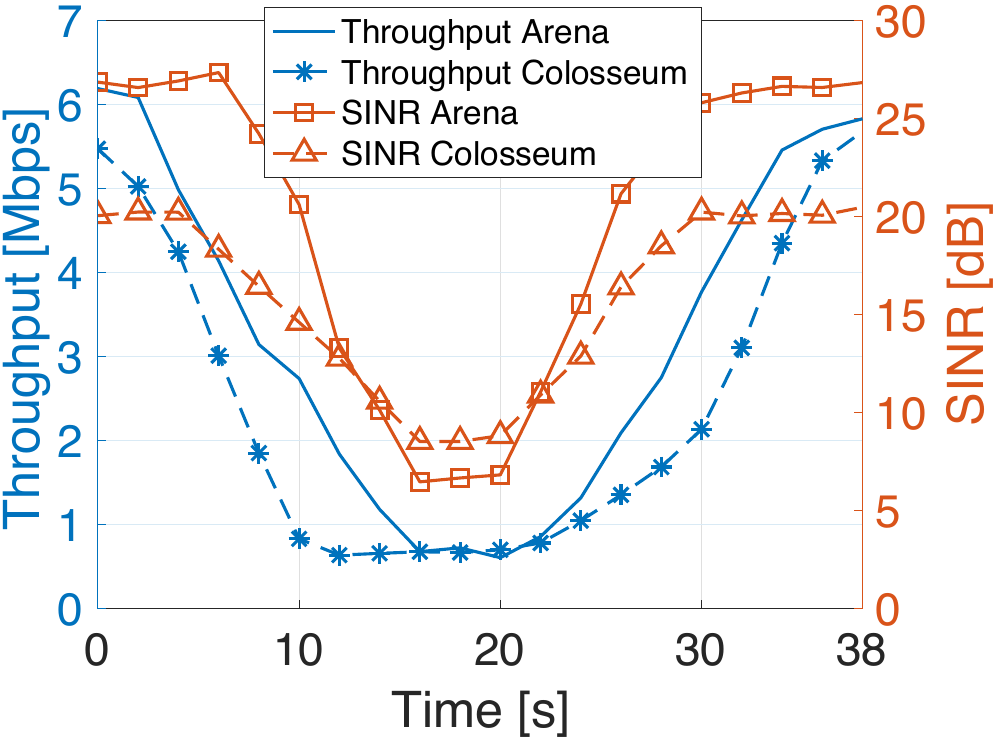}
    \setlength\abovecaptionskip{0cm}
    \caption{\centering Narrowband}
    \label{chap2-fig:jamming-narrowband-mobile-results}
\end{subfigure}
    \hfill
\begin{subfigure}{0.49\linewidth}
    \centering
    \includegraphics[width=\linewidth]{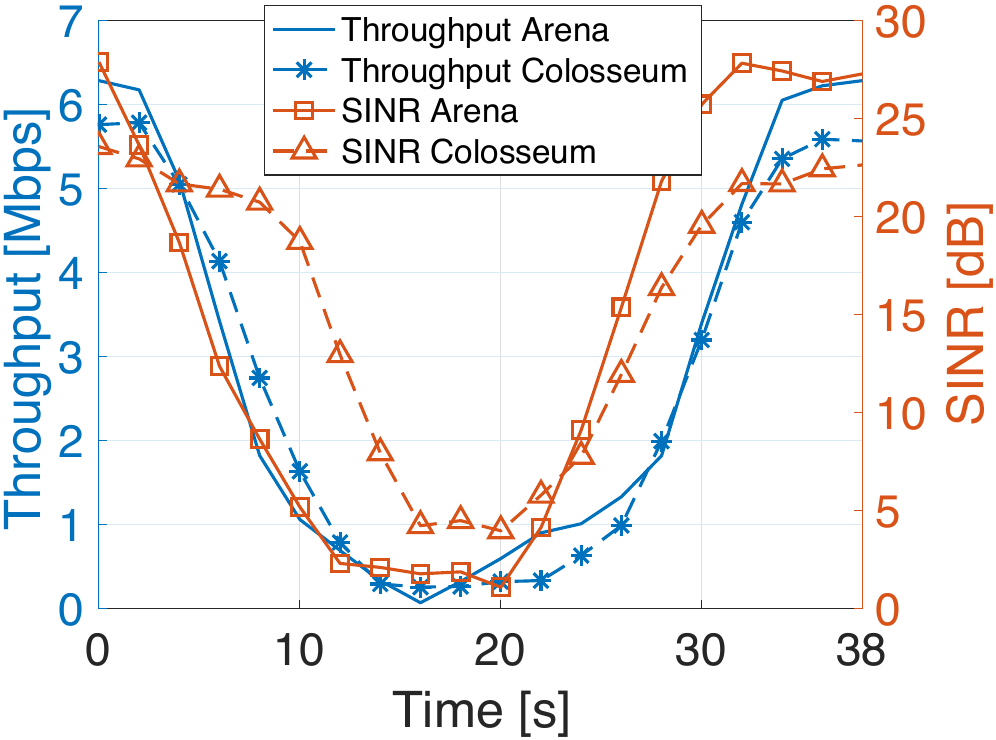}
    \setlength\abovecaptionskip{0cm}
    \caption{\centering Wideband}
    \label{chap2-fig:jamming-wideband-mobile-results}
\end{subfigure}
    \caption{\justifying \blue{Impact of a moving jammer for narrowband (a) and wideband (b) use cases on the throughput and \acrshort{sinr} of Wi-Fi nodes on Arena and Colosseum testbed.}}
    \label{chap2-fig:jamming-mobile-results}
\end{figure}

\blue{\textbf{Mobile Jamming.}} Now, we evaluate the impact that a \blue{narrowband and wideband} mobile jammer (node~41 in Figure~\ref{chap2-fig:jamming-map}) moving at pedestrian speed has on the Wi-Fi throughput.
Wi-Fi nodes are located as in the previous case, i.e., nodes~8 and~9 in the figure.
%
%
Results are shown in Figure~\ref{chap2-fig:jamming-mobile-results}.
As expected, the impact of the jamming signal on the Wi-Fi throughput varies as the jammer moves closer or farther from the Wi-Fi receiver, \blue{and it also depends on the type of jamming, i.e., narrowband vs. wideband}.
Specifically, as the jammer gets closer to the Wi-Fi nodes (i.e., seconds $5$ to $30$) \blue{in the narrowband case}, we observe a $\sim$90\% decrease in the Wi-Fi throughput in both real-world and \gls{dt} scenarios (see Figure~\ref{chap2-fig:jamming-narrowband-mobile-results}).
\blue{A comparable decrease can be observed in the \gls{sinr} as well, where we notice a clear drop in both the over-the-air Arena case and the \gls{dt} of about $\sim$65\% in a lookalike trend}.
\blue{Following a similar pattern, in the wideband case, we observe a more pronounced drop as the jammer approaches (i.e., seconds $5$ to $30$), reaching peaks near 100\% drops in throughput and \gls{sinr} in the closest locations with the nodes at around second $15$.}

\textbf{\blue{Comparison.}} As for the cellular experiment, Table~\ref{chap2-table:jamming-corr-results} shows the normalized cross-correlation results and their averages, considering the maximum values between 10 lags, for each jamming experiment.
We observe a strong similarity between the two testbeds in both static and mobile experiments, with individual values consistently above \blue{0.93}.
\begin{table}[htbp]
    \centering
    \caption{Normalized cross-correlation results, and their averages, considering the maximum values between 10 lags for each jamming experiment.}
    \label{chap2-table:jamming-corr-results}
    \scriptsize
    \setlength{\tabcolsep}{4pt}
    \begin{tabular}{lccc}
        \toprule
        \textbf{Metric} & \textbf{Narrowband} & \textbf{Wideband} & \textbf{Average} \\
        \midrule
        Static Throughput & 0.996 & 0.982 & 0.989 \\
        Static \gls{sinr} & 0.986 & 0.984 & 0.985 \\
        Mobile Throughput & 0.982 & \blue{0.993} & \blue{0.988} \\
        Mobile \gls{sinr} & 0.993 & \blue{0.935} & \blue{0.964} \\
        \bottomrule
    \end{tabular}
\end{table}

Overall, considering all sample experiments in this work, our \gls{dtmn} is able to achieve an average similarity of \blue{0.987} in throughput and \blue{0.982} in \gls{sinr}.
These results prove the ability of our system to properly emulate various use case experiments with different protocol stacks and scenarios.

\section{Related Work}
\label{chap2-sec:relatedwork}

The concept of \gls{dt} is rapidly gaining momentum in both industry and academia.
Initial approaches showcase the use of \gls{dt}s for industry~4.0~\cite{rolle2020architecture}, and to assist design, assembly, and production operations in the manufacturing process~\cite{Tao2018dt}.
A comprehensive literature review on \gls{dt}-related applications in manufacturing is provided by Kritzinger et al.\ in~\cite{kritzinger2018dt}.

Recently, researchers and practitioners have started to apply the concept of \gls{dt} to the wireless ecosystem due to the potential of digitalization processes, and easier integration and monitoring of interconnected intelligent components, as Zeb et al.\ discuss in~\cite{zeb2022industrial}.
Nguyen et al.\ theoretically discuss how \gls{dt}s can enable swift testing and validation on real-time digital replicas of real-world 5G~cellular networks~\cite{nguyen2021dt}, while Khan et al.\ provide the architectural requirements for 5G-oriented \gls{dt}s, mentioning them as key components for the development of 6G~networks~\cite{latif2022dte}.
He et al.\ leverage the \gls{dt}s and mobile edge computing in cellular networks to enhance the creation of digital models affected by the straggler effect of user devices in a \gls{fl} process~\cite{he2022resource}.
%
Lu et al.\ incorporate \gls{dt}s into wireless networks to mitigate long and unreliable communications among users and \gls{bs} and define a permissioned blockchain-based \gls{fl} framework for edge computing~\cite{lu2021low}.
Zhao et al.\ combine 
\gls{dt}s with software-defined vehicular networks to learn, update, and verify physical environments to foresee future states of the system while improving the network performance~\cite{zhao2020idt}.

Overall, the above works agree on the potential of \gls{dt}s in: (i)~assessing the network performance; (ii)~creating realistic and accurate system models; (iii)~predicting the impact of changes in the deployment environment; and (iv)~reacting and optimizing the performance of the network.

The works most similar to our \gls{cast} toolchain in modeling and simulating channel characteristics are those of Patnaik et al.~\cite{patnaik2014implementation}, Ju and Rappaport~\cite{ju2018simulating}, Bilibashi et al.~\cite{bilibashi2020dynamic}, and Oliveira et al.~\cite{oliveria2019ray}.
Specifically, Patnaik et al.\ compare the response of \gls{fir} filters with their simulated counterpart~\cite{patnaik2014implementation}, while Ju and Rappaport devise a technique to improve the representation of channel impairments and variations for adaptive antenna algorithms in a mmWave channel simulator~\cite{ju2018simulating}.
Bilibashi et al., and Oliveira et al., instead, leverage ray-tracing approaches to include mobility in the emulated channels in~\cite{bilibashi2020dynamic} and~\cite{oliveria2019ray}, respectively.
However, these works only target specific use cases, and they cannot model generic scenarios and deployments, as our \gls{cast} toolchain does.

Finally, to the best of our knowledge, there are no practical works that encompass all the various building blocks of a \gls{dt} system, from channel characterization and modeling to large-scale experimentation on a \gls{dt}, to real-world validation on an \gls{ota} testbed, as we carry out in this work.

\section{Conclusions}
\label{chap2-sec:conclusion}



In this chapter, we applied the concept of \glspl{dt} to wireless communication systems, presented Colosseum---the world's largest wireless network emulator---as an ideal candidate for realizing \glspl{dtmn}, and introduced \gls{cast}, an open toolchain to create and validate realistic channel scenarios.
We validated \gls{cast} through experiments on a lab setup and on Colosseum, showing that it achieves up to $20\:\mathrm{ns}$ accuracy in sounding the \gls{cir} tap delays and $0.5\:\mathrm{dB}$ accuracy in measuring the tap gains.
We then used \gls{cast} to digitize and characterize a variety of emulation scenarios, including a ray-tracing-based \gls{v2x} deployment and the Arena testbed, and validated the twinning process through cellular, Wi-Fi, and jamming experiments, achieving up to $0.987$ similarity in throughput and $0.982$ in \gls{sinr} between emulated and real-world results. This proves that \glspl{dt} can closely reproduce real-world setups, enabling users to run meaningful experiments in a highly realistic environment.

Connecting to the four key research challenges outlined in Chapter~\ref{chap:intro}, this chapter: (i) demonstrated automated scenario creation and validation through the \gls{cast} toolchain (\emph{deployment}); (ii) validated \glspl{dt} that closely reproduce real setups via paired experiments between physical and digital platforms (\emph{realism}); and (iii) released open-source tools (e.g., \gls{cast}), methodologies, and scenarios to ease experimentation by other researchers (\emph{usability}).
The fourth challenge, namely \emph{use cases}, is addressed in Chapter~\ref{chap:3}, where we show how Colosseum and \glspl{dt} enable research across various areas, such as spectrum sharing, \gls{ai}-driven modeling, and synthetic data generation.

\chapter{Spectrum Sharing and AI-Driven Modeling on Digital Platforms}
\label{chap:3}

\section{Introduction}


Chapter~\ref{chap:2} has established \glspl{dt} and large-scale wireless emulation platforms, such as Colosseum, together with \gls{cast} for the creation and validation of these \glspl{dt}, as valuable tools for the research community.
They provide a controlled, repeatable, and reproducible environment for wireless experimentation in which to explore complex behaviors that would otherwise be impractical, expensive, unsafe, or constrained by regulatory requirements for \gls{ota} testing in real-world settings.
This chapter demonstrates the \emph{use case} dimension of \gls{dt} platforms, addressing the fourth research challenge introduced in Chapter~\ref{chap:intro} and presenting three classes of problems enabled by Colosseum, building on~\cite{villa2023wintech,villa2023demointelligent,saeizadeh2024camad,saeizadeh2025airmap,basaran2025gentwin}: (i) spectrum sharing and incumbent signal detection; (ii) \gls{ai}-driven radio-map generation; and (iii) generative models for synthetic \gls{rf} data.

Section~\ref{chap3.1-sec:spectrum} presents a framework for spectrum sharing between cellular networks and radar systems in the \gls{cbrs} band, showing how commercial waveforms can be twinned on Colosseum and how data can be collected on the emulator to train a \gls{cnn}-based detector for incumbent radar transmissions.
Section~\ref{chap3.2-sec:airmap} addresses the challenge of real-time channel modeling for \gls{dt}-based applications, presenting AIRMap, a deep-learning framework for ultra-fast radio-map estimation, and demonstrating its integration with Colosseum for validation.
Section~\ref{chap3.3-sec:gentwin} explores the use of generative \gls{ai} to augment \gls{rf} datasets for O-RAN applications, describing Gen-TWIN and its dataset generation methodology using Colosseum and another testbed to enable model training that generalizes across diverse channel conditions.
Finally, Section~\ref{chap3.4-sec:conclusion} summarizes the research impact enabled by \gls{dt} platforms and discusses their limitations, motivating the transition to \gls{ota} testing in the following chapters.

\section{Spectrum Sharing and Incumbent Signal Detection}
\label{chap3.1-sec:spectrum}

The increasing demand for wireless spectrum is driving the adoption of spectrum-sharing techniques, in which multiple wireless systems from different vendors and with different objectives coexist in the same frequency bands.
This presents significant challenges, particularly when commercial networks must share the spectrum with incumbent systems operating mission-critical communication links, such as nautical and aerial fleet radars~\cite{guiyang2022artificial}.
One prominent example is the potential interference of \gls{5g} \glspl{ran} on incumbent radar communications in the \gls{cbrs} band, ranging from $3.55$~GHz to $3.7$~GHz, where U.S. military radar systems operate along coastal areas. According to federal regulations, dynamic access to the \gls{cbrs} spectrum is permitted, but it requires thorough study, research, and the development of mitigation strategies to ensure the reliable operation of both systems, i.e., seamless communication in cellular networks while avoiding harmful interference to incumbent radar systems~\cite{caromi2018detection}.

\gls{ai}/\gls{ml}-based solutions have emerged as promising tools to address this challenge, thanks to their ability to learn complex patterns, generalize across diverse channel conditions, and make real-time decisions~\cite{baldesi2022charm}.
In the context of Open \gls{ran}, these \gls{ai}/\gls{ml} agents can be deployed in the \glspl{ric} proposed by O-RAN~\cite{polese2023understanding}, i.e., as xApps and rApps, or at the \gls{bs} directly as dApps~\cite{doro2022dapps}.
However, these methods require abundant high-quality data to train effective \gls{ml} models. Gathering diverse, representative datasets that capture real-world scenarios can be time-consuming and resource-intensive, but it is crucial for achieving accurate, robust performance of \gls{ai} algorithms.
Prior work generates datasets that are not always able to capture high-fidelity, diverse environments or leverage synthetic waveforms that are not representative of commercial radios, resulting in impractical \gls{ai} models whose performance substantially degrades when deployed in the real world~\cite{gao2019deep,wang2019data,zheng2020spectrum}.
In the context of incumbent radios and communication links, such as nautical and aerial fleets, safety is of paramount importance, and non-carefully planned real-world experiments can endanger critical operations.
As discussed in Chapter~\ref{chap:2}, high-fidelity emulation platforms and \glspl{dt}, such as Colosseum, allow researchers to simulate interference scenarios realistically and evaluate their impact without posing any actual risk to operational systems.

In this section, we develop a framework to emulate a spectrum-sharing scenario with cellular and radar nodes implemented in a high-fidelity \gls{dt} system on Colosseum~\cite{bonati2021colosseum}.
We create a scenario emulating the coastal environment of Waikiki Beach in Honolulu, Hawaii, where a \gls{bs} operating in the \gls{cbrs} band must detect radar transmissions from a ship moving in the North Pacific Ocean.
We collect \gls{iq} samples from radar and cellular communications and train a \gls{cnn} that can be deployed as a dApp to detect the presence of radar signals and notify the \gls{ran} to vacate the bandwidth and eliminate interference with the incumbent radar communications.
Our experimental results show an average accuracy of 88\%, with accuracy above 90\% in \gls{snr} regimes above $ 0$~dB and in \gls{sinr} regimes above $-20$~dB. Through timing experiments, we also observe an average detection time of $137$~ms.
This demonstrates the effectiveness of our system in detecting in-band interference in the \gls{cbrs} band, as it complies with both maximum timing requirements ($60$~s) and accuracy (99\% within the $60$~s time window)~\cite{goldreich2016requirements}.

The remainder of this section is organized as follows.
Section~\ref{chap3.1-subsec:twinning} details the waveform twinning methodology and radar characterization.
Section~\ref{chap3.1-subsec:rf-scenario} describes the spectrum-sharing scenario for the coexistence of cellular and radar signals.
Section~\ref{chap3.1-sec:radardetection} presents the intelligent radar detection approach, while Section~\ref{chap3.1-sec:results} discusses the experimental results.

\subsection{Waveform Twinning and Radar Characterization}
\label{chap3.1-subsec:twinning}

\subsubsection{Waveform Twinning}
\begin{figure}[htb]
    \centering
    \includegraphics[width=\columnwidth]{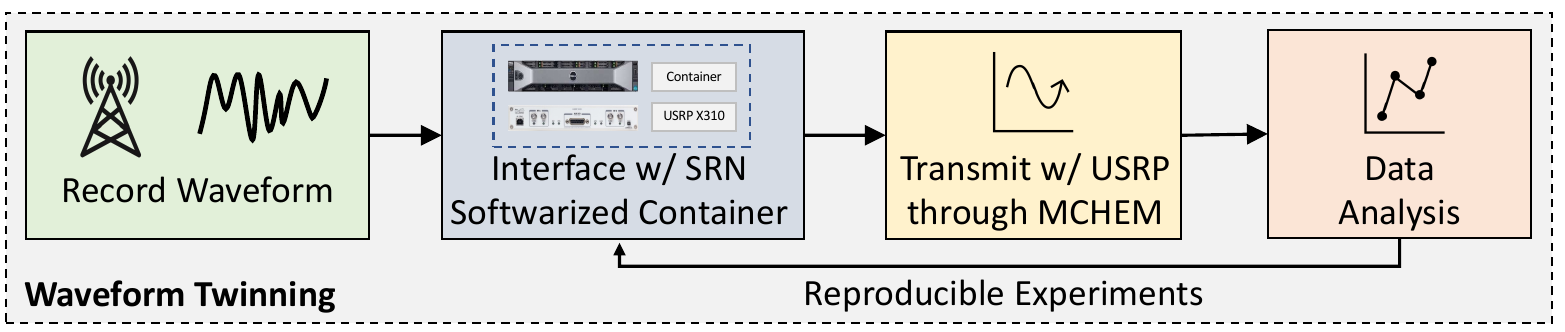}
    \caption{Waveform twinning on Colosseum.}
    \label{chap3.1-fig:waveformblock}
\end{figure}

As introduced in Chapter~\ref{chap:2}, Colosseum enables the deployment of softwarized protocol stacks through \gls{lxc} containers.
This capability extends to twinning arbitrary waveforms, including commercial radar signals, allowing researchers to experiment with spectrum-sharing scenarios in a controlled environment.
This process is shown at a high level in Figure~\ref{chap3.1-fig:waveformblock}. First, the waveform is either recorded from a real-world transmission (e.g., radar, Wi-Fi, or cellular transmission) or synthetically generated.
The waveform is imported on Colosseum and interfaced with the softwarized \gls{lxc} container running on Colosseum \gls{srn}.
It is then transmitted by the \gls{srn} USRP to the other nodes of the experiment through \gls{mchem}.
At the end of the experiment, data is collected and analyzed for post-processing purposes.
Finally, it is worth noting that this procedure can be repeated on the Colosseum for fine-tuning designed user solutions and their validation, thus allowing reproducible experiments to be carried out on the testbed.

\subsubsection{Radar Characterization}
\label{chap3.1-sec:radar}

Radar systems leverage reflections of \gls{rf} electromagnetic signals from a target to infer information on such a target~\cite{mahafza2017introduction}.
Typical information may include detection, tracking, localization, recognition, and composition of the target, which may include aircraft, ships, spacecraft, vehicles, astronomical bodies, animals, and weather phenomena.
Even though radar's primary uses were mainly related to military applications, nowadays this technology is commonly used in other areas, such as weather forecasting and automotive applications.

%

\begin{figure}[htb]
    \centering
    \begin{subfigure}{0.49\linewidth}
        \includegraphics[width=\linewidth]{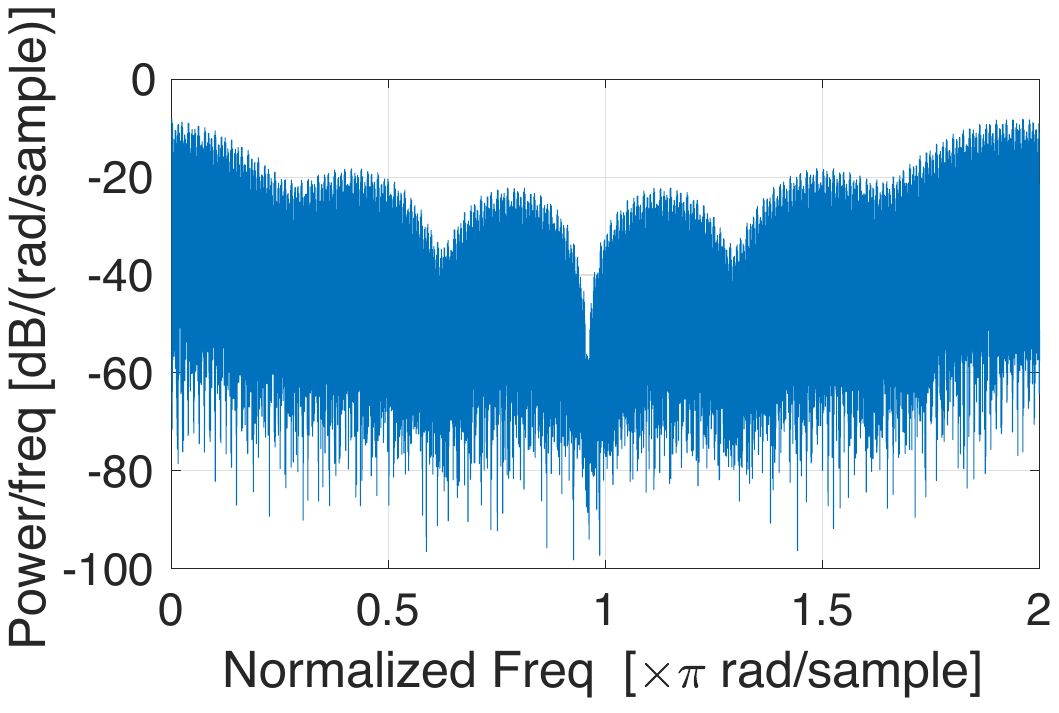}
        \caption{Radar \acrshort{psd}}
        \label{chap3.1-fig:radarpsd}
    \end{subfigure}
    \begin{subfigure}{0.49\linewidth}
        \includegraphics[width=\linewidth]{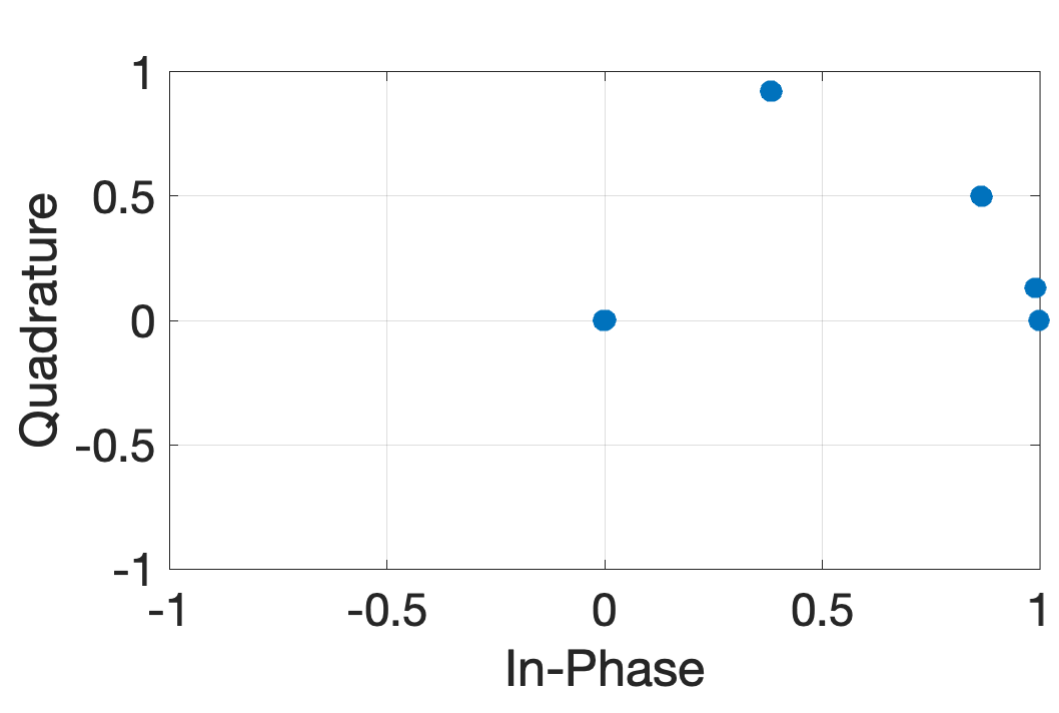}
        \caption{Radar constellation}
        \label{chap3.1-fig:radarconst}
    \end{subfigure}
    \caption{Radar characterization with PSD and constellation plots.}
    \label{chap3.1-fig:radarchar}
\end{figure}

In this work, we leverage a weather radar that combines techniques typical of continuous-wave radars, e.g., pulse-timing to compute the distance of the target, and of pulse radars, like the Doppler effect of the returned signal to establish the velocity of the moving target~\cite{doviak2006doppler}.
Note that similar considerations can be applied to any other radar or waveform type, and the radar signal considered in this work is a use-case study (without loss of generality) to showcase Colosseum capabilities.
Our radar operates in the S-Band, typically located within the $[3.0, 3.8]$~GHz frequency range.
The signal has been synthetically generated as a collection of \gls{iq} samples and timestamps, with a sampling rate of $6$\:MS/s and 106657 sampling points for a total duration of $17.8$\:ms. Figure~\ref{chap3.1-fig:radarchar} shows some characterization of the radar signal.
Figure~\ref{chap3.1-fig:radarpsd} depicts the \gls{psd} of the radar.
%
Figure~\ref{chap3.1-fig:radarconst} displays the constellation diagram of the transmitted signal. We notice that the signal lies only in the first quadrant of the \gls{iq}-plane, which is typical of some radars.

\begin{figure}[ht]
    \centering
    \includegraphics[width=0.7\linewidth]{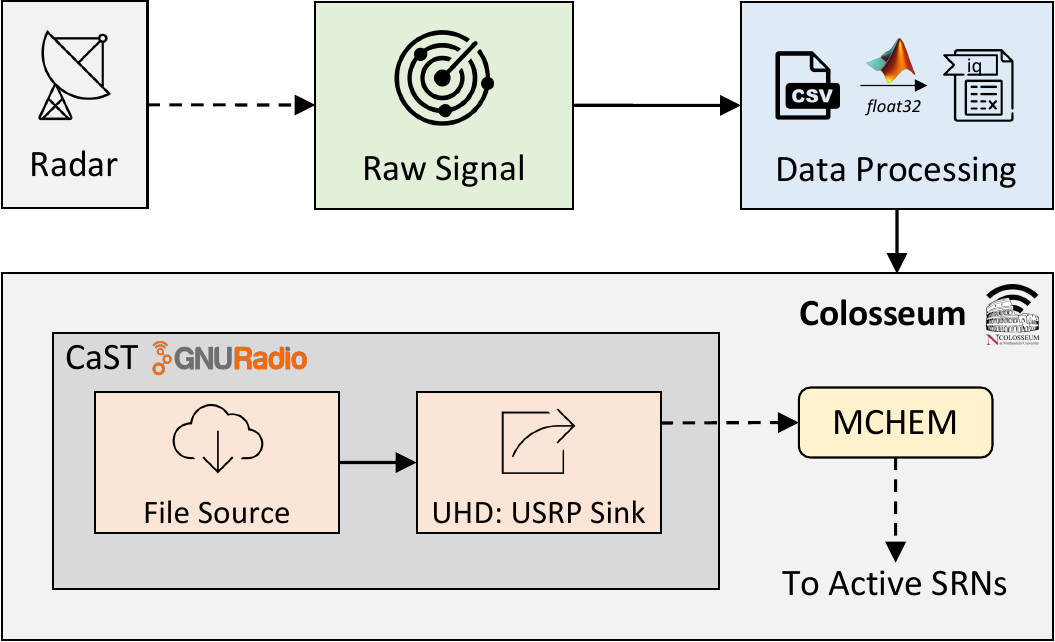}
    \caption{Block diagram of the operations needed to integrate the radar signal in Colosseum.}
    \label{chap3.1-fig:radarblock}
\end{figure}

Figure~\ref{chap3.1-fig:radarblock} shows the various operations that we developed to integrate an arbitrary waveform (radar in this case) into the Colosseum environment.
%
%
In the first step, the radar signal is generated either through a hardware device or in a synthetic manner. The output of this step is a raw signal formed of \gls{iq} samples for given time instances, which in our case is stored in a \texttt{.csv} file.
The raw data is then processed to convert it into a format that can be interpreted by Colosseum. We use MATLAB to read the \texttt{.csv} raw signal and generate a \texttt{.iq} file with the array of \gls{iq} values sequentially saved in a \texttt{float32} format.
Finally, the newly created \texttt{.iq} file is transmitted in the Colosseum environment by leveraging the open-source \gls{cast} framework, which is based on GNU Radio~\cite{villa2024dt, gnuradio}.
For the purpose of this work, \gls{cast} has been modified to include a \textit{File Source} block that allows us to load the \texttt{.iq} file on the Colosseum system. The signal is then passed through an \textit{UHD: USRP Sink} block, which connects to the USRP \gls{sdr} in Colosseum, and transmits the signal over \gls{mchem} to the other \glspl{srn}.

\begin{figure}[htp]
    \centering
    \begin{subfigure}{0.49\linewidth}
        \includegraphics[width=\linewidth]{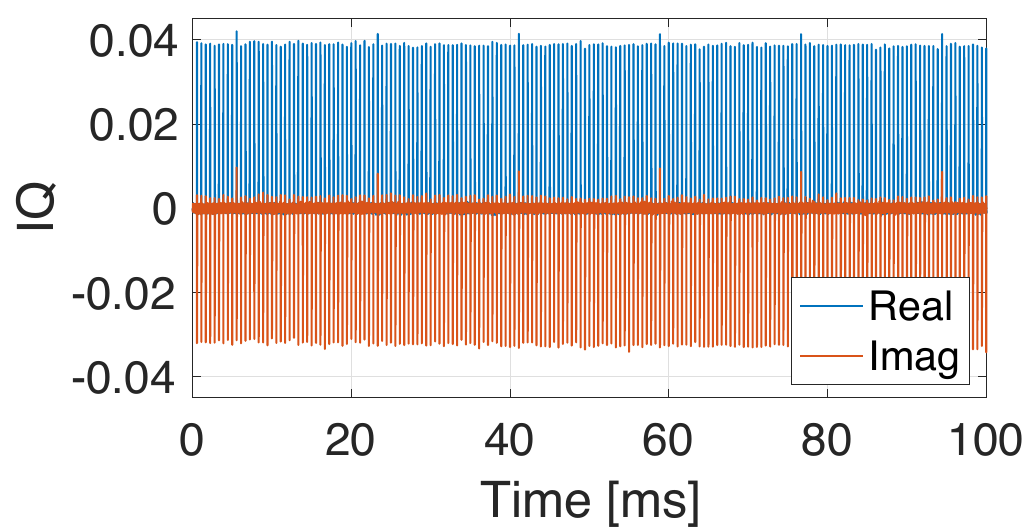}
        \caption{Received radar waveform}
        \label{chap3.1-fig:rxplainiq}
    \end{subfigure}
    \begin{subfigure}{0.49\linewidth}
        \includegraphics[width=\linewidth]{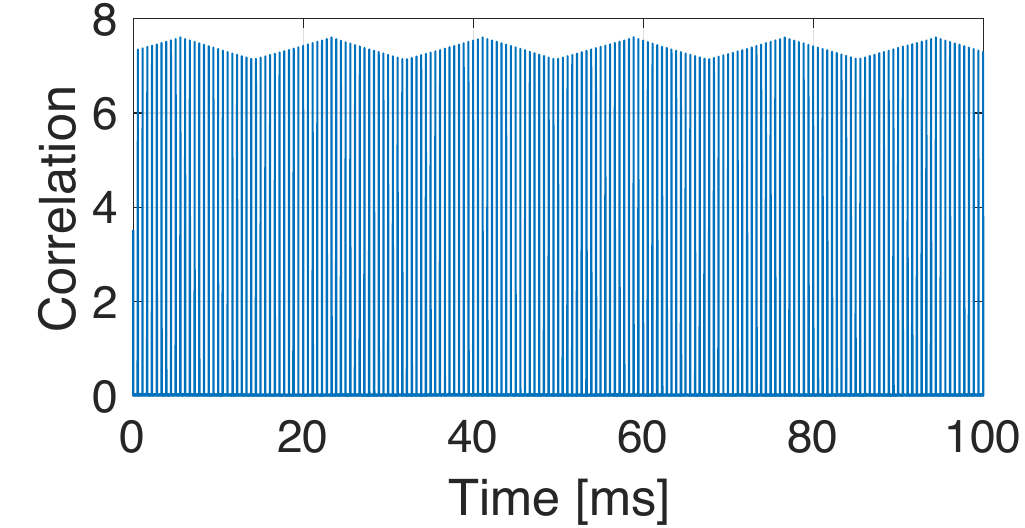}
        \caption{Radar correlation}
        \label{chap3.1-fig:rxcorr}
    \end{subfigure}
    \caption{Results of a radar transmission through \acrshort{cast}.}
    \label{chap3.1-fig:rxradarchar}
\end{figure}

%
Figure~\ref{chap3.1-fig:rxradarchar} shows the radar signal transmitted on the Colosseum wireless network emulator through \gls{cast} and received by another \gls{srn}.
Figure~\ref{chap3.1-fig:rxplainiq} displays real and imaginary parts of the raw radar waveform at the receiver node, while Figure~\ref{chap3.1-fig:rxcorr} displays the correlation between the original radar signal and the received waveform. We notice that the correlation values are clearly visible, meaning that the radar signal is correctly transmitted and detected at the receiver side.
We also notice a periodic jigsaw trend in the correlation results. This behavior is due to the large length of the transmitted radar \gls{iq} sequence (106657 complex points).
Upon performing the correlation operation, the length of the sequence causes peaks at the beginning of the sequence, as well as valleys, because of leftover samples from the correlation.
%

\subsection{Spectrum-Sharing Waikiki Beach Scenario}
\label{chap3.1-subsec:rf-scenario}

To evaluate the coexistence of cellular and radar technologies, we consider a 4G/5G \gls{ran} operating in the \gls{cbrs} band that must vacate the spectrum upon detection of an incumbent radar transmission.

\gls{cbrs} regulations allow commercial broadband access to the \gls{rf} spectrum ranging from $3.55$\:GHz to $3.7$\:GHz, as depicted in the \gls{cfr}~\cite{cfr2016}. This spectrum is shared with various incumbents, including the U.S. military, which operates radar systems in this frequency range, e.g., shipborne radars along the U.S. coasts. According to the regulations, dynamic access to the spectrum is permitted as long as the network can detect the presence of the radar and activate interference mitigation measures when necessary~\cite{caromi2018detection}.
Once detected, \glspl{bs} can either move to an unused bandwidth, if any, or terminate any ongoing communication to give priority to the radar.

\begin{figure}[t]
    \centering
    \includegraphics[width=0.7\linewidth]{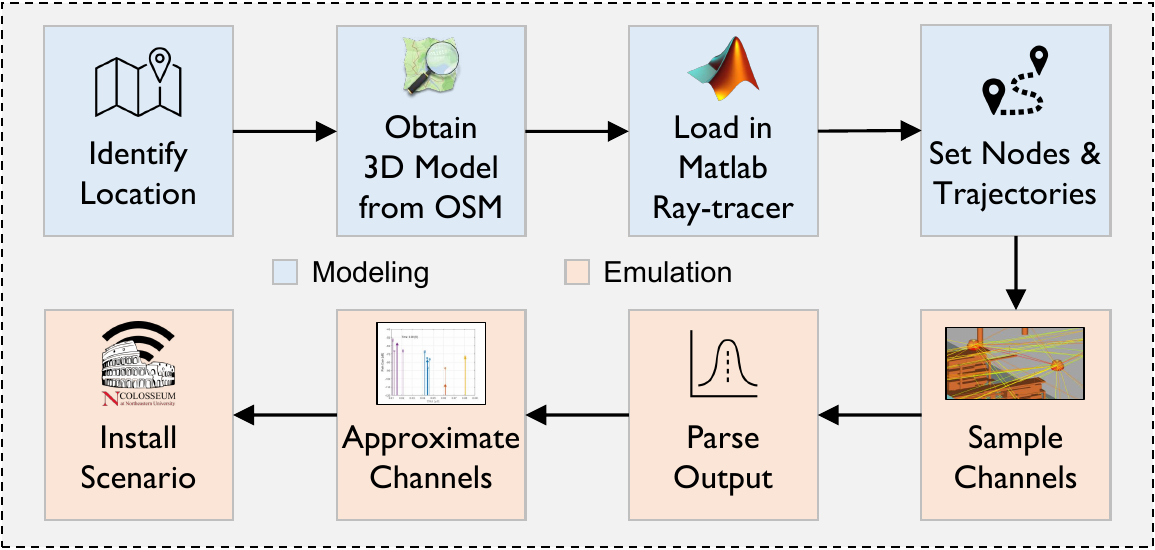}
    \caption{\acrshort{cast} scenario creation toolchain blocks diagram. Figure adapted from~\cite{villa2024dt}.}
    \label{chap3.1-fig:castscenarioblocks}
\end{figure}

To effectively study this use case on Colosseum, we created a novel \gls{rf} scenario that emulates the propagation environment of Waikiki Beach in Honolulu, Hawaii. This scenario involves a \gls{bs}---whose location was taken from the OpenCelliD database~\cite{opencellid} of real-world cellular deployments---that serves 6~\glspl{ue}, and a radar-equipped ship that moves in the North Pacific Ocean.
This scenario was created with the \gls{cast} toolchain~\cite{villa2024dt} following the steps of Figure~\ref{chap3.1-fig:castscenarioblocks}.
%

\begin{figure}[hb]
    \centering
    \includegraphics[width=0.7\linewidth]{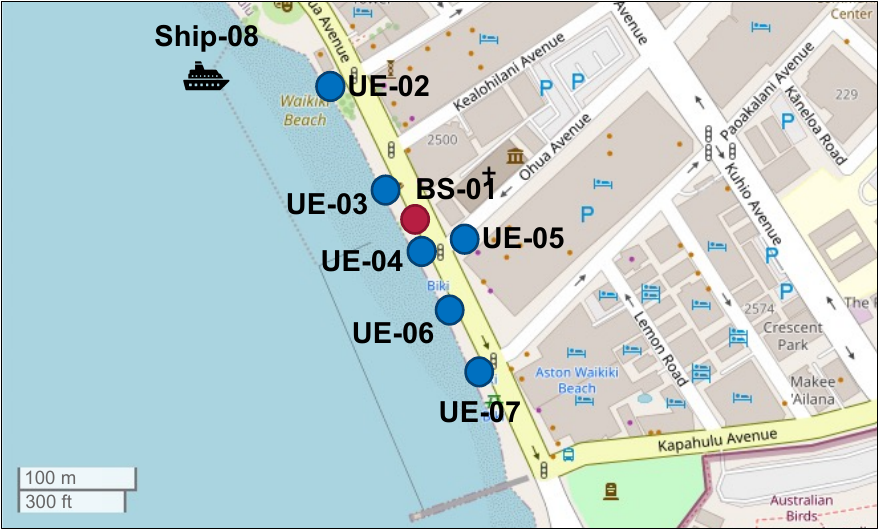}
    \caption{Location of the nodes in the Waikiki Beach scenario.}
    \label{chap3.1-fig:scenarionodes}
\end{figure}

In the first step, we identify the scenario location. Since we are considering a ship node for the radar, we chose the coastal area of Waikiki Beach in Honolulu, Hawaii.
Next, we obtain the 3D model of the selected location through the \gls{osm} tool~\cite{openstreetmap}. We generate an \texttt{.osm} file of a rectangular area of about $700 \times 800\:\mathrm{m}^2$, which includes Waikiki Beach, nearby buildings, and skyscrapers, as well as a portion of the ocean.
We then load the 3D model into the MATLAB ray-tracer,
and define the nodes of our scenario (shown in Figure~\ref{chap3.1-fig:scenarionodes}) as well as their trajectories.

\noindent
Our nodes are as follows.
\begin{itemize}
    \item One cellular \gls{bs} (red circle in Figure~\ref{chap3.1-fig:scenarionodes}), whose antennas are located at $3$\:m from the ground.

    \item Six static \glspl{ue} (blue circles in Figure~\ref{chap3.1-fig:scenarionodes}) uniformly distributed in the surroundings of the \gls{bs}. \glspl{ue} are located at $1$\:m from the ground level to emulate hand-held devices.

    \item One ship (shown in black in Figure~\ref{chap3.1-fig:scenarionodes}) equipped with a radar, whose antennas are located at a height of $3$\:m.
    The ship moves following a North-South linear trajectory along Waikiki beach at a constant speed of $20$\:knots ($\sim\!\!10$\:m/s).
    This speed was derived as the average between the speed typical of civilian container ships---which travel at around $10$\:knots ($\sim\!\!5$\:m/s)---and that of aircraft carriers---which reach speeds of around $30$\:knots ($\sim\!\!15$\:m/s)~\cite{taylor2013speed}.
    
\end{itemize}
%

%
\begin{table}[htbp]
    \centering
    \caption{Parameters of the Waikiki Beach scenario.}
    \label{chap3.1-table:scenarioparameters}
    \scriptsize
    \setlength{\tabcolsep}{4pt}
    \begin{tabular}{lc}
        \toprule
        \textbf{Parameter} & \textbf{Value} \\
        \midrule
        Signal bandwidth & $20$\:MHz \\
        Transmit power (BS and ship) & $30$\:dBm \\
        Transmit power (\glspl{ue}) & $20$\:dBm \\
        Antenna height (BS and ship) & $3$\:m \\
        Antenna height (\glspl{ue}) & $1$\:m \\
        Building material & Concrete \\
        Max number of reflections & $3$ \\
        Sampling time & $1$\:second \\ 
        Ship speed & $10$\:m/s \\
        Emulation area & $700x800\:\mathrm{m}^2$ \\
        \bottomrule
    \end{tabular}
\end{table}
%
%

\begin{figure}[hb]
    \centering
    \includegraphics[width=0.7\columnwidth]{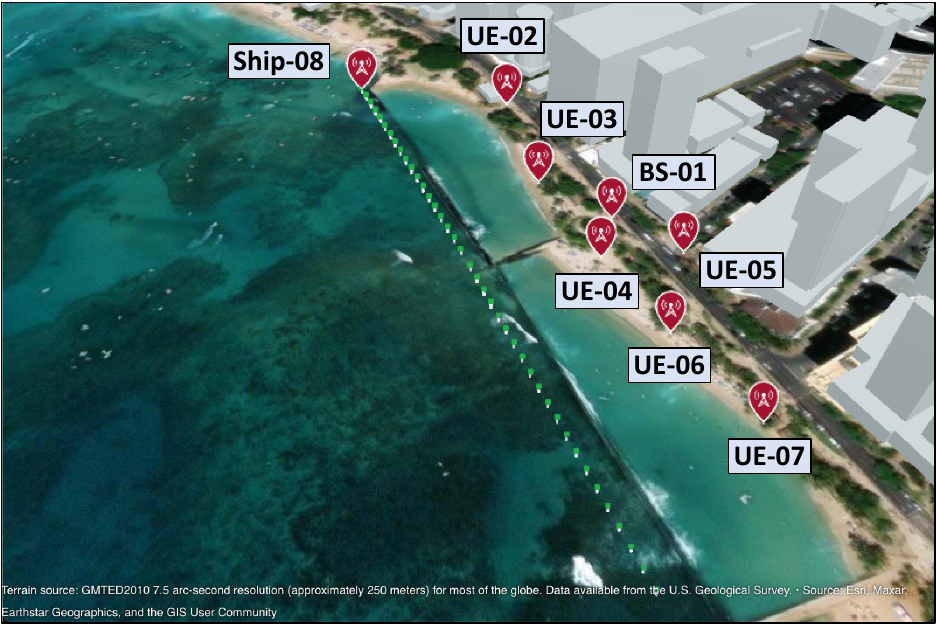}
    \caption{Layout of the scenario loaded in the MATLAB ray-tracer and visualized with Site Viewer.}
    \label{chap3.1-fig:scenariositeviewer}
\end{figure}

Table~\ref{chap3.1-table:scenarioparameters} summarizes the wireless parameters defined for the designed Colosseum \gls{rf} emulation scenario.
Figure~\ref{chap3.1-fig:scenariositeviewer} shows the layout of the scenario loaded in the MATLAB ray-tracer. We notice the 3D model of the environment (white building blocks in the figure), together with the radio node locations (red icons), and the trajectory of the ship (green dots). In this step, we perform ray-tracing to characterize the environment of interest and derive the channel taps between each pair of the nodes of our scenario.

\begin{figure}[ht]
    \centering
    \includegraphics[width=0.7\columnwidth]{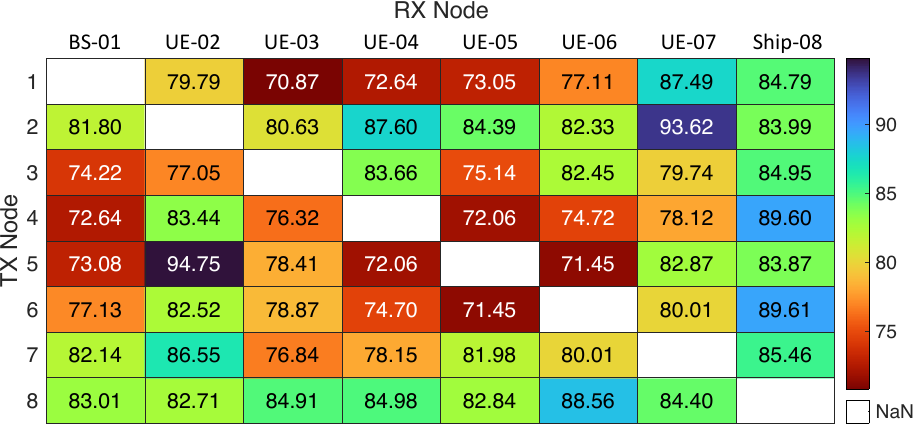}
    \caption{Heat map of the path loss among the nodes of the Waikiki scenario in Figure~\ref{chap3.1-fig:scenariositeviewer}. The mobile ship node is considered in its starting position at the top.}
    \label{chap3.1-fig:scenarioheatmap}
\end{figure}

\noindent After these operations are completed, the next step involves approximating the channel taps returned by the ray-tracer. This step is required to install the scenario in Colosseum, since \gls{mchem} supports a maximum of 4 non-zero channel taps, with a maximum delay spread
of $5.12\:\mu\mathrm{s}$.
This is performed through a k-means clustering algorithm that we previously developed~\cite{tehrani2021creating}.
The heat map of the path loss among each pair of nodes after this channel approximation step is depicted in Figure~\ref{chap3.1-fig:scenarioheatmap}. (The ship node is considered to be in the top position at the beginning of the scenario.)
As expected, closer nodes experience lower path loss values.
%


As a final step, the channel taps are converted into \gls{fpga}-readable format, and the scenario is installed in Colosseum.
%
Generating the channel taps with the ray-tracing software, and approximating them to the four non-zero taps
required around $7$\:hours by using a 2021 MacBook Pro M1 with 10 cores and $16$\:GB of RAM.
Installing the scenario on the Colosseum system required around $50$\:minutes by leveraging a virtual machine hosted on a Dell PowerEdge M630 Server with 24 CPU cores and $96$\:GB of RAM.

\subsection{Intelligent Radar Detection}
\label{chap3.1-sec:radardetection}

The \gls{bs} leverages an \gls{ai} model
to detect radar signals during or before cellular communications. This section explains how we collect and pre-process the data before feeding it into our model, as well as the structure of the model itself.

\subsubsection{Data Collection}

By using the scenario of Section~\ref{chap3.1-subsec:rf-scenario}, radar and cellular signals are transmitted in different combinations and varying reception gain. Specifically, we collect \gls{iq} samples when only the radar is present, only the cellular signal is present, both are present, and neither is present (empty channel). These combinations encompass all the possible real scenarios that the intelligent radar detector might come across.

We pre-process these recordings by first breaking them into smaller samples of 1024 \glspl{iq}, as this is the input size to the \gls{ml} agent. This input size was chosen as we have found it to be the smallest size that still offers high classification performance. We then convert each sample to its frequency domain representation. Finally, we offer a binary label to each sample:~\texttt{1} if radar exists in the sample, and \texttt{0} otherwise. In this way, the model groups empty channels and un-interfered cellular transmissions as \texttt{0} and therefore is free to communicate in the given band.

\subsubsection{Model Design and Training}

We use a lightweight, modified \gls{cnn} for radar detection. Specifically, we use a smaller version of VGG16 \cite{simonyan2014very}. We chose this structure as it is commonly used in wireless applications and can adequately show the capabilities of our framework.\footnote{More complex \gls{ai} algorithms can be used for this task. However, this is out of the scope of this work.} 
\begin{figure}[htb]
\setlength\abovecaptionskip{1pt}
\setlength\belowcaptionskip{-5pt}
    \centering
    \includegraphics[width=0.8\columnwidth]{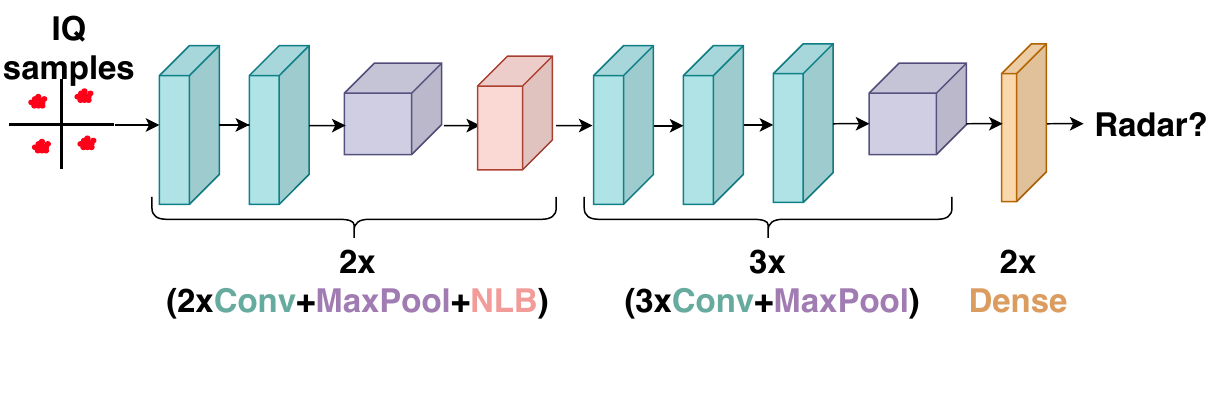}
    \caption{\acrshort{cnn} model used to train the radar detector.}
    \label{chap3.1-fig:cnn}
\end{figure}
The model architecture can be seen in Figure~\ref{chap3.1-fig:cnn}.
For the first two convolutional blocks, we append a \gls{nlb}. 
These blocks help the \gls{cnn} achieve self-attention and focus on spatially distant information.
More on these blocks can be read in \cite{wang2018non}.
This is a desirable trait as it can help identify long-range dependencies that may be present in the wireless signal rather than only focusing on adjacent \glspl{iq}.
The model takes as input \glspl{iq} in the shape of $(batch\_size, 1024, 2)$ where the last dimension is the real and complex part of the \gls{iq} separated into two distinct channels.

\subsection{Performance Evaluation}
\label{chap3.1-sec:results}


In this section, we present results on the effect of a radar signal on a cellular network deployed on the Colosseum wireless network emulator by showing: (i) the performance and accuracy of the \gls{ml} intelligent detector model; and (ii) the \glspl{kpi}, e.g., throughput, \gls{cqi}, and computation time of real-time experiments
with and without radar transmissions and the intelligent detector.

\subsubsection{Intelligent Detector Results}


We test our radar detector on data withheld during training and observe an accuracy of 88\%, a precision of 94\%, and a recall of 79\%. This tells us that our model is not susceptible to false positives (misclassifying empty channels or cellular signals as radar), but is susceptible to false negatives (misclassifying radar as an empty channel or a cellular signal). 

\begin{figure}[htbp]
    \centering
    \begin{subfigure}{0.49\linewidth}
        \includegraphics[width=\linewidth]{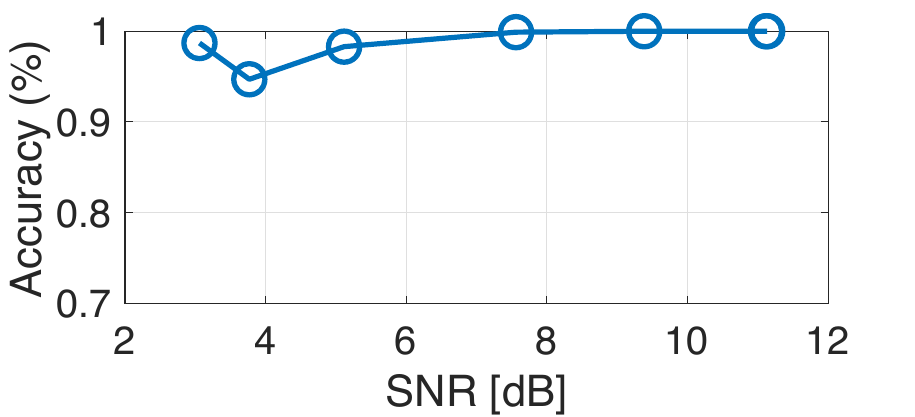}
        \caption{SNR accuracy}
        \label{chap3.1-fig:snracc}
    \end{subfigure}
       \hfill
    \begin{subfigure}{0.49\linewidth}
        \includegraphics[width=\linewidth]{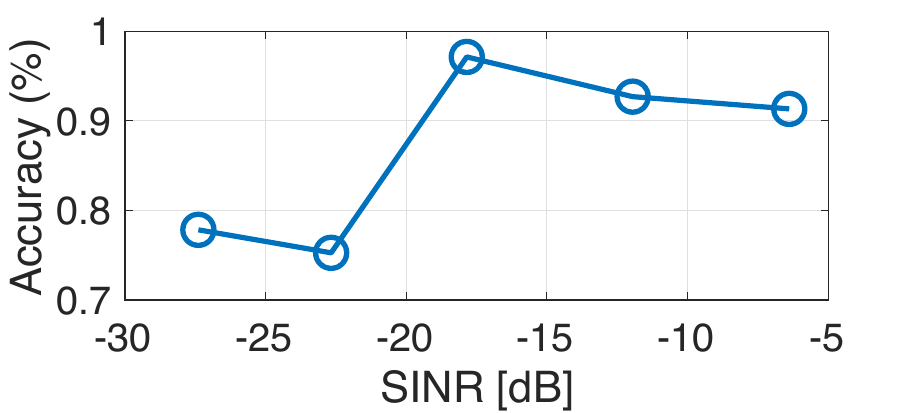}
        \caption{SINR accuracy}
        \label{chap3.1-fig:sinracc}
    \end{subfigure}
    \caption{\acrshort{cnn} radar detection accuracy with varying 
    \acrshort{snr} of the radar signal, and varying \acrshort{sinr} where the cellular signal is considered to be interference.}
    \label{chap3.1-fig:snrsinracc}
\end{figure}

To delve deeper into these results, we plot the accuracy as a function of \gls{snr} in Figure~\ref{chap3.1-fig:snracc}. For this plot, we keep the cellular nodes static in the scenario and only vary the radar gain. Here we can see that through varying \glspl{snr} we have very high and consistent performance in detecting radar signals, above 90\%. However, when we add cellular signals into the channel, this makes the detection of radar more difficult, as can be seen in Figure~\ref{chap3.1-fig:sinracc}, where we plot accuracy as a function of \gls{sinr}. The cellular signal is considered interference in this plot. Indeed, with high cellular signal gain and low radar gain, we see that our classification accuracy decreases to about 75\%.  

\subsubsection{Experimental Results}

To properly study the network performance with and without the presence of radar transmissions, we leverage Colosseum and the newly created scenario described in Section~\ref{chap3.1-subsec:rf-scenario} to deploy a cellular network and run traffic analysis.
The parameters of the experiments are summarized
\begin{table}[htbp]
    \centering
    \caption{Parameters of the experiments.}
    \label{chap3.1-table:emulationparameters}
    \scriptsize
    \setlength{\tabcolsep}{4pt}
    \begin{tabular}{lc}
        \toprule
        \textbf{Parameter} & \textbf{Value} \\
        \midrule
        Center frequency & $3.6$\:GHz \\
        Signal bandwidth (radar) & $20$\:MHz \\
        Signal bandwidth (cellular) & $10$\:MHz \\
        Number of \glspl{bs} & 1 \\
        Number of \glspl{ue} & 6 \\
        USRP BS gains (Tx and Rx) & $[10, 30]$\:dB \\
        USRP \gls{ue} gains (Tx and Rx) & $20$\:dB \\
        USRP radar Tx gain & $20$\:dB \\
        Scenario Duration & $40$\:s \\
        Traffic type & UDP Downlink \\
        Traffic rate & $10$\:Mbps \\
        Scheduling policy & Round-robin \\
        \bottomrule
    \end{tabular}
\end{table}
in Table~\ref{chap3.1-table:emulationparameters}.
The center frequency is set to $3.6$\:GHz in the newly opened \gls{cbrs} band, which is also used by S-Band-type radars. It is worth noticing that, even though characterized at $3.6$\:GHz, the scenario has been installed in Colosseum at a center frequency of $1$\:GHz, at which \gls{mchem} is optimized to work.
%
%
The scenario duration is set to $40$\:s, which is the time needed by the ship to travel the planned $400$\:m trajectory at the constant speed of $10$\:m/s.
Then, the scenario repeats cyclically from the beginning indefinitely.

We leverage the open-source SCOPE framework to deploy a twinned srsRAN protocol stack with one \gls{bs} and six \glspl{ue}~\cite{bonati2021scope,gomez2016srslte}. Additionally, the radar signal is transmitted through the use of the \gls{cast} transmit node~\cite{villa2022cast}, which has been modified to support custom radio waveforms.
In the following subsections, we show the results for three main cellular network use cases: (i) no radar transmission; (ii) with radar signal interference; and (iii) with radar and an intelligent detector.
In all experiments, a \gls{udp} downlink traffic of $10$\:Mbps among \gls{bs} and \glspl{ue} is generated through iPerf, a tool to benchmark the performance of \gls{ip} networks.

\begin{figure}[tb]
    \centering
    \begin{subfigure}{0.49\linewidth}
        \includegraphics[width=\linewidth]{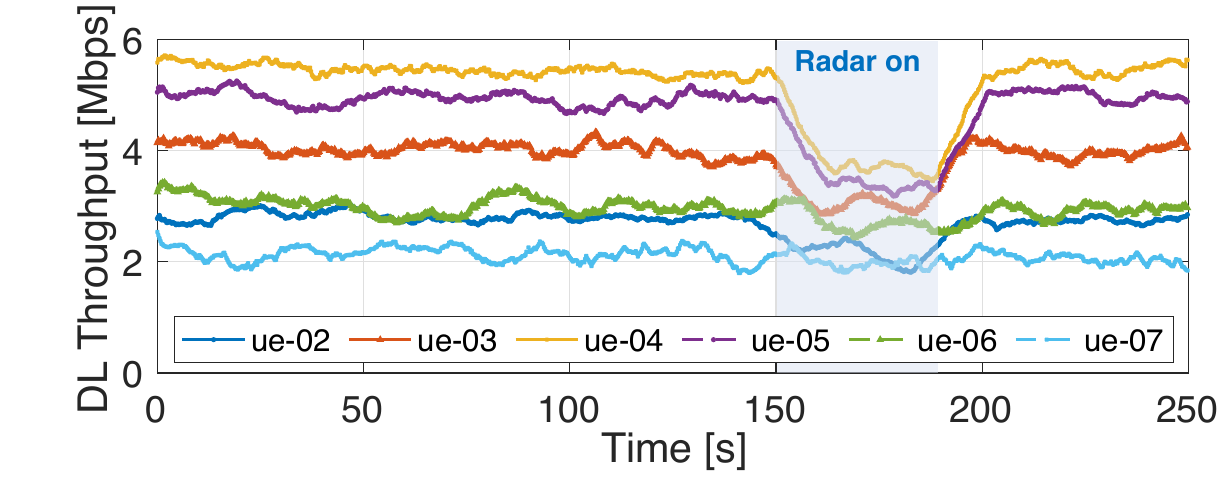}
        \caption{Downlink throughput}
        \label{chap3.1-fig:1resthroughput}
    \end{subfigure}
    \begin{subfigure}{0.49\linewidth}
        \includegraphics[width=\linewidth]{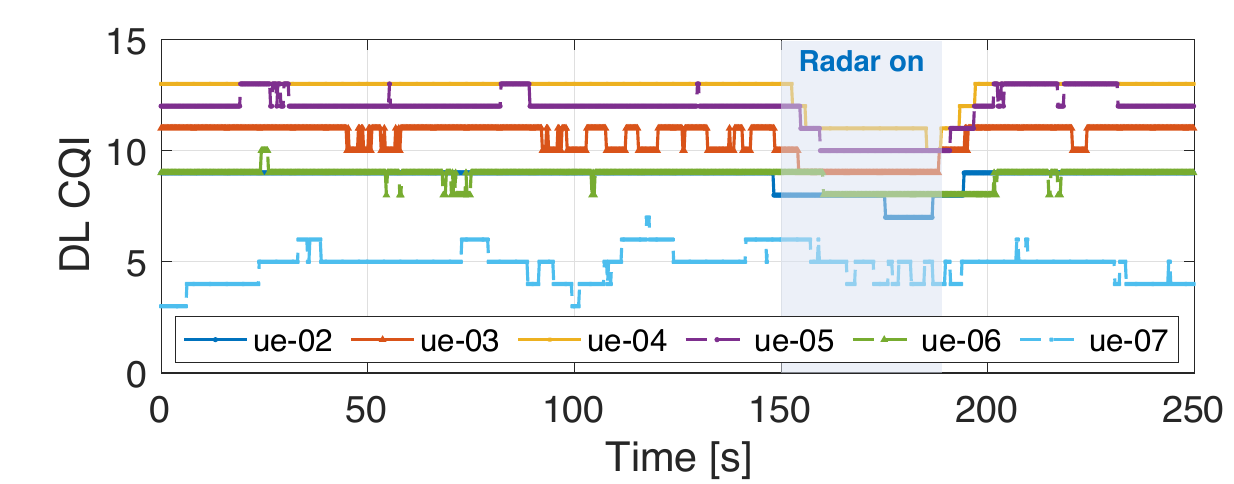}
        \caption{Downlink \acrshort{cqi}}
        \label{chap3.1-fig:1rescqi}
    \end{subfigure}
    \caption{Moving average of cellular network downlink throughput and CQI. A radar transmission is ongoing from second 150 to second 190, highlighted with a blue shade.}
    \label{chap3.1-fig:resultsnoradar}
\end{figure}

\textbf{No Radar}
\label{chap3.1-sec:expnoradar}
%
%
The performance of the cellular network, in terms of downlink throughput and \gls{cqi}, without radar transmissions is shown in Figure~\ref{chap3.1-fig:resultsnoradar}, from second $0$ to $150$.
The gains of the \gls{bs} USRP are set to $25$\:dB.%
As expected, the throughput, shown in Figure~\ref{chap3.1-fig:1resthroughput}, decreases with the increase of the distance between \glspl{ue} and \gls{bs}.
The best performance is achieved by UE-04, with values between $5.22$ and $5.71$\:Mbps, while the other \glspl{ue} achieve between $1.82$ and $5.25$\:Mbps.
UE-07 achieves the worst throughput due to its large distance from the \gls{bs}, environmental conditions, and interference with the other nodes.
The \gls{cqi}, shown in Figure~\ref{chap3.1-fig:1rescqi}, follows a similar trend. Best values are reported by UE-04, with a stable \gls{cqi} of $13$.
The other \glspl{ue} show \gls{cqi} values between $2$ and $13$, with UE-07 reporting the lowest \gls{cqi} values (between $2$ and $7$).

\textbf{Radar}
\label{chap3.1-sec:exponlyradar}
The impact of radar transmissions on the cellular performance is shown in Figure~\ref{chap3.1-fig:resultsnoradar}, from seconds $150$ to $190$.
As expected, we notice a drop in the throughput (Figure~\ref{chap3.1-fig:1resthroughput}) and \gls{cqi} values reported by the \glspl{ue} (Figure~\ref{chap3.1-fig:1rescqi}).
This is more visible for the nodes closer to the \gls{bs}, e.g., UE-03, UE-04, and UE-05, since they
get more affected by the radar transmission.
When the radar stops transmitting, i.e., at around second $190$, the performance of the \glspl{ue} goes back to the initial values, i.e., to the values in the $[0, 150]$\:s window.

\begin{figure}[t]
    \centering
    \includegraphics[width=0.8\columnwidth]{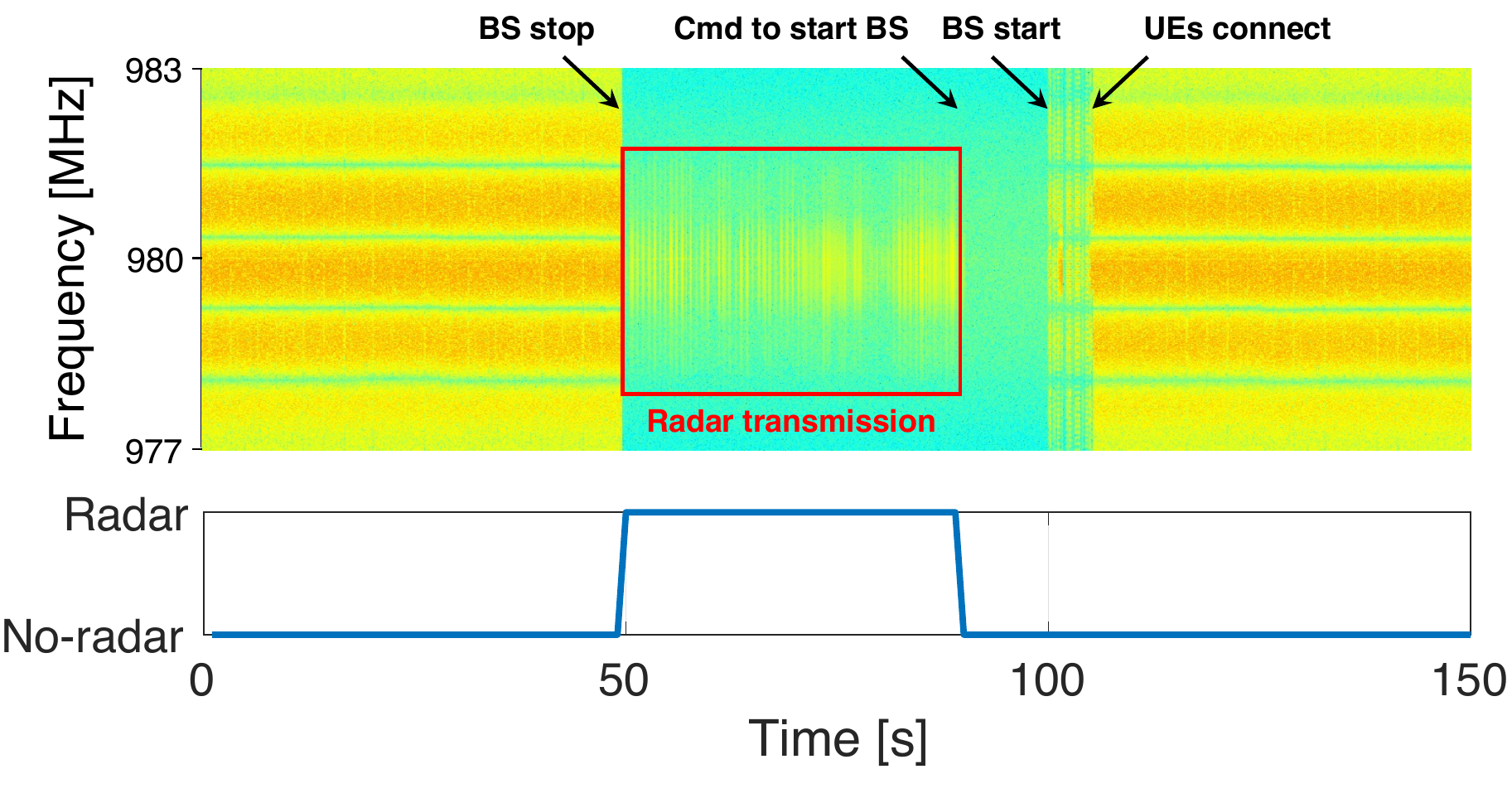}
    \caption{(top) Downlink cellular spectrogram; (bottom) radar detection system. The BS is shut down when a radar transmission is detected and resumes normal operations after no radar is detected.}
    \label{chap3.1-fig:expspectr}
\end{figure}

\textbf{Intelligent Radar Detection}
\label{chap3.1-sec:expradardetector}
In this last use case, we evaluate the effectiveness of our intelligent detector in understanding the presence of the radar signal, as shown in Figure~\ref{chap3.1-fig:expspectr}.
%
The top portion of the figure shows the downlink cellular spectrogram centered at $980$\:MHz (i.e., the downlink center frequency we use for srsRAN in Colosseum) with a $6$\:MHz span; the bottom one displays the results of the radar detection system.
At the beginning of the experiment---from second $0$ to second $50$---the \gls{bs} is serving the \glspl{ue} through \gls{udp} downlink traffic (Figure~\ref{chap3.1-fig:expspectr}, top), as we notice from the orange and yellow stripes.
Then, at second $50$, a radar transmission is detected by the intelligent detector (Figure~\ref{chap3.1-fig:expspectr}, bottom) described in Section~\ref{chap3.1-sec:radardetection}, and the \gls{bs} is shut down accordingly.
After the radar transmission ends, i.e., at second $90$, the \gls{bs} receives the command to power back on and, after around $10$\:seconds, it resumes its operations (second $100$).
Finally, at second $110$, the \glspl{ue} reconnect to the \gls{bs}, and the downlink transmissions are restarted.
Overall, this demonstrates the effectiveness of our intelligent detector in identifying radar signals and vacating the cellular bandwidth.
Note that even if we have not tested our \gls{ml} agent with different radar signal types, changes in the radar waveform that impact its frequency domain representation might
require a re-training of the model to achieve similar performance.

\begin{figure}[b]
    \centering
    \includegraphics[width=\columnwidth]{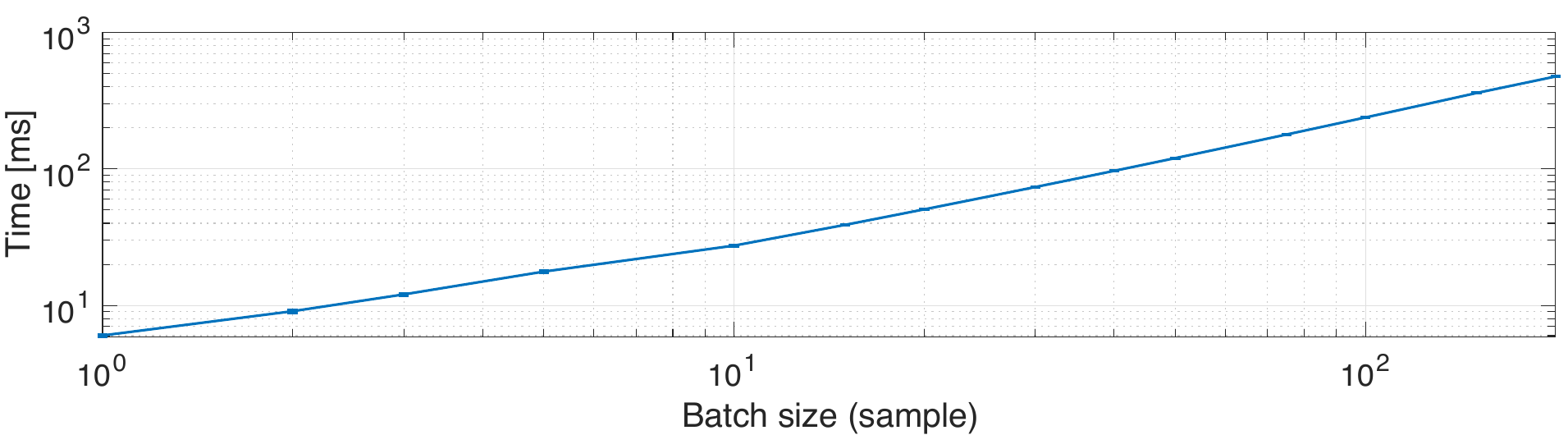}
    \caption{Computation time required for a radar classification, with different batch sizes, run on CPU on a Colosseum \acrshort{srn} with 48-cores Intel Xeon processor and 126~GB of RAM.}
    \label{chap3.1-fig:expbatch}
\end{figure}

Figure~\ref{chap3.1-fig:expbatch} shows the required computation time for the classification, performed on CPU, with different batch sizes in a Colosseum \gls{srn}.
We notice that values grow linearly with the batch size, e.g., $6$\:ms for a batch size of~1 sample, $27.4$\:ms for 10, $239$\:ms for 100.
This can be traced back to the fact that these operations run on a CPU, so there is not much parallelization of the processes as there would be in a GPU.
However, even in the case of a batch size of 100 samples, the maximum time of $60$\:seconds required for the detection of commercial transmissions in the \gls{cbrs} band is satisfied~\cite{goldreich2016requirements}.
To avoid false positives and false negatives, we leverage a voting system of 100 samples, in which the signal to shut down the \gls{bs} is sent only when more than $50$\% of the outputs are detecting a radar.
Moreover, we use a batch size of 10 samples, which gives us a good tradeoff between computation time and granularity of the output samples needed for the voting system.
In these conditions, our intelligent detector is able to detect an incumbent radar transmission and vacate the cellular bandwidth within $137$\:ms---which is the average time for the \gls{ml} model to generate 50 new outputs with a batch size of 10 samples---and with an accuracy of $88$\%.
%


\subsection{Conclusions}
\label{chap3.1-sec:conclusion}
In this section, we developed a framework for high-fidelity emulation-based spectrum-sharing scenarios with cellular and radar nodes implemented as a \gls{dt} system on the Colosseum wireless network emulator.
%
First, we twinned the radar waveform on Colosseum, then we collected \gls{iq} samples of radar and cellular communications in different conditions.
Finally, we trained a \gls{cnn} network ---that can run as a dApp--- to detect the presence of the radar signal and halt the cellular network to eliminate the undesired interference on the incumbent radar communications.
Our experimental results show that our detector obtains an average accuracy of 88\% (above 90\% when \gls{snr} and \gls{sinr} are greater than $0$\:dB and $-20$\:dB respectively), and requires an average time of $137$\:ms to detect ongoing radar transmissions.
This work demonstrates how \gls{dt} platforms like Colosseum enable the study of spectrum-sharing scenarios that would be difficult or unsafe to conduct in real-world environments, thereby showcasing the platform use-case key challenge dimension for \gls{ai}-driven wireless research.

\section{AI-Driven Radio Map Generation}
\label{chap3.2-sec:airmap}




Accurate, low-latency channel modeling is essential for effective real‑time wireless network simulation and \gls{dt} applications to precisely and swiftly characterize the radio signal propagation through a dynamic environment.
Channel modeling for \gls{rf} scenarios traditionally falls into three main categories: measurement-based models, statistical models, and deterministic methods (e.g., ray‑tracing).
Measurement‑based approaches capture site‑specific phenomena with high accuracy but are costly, labor‑intensive, and quickly outdated in dynamic environments~\cite{zhao2020playback}.
Statistical channel models employ stochastic or deterministic mathematical equations to characterize wireless propagation~\cite{3gpp38901}. However, these models often fail to represent all scenarios or capture environmental intricacies. Their reliance on simplified environmental assumptions leads to prediction inaccuracies, particularly in site-specific scenarios~\cite{zemen_site-specific_2025}.
Ray tracing provides a deterministic means of modeling wireless channels by launching rays and simulating their interactions—reflection, diffraction, and transmission—with environmental geometries using material-specific parameters~\cite{fuschini_ray_2015}. Although it achieves higher site-specific accuracy than empirical models---especially in architecturally complex settings---it remains impractical for real‑time use. Even with GPU‑accelerated implementations, the computational burden of dynamic mobility modeling and scenario adaptation is prohibitive, and the exhaustive pre-computation generally demands extensive storage. Furthermore, its fidelity depends on highly detailed 3D maps with accurate material assignments and accurate antenna patterns, since electromagnetic responses vary noticeably across surfaces.
In a trade-off against computational burden, ray tracing remains an approximation—constrained by finite ray sampling and often incomplete modeling of propagation phenomena like diffraction, scattering, and reflection for all possible points. These factors limit ray tracing’s suitability for high‑fidelity channel modeling in \glspl{dt}.

To address these limitations, advanced \gls{ai}-based techniques have been proposed. 
These methods leverage ground truth data from measurements or simulations to train data-driven models, enabling rapid and precise predictions of channel properties. 
By establishing a mapping function between the wireless environment and channel parameters, \gls{ai} tools facilitate proactive network design. 
They effectively tackle challenges such as resource allocation, user mobility analysis, localization, and radio propagation modeling. 
AI-based techniques offer greater flexibility, scalability, and reduced computational complexity, thus enabling real-time propagation modeling in complex urban environments

Radio (environment) maps provide a two‑dimensional representation of averaged statistics of channel characteristics (e.g., received signal power, interference power, power spectral density, delay spread, and channel gain) over a geographic region~\cite{el-friakh_crowdsourced_2018}. Unlike individual point-wise channel estimates, radio maps capture spatial relationships and large-scale propagation patterns, reflecting how neighboring locations influence one another. This inherent spatial structure makes radio maps a natural and effective output format for \gls{dl} models, which can leverage locality and spatial dependencies. For instance, \gls{cnn} architectures can process environmental inputs such as terrain and building layouts to generate channel predictions for an entire area in a single inference step, rather than predicting each point independently. This structured approach not only improves scalability and inference speed but also provides the spatial context essential for real-time digital twin applications.

This section presents work on \gls{dl}-based frameworks for real-time, high-fidelity radio map estimation, developed in collaboration with the authors of~\cite{saeizadeh2024camad, saeizadeh2025airmap}.
We first describe the initial approach introduced in~\cite{saeizadeh2024camad}, which demonstrated real-time path gain prediction using a U-Net architecture with elevation maps and rough propagation estimates as inputs, achieving inference times of $46$~ms on GPU.
We then mention its extension, AIRMap~\cite{saeizadeh2025airmap}, which simplifies the input to a single-channel elevation map and further reduces inference time to under $4$~ms, enabling integration with the Colosseum emulator for real-time channel emulation.


The remainder of this section is organized as follows.
Section~\ref{chap3.2-subsec:overview} provides a general overview of the framework.
Section~\ref{chap3.2-subsec:airmap-data} describes the data collection process, including ray-tracing simulations and the measurement campaign.
Section~\ref{chap3.2-subsec:dl} details the \gls{dl}-based propagation model.
Section~\ref{chap3.2-subsec:airmap-results} presents the performance results.
Section~\ref{chap3.2-subsec:airmap-colosseum} demonstrates the integration with the Colosseum emulator.
Finally, Section~\ref{chap3.2-subsec:airmap-conclusions} concludes this section.

\subsection{Framework Overview}
\label{chap3.2-subsec:overview}

\begin{figure}[htb]
    \centering
    \includegraphics[width=0.6\linewidth]{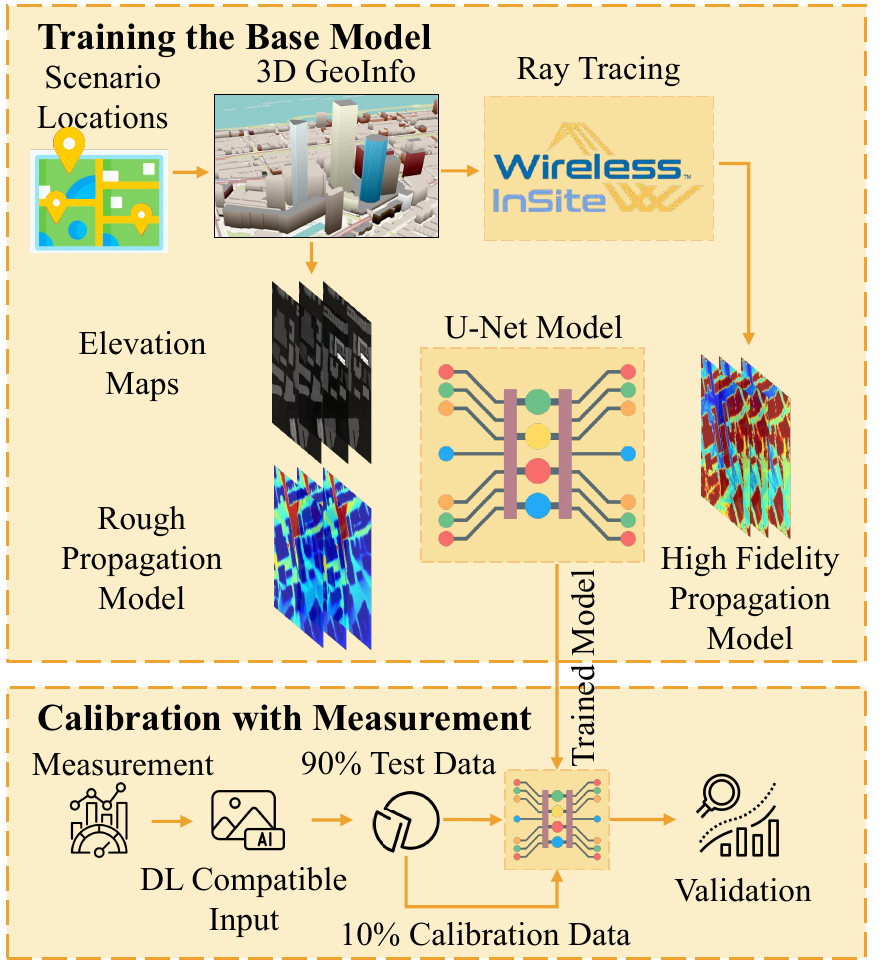}
    \caption{Training process, data calibration, and model validation.}
    \label{chap3.2-fig:diag}
\end{figure}

The overall training process, data calibration, and model validation pipeline is illustrated in Figure~\ref{chap3.2-fig:diag}.
To achieve this, we employ a U-Net structure inspired by Lee and Molish~\cite{lee2023scalable}, which allows the model to cover an entire area with a single inference while capturing spatial features effectively. 
We leverage two key inputs: an elevation map to accurately convey 3D geographical information to the model, and a rough estimation of the propagation model to maintain generalizability. 
This rough estimation is subsequently upsampled and refined into a high-fidelity propagation model. 
Unlike other approaches, we place the \gls{tx} at the center to reduce the model's complexity. 
This improves model performance compared to using an additional channel for the \gls{tx} location.
By implementing these changes, we address the generalizability issues of previous works, enabling real-time propagation modeling for any environment with available 3D geographical data. Our approach significantly improves both accuracy and computational efficiency. Specifically, we achieve a normalized \gls{rmse} of less than $0.035$~dB over a 37,210 square meter area, with processing times of just $46$~ms on a GPU. These results demonstrate the model's ability to rapidly provide high-fidelity propagation predictions, surpassing traditional ray tracing methods that require over $387.6$~seconds. 
Additionally, refining the model with a small amount of measurement data shows an \gls{rmse} of $0.0113$, demonstrating its adaptability to real-world data.
%
%
The remarkable performance and efficiency of AI-driven techniques underscore their potential to revolutionize wireless network design and optimization, enabling real-time adaptation to dynamic and complex telecommunication environments.

\subsection{Training the Model: Data Collection}
\label{chap3.2-subsec:airmap-data}

To construct a comprehensive and precise dataset for predicting path gain (expressed in dB throughout this study), we use both ray tracing simulations and empirical measurements. Our objective is to forecast the Path Gain ($\mathrm{PG}$), as delineated in equation (\ref{chap3.2-eqn:pg_general}), by leveraging the received power at the receiver ($\text{P}_{\text{RX}}$) and the transmitted power ($\text{P}_{\text{TX}}$).
\begin{equation}
    \mathrm{PG}(t)
    = P_{\mathrm{RX}}(t) 
    - P_{\mathrm{TX}}(t),
    \label{chap3.2-eqn:pg_general}
\end{equation}
The \gls{cir} can be defined as:
\begin{equation}
h(t, \tau) = \sum_{i=1}^{N} \alpha_i(t) \delta\left(\tau - \tau_i(t)\right),
\label{chap3.2-eqn:cir}
\end{equation}
\noindent where \(\alpha_i(t)\) is the time-varying complex amplitude of the \(i\)-th path and \(\tau_i(t)\) is the time-varying delay of the \(i\)-th path.
To extract the received power from the channel, we use the \gls{cir} as shown in (\ref{chap3.2-eqn:p_rx})
\begin{equation}
\begin{aligned}
    P_{\text{RX}}(t) 
    &= 10 \log_{10} \left( \int_{-\infty}^{\infty}|h(t, \tau)|^2 d \tau \right) \\
    &= 10 \log_{10} \left( \sum_{i} \left|\alpha_i(t)\right|^2 \right).
\label{chap3.2-eqn:p_rx}
\end{aligned}
\end{equation}
In a stationary environment with a mobile receiver, $PG(t)$ and $\text{PG}(\boldsymbol{q}_{\mathrm{RX}})$, where $\boldsymbol{q}_{\mathrm{RX}}$ is the location of \gls{rx} can be considered equivalent since the position \(\boldsymbol{q}_{\mathrm{RX}}\) of the receiver varies with time \(t\). Thus, modeling \(\text{PG}(\boldsymbol{q}_{\mathrm{RX}})\) effectively captures the time-varying nature of the path gain \(\text{PG}(t)\) as the receiver moves through different locations \(\boldsymbol{q}_{\mathrm{RX}}\).

\subsubsection{Ray Tracing}

Urban environments pose challenges for conventional channel models, which often fail to accurately characterize channel properties. Ray tracing methods offer a potential solution for these complex scenarios. To build a comprehensive dataset, we employed Wireless InSite Ray Tracing software~\cite{remcom_wireless_insite} by RemCom to collect high-fidelity data. The ray tracing model is configured by one diffraction and four reflections. In the scenario described below, simulating one transmitter takes 1:25:12 using CPU or 0:03:16 using GPU on a machine with two Intel Xeon E5-2660 processors with 28 cores and one Nvidia Tesla K40c GPU with 2880 CUDA cores.

\begin{figure}
    \centering
    \includegraphics[width=.7\linewidth]{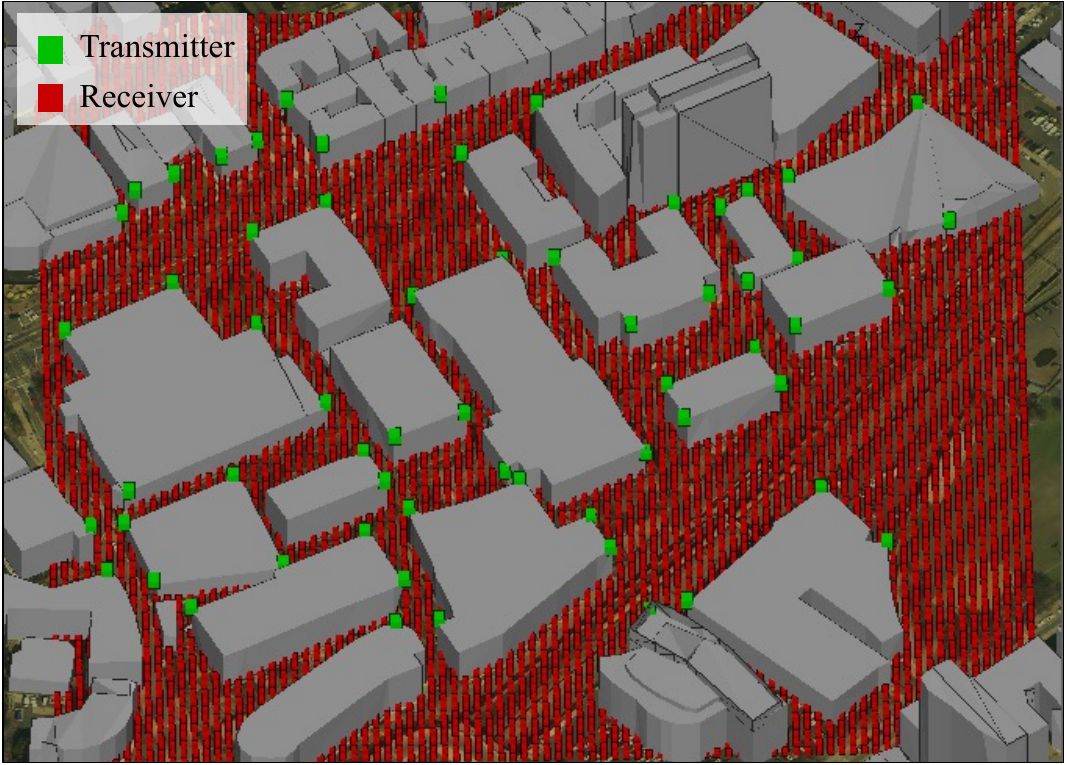}
    \caption{Northeastern University, Boston Campus, used for generating the main dataset consisting of 61 \acrshort{tx} and 7,569 \acrshort{rx}. Location: 42°20'22"N, 71°05'14"W.}
    \label{chap3.2-fig:neu}
\end{figure}

In this study, we focus on the Northeastern University campus in Boston, shown in Figure~\ref{chap3.2-fig:neu}, as an urban use case scenario to train and test the model. We consider a grid of potential \gls{rx} locations, comprising 7,569 points (red points) spread over a $435 \times 435$ square meters area, and 61 \gls{tx} locations (green points) situated at the corners of building rooftops for potential \gls{bs}. Additionally, we examine Fenway Park (42°20'26"N 71°05'38"W), as a separate location with a lower density of buildings (and not seen in the \gls{dl} training phase) with another 16 \glspl{tx} to evaluate the model's generalization capabilities. To ensure accuracy, we import a precise 3D model of the area using data gathered by Boston Planning and Development Agency~\cite{BPDA_3D_Data_Maps}.

\begin{figure}[htb]
    \begin{subfigure}{0.23\linewidth}
        \includegraphics[width=.99\linewidth]{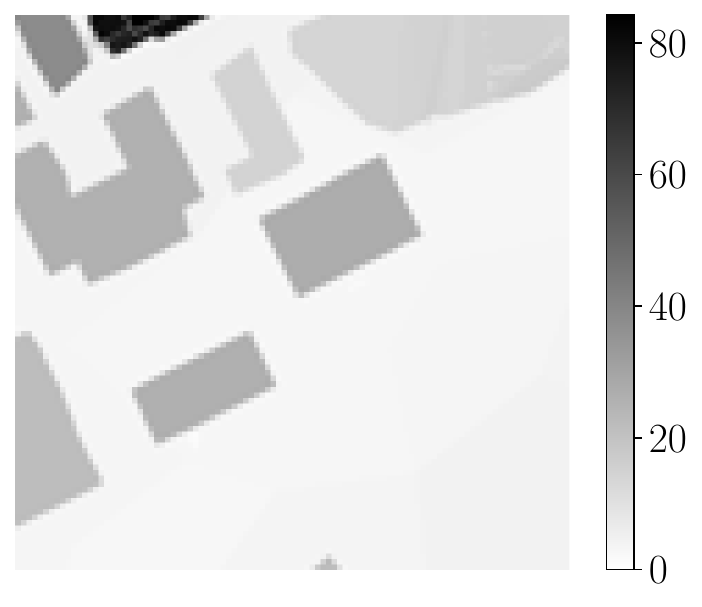}
        \caption{Elevation Map [m]}
        \label{chap3.2-fig:building_maps}
    \end{subfigure}
    \hfill
    \begin{subfigure}{0.24\linewidth}
        \includegraphics[width=.99\linewidth]{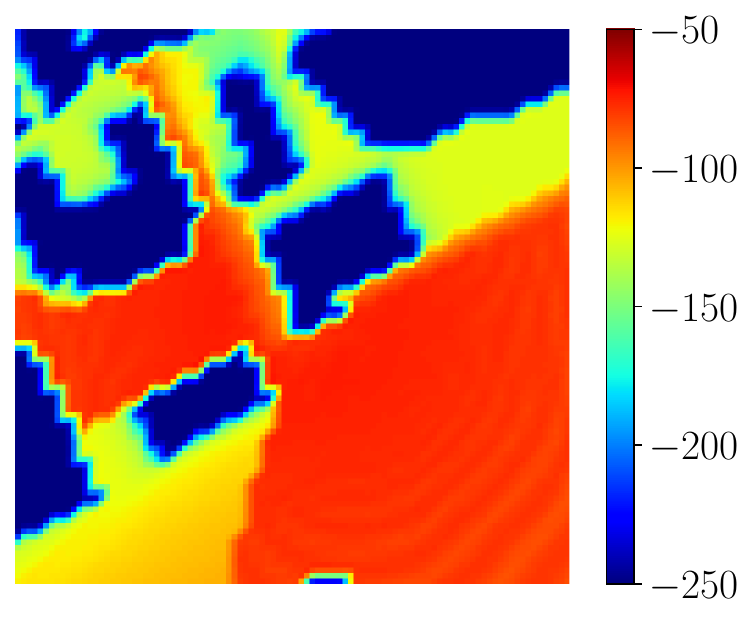}
        \caption{Propagat. Est. [dB]}
        \label{chap3.2-fig:rough}
    \end{subfigure}
    \begin{subfigure}{0.24\linewidth}
        \includegraphics[width=.99\linewidth]{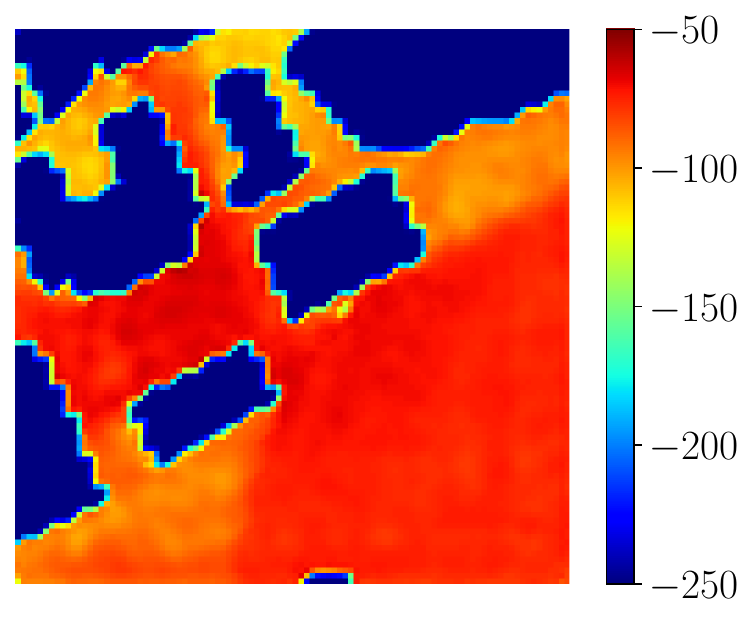}
        \caption{Model's Output [dB]}
        \label{chap3.2-fig:output}
    \end{subfigure}
    \hfill
    \begin{subfigure}{0.24\linewidth}
        \includegraphics[width=.99\linewidth]{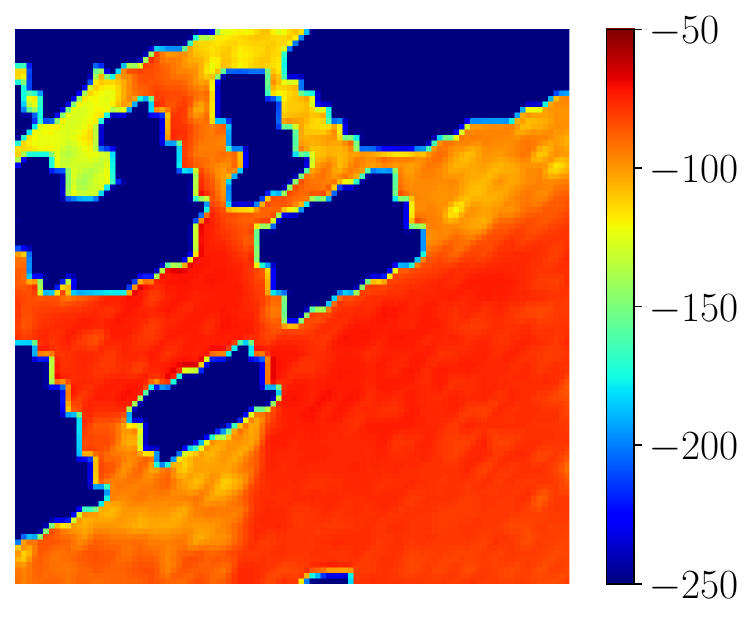}
        \caption{Ray Tracing [dB]}
        \label{chap3.2-fig:ground_truth}
    \end{subfigure}
    \caption{A comparison of the model's output (Figure~\ref{chap3.2-fig:output}) with the ground truth heatmap generated by Wireless InSite (Figure~\ref{chap3.2-fig:ground_truth}), showing its superior performance compared to (Figure~\ref{chap3.2-fig:rough}).}
    \label{chap3.2-fig:out}
\end{figure}

To prepare the input data for the model, we simulate ray tracing for all transmitter locations in the Northeastern University scenario, as shown in Figure~\ref{chap3.2-fig:neu}, to obtain path gain heat maps. These heat maps serve as the ground truth for training the model, with Figure~\ref{chap3.2-fig:out} presenting an example. We convert these maps into gray-scale single-channel images to reduce the data requirements and mitigate overfitting. Since \gls{dl} models require fixed input sizes, we crop the images to ensure uniform input sizes across different scenarios. Each image in our dataset measures $100\times100$ pixels, representing approximately a $1.929\times1.929$ meter area per pixel. Generating data for numerous scenarios is impractical, so we employ data augmentation techniques to create an augmented dataset from a limited synthetic dataset. Specifically, we use random rotations to incorporate the transmitter location into various input configurations, ensuring a substantial and diverse training dataset to enhance the model's robustness and accuracy.

\subsubsection{Measurement Campaign}

\begin{figure}[htb]
    \centering
    \begin{subfigure}{0.49\textwidth}
        \centering
        \includegraphics[width=\textwidth]{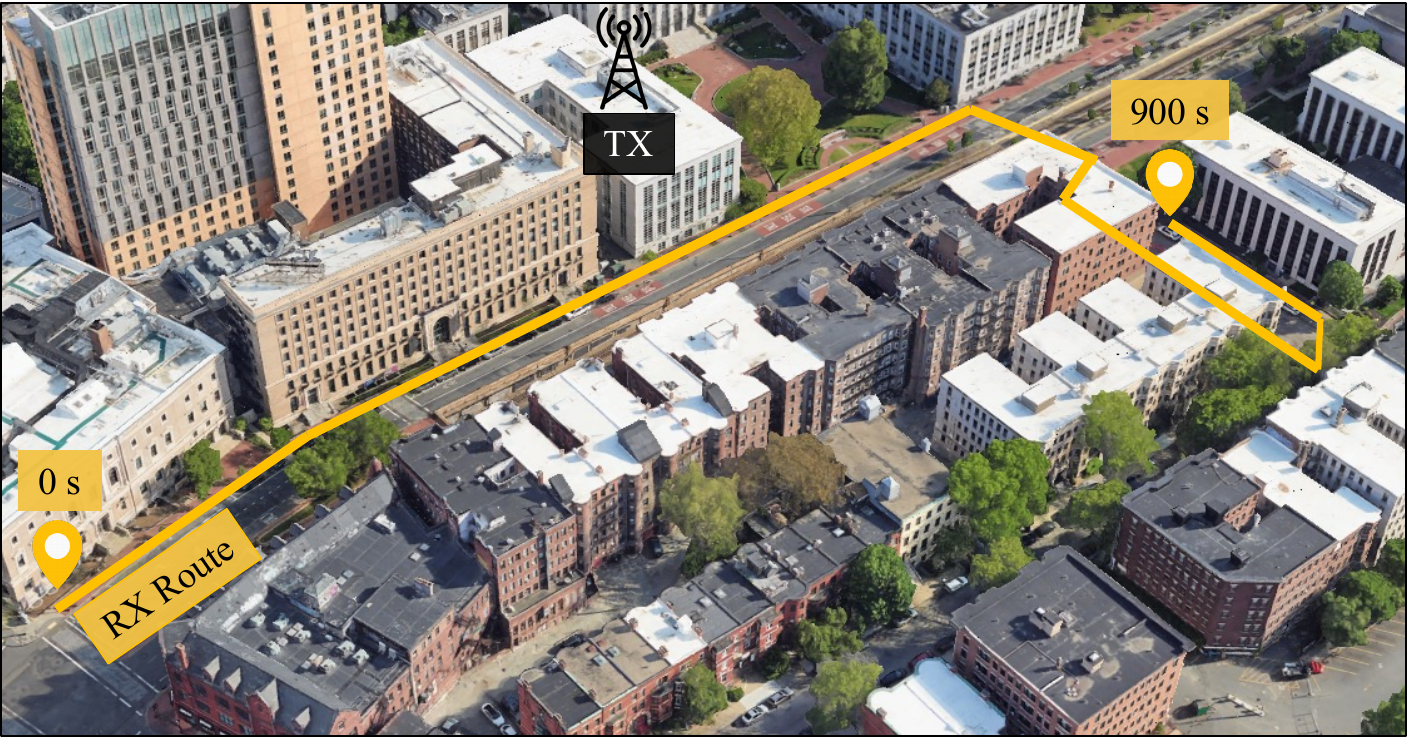}
        \caption{Path followed during the measurement campaign.}
        \label{chap3.2-fig:meas}
    \end{subfigure}
    \hfill
    \begin{subfigure}{0.49\textwidth}
        \centering
        \includegraphics[width=\textwidth]{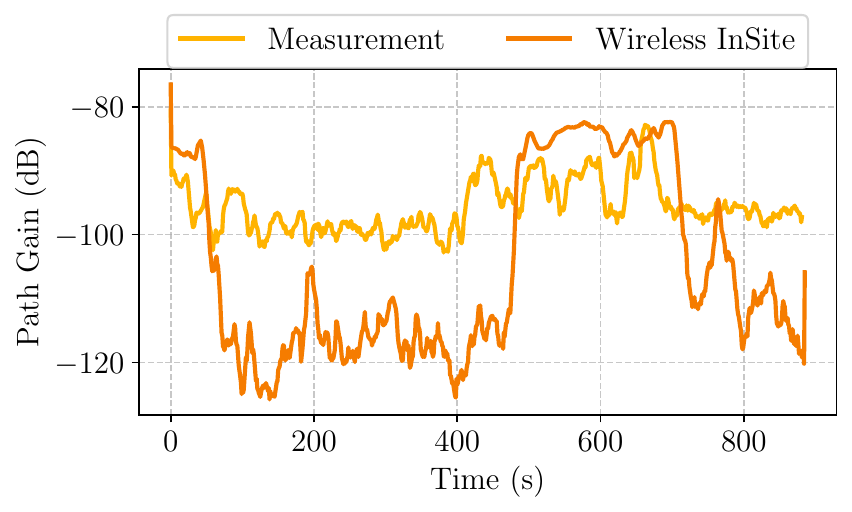}
        \caption{The path gain derived from the measurement campaign with moving average applied.}
        \label{chap3.2-fig:path_gain}
    \end{subfigure}
    \caption{Measurement campaign at Northeastern University Campus in collaboration with VIAVI to gather real-world data for model validation and refinement.}
    \label{chap3.2-fig:combined}
\end{figure}

Although ray tracing is a powerful tool for modeling channels, real-world scenarios present unique variations that can significantly alter the channel characteristics (see Figure~\ref{chap3.2-fig:path_gain}). Three main factors contribute to these variations: (1)~environmental dynamics, such as moving vehicles, constantly change the channel conditions; (2)~the precise shapes and materials of buildings, which 3D models may not accurately capture, can affect channel behavior, as materials are often not modeled with exact fidelity; (3)~interference from other users and background noise introduce additional environment noise into the channel, complicating accurate modeling.

To refine, calibrate, and validate the \gls{dl} model in a real-world scenario, we conducted a measurement campaign in collaboration with VIAVI Solutions around the Northeastern University campus, as depicted in Figure~\ref{chap3.2-fig:meas}. The details of the measurements are provided in Table \ref{chap3.2-tab:meas-setup}.
\begin{table}[htb]
    \centering
    \caption{Measurement Setup and Equipment}
    \label{chap3.2-tab:meas-setup}
    \begin{tabular}{p{2.5cm}p{4.5cm}}
    \toprule
    \textbf{\textit{\gls{tx}}} & \\ 
    \midrule
    \gls{ru} & Ettus USRP X410 \\ 
    Amplifier & Minicircuits ZHL-1000-3W+ (38 dB) \\ 
    Antenna & Pasternack PE51OM1014 (6 dBi) \\ 
    Location & 42°20'25"N 71°05'16"W \\
    \toprule
    \textbf{\textit{\gls{rx}}} & \\ 
    \midrule
    \gls{ru} & Ranger provided by VIAVI Solutions \\
    Antenna & Waveform directional antenna (10 dBi) \\ 
    Location & Mobile (see Figure~\ref{chap3.2-fig:meas}) \\
    \toprule
    \textbf{\textit{Measurement Details}} & \\
    \midrule
    Frequency & 910 MHz \\ 
    Bandwidth & 122.88 MHz \\ 
    Codeword & GLFSR-14 \\ 
    Synchronization & GPS clock for both \gls{tx} and \gls{rx} \\ 
    \bottomrule
    \end{tabular}
\end{table}

For channel sounding, the VIAVI Solutions Ranger, an \gls{rf} waveform generator and capture platform, was used. It supports two full-duplex channels with $200$\:MHz bandwidth up to $6$\:GHz and captures three and one-half hours of recordings. Analysis and waveform modifications are done using the Signal Workshop application, which can be run locally or remotely.

The recorded \gls{iq} samples by the Ranger were processed in two steps: (1) extracting \gls{cir} ($h(\tau,t)$) from the raw \gls{iq} samples ($R(\tau, t)$); and (2) calibration to determine the actual Path Gain ($PG(\mathbf{q}_{\mathrm{RX}})$). For the first step, we utilized a Galois Linear Feedback Shift Register 14 (GLFSR-14) codeword, modulated it using \gls{bpsk}, resampled it to match the \gls{rx} sampling frequency ($\mathbf{s}(t)$), and then correlated it with the received data. This process was repeated for each second of the data. Thus, \gls{cir} at the receiver $h[\tau,t=n]$ at the $n$-th second at location $\mathbf{q}_{\mathrm{RX}}$ using GPS logs is
\begin{equation}
    h(\tau,t) = R(\tau, t) \ast \mathbf{s}(t).
\end{equation}

To remove background noise from the channel, we identified the significant peaks in the channel for every length of the codeword:
\begin{equation}
    \mathbf{\alpha} = h(\tau, t) \cdot \mathbf{1}_{\{\text{peaks}(h(\tau, t))\}}.
    \label{chap3.2-eqn:peaks}
\end{equation}

The indicator function \(\mathbf{1}_{\{\text{peaks}(h(\tau, t))\}}\) is defined as
\begin{equation}
\mathbf{1}_{\{\text{peaks}(h(\tau, t))\}} =
\begin{cases} 
1 & \text{peaks of } h(\tau, t)  \\
0 & \text{otherwise.}
\end{cases}
\end{equation}
This function is $1$ at the peak (threshold of $+3 dB$ of the neighboring points) positions of \(h(\tau, t)\) and $0$ elsewhere, effectively isolating the peak values of the \gls{cir}. After extracting the sparse channel, we sum the values like in~(\ref{chap3.2-eqn:p_rx}).

To accurately determine the path gain, a conducted calibration test \gls{otc} was performed with all equipment in the loop except for the antennas. Instead of antennas, 60\,dB attenuators ($L_{\text{ATT}}$) were placed between \gls{rx} and \gls{tx} to prevent \glspl{adc} saturation, and the cable loss ($L_{\text{c}}$) was measured.
\begin{equation}
    P_{\text{\gls{rx}}}^{\text{\gls{otc}}}
    = P_{\text{\gls{tx}}} 
    + G_{\text{AMP}} 
    - L_{\text{c}} 
    - L_{\text{ATT}}
\end{equation}

After field measurements, the average received power from the calibration data ($P_{\text{\gls{rx}}}^{\text{\gls{otc}}}$) was subtracted from the received power in the actual measurement ($P_{\text{\gls{rx}}}^{\text{\gls{ota}}}$) to compensate for amplifiers ($G_{\text{AMP}}$), cable losses ($L_{\text{c}}$), and \gls{tx} power ($P_{\text{\gls{tx}}}$).
\begin{equation}
    P_{\text{\gls{rx}}}^{\text{\gls{ota}}}
    = P_{\text{\gls{tx}}}
    + G_{\text{AMP}}
    + G_{\text{ANT}}
    + \mathrm{PG}^{\text{\gls{ota}}}
\end{equation}

The nominal gain of the antennas ($G_{\text{ANT}}$) was also considered in $\mathrm{PG}^{\text{\gls{ota}}}$:
\begin{equation}
    \mathrm{PG}^{\text{\gls{ota}}}
    = P_{\text{RX}}^{\text{OTA}}
    - P_{\text{RX}}^{\text{\gls{otc}}}
    - L_{\text{c}}
    - L_{\text{ATT}}
    - G_{\text{ANT}}
\end{equation}

\subsection{Deep Learning-Based Propagation Model}
\label{chap3.2-subsec:dl}
\begin{figure}[htb]
    \centering
    \includegraphics[width=0.8\textwidth]{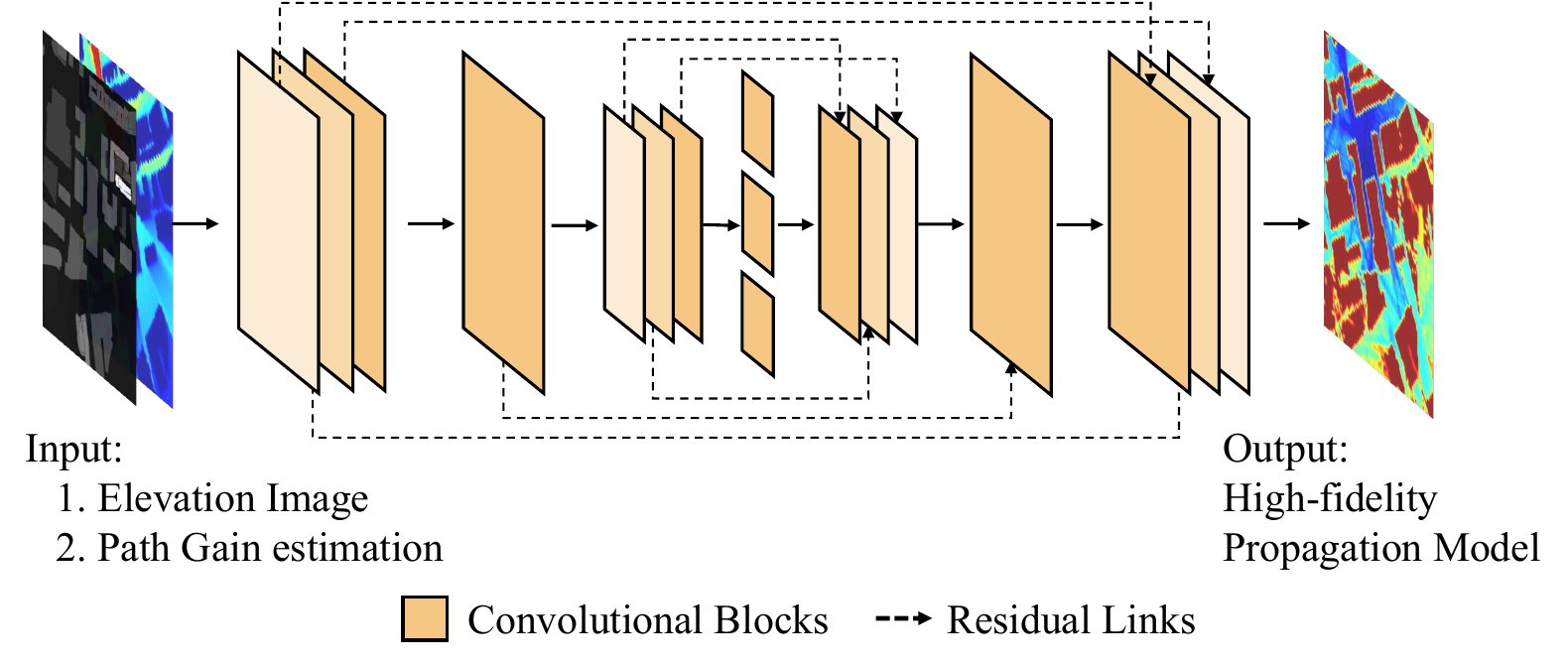}
    \caption{U-Net architecture adapted from PMNet~\cite{lee2023scalable}. We use elevation maps and propagation estimates as inputs to achieve an accurate propagation model.}
    \label{chap3.2-fig:dl_arch}
\end{figure}

We propose a novel model shown in Figure~\ref{chap3.2-fig:dl_arch} inspired by PMNet~\cite{lee2023scalable}, which originally takes two inputs: (1) a building map, indicating building locations with binary values (ones and zeros); and (2) a one-hot encoded \gls{tx} location. We have modified these inputs for improved performance. Instead of using a binary building map, we now input an elevation map ($\textbf{I}_{el}$) that shows building heights (Figure~\ref{chap3.2-fig:building_maps}). Additionally, rather than a one-hot encoded \gls{tx} location, we use a rough estimation of the propagation model derived from real-time ray tracing by Wireless InSite ($\textbf{I}_{est}$), which runs in approximately $30$\:ms, as illustrated in Figure~\ref{chap3.2-fig:rough}. The \gls{tx} location is always at the center of the image, so there is no need to include the \gls{tx} location in another channel. These modifications to the input enable our model to deliver high-fidelity propagation predictions in real time for different scenarios.

The \gls{dl} model is designed to predict the path gain ($PG_{dB}(\boldsymbol{q}_{\mathrm{RX}})$) given the input $\textbf{x}$ and the model parameters $\theta$ (weights and biases). However, the model's output is not limited to the path gain for a specific location, i.e., pixels, but rather it provides the path gain for an entire area, i.e., a heatmap. Mathematically, this can be represented as \(P(PG_{dB}(\boldsymbol{q}_{\mathrm{RX}}) | \textbf{x}, \mathbf{\theta})\), where $\textbf{x}$ is the concatenated input $[\textbf{I}_{el}, \textbf{I}_{est}]$, and $\theta$ encompasses all the parameters of the model. The objective is to learn the distribution of path gain conditioned on the input data and model parameters. This is achieved by optimizing the model parameters $\theta$ during training to minimize the prediction error.

As mentioned in Section~\ref{chap3.2-subsec:airmap-data}, the main dataset consists of $61$~different scenarios (\gls{tx} locations), each with $100$~augmented scenarios, resulting in a total of $6,100$~images. To validate the robustness and ensure a fair comparison of the model, we randomly split the data into different ratios ten times. This approach helps to: (1) ensure there is no bias towards specific scenarios; and (2) evaluate the model's generalization across various test ratios. Additionally, the trained model is tested in a different type of environment at Fenway Park with 16 \glspl{tx} to further assess its performance. Besides ray tracing, measurement data is also used to validate the model. The entire process is illustrated in the diagram in Figure~\ref{chap3.2-fig:diag}.

\subsection{Performance Results}
\label{chap3.2-subsec:airmap-results}

\begin{figure}[htb]
    \centering
    \begin{subfigure}{0.49\linewidth}
        \centering
        \includegraphics[width=\linewidth]{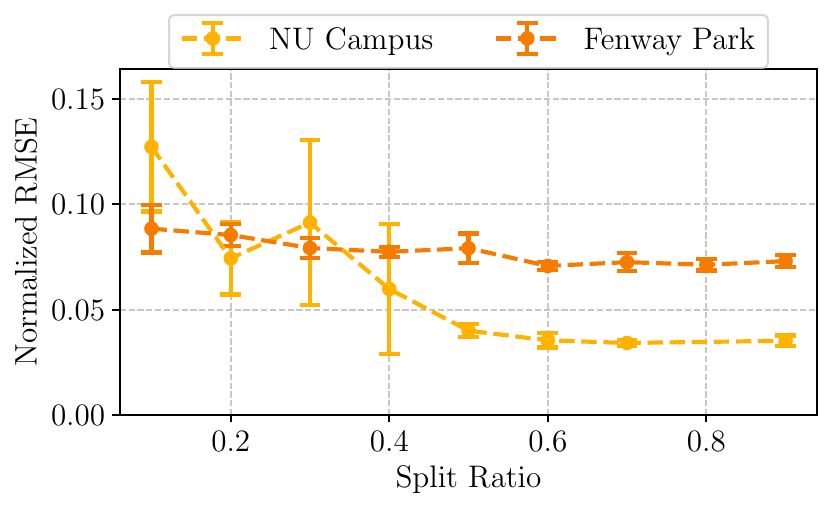}
        \caption{\acrshort{rmse} versus split ratio to evaluate the model's generalizability.}
        \label{chap3.2-fig:split}
    \end{subfigure}
    \hfill
    \begin{subfigure}{0.49\linewidth}
        \centering
        \includegraphics[width=\linewidth]{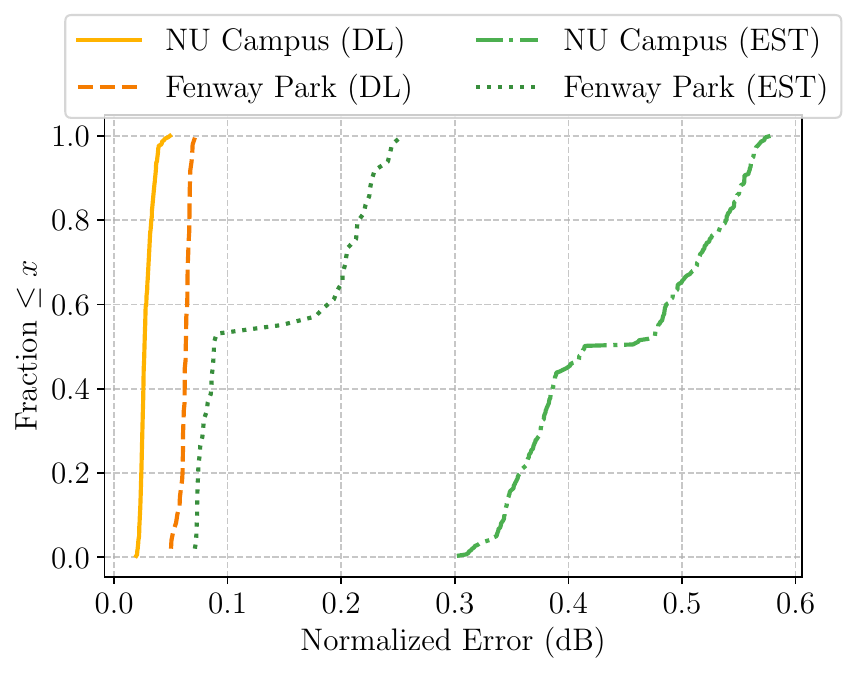}
        \caption{\acrshort{ecdf} for the \acrshort{dl} model compared to real-time propagation modeling from the WI model (EST).}
        \label{chap3.2-fig:ecdf}
    \end{subfigure}
    \caption{Model performance comparison at Northeastern University and Fenway Park.}
    \label{chap3.2-fig:result}
\end{figure}

As described in Section~\ref{chap3.2-subsec:airmap-data}, the model is trained and tested with different split ratios using 10-fold cross-validation to find the optimal configuration. The results are shown in Figure~\ref{chap3.2-fig:split}. As observed, the model stabilizes after a $0.6$~split ratio, indicating that no further training is required for effective inference. The error has been normalized across all plots, with the path gain range spanning from $-50$ to $-250$\:dB.

Also, the \gls{ecdf} of error for the best-performing model across 10-fold cross-validation has been plotted using two different test datasets (unseen by the model): (1) the same area at Northeastern; and (2) a different area at Fenway Park (42°20'26"N, 71°05'38"W), where the environment is more open compared to the dense structure of the Northeastern Campus. 

Results in Figure~\ref{chap3.2-fig:ecdf} show that the model outperform traditional propagation estimation. Specifically, the median error for \gls{dl} at Northeastern Campus is $0.0268$, and at Fenway Park it is $0.0484$. However, for the traditional model, the median error at Northeastern Campus is $0.4146$, and at Fenway Park it is $0.0778$. These results highlight the model's robustness and adaptability to different environments, maintaining superior performance in both familiar and new scenarios.

\begin{figure}[htb]
    \centering
    \includegraphics[width=0.6\textwidth]{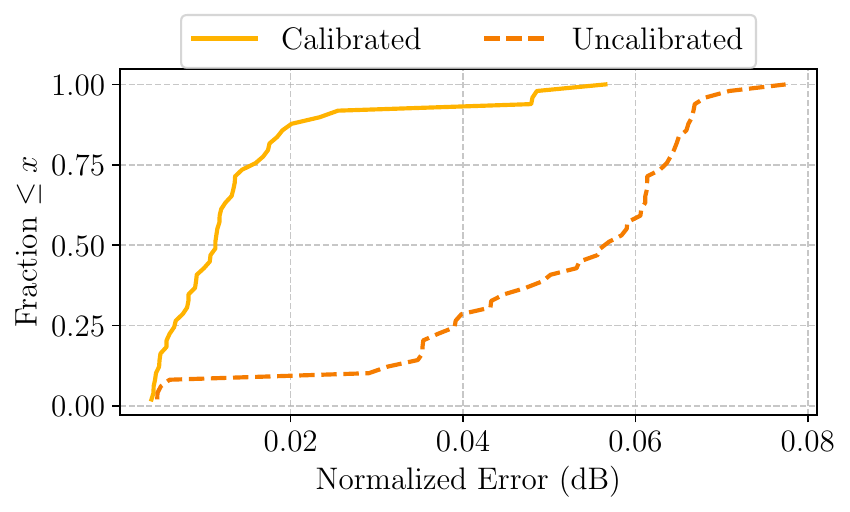}
    \caption{\acrshort{ecdf} of error w/ and w/o calibration using the measurement data (10\% of the dataset).}
    \label{chap3.2-fig:ecdf-meas}
\end{figure}

In addition to ray tracing, we evaluate the model's performance after adding measurement data in the training phase (Figure~\ref{chap3.2-fig:ecdf-meas}). For this, we use 90\% of the points for testing. Initially, the error is relatively high (median of $0.0569$\:dB), but after refining the model with a small amount of measurement data, the performance improves significantly, reducing the median error to $0.0113$\:dB. This demonstrates that the \gls{dl} model can be calibrated with measurement data, unlike common ray tracing software.

Besides the low overall error of the propagation model, using \gls{dl} models enables near real-time accurate propagation modeling. From our test benches, real-time ray tracing by Wireless Insite takes approximately $30$\:ms to run, while high-fidelity ray tracing takes over $387.6$\:s to complete. However, 
our model requires only $46$\:ms on a GPU and $183$\:ms on a CPU to run with minimal error.

\subsection{Integration with the Colosseum Emulator}
\label{chap3.2-subsec:airmap-colosseum}

\begin{figure}[hbt]
    \centering
    \includegraphics[width=0.7\linewidth]{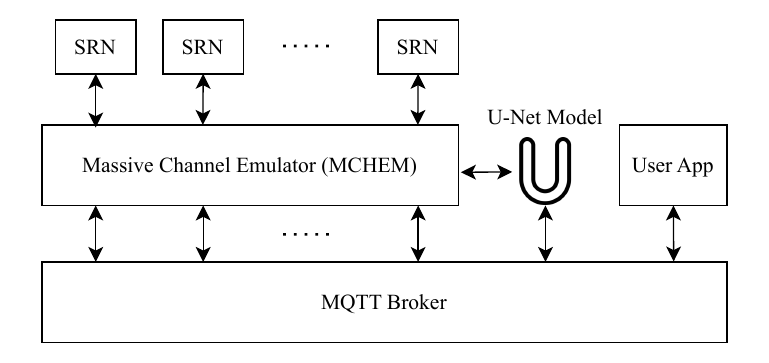}
    \caption{Real-time channel emulation pipeline on Colosseum. User-selected locations trigger U-Net inference, with predicted taps sent via MQTT to \acrshort{mchem} for emulation.}
    \label{chap3.2-fig:mqtt}
\end{figure}

To enable real-time channel emulation as well as simulation using our U-Net model, we integrated it into the Colosseum platform. We followed the approach introduced in ColosSUMO~\cite{gemmi_colossumo_2024}, which extends Colosseum with real-time scenario generation capabilities. As shown in Fig.~\ref{chap3.2-fig:mqtt}, when a user selects or updates a location, the corresponding building elevation map is loaded, transformed into the appropriate input shape and format within $20$~ms, and passed to our U-Net model. The model performs inference in approximately $4$~ms, producing a path gain prediction. This result is then sent to an MQTT broker, which forwards it to Colosseum's \gls{mchem}, the FPGA-based Massive Channel Emulator. \gls{mchem} then uses these taps to emulate the corresponding wireless channel in real time.

\begin{figure}[htb]
    \centering
    \includegraphics[width=0.7\textwidth]{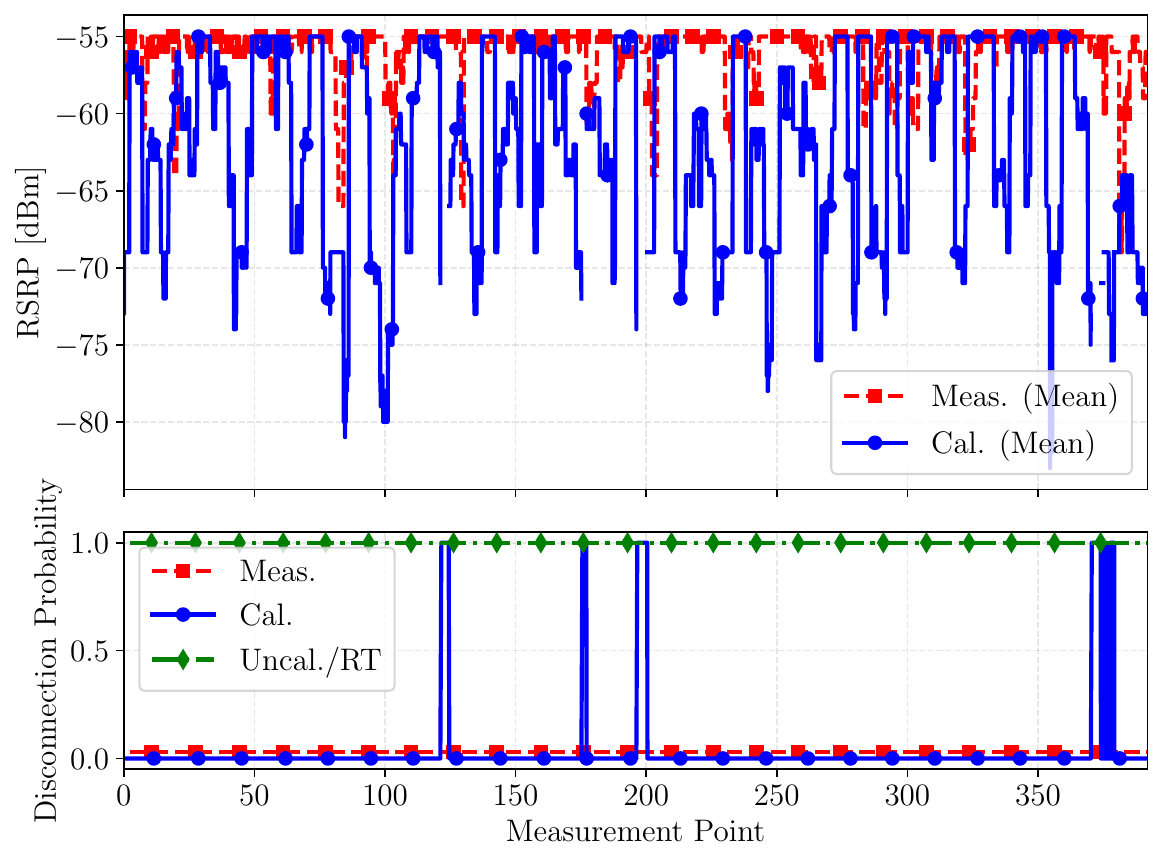}
    \caption{Comparison of \acrshort{rsrp} reports from \acrshort{oai} on Colosseum (10 runs per experiment). The Calibrated U-Net closely matches measurements, while Ray Tracing and Uncalibrated U-Net failed to establish a connection in this setup.}
    \label{chap3.2-fig:rsrp_oai}
\end{figure}

In Figure~\ref{chap3.2-fig:rsrp_oai}, we show results from running \gls{oai} on the Colosseum testbed using single-tap time-domain channels with simple propagation delays to simulate mobile scenarios across four channel models: Ray Tracing, Calibrated U-Net, Uncalibrated U-Net, and actual measurements. Each experiment was repeated 10 times. Both the Ray Tracing and Uncalibrated U-Net models failed to establish a connection under the same configuration, resulting in no \gls{rsrp} data. In contrast, the Calibrated U-Net model successfully produced \gls{rsrp} patterns closely matching empirical measurements, with an \gls{rmse} of $8.86$~dBm.
%

\subsection{Conclusions}
\label{chap3.2-subsec:airmap-conclusions}

This section presented a real-time path gain estimator leveraging advanced \gls{dl} techniques for radio map creation. 
The initial approach~\cite{saeizadeh2024camad} integrates elevation maps and rough propagation model estimations as inputs to a U-Net architecture, achieving a normalized \gls{rmse} of less than $0.035$~dB across a $37{,}210$ square meter area with inference times of $46$~ms on GPU and $183$~ms on CPU---a significant advancement over traditional high-fidelity ray tracing methods, which typically require over $387.6$~seconds.
The measurement campaign conducted at Northeastern University enabled model calibration through transfer learning, reducing the median error to $0.0113$~dB with only 10\% of the data used for calibration.

The extended framework, AIRMap~\cite{saeizadeh2025airmap}, further simplifies the approach by relying solely on a single-channel elevation map as input, reducing inference time to under $4$~ms while maintaining sub-$5$~dB \gls{rmse}.
This resolution-adaptive design supports radio-map estimation across physical areas ranging from $500$~m to $3$~km per side without altering the architecture.
Integration into the Colosseum emulator demonstrated the practical feasibility of \gls{ai}-driven channel models for real-time wireless network emulation, with the calibrated model successfully producing \gls{rsrp} patterns matching empirical measurements (\gls{rmse} of $8.86$~dBm) while ray tracing and uncalibrated models failed to establish connections.

This work demonstrates how \gls{dt} platforms like Colosseum can serve both as data collection environments for training \gls{ai} models and as deployment systems for real-time channel emulation.
The ability to generate accurate radio maps in milliseconds---rather than minutes or hours---enables new use cases such as dynamic scenario adaptation, real-time mobility modeling, and closed-loop network optimization that were previously impractical with traditional ray-tracing methods.

\section{Generative Models for Synthetic RF Data}
\label{chap3.3-sec:gentwin}

The realization of efficient \gls{ai} solutions for the optimization of next-generation \gls{ran} relies on the availability of expansive, high-quality datasets that accurately capture nuanced, site-specific conditions.
However, obtaining such abundant, domain-specific measurements poses a significant challenge, especially as network complexity and energy efficiency demand surge toward \gls{6g}.
As discussed in Section~\ref{chap3.2-sec:airmap}, creating accurate \gls{dt} representations requires substantial data, and field measurements are costly, time-consuming, and quickly become outdated in dynamic environments.

To address these challenges, this section presents \gls{gentwin}, a \gls{genai}-enabled \gls{dt} platform that leverages a novel \textit{soft}-\gls{gan} based on \gls{lstm} layers (soft-GAN).
This model is designed to generate synthetic \gls{dtran} \gls{rf} data for both transmitter and receiver ends, augmenting the limited datasets collected through field measurements by synthetically replicating key features of \gls{rf} signals.
Building upon the Colosseum \gls{dt} platform described in Chapter~\ref{chap:2}, \gls{gentwin} demonstrates how testbeds can serve not only as testing environments but also as data sources for training generative models.

The key contributions of this section include:
\begin{itemize}
    \item The \gls{gentwin} platform, integrating \glsentrylong{dt} and \glsentrylong{genai} layers.
    \item A soft-attention \gls{lstm}-based time-series \gls{gan} model, named \textit{soft}-\gls{gan}, specifically for \glspl{dtran}.
    \item Performance evaluation comparing \textit{soft}-\gls{gan} against various baseline generative models, demonstrating up to 19\% improvement in accuracy.
\end{itemize}

The remainder of this section is organized as follows.
Section~\ref{chap3.3-subsec:gentwin-platform} presents the Gen-TWIN platform architecture, describing both the \gls{dt} layer built on Colosseum and the \gls{genai} layer.
Section~\ref{chap3.3-subsec:gentwin-softgan} and \ref{chap3.3-subsec:gentwin-data} detail the soft-GAN model and the data collection methodology, respectively.
Section~\ref{chap3.3-subsec:gentwin-results} presents the experimental evaluation, comparing soft-GAN against baseline generative models.
Finally, Section~\ref{chap3.3-subsec:gentwin-conclusions} concludes this section.

\subsection{Gen-TWIN Platform}
\label{chap3.3-subsec:gentwin-platform}

\begin{figure*}[htb]
  \centering
  \includegraphics[width=\textwidth]{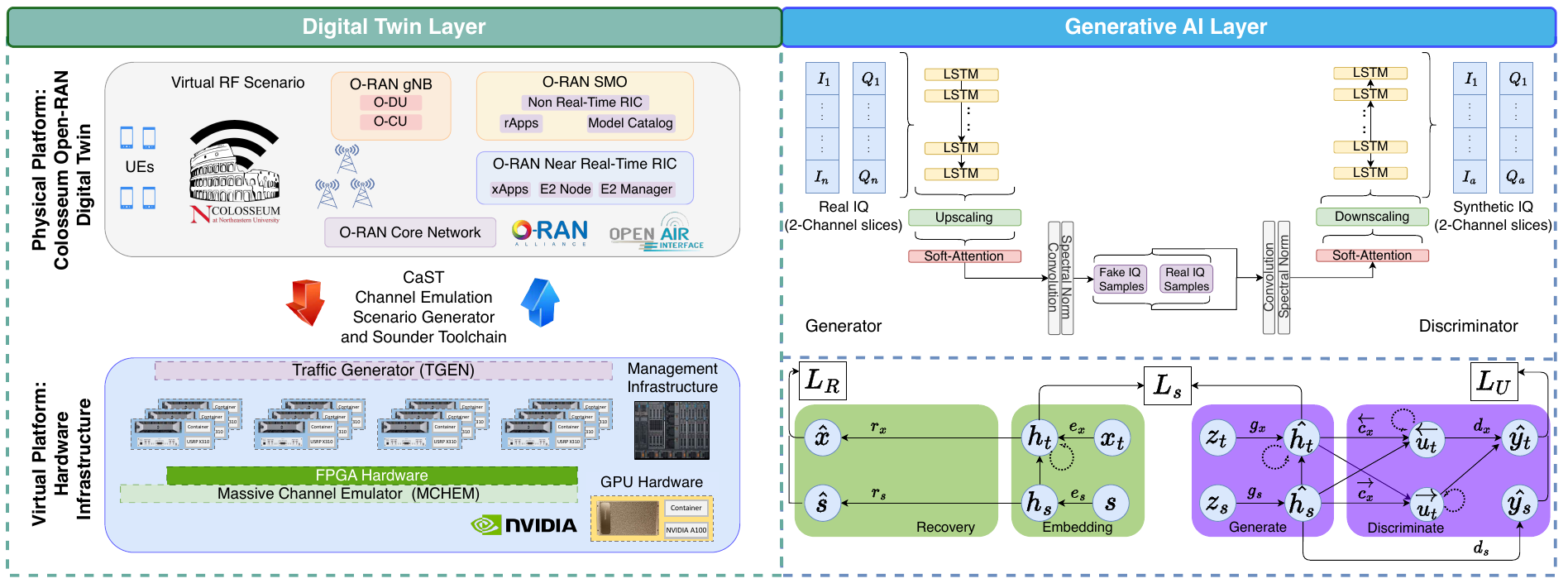}
  \caption{Left: O-RAN Digital Twin Layer; Colosseum—the world's largest wireless network emulator with hardware in the loop~\cite{bonati2021colosseum} to create a comprehensive twin of the O-RAN network. Right: Generative AI Layer; Proposed \textit{soft}-GAN Model --- soft-attention-LSTM based Generative Adversarial Network.}
  \label{chap3.3-fig_gentwin}
\end{figure*}

Our proposed platform comprises two integral components, \gls{dt} and \gls{genai} layers, as shown in Figure~\ref{chap3.3-fig_gentwin}.

\textbf{Digital Twin Layer:}
The \gls{dt} layer leverages Colosseum, as the O-RAN \gls{dt} for the network emulator part, and \gls{cast}~\cite{villa2022cast} for data collection, both described in Chapter~\ref{chap:2}.

\textbf{Generative AI Layer:} Our \gls{gan} framework can be described by the following minmax game:
\begin{equation}
    \min_G \max_D \, \mathbb{E}_{\mathbf{x} \sim p_{\text{data}}(\mathbf{x})} [\log D(\mathbf{x})] + \mathbb{E}_{\mathbf{z} \sim p_{\mathbf{z}}(\mathbf{z})} [\log (1 - D(G(\mathbf{z})))],
\end{equation}
where:
\begin{itemize}
    \item $\mathbf{x}$ represents real \gls{iq} data samples from \gls{dt} layer.
    \item $p_{\text{data}}(\mathbf{x})$ denotes the distribution of real time series data.
    \item $\mathbf{z}$ represents random noise vectors sampled from a prior distribution $p_{\mathbf{z}}(\mathbf{z})$ (e.g., a Gaussian distribution).
    \item $G(\mathbf{z})$ generates synthetic \gls{iq} data samples from the noise vector $\mathbf{z}$.
    \item $D(\mathbf{x})$ represents the probability that $\mathbf{x}$ originates from the real data distribution rather than from $G$.
\end{itemize}
In the context of \gls{iq} data augmentation, the generator $G$ learns to produce realistic \gls{iq} data sequences, while the discriminator $D$ tries to distinguish between real and synthetic \gls{iq} data. 

The training process iteratively updates the parameters of \(G\) and \(D\) using gradient-based optimization. Specifically, the discriminator is trained to maximize the likelihood of correctly distinguishing between real and synthetic samples:
\begin{equation}
    \mathcal{L}_D = -\mathbb{E}_{x \sim p_{\text{data}}(x)}[\log D(x)] - \\ \mathbb{E}_{z \sim p_z(z)}[\log (1 - D(G(z)))].
\end{equation}

The generator is trained to minimize the likelihood of the discriminator correctly distinguishing synthetic from real samples:
\begin{equation}
    \mathcal{L}_G = -\mathbb{E}_{z \sim p_z(z)}[\log D(G(z))].
\end{equation}
These loss functions are alternately optimized at each iteration, leading to gradual improvement in both networks. Once trained, the generator can produce synthetic univariate time series that augment the original dataset.
Given a noise vector \(z\), the generator produces a synthetic time series $\hat{x} = G(z)$.
These synthetic samples can be used to expand the training dataset, thereby improving the performance and generalization of machine learning models.
The augmented dataset \( \mathbf{X}' \) can be represented as $\mathbf{X}' = \mathbf{X} \cup \{G(z_i)\}_{i=1}^{N}$ where \( \mathbf{X} \) is the original dataset and \( N \) is the number of synthetic generated samples.

\subsection{Data Augmentation with \textit{soft}-GAN}
\label{chap3.3-subsec:gentwin-softgan}

\subsubsection{Fundamentals of \textit{soft}-GAN Model}

The values of the time series data are denoted as $m_{i,t} \in \mathbb{R}$, where $i \in \{1, 2, 3, \dots, N\}$ represents the index of individual samples and $t \in \{1, 2, 3, \dots, T\}$ indicates the time points.
The associated matrix of time feature vectors is $X_{1,T} = (x_1, \dots, x_T)$ in $\mathbb{R}^{D \times T}$, where $D$ represents the number of time features.
Consequently, we take the time series to have a fixed length $\zeta$, denoted as $\hat{M}_{i,t,\zeta} = (\hat{m}_{i,t}, \dots, \hat{m}_{i,t,\zeta})$.
Then, using a generator function $G$ and a fixed time sequence $t$, we can model $\hat{M}_{i,t,\zeta} = G(n, \phi(i), X_{t:t+\zeta})$, where $n$ is a noise vector and $\phi$ is an embedding function that maps the index of the time series to a vector representation.
In the generator $G$ and discriminator $D$, we use a soft attention mechanism along with convolutional spectral normalization $SN$ as described in~\cite{miyato2018spectral}.

For a sequence of length $l$, we can write:
\begin{equation}
\begin{split}
    p: \mathbb{R}^{n_{f} \times l} &\rightarrow \mathbb{R}^{n_{f}^{'} \times l} \\
    x &\longmapsto \gamma \, \text{softA}(f(x) + f(x))
\end{split}
\label{mainfunc}
\end{equation}
\begin{equation}
    f(x) = SN(LeakyReLU(c(x)))
    \label{relu},
\end{equation}
where $n_{f}^{'}$ is the number of output features, $c$ is the convolution operator, and \textit{softA} is the soft-attention mechanism.
We use \textit{LeakyReLU}~\cite{xu2015empirical} to avoid the well-known dying ReLU problem, where neurons can become inactive and always output zero for any input.
On the generator side, we add a spectral normalization layer to further stabilize the training process and prevent sudden escalation of the gradient magnitude~\cite{zhang2019self}.
$\phi$ is a learnable parameter that, when indexed to 0, influences the network's ability to learn local features through the function $p$.


\textbf{soft-GAN Generator:} The first layer of the generator, $G1$, uses the main function block $p$:
\begin{equation}
\begin{split}
    G1: \mathbb{R}^{n_{f}^{'} \times 2^3} &\rightarrow \mathbb{R}^{n_{f} \times 2^3} \\
    \hat{M}_0 &\longmapsto \hat{M}_1 = p(\hat{M}_0),
\end{split}
\label{gen1}
\end{equation}
where $i \in [2, L]$. $G1$ maps two channel-separated \gls{iq} sequences $\hat{M}_{i-1}$ to an output $\hat{M}_{i}$, then applies an upscale to the function block $p$:
\begin{equation}
\begin{split}
    G_i: \mathbb{R}^{n_{f} \times 2^{i+2}} &\rightarrow \mathbb{R}^{n_{f} \times 2^{i+3}} \\
    \hat{M}_{i-1} &\longmapsto \hat{M}_i = p(\text{UP}(\hat{M}_{i-1})).
\end{split}
\label{gen2}
\end{equation}
In the last layer of the generator, a multivariate sequence is obtained univariately as $\hat{M}_{i,t,\zeta}$ through a one-dimensional convolution operation.

\begin{algorithm}[htb]
\caption{\textit{soft}-GAN Generator}
\begin{algorithmic}[1]
\REQUIRE Noise vector \( z \in \mathbb{R}^{(batch\_size, seq\_len, input\_dim)} \)
\STATE \textbf{Input:} Real IQ Samples
\STATE \textbf{First LSTM Layer:} Process IQ and \( z \) with an LSTM layer having 256 units, return sequences, dropout of 0.3, and recurrent dropout of 0.3.
\STATE \( h_1 = \text{LSTM}_{256}(z) \)
\STATE \textbf{Batch Normalization:} Normalize the output.
\STATE \( h_1 = \text{BatchNorm}(h_1) \)
\STATE \textbf{Second LSTM Layer:} Process the output with another LSTM layer having 256 units, return sequences, dropout of 0.3, and recurrent dropout of 0.3.
\STATE \( h_2 = \text{LSTM}_{256}(h_1) \)
\STATE \textbf{Batch Normalization:} Normalize the output.
\STATE \( h_2 = \text{BatchNorm}(h_2) \)
\STATE \textbf{Third LSTM Layer:} Process the output with another LSTM layer having 128 units, return sequences, dropout of 0.3, and recurrent dropout of 0.3.
\STATE \( h_3 = \text{LSTM}_{128}(h_2) \)
\STATE \textbf{Spectral Normalization:} Normalize the output.
\STATE \( h_3 = \text{SpectNorm}(h_3) \)
\STATE \textbf{Attention Mechanism:} Apply soft-attention to the output and concatenate it with the LSTM output.
\STATE \( \text{attention\_out} = \text{Attention}(h_3, h_3) \)
\STATE \( \text{context\_vector} = \text{Concat}(h_3, \text{attention\_out}) \)
\STATE \textbf{Fully Connected Layer:} Apply a dense layer with 128 units and LeakyReLU activation.
\STATE \( d_1 = \text{Dense}_{128}(\text{context\_vector}, \text{activation='LeakyReLU'}) \)
\STATE \textbf{Dropout:} Apply dropout with a rate of 0.3.
\STATE \( d_1 = \text{Dropout}_{0.3}(d_1) \)
\STATE \textbf{Output:} Apply a final dense layer with 2 units to produce the synthetic IQ data.
\STATE \( \text{output} = \text{Dense}_{2}(d_1) \)
\STATE \textbf{Return:}  \( \text{Samples} \)
\end{algorithmic}
\end{algorithm}

\textbf{soft-GAN Discriminator:} The discriminator design is the same as the generator.
The discriminator maps the generator's output $\hat{M}_{i,t,\zeta}$ and $X_{t:t+\zeta}$ to a discriminator score $d$.
The \textit{LeakyReLU} activation function is then used as follows:
\begin{equation}
    d_{L+1}: \mathbb{R}^{1+D, \zeta} \rightarrow \mathbb{R}^{n_{f}, \zeta}
\end{equation}  
\begin{equation}
    (\hat{M}_{L+1}, X_{t:t+\zeta}) \longmapsto \hat{M}_L = \text{LR}(c_{1}(\hat{M}_{L+1}, X_{t:t+\zeta})),
\label{disc}
\end{equation}
where $i \in [2, L]$.
Downscaling is then applied to the main function block:
\begin{equation}
\begin{split}
    d_{i}: \mathbb{R}^{n_{f} \times 2^{i+3}} &\rightarrow \mathbb{R}^{n_{f} \times 2^{i+2}} \\
    Y_{i} &\longmapsto Y_{i-1} = \text{DOWN}(m(Y_i)).
\end{split}
\label{down}
\end{equation}    
The final operation is handled by a fully connected layer:
\begin{equation}
\begin{split}
    d_{1}: \mathbb{R}^{n_{f}} &\rightarrow \mathbb{R} \\
    Y_{i} &\longmapsto Y_{0} = \text{SN}(\text{FC}(\text{LeakyReLU}(\text{softA}(c(Y_1)))))
\end{split}
\end{equation}

\subsection{Data Collection}
\label{chap3.3-subsec:gentwin-data}

\begin{figure}
    \centering
    \includegraphics[width=1\linewidth]{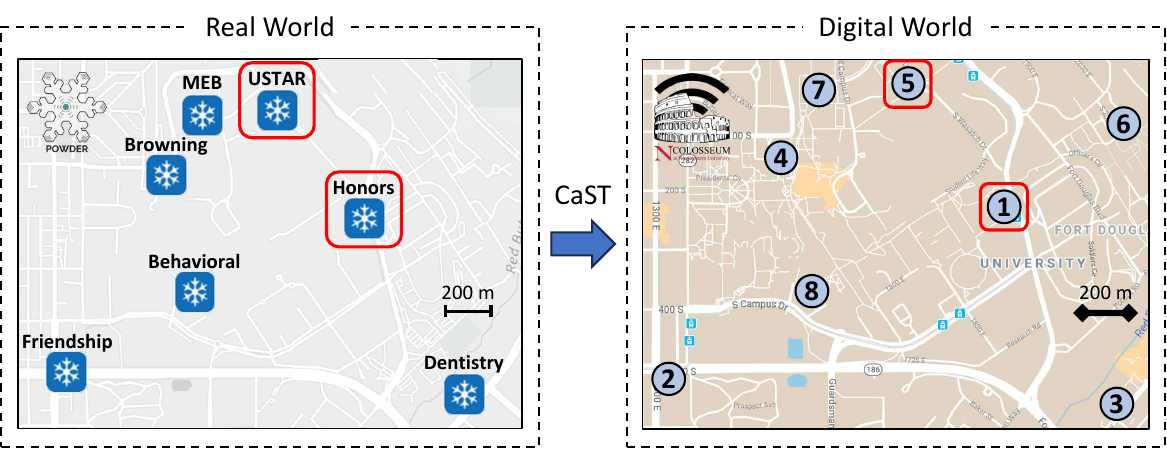}
    \caption{Data collection scenarios for the real world (POWDER, left picture) and digital world (Colosseum, right picture). The red circles indicate the actual nodes used for data collection. Colosseum node 6 is a remnant of the now-decommissioned POWDER Medical Tower Rooftop node.}
    \label{chap3.3-fig:collection-scenario}
\end{figure}

The data collection operation aims to collect \gls{iq} samples that can serve as a baseline for the data augmentation step.
We collect \gls{iq} samples through two different testbeds: (i) POWDER for real-world data, and (ii) Colosseum for emulated data.
A summary of the parameters for the data collection is shown in Table~\ref{chap3.3-table:data-collection}.
Additionally, Figure~\ref{chap3.3-fig:collection-scenario} depicts the data collection scenarios for the real world (POWDER) and the digital world (Colosseum).
The POWDER testbed~\cite{breen2020powder} is a city-scale facility located at the University of Utah in Salt Lake City, UT.
The platform supports a wide range of hardware and software configurations, enabling experimentation with real-world \gls{ota} scenarios and the development of new wireless technologies.
It provides various types of end-to-end software-defined nodes with different capabilities, deployed across the university campus.

In our real-world \gls{ota} data collection, we leverage two of its rooftop base station nodes, as shown by the red circles in Figure~\ref{chap3.3-fig:collection-scenario}: Honors and USTAR.
For the emulation data, we utilize Colosseum's capabilities for digitizing the real world~\cite{villa2024dt}.
We use a \gls{dt} representation of the POWDER platform to recreate a similar real-world environment in an emulated one.
Thus, we use the POWDER scenario described in~\cite{bonati2021scope} to collect data between the same digitized nodes for Honors and USTAR rooftop base stations, digitized as nodes 1 and 5, respectively.

\begin{table}
    \centering
    \footnotesize
    \caption{Data collection parameters summary between the two testbeds: POWDER for OTA data, and Colosseum for emulated data.}
    \label{chap3.3-table:data-collection}
    \begin{tabular}{@{}lll@{}}
        \toprule
        Parameter & POWDER & Colosseum \\
        \midrule
        SDR & USRP X310 & USRP X310 \\
        Nodes & Honors and USTAR Rooftop BS & Digitized Honors and USTAR Nodes \\
        Frequency & 3.46-3.465 GHz & 1 GHz \\
        Bandwidth & 1-10 MHz & 80 MHz \\
        Time Capture & 30 seconds & 30 seconds \\
        Capture Size & 2,57 GB& 2,96 GB \\
        Sample Time & 10 MS/s & 10 MS/s \\
        Tx gain & 25 dB & 15-25 dB \\
        Rx gain & 25 dB & 15 dB \\
        \bottomrule
    \end{tabular}
\end{table}

As shown in Table~\ref{chap3.3-table:data-collection}, in both cases, we leverage USRP X310 as \gls{sdr} hardware.
The frequencies used for POWDER fall in the CBRS band between 3.46 and 3.465 GHz with bandwidths of 1 MHz and 10 MHz, respectively.
In contrast, Colosseum uses its typical emulation of the baseband signal with a frequency of 1 GHz and a bandwidth of 80 MHz.
The data capture duration for both platforms is 30 seconds per use case, with a sampling frequency of 10 megasamples per second.
The transmission and reception gains of each \gls{sdr} vary between 15 and 25 dB to compensate for the long distance of about 800 meters between the two nodes. 

In each scenario, we leverage CaST~\cite{villa2022cast}, which is based on the GNU Radio software development toolkit, to send a \gls{bpsk}-modulated \gls{glfsr} code sequence from the sender node (Honors) and to store the raw received \gls{iq} samples without any processing or equalization on the receiver side (USTAR).

\subsection{Experiment Evaluation}
\label{chap3.3-subsec:gentwin-results}

\subsubsection{Hyper-parameter Tuning and Training}

In our training stage, the initial step in handling the dataset involves splitting it into three distinct parts: 80\% training set, 10\% validation set, and 10\% test set.
This division ensures that the model has adequate data for training, tuning, and evaluation, ultimately leading to better performance and generalization.
Preprocessing is a critical step in preparing the data for training.
The main goal is to normalize the \gls{iq} data to ensure consistency and stability during training.
Normalization involves scaling the data so that its values fall within a specific range, -1 to 1.
We achieve this by computing the mean ($\mu$) and standard deviation ($\sigma$) of the training data and then normalizing each data point accordingly.

\begin{table}[htb]
    \centering
    \footnotesize
    \caption{Hyper-parameter tuning stage of \textit{soft}-GAN}
    \label{chap3.3-table:hyperparam}
    \begin{tabular}{@{}lll@{}}
    \toprule
        Hyper-param. & Search Space & Selected Parameter \\
         \midrule
        Input size  & [2, 4, 8, 16, 32] & 8 \\
        Hidden layers & [1, 2, 3] & 3 \\
        Hidden layers & [0.5, 1.5, 2.0, 3.0, 5.0]	 & 2.0 $x$ Input Size\\
        Learning rate & [0.0001, 0.0005, 0.001, 0.005, 0.01] & 0.005\\
        Batch size & [64, 128, 256, 512, 1024] & 256 \\
        Loss Function & [MAE, MSE, sMAPE] & MAE \\
        Training steps & [1000, 5000, 10,000, 20,000, 40,000] & 20,000 \\
        Optimizer & [RMSprop, SGD, Adam] & Adam \\
    \bottomrule
    \end{tabular}
\end{table}

\begin{figure}[htb]
    \centering
    \includegraphics[width=1\linewidth]{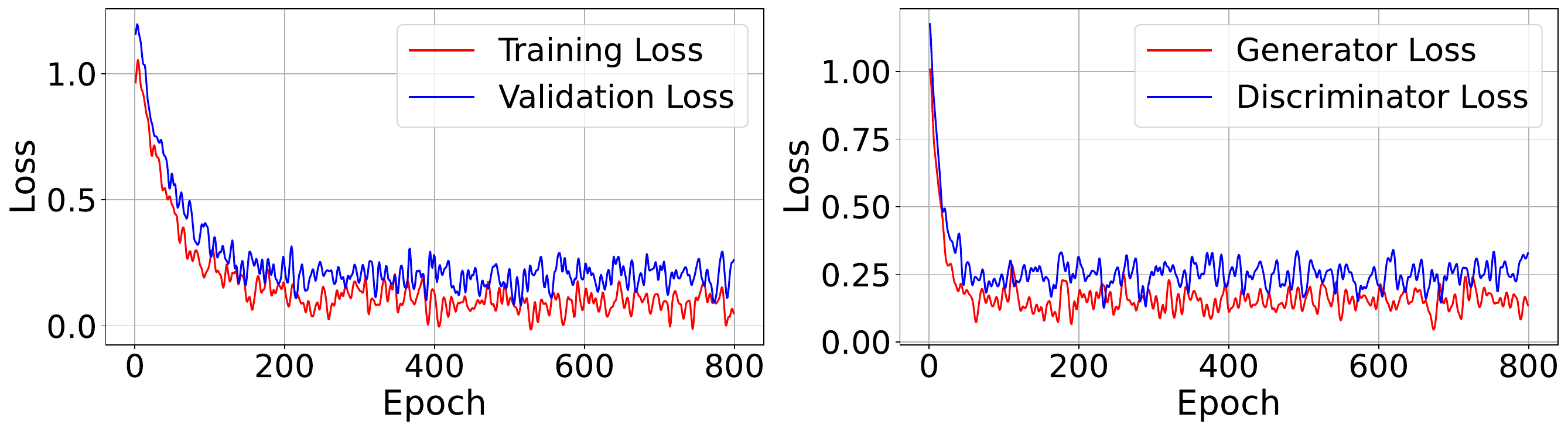}
    \caption{\textit{soft}-GAN training and validation loss over epochs (left picture). Generator and discriminator loss over epochs (right picture).}
    \label{chap3.3-fig:losses}
\end{figure}

Table~\ref{chap3.3-table:hyperparam} shows our parameter space.
During the hyperparameter tuning stage, an input size of 8 achieves a balance between model complexity and performance.
Larger input sizes increase the model's complexity without significantly improving performance.
Having three hidden layers allows the model to capture more complex patterns in the data.
Setting the hidden layer size to $2 \times Input Size$ ensures that the network has sufficient neurons to learn intricate features of \gls{iq} samples without overcomplicating the model.
A learning rate of 0.005 is a moderate choice that facilitates faster convergence while avoiding the pitfalls of overshooting the optimal point, which can occur with a higher learning rate.
A batch size of 256 provides a good compromise between training stability and computational efficiency.
We found that larger batch sizes require more memory and longer computation times per update, while smaller batch sizes lead to noisy gradient estimates.

As shown in Figure~\ref{chap3.3-fig:losses} (left), both training and validation losses start at a relatively high value, around 1.2.
This is expected as the model begins learning the data patterns from scratch.
As training progresses, both losses decrease significantly, stabilizing at lower values, around 0.2.
The training loss declines more smoothly than the validation loss, which shows more fluctuations.
However, these fluctuations are within a narrow range, and both losses remain close to each other throughout the training process.
This behavior suggests that the model is learning effectively without overfitting.
If the model were overfitting, we would observe the validation loss diverging from the training loss, with the validation loss increasing while the training loss continues to decrease.
Conversely, underfitting would be indicated by both losses remaining high and not decreasing significantly.
The close alignment between the training and validation losses in our model demonstrates a good balance, indicating that the \textit{soft}-GAN generalizes well to unseen data.

\renewcommand{\baselinestretch}{1}
\begin{table}[htb]
\centering
\caption{NRMSE and Context FID-Scores (Lower values in these metrics indicate better model performance)}
    \label{chap3.3-tableIII}
    \begin{tabular}{*{8}{l}}
    \toprule
    &   \multicolumn{2}{c}{NRMSE} 
    &   \multicolumn{2}{c}{Context FID-scores} 
    \\
    \cmidrule(r) {2-3}
    \cmidrule(lr){4-5}
    \cmidrule(lr){6-7}

    Models                          
      &   128    &   256    
          & 128      &   256      \\
    \midrule
TimeGAN   & $0.27\pm 0.06 $        & $0.26\pm 0.02$     &  $0.23\pm 0.04 $  &  $0.63\pm 0.10 $  
     \\
Autoencoder & $0.45\pm 0.07 $    & $0.43\pm 0.01 $     & $0.58\pm 0.03 $  & $0.42\pm 0.10 $  
  \\
BiLSTM & $1.03\pm 0.05 $5    & $0.97\pm 0.003 $     & $1.07\pm 0.20 $       & $0.92\pm 0.07 $  
     \\
BiLSTMGAN &  $0.95\pm 0.06 $   &  $1.13\pm 0.04 $     & $0.71\pm 0.01 $    & $0.23\pm 0.05 $    \\
CnnGAN &  $1.87\pm 0.07 $  &  $2.73\pm 0.02 $  & $1.87\pm 0.09 $ & $1.63\pm 0.06 $    \\
\textbf{soft-GAN (ours)} & $0.23\pm 0.02 $   & \textbf{$0.21\pm 0.07$}  &   $0.15\pm 0.01 $ &  $0.09\pm 0.04 $ \\
    \bottomrule
    \end{tabular}
\end{table}

\renewcommand{\baselinestretch}{1}
\begin{table}[htb]
\centering
\caption{Discriminative and Predictive Score Results (Lower values in these metrics indicate better model performance)}
    \label{chap3.3-tableIV}
    \begin{tabular}{*{8}{l}}
    \toprule
    &   \multicolumn{2}{c}{Discriminative Score} 
    &   \multicolumn{2}{c}{Predictive Score} 
    \\
    \cmidrule(r) {2-3}
    \cmidrule(lr){4-5}
    \cmidrule(lr){6-7}

Models                          
    &   128    &   256    
        & 128      &   256      \\
    \midrule
TimeGAN & $0.028\pm 0.003 $        & $0.013\pm 0.01 $   &  $0.019\pm 0.08 $  &  $0.17\pm 0.03 $  
     \\
Autoencoder & $0.15\pm 0.03 $    & $0.17\pm 0.08 $   & $0.24\pm 0.09 $  & $0.23\pm 0.01 $  
  \\
BiLSTM & $0.04\pm 0.03 $& $0.053\pm 0.002 $   & $0.18\pm 0.04 $       & $0.13\pm 0.04 $  
     \\
BiLSTMGAN &  $0.113\pm 0.09 $   &  $0.088\pm 0.02 $ & $0.27\pm 0.017 $    & $0.09\pm 0.03 $    \\
CnnGAN &  $0.43\pm 0.08 $  &  $0.39\pm 0.01 $ & $0.36\pm 0.008 $ & $0.33\pm 0.004 $  \\
\textbf{soft-GAN (ours)} & $0.017\pm 0.02 $   & $0.013\pm 0.06$  &  $0.016\pm 0.004 $ &  $0.009\pm 0.006 $  \\
    \bottomrule
    \end{tabular}
\end{table}

\renewcommand{\baselinestretch}{1}
\begin{table}[htb]
    \centering
    \caption{Comparison of classification performance of baseline models}
    \label{chap3.3-tableV}
    \begin{tabular}{lcccc}
    \toprule
    Models & Accuracy & Precision & Recall & F1-score \\
    \midrule
    TimeGAN        & 0.83 & 0.68 & 0.84 & 0.87 \\
    Autoencoder    & 0.88 & 0.63 & 0.82 & 0.85 \\
    BiLSTM         & 0.78 & 0.70 & 0.80 & 0.81 \\
    BiLSTMGAN      & 0.83 & 0.73 & 0.81 & 0.83 \\
    CnnGAN         & 0.77 & 0.61 & 0.79 & 0.73 \\
    \textbf{soft-GAN (ours)} & \textbf{0.93} & \textbf{0.78} & \textbf{0.91} & \textbf{0.89} \\
    \bottomrule
    \end{tabular}
\end{table}

Figure~\ref{chap3.3-fig:losses} (right) plot shows the generator and discriminator losses over the epochs.
Initially, both losses start high, around 1.0, which is typical at the beginning of \gls{gan} training.
As training proceeds, the generator loss decreases steadily, stabilizing around 0.1 to 0.2.
The discriminator loss also decreases but at a slower rate and stabilizes around 0.2 to 0.3.
This dynamic is crucial in \gls{gan} training.
The generator aims to produce data that the discriminator cannot distinguish from real data, while the discriminator aims to correctly identify real and generated data.
The observed balance between these losses indicates that the training process is stable.
If the discriminator loss were to decrease too rapidly, it would suggest that the discriminator is too powerful, making it difficult for the generator to improve.
On the other hand, if the generator loss were too low compared to the discriminator loss, it might indicate mode collapse, where the generator produces limited diversity in the generated data.
The balance observed here suggests that both the generator and discriminator are improving in tandem, contributing to the overall performance of the model.
The stability and convergence of these loss curves provide a numerical testament to the model's performance.
The training and validation losses stabilize around 0.2, and the generator and discriminator losses around 0.1 to 0.3, indicating that the model has learned the underlying data distribution effectively.

\subsubsection{Performance Comparison with Baseline Models}

This section provides a comprehensive evaluation of the proposed \textit{soft}-GAN architecture against baseline models for time-series data augmentation, all trained on our dataset, including TimeGAN~\cite{yoon2019time-series}, Autoencoder~\cite{bank2021autoencoders}, BiLSTM~\cite{huang2015bidirectional}, BiLSTMGAN~\cite{benchama2024novel}, and CNNGAN~\cite{radford2016unsupervised}.
The performance metrics presented in Table~\ref{chap3.3-tableIII} include \gls{nrmse} and Context \gls{fid}-scores, which are widely used in computer vision to evaluate the quality of synthetic data~\cite{heusel2018gans}.
Table~\ref{chap3.3-tableIV} includes Discriminative and Predictive Scores, derived as evaluation criterion from baseline work~\cite{yoon2019time-series}, where a post-hoc time-series classification model (2-Layer \gls{lstm}) is trained to distinguish between original and generated sequences, labeled as "real" and "generated," respectively.
The classification error on a held-out test set is reported as a quantitative metric for similarity assessment.
In Table~\ref{chap3.3-tableV}, the results are presented using well-known classification metrics, including accuracy, precision, recall, and F1-score.

\begin{figure}[htb]
    \centering
    \includegraphics[width=0.6\linewidth]{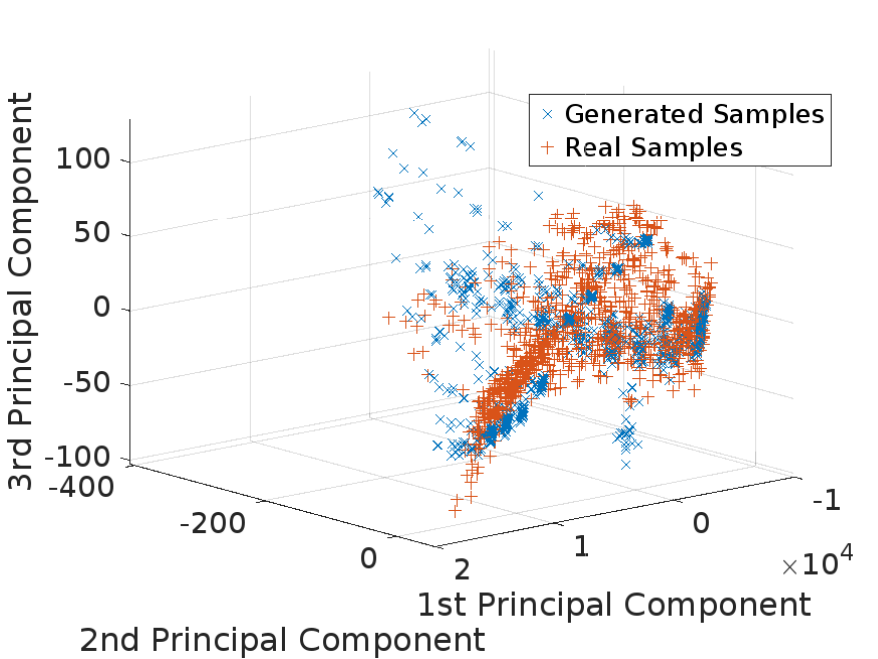}
    \caption{Components of real and generated samples in the PCA subspace;  Generated samples (marked as “×”) exhibit a substantial overlap with the real samples (marked as “+”).}
    \label{chap3.3-fig:pca}
\end{figure}

\begin{figure}[htb]
    \centering
    \includegraphics[width=0.7\linewidth]{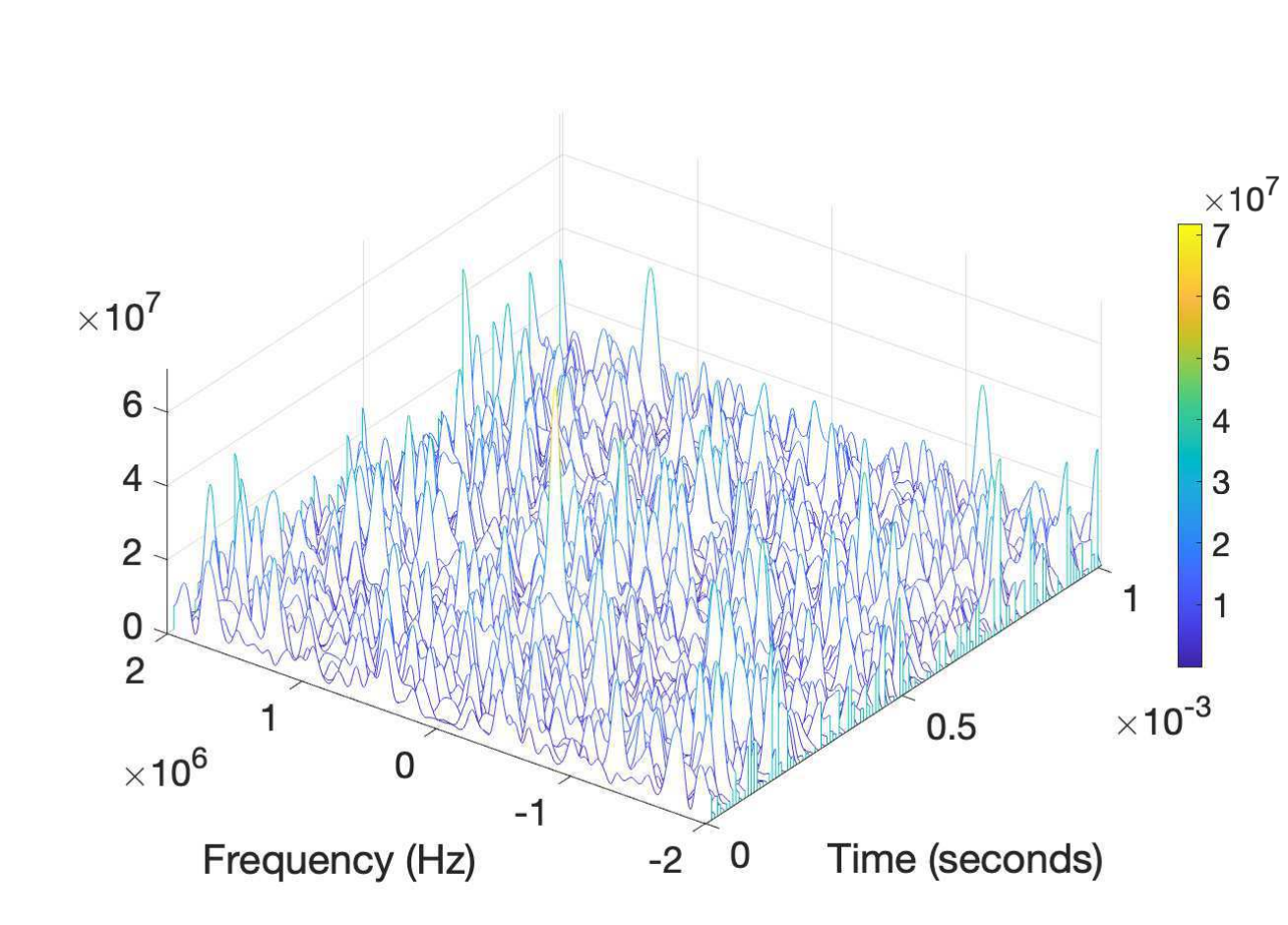}
    \caption{Spectrogram of \textit{soft}-GAN generated synthetic \acrshort{rf} (25 dB gain) Signals}
    \label{chap3.3-fig:spect}
\end{figure}

Performance results show that the use of a soft-attention mechanism allows the model to focus on the most relevant parts of the input sequence, enhancing its ability to capture long-range dependencies and intricate patterns in the time-series data.
Table~\ref{chap3.3-tableIII} shows how \textit{soft}-GAN achieves an \gls{nrmse} of approximately 0.23 and 0.21, outperforming the best-performing baseline, TimeGAN, whose \gls{nrmse} ranges from 0.26 to 0.27.
Statistically, this equates to about an 11–19\% reduction in the \gls{nrmse} relative to strong baselines.
Lower \gls{nrmse} signifies a tighter fit of the generated time series to the real data distribution, indicating that \textit{soft}-GAN's internal soft-attention and recurrent mechanisms effectively preserve key temporal dependencies.
Decreasing the \gls{fid}-score relative to leading baselines suggests that \textit{soft}-GAN maintains more accurate local and global temporal structures.
Table~\ref{chap3.3-tableIV} shows that for 128-length sequences, \textit{soft}-GAN achieves a Discriminative Score approximately 39\% lower than TimeGAN (0.017 vs. 0.028) and significantly lower than all other baselines.
This improvement at the 256-length scale is also notable, matching TimeGAN at approximately 0.013 but with a more stable variance.
Such a low Discriminative Score suggests that the synthetic sequences from \textit{soft}-GAN are statistically indistinguishable from real data, reinforcing that our generative model is not merely capturing coarse trends but rather the nuanced statistical behavior observed in authentic RAN measurements.

To further validate the utility of generated data, Table~\ref{chap3.3-tableV} presents results from a classification task trained on synthetic samples.
\textit{soft}-GAN achieves an accuracy of 0.93, outperforming all baselines by at least 5 percentage points.
Its F1-score of 0.89 represents a $\sim9\%$ improvement over the closest competitor.
This higher classification metric confirms that \textit{soft}-GAN does not just produce visually or statistically similar data, but also preserves class-discriminative features.
Higher precision results of our model indicate that \textit{soft}-GAN has a lower false positive rate.
\textit{Soft}-GAN improves accuracy by 5-11\% over TimeGAN and 7-19\% over BiLSTM. 

The \gls{pca} projection in Figure~\ref{chap3.3-fig:pca} offers a reduced dimensionality view of both real and \textit{soft}-GAN generated samples.
If we consider the distributions of points along each principal component axis, the generated samples do not form a distinctly separate cluster; rather, they integrate into the same regions occupied by the real samples.
This suggests that the generative model has learned not only the mean and covariance structure of the underlying data but also the higher-order moments that define the geometry of the data manifold in a lower-dimensional space.
The close spatial proximity and pattern similarity between real and synthetic points imply that \textit{soft}-GAN captured complex temporal correlations, amplitude variations, and other spectral signatures pertinent to RAN signals.
The spectrogram in Figure~\ref{chap3.3-fig:spect} provides a joint time-frequency representation of a \textit{soft}-GAN generated RF signal at $25$~dB signal power level.
By applying a time-frequency decomposition—commonly via \gls{stft}—we visualize how signal energy is distributed across frequencies as a function of time.
The generated spectrogram exhibits key spectral and temporal patterns that mirror realistic RF conditions.

\subsection{Conclusions}
\label{chap3.3-subsec:gentwin-conclusions}


This section introduced \gls{gentwin}, a synthetic data generation \gls{dt} platform, that fulfills the data-intensive demands of advanced \gls{dtran} operations.
By producing high-fidelity \gls{rf} signals through the proposed soft-GAN model, \gls{gentwin} empowers \gls{ai}-driven network management with richer, more representative training sets.
The experimental evaluation demonstrated that soft-GAN achieves up to 19\% improvement in \gls{nrmse} compared to baseline models, with generated samples that are statistically indistinguishable from real data.

The data collection methodology presented in this section further demonstrates the potential connections and use cases between real-world testbeds (POWDER) and \gls{dt} platforms (Colosseum).
We demonstrated that the tools developed in this dissertation, such as \gls{cast}, are portable and can be used to collect \gls{iq} samples across both environments. Additionally, it showed how digitized scenarios can complement physical measurements to create comprehensive datasets for training generative models.
This approach also highlighted one of the key enablers of testbeds and platforms, high-volume data generation for \gls{ml} training. It positions testbeds and \glspl{dt} not only as environments for validation and experimentation, but also as critical data sources for developing and evaluating \gls{ai}/\gls{ml} solutions for next-generation wireless networks.

\section{Summary and Discussion}
\label{chap3.4-sec:conclusion}


This chapter demonstrated the \emph{use case} dimension of \glspl{dt} and emulation platforms, showing how they enable \gls{e2e} pipelines for three specific applications: spectrum sharing, \gls{ai}-driven channel modeling, and synthetic data generation. Each use case highlighted a different aspect and capability of the Colosseum emulation platform, the main experimental platform in this dissertation.
First, we demonstrated how spectrum-sharing \gls{cbrs} scenarios between \gls{4g} and \gls{5g} cellular and incumbent radar systems can be emulated safely on Colosseum using waveform twinning and dApp-based detection to study interference management in conditions that would be difficult, expensive, or unsafe to reproduce \gls{ota} in the real world.
Then, we showed the AIRMap framework and how \gls{ai}-driven propagation modeling can create ultra-fast radio maps using ray-tracing datasets, real-world measurements, and a simple elevation map, which can then be fed back into the emulator for real-time dynamic scenario adaptation.
Finally, we introduced \gls{gentwin}, in which Colosseum and other \gls{ota} testbeds serve not only as execution environments but also as data factories for training generative models, with the objective of augmenting \gls{rf} data for \gls{ai}-native and AI-RAN-related tasks.

These results highlight the broader impact that \gls{dt} platforms, such as Colosseum, have on the research community. They provide a controlled, repeatable, and reproducible environment in which to prototype algorithms, collect large and diverse datasets, and test and validate these solutions safely in a realistic setting before committing to more expensive field trials.
Additionally, the tools we developed, such as \gls{cast}, were used to create and validate scenarios and to collect data across various platforms, demonstrating their portability across different experimental environments. This is key to following the experimental workflow envisioned in this work for transitioning from emulation to the real world.

However, while \gls{dt} platforms like Colosseum offer significant advantages thanks to their controlled environments, they also have inherent limitations.
Channel emulation, regardless of its fidelity, remains an approximation of reality. \glspl{dt} are derived from ray-tracing simulations or measurement campaigns, but they cannot capture the full complexity of dynamic environments and channel effects, such as scattering, unwanted interference, hardware impairments in actual deployments, and environmental imperfections.
These limitations motivate the transition from emulated to \gls{ota} experimentation. Once algorithms have been sufficiently validated in controlled emulation scenarios, the next step is to deploy these solutions on a physical testbed and test them under real \gls{rf} propagation, hardware constraints, and environmental conditions.
The following chapters address this next phase by designing, deploying, and experimenting with an \gls{ota} platform, namely X5G, to continue the experimental journey and move from \glspl{dt} to physical wireless networks.
\chapter[Design and Deployment of a Private 5G O-RAN Testbed]{Design and Deployment of an Open and Programmable Private 5G O-RAN Testbed}
\label{chap:4}



The \gls{dt} platforms presented in Chapters~\ref{chap:2} and \ref{chap:3} have shown how large-scale emulation, such as Colosseum, can provide a controlled, repeatable, and reproducible environment for wireless network experimentation. However, as discussed in Section~\ref{chap3.4-sec:conclusion}, while these platforms provide control over channel conditions, topology, and traffic, they inherently approximate real-world effects and abstract away some important aspects of real deployments, including hardware impairments, dynamic wireless propagation, and unexpected real-world interference and situations. To complete the experimental journey outlined in Chapter~\ref{chap:intro} and validate the solutions under realistic conditions, an \gls{ota} testbed with physical, commercial-grade hardware and software components is an essential step.

This chapter addresses the \emph{deployment} and \emph{usability} dimensions of experimental wireless platforms through the design and implementation of X5G, an open, programmable, multi-vendor private 5G O-RAN-compliant testbed~\cite{villa2024x5g,villa2025tmc,villa2025cusense}. While \glspl{dt} like Colosseum excel at providing repeatable experimentation at scale, \gls{ota} testbeds provide the necessary \emph{realism} for final validation before moving to a large-scale production deployment.

\section{Introduction}
\label{chap4-sec:intro}

The evolution of the \gls{ran} in \gls{5g} networks has led to key performance improvements in cell and user data rates, now averaging hundreds of Mbps, and in air-interface latency~\cite{aarayanan2022comparative}, thanks to specifications developed within \gls{3gpp}. 
From an architectural point of view, 
\gls{5g} deployments are also becoming more open, intelligent, programmable, and software-based~\cite{polese2023understanding}, through activities led by the O-RAN ALLIANCE, which is developing the network architecture for Open \gls{ran}.
These elements have the potential to transform how we deploy and manage wireless mobile networks~\cite{bonati2023neutran}, leveraging intelligent control, with \gls{ran} optimization and automation exercised via closed-loop data-driven control; softwarization, with the components of the end-to-end protocol stack defined through software rather than dedicated hardware; and disaggregation, with the \gls{5g} \gls{ran} layers distributed across different network functions, i.e., the \gls{cu}, the \gls{du}, and the \gls{ru}. 

Open and programmable networks are often associated with lower capital and operational expenditures, facilitated by the increasing robustness and diversity of the telecom supply chain~\cite{dellOroRAN}, now also including open-source projects~\cite{kaltenberger2024driving,gomez2016srslte} and vendors focused on specific components of the disaggregated \gls{ran}. This, along with increased spectrum availability in dedicated or shared bands, has opened opportunities to deploy private \gls{5g} systems, complementing public \gls{5g} networks with more agile, dynamic deployments for site-specific use cases (e.g., events, warehouse automation, industrial control).

While the transition to disaggregated, software-based, and programmable networks offers significant benefits, several challenges must be addressed before Open \gls{ran} systems can match or exceed the performance of traditional cellular systems and translate the potential cost savings into actual deployments.
First, the radio domain still exhibits a low degree of automation and zero-touch provisioning for the \gls{ran} configuration, complicating the successful \emph{deployment} of end-to-end cellular systems. Second, the diverse vendor ecosystem poses challenges related to interoperability and end-to-end integration across multiple products, potentially from different vendors~\cite{5gtesting,bahl2023accelerating,tang2023ai} (\emph{usability} dimension). Third, the \gls{dsp} at the \gls{phy} layer of the stack is a computationally complex element, using about $90$\% of the available compute when run on general-purpose CPUs, and thus introducing a burden on the software-based and virtualized \gls{5g} stack components. Finally, there are still open questions in terms of how the intelligent and data-driven control loops can be implemented with \gls{ai} and \gls{ml} solutions that generalize well across a multitude of cellular network scenarios~\cite{fiandrino2023explora}.
These challenges call for a concerted effort across different communities (including hardware, \gls{dsp}, software, DevOps, \gls{ai}/\gls{ml}) that aims to design and deploy open, programmable, multi-vendor cellular networks and testbeds that can support private \gls{5g} requirements and use cases with the stability and performance of production-level systems.

Beyond the challenges of deploying and integrating Open RAN systems, enabling real-time, \gls{ai}-native applications on the \gls{ran} infrastructure presents additional requirements. Current \gls{ran} control and observability (e.g., with O-RAN interfaces)
remain limited to near- or non-real-time scales (i.e., above $10$~ms), and focused on the control plane. O-RAN programmable applications, xApps and rApps, are deployed in dedicated controllers that aggregate data from tens of \glspl{gnb}. They typically operate with timescales of tens of milliseconds to seconds and observe the \gls{ran} through aggregated \glspl{kpi}, per-device and per-cell statistics computed over many slots and subcarriers. This design is intentional, as sending granular control- or user-plane data---for example, \acrshort{iq} samples or full \gls{csi} matrices---to a remote controller would incur prohibitive (Gbps-scale) data rates and raise security and privacy concerns. Today's programmable \glspl{ran} cannot thus leverage fine-grained \gls{phy}/\gls{mac} telemetry for tasks that require per-slot accuracy, such as fast link adaptation, beam management, or \gls{isac}.

\begin{figure}[htb]
    \centering
    \includegraphics[width=\linewidth]{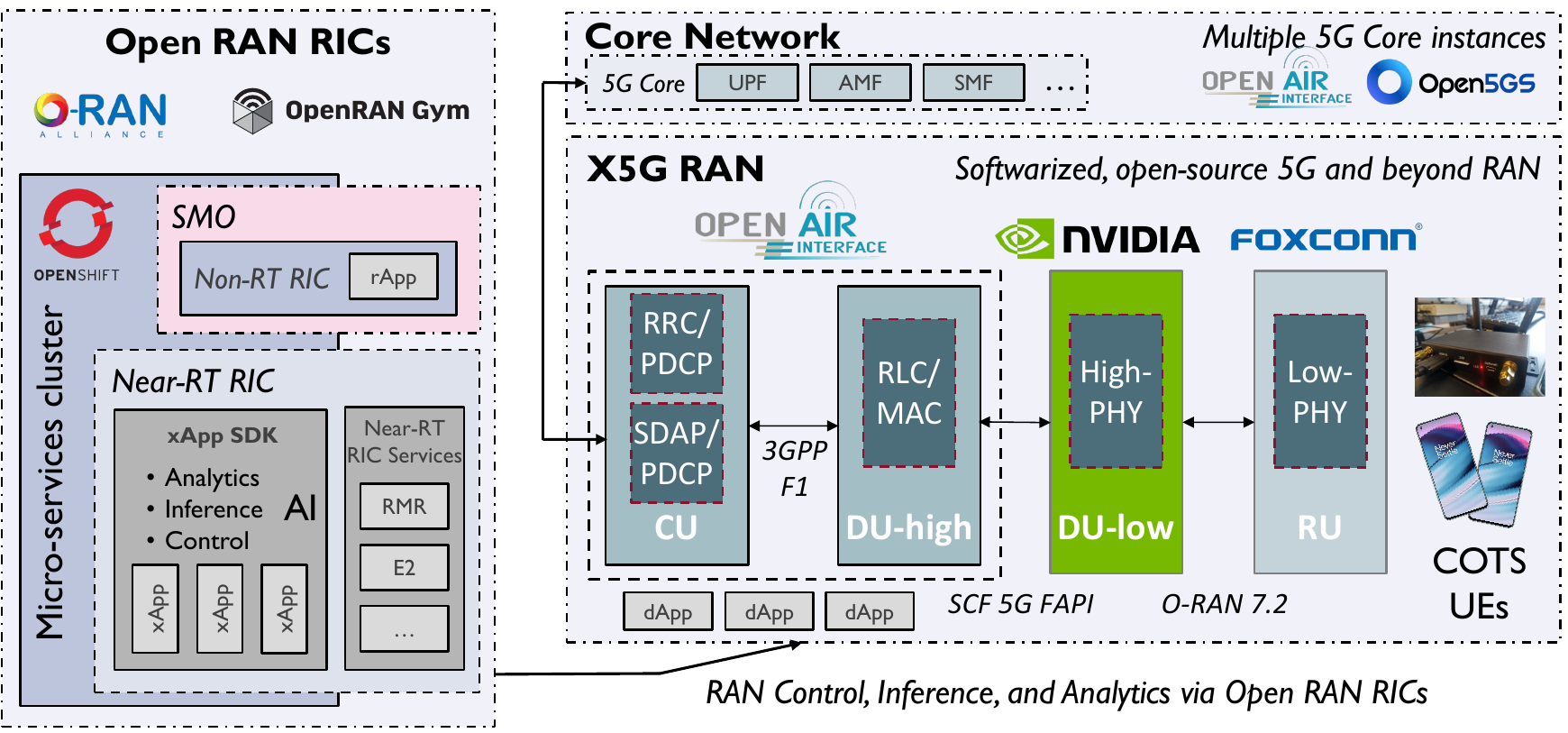}
    \caption{X5G end-to-end programmable testbed overview.}
    \label{chap4-fig:x5g-e2e-overview}
\end{figure}

This chapter presents the design, deployment, and experimental evaluation of X5G, a private \gls{5g} network testbed deployed at Northeastern University in Boston, MA, and based on multiple programmable and open-source components from the physical layer all the way up to the \gls{cn}, as shown in Figure~\ref{chap4-fig:x5g-e2e-overview}.
X5G integrates a \gls{phy} layer implemented on \gls{gpu} (i.e., NVIDIA Aerial) with \gls{oai} for the higher layers of the 5G stack through the \gls{scf} \gls{fapi}, which regulates the interaction between the \gls{phy} and \gls{mac} layers.
The testbed also connects to a near-real-time \gls{ric} from the \gls{osc} running on an OpenShift-based cluster, and supports multiple \gls{5g} \gls{cn} implementations, multiple \glspl{ru}, and commercial smartphones and programmable devices.

Beyond the core platform infrastructure, a distinctive feature of X5G is its GPU-accelerated \gls{ran} architecture, which not only offloads the \gls{phy} processing to NVIDIA \glspl{gpu} but also allows reuse of spare resources for \gls{ai}/\gls{ml} workloads co-located with the \gls{gnb}, enabling the execution of distributed applications, or dApps~\cite{doro2022dapps}, timescales in the real-time domain.
In this model, these dApps can run on the same accelerator used for baseband processing, access low-level \gls{phy} measurements, and close control loops at timescales in the real-time domain, complementing xApps and rApps near-real-time and non-real-time cycles. Therefore, we also focus on the design and integration of this dApp framework within a \gls{gpu}-accelerated \gls{gnb}, such as X5G.

The key contributions of this chapter include:
\begin{itemize}
    \item \textbf{Multi-vendor 5G O-RAN integration:} The integration of NVIDIA Aerial GPU-accelerated \gls{phy} with \gls{oai} higher layers through the \gls{scf} \gls{fapi} interface, demonstrating interoperability across open-source and commercial components.
    \item \textbf{End-to-end infrastructure deployment:} A comprehensive deployment including synchronization and networking infrastructure, multiple \glspl{ru}, and integration with various \glspl{cn}.
    \item \textbf{Digital twin-based RF planning:} A ray-tracing-based methodology using a \gls{dt} framework to optimize \gls{ru} placement in an indoor space, combining \gls{dt} and \gls{ota} experimentation.
    \item \textbf{GPU-accelerated dApp framework:} A real-time framework for distributed applications that exposes PHY/MAC telemetry via shared memory and an E3-based interface, enabling \gls{e2e} control loops with sub-millisecond latency.
    \item \textbf{Experimental validation:} Performance profiling with commercial smartphones and emulated \glspl{ue}, demonstrating cell throughput up to $1.65$~Gbps in downlink and system stability spanning multiple days.
\end{itemize}
The rest of the chapter is organized as follows. Section~\ref{chap4-sec:software} introduces the X5G software architecture. 
Section~\ref{chap4-sec:arc-hardware} concerns the deployment and configuration of the X5G physical hardware infrastructure. 
Section~\ref{chap4-sec:ray-tracing} describes an \acrshort{rf} planning study to determine an optimal location for deploying the \glspl{ru} using a \gls{dt} framework.
System performance is evaluated in Section~\ref{chap4-sec:exp-results} through various use case scenarios with multiple \gls{cots} \glspl{ue} and applications.
Section~\ref{chap4-sec:dapp-framework} introduces the GPU-accelerated dApp framework for real-time \gls{ai} applications on the \gls{ran}.
Section~\ref{chap4-sec:related-work} compares X5G with the state of the art.
Section~\ref{chap4-sec:conclusions} summarizes the main takeaways and connects the X5G platform to the \gls{ota} use cases presented in Chapter~\ref{chap:5}.

\glsreset{cn}
\section{X5G Software}
\label{chap4-sec:software}

This section describes the software components of X5G, also shown in Figure~\ref{chap4-fig:x5g-e2e-overview}.
These components can be divided into three main groups: (i) a full-stack programmable \gls{gnb} (X5G \gls{ran})\blue{;} (ii) the Open \gls{ran} \glspl{ric} deployed on a micro-services cluster based on OpenShift\blue{;} and (iii) various \glspl{cn} deployed in a micro-services-based architecture essential for the effective functioning of the \gls{5g} network.

\glsunset{gnb}
\subsection{Full-stack Programmable RAN with NVIDIA Aerial and OpenAirInterface}
\label{chap4-sec:arc}


\glsreset{arc}





\begin{figure}[htb]
    \centering
    \includegraphics[width=0.6\linewidth]{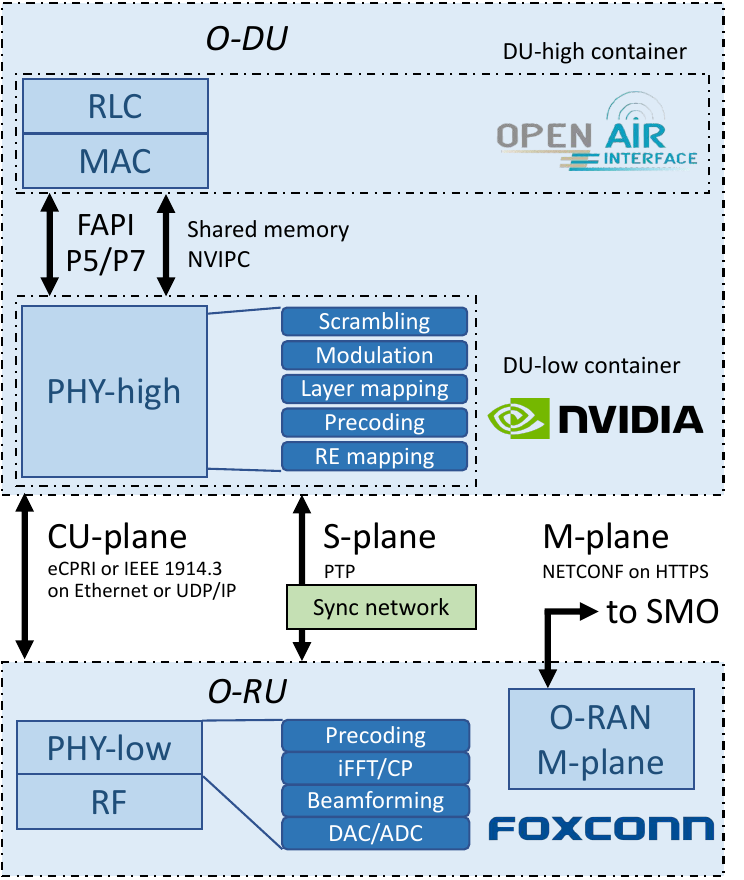}
    \caption{Architecture of the lower layers of the X5G \acrshort{ran} following O-RAN specifications and consisting of: (i) a Foxconn O-RU; (ii) an O-DU-low based on NVIDIA Aerial SDK; (iii) an O-DU-high based on OpenAirInterface with their corresponding interfaces.}
    \label{chap4-fig:ARC_archi}
\end{figure}

The right part of Figure~\ref{chap4-fig:x5g-e2e-overview} shows a detailed breakdown of the architecture of the X5G \gls{ran}, which follows the basic O-RAN architecture split into \gls{cu}, \gls{du}, and \gls{ru}. The \gls{du} is further split into a DU-low, implementing Layer 1 (\gls{phy}, or L1) functionalities, and into a DU-high, implementing Layer 2 (\gls{mac} and \gls{rlc}, or L2) ones. As shown in Figure~\ref{chap4-fig:ARC_archi}, DU-low and DU-high communicate over the 5G \gls{fapi} interface specified by the \gls{scf}~\cite{SCF2020FAPI}. The DU-low is implemented using the NVIDIA Aerial \gls{sdk}~\cite{aerialsdk} on in-line \gls{gpu} accelerator cards, whereas DU-high and CU are implemented by \gls{oai} on general-purpose \glspl{cpu}. We deploy each function in separate Docker containers, sharing a dedicated memory space for the inter-process communication library that enables the \gls{fapi} interface. In our setup, we also combine the \gls{cu} and the \gls{du}-high into a combined L2/L3 \gls{gnb} Docker container, but the F1 split has also been deployed and tested.

The \gls{fapi} interface between the DU-high and DU-low defines two sets of procedures: configuration and slot procedures. Configuration procedures handle the management of the \gls{phy} layer and happen infrequently, e.g., when the \gls{gnb} stack is bootstrapped or reconfigured. On the contrary, slot procedures happen in every slot (i.e., every $500$\:$\mu$s for a $30$\:kHz subcarrier spacing) and determine the structure of each \gls{dl} and \gls{ul} slot. In our case, L1 serves as the primary and L2 as the subordinate. Upon the reception of a slot indication message from L1, L2 sends either an \gls{ul} or \gls{dl} request to dictate the required actions for the \gls{phy} layer in each slot. Additionally, L1 might transmit other indicators to L2, signaling the receipt of data related to \gls{rach}, \gls{uci}, \gls{srs}, checksums, or user plane activities.

In our implementation, we use \gls{fapi} version 222.10.02 with a few exceptions as outlined in the NVIDIA Aerial release notes \cite{aerialsdk-website}. 
The transport mechanism for \gls{fapi} messages is specified in the networked \gls{fapi} (nFAPI) specification~\cite{SCF2021nFAPI}, which assumes that messages are transported over a network. However, in our implementation, the L1 and L2 Docker containers communicate through the \gls{nvipc} library. This tool provides a robust shared memory communication framework specifically designed to meet the real-time performance demand of the data exchanges between \gls{mac} and \gls{phy} layers. In our implementation, we choose to transport the messages using little-endian with zero padding to $32$\:bits. The \gls{nvipc} library is also capable of tracing the \gls{fapi} messages and exporting them to a \texttt{pcap} file that can be analyzed offline with tools such as Wireshark.

The NVIDIA physical layer in the DU-low implements the O-RAN Open Fronthaul interface, also known as the O-RAN 7.2 interface~\cite{ORA2023}, to communicate directly with the O-RU, in our case manufactured by Foxconn. This interface transports frequency domain \gls{iq} samples (with optional block floating point compression) over a switched network, allowing for flexible deployments. The interface includes synchronization, control, and user planes. The synchronization plane, or S-plane, is based on PTPv2. We use synchronization architecture option 3~\cite{oran-wg4-fronthaul-cus}, where the fronthaul switch provides timing to both DU and RU. The interface also includes a management plane, although \blue{our system currently does not support it}.

\begin{table}[htbp]
    \centering
    \caption{X5G \acrshort{arc} deployment main features.}
    \label{chap4-table:testbeds-features}
    \scriptsize
    \setlength{\tabcolsep}{4pt}
    \begin{tabular}{ll}
        \toprule
        \textbf{Feature} & \textbf{Description} \\
        \midrule
        3GPP Release & 15 \\
        Frequency Band & n78 (FR1, TDD) \\
        Carrier Frequency & $3.75$\:GHz \\
        Bandwidth \blue{($\beta$)} & $100$\:MHz \\
        Subcarrier spacing \blue{($\Delta f$)} & $30$\:kHz \\
        \blue{Resource Block size ($\chi$)} & \blue{12 subcarriers} \\
        \blue{Modulation order ($Q_{m}$)} & \blue{8 (256-QAM)} \\
        TDD config & DDDSU, DDDDDDSUUU$^*$ \\
        Number of antennas used & \blue{4 TX, 4 RX} \\
        MIMO config \blue{($L_{DL}, L_{UL}$)} & \blue{4 layers DL, 1 layer UL} \\
        Max theoretical cell throughput$^{**}$ \blue{($T_{DL}, T_{UL}$)} & \blue{$1.64$\:Gbps DL, $204$\:Mbps UL} \\
        \bottomrule
    \end{tabular}
    
    \raggedright
    \footnotesize
    
    $^*$Currently the special slot is unused due to limitations in Foxconn radios.
    
    $^{**}$The single-user maximum theoretical DL throughput can currently only be reached in the DDDDDDSUUU TDD configuration. In the DDDSU TDD configuration, it is limited to $350$\:Mbps since we can schedule a maximum of 2~DL slots per user in one TDD period, as only 2~ACK/NACK feedback bits are available per user.
\end{table}
\glsunset{arc}

Table~\ref{chap4-table:testbeds-features} summarizes the main features and operational parameters of the \gls{arc-ota} deployment in the X5G testbed. The protocol stack is aligned with 3GPP Release 15 and uses the 5G n78 \gls{tdd} band and numerology~1. The DDDSU \gls{tdd} pattern, which repeats every $2.5$\:ms, includes three downlink slots, one special slot (which is not used due to limitations in the Foxconn \glspl{ru}), and an uplink slot. The uplink slot format implemented in \gls{oai} carries only two feedback bits for ACK/NACK per \gls{ue}, thus allowing only the scheduling of two downlink slots per \gls{ue}, eventually limiting the single \gls{ue} throughput. \blue{Alternative \gls{tdd} patterns, including DDDDDDSUUU and DDDDDDDSUU, repeating every $5$\:ms, are also already in use} to provide additional ACK/NACK bits for reporting from the \glspl{ue} \blue{and mitigate this limitation}.

\blue{To compute the maximum theoretical cell throughput in downlink ($T_{DL}$) and uplink ($T_{UL}$), we first derive a few additional parameters from Table~\ref{chap4-table:testbeds-features}. The number of resource blocks ($N_{RB}$) is computed using
}
\begin{align}\label{chap4-eq:nrb}
    \blue{N_{RB} = \frac{\beta}{\chi \cdot \Delta f} = 273.}
\end{align}
\blue{By default, the number of \gls{ofdm} symbols per slot ($N_{sym}$) is 14. The number of slots per second ($N_{slot}$) is inversely proportional to the slot duration, which for numerology $\mu = 1$ is  $0.5$~ms. Hence, $N_{slot} = 1s/0.5$ms$ = 2000$ slots/second.
The maximum theoretical cell throughput for downlink and uplink is given by
}
\begin{align}\label{chap4-eq:maxth}
    \blue{T_{DL,UL} = N_{RB} \cdot \chi \cdot N_{sym} \cdot N_{slot} \cdot Q_m \cdot R \cdot L_{DL,UL} \cdot \eta,}
\end{align}
\blue{where $R$ is the effective code rate, which can approach 0.93 (as specified in the 3GPP standard~\cite{3gpp5gnr}), and $\eta$ is the fraction of time allocated for downlink or uplink operations based on the chosen TDD pattern.
Considering the DDDDDDSUUU pattern to circumvent current \gls{oai} limitations on the ACK/NACK feedback bits, 60\% of time is allocated for downlink and 30\% for uplink since the special slot is unused due to Foxconn \gls{ru} constraints.
Consequently, the resulting theoretical peak cell throughput is $1.64$~Gbps for downlink ($T_{DL}$) and $204$~Mbps for uplink ($T_{UL}$).
These values do not account for overheads typical of real networks---such as \gls{dmrs}, \gls{pucch}, and \gls{pdcch}---which may further reduce net throughput.
As shown by the experimental results in Section~\ref{chap4-sec:exp-peak}, X5G peak performance nearly reaches the theoretical downlink throughput, while the uplink is still under improvement.
}

\textbf{Core Network.} The X5G testbed facilitates the integration and testing of different \glspl{cn} from various vendors and projects. We leverage virtualization to deploy all the necessary micro-services, e.g., \gls{amf}, \gls{smf}, \gls{upf}, in the OpenShift cluster that also supports the Near-RT \gls{ric}.
We have successfully tested and integrated the X5G \gls{ran} with two open-source core network implementations, i.e., the \gls{5g} \glspl{cn} from \gls{oai}~\cite{kaltenberger2024driving} and Open5GS~\cite{open5gs_website}, \blue{as well as the CoreSIM software from Keysight~\cite{keysight_coresim}}. As part of our ongoing efforts, we plan to incorporate additional cores, including the commercial core from A5G~\cite{a5gnetworks}.


\subsection{Integration with the OSC Near-RT RIC}
\label{chap4-sec:e2}

%
One of the key components of an O-RAN deployment is the Near-Real-Time (or Near-RT) \gls{ric}, and the intelligent applications hosted therein, namely xApps. These can implement closed-control loops at timescales between $10$\:ms and $1$\:s to provide optimization and monitoring of the \gls{ran}~\cite{abdalla2022toward,mungari2021rl}. In the \blue{current} X5G setup, we deploy the ``E'' release of the \gls{osc} Near-RT \gls{ric} on a RedHat OpenShift cluster~\cite{bonati2023neutran}, which manages the lifecycle of edge-computing workloads instantiated as containerized applications.
%
\begin{figure}[bth]
    \centering
    \includegraphics[width=0.6\linewidth]{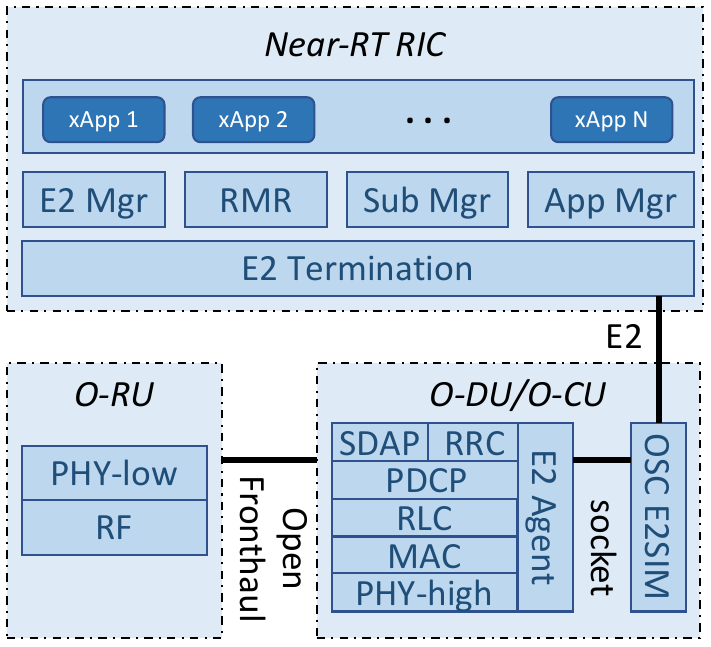}
    \caption{Integration of the \acrshort{osc} Near-RT \acrshort{ric} in the OpenShift cluster with the X5G \acrshort{ran}.}
    \label{chap4-fig:ARC_e2if}
\end{figure}
%
The Near-RT \gls{ric} and the \gls{arc-ota} \gls{ran} are connected through the O-RAN E2 interface (see Figure~\ref{chap4-fig:ARC_e2if}), based on \gls{sctp} and an O-RAN-defined application protocol (E2AP) with multiple service models implementing the semantics of the interface (e.g., control, reporting, etc)~\cite{polese2023understanding}.
%
%
On the \gls{gnb} side, we integrate an E2 agent based on the \textit{e2sim} software library~\cite{moro2023nfv,e2sim}, which is used to transmit the metrics collected by the \gls{oai}~\gls{gnb} to the \gls{ric} via the \gls{kpm} E2 service model.
These metrics are then processed by xApps deployed on the \gls{ric}, and used to compute some control action (e.g., through \gls{ai}/\gls{ml} agents) that is sent to the \gls{ran} through the E2 interface and processed by the \textit{e2sim} agent.

\begin{figure}[htb]
    \centering
    \includegraphics[width=0.9\linewidth]{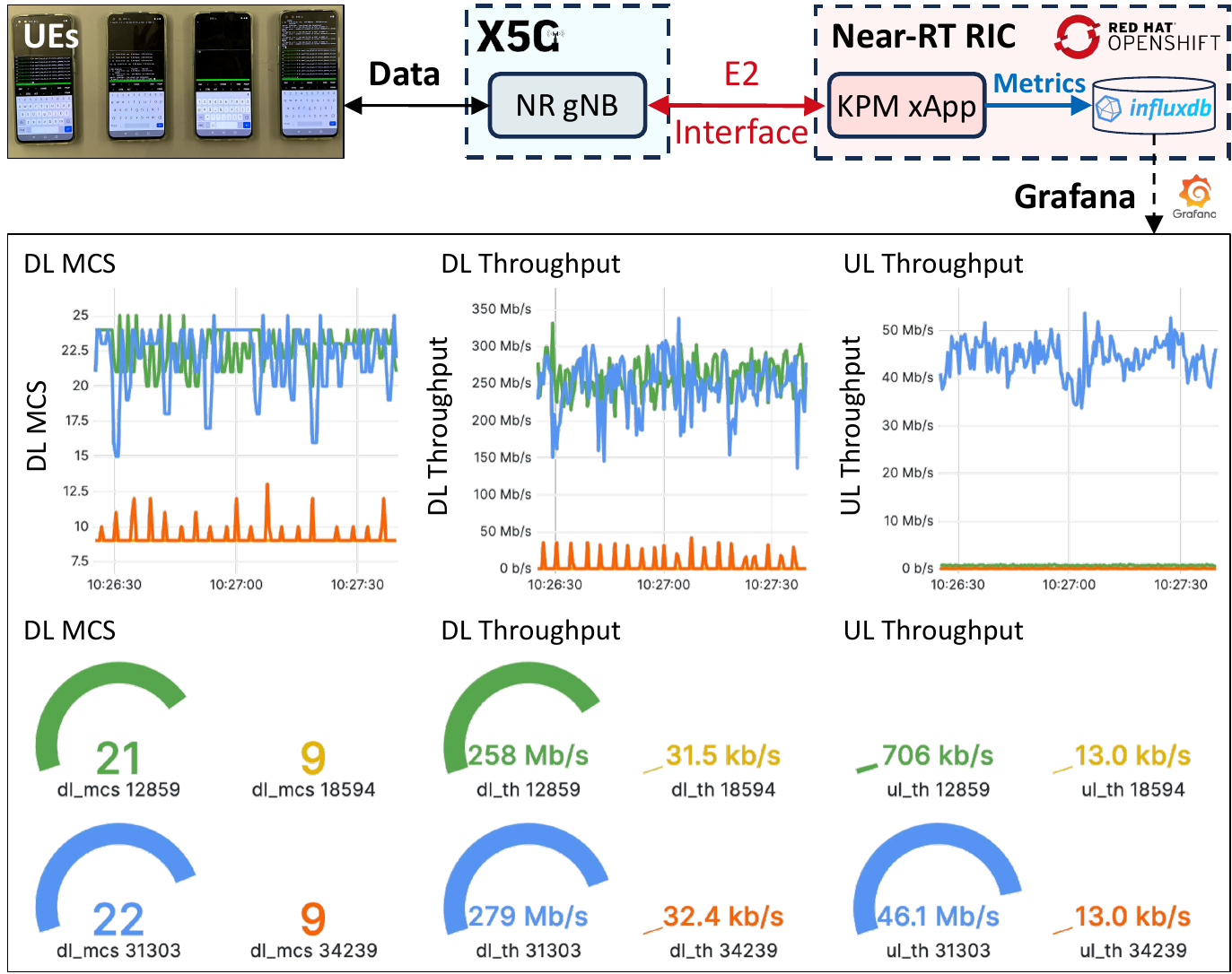}
    \caption{\acrshort{kpm} xApp example architecture \blue{including an X5G \acrshort{gnb} with four connected \acrshortpl{ue}, each performing a different operation (ping, video streaming, \acrshort{dl} test, and \acrshort{dl}/\acrshort{ul} tests), and a \acrshort{kpm} xApp that pushes \acrshort{ue} metrics into an Influx database, which are then visualized in a Grafana dashboard.}}
    \label{chap4-fig:kpm_xapp}
\end{figure}

As an example, Figure~\ref{chap4-fig:kpm_xapp} shows the architecture of a \gls{kpm} xApp integrated with the X5G testbed.
%
%
This xApp receives metrics from the E2 agent in the \gls{gnb}, including throughput, number of \glspl{ue}, and \gls{rsrp}, and stores them in an InfluxDB database~\cite{influxdb}. The database is then queried to display the \gls{ran} performance on a Grafana dashboard~\cite{grafana} (see Figure~\ref{chap4-fig:kpm_xapp}).
This setup creates a user-friendly observation point for monitoring network performance and demonstrates the effective integration of the near-RT \gls{ric} in our configuration.
A tutorial on how to deploy and run this xApp in X5G or on a similar testbed can be found on the OpenRAN Gym website~\cite{openrangymwebsite}, which hosts an open-source project and framework for collaborative research in the O-RAN ecosystem~\cite{bonati2022openrangym-pawr}.

\begin{figure}[htb]
\centering
    \subfloat[DL Throughput]{
    \label{chap4-fig:slicing_xapp_th}
    \centering
    \setlength\fwidth{.8\linewidth}
    \setlength\fheight{.25\linewidth}
    \input{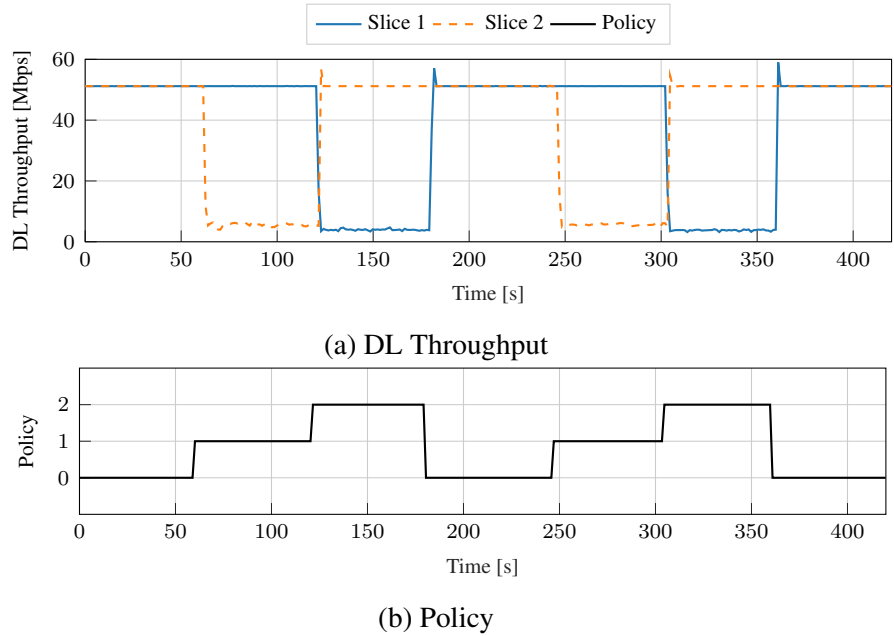}
    \setlength\abovecaptionskip{.05cm}}
    \hfill    
    \subfloat[Policy]{
    \label{chap4-fig:slicing_xapp_policy}
    \centering
    \setlength\fwidth{.8\linewidth}
    \setlength\fheight{.22\linewidth}
\begin{tikzpicture}
\pgfplotsset{every tick label/.append style={font=\scriptsize}}

\definecolor{darkslategray38}{RGB}{38,38,38}
\definecolor{lightgray204}{RGB}{204,204,204}
\definecolor{steelblue31119180}{RGB}{31,119,180}
\definecolor{darkorange25512714}{RGB}{255,127,14}

\begin{axis}[
width=1\fwidth, 
height=1.05\fheight, 
axis line style={color=black},
legend cell align={center},
legend style={
  fill opacity=0.8,
  draw opacity=1,
  text opacity=1,
  at={(0.5, 1.08)},
  anchor=south,
  draw=lightgray204,
  font=\scriptsize,
},
tick align=inside,
tick pos=left,
x grid style={lightgray204},
xlabel=\textcolor{darkslategray38}{Time [s]},
label style={font=\scriptsize},
xmajorgrids,
xmin=0, xmax=420,
xtick style={color=darkslategray38},
y grid style={lightgray204},
ymajorgrids,
ymin=-1, ymax=3,
ytick={0,1,2},
ytick style={color=darkslategray38},
ylabel={Policy},
legend columns=2,
]

\addplot [thick, black]
table {%
0         0
1.28      0
2.56      0
3.84      0
5.12      0
7.68      0
8.96      0
10.24     0
11.52     0
12.80     0
14.08     0
15.36     0
16.64     0
17.92     0
19.20     0
20.48     0
21.76     0
23.04     0
24.32     0
25.60     0
26.88     0
28.16     0
29.44     0
30.72     0
32.00     0
33.28     0
34.56     0
35.84     0
37.12     0
38.40     0
39.68     0
40.96     0
42.24     0
43.52     0
44.80     0
46.08     0
47.36     0
48.64     0
49.92     0
51.20     0
52.48     0
53.76     0
55.04     0
56.32     0
57.60     0
58.88     0
60.16     1
61.44     1
62.72     1
64.00     1
65.28     1
66.56     1
67.84     1
69.12     1
70.40     1
71.68     1
72.96     1
74.24     1
75.52     1
76.80     1
78.08     1
79.36     1
80.64     1
81.92     1
83.20     1
84.48     1
85.76     1
87.04     1
88.32     1
89.60     1
90.88     1
92.16     1
93.44     1
94.72     1
96.00     1
97.28     1
98.56     1
99.84     1
101.12    1
102.40    1
103.68    1
104.96    1
106.24    1
107.52    1
108.80    1
110.08    1
111.36    1
112.64    1
113.92    1
115.20    1
116.48    1
117.76    1
119.04    1
120.32    1
121.60    2
122.88    2
124.16    2
125.44    2
126.72    2
128.00    2
129.28    2
130.56    2
131.84    2
133.12    2
134.40    2
135.68    2
136.96    2
138.24    2
139.52    2
140.80    2
142.08    2
143.36    2
144.64    2
145.92    2
147.20    2
148.48    2
149.76    2
151.04    2
152.32    2
153.60    2
154.88    2
156.16    2
157.44    2
158.72    2
160.00    2
161.28    2
162.56    2
163.84    2
165.12    2
166.40    2
167.68    2
168.96    2
170.24    2
171.52    2
172.80    2
174.08    2
175.36    2
176.64    2
177.92    2
179.20    2
180.48    0
181.76    0
183.04    0
184.32    0
185.60    0
186.88    0
188.16    0
189.44    0
190.72    0
192.00    0
193.28    0
194.56    0
195.84    0
197.12    0
198.40    0
199.68    0
200.96    0
202.24    0
203.52    0
204.80    0
206.08    0
207.36    0
208.64    0
209.92    0
211.20    0
212.48    0
213.76    0
215.04    0
216.32    0
217.60    0
218.88    0
220.16    0
221.44    0
222.72    0
224.00    0
225.28    0
226.56    0
227.84    0
229.12    0
230.40    0
231.68    0
232.96    0
234.24    0
235.52    0
236.80    0
238.08    0
239.36    0
240.64    0
241.92    0
243.20    0
244.48    0
245.76    0
247.04    1
248.32    1
249.60    1
250.88    1
252.16    1
253.44    1
254.72    1
256.00    1
257.28    1
258.56    1
259.84    1
261.12    1
262.40    1
263.68    1
264.96    1
266.24    1
267.52    1
268.80    1
270.08    1
271.36    1
272.64    1
273.92    1
275.20    1
276.48    1
277.76    1
279.04    1
280.32    1
281.60    1
282.88    1
284.16    1
285.44    1
286.72    1
288.00    1
289.28    1
290.56    1
291.84    1
293.12    1
294.40    1
295.68    1
296.96    1
298.24    1
299.52    1
300.80    1
302.08    1
303.36    1
304.64    2
305.92    2
307.20    2
308.48    2
309.76    2
311.04    2
312.32    2
313.60    2
314.88    2
316.16    2
317.44    2
318.72    2
320.00    2
321.28    2
322.56    2
323.84    2
325.12    2
326.40    2
327.68    2
328.96    2
330.24    2
331.52    2
332.80    2
334.08    2
335.36    2
336.64    2
337.92    2
339.20    2
340.48    2
341.76    2
343.04    2
344.32    2
345.60    2
346.88    2
348.16    2
349.44    2
350.72    2
352.00    2
353.28    2
354.56    2
355.84    2
357.12    2
358.40    2
359.68    2
360.96    0
362.24    0
363.52    0
364.80    0
366.08    0
367.36    0
368.64    0
369.92    0
371.20    0
372.48    0
373.76    0
375.04    0
376.32    0
377.60    0
378.88    0
380.16    0
381.44    0
382.72    0
384.00    0
385.28    0
386.56    0
387.84    0
389.12    0
390.40    0
391.68    0
392.96    0
394.24    0
395.52    0
396.80    0
398.08    0
399.36    0
400.64    0
401.92    0
403.20    0
404.48    0
405.76    0
407.04    0
408.32    0
409.60    0
410.88    0
412.16    0
413.44    0
414.72    0
416.00    0
417.28    0
418.56    0
419.84    0
};

\end{axis}

\end{tikzpicture}
    \setlength\abovecaptionskip{.05cm}}
\caption{\blue{Slicing xApp example showing: (a) \acrshort{dl} throughput for two different slices, each with a single \acrshort{ue} connected and pushing $50$\:Mbps of \acrshort{udp} traffic; (b) the network policy applied by the slicing xApp, switching between no-priority (0), prioritize slice 1 (1), and prioritize slice 2 (2).}}
\label{chap4-fig:slicing_xapp_demo}
\end{figure}

\blue{The metrics collected by a \gls{kpm} xApp can then be leveraged by a second xApp or an rApp to perform smart closed-loop \gls{ran} controls at runtime, based on an arbitrary optimization strategy or specific requirements.
ORANSlice---an open-source, network-slicing-enabled Open \gls{ran} system that leverages open-source \gls{ran} frameworks such as \gls{oai}~\cite{Cheng2024oranslice}---was successfully integrated and tested in X5G, enabling near-real-time slicing control of the resources allocated by a \gls{gnb} to multiple slices of the network, according to different policies set by the network manager.
Figure~\ref{chap4-fig:slicing_xapp_demo} presents the effects of various network policies applied by an ORANSlice slicing xApp in a X5G\ \gls{gnb}. Figure~\ref{chap4-fig:slicing_xapp_th} shows the \gls{dl} throughput results for the two slices (slice 1 in blue and slice 2 in orange), with a single \gls{ue} per slice connected and transmitting $50$\:Mbps of \gls{dl} \gls{udp} data, according to the policy shown in Figure~\ref{chap4-fig:slicing_xapp_policy}. The slicing xApp switches between three policies: (0) no-priority, where all slices share all resources, so both \glspl{ue} achieve the target throughput of $50$\:Mbps; (1) prioritize slice 1: where 98\% of resources are reserved for the first slice and 2\% for the second one, causing the latter performance to drop to only $6$\:Mbps; (2) prioritize slice 2, where the opposite behavior of policy 1 is observed, with slice 1 now unable to achieve the target throughput.
In this example, the policy is applied arbitrarily as a proof-of-concept for the network slicing control capabilities of X5G, while more intelligent strategies employing \gls{ai}/\gls{ml} components can be easily integrated into the decision process.
Additional applications, including the emerging dApps~\cite{lacava2025dapps}, are currently being integrated into X5G to fully leverage its openness and programmability, further demonstrating the benefits of smart closed-loop control within the O-RAN ecosystem.}


\subsection{X5G Software Licensing and Tutorials}

X5G, including the Aerial \gls{phy}, the \gls{oai} higher layers, as well as the \gls{osc} \gls{ric}, is open and can be extended with custom features and functionalities. The NVIDIA \gls{arc-ota} framework is documented on the NVIDIA portal~\cite{aerialsdk-website}, which is accessible through NVIDIA's 6G developer program. As mentioned in Section~\ref{chap4-sec:e2}, the step-by-step integration between the \gls{osc} \gls{ric} and the \gls{arc-ota} stack through the X5G E2 agent is discussed in a tutorial on the OpenRAN Gym website~\cite{openrangymwebsite,bonati2022openrangym-pawr}.

The components implemented by \gls{oai} are published under the \gls{oai} public license v1.1  created by the \gls{oai} Software Alliance (OSA) in 2017~\cite{oai}.
This license is a modified Apache v2.0 License, with an additional clause that allows contributors to make patent licenses available to third parties under \gls{frand} terms, similar to \gls{3gpp} for commercial exploitation, to allow contributions from companies holding intellectual property in related areas. The usage of \gls{oai} code is free for non-commercial/academic research purposes. The Aerial \gls{sdk} is available through an early adopter program~\cite{aerialsdk-website}. The \gls{osc} software is published under the Apache v2.0 License.

\section{X5G Infrastructure}
\label{chap4-sec:arc-hardware}

\begin{figure*}[t]
    \centering
    \includegraphics[width=1\linewidth]{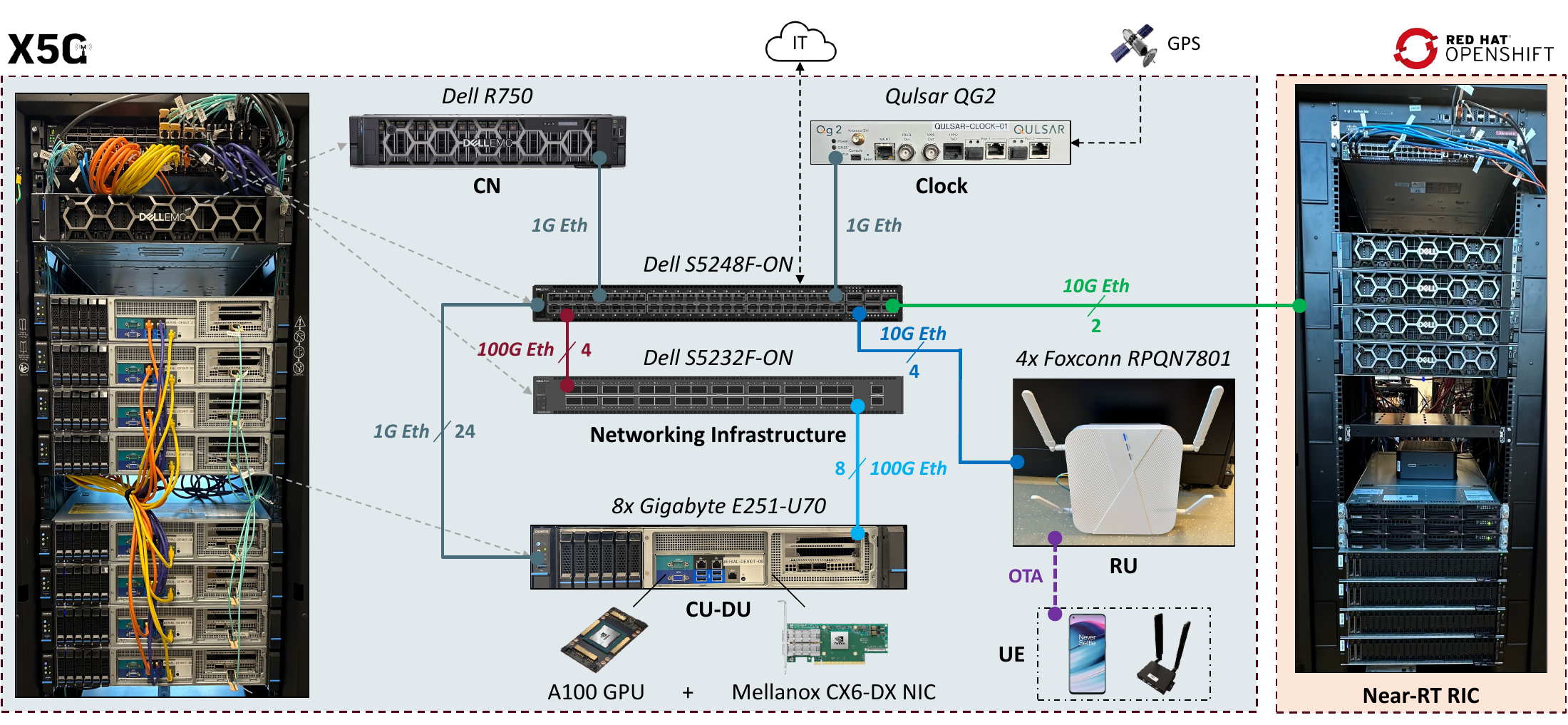}
    \caption{Hardware architecture infrastructure of the X5G deployment at Northeastern University.}
    \label{chap4-fig:arc-hardware}
\end{figure*}

This section describes the X5G physical deployment that is currently located on the Northeastern University campus in Boston, MA.\footnote{X5G website: \url{https://x5g.org}.} The deployment includes a server room with a dedicated rack for the private 5G system and an indoor laboratory open space area with benches and experimental equipment that provide a realistic \gls{rf} environment with rich scattering and obstacles.
Figure~\ref{chap4-fig:arc-hardware} illustrates the hardware infrastructure that we deployed to support the X5G operations. This includes synchronization and networking infrastructures, radio nodes, eight \gls{arc-ota} servers with integrated \gls{du} and \gls{cu}, and additional compute infrastructure for the \gls{ric} and \gls{cn} deployments. This infrastructure, which will be described next, has been leveraged to provide connectivity for up to eight concurrent \gls{cots} \glspl{ue}, such as OnePlus smartphones (AC Nord 2003) and Sierra Wireless boards (EM9191)~\cite{sierrawireless}.

\textbf{Synchronization Infrastructure.} The synchronization infrastructure consists of a Qulsar (now VIAVI) QG-2 device acting as grandmaster clock. The QG-2 unit is connected to a GPS antenna for precise class-6 timing and generates both \gls{ptp} and \gls{synce} signals to provide frequency, phase, and time synchronization compliant with the ITU-T G.8265.1, G.8275.1, and G.8275.2 profiles. It sends the synchronization packets to the networking infrastructure through a $1$\:Gbps Ethernet connection. The networking infrastructure then offers full on-path support, which is necessary to distribute phase synchronization throughout the X5G platform.

\textbf{Networking Infrastructure.} The networking infrastructure provides connectivity between all the components of the X5G platform. It features fronthaul and backhaul capabilities through the use of two Dell switches (S5248F-ON and S5232F-ON) interconnected via four $100$\:Gbps cables in a port channel configuration. This configuration allows for the aggregation of multiple physical links into a single logical one to increase bandwidth and provide redundancy in case some of them fail. All switch ports are sliced into different \glspl{vlan} to allow the proper coexistence of the various types of traffic (i.e., fronthaul, backhaul, management).
The Dell S5248F-ON switch primarily provides backhaul capabilities to the network and acts as a boundary clock in the synchronization plane, receiving \gls{ptp} signals from the synchronization infrastructure. This switch includes 48~\glsname{sfp+} ports: 12 ports are dedicated to the \glspl{ru} and receive \gls{ptp} synchronization packets, 10 are used to connect to the OpenShift cluster and service network, 10 are used for the out-of-band management network, and 16 connect to the \gls{cn} and the Internet. Additionally, the switch includes 6~\glsname{qsfp28} ports, 4 of which interconnect with the second switch.
The Dell S5232F-ON switch mainly provides fronthaul connectivity to the \glspl{gnb}. It includes 32~\glsname{qsfp28} ports: 8 ports connect to the Mellanox cards of the \gls{arc-ota} nodes via $100$\:Gbps fiber links, and 4 connect to the Dell S5248F-ON switch. The latter also acts as a boundary clock, receiving the synchronization messages from the S5232F-ON and delivering them to the \glspl{gnb}.


\textbf{\gls{ru}.} We deployed \blue{eight} Foxconn RPQN 4T4R \glspl{ru}, operating in the \blue{n78} band, with additional units being tested in the lab, and the Keysight RuSIM emulator. The Foxconn units have 4 externally mounted antennas, each antenna with a $5$\:dBi gain, and $24$\:dBm of transmit power. The \gls{ota} transmissions are regulated as part of the Northeastern University \gls{fcc} Innovation Zone~\cite{FCC-IZ-Boston}, with an additional transmit attenuation of $20$\:dB per port to comply with transmit power limits and guarantee the coexistence of multiple in-band \glspl{ru} in the same environment. As we will discuss in Section~\ref{chap4-sec:exp-results}, we leverage two of these \glspl{ru} for the experimental analysis presented in this work. These \glspl{ru} are deployed following \gls{rf} planning procedures discussed in Section~\ref{chap4-sec:ray-tracing}. 
Plans are in place to procure \gls{cbrs} \glspl{ru} and deploy them in outdoor locations. 
\glspl{ru} from additional vendors are being tested and integrated as part of our future works. Finally, we also tested and integrated the \gls{arc-ota} stack with the Keysight RuSIM emulator, which supports the termination of the fronthaul interface on the \gls{ru} side and exposes multiple \glspl{ru} to the \gls{ran} stack, for troubleshooting, conformance testing, and performance testing~\cite{keysight_rusim}. 


\textbf{\gls{cu} and \gls{du}.} The 8~\gls{arc-ota} nodes that execute the containerized \gls{cu}/\gls{du} workloads are deployed on Gigabyte E251-U70 servers with 24-core Intel Xeon Gold 6240R CPU and $96$\:GB of RAM.
The servers---which come in a half rack chassis for deployment in \gls{ran} and edge scenarios---are equipped with a Broadcom PEX 8747 \gls{pci} switch that 
enables direct connectivity between cards installed in two dedicated \gls{pci} slots without the need for interactions with the CPU.
Specifically, the two \gls{pci} slots host an NVIDIA A100 \gls{gpu}, which supports the computational operations of the NVIDIA Aerial \gls{phy} layer, as well as a Mellanox ConnectX-6 Dx \gls{nic}. The latter, which is used for the fronthaul interface, connects to the fronthaul part of the networking infrastructure via a \acrshort{qsfp28} port and $100$\:Gbps fiber-optic cable.
In this way, the \gls{nic} can offload or receive packets directly from the GPU, thus enabling low-latency packet processing.
Finally, each server is connected to the backhaul part of the networking infrastructure through three $1$\:Gbps Ethernet links, which provide connectivity with the OpenShift cluster (and thus the Near-RT \gls{ric} and the core networks), the management infrastructure, and the Internet.
\blue{In addition to the Gigabyte servers, seven \gls{gh} machines---one of the latest NVIDIA high-computing ARM-based devices---are currently being integrated into X5G to run the \gls{arc-ota} \gls{cu}/\gls{du} worloads. Each \gls{gh} combines a 72-core NVIDIA Grace CPU Superchip and an NVIDIA H100 Tensor Core GPU, linked through NVIDIA NVLink-C2C technology, which ensures seamless data sharing with up to 900 GB/s of bandwidth. It also features $480$\:GB of RAM, two BlueField-3 \glspl{dpu}, and ConnectX-7 \glspl{nic}. This configuration provides significantly higher computational capabilities compared to the Gigabyte servers, enabling the efficient support of concurrent \gls{ran} and \gls{ai}/\gls{ml} workloads.}

\textbf{Additional Compute.} We leverage additional servers that are part of the OpenShift cluster and are used to instantiate the various \glspl{cn} and the Near-RT \gls{ric}. 
The OpenShift cluster includes three Dell R740 servers acting as control-plane nodes and two Microway Navion Dual servers as worker nodes. The OpenShift rack is linked to the X5G rack through two $10$ Gbps connections, one dedicated to OpenShift operations, and the other for the out-of-band management.
Additionally, a Dell R750 server with $56$\:cores and $256$\:GB RAM is available for the deployment and testing of additional core network elements. This server connects to the networking infrastructure via a $1$\:Gbps Ethernet link and has access to the Internet through the Northeastern University network.




\section{RF Planning with Ray-tracing}
\label{chap4-sec:ray-tracing}


In this section, we present \gls{rf} planning procedures to identify suitable locations for the \gls{ru} deployment. 
\blue{This approach leverages an exhaustive search within a ray-tracing-based digital twin framework, with the objective of maximizing the \glspl{ru} coverage while minimizing the overall interference.
The study is conducted only once during the system deployment phase and remains valid as long as no significant changes occur in the environment.}
%
%
We perform ray-tracing in a detailed \blue{digitized} representation of our indoor laboratory space in the Northeastern University ISEC building in Boston, MA, to achieve high fidelity between the real-world environment and the \blue{digital} one. We carefully study how to deploy 2 \glspl{ru} by considering the \gls{sinr} between the \glspl{ru} and the \glspl{ue} as the objective function in the optimization problem. We \blue{limit} the optimization space by using a grid of 24 possible \gls{ru} locations and 52 \gls{ue} test points\blue{, enabling an exhaustive search approach instead of a formal integer optimization problem, since these constraints keep the computation manageable.} 

\begin{figure}[htb]
    \centering
    \subfloat[Site viewer top view.]
    {\label{chap4-fig:siteviewertop}
    \includegraphics[width=0.9\linewidth]{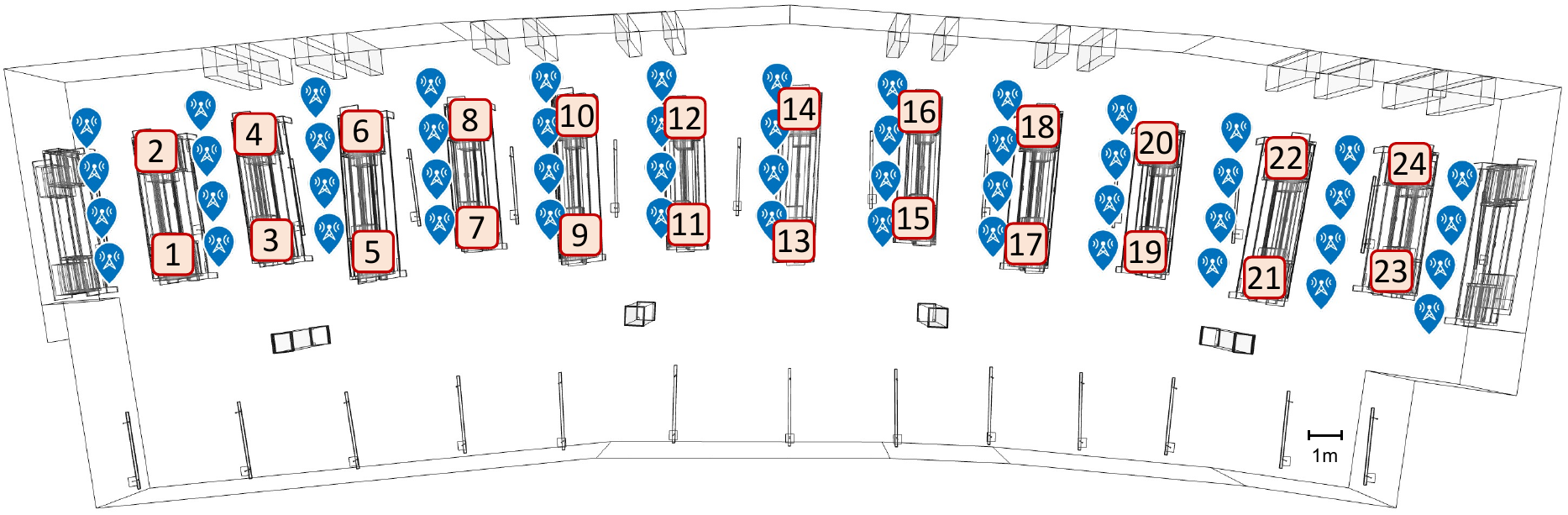}}
    \hfill
    \subfloat[Site viewer side view.]
    {
    \label{chap4-fig:siteviewerside}
    \includegraphics[width=0.9\linewidth]{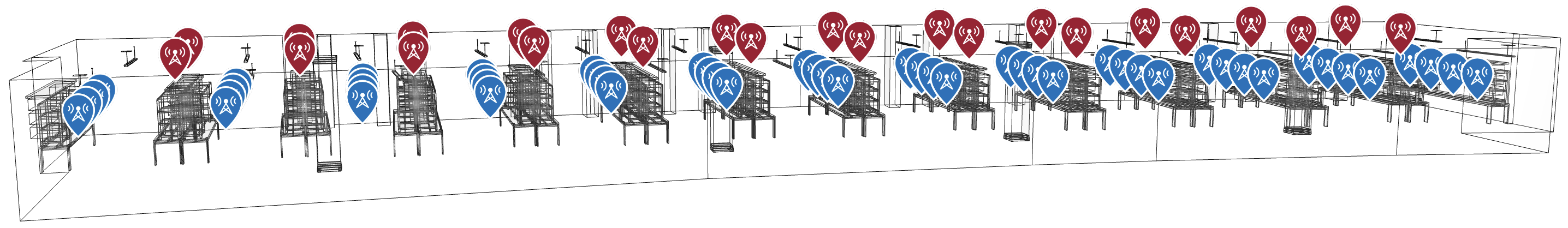}}        
    \caption{Site viewer with \acrshort{ru} (red squares) and \acrshort{ue} (blue icons) locations.}
    \label{chap4-fig:siteviewer}
\end{figure}

First, we leverage the 3D representation of our laboratory space, \blue{created as part of the digital twin framework developed in~\cite{villa2024dt} using the SketchUp modeling software.}
We then import the model in the MATLAB ray-tracing software and define the locations of \glspl{ru} and \glspl{ue} as shown in Figure~\ref{chap4-fig:siteviewertop} (from a top perspective) and in Figure~\ref{chap4-fig:siteviewerside} (from a side view). The 24 possible \glspl{ru} locations (2 for each bench) are shown in red, while the 52 test points for the \glspl{ue} (arranged in a $4 \times 13$ grid) are in blue.
%
%
Tables~\ref{chap4-table:testbeds-features} and~\ref{chap4-table:raytracing-parameters} summarize the parameters used in our ray-tracing model. For the deployment planning purpose, we consider the \glspl{ru} as transmitter nodes (TX) and the \glspl{ue} as receiver ones (RX), i.e., we tailor our deployment to downlink transmissions.
%

\begin{table}[htbp]
    \centering
    \caption{Parameters of the MATLAB ray-tracing study to determine \acrshort{ru} locations.}
    \label{chap4-table:raytracing-parameters}
    \scriptsize
    \setlength{\tabcolsep}{4pt}
    \begin{tabular}{ll}
        \toprule
        \textbf{Parameter} & \textbf{Value} \\
        \midrule
        \gls{ru} antenna spacing & $0.25$\:m \\
        \gls{ru} antenna TX power ($P_{RU}$) & $24$\:dBm \\
        \gls{ru} antenna gain ($G_{RU}$) & $5$\:dBi \\
        \gls{ru} antenna pattern & Isotropic \\
        \gls{ru} TX attenuation ($A_{RU}$) & $[0-50]$\:dB \\
        \blue{Set} of \gls{ru} locations (\blue{$\mathcal{R}$}) & $24$ in a $2 \times 12$ grid \\
        \gls{ru} height & $2.2$\:m \\
        \gls{ue} number of antennas & 2 \\
        \gls{ue} antenna spacing & $0.07$\:m \\
        \gls{ue} antenna gain ($G_{UE}$) & $1.1$\:dBi \\
        \gls{ue} noise figure ($F_{UE}$) & $5$\:dB \\
        \blue{Set} of \glspl{ue} locations (\blue{$\mathcal{U}$}) & $52$ in a $4 \times 13$ grid \\
        \gls{ue} height & $0.8$\:m \\
        Environment material & Wood \\
        Max number of reflections & $3$ \\
        Max diffraction order & $1$ \\
        Ray-tracing method & Shooting and bouncing rays \\
        \bottomrule
    \end{tabular}
\end{table}

The ray-tracer generates a $24 \times 52$ matrix $\mathbf{C}$ where each entry $c_{i,j}$ corresponds to the channel information between $RU_i$ with \blue{$i \in \mathcal{R}$, $ \mathcal{R} = {1,...,24}$, and $UE_j$ with $j \in \mathcal{U}$, $ \mathcal{U} = {1,...,52}$}. We use this to derive relevant parameters such as the thermal noise (\blue{$\mathcal{N}$}) and the path loss (\blue{$\mathcal{L}$}) to compute 
%
the \gls{rssi} $\blue{\mathcal{S}}_{i,j}$ for $UE_j$ connected to $RU_i$, as follows:
%
%
\begin{align}\label{chap4-eq:rssi}\small
    \blue{\mathcal{S}}_{i,j} = P_{RU,i} + G_{RU,i} - A_{RU,i} - \blue{\mathcal{L}}_{i,j} + G_{UE,j},
\end{align}
where $P_{RU,i}$, $G_{RU,i}$, and $A_{RU,i}$ are the antenna TX power, gain, and attenuation of $RU_i$, respectively.
%
Then, considering the linear representation of $\hat{\blue{\mathcal{S}}}_{i,j}$, the \gls{sinr} $\Gamma_{i,j}$ is
\begin{align}\label{chap4-eq:sinr}
    \Gamma_{i,j} = \frac{\hat{\blue{\mathcal{S}}}_{i,j}}{\blue{\mathcal{N}} F_{UE,i} + \sum\limits_{u=1, u \neq i}^M \hat{\blue{\mathcal{S}}}_{u,j}},
\end{align}
%
%
%
where 
$M$ is the number of \glspl{ru} being deployed, $\blue{\mathcal{N}}$ is the thermal noise\blue{, and $F_{UE,i}$ is the noise figure of $UE_i$}.
%
The \gls{sinr} $\Gamma_{i,j}$ considers the interference to the signal from $RU_i$ to $UE_j$ due to downlink transmissions of all other $M - 1$ \glspl{ru} being deployed.

In our \gls{rf} planning, we deploy two \glspl{ru} (i.e., $M=2$). In the following study, we consider scenarios where the first \gls{ru} serves one \gls{ue} from the test locations, while we assume that the second \gls{ru} creates interference with the first, even without being assigned any \gls{ue} from the list.
%
%
We test all possible combinations of the 24~\gls{ru} test locations, which, following the combinatorial equation of choosing 24 elements ($n$) in groups of~2 ($r$) as $C(n, r) = \frac{n!}{r!(n-r)!}$, results in a total of 276 pairs.
\blue{The proposed approach for determining the optimal \gls{ru} locations and the maximum average \gls{sinr} ($\Phi_{\max}(\Gamma)$), called score, is presented in Algorithm~\ref{chap4-algo:rfplanning}.
It takes as input the set of \gls{ru} locations ($\mathcal{R}$), the set of \gls{ue} test points ($\mathcal{U}$), and the \gls{sinr} matrix $\mathbf{\Gamma}$. Then, it performs an exhaustive search, testing all pairs of \gls{ru} against all \glspl{ue} to determine the optimal \gls{ru} pair $(p^*, q^*)$ with the best maximum average \gls{sinr} $\Phi_{\max}(\Gamma)$.}
%
%
%
%
%
\begin{algorithm}[htb]
\caption{\blue{Exhaustive Search Algorithm for RF Planning}}
\begin{algorithmic}[1]\label{chap4-algo:rfplanning}
\REQUIRE \blue{Set of RU locations ($\mathcal{R}$), set of UE test points ($\mathcal{U}$), precomputed SINR matrix $\mathbf{\Gamma}$}
\ENSURE \blue{Optimal RU pair $(p^*, q^*)$ and maximum average SINR $\Phi_{\max}(\Gamma)$}
\blue{
\STATE Initialize $bestScore \gets -\infty$ and $bestPair \gets \text{None}$
\FORALL{RU pairs $(p, q) \in \binom{\mathcal{R}}{2}$}
    \STATE $sumSINR \gets 0$
    \FORALL{UE $j \in \mathcal{U}$}
        \STATE Compute $\Gamma_{p,j}$ (RU $p$ serving, RU $q$ interfering)
        \STATE Compute $\Gamma_{q,j}$ (RU $q$ serving, RU $p$ interfering)
        \STATE $sinrMax \gets \max(\Gamma_{p,j}, \Gamma_{q,j})$
        \STATE $sumSINR \gets sumSINR + sinrMax$
    \ENDFOR
    \STATE $avgSINR \gets sumSINR / |\mathcal{U}|$
    \IF{$avgSINR > bestScore$}
        \STATE $bestScore \gets avgSINR$
        \STATE $bestPair \gets (p, q)$
    \ENDIF
\ENDFOR
\RETURN $bestPair, bestScore$
}
\end{algorithmic}
\end{algorithm}

We test this algorithm with different values of the attenuation $A_{RU}$, from $0$ to $50$\:dB in $10$\:dB increments.
%
%
\begin{figure}[htb]
  \centering
  \subfloat[$0$\:dB attenuation]{\label{chap4-fig:heatmap-scores-0db}\includegraphics[width=0.33\linewidth]{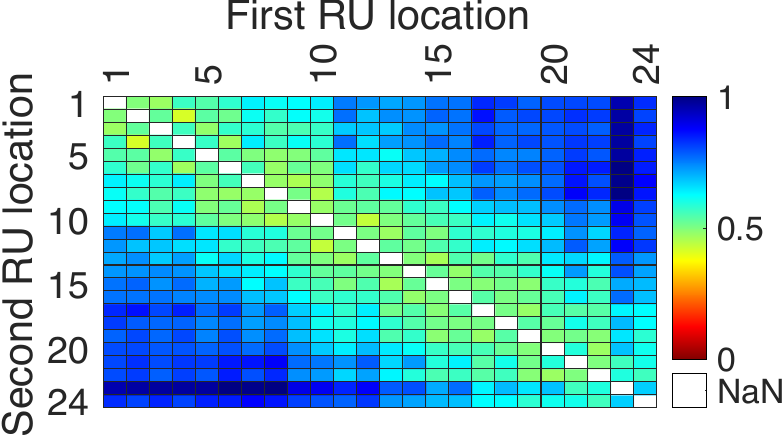}}
  \hfill
  \subfloat[$10$\:dB attenuation]{\label{chap4-fig:heatmap-scores-10db}\includegraphics[width=0.33\linewidth]{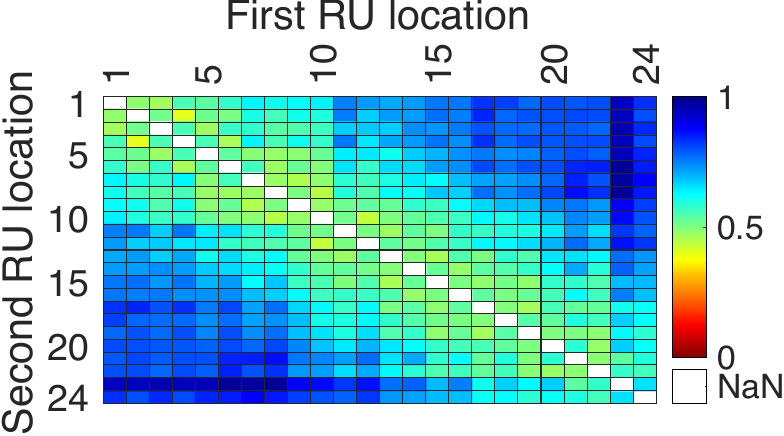}}
  \hfill
  \subfloat[$20$\:dB attenuation]
  {\label{chap4-fig:heatmap-scores-20db}\includegraphics[width=0.33\linewidth]{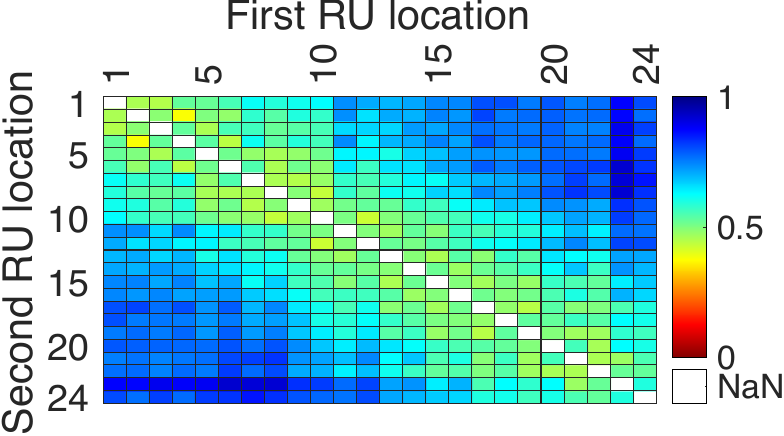}}
  \hfill
  \subfloat[$30$\:dB attenuation]
  {\label{chap4-fig:heatmap-scores-30db}\includegraphics[width=0.33\linewidth]{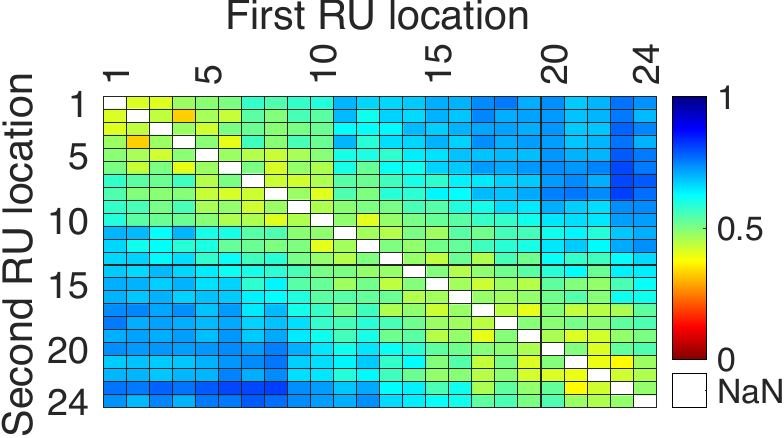}}
  \hfill
  \subfloat[$40$\:dB attenuation]{\label{chap4-fig:heatmap-scores-40db}\includegraphics[width=0.33\linewidth]{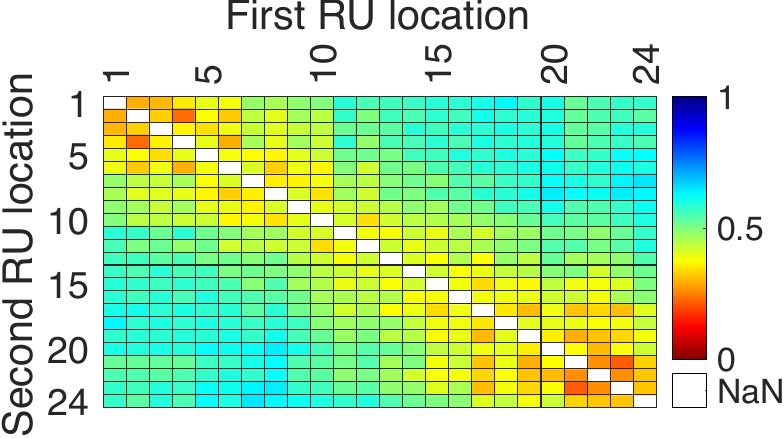}}
  \hfill
  \subfloat[$50$\:dB attenuation]{\label{chap4-fig:heatmap-scores-50db}    \includegraphics[width=0.33\linewidth]{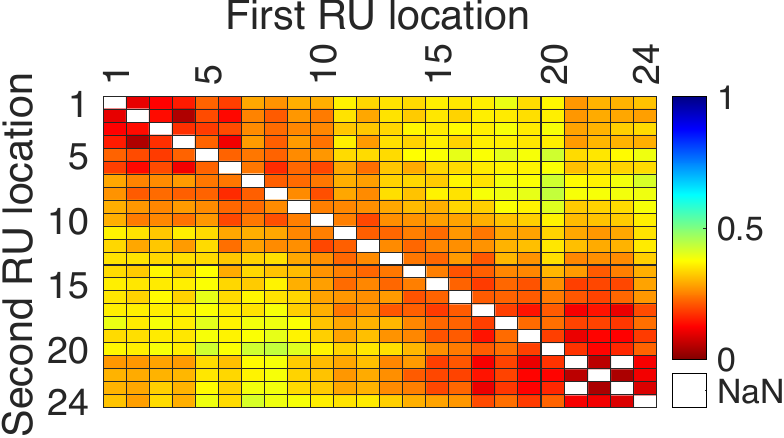}}
  \caption{Heatmap results of the normalized average \acrshort{sinr} $\blue{\Phi}(\Gamma)$ with 2 \acrshortpl{ru}.}
  \label{chap4-fig:heatmap-scores}
\end{figure}
\begin{table}[htbp]
    \centering
    \caption{Best \acrshortpl{ru} and average \acrshort{sinr} $\blue{\Phi}(\Gamma)$ range values.}
    \label{chap4-table:score-results}
    \scriptsize
    \setlength{\tabcolsep}{4pt}
    \begin{tabular}{lcc}
        \toprule
        \textbf{$A_{RU}$ [dB]} & \textbf{\acrshort{ru} locations with best \acrshort{sinr}} & \textbf{[Min, Max] $\blue{\Phi}(\Gamma)$ [dB]} \\
        \midrule
        0 & [8, 23] & [6.08, 23.33] \\
        10 & [6, 23] & [5.71, 22.66] \\
        20 & [6, 23] & [5.00, 21.03] \\
        30 & [8, 23] & [3.58, 17.82] \\
        40 & [7, 24] & [0.19, 12.94] \\
        50 & [8, 20] & [-6.23, 6.63] \\
        \bottomrule
    \end{tabular}
\end{table}
Figure~\ref{chap4-fig:heatmap-scores} visualizes the normalized values of \blue{the score $\Phi(\Gamma)$} for all possible combinations of \gls{ru} pairs and different attenuation values. Additionally, Table~\ref{chap4-table:score-results} provides the best \gls{ru} locations including the minimum and maximum values of $\blue{\Phi}(\Gamma)$ for the corresponding combinations.
As expected, locations with further \glspl{ru} exhibit higher average \gls{sinr} values, as they are less affected by interference. However, it is important to note that the score also considers coverage, \blue{as it is computed based on the \gls{sinr}}. Consequently, the optimal combination of locations identifies \glspl{ru} that are further apart but not necessarily the furthermost pair.
Considering these results, for the experiments in Section~\ref{chap4-sec:exp-results}, we select a TX attenuation of $20$\:dB, which exhibits a good trade-off between coverage and average \gls{sinr} values.
Moreover, during our real-world experiments, we observed that a $20$\:dB attenuation leads to increased system stability and reduced degradation compared to lower attenuation values, resulting in improved overall performance, as it reduces the likelihood of saturation at the \gls{ue} antenna.
Therefore, we select locations [6,23] for our \glspl{ru} deployment.

\section{Experiment Results}
\label{chap4-sec:exp-results}



In this section, we describe the design and execution of a comprehensive set of experiments that illustrate the capabilities of the X5G infrastructure in a variety of operational scenarios.
We assess the adaptability of the testbed through rigorous testing, utilizing iPerf to measure network throughput and MPEG-DASH to gauge video streaming quality. 
The experiments are \blue{mainly} conducted in the same indoor laboratory area modeled in Section~\ref{chap4-sec:ray-tracing}. They include static configurations with a single \gls{ue} as well as more complex setups with multiple \glspl{ue} and \glspl{ru}, \blue{leveraging the Keysight RuSIM emulator}, and scenarios with \gls{ue} mobility.

\subsection{Setup Overview}
\label{chap4-sec:exp-setup-overview}
%
\blue{We consider two different setups: (i)~Gigabyte \gls{ran} servers with a 2x2~\gls{mimo} configuration ($L_{DL}, L_{UL}$), 2~layers \gls{dl}, 1~layer \gls{ul}, a DDDSU \gls{tdd} pattern, and a modulation order ($Q_{m}$) up to 64-\gls{qam} (results for this setup are shown in Sections~\ref{chap4-sec:exp-static}, \ref{chap4-sec:exp-mobile}, and \ref{chap4-sec:exp-video}); and (ii)~\gls{gh} \gls{ran} servers with a 4x4~\gls{mimo} configuration, 4~layers \gls{dl}, 1~layer \gls{ul}, a DDDDDDSUUU \gls{tdd} pattern, and a $Q_{m}$ up to 256-\gls{qam} (Sections~\ref{chap4-sec:exp-peak} and \ref{chap4-sec:exp-long}).
All experiments utilize a carrier frequency of $3.75$\:GHz with a bandwidth ($\beta$) of $100$\:MHz.
}

%
%

\begin{figure}[htb]
    \centering
    \includegraphics[width=.95\linewidth]{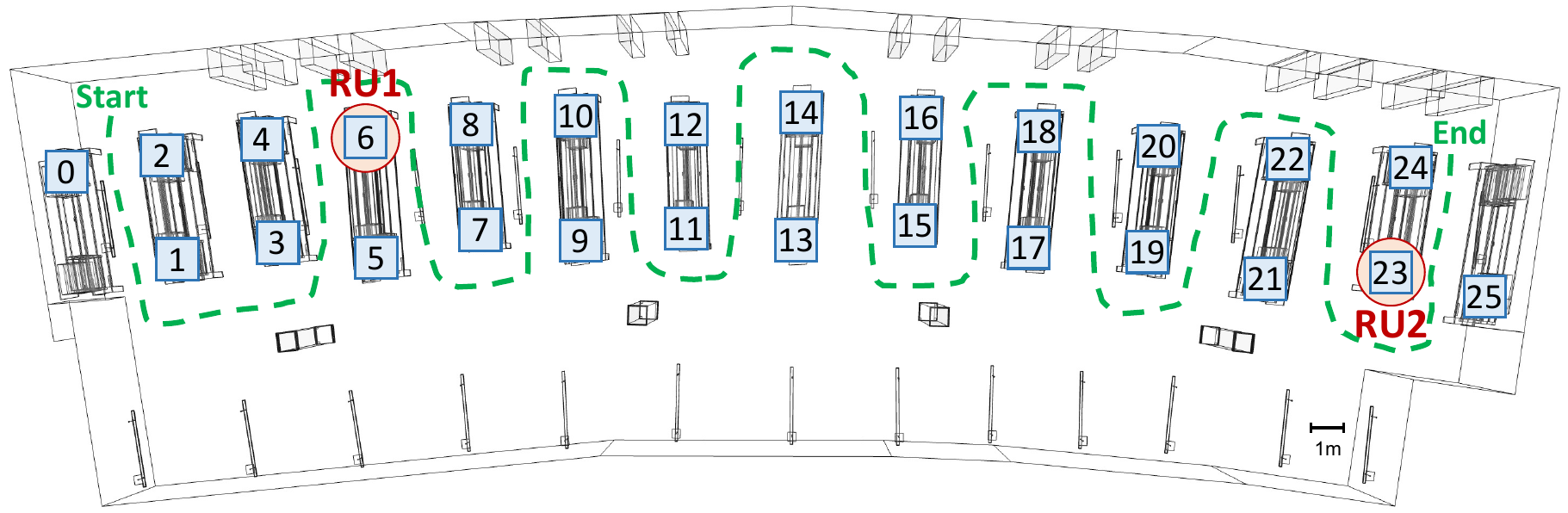}
    \setlength\belowcaptionskip{-.5cm}
    \caption{Node locations considered in our experiments: \acrshortpl{ru} (red circles in 6 and 23); possible static \acrshortpl{ue} (blue squares); and mobile \acrshortpl{ue} (green dashed line).}
    \label{chap4-fig:node-locations}
\end{figure}

The experiments are \blue{mainly} conducted in the laboratory area shown in Figure~\ref{chap4-fig:node-locations}, which highlights the \gls{ru} locations (outcome of the ray-tracing study discussed in Section~\ref{chap4-sec:ray-tracing}), as well as the \gls{ue} locations and the mobility pattern for the non-static experiments. All tests involving a single \gls{ru} are conducted at location~6, as illustrated in Figure~\ref{chap4-fig:node-locations}.
%
%
%
%
%
An edge server, configured to support the iPerf and MPEG-DASH applications, is deployed within the campus network to ensure minimal latency, ranging from $1$ to $2$\:ms. During static throughput tests, \gls{tcp} backlogged traffic is transmitted first in the downlink and then in the uplink directions for $40$\:seconds each across different \gls{ue} configurations.
For video streaming, the server employs FFmpeg~\cite{ffmpeg} to deliver five distinct profiles simultaneously at various resolutions—ranging from 1080P at $250$~Mbps to 540P at $10$~Mbps—to the \glspl{ue}. On the device side, we leverage a pre-compiled iPerf3 binary for Android to generate \gls{tcp} traffic, and Google’s ExoPlayer for client-side video playback. 
Each set of experiments is replicated five times to ensure data reliability, with results including mean values and 95\% confidence intervals of the metrics plotted. These metrics encompass application layer measurements such as throughput, bitrate, and rebuffer ratio, alongside \gls{mac} layer metrics like \gls{sinr}, \gls{rsrp}, and \gls{mcs}, collected at the \gls{oai} \gls{gnb} level.



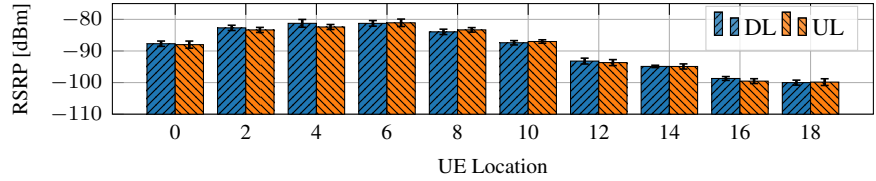
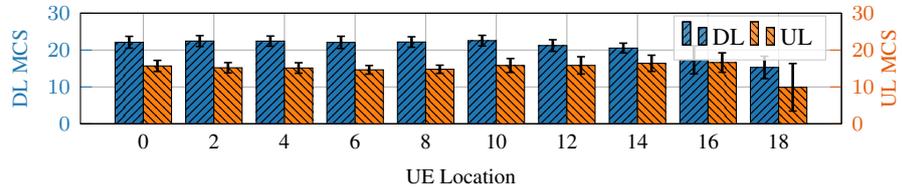
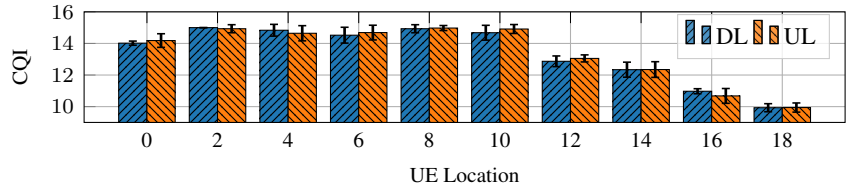
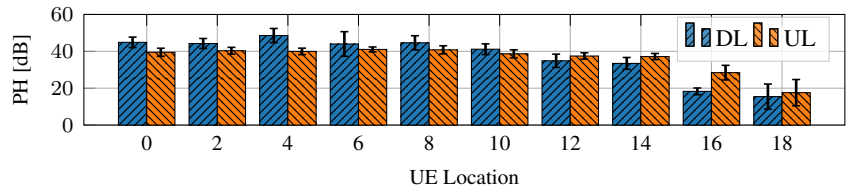
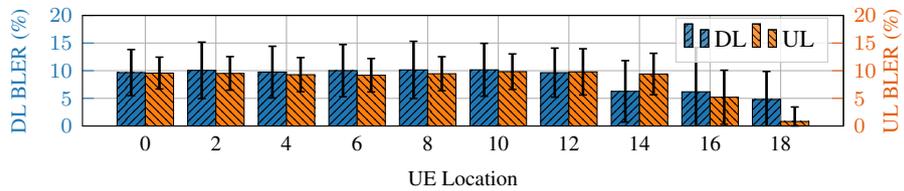
\begin{figure}[htbp]
\centering
    \subfloat[Throughput]{
    \label{chap4-fig:1ru1ue_static_iperf_th}
    \centering
    \setlength\fwidth{0.8\linewidth}
    \setlength\fheight{.2\linewidth}
\begin{tikzpicture}
\pgfplotsset{every tick label/.append style={font=\scriptsize}}

\definecolor{darkgray176}{RGB}{176,176,176}
\definecolor{darkorange25512714}{RGB}{255,127,14}
\definecolor{darkorange2309111}{RGB}{230,91,11}
\definecolor{lightgray204}{RGB}{204,204,204}
\definecolor{steelblue31119180}{RGB}{31,119,180}


\begin{axis}[
width=0.951\fwidth,
height=\fheight,
at={(0\fwidth,0\fheight)},
legend cell align={left},
legend style={fill opacity=0.8, draw opacity=1, text opacity=1, draw=lightgray204, font=\footnotesize},
legend columns=2,
x grid style={darkgray176},
xmajorticks=false,
xmin=-0.69, xmax=10.09,
xtick={0.2,1.2,2.2,3.2,4.2,5.2,6.2,7.2,8.2,9.2},
xticklabel style={rotate=45.0},
y grid style={darkgray176},
ylabel=\textcolor{steelblue31119180}{DL TH [Mbps]},
ylabel style={font=\scriptsize},
xlabel style={font=\scriptsize},
ymin=0, ymax=350,
ytick pos=left,
ytick style={color=steelblue31119180},
yticklabel style={color=steelblue31119180},
xmajorgrids,
ymajorgrids
]


\addlegendimage{ybar,ybar legend,draw=black,fill=steelblue31119180,postaction={pattern=north east lines,pattern color=black}}
\addlegendentry{DL}
\addlegendimage{ybar,ybar legend,draw=black,fill=darkorange25512714,postaction={pattern=north west lines,pattern color=black}}
\addlegendentry{UL}


\draw[draw=black,fill=steelblue31119180,postaction={pattern=north east lines,pattern color=black}] (axis cs:-0.2,0) rectangle (axis cs:0.2,297.779273813333);
\draw[draw=black,fill=steelblue31119180,postaction={pattern=north east lines,pattern color=black}] (axis cs:0.8,0) rectangle (axis cs:1.2,278.951601066667);
\draw[draw=black,fill=steelblue31119180,postaction={pattern=north east lines,pattern color=black}] (axis cs:1.8,0) rectangle (axis cs:2.2,291.50394624);
\draw[draw=black,fill=steelblue31119180,postaction={pattern=north east lines,pattern color=black}] (axis cs:2.8,0) rectangle (axis cs:3.2,281.087304533333);
\draw[draw=black,fill=steelblue31119180,postaction={pattern=north east lines,pattern color=black}] (axis cs:3.8,0) rectangle (axis cs:4.2,286.19221888);
\draw[draw=black,fill=steelblue31119180,postaction={pattern=north east lines,pattern color=black}] (axis cs:4.8,0) rectangle (axis cs:5.2,293.477207893333);
\draw[draw=black,fill=steelblue31119180,postaction={pattern=north east lines,pattern color=black}] (axis cs:5.8,0) rectangle (axis cs:6.2,281.056259413333);
\draw[draw=black,fill=steelblue31119180,postaction={pattern=north east lines,pattern color=black}] (axis cs:6.8,0) rectangle (axis cs:7.2,282.361967786667);
\draw[draw=black,fill=steelblue31119180,postaction={pattern=north east lines,pattern color=black}] (axis cs:7.8,0) rectangle (axis cs:8.2,226.071218773333);
\draw[draw=black,fill=steelblue31119180,postaction={pattern=north east lines,pattern color=black}] (axis cs:8.8,0) rectangle (axis cs:9.2,177.708098986667);


\path [draw=black, line width=1pt]
(axis cs:0,293.561317740977)
--(axis cs:0,301.997229885689);
\path [draw=black, line width=1pt]
(axis cs:1,274.913928006811)
--(axis cs:1,282.989274126523);
\path [draw=black, line width=1pt]
(axis cs:2,285.413594787131)
--(axis cs:2,297.594297692869);
\path [draw=black, line width=1pt]
(axis cs:3,274.57133522018)
--(axis cs:3,287.603273846487);
\path [draw=black, line width=1pt]
(axis cs:4,280.463374578279)
--(axis cs:4,291.921063181721);
\path [draw=black, line width=1pt]
(axis cs:5,289.103429270324)
--(axis cs:5,297.850986516343);
\path [draw=black, line width=1pt]
(axis cs:6,276.853204662378)
--(axis cs:6,285.259314164289);
\path [draw=black, line width=1pt]
(axis cs:7,278.164506604369)
--(axis cs:7,286.559428968964);
\path [draw=black, line width=1pt]
(axis cs:8,215.497162859975)
--(axis cs:8,236.645274686691);
\path [draw=black, line width=1pt]
(axis cs:9,163.498710494908)
--(axis cs:9,191.917487478425);


\addplot [semithick, black, mark=-, mark size=1.5, mark options={solid}, only marks]
table {%
0 293.561317740977
1 274.913928006811
2 285.413594787131
3 274.57133522018
4 280.463374578279
5 289.103429270324
6 276.853204662378
7 278.164506604369
8 215.497162859975
9 163.498710494908
};


\addplot [semithick, black, mark=-, mark size=1.5, mark options={solid}, only marks]
table {%
0 301.997229885689
1 282.989274126523
2 297.594297692869
3 287.603273846487
4 291.921063181721
5 297.850986516343
6 285.259314164289
7 286.559428968964
8 236.645274686691
9 191.917487478425
};
\end{axis}


\begin{axis}[
width=0.951\fwidth,
height=\fheight,
at={(0\fwidth,0\fheight)},
axis y line*=right,
legend cell align={left},
legend style={fill opacity=0.8, draw opacity=1, text opacity=1, draw=lightgray204, font=\footnotesize},
legend columns=2,
x grid style={darkgray176},
xmin=-0.69, xmax=10.09,
xtick pos=left,
ytick style={color=darkorange2309111},
xtick={0.2,1.2,2.2,3.2,4.2,5.2,6.2,7.2,8.2,9.2},
xticklabels={0,2,4,6,8,10,12,14,16,18},
y grid style={darkgray176},
ylabel=\textcolor{darkorange2309111}{UL TH [Mbps]},
ylabel style={font=\scriptsize},
xlabel style={font=\scriptsize},
xlabel={UE Location},
ymin=0, ymax=50,
ytick pos=right,
ytick style={color=darkorange2309111},
yticklabel style={anchor=west, color=darkorange2309111},
ylabel shift=-5pt
]


\draw[draw=black,fill=darkorange25512714,postaction={pattern=north west lines,pattern color=black}] (axis cs:0.2,0) rectangle (axis cs:0.6,29.4579950933333);
\draw[draw=black,fill=darkorange25512714,postaction={pattern=north west lines,pattern color=black}] (axis cs:1.2,0) rectangle (axis cs:1.6,34.2534826666667);
\draw[draw=black,fill=darkorange25512714,postaction={pattern=north west lines,pattern color=black}] (axis cs:2.2,0) rectangle (axis cs:2.6,32.65265664);
\draw[draw=black,fill=darkorange25512714,postaction={pattern=north west lines,pattern color=black}] (axis cs:3.2,0) rectangle (axis cs:3.6,36.9588087466667);
\draw[draw=black,fill=darkorange25512714,postaction={pattern=north west lines,pattern color=black}] (axis cs:4.2,0) rectangle (axis cs:4.6,34.6589320533333);
\draw[draw=black,fill=darkorange25512714,postaction={pattern=north west lines,pattern color=black}] (axis cs:5.2,0) rectangle (axis cs:5.6,25.6132164266667);
\draw[draw=black,fill=darkorange25512714,postaction={pattern=north west lines,pattern color=black}] (axis cs:6.2,0) rectangle (axis cs:6.6,19.6153617066667);
\draw[draw=black,fill=darkorange25512714,postaction={pattern=north west lines,pattern color=black}] (axis cs:7.2,0) rectangle (axis cs:7.6,16.84013056);
\draw[draw=black,fill=darkorange25512714,postaction={pattern=north west lines,pattern color=black}] (axis cs:8.2,0) rectangle (axis cs:8.6,2.56551594666667);
\draw[draw=black,fill=darkorange25512714,postaction={pattern=north west lines,pattern color=black}] (axis cs:9.2,0) rectangle (axis cs:9.6,1.18139562666667);


\path [draw=black, line width=1pt]
(axis cs:0.4,28.0220236375514)
--(axis cs:0.4,30.8939665491152);
\path [draw=black, line width=1pt]
(axis cs:1.4,33.0880112122231)
--(axis cs:1.4,35.4189541211102);
\path [draw=black, line width=1pt]
(axis cs:2.4,31.3074846346825)
--(axis cs:2.4,33.9978286453175);
\path [draw=black, line width=1pt]
(axis cs:3.4,36.5276271183574)
--(axis cs:3.4,37.3899903749759);
\path [draw=black, line width=1pt]
(axis cs:4.4,33.8209615410315)
--(axis cs:4.4,35.4969025656351);
\path [draw=black, line width=1pt]
(axis cs:5.4,24.0539830782761)
--(axis cs:5.4,27.1724497750572);
\path [draw=black, line width=1pt]
(axis cs:6.4,18.2651274938699)
--(axis cs:6.4,20.9655959194635);
\path [draw=black, line width=1pt]
(axis cs:7.4,15.5330246472272)
--(axis cs:7.4,18.1472364727728);
\path [draw=black, line width=1pt]
(axis cs:8.4,2.16213861293845)
--(axis cs:8.4,2.96889328039488);
\path [draw=black, line width=1pt]
(axis cs:9.4,0.995143241485241)
--(axis cs:9.4,1.36764801184809);


\addplot [semithick, black, mark=-, mark size=1.5, mark options={solid}, only marks]
table {%
0.4 28.0220236375514
1.4 33.0880112122231
2.4 31.3074846346825
3.4 36.5276271183574
4.4 33.8209615410315
5.4 24.0539830782761
6.4 18.2651274938699
7.4 15.5330246472272
8.4 2.16213861293845
9.4 0.995143241485241
};


\addplot [semithick, black, mark=-, mark size=1.5, mark options={solid}, only marks]
table {%
0.4 30.8939665491152
1.4 35.4189541211102
2.4 33.9978286453175
3.4 37.3899903749759
4.4 35.4969025656351
5.4 27.1724497750572
6.4 20.9655959194635
7.4 18.1472364727728
8.4 2.96889328039488
9.4 1.36764801184809
};

\end{axis}

\end{tikzpicture}}
    
    \hfill
    
    \subfloat[\gls{rsrp}]{
    \label{chap4-fig:1ru1ue_static_iperf_rsrp}
    \centering
    \setlength\fwidth{0.8\linewidth}
    \setlength\fheight{.2\linewidth}
\begin{tikzpicture}
\pgfplotsset{every tick label/.append style={font=\scriptsize}}

\definecolor{darkgray176}{RGB}{176,176,176}
\definecolor{darkorange25512714}{RGB}{255,127,14}
\definecolor{lightgray204}{RGB}{204,204,204}
\definecolor{steelblue31119180}{RGB}{31,119,180}


\begin{axis}[
width=0.951\fwidth,
height=\fheight,
at={(0\fwidth,0\fheight)},
legend cell align={left},
legend style={fill opacity=0.8, draw opacity=1, text opacity=1, draw=lightgray204, font=\footnotesize},
legend columns=2,
x grid style={darkgray176},
xmin=-0.69, xmax=10.09,
xtick style={color=black},
xtick={0.2,1.2,2.2,3.2,4.2,5.2,6.2,7.2,8.2,9.2},
xticklabels={0,2,4,6,8,10,12,14,16,18},
y grid style={darkgray176},
ylabel={RSRP [dBm]},
xlabel={UE Location},
ylabel style={font=\scriptsize},
xlabel style={font=\scriptsize},
ymin=-110, ymax=-75,
ytick pos=left,
xmajorgrids,
ymajorgrids
]


\addlegendimage{ybar,ybar legend,draw=black,fill=steelblue31119180,postaction={pattern=north east lines,pattern color=black}}
\addlegendentry{DL}


\draw[draw=black,fill=steelblue31119180,postaction={pattern=north east lines,pattern color=black}] (axis cs:-0.2,-110) rectangle (axis cs:0.2,-87.704347826087);
\draw[draw=black,fill=steelblue31119180,postaction={pattern=north east lines,pattern color=black}] (axis cs:0.8,-110) rectangle (axis cs:1.2,-82.6869565217391);
\draw[draw=black,fill=steelblue31119180,postaction={pattern=north east lines,pattern color=black}] (axis cs:1.8,-110) rectangle (axis cs:2.2,-81.2521739130435);
\draw[draw=black,fill=steelblue31119180,postaction={pattern=north east lines,pattern color=black}] (axis cs:2.8,-110) rectangle (axis cs:3.2,-81.2608695652174);
\draw[draw=black,fill=steelblue31119180,postaction={pattern=north east lines,pattern color=black}] (axis cs:3.8,-110) rectangle (axis cs:4.2,-83.9478260869565);
\draw[draw=black,fill=steelblue31119180,postaction={pattern=north east lines,pattern color=black}] (axis cs:4.8,-110) rectangle (axis cs:5.2,-87.4);
\draw[draw=black,fill=steelblue31119180,postaction={pattern=north east lines,pattern color=black}] (axis cs:5.8,-110) rectangle (axis cs:6.2,-93.2);
\draw[draw=black,fill=steelblue31119180,postaction={pattern=north east lines,pattern color=black}] (axis cs:6.8,-110) rectangle (axis cs:7.2,-94.8869565217391);
\draw[draw=black,fill=steelblue31119180,postaction={pattern=north east lines,pattern color=black}] (axis cs:7.8,-110) rectangle (axis cs:8.2,-98.704347826087);
\draw[draw=black,fill=steelblue31119180,postaction={pattern=north east lines,pattern color=black}] (axis cs:8.8,-110) rectangle (axis cs:9.2,-100.017391304348);


\addlegendimage{ybar,ybar legend,draw=black,fill=darkorange25512714,postaction={pattern=north west lines,pattern color=black}}
\addlegendentry{UL}


\draw[draw=black,fill=darkorange25512714,postaction={pattern=north west lines,pattern color=black}] (axis cs:0.2,-110) rectangle (axis cs:0.6,-87.9913043478261);
\draw[draw=black,fill=darkorange25512714,postaction={pattern=north west lines,pattern color=black}] (axis cs:1.2,-110) rectangle (axis cs:1.6,-83.3391304347826);
\draw[draw=black,fill=darkorange25512714,postaction={pattern=north west lines,pattern color=black}] (axis cs:2.2,-110) rectangle (axis cs:2.6,-82.4086956521739);
\draw[draw=black,fill=darkorange25512714,postaction={pattern=north west lines,pattern color=black}] (axis cs:3.2,-110) rectangle (axis cs:3.6,-81.0869565217391);
\draw[draw=black,fill=darkorange25512714,postaction={pattern=north west lines,pattern color=black}] (axis cs:4.2,-110) rectangle (axis cs:4.6,-83.3217391304348);
\draw[draw=black,fill=darkorange25512714,postaction={pattern=north west lines,pattern color=black}] (axis cs:5.2,-110) rectangle (axis cs:5.6,-87);
\draw[draw=black,fill=darkorange25512714,postaction={pattern=north west lines,pattern color=black}] (axis cs:6.2,-110) rectangle (axis cs:6.6,-93.6782608695652);
\draw[draw=black,fill=darkorange25512714,postaction={pattern=north west lines,pattern color=black}] (axis cs:7.2,-110) rectangle (axis cs:7.6,-94.904347826087);
\draw[draw=black,fill=darkorange25512714,postaction={pattern=north west lines,pattern color=black}] (axis cs:8.2,-110) rectangle (axis cs:8.6,-99.5391304347826);
\draw[draw=black,fill=darkorange25512714,postaction={pattern=north west lines,pattern color=black}] (axis cs:9.2,-110) rectangle (axis cs:9.6,-99.8550724637681);


\path [draw=black, line width=1pt]
(axis cs:0,-88.5521924932933)
--(axis cs:0,-86.8565031588807);
\path [draw=black, line width=1pt]
(axis cs:1,-83.5179907261917)
--(axis cs:1,-81.8559223172865);
\path [draw=black, line width=1pt]
(axis cs:2,-82.4865036072704)
--(axis cs:2,-80.0178442188166);
\path [draw=black, line width=1pt]
(axis cs:3,-82.1502509441677)
--(axis cs:3,-80.3714881862671);
\path [draw=black, line width=1pt]
(axis cs:4,-84.794320180374)
--(axis cs:4,-83.1013319935389);
\path [draw=black, line width=1pt]
(axis cs:5,-88.085693399849)
--(axis cs:5,-86.714306600151);
\path [draw=black, line width=1pt]
(axis cs:6,-94.1290629121523)
--(axis cs:6,-92.2709370878477);
\path [draw=black, line width=1pt]
(axis cs:7,-95.2560525581009)
--(axis cs:7,-94.5178604853773);
\path [draw=black, line width=1pt]
(axis cs:8,-99.3109476832644)
--(axis cs:8,-98.0976480689096);
\path [draw=black, line width=1pt]
(axis cs:9,-100.811918734186)
--(axis cs:9,-99.22286387451);


\addplot [semithick, black, mark=-, mark size=1.5, mark options={solid}, only marks]
table {%
0 -88.5521924932933
1 -83.5179907261917
2 -82.4865036072704
3 -82.15025094416779
4 -84.794320180374
5 -88.085693399849
6 -94.1290629121523
7 -95.2560525581009
8 -99.3109476832644
9 -100.811918734186
};


\addplot [semithick, black, mark=-, mark size=1.5, mark options={solid}, only marks]
table {%
0 -86.8565031588807
1 -81.8559223172865
2 -80.0178442188166
3 -80.3714881862671
4 -83.1013319935389
5 -86.714306600151
6 -92.2709370878477
7 -94.5178604853773
8 -98.0976480689096
9 -99.22286387451
};


\path [draw=black, line width=1pt]
(axis cs:0.4,-89.1190691933993)
--(axis cs:0.4,-86.8635395022529);
\path [draw=black, line width=1pt]
(axis cs:1.4,-84.1435803015243)
--(axis cs:1.4,-82.5346805670409);
\path [draw=black, line width=1pt]
(axis cs:2.4,-83.2022624633202)
--(axis cs:2.4,-81.6151288400276);
\path [draw=black, line width=1pt]
(axis cs:3.4,-82.2684306039206)
--(axis cs:3.4,-79.9054824395579);
\path [draw=black, line width=1pt]
(axis cs:4.4,-84.0417272649247)
--(axis cs:4.4,-82.6017509959449);
\path [draw=black, line width=1pt]
(axis cs:5.4,-87.5461186812728)
--(axis cs:5.4,-86.4538813187272);
\path [draw=black, line width=1pt]
(axis cs:6.4,-94.6292329123125)
--(axis cs:6.4,-92.7272888268179);
\path [draw=black, line width=1pt]
(axis cs:7.4,-95.7205640969140)
--(axis cs:7.4,-94.0881315552600);
\path [draw=black, line width=1pt]
(axis cs:8.4,-100.303317383006)
--(axis cs:8.4,-98.7749434865592);
\path [draw=black, line width=1pt]
(axis cs:9.4,-100.919087203785)
--(axis cs:9.4,-98.7910577237512);


\addplot [semithick, black, mark=-, mark size=1.5, mark options={solid}, only marks]
table {%
0.4 -89.1190691933993
1.4 -84.1435803015243
2.4 -83.2022624633202
3.4 -82.2684306039206
4.4 -84.0417272649247
5.4 -87.5461186812728
6.4 -94.6292329123125
7.4 -95.7205640969140
8.4 -100.303317383006
9.4 -100.919087203785
};


\addplot [semithick, black, mark=-, mark size=1.5, mark options={solid}, only marks]
table {%
0.4 -86.8635395022529
1.4 -82.5346805670409
2.4 -81.6151288400276
3.4 -79.9054824395579
4.4 -82.6017509959449
5.4 -86.4538813187272
6.4 -92.7272888268179
7.4 -94.0881315552600
8.4 -98.7749434865592
9.4 -98.7910577237512
};

\end{axis}

\end{tikzpicture}}

    \hfill
    
    \subfloat[\gls{mcs}]{
    \label{chap4-fig:1ru1ue_static_iperf_mcs}
    \centering
    \setlength\fwidth{0.8\linewidth}
    \setlength\fheight{.2\linewidth}
\begin{tikzpicture}
\pgfplotsset{every tick label/.append style={font=\scriptsize}}

\definecolor{darkgray176}{RGB}{176,176,176}
\definecolor{darkorange25512714}{RGB}{255,127,14}
\definecolor{darkorange2309111}{RGB}{230,91,11}
\definecolor{lightgray204}{RGB}{204,204,204}
\definecolor{steelblue31119180}{RGB}{31,119,180}


\begin{axis}[
width=0.951\fwidth,
height=\fheight,
at={(0\fwidth,0\fheight)},
legend cell align={left},
legend style={fill opacity=0.8, draw opacity=1, text opacity=1, draw=lightgray204, font=\footnotesize},
legend columns=2,
x grid style={darkgray176},
xmajorticks=false,
xmin=-0.69, xmax=10.09,
xtick={0.2,1.2,2.2,3.2,4.2,5.2,6.2,7.2,8.2,9.2},
xticklabel style={rotate=45.0},
y grid style={darkgray176},
ylabel=\textcolor{steelblue31119180}{DL MCS},
ylabel style={font=\scriptsize},
xlabel style={font=\scriptsize},
ymin=0, ymax=30,
ytick pos=left,
ytick style={color=steelblue31119180},
yticklabel style={color=steelblue31119180},
xmajorgrids,
ymajorgrids
]


\addlegendimage{ybar,ybar legend,draw=black,fill=steelblue31119180,postaction={pattern=north east lines,pattern color=black}}
\addlegendentry{DL}
\addlegendimage{ybar,ybar legend,draw=black,fill=darkorange25512714,postaction={pattern=north west lines,pattern color=black}}
\addlegendentry{UL}


\draw[draw=black,fill=steelblue31119180,postaction={pattern=north east lines,pattern color=black}] (axis cs:-0.2,0) rectangle (axis cs:0.2,22.1130434782609);
\draw[draw=black,fill=steelblue31119180,postaction={pattern=north east lines,pattern color=black}] (axis cs:0.8,0) rectangle (axis cs:1.2,22.4434782608696);
\draw[draw=black,fill=steelblue31119180,postaction={pattern=north east lines,pattern color=black}] (axis cs:1.8,0) rectangle (axis cs:2.2,22.4260869565217);
\draw[draw=black,fill=steelblue31119180,postaction={pattern=north east lines,pattern color=black}] (axis cs:2.8,0) rectangle (axis cs:3.2,22.0782608695652);
\draw[draw=black,fill=steelblue31119180,postaction={pattern=north east lines,pattern color=black}] (axis cs:3.8,0) rectangle (axis cs:4.2,22.1826086956522);
\draw[draw=black,fill=steelblue31119180,postaction={pattern=north east lines,pattern color=black}] (axis cs:4.8,0) rectangle (axis cs:5.2,22.5565217391304);
\draw[draw=black,fill=steelblue31119180,postaction={pattern=north east lines,pattern color=black}] (axis cs:5.8,0) rectangle (axis cs:6.2,21.2260869565217);
\draw[draw=black,fill=steelblue31119180,postaction={pattern=north east lines,pattern color=black}] (axis cs:6.8,0) rectangle (axis cs:7.2,20.5391304347826);
\draw[draw=black,fill=steelblue31119180,postaction={pattern=north east lines,pattern color=black}] (axis cs:7.8,0) rectangle (axis cs:8.2,17.3304347826087);
\draw[draw=black,fill=steelblue31119180,postaction={pattern=north east lines,pattern color=black}] (axis cs:8.8,0) rectangle (axis cs:9.2,15.2956521739130);


\path [draw=black, line width=1pt]
(axis cs:0,20.5029589354697)
--(axis cs:0,23.7231280210521);
\path [draw=black, line width=1pt]
(axis cs:1,20.96147331055099)
--(axis cs:1,23.92548321118821);
\path [draw=black, line width=1pt]
(axis cs:2,21.04920470693957)
--(axis cs:2,23.80296920610383);
\path [draw=black, line width=1pt]
(axis cs:3,20.41257084297458)
--(axis cs:3,23.74395089615582);
\path [draw=black, line width=1pt]
(axis cs:4,20.75861253525148)
--(axis cs:4,23.60660485605292);
\path [draw=black, line width=1pt]
(axis cs:5,21.11654547015059)
--(axis cs:5,23.99649800811021);
\path [draw=black, line width=1pt]
(axis cs:6,19.64738623843593)
--(axis cs:6,22.80478767460747);
\path [draw=black, line width=1pt]
(axis cs:7,19.20610224696953)
--(axis cs:7,21.87215862259567);
\path [draw=black, line width=1pt]
(axis cs:8,13.53357149010824)
--(axis cs:8,21.12729807510916);
\path [draw=black, line width=1pt]
(axis cs:9,12.26084268238217)
--(axis cs:9,18.33046166544383);


\addplot [semithick, black, mark=-, mark size=1.5, mark options={solid}, only marks]
table {%
0 20.5029589354697
1 20.96147331055099
2 21.04920470693957
3 20.41257084297458
4 20.75861253525148
5 21.11654547015059
6 19.64738623843593
7 19.20610224696953
8 13.53357149010824
9 12.26084268238217
};


\addplot [semithick, black, mark=-, mark size=1.5, mark options={solid}, only marks]
table {%
0 23.7231280210521
1 23.92548321118821
2 23.80296920610383
3 23.74395089615582
4 23.60660485605292
5 23.99649800811021
6 22.80478767460747
7 21.87215862259567
8 21.12729807510916
9 18.33046166544383
};
\end{axis}


\begin{axis}[
width=0.951\fwidth,
height=\fheight,
at={(0\fwidth,0\fheight)},
axis y line*=right,
legend cell align={left},
legend style={fill opacity=0.8, draw opacity=1, text opacity=1, draw=lightgray204, font=\footnotesize},
legend columns=2,
x grid style={darkgray176},
xmin=-0.69, xmax=10.09,
xtick pos=left,
ytick style={color=darkorange2309111},
xtick={0.2,1.2,2.2,3.2,4.2,5.2,6.2,7.2,8.2,9.2},
xticklabels={0,2,4,6,8,10,12,14,16,18},
y grid style={darkgray176},
ylabel=\textcolor{darkorange2309111}{UL MCS},
ylabel style={font=\scriptsize},
xlabel style={font=\scriptsize},
xlabel={UE Location},
ymin=0, ymax=30,
ytick pos=right,
ytick style={color=darkorange2309111},
yticklabel style={anchor=west, color=darkorange2309111},
ylabel shift=-5pt
]


\draw[draw=black,fill=darkorange25512714,postaction={pattern=north west lines,pattern color=black}] (axis cs:0.2,0) rectangle (axis cs:0.6,15.6956521739130);
\draw[draw=black,fill=darkorange25512714,postaction={pattern=north west lines,pattern color=black}] (axis cs:1.2,0) rectangle (axis cs:1.6,15.2000000000000);
\draw[draw=black,fill=darkorange25512714,postaction={pattern=north west lines,pattern color=black}] (axis cs:2.2,0) rectangle (axis cs:2.6,15.1391304347826);
\draw[draw=black,fill=darkorange25512714,postaction={pattern=north west lines,pattern color=black}] (axis cs:3.2,0) rectangle (axis cs:3.6,14.6434782608696);
\draw[draw=black,fill=darkorange25512714,postaction={pattern=north west lines,pattern color=black}] (axis cs:4.2,0) rectangle (axis cs:4.6,14.7913043478261);
\draw[draw=black,fill=darkorange25512714,postaction={pattern=north west lines,pattern color=black}] (axis cs:5.2,0) rectangle (axis cs:5.6,15.8173913043478);
\draw[draw=black,fill=darkorange25512714,postaction={pattern=north west lines,pattern color=black}] (axis cs:6.2,0) rectangle (axis cs:6.6,15.8434782608696);
\draw[draw=black,fill=darkorange25512714,postaction={pattern=north west lines,pattern color=black}] (axis cs:7.2,0) rectangle (axis cs:7.6,16.4086956521739);
\draw[draw=black,fill=darkorange25512714,postaction={pattern=north west lines,pattern color=black}] (axis cs:8.2,0) rectangle (axis cs:8.6,16.6173913043478);
\draw[draw=black,fill=darkorange25512714,postaction={pattern=north west lines,pattern color=black}] (axis cs:9.2,0) rectangle (axis cs:9.6,9.89855072463768);


\path [draw=black, line width=1pt]
(axis cs:0.4,14.1844431015435)
--(axis cs:0.4,17.2068612462825);
\path [draw=black, line width=1pt]
(axis cs:1.4,13.79699570469953)
--(axis cs:1.4,16.60300429530047);
\path [draw=black, line width=1pt]
(axis cs:2.4,13.71326070006579)
--(axis cs:2.4,16.56500016949941);
\path [draw=black, line width=1pt]
(axis cs:3.4,13.47224930406393)
--(axis cs:3.4,15.81470721767527);
\path [draw=black, line width=1pt]
(axis cs:4.4,13.67920460966746)
--(axis cs:4.4,15.90340408698474);
\path [draw=black, line width=1pt]
(axis cs:5.4,13.92288118993618)
--(axis cs:5.4,17.71190141875942);
\path [draw=black, line width=1pt]
(axis cs:6.4,13.49419205218562)
--(axis cs:6.4,18.19276446955358);
\path [draw=black, line width=1pt]
(axis cs:7.4,14.21262399125452)
--(axis cs:7.4,18.60476731309328);
\path [draw=black, line width=1pt]
(axis cs:8.4,13.98299464351815)
--(axis cs:8.4,19.25178796517745);
\path [draw=black, line width=1pt]
(axis cs:9.4,3.45249171805185)
--(axis cs:9.4,16.34460973122351);


\addplot [semithick, black, mark=-, mark size=1.5, mark options={solid}, only marks]
table {%
0.4 14.1844431015435
1.4 13.79699570469953
2.4 13.71326070006579
3.4 13.47224930406393
4.4 13.67920460966746
5.4 13.92288118993618
6.4 13.49419205218562
7.4 14.21262399125452
8.4 13.98299464351815
9.4 3.45249171805185
};


\addplot [semithick, black, mark=-, mark size=1.5, mark options={solid}, only marks]
table {%
0.4 17.2068612462825
1.4 16.60300429530047
2.4 16.56500016949941
3.4 15.81470721767527
4.4 15.90340408698474
5.4 17.71190141875942
6.4 18.19276446955358
7.4 18.60476731309328
8.4 19.25178796517745
9.4 16.34460973122351
};

\end{axis}

\end{tikzpicture}}

    \hfill
    
    \subfloat[\gls{cqi}]{
    \label{chap4-fig:1ru1ue_static_iperf_cqi}
    \centering
    \setlength\fwidth{0.8\linewidth}
    \setlength\fheight{.2\linewidth}
\begin{tikzpicture}
\pgfplotsset{every tick label/.append style={font=\scriptsize}}

\definecolor{darkgray176}{RGB}{176,176,176}
\definecolor{darkorange25512714}{RGB}{255,127,14}
\definecolor{lightgray204}{RGB}{204,204,204}
\definecolor{steelblue31119180}{RGB}{31,119,180}


\begin{axis}[
width=0.951\fwidth,
height=\fheight,
at={(0\fwidth,0\fheight)},
legend cell align={left},
legend style={fill opacity=0.8, draw opacity=1, text opacity=1, draw=lightgray204, font=\footnotesize},
legend columns=2,
x grid style={darkgray176},
xmin=-0.69, xmax=10.09,
xtick style={color=black},
xtick={0.2,1.2,2.2,3.2,4.2,5.2,6.2,7.2,8.2,9.2},
xticklabels={0,2,4,6,8,10,12,14,16,18},
y grid style={darkgray176},
ylabel={CQI},
xlabel={UE Location},
ylabel style={font=\scriptsize},
xlabel style={font=\scriptsize},
ymin=9, ymax=16,
ytick pos=left,
xmajorgrids,
ymajorgrids
]


\addlegendimage{ybar,ybar legend,draw=black,fill=steelblue31119180,postaction={pattern=north east lines,pattern color=black}}
\addlegendentry{DL}


\draw[draw=black,fill=steelblue31119180,postaction={pattern=north east lines,pattern color=black}] (axis cs:-0.2,9) rectangle (axis cs:0.2,14.0173913043478);
\draw[draw=black,fill=steelblue31119180,postaction={pattern=north east lines,pattern color=black}] (axis cs:0.8,9) rectangle (axis cs:1.2,15.0000000000000);
\draw[draw=black,fill=steelblue31119180,postaction={pattern=north east lines,pattern color=black}] (axis cs:1.8,9) rectangle (axis cs:2.2,14.8347826086957);
\draw[draw=black,fill=steelblue31119180,postaction={pattern=north east lines,pattern color=black}] (axis cs:2.8,9) rectangle (axis cs:3.2,14.5217391304348);
\draw[draw=black,fill=steelblue31119180,postaction={pattern=north east lines,pattern color=black}] (axis cs:3.8,9) rectangle (axis cs:4.2,14.9304347826087);
\draw[draw=black,fill=steelblue31119180,postaction={pattern=north east lines,pattern color=black}] (axis cs:4.8,9) rectangle (axis cs:5.2,14.6782608695652);
\draw[draw=black,fill=steelblue31119180,postaction={pattern=north east lines,pattern color=black}] (axis cs:5.8,9) rectangle (axis cs:6.2,12.8695652173913);
\draw[draw=black,fill=steelblue31119180,postaction={pattern=north east lines,pattern color=black}] (axis cs:6.8,9) rectangle (axis cs:7.2,12.3391304347826);
\draw[draw=black,fill=steelblue31119180,postaction={pattern=north east lines,pattern color=black}] (axis cs:7.8,9) rectangle (axis cs:8.2,10.9739130434783);
\draw[draw=black,fill=steelblue31119180,postaction={pattern=north east lines,pattern color=black}] (axis cs:8.8,9) rectangle (axis cs:9.2,9.93043478260870);


\addlegendimage{ybar,ybar legend,draw=black,fill=darkorange25512714,postaction={pattern=north west lines,pattern color=black}}
\addlegendentry{UL}


\draw[draw=black,fill=darkorange25512714,postaction={pattern=north west lines,pattern color=black}] (axis cs:0.2,9) rectangle (axis cs:0.6,14.1826086956522);
\draw[draw=black,fill=darkorange25512714,postaction={pattern=north west lines,pattern color=black}] (axis cs:1.2,9) rectangle (axis cs:1.6,14.9304347826087);
\draw[draw=black,fill=darkorange25512714,postaction={pattern=north west lines,pattern color=black}] (axis cs:2.2,9) rectangle (axis cs:2.6,14.6434782608696);
\draw[draw=black,fill=darkorange25512714,postaction={pattern=north west lines,pattern color=black}] (axis cs:3.2,9) rectangle (axis cs:3.6,14.6869565217391);
\draw[draw=black,fill=darkorange25512714,postaction={pattern=north west lines,pattern color=black}] (axis cs:4.2,9) rectangle (axis cs:4.6,14.9739130434783);
\draw[draw=black,fill=darkorange25512714,postaction={pattern=north west lines,pattern color=black}] (axis cs:5.2,9) rectangle (axis cs:5.6,14.9130434782609);
\draw[draw=black,fill=darkorange25512714,postaction={pattern=north west lines,pattern color=black}] (axis cs:6.2,9) rectangle (axis cs:6.6,13.0521739130435);
\draw[draw=black,fill=darkorange25512714,postaction={pattern=north west lines,pattern color=black}] (axis cs:7.2,9) rectangle (axis cs:7.6,12.3478260869565);
\draw[draw=black,fill=darkorange25512714,postaction={pattern=north west lines,pattern color=black}] (axis cs:8.2,9) rectangle (axis cs:8.6,10.6782608695652);
\draw[draw=black,fill=darkorange25512714,postaction={pattern=north west lines,pattern color=black}] (axis cs:9.2,9) rectangle (axis cs:9.6,9.94202898550725);


\path [draw=black, line width=1pt]
(axis cs:0,13.88609488761631)
--(axis cs:0,14.14868772107929);
\path [draw=black, line width=1pt]
(axis cs:1,14.9999)
--(axis cs:1,15.0001);
\path [draw=black, line width=1pt]
(axis cs:2,14.461780685399575)
--(axis cs:2,15.207784532991825);
\path [draw=black, line width=1pt]
(axis cs:3,14.020025818732202)
--(axis cs:3,15.023452442137398);
\path [draw=black, line width=1pt]
(axis cs:4,14.674908523772164)
--(axis cs:4,15.185961041445236);
\path [draw=black, line width=1pt]
(axis cs:5,14.209072954363661)
--(axis cs:5,15.147448784766739);
\path [draw=black, line width=1pt]
(axis cs:6,12.53131017164543)
--(axis cs:6,13.20782026313717);
\path [draw=black, line width=1pt]
(axis cs:7,11.86364441340896)
--(axis cs:7,12.81461645615624);
\path [draw=black, line width=1pt]
(axis cs:8,10.813821536464814)
--(axis cs:8,11.134004550491786);
\path [draw=black, line width=1pt]
(axis cs:9,9.674908523772164)
--(axis cs:9,10.185961041445236);


\addplot [semithick, black, mark=-, mark size=1.5, mark options={solid}, only marks]
table {%
0 13.88609488761631
1 14.9999
2 14.461780685399575
3 14.020025818732202
4 14.674908523772164
5 14.209072954363661
6 12.53131017164543
7 11.86364441340896
8 10.813821536464814
9 9.674908523772164
};


\addplot [semithick, black, mark=-, mark size=1.5, mark options={solid}, only marks]
table {%
0 14.14868772107929
1 15.0001
2 15.207784532991825
3 15.023452442137398
4 15.185961041445236
5 15.147448784766739
6 13.20782026313717
7 12.81461645615624
8 11.134004550491786
9 10.185961041445236
};


\path [draw=black, line width=1pt]
(axis cs:0.4,13.751726111091036)
--(axis cs:0.4,14.613491280213364);
\path [draw=black, line width=1pt]
(axis cs:1.4,14.674908523772165)
--(axis cs:1.4,15.185961041445235);
\path [draw=black, line width=1pt]
(axis cs:2.4,14.162410290550307)
--(axis cs:2.4,15.124546231188893);
\path [draw=black, line width=1pt]
(axis cs:3.4,14.221195168207128)
--(axis cs:3.4,15.152717875271072);
\path [draw=black, line width=1pt]
(axis cs:4.4,14.813821536464814)
--(axis cs:4.4,15.134004550491786);
\path [draw=black, line width=1pt]
(axis cs:5.4,14.630039002711876)
--(axis cs:5.4,15.196047953809924);
\path [draw=black, line width=1pt]
(axis cs:6.4,12.828823104842533)
--(axis cs:6.4,13.275524721244467);
\path [draw=black, line width=1pt]
(axis cs:7.4,11.851462498853784)
--(axis cs:7.4,12.844189675059216);
\path [draw=black, line width=1pt]
(axis cs:8.4,10.209072954363661)
--(axis cs:8.4,11.147448784766739);
\path [draw=black, line width=1pt]
(axis cs:9.4,9.651846160125718)
--(axis cs:9.4,10.232211810888782);


\addplot [semithick, black, mark=-, mark size=1.5, mark options={solid}, only marks]
table {%
0.4 13.751726111091036
1.4 14.674908523772165
2.4 14.162410290550307
3.4 14.221195168207128
4.4 14.813821536464814
5.4 14.630039002711876
6.4 12.828823104842533
7.4 11.851462498853784
8.4 10.209072954363661
9.4 9.651846160125718
};


\addplot [semithick, black, mark=-, mark size=1.5, mark options={solid}, only marks]
table {%
0.4 14.613491280213364
1.4 15.185961041445235
2.4 15.124546231188893
3.4 15.152717875271072
4.4 15.134004550491786
5.4 15.196047953809924
6.4 13.275524721244467
7.4 12.844189675059216
8.4 11.147448784766739
9.4 10.232211810888782
};

\end{axis}

\end{tikzpicture}}

    \hfill
    
    \subfloat[\gls{ph}]{
    \label{chap4-fig:1ru1ue_static_iperf_ph}
    \centering
    \setlength\fwidth{0.8\linewidth}
    \setlength\fheight{.2\linewidth}
\begin{tikzpicture}
\pgfplotsset{every tick label/.append style={font=\scriptsize}}

\definecolor{darkgray176}{RGB}{176,176,176}
\definecolor{darkorange25512714}{RGB}{255,127,14}
\definecolor{lightgray204}{RGB}{204,204,204}
\definecolor{steelblue31119180}{RGB}{31,119,180}


\begin{axis}[
width=0.951\fwidth,
height=\fheight,
at={(0\fwidth,0\fheight)},
legend cell align={left},
legend style={fill opacity=0.8, draw opacity=1, text opacity=1, draw=lightgray204, font=\footnotesize},
legend columns=2,
x grid style={darkgray176},
xmin=-0.69, xmax=10.09,
xtick style={color=black},
xtick={0.2,1.2,2.2,3.2,4.2,5.2,6.2,7.2,8.2,9.2},
xticklabels={0,2,4,6,8,10,12,14,16,18},
y grid style={darkgray176},
ylabel={PH [dB]},
xlabel={UE Location},
ylabel style={font=\scriptsize},
xlabel style={font=\scriptsize},
ymin=0, ymax=60,
ytick pos=left,
xmajorgrids,
ymajorgrids
]


\addlegendimage{ybar,ybar legend,draw=black,fill=steelblue31119180,postaction={pattern=north east lines,pattern color=black}}
\addlegendentry{DL}


\draw[draw=black,fill=steelblue31119180,postaction={pattern=north east lines,pattern color=black}] (axis cs:-0.2,0) rectangle (axis cs:0.2,44.8782608695652);
\draw[draw=black,fill=steelblue31119180,postaction={pattern=north east lines,pattern color=black}] (axis cs:0.8,0) rectangle (axis cs:1.2,44.2434782608696);
\draw[draw=black,fill=steelblue31119180,postaction={pattern=north east lines,pattern color=black}] (axis cs:1.8,0) rectangle (axis cs:2.2,48.5565217391304);
\draw[draw=black,fill=steelblue31119180,postaction={pattern=north east lines,pattern color=black}] (axis cs:2.8,0) rectangle (axis cs:3.2,43.9826086956522);
\draw[draw=black,fill=steelblue31119180,postaction={pattern=north east lines,pattern color=black}] (axis cs:3.8,0) rectangle (axis cs:4.2,44.6347826086957);
\draw[draw=black,fill=steelblue31119180,postaction={pattern=north east lines,pattern color=black}] (axis cs:4.8,0) rectangle (axis cs:5.2,41.1478260869565);
\draw[draw=black,fill=steelblue31119180,postaction={pattern=north east lines,pattern color=black}] (axis cs:5.8,0) rectangle (axis cs:6.2,34.8608695652174);
\draw[draw=black,fill=steelblue31119180,postaction={pattern=north east lines,pattern color=black}] (axis cs:6.8,0) rectangle (axis cs:7.2,33.4521739130435);
\draw[draw=black,fill=steelblue31119180,postaction={pattern=north east lines,pattern color=black}] (axis cs:7.8,0) rectangle (axis cs:8.2,18.2695652173913);
\draw[draw=black,fill=steelblue31119180,postaction={pattern=north east lines,pattern color=black}] (axis cs:8.8,0) rectangle (axis cs:9.2,15.4434782608696);


\addlegendimage{ybar,ybar legend,draw=black,fill=darkorange25512714,postaction={pattern=north west lines,pattern color=black}}
\addlegendentry{UL}


\draw[draw=black,fill=darkorange25512714,postaction={pattern=north west lines,pattern color=black}] (axis cs:0.2,0) rectangle (axis cs:0.6,39.4869565217391);
\draw[draw=black,fill=darkorange25512714,postaction={pattern=north west lines,pattern color=black}] (axis cs:1.2,0) rectangle (axis cs:1.6,40.3130434782609);
\draw[draw=black,fill=darkorange25512714,postaction={pattern=north west lines,pattern color=black}] (axis cs:2.2,0) rectangle (axis cs:2.6,39.9565217391304);
\draw[draw=black,fill=darkorange25512714,postaction={pattern=north west lines,pattern color=black}] (axis cs:3.2,0) rectangle (axis cs:3.6,41.0347826086957);
\draw[draw=black,fill=darkorange25512714,postaction={pattern=north west lines,pattern color=black}] (axis cs:4.2,0) rectangle (axis cs:4.6,40.8000000000000);
\draw[draw=black,fill=darkorange25512714,postaction={pattern=north west lines,pattern color=black}] (axis cs:5.2,0) rectangle (axis cs:5.6,38.6434782608696);
\draw[draw=black,fill=darkorange25512714,postaction={pattern=north west lines,pattern color=black}] (axis cs:6.2,0) rectangle (axis cs:6.6,37.4521739130435);
\draw[draw=black,fill=darkorange25512714,postaction={pattern=north west lines,pattern color=black}] (axis cs:7.2,0) rectangle (axis cs:7.6,37.1739130434783);
\draw[draw=black,fill=darkorange25512714,postaction={pattern=north west lines,pattern color=black}] (axis cs:8.2,0) rectangle (axis cs:8.6,28.4347826086957);
\draw[draw=black,fill=darkorange25512714,postaction={pattern=north west lines,pattern color=black}] (axis cs:9.2,0) rectangle (axis cs:9.6,17.5797101449275);


\path [draw=black, line width=1pt]
(axis cs:0,42.036999023357)
--(axis cs:0,47.7195227157734);
\path [draw=black, line width=1pt]
(axis cs:1,41.520633694535)
--(axis cs:1,46.9663228272042);
\path [draw=black, line width=1pt]
(axis cs:2,44.7126583126817)
--(axis cs:2,52.4003851655791);
\path [draw=black, line width=1pt]
(axis cs:3,37.2966944518586)
--(axis cs:3,50.6685239394458);
\path [draw=black, line width=1pt]
(axis cs:4,40.7951681480093)
--(axis cs:4,48.4743970693821);
\path [draw=black, line width=1pt]
(axis cs:5,38.2361331500274)
--(axis cs:5,44.0595190238856);
\path [draw=black, line width=1pt]
(axis cs:6,31.2531610624788)
--(axis cs:6,38.468578067956);
\path [draw=black, line width=1pt]
(axis cs:7,30.2315074290579)
--(axis cs:7,36.6728403970291);
\path [draw=black, line width=1pt]
(axis cs:8,16.4327022397091)
--(axis cs:8,20.1064281950735);
\path [draw=black, line width=1pt]
(axis cs:9,8.65572961678848)
--(axis cs:9,22.2312269049507);


\addplot [semithick, black, mark=-, mark size=1.5, mark options={solid}, only marks]
table {%
0 42.036999023357
1 41.520633694535
2 44.7126583126817
3 37.2966944518586
4 40.7951681480093
5 38.2361331500274
6 31.2531610624788
7 30.2315074290579
8 16.4327022397091
9 8.65572961678848
};


\addplot [semithick, black, mark=-, mark size=1.5, mark options={solid}, only marks]
table {%
0 47.7195227157734
1 46.9663228272042
2 52.4003851655791
3 50.6685239394458
4 48.4743970693821
5 44.0595190238856
6 38.468578067956
7 36.6728403970291
8 20.1064281950735
9 22.2312269049507
};


\path [draw=black, line width=1pt]
(axis cs:0.4,37.2970065207425)
--(axis cs:0.4,41.6769065227357);
\path [draw=black, line width=1pt]
(axis cs:1.4,38.461703435418)
--(axis cs:1.4,42.1643835211038);
\path [draw=black, line width=1pt]
(axis cs:2.4,38.2023722508727)
--(axis cs:2.4,41.7106712273881);
\path [draw=black, line width=1pt]
(axis cs:3.4,39.6909831067427)
--(axis cs:3.4,42.3785821106487);
\path [draw=black, line width=1pt]
(axis cs:4.4,38.5868019487549)
--(axis cs:4.4,43.0131980512451);
\path [draw=black, line width=1pt]
(axis cs:5.4,36.4462259721833)
--(axis cs:5.4,40.8407305495559);
\path [draw=black, line width=1pt]
(axis cs:6.4,35.7239587387127)
--(axis cs:6.4,39.1803890873743);
\path [draw=black, line width=1pt]
(axis cs:7.4,35.5076278087119)
--(axis cs:7.4,38.8401982782447);
\path [draw=black, line width=1pt]
(axis cs:8.4,24.4608976061157)
--(axis cs:8.4,32.4086676112757);
\path [draw=black, line width=1pt]
(axis cs:9.4,10.4461443816284)
--(axis cs:9.4,24.7132759082266);


\addplot [semithick, black, mark=-, mark size=1.5, mark options={solid}, only marks]
table {%
0.4 37.2970065207425
1.4 38.461703435418
2.4 38.2023722508727
3.4 39.6909831067427
4.4 38.5868019487549
5.4 36.4462259721833
6.4 35.7239587387127
7.4 35.5076278087119
8.4 24.4608976061157
9.4 10.4461443816284
};


\addplot [semithick, black, mark=-, mark size=1.5, mark options={solid}, only marks]
table {%
0.4 41.6769065227357
1.4 42.1643835211038
2.4 41.7106712273881
3.4 42.3785821106487
4.4 43.0131980512451
5.4 40.8407305495559
6.4 39.1803890873743
7.4 38.8401982782447
8.4 32.4086676112757
9.4 24.7132759082266
};

\end{axis}

\end{tikzpicture}}

    \hfill
    
    \subfloat[\gls{bler}]{
    \label{chap4-fig:1ru1ue_static_iperf_bler}
    \centering
    \setlength\fwidth{0.8\linewidth}
    \setlength\fheight{.2\linewidth}
\begin{tikzpicture}
\pgfplotsset{every tick label/.append style={font=\scriptsize}}

\definecolor{darkgray176}{RGB}{176,176,176}
\definecolor{darkorange25512714}{RGB}{255,127,14}
\definecolor{darkorange2309111}{RGB}{230,91,11}
\definecolor{lightgray204}{RGB}{204,204,204}
\definecolor{steelblue31119180}{RGB}{31,119,180}


\begin{axis}[
width=0.951\fwidth,
height=\fheight,
at={(0\fwidth,0\fheight)},
legend cell align={left},
legend style={fill opacity=0.8, draw opacity=1, text opacity=1, draw=lightgray204, font=\footnotesize},
legend columns=2,
x grid style={darkgray176},
xmajorticks=false,
xmin=-0.69, xmax=10.09,
xtick={0.2,1.2,2.2,3.2,4.2,5.2,6.2,7.2,8.2,9.2},
xticklabel style={rotate=45.0},
y grid style={darkgray176},
ylabel=\textcolor{steelblue31119180}{DL BLER (\%)},
ylabel style={font=\scriptsize},
xlabel style={font=\scriptsize},
ymin=0, ymax=20,
ytick pos=left,
ytick style={color=steelblue31119180},
yticklabel style={color=steelblue31119180},
xmajorgrids,
ymajorgrids
]


\addlegendimage{ybar,ybar legend,draw=black,fill=steelblue31119180,postaction={pattern=north east lines,pattern color=black}}
\addlegendentry{DL}
\addlegendimage{ybar,ybar legend,draw=black,fill=darkorange25512714,postaction={pattern=north west lines,pattern color=black}}
\addlegendentry{UL}


\draw[draw=black,fill=steelblue31119180,postaction={pattern=north east lines,pattern color=black}] (axis cs:-0.2,0) rectangle (axis cs:0.2,9.65520869565217);
\draw[draw=black,fill=steelblue31119180,postaction={pattern=north east lines,pattern color=black}] (axis cs:0.8,0) rectangle (axis cs:1.2,10.0563826086957);
\draw[draw=black,fill=steelblue31119180,postaction={pattern=north east lines,pattern color=black}] (axis cs:1.8,0) rectangle (axis cs:2.2,9.74211304347826);
\draw[draw=black,fill=steelblue31119180,postaction={pattern=north east lines,pattern color=black}] (axis cs:2.8,0) rectangle (axis cs:3.2,10.0126173913043);
\draw[draw=black,fill=steelblue31119180,postaction={pattern=north east lines,pattern color=black}] (axis cs:3.8,0) rectangle (axis cs:4.2,10.1451913043478);
\draw[draw=black,fill=steelblue31119180,postaction={pattern=north east lines,pattern color=black}] (axis cs:4.8,0) rectangle (axis cs:5.2,10.1578521739130);
\draw[draw=black,fill=steelblue31119180,postaction={pattern=north east lines,pattern color=black}] (axis cs:5.8,0) rectangle (axis cs:6.2,9.64549565217391);
\draw[draw=black,fill=steelblue31119180,postaction={pattern=north east lines,pattern color=black}] (axis cs:6.8,0) rectangle (axis cs:7.2,6.26393043478261);
\draw[draw=black,fill=steelblue31119180,postaction={pattern=north east lines,pattern color=black}] (axis cs:7.8,0) rectangle (axis cs:8.2,6.18401739130435);
\draw[draw=black,fill=steelblue31119180,postaction={pattern=north east lines,pattern color=black}] (axis cs:8.8,0) rectangle (axis cs:9.2,4.79395652173913);


\path [draw=black, line width=1pt]
(axis cs:0,5.48903857823112)
--(axis cs:0,13.82137881307322);
\path [draw=black, line width=1pt]
(axis cs:1,4.94947400010116)
--(axis cs:1,15.16329121729024);
\path [draw=black, line width=1pt]
(axis cs:2,5.06530920953951)
--(axis cs:2,14.41891687741701);
\path [draw=black, line width=1pt]
(axis cs:3,5.29759739435575)
--(axis cs:3,14.72763738825285);
\path [draw=black, line width=1pt]
(axis cs:4,4.97515481206821)
--(axis cs:4,15.31522779662739);
\path [draw=black, line width=1pt]
(axis cs:5,5.37444307204324)
--(axis cs:5,14.94126127578276);
\path [draw=black, line width=1pt]
(axis cs:6,5.21040864502897)
--(axis cs:6,14.08058265931885);
\path [draw=black, line width=1pt]
(axis cs:7,0.71300871944064)
--(axis cs:7,11.81485215012458);
\path [draw=black, line width=1pt]
(axis cs:8,0.000000)
--(axis cs:8,14.19494521607783);
\path [draw=black, line width=1pt]
(axis cs:9,0.000000)
--(axis cs:9,9.86617351753883);


\addplot [semithick, black, mark=-, mark size=1.5, mark options={solid}, only marks]
table {%
0 5.48903857823112
1 4.94947400010116
2 5.06530920953951
3 5.29759739435575
4 4.97515481206821
5 5.37444307204324
6 5.21040864502897
7 0.71300871944064
8 -1.82691043346913
9 -0.27826047406054
};


\addplot [semithick, black, mark=-, mark size=1.5, mark options={solid}, only marks]
table {%
0 13.82137881307322
1 15.16329121729024
2 14.41891687741701
3 14.72763738825285
4 15.31522779662739
5 14.94126127578276
6 14.08058265931885
7 11.81485215012458
8 14.19494521607783
9 9.86617351753883
};
\end{axis}


\begin{axis}[
width=0.951\fwidth,
height=\fheight,
at={(0\fwidth,0\fheight)},
axis y line*=right,
legend cell align={left},
legend style={fill opacity=0.8, draw opacity=1, text opacity=1, draw=lightgray204, font=\footnotesize},
legend columns=2,
x grid style={darkgray176},
xmin=-0.69, xmax=10.09,
xtick pos=left,
ytick style={color=darkorange2309111},
xtick={0.2,1.2,2.2,3.2,4.2,5.2,6.2,7.2,8.2,9.2},
xticklabels={0,2,4,6,8,10,12,14,16,18},
y grid style={darkgray176},
ylabel=\textcolor{darkorange2309111}{UL BLER (\%)},
ylabel style={font=\scriptsize},
xlabel style={font=\scriptsize},
xlabel={UE Location},
ymin=0, ymax=20,
ytick pos=right,
ytick style={color=darkorange2309111},
yticklabel style={anchor=west, color=darkorange2309111},
ylabel shift=-5pt
]


\draw[draw=black,fill=darkorange25512714,postaction={pattern=north west lines,pattern color=black}] (axis cs:0.2,0) rectangle (axis cs:0.6,9.56812173913044);
\draw[draw=black,fill=darkorange25512714,postaction={pattern=north west lines,pattern color=black}] (axis cs:1.2,0) rectangle (axis cs:1.6,9.50712173913043);
\draw[draw=black,fill=darkorange25512714,postaction={pattern=north west lines,pattern color=black}] (axis cs:2.2,0) rectangle (axis cs:2.6,9.27707826086957);
\draw[draw=black,fill=darkorange25512714,postaction={pattern=north west lines,pattern color=black}] (axis cs:3.2,0) rectangle (axis cs:3.6,9.17420000000000);
\draw[draw=black,fill=darkorange25512714,postaction={pattern=north west lines,pattern color=black}] (axis cs:4.2,0) rectangle (axis cs:4.6,9.45190434782609);
\draw[draw=black,fill=darkorange25512714,postaction={pattern=north west lines,pattern color=black}] (axis cs:5.2,0) rectangle (axis cs:5.6,9.82486956521739);
\draw[draw=black,fill=darkorange25512714,postaction={pattern=north west lines,pattern color=black}] (axis cs:6.2,0) rectangle (axis cs:6.6,9.78168695652174);
\draw[draw=black,fill=darkorange25512714,postaction={pattern=north west lines,pattern color=black}] (axis cs:7.2,0) rectangle (axis cs:7.6,9.39971304347826);
\draw[draw=black,fill=darkorange25512714,postaction={pattern=north west lines,pattern color=black}] (axis cs:8.2,0) rectangle (axis cs:8.6,5.20550434782609);
\draw[draw=black,fill=darkorange25512714,postaction={pattern=north west lines,pattern color=black}] (axis cs:9.2,0) rectangle (axis cs:9.6,0.81201449275362);


\path [draw=black, line width=1pt]
(axis cs:0.4,6.68591234005179)
--(axis cs:0.4,12.45033113820909);
\path [draw=black, line width=1pt]
(axis cs:1.4,6.4711601927064)
--(axis cs:1.4,12.54308328555446);
\path [draw=black, line width=1pt]
(axis cs:2.4,6.18722643012966)
--(axis cs:2.4,12.36693009160948);
\path [draw=black, line width=1pt]
(axis cs:3.4,6.1539307335098)
--(axis cs:3.4,12.1944692664902);
\path [draw=black, line width=1pt]
(axis cs:4.4,6.37777071442081)
--(axis cs:4.4,12.52603798123137);
\path [draw=black, line width=1pt]
(axis cs:5.4,6.61743310313804)
--(axis cs:5.4,13.03230602729674);
\path [draw=black, line width=1pt]
(axis cs:6.4,5.5995890267589)
--(axis cs:6.4,13.96378488628458);
\path [draw=black, line width=1pt]
(axis cs:7.4,5.65376859843443)
--(axis cs:7.4,13.14565748852209);
\path [draw=black, line width=1pt]
(axis cs:8.4,0.32297303824742)
--(axis cs:8.4,10.08803565740476);
\path [draw=black, line width=1pt]
(axis cs:9.4,0.000000)
--(axis cs:9.4,3.43483478899158);


\addplot [semithick, black, mark=-, mark size=1.5, mark options={solid}, only marks]
table {%
0.4 6.68591234005179
1.4 6.4711601927064
2.4 6.18722643012966
3.4 6.1539307335098
4.4 6.37777071442081
5.4 6.61743310313804
6.4 5.5995890267589
7.4 5.65376859843443
8.4 0.32297303824742
9.4 0.000000
};


\addplot [semithick, black, mark=-, mark size=1.5, mark options={solid}, only marks]
table {%
0.4 12.45033113820909
1.4 12.54308328555446
2.4 12.36693009160948
3.4 12.1944692664902
4.4 12.52603798123137
5.4 13.03230602729674
6.4 13.96378488628458
7.4 13.14565748852209
8.4 10.08803565740476
9.4 3.43483478899158
};

\end{axis}

\end{tikzpicture}}
    
\caption{Performance profiling with one \acrshort{ue} and single \acrshort{ru} for the static iPerf use case during \acrshort{dl} (blue bars) and \acrshort{ul} (orange bars) data transmissions.}
\label{chap4-fig:1ru1ue_static_iperf}
\vspace{-10pt}
\end{figure}

\subsection{Static Experiments}
\label{chap4-sec:exp-static}
\textbf{1 \gls{ue}, static, iPerf.} In the initial series of tests, we analyze the performance of a single \gls{ue} in ten static locations at varying distances from the \gls{ru}, as
shown in Figure~\ref{chap4-fig:1ru1ue_static_iperf}, \blue{using the first configuration setup of 2x2 \gls{mimo}, 2~layers \gls{dl}, 1~layer \gls{ul}, a DDDSU \gls{tdd} pattern, and a $Q_{m}$ up to 64-\gls{qam}}.

The iPerf throughput results in Figure~\ref{chap4-fig:1ru1ue_static_iperf_th} highlight the upper layer's responsiveness, showing a significant reduction from an average downlink throughput of $300$~Mbps and an uplink one of $38$~Mbps at locations near the \gls{ru}, to significantly lower rates of $177$~Mbps in downlink and $1.5$~Mbps in uplink at the most remote point, i.e., location 18. This high-level data throughput behavior is supported by corresponding shifts in the lower layers.

For example, this trend is clearly noticeable from the results of Figure~\ref{chap4-fig:1ru1ue_static_iperf_rsrp}, which shows the \gls{rsrp} values reported by the \gls{ue} to the \gls{gnb} during \gls{dl} (blue bars) and \gls{ul} (orange bars) data transmissions. As the distance from the \gls{ru} initially decreases starting from location~0 to location~6, the \gls{rsrp} values peak at around $-80$\:dBm. Subsequently, as the distance starts to increase again, moving from location~6 towards location~18, the \gls{rsrp} values begin to decrease, reaching as low as $-100$\:dBm.

In the same way, the \gls{mcs} values for both downlink and uplink initially mirror each other, then begin to diverge from around location~10, as shown in Figure~\ref{chap4-fig:1ru1ue_static_iperf_mcs}. In uplink, the \gls{mcs} values tend to remain stable or slightly increase, whereas downlink \gls{mcs} values experience a slight decline possibly due to power control strategies that better favor the uplink at more distant locations.

Similarly, the \gls{cqi} (Figure~\ref{chap4-fig:1ru1ue_static_iperf_cqi}) for both downlink and uplink starts closely matched but begins to show variation past the midpoint of the locations. The downlink \gls{cqi} experiences a modest decline, while the uplink \gls{cqi} sustains higher values. This suggests better channel conditions or more effective adaptation mechanisms in the uplink, due to adaptive power adjustments in uplink transmissions that maintain signal quality over distance.

\gls{ph} metrics, shown in Figure~\ref{chap4-fig:1ru1ue_static_iperf_ph}, reveal that downlink power headroom remains relatively stable across all locations, indicating a consistent application of power levels for downlink transmissions. In contrast, the uplink displays greater variability and generally higher values in distant locations to compensate for potential path loss and ensure that the transmit power remains adequate to maintain quality of service as the \gls{ue} moves further from the \gls{ru}.

Finally, the \gls{bler} (Figure~\ref{chap4-fig:1ru1ue_static_iperf_bler}) for downlink remains below 10\% for most locations, pointing to good reliability and effective adaptation of the \gls{mcs}. However, while the uplink \gls{bler} is generally low, it exhibits some peaks, especially around mid-range locations which may be related to specific lab obstacles or multipath effects and slow adaptation loops. 

\begin{figure}[htb]
\centering
    \subfloat[Downlink]{
    \label{chap4-fig:dl_1ru2ue}
    \centering
    \setlength\fwidth{0.8\linewidth}
    \setlength\fheight{.25\linewidth}
\begin{tikzpicture}
\pgfplotsset{every tick label/.append style={font=\scriptsize}}

\definecolor{darkgray176}{RGB}{176,176,176}
\definecolor{darkorange25512714}{RGB}{255,127,14}
\definecolor{lightgray204}{RGB}{204,204,204}
\definecolor{steelblue31119180}{RGB}{31,119,180}


\begin{axis}[
width=0.951\fwidth,
height=\fheight,
at={(0\fwidth,0\fheight)},
legend cell align={left},
legend columns=2,
legend style={fill opacity=0.8, draw opacity=1, text opacity=1, draw=lightgray204, at={(0.5,1.02)}, anchor=south, font =\footnotesize},
x grid style={darkgray176},
xmin=-0.69, xmax=10.09,
xtick style={color=black},
xtick={0.2,1.2,2.2,3.2,4.2,5.2,6.2,7.2,8.2,9.2},
xticklabels={0,2,4,6,8,10,12,14,16,18},
y grid style={darkgray176},
ylabel={DL Throughput [Mbps]},
xlabel={Location},
ylabel style={font=\scriptsize},
xlabel style={font=\scriptsize},
ymin=0, ymax=250,
ytick pos=left,
xmajorgrids,
ymajorgrids
]


\addlegendimage{ybar,ybar legend,draw=black,fill=steelblue31119180,postaction={pattern=north east lines,pattern color=black}}
\addlegendentry{UE1}


\addlegendimage{ybar,ybar legend,draw=black,fill=darkorange25512714,postaction={pattern=north west lines,pattern color=black}}
\addlegendentry{UE2}

\draw[draw=black,fill=steelblue31119180,postaction={pattern=north east lines,pattern color=black}] (axis cs:-0.2,0) rectangle (axis cs:0.2,209.31375104);
\draw[draw=black,fill=steelblue31119180,postaction={pattern=north east lines,pattern color=black}] (axis cs:0.8,0) rectangle (axis cs:1.2,186.617992838095);
\draw[draw=black,fill=steelblue31119180,postaction={pattern=north east lines,pattern color=black}] (axis cs:1.8,0) rectangle (axis cs:2.2,219.202990933333);
\draw[draw=black,fill=steelblue31119180,postaction={pattern=north east lines,pattern color=black}] (axis cs:2.8,0) rectangle (axis cs:3.2,217.906626986667);
\draw[draw=black,fill=steelblue31119180,postaction={pattern=north east lines,pattern color=black}] (axis cs:3.8,0) rectangle (axis cs:4.2,199.6595072);
\draw[draw=black,fill=steelblue31119180,postaction={pattern=north east lines,pattern color=black}] (axis cs:4.8,0) rectangle (axis cs:5.2,191.50603008);
\draw[draw=black,fill=steelblue31119180,postaction={pattern=north east lines,pattern color=black}] (axis cs:5.8,0) rectangle (axis cs:6.2,188.67632768);
\draw[draw=black,fill=steelblue31119180,postaction={pattern=north east lines,pattern color=black}] (axis cs:6.8,0) rectangle (axis cs:7.2,214.866522026667);
\draw[draw=black,fill=steelblue31119180,postaction={pattern=north east lines,pattern color=black}] (axis cs:7.8,0) rectangle (axis cs:8.2,202.916541866667);
\draw[draw=black,fill=steelblue31119180,postaction={pattern=north east lines,pattern color=black}] (axis cs:8.8,0) rectangle (axis cs:9.2,175.75314304);


\draw[draw=black,fill=darkorange25512714,postaction={pattern=north west lines,pattern color=black}] (axis cs:0.2,0) rectangle (axis cs:0.6,197.091720533333);
\draw[draw=black,fill=darkorange25512714,postaction={pattern=north west lines,pattern color=black}] (axis cs:1.2,0) rectangle (axis cs:1.6,178.269318095238);
\draw[draw=black,fill=darkorange25512714,postaction={pattern=north west lines,pattern color=black}] (axis cs:2.2,0) rectangle (axis cs:2.6,167.423474773333);
\draw[draw=black,fill=darkorange25512714,postaction={pattern=north west lines,pattern color=black}] (axis cs:3.2,0) rectangle (axis cs:3.6,178.305252693333);
\draw[draw=black,fill=darkorange25512714,postaction={pattern=north west lines,pattern color=black}] (axis cs:4.2,0) rectangle (axis cs:4.6,197.397266346667);
\draw[draw=black,fill=darkorange25512714,postaction={pattern=north west lines,pattern color=black}] (axis cs:5.2,0) rectangle (axis cs:5.6,206.263659946667);
\draw[draw=black,fill=darkorange25512714,postaction={pattern=north west lines,pattern color=black}] (axis cs:6.2,0) rectangle (axis cs:6.6,201.951863466667);
\draw[draw=black,fill=darkorange25512714,postaction={pattern=north west lines,pattern color=black}] (axis cs:7.2,0) rectangle (axis cs:7.6,201.720223573333);
\draw[draw=black,fill=darkorange25512714,postaction={pattern=north west lines,pattern color=black}] (axis cs:8.2,0) rectangle (axis cs:8.6,200.045983573333);
\draw[draw=black,fill=darkorange25512714,postaction={pattern=north west lines,pattern color=black}] (axis cs:9.2,0) rectangle (axis cs:9.6,187.889971626667);


\path [draw=black, line width=1pt]
(axis cs:0,201.847342649942)
--(axis cs:0,216.780159430058);
\path [draw=black, line width=1pt]
(axis cs:1,177.31164661874)
--(axis cs:1,195.924339057451);
\path [draw=black, line width=1pt]
(axis cs:2,214.074751221685)
--(axis cs:2,224.331230644982);
\path [draw=black, line width=1pt]
(axis cs:3,213.073034039615)
--(axis cs:3,222.740219933718);
\path [draw=black, line width=1pt]
(axis cs:4,189.806104456157)
--(axis cs:4,209.512909943843);
\path [draw=black, line width=1pt]
(axis cs:5,186.048853581522)
--(axis cs:5,196.963206578478);
\path [draw=black, line width=1pt]
(axis cs:6,180.538676605231)
--(axis cs:6,196.813978754769);
\path [draw=black, line width=1pt]
(axis cs:7,210.616087254379)
--(axis cs:7,219.116956798954);
\path [draw=black, line width=1pt]
(axis cs:8,200.64366464943)
--(axis cs:8,205.189419083904);
\path [draw=black, line width=1pt]
(axis cs:9,172.587760013351)
--(axis cs:9,178.918526066649);


\addplot [semithick, black, mark=-, mark size=1.5, mark options={solid}, only marks]
table {%
0 201.847342649942
1 177.31164661874
2 214.074751221685
3 213.073034039615
4 189.806104456157
5 186.048853581522
6 180.538676605231
7 210.616087254379
8 200.64366464943
9 172.587760013351
};


\addplot [semithick, black, mark=-, mark size=1.5, mark options={solid}, only marks]
table {%
0 216.780159430058
1 195.924339057451
2 224.331230644982
3 222.740219933718
4 209.512909943843
5 196.963206578478
6 196.813978754769
7 219.116956798954
8 205.189419083904
9 178.918526066649
};


\path [draw=black, line width=1pt]
(axis cs:0.4,189.979224182537)
--(axis cs:0.4,204.20421688413);
\path [draw=black, line width=1pt]
(axis cs:1.4,168.810219824794)
--(axis cs:1.4,187.728416365682);
\path [draw=black, line width=1pt]
(axis cs:2.4,160.967087065603)
--(axis cs:2.4,173.879862481064);
\path [draw=black, line width=1pt]
(axis cs:3.4,172.736335356394)
--(axis cs:3.4,183.874170030273);
\path [draw=black, line width=1pt]
(axis cs:4.4,190.338614817949)
--(axis cs:4.4,204.455917875384);
\path [draw=black, line width=1pt]
(axis cs:5.4,201.296433481144)
--(axis cs:5.4,211.23088641219);
\path [draw=black, line width=1pt]
(axis cs:6.4,197.585972034881)
--(axis cs:6.4,206.317754898452);
\path [draw=black, line width=1pt]
(axis cs:7.4,195.473528009698)
--(axis cs:7.4,207.966919136969);
\path [draw=black, line width=1pt]
(axis cs:8.4,196.563119925935)
--(axis cs:8.4,203.528847220731);
\path [draw=black, line width=1pt]
(axis cs:9.4,183.098790508473)
--(axis cs:9.4,192.681152744861);


\addplot [semithick, black, mark=-, mark size=1.5, mark options={solid}, only marks]
table {%
0.4 189.979224182537
1.4 168.810219824794
2.4 160.967087065603
3.4 172.736335356394
4.4 190.338614817949
5.4 201.296433481144
6.4 197.585972034881
7.4 195.473528009698
8.4 196.563119925935
9.4 183.098790508473
};


\addplot [semithick, black, mark=-, mark size=1.5, mark options={solid}, only marks]
table {%
0.4 204.20421688413
1.4 187.728416365682
2.4 173.879862481064
3.4 183.874170030273
4.4 204.455917875384
5.4 211.23088641219
6.4 206.317754898452
7.4 207.966919136969
8.4 203.528847220731
9.4 192.681152744861
};

\end{axis}

\end{tikzpicture}
    \setlength\abovecaptionskip{.05cm}}
    \hfill    
    \subfloat[Uplink]{
    \label{chap4-fig:ul_1ru2ue}
    \centering
        \setlength\fwidth{0.8\linewidth}
        \setlength\fheight{.25\linewidth}
\begin{tikzpicture}
\pgfplotsset{every tick label/.append style={font=\scriptsize}}

\definecolor{darkgray176}{RGB}{176,176,176}
\definecolor{darkorange25512714}{RGB}{255,127,14}
\definecolor{lightgray204}{RGB}{204,204,204}
\definecolor{steelblue31119180}{RGB}{31,119,180}

\begin{axis}[
width=0.951\fwidth,
height=\fheight,
at={(0\fwidth,0\fheight)},
legend cell align={left},
legend columns=2,
legend style={fill opacity=0.8, draw opacity=1, text opacity=1, draw=lightgray204, at={(0.5,1.02)}, anchor=south, font =\footnotesize},
x grid style={darkgray176},
xmin=-0.69, xmax=10.09,
xtick style={color=black},
xtick={0.2,1.2,2.2,3.2,4.2,5.2,6.2,7.2,8.2,9.2},
xticklabels={0,2,4,6,8,10,12,14,16,18},
y grid style={darkgray176},
ylabel={UL Throughput [Mbps]},
xlabel={Location},
ylabel style={font=\scriptsize},
xlabel style={font=\scriptsize},
ymin=0, ymax=30,
ytick pos=left,
xmajorgrids,
ymajorgrids
]


\addlegendimage{ybar,ybar legend,draw=black,fill=steelblue31119180,postaction={pattern=north east lines,pattern color=black}}
\addlegendentry{UE1}


\addlegendimage{ybar,ybar legend,draw=black,fill=darkorange25512714,postaction={pattern=north west lines,pattern color=black}}
\addlegendentry{UE2}

\draw[draw=black,fill=steelblue31119180,postaction={pattern=north east lines,pattern color=black}] (axis cs:-0.2,0) rectangle (axis cs:0.2,15.9243741866667);
\draw[draw=black,fill=steelblue31119180,postaction={pattern=north east lines,pattern color=black}] (axis cs:0.8,0) rectangle (axis cs:1.2,16.58847232);
\draw[draw=black,fill=steelblue31119180,postaction={pattern=north east lines,pattern color=black}] (axis cs:1.8,0) rectangle (axis cs:2.2,22.79604224);
\draw[draw=black,fill=steelblue31119180,postaction={pattern=north east lines,pattern color=black}] (axis cs:2.8,0) rectangle (axis cs:3.2,21.8173713066667);
\draw[draw=black,fill=steelblue31119180,postaction={pattern=north east lines,pattern color=black}] (axis cs:3.8,0) rectangle (axis cs:4.2,20.69889024);
\draw[draw=black,fill=steelblue31119180,postaction={pattern=north east lines,pattern color=black}] (axis cs:4.8,0) rectangle (axis cs:5.2,20.3074218666667);
\draw[draw=black,fill=steelblue31119180,postaction={pattern=north east lines,pattern color=black}] (axis cs:5.8,0) rectangle (axis cs:6.2,15.2113425066667);
\draw[draw=black,fill=steelblue31119180,postaction={pattern=north east lines,pattern color=black}] (axis cs:6.8,0) rectangle (axis cs:7.2,9.48611754666667);
\draw[draw=black,fill=steelblue31119180,postaction={pattern=north east lines,pattern color=black}] (axis cs:7.8,0) rectangle (axis cs:8.2,11.6042410666667);
\draw[draw=black,fill=steelblue31119180,postaction={pattern=north east lines,pattern color=black}] (axis cs:8.8,0) rectangle (axis cs:9.2,4.04051285333333);
\draw[draw=black,fill=darkorange25512714,postaction={pattern=north west lines,pattern color=black}] (axis cs:0.2,0) rectangle (axis cs:0.6,24.30599168);

\draw[draw=black,fill=darkorange25512714,postaction={pattern=north west lines,pattern color=black}] (axis cs:1.2,0) rectangle (axis cs:1.6,23.6838365866667);
\draw[draw=black,fill=darkorange25512714,postaction={pattern=north west lines,pattern color=black}] (axis cs:2.2,0) rectangle (axis cs:2.6,21.3559978666667);
\draw[draw=black,fill=darkorange25512714,postaction={pattern=north west lines,pattern color=black}] (axis cs:3.2,0) rectangle (axis cs:3.6,21.1952162133333);
\draw[draw=black,fill=darkorange25512714,postaction={pattern=north west lines,pattern color=black}] (axis cs:4.2,0) rectangle (axis cs:4.6,20.5381085866667);
\draw[draw=black,fill=darkorange25512714,postaction={pattern=north west lines,pattern color=black}] (axis cs:5.2,0) rectangle (axis cs:5.6,20.1606212266667);
\draw[draw=black,fill=darkorange25512714,postaction={pattern=north west lines,pattern color=black}] (axis cs:6.2,0) rectangle (axis cs:6.6,18.1473553066667);
\draw[draw=black,fill=darkorange25512714,postaction={pattern=north west lines,pattern color=black}] (axis cs:7.2,0) rectangle (axis cs:7.6,10.0313770666667);
\draw[draw=black,fill=darkorange25512714,postaction={pattern=north west lines,pattern color=black}] (axis cs:8.2,0) rectangle (axis cs:8.6,5.48055722666667);
\draw[draw=black,fill=darkorange25512714,postaction={pattern=north west lines,pattern color=black}] (axis cs:9.2,0) rectangle (axis cs:9.6,3.08980394666667);
\path [draw=black, line width=1pt]
(axis cs:0,15.1854121927106)
--(axis cs:0,16.6633361806227);

\path [draw=black, line width=1pt]
(axis cs:1,15.9339611695867)
--(axis cs:1,17.2429834704133);

\path [draw=black, line width=1pt]
(axis cs:2,22.4809501886116)
--(axis cs:2,23.1111342913884);

\path [draw=black, line width=1pt]
(axis cs:3,21.4878175036607)
--(axis cs:3,22.1469251096727);

\path [draw=black, line width=1pt]
(axis cs:4,20.4020033048218)
--(axis cs:4,20.9957771751782);

\path [draw=black, line width=1pt]
(axis cs:5,19.6595932286659)
--(axis cs:5,20.9552505046674);

\path [draw=black, line width=1pt]
(axis cs:6,14.1801751599062)
--(axis cs:6,16.2425098534271);

\path [draw=black, line width=1pt]
(axis cs:7,8.36420179649132)
--(axis cs:7,10.608033296842);

\path [draw=black, line width=1pt]
(axis cs:8,10.3518558129289)
--(axis cs:8,12.8566263204044);

\path [draw=black, line width=1pt]
(axis cs:9,3.34300492484811)
--(axis cs:9,4.73802078181856);

\addplot [semithick, black, mark=-, mark size=1.5, mark options={solid}, only marks]
table {%
0 15.1854121927106
1 15.9339611695867
2 22.4809501886116
3 21.4878175036607
4 20.4020033048218
5 19.6595932286659
6 14.1801751599062
7 8.36420179649132
8 10.3518558129289
9 3.34300492484811
};
\addplot [semithick, black, mark=-, mark size=1.5, mark options={solid}, only marks]
table {%
0 16.6633361806227
1 17.2429834704133
2 23.1111342913884
3 22.1469251096727
4 20.9957771751782
5 20.9552505046674
6 16.2425098534271
7 10.608033296842
8 12.8566263204044
9 4.73802078181856
};
\path [draw=black, line width=1pt]
(axis cs:0.4,23.6666234045624)
--(axis cs:0.4,24.9453599554376);

\path [draw=black, line width=1pt]
(axis cs:1.4,23.0062202721671)
--(axis cs:1.4,24.3614529011663);

\path [draw=black, line width=1pt]
(axis cs:2.4,21.0196821373176)
--(axis cs:2.4,21.6923135960157);

\path [draw=black, line width=1pt]
(axis cs:3.4,20.8733946143571)
--(axis cs:3.4,21.5170378123096);

\path [draw=black, line width=1pt]
(axis cs:4.4,20.2846734632244)
--(axis cs:4.4,20.7915437101089);

\path [draw=black, line width=1pt]
(axis cs:5.4,19.5017001671464)
--(axis cs:5.4,20.8195422861869);

\path [draw=black, line width=1pt]
(axis cs:6.4,17.1807637278958)
--(axis cs:6.4,19.1139468854375);

\path [draw=black, line width=1pt]
(axis cs:7.4,8.89414285024027)
--(axis cs:7.4,11.1686112830931);

\path [draw=black, line width=1pt]
(axis cs:8.4,4.66292328647226)
--(axis cs:8.4,6.29819116686107);

\path [draw=black, line width=1pt]
(axis cs:9.4,2.62564703322214)
--(axis cs:9.4,3.55396086011119);

\addplot [semithick, black, mark=-, mark size=1.5, mark options={solid}, only marks]
table {%
0.4 23.6666234045624
1.4 23.0062202721671
2.4 21.0196821373176
3.4 20.8733946143571
4.4 20.2846734632244
5.4 19.5017001671464
6.4 17.1807637278958
7.4 8.89414285024027
8.4 4.66292328647226
9.4 2.62564703322214
};
\addplot [semithick, black, mark=-, mark size=1.5, mark options={solid}, only marks]
table {%
0.4 24.9453599554376
1.4 24.3614529011663
2.4 21.6923135960157
3.4 21.5170378123096
4.4 20.7915437101089
5.4 20.8195422861869
6.4 19.1139468854375
7.4 11.1686112830931
8.4 6.29819116686107
9.4 3.55396086011119
};
\end{axis}

\end{tikzpicture}
        \setlength\abovecaptionskip{.05cm}}
\caption{Performance profiling for one \acrshort{ru} and two static \acrshortpl{ue} for the static iPerf use case.}
\label{chap4-fig:1RU2UE_Static}
\end{figure}
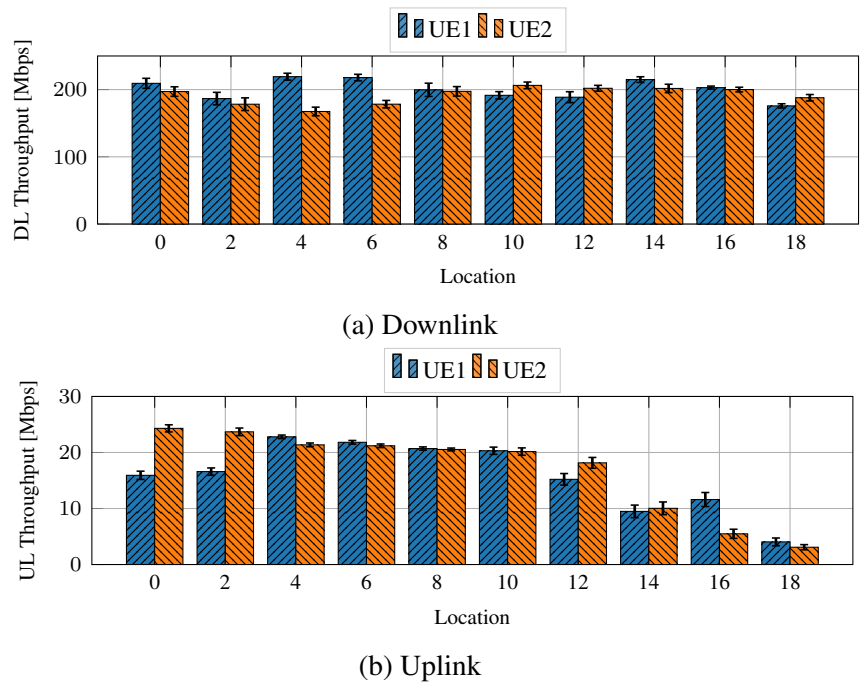

\textbf{2 \glspl{ue}, static, iPerf.} We test the performance of 2 \glspl{ue} for a single \gls{ru}. We position the \glspl{ue} at the same static locations as in the previous single \gls{ue} static case. The results are plotted in Figure~\ref{chap4-fig:1RU2UE_Static} for both downlink and uplink transmissions. We observe that, in most cases, the \glspl{ue} are able to share bandwidth fairly. The best achievable average aggregate throughput from both \glspl{ue} is around $400$\:Mbps in \gls{dl} and $44$\:Mbps in \gls{ul}. This shows that the total cell throughput can be higher than the single \gls{ue} throughput. As discussed in Table~\ref{chap4-table:testbeds-features}, this is due to a limitation in the number of transport blocks that can be acknowledged in a single slot for a single \gls{ue} \blue{in the case of the DDDSU \gls{tdd} pattern used in this first set of experiments}. Therefore, scheduling multiple \glspl{ue} improves the resource utilization of the system.


\begin{figure}[htb]
\centering
    \setlength\fwidth{0.7\linewidth}
    \setlength\fheight{.2\linewidth}
    \input{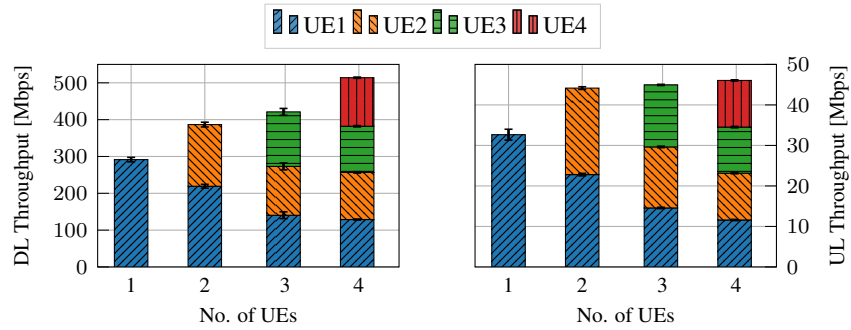}
\vspace{-0.10in}
\caption{Performance profiling with multiple \acrshortpl{ue} at fixed location~4 for the static iPerf use case \blue{using a DDDSU \acrshort{tdd} pattern, a 2x2 MIMO configuration, 2~layers DL and 1~layer UL}.}
\label{chap4-fig:4ue_static}
\end{figure}

\textbf{1-4 \acrshortpl{ue}, static, iPerf.}
We further extend our evaluation to include additional tests with multiple \glspl{ue}. At fixed location~4, we compare system performance with varying numbers of \glspl{ue} (from 1 to 4) connected to our network. The average throughput and $95$\% confidence intervals are plotted in Figure~\ref{chap4-fig:4ue_static}. We observe that the \glspl{ue} achieve steady throughput in all the cases, as indicated by the small confidence interval values. Additionally, the combined throughput increases with the number of \glspl{ue} connected: with four \glspl{ue}, the aggregate throughput reaches $512$\:Mbps in \gls{dl} and $46$\:Mbps in \gls{ul}.
This scenario highlights the maximum throughput performance that X5G is able to achieve with the current \blue{2x2 \gls{mimo} configuration, featuring 2~layers in \gls{dl} and a DDDSU \gls{tdd} pattern, ensuring a fair distribution of resources among all \glspl{ue}.} It is worth noting that \blue{the peak performance of X5G is detailed in Section~\ref{chap4-sec:exp-peak}}.

\textbf{2 \glspl{ru}, 1 \gls{ue} per \gls{ru}, static, iPerf.}
Finally, we evaluate X5G performance with two \glspl{ru} by connecting \gls{ue}1 to \gls{ru}1 and \gls{ue}2 to \gls{ru}2. \gls{ru}1 is located at position~6, and \gls{ru}2 is at position~23. We select six pairs of locations---{(0,25), (2,23), (4,21), (6,19), (8,17), (10,15)}---for the \glspl{ue} to ensure different distances among them and the \gls{ru}.

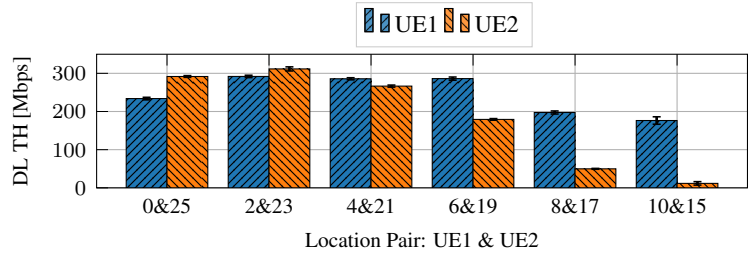
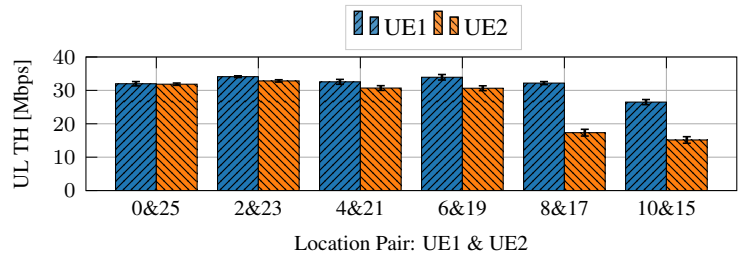
\begin{figure}[htb]
\centering
    \subfloat[Downlink]{
    \label{chap4-fig:dl_2ru2ue_new}
    \centering
    \setlength\fwidth{.7\linewidth}
    \setlength\fheight{.22\linewidth}
\begin{tikzpicture}
\pgfplotsset{every tick label/.append style={font=\scriptsize}}

\definecolor{darkgray176}{RGB}{176,176,176}
\definecolor{darkorange25512714}{RGB}{255,127,14}
\definecolor{lightgray204}{RGB}{204,204,204}
\definecolor{steelblue31119180}{RGB}{31,119,180}

\begin{axis}[
width=0.951\fwidth,
height=\fheight,
at={(0\fwidth,0\fheight)},
legend cell align={left},
legend columns=2,
legend style={fill opacity=0.8, 
draw opacity=1, 
text opacity=1, 
draw=lightgray204, 
font=\footnotesize,
at={(0.67, 1.39)}},
x grid style={darkgray176},
xmin=-0.49, xmax=5.89,
xtick style={color=black},
xtick={0.2,1.2,2.2,3.2,4.2,5.2},
xticklabels={0\&25,2\&23,4\&21,6\&19,8\&17,10\&15},
y grid style={darkgray176},
ylabel={DL TH [Mbps]},
xlabel={Location Pair: UE1 \& UE2},
ylabel style={font=\scriptsize},
xlabel style={font=\scriptsize},
ymin=0, ymax=350,
ytick pos=left,
ytick style={color=black},
xmajorgrids,
ymajorgrids
]

\addlegendimage{ybar,ybar legend,draw=black,fill=steelblue31119180,postaction={pattern=north east lines,pattern color=black}}
\addlegendentry{UE1}

\draw[draw=black,fill=steelblue31119180,postaction={pattern=north east lines,pattern color=black}] (axis cs:-0.2,0) rectangle (axis cs:0.2,233.86456832);
\draw[draw=black,fill=steelblue31119180,postaction={pattern=north east lines,pattern color=black}] (axis cs:0.8,0) rectangle (axis cs:1.2,291.89891328);
\draw[draw=black,fill=steelblue31119180,postaction={pattern=north east lines,pattern color=black}] (axis cs:1.8,0) rectangle (axis cs:2.2,285.713666133333);
\draw[draw=black,fill=steelblue31119180,postaction={pattern=north east lines,pattern color=black}] (axis cs:2.8,0) rectangle (axis cs:3.2,286.180349866667);
\draw[draw=black,fill=steelblue31119180,postaction={pattern=north east lines,pattern color=black}] (axis cs:3.8,0) rectangle (axis cs:4.2,197.37538432);
\draw[draw=black,fill=steelblue31119180,postaction={pattern=north east lines,pattern color=black}] (axis cs:4.8,0) rectangle (axis cs:5.2,176.3935744);

\addlegendimage{ybar,ybar legend,draw=black,fill=darkorange25512714,postaction={pattern=north west lines,pattern color=black}}
\addlegendentry{UE2}

\draw[draw=black,fill=darkorange25512714,postaction={pattern=north west lines,pattern color=black}] (axis cs:0.2,0) rectangle (axis cs:0.6,291.741805226667);
\draw[draw=black,fill=darkorange25512714,postaction={pattern=north west lines,pattern color=black}] (axis cs:1.2,0) rectangle (axis cs:1.6,311.55290112);
\draw[draw=black,fill=darkorange25512714,postaction={pattern=north west lines,pattern color=black}] (axis cs:2.2,0) rectangle (axis cs:2.6,266.541028693333);
\draw[draw=black,fill=darkorange25512714,postaction={pattern=north west lines,pattern color=black}] (axis cs:3.2,0) rectangle (axis cs:3.6,179.292514986667);
\draw[draw=black,fill=darkorange25512714,postaction={pattern=north west lines,pattern color=black}] (axis cs:4.2,0) rectangle (axis cs:4.6,49.80736);
\draw[draw=black,fill=darkorange25512714,postaction={pattern=north west lines,pattern color=black}] (axis cs:5.2,0) rectangle (axis cs:5.6,11.7059211636364);

\path [draw=black, line width=1pt]
(axis cs:0,230.547928827315)
--(axis cs:0,237.181207812685);

\path [draw=black, line width=1pt]
(axis cs:1,288.8207988582)
--(axis cs:1,294.9770277018);

\path [draw=black, line width=1pt]
(axis cs:2,282.98840323712)
--(axis cs:2,288.438929029546);

\path [draw=black, line width=1pt]
(axis cs:3,282.045078669539)
--(axis cs:3,290.315621063794);

\path [draw=black, line width=1pt]
(axis cs:4,193.352877110513)
--(axis cs:4,201.397891529487);

\path [draw=black, line width=1pt]
(axis cs:5,166.642034928531)
--(axis cs:5,186.145113871469);

\addplot [semithick, black, mark=-, mark size=1.5, mark options={solid}, only marks]
table {%
0 230.547928827315
1 288.8207988582
2 282.98840323712
3 282.045078669539
4 193.352877110513
5 166.642034928531
};
\addplot [semithick, black, mark=-, mark size=1.5, mark options={solid}, only marks]
table {%
0 237.181207812685
1 294.9770277018
2 288.438929029546
3 290.315621063794
4 201.397891529487
5 186.145113871469
};

\path [draw=black, line width=1pt]
(axis cs:0.4,289.430103281427)
--(axis cs:0.4,294.053507171906);

\path [draw=black, line width=1pt]
(axis cs:1.4,306.152562756234)
--(axis cs:1.4,316.953239483766);

\path [draw=black, line width=1pt]
(axis cs:2.4,263.856146802523)
--(axis cs:2.4,269.225910584144);

\path [draw=black, line width=1pt]
(axis cs:3.4,177.173672478772)
--(axis cs:3.4,181.411357494562);

\path [draw=black, line width=1pt]
(axis cs:4.4,48.6648455686547)
--(axis cs:4.4,50.9498744313453);

\path [draw=black, line width=1pt]
(axis cs:5.4,7.21392651935996)
--(axis cs:5.4,16.1979158079128);

\addplot [semithick, black, mark=-, mark size=1.5, mark options={solid}, only marks]
table {%
0.4 289.430103281427
1.4 306.152562756234
2.4 263.856146802523
3.4 177.173672478772
4.4 48.6648455686547
5.4 7.21392651935996
};
\addplot [semithick, black, mark=-, mark size=1.5, mark options={solid}, only marks]
table {%
0.4 294.053507171906
1.4 316.953239483766
2.4 269.225910584144
3.4 181.411357494562
4.4 50.9498744313453
5.4 16.1979158079128
};

\end{axis}

\end{tikzpicture}
    \setlength\abovecaptionskip{.05cm}}
    \hfill    
    \subfloat[Uplink]{
    \label{chap4-fig:ul_2ru2ue_new}
    \centering
        \setlength\fwidth{.7\linewidth}
        \setlength\fheight{.22\linewidth}
\begin{tikzpicture}
\pgfplotsset{every tick label/.append style={font=\scriptsize}}

\definecolor{darkgray176}{RGB}{176,176,176}
\definecolor{darkorange25512714}{RGB}{255,127,14}
\definecolor{lightgray204}{RGB}{204,204,204}
\definecolor{steelblue31119180}{RGB}{31,119,180}

\begin{axis}[
width=0.951\fwidth,
height=\fheight,
at={(0\fwidth,0\fheight)},
legend cell align={left},
legend columns=2,
legend style={fill opacity=0.8, draw opacity=1, text opacity=1, draw=lightgray204, font=\footnotesize,at={(0.67, 1.39)}},
x grid style={darkgray176},
xmin=-0.49, xmax=5.89,
xtick style={color=black},
xtick={0.2,1.2,2.2,3.2,4.2,5.2},
xticklabels={0\&25,2\&23,4\&21,6\&19,8\&17,10\&15},
y grid style={darkgray176},
ylabel={UL TH [Mbps]},
xlabel={Location Pair: UE1 \& UE2},
ylabel style={font=\scriptsize},
xlabel style={font=\scriptsize},
ymin=0, ymax=40,
ytick pos=left,
ytick style={color=black},
xmajorgrids,
ymajorgrids
]

\addlegendimage{ybar,ybar legend,draw=black,fill=steelblue31119180,postaction={pattern=north east lines,pattern color=black}}
\addlegendentry{UE1}

\addlegendimage{ybar,ybar legend,draw=black,fill=darkorange25512714,postaction={pattern=north west lines,pattern color=black}}
\addlegendentry{UE2}

\draw[draw=black,fill=steelblue31119180,postaction={pattern=north east lines,pattern color=black}] (axis cs:-0.2,0) rectangle (axis cs:0.2,31.9885585066667);
\draw[draw=black,fill=steelblue31119180,postaction={pattern=north east lines,pattern color=black}] (axis cs:0.8,0) rectangle (axis cs:1.2,34.1066820266667);
\draw[draw=black,fill=steelblue31119180,postaction={pattern=north east lines,pattern color=black}] (axis cs:1.8,0) rectangle (axis cs:2.2,32.5617800533333);
\draw[draw=black,fill=steelblue31119180,postaction={pattern=north east lines,pattern color=black}] (axis cs:2.8,0) rectangle (axis cs:3.2,33.93191936);
\draw[draw=black,fill=steelblue31119180,postaction={pattern=north east lines,pattern color=black}] (axis cs:3.8,0) rectangle (axis cs:4.2,32.1842926933333);
\draw[draw=black,fill=steelblue31119180,postaction={pattern=north east lines,pattern color=black}] (axis cs:4.8,0) rectangle (axis cs:5.2,26.4940202666667);

\draw[draw=black,fill=darkorange25512714,postaction={pattern=north west lines,pattern color=black}] (axis cs:0.2,0) rectangle (axis cs:0.6,31.85573888);
\draw[draw=black,fill=darkorange25512714,postaction={pattern=north west lines,pattern color=black}] (axis cs:1.2,0) rectangle (axis cs:1.6,32.8344098133333);
\draw[draw=black,fill=darkorange25512714,postaction={pattern=north west lines,pattern color=black}] (axis cs:2.2,0) rectangle (axis cs:2.6,30.70230528);
\draw[draw=black,fill=darkorange25512714,postaction={pattern=north west lines,pattern color=black}] (axis cs:3.2,0) rectangle (axis cs:3.6,30.6463812266667);
\draw[draw=black,fill=darkorange25512714,postaction={pattern=north west lines,pattern color=black}] (axis cs:4.2,0) rectangle (axis cs:4.6,17.3364565333333);
\draw[draw=black,fill=darkorange25512714,postaction={pattern=north west lines,pattern color=black}] (axis cs:5.2,0) rectangle (axis cs:5.6,15.1554184533333);

\path [draw=black, line width=1pt]
(axis cs:0.4,31.513056012484)
--(axis cs:0.4,32.198421747516);

\path [draw=black, line width=1pt]
(axis cs:1.4,32.4809834366442)
--(axis cs:1.4,33.1878361900225);

\path [draw=black, line width=1pt]
(axis cs:2.4,29.9925850167168)
--(axis cs:2.4,31.4120255432832);

\path [draw=black, line width=1pt]
(axis cs:3.4,29.9176908246676)
--(axis cs:3.4,31.3750716286657);

\path [draw=black, line width=1pt]
(axis cs:4.4,16.3248632855358)
--(axis cs:4.4,18.3480497811309);

\path [draw=black, line width=1pt]
(axis cs:5.4,14.1746933344082)
--(axis cs:5.4,16.1361435722584);

\path [draw=black, line width=1pt]
(axis cs:0,31.3367893521494)
--(axis cs:0,32.6403276611839);

\path [draw=black, line width=1pt]
(axis cs:1,33.8468768830825)
--(axis cs:1,34.3664871702508);

\path [draw=black, line width=1pt]
(axis cs:2,31.8263617954621)
--(axis cs:2,33.2971983112046);

\path [draw=black, line width=1pt]
(axis cs:3,33.1077024863186)
--(axis cs:3,34.7561362336814);

\path [draw=black, line width=1pt]
(axis cs:4,31.7343581731514)
--(axis cs:4,32.6342272135153);

\path [draw=black, line width=1pt]
(axis cs:5,25.7266767289913)
--(axis cs:5,27.2613638043421);

\addplot [semithick, black, mark=-, mark size=1.5, mark options={solid}, only marks]
table {%
0.4 31.513056012484
1.4 32.4809834366442
2.4 29.9925850167168
3.4 29.9176908246676
4.4 16.3248632855358
5.4 14.1746933344082
};

\addplot [semithick, black, mark=-, mark size=1.5, mark options={solid}, only marks]
table {%
0.4 32.198421747516
1.4 33.1878361900225
2.4 31.4120255432832
3.4 31.3750716286657
4.4 18.3480497811309
5.4 16.1361435722584
};

\addplot [semithick, black, mark=-, mark size=1.5, mark options={solid}, only marks]
table {%
0 31.3367893521494
1 33.8468768830825
2 31.8263617954621
3 33.1077024863186
4 31.7343581731514
5 25.7266767289913
};

\addplot [semithick, black, mark=-, mark size=1.5, mark options={solid}, only marks]
table {%
0 32.6403276611839
1 34.3664871702508
2 33.2971983112046
3 34.7561362336814
4 32.6342272135153
5 27.2613638043421
};

\end{axis}

\end{tikzpicture}
        \setlength\abovecaptionskip{.05cm}}
\caption{Performance profiling for two \acrshortpl{ru} in the static iPerf use case, each with one assigned \acrshort{ue}: \acrshort{ue}1 to \acrshort{ru}1 and \acrshort{ue}2 to \acrshort{ru}2.}
\label{chap4-fig:RF2RU2UE_Static_th_New}
\end{figure}

From Figure~\ref{chap4-fig:dl_2ru2ue_new}, we observe that the \gls{dl} throughput is significantly impacted by interference, particularly at cell edge locations. The throughput for \gls{ue}1 shows a reduction of up to 90\% as the \glspl{ue} approach each other, while \gls{ue}2 throughput decreases by up to 50\%. These observations indicate that interference predominantly affects the \gls{dl} direction. Conversely, as depicted in Figure~\ref{chap4-fig:ul_2ru2ue_new}, \gls{ul} throughput remains relatively stable across different location pairs, suggesting that \gls{ul} is less susceptible to the types of interference affecting \gls{dl} throughput.

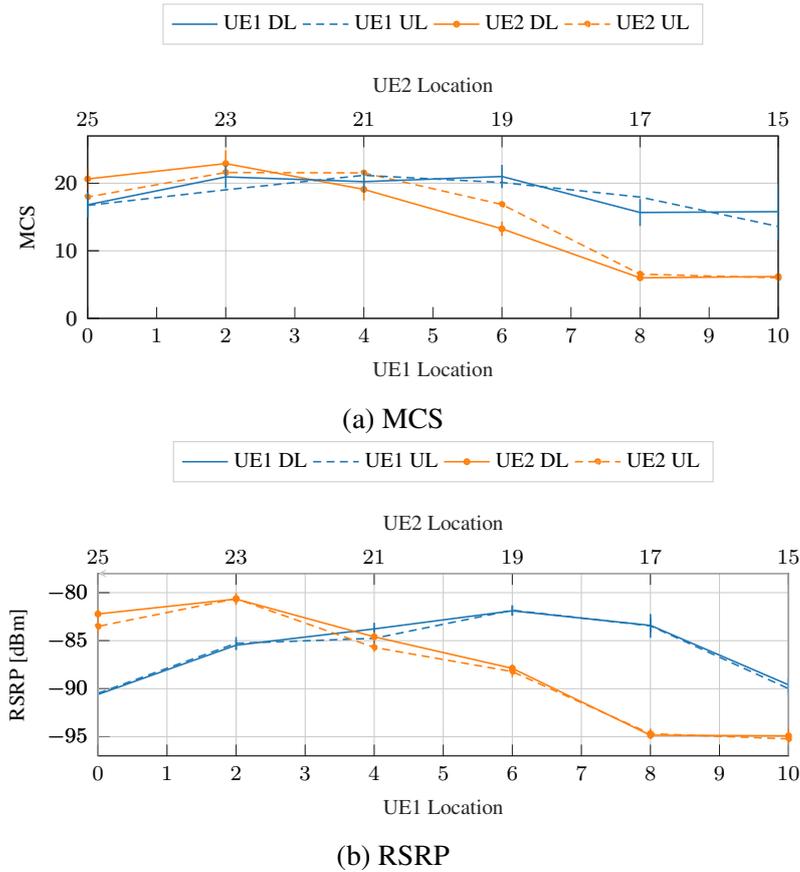
\begin{figure}[htb]
\centering
    \subfloat[MCS]{
    \label{chap4-fig:mcs_2ru2ue_new}
    \centering
    \setlength\fwidth{.7\linewidth}
    \setlength\fheight{.25\linewidth}
\begin{tikzpicture}
\pgfplotsset{every tick label/.append style={font=\scriptsize}}

\definecolor{black}{RGB}{38,38,38}
\definecolor{lightgray204}{RGB}{204,204,204}
\definecolor{steelblue31119180}{RGB}{31,119,180}
\definecolor{darkorange25512714}{RGB}{255,127,14}

\begin{axis}[
width=1\fwidth, 
height=1.05\fheight, 
at={(0\fwidth,0\fheight)},
axis line style={color=black},
legend cell align={center},
legend style={
  fill opacity=0.8,
  draw opacity=1,
  text opacity=1,
  at={(0.5, 1.5)},
  anchor=south,
  draw=lightgray204,
  font=\scriptsize,
},
axis line style={color=black},
tick align=inside,
x dir=reverse,
x grid style={lightgray204},
xlabel=\textcolor{black}{UE2 Location},
label style={font=\scriptsize},
xmajorgrids,
xmin=15, xmax=25,
xtick={15, 17, 19, 21, 23, 25},
xtick pos=right,
xtick style={color=black},
y grid style={lightgray204},
ymajorgrids,
ymin=0, ymax=27,
ytick pos=left,
ytick style={color=black},
ylabel={MCS},
legend columns=4,
]

\path [draw=darkorange25512714, semithick]
(axis cs:15,5.16716845905602)
--(axis cs:15,7.22172042983287);
\path [draw=darkorange25512714, semithick]
(axis cs:17,6)
--(axis cs:17,6);
\path [draw=darkorange25512714, semithick]
(axis cs:19,12.1842084276582)
--(axis cs:19,14.3248824814328);
\path [draw=darkorange25512714, semithick]
(axis cs:21,17.4511202560732)
--(axis cs:21,20.730697925745);
\path [draw=darkorange25512714, semithick]
(axis cs:23,20.9435835065462)
--(axis cs:23,24.8745983116356);
\path [draw=darkorange25512714, semithick]
(axis cs:25,19.0582607645917)
--(axis cs:25,22.2144665081356);

\addplot [semithick, darkorange25512714, mark=*, mark size=1pt]
table {%
15 6.19444444444444
17 6
19 13.2545454545455
21 19.0909090909091
23 22.9090909090909
25 20.6363636363636
};

\path [draw=steelblue31119180, semithick]
(axis cs:0,14.8066916065866)
--(axis cs:0,18.629672029777);
\path [draw=steelblue31119180, semithick]
(axis cs:2,17.6831786740086)
--(axis cs:2,20.3713667805369);
\path [draw=steelblue31119180, semithick]
(axis cs:4,18.4606777382919)
--(axis cs:4,23.8847768071626);
\path [draw=steelblue31119180, semithick]
(axis cs:6,17.5306145275091)
--(axis cs:6,22.6875672906727);
\path [draw=steelblue31119180, semithick]
(axis cs:8,15.5148644570635)
--(axis cs:8,20.3942264520274);
\path [draw=steelblue31119180, semithick]
(axis cs:10,10.0883025514888)
--(axis cs:10,17.1227066228231);

\addplot [semithick, darkorange25512714, densely dashed, mark=*, mark size=1pt]
table {%
15 6
17 6.54545454545455
19 16.8545454545455
21 21.5272727272727
23 21.5909090909091
25 17.9636363636364
};

\end{axis}

\begin{axis}[
width=1\fwidth, 
height=1.05\fheight, 
axis line style={color=black},
at={(0\fwidth,0\fheight)},
legend cell align={center},
legend columns=4,
legend style={
  fill opacity=0.8,
  draw opacity=1,
  text opacity=1,
  at={(0.5, 1.5)},
  anchor=south,
  draw=lightgray204,
  font=\scriptsize,
},
tick align=inside,
tick pos=left,
x grid style={lightgray204},
xlabel=\textcolor{black}{UE1 Location},
label style={font=\scriptsize},
xmin=0, xmax=10,
xtick style={color=black},
y grid style={lightgray204},
ymin=0, ymax=27,
ytick style={color=black}, 
yticklabel style={text=black} 
]
\path [draw=steelblue31119180, semithick]
(axis cs:0,14.954290766315)
--(axis cs:0,18.6638910518668);
\path [draw=steelblue31119180, semithick]
(axis cs:2,19.2813613623151)
--(axis cs:2,22.5913659104122);
\path [draw=steelblue31119180, semithick]
(axis cs:4,18.6056533575758)
--(axis cs:4,21.8670739151515);
\path [draw=steelblue31119180, semithick]
(axis cs:6,19.2865882524875)
--(axis cs:6,22.7134117475125);
\path [draw=steelblue31119180, semithick]
(axis cs:8,13.6759077642524)
--(axis cs:8,17.663714877257);
\path [draw=steelblue31119180, semithick]
(axis cs:10,11.6354323899127)
--(axis cs:10,19.9533526568163);

\addplot [semithick, steelblue31119180, forget plot]
table {%
0 16.8090909090909
2 20.9363636363636
4 20.2363636363636
6 21
8 15.6698113207547
10 15.7943925233645
};

\path [draw=darkorange25512714, semithick]
(axis cs:15,6)
--(axis cs:15,6);
\path [draw=darkorange25512714, semithick]
(axis cs:17,5.16647497567179)
--(axis cs:17,7.9244341152373);
\path [draw=darkorange25512714, semithick]
(axis cs:19,14.8461394334378)
--(axis cs:19,18.8629514756531);
\path [draw=darkorange25512714, semithick]
(axis cs:21,20.0328851173779)
--(axis cs:21,23.0216603371676);
\path [draw=darkorange25512714, semithick]
(axis cs:23,19.8284282781021)
--(axis cs:23,23.3533899037161);
\path [draw=darkorange25512714, semithick]
(axis cs:25,15.2547408145402)
--(axis cs:25,20.6725319127326);

\addplot [semithick, steelblue31119180, densely dashed, forget plot]
table {%
0 16.7181818181818
2 19.0272727272727
4 21.1727272727273
6 20.1090909090909
8 17.9545454545455
10 13.605504587156
};

\addlegendimage{semithick, steelblue31119180}
\addlegendentry{UE1 DL}

\addlegendimage{semithick, steelblue31119180, densely dashed}
\addlegendentry{UE1 UL}

\addlegendimage{semithick, darkorange25512714, mark=*, mark size=1pt}
\addlegendentry{UE2 DL}

\addlegendimage{semithick, darkorange25512714, densely dashed, mark=*, mark size=1pt}
\addlegendentry{UE2 UL}

\end{axis}

\end{tikzpicture}
    \setlength\abovecaptionskip{.05cm}}
    \hfill    
    \subfloat[RSRP]{
    \label{chap4-fig:rsrp_2ru2ue_new}
    \centering
    \setlength\fwidth{.7\linewidth}
    \setlength\fheight{.25\linewidth}
\begin{tikzpicture}
\pgfplotsset{every tick label/.append style={font=\scriptsize}}

\definecolor{darkslategray38}{RGB}{38,38,38}
\definecolor{lightgray204}{RGB}{204,204,204}
\definecolor{steelblue31119180}{RGB}{31,119,180}
\definecolor{darkorange25512714}{RGB}{255,127,14}

\begin{axis}[
width=1\fwidth, 
height=1.05\fheight, 
axis line style={color=black},
legend cell align={center},
legend style={
  fill opacity=0.8,
  draw opacity=1,
  text opacity=1,
  at={(0.5, 1.5)},
  anchor=south,
  draw=lightgray204,
  font=\scriptsize,
},
tick align=inside,
tick pos=left,
x grid style={lightgray204},
xlabel=\textcolor{darkslategray38}{UE1 Location},
label style={font=\scriptsize},
xmajorgrids,
xmin=0, xmax=10,
xtick style={color=darkslategray38},
y grid style={lightgray204},
ymajorgrids,
ymin=-97, ymax=-78,
ytick style={color=darkslategray38},
ylabel={RSRP [dBm]},
legend columns=4,
]

\path [draw=steelblue31119180, semithick]
(axis cs:0,-91.1617796100565)
--(axis cs:0,-90.0382203899435);
\path [draw=steelblue31119180, semithick]
(axis cs:2,-85.9742678518038)
--(axis cs:2,-84.9711866936508);
\path [draw=steelblue31119180, semithick]
(axis cs:4,-84.4175242453835)
--(axis cs:4,-83.1279303000711);
\path [draw=steelblue31119180, semithick]
(axis cs:6,-82.4039581621078)
--(axis cs:6,-81.3778600197104);
\path [draw=steelblue31119180, semithick]
(axis cs:8,-84.1587910102666)
--(axis cs:8,-82.6525297444503);
\path [draw=steelblue31119180, semithick]
(axis cs:10,-90.373455643225)
--(axis cs:10,-88.822806039018);

\addplot [semithick, steelblue31119180]
table {%
0 -90.6
2 -85.4727272727273
4 -83.7727272727273
6 -81.8909090909091
8 -83.4056603773585
10 -89.5981308411215
};

\path [draw=steelblue31119180, semithick]
(axis cs:0,-90.9922381737809)
--(axis cs:0,-89.9532163716736);
\path [draw=steelblue31119180, semithick]
(axis cs:2,-85.9350707755739)
--(axis cs:2,-84.6103837698806);
\path [draw=steelblue31119180, semithick]
(axis cs:4,-85.8730828648639)
--(axis cs:4,-83.636008044227);
\path [draw=steelblue31119180, semithick]
(axis cs:6,-82.3494258368982)
--(axis cs:6,-81.3051196176472);
\path [draw=steelblue31119180, semithick]
(axis cs:8,-84.6933401012945)
--(axis cs:8,-82.2157508077964);
\path [draw=steelblue31119180, semithick]
(axis cs:10,-90.6362571091663)
--(axis cs:10,-89.3453942669805);

\addplot [semithick, steelblue31119180, densely dashed]
table {%
0 -90.4727272727273
2 -85.2727272727273
4 -84.7545454545455
6 -81.8272727272727
8 -83.4545454545455
10 -89.9908256880734
};

\end{axis}

\begin{axis}[
width=1\fwidth, 
height=1.05\fheight, 
axis line style={lightgray204},
axis x line=top,
legend cell align={left},
legend columns=4,
legend style={
  fill opacity=0.8,
  draw opacity=1,
  text opacity=1,
  at={(0.5, 1.5)},
  anchor=south,
  draw=lightgray204,
  font=\scriptsize,
},
tick align=inside,
x dir=reverse,
x grid style={lightgray204},
xlabel=\textcolor{darkslategray38}{UE2 Location},
label style={font=\scriptsize},
xmin=15, xmax=25,
xtick={15, 17, 19, 21, 23, 25},
xtick pos=right,
xtick style={color=darkslategray38},
y grid style={lightgray204},
ymin=-97, ymax=-78,
ytick pos=left,
]
\path [draw=darkorange25512714, semithick]
(axis cs:15,-95.6116938433622)
--(axis cs:15,-94.2031209714526);
\path [draw=darkorange25512714, semithick]
(axis cs:17,-95.2085797073974)
--(axis cs:17,-94.4823293835117);
\path [draw=darkorange25512714, semithick]
(axis cs:19,-88.3043882602975)
--(axis cs:19,-87.4592481033389);
\path [draw=darkorange25512714, semithick]
(axis cs:21,-85.0848253737061)
--(axis cs:21,-84.0969928081121);
\path [draw=darkorange25512714, semithick]
(axis cs:23,-81.2736036168899)
--(axis cs:23,-80.0536691103828);
\path [draw=darkorange25512714, semithick]
(axis cs:25,-82.7318807301365)
--(axis cs:25,-81.7044829062271);

\addplot [semithick, darkorange25512714, forget plot, mark=*, mark size=1pt]
table {%
15 -94.9074074074074
17 -94.8454545454545
19 -87.8818181818182
21 -84.5909090909091
23 -80.6636363636364
25 -82.2181818181818
};

\path [draw=darkorange25512714, semithick]
(axis cs:15,-95.6695166076239)
--(axis cs:15,-94.7850288469215);
\path [draw=darkorange25512714, semithick]
(axis cs:17,-95.2284101819335)
--(axis cs:17,-94.1534079998847);
\path [draw=darkorange25512714, semithick]
(axis cs:19,-88.7989408624991)
--(axis cs:19,-87.6374227738646);
\path [draw=darkorange25512714, semithick]
(axis cs:21,-86.1603548691006)
--(axis cs:21,-85.2396451308994);
\path [draw=darkorange25512714, semithick]
(axis cs:23,-81.1997146382569)
--(axis cs:23,-80.0548308162886);
\path [draw=darkorange25512714, semithick]
(axis cs:25,-84.3737428732838)
--(axis cs:25,-82.644438944898);

\addplot [semithick, darkorange25512714, densely dashed, forget plot, mark=*, mark size=1pt]
table {%
15 -95.2272727272727
17 -94.6909090909091
19 -88.2181818181818
21 -85.7
23 -80.6272727272727
25 -83.5090909090909
};

\addlegendimage{semithick, steelblue31119180}
\addlegendentry{UE1 DL}

\addlegendimage{semithick, steelblue31119180, densely dashed}
\addlegendentry{UE1 UL}

\addlegendimage{semithick, darkorange25512714, mark=*, mark size=1pt}
\addlegendentry{UE2 DL}

\addlegendimage{semithick, darkorange25512714, densely dashed, mark=*, mark size=1pt}
\addlegendentry{UE2 UL}

\end{axis}

\end{tikzpicture}
    \setlength\abovecaptionskip{.05cm}}
\caption{\acrshort{mac} \acrshortpl{kpi} in the two \acrshortpl{ru} iPerf use case, each with one static \acrshort{ue} (\acrshort{ue}1 assigned to \acrshort{ru}1 and \acrshort{ue}2 to \acrshort{ru}2): (a) averages and confidence intervals for \acrshort{dl} \acrshort{mcs} (solid lines) during \acrshort{dl} data transmissions, and \acrshort{ul} \acrshort{mcs} (dashed lines) during \acrshort{ul} transmissions, for \acrshort{ue}1 (blue) and \acrshort{ue}2 (orange); (b) averages and confidence intervals of \acrshort{rsrp} reported by \acrshort{ue}1 (blue) and \acrshort{ue}2 (orange) during \acrshort{dl} (solid lines) and \acrshort{ul} (dashed lines) transmissions.}
\label{chap4-fig:RF2RU2UE_Static_New}
\end{figure}

Figure~\ref{chap4-fig:RF2RU2UE_Static_New} further supports these observations by presenting additional \glspl{kpi} from the \gls{mac} layer. Figure~\ref{chap4-fig:mcs_2ru2ue_new} shows that the \gls{mcs} for both \glspl{ue} decreases as the distance between the \glspl{ue} diminishes, indicative of increasing interference levels. 
Figure~\ref{chap4-fig:rsrp_2ru2ue_new} illustrates the \gls{rsrp}, which varies in response to the \glspl{ue} locations. Notably, despite adequate \gls{rsrp} levels, the throughput remains low, highlighting the significant impact of interference, particularly in the \gls{dl} direction.

\subsection{Mobile Experiments}
\label{chap4-sec:exp-mobile}

\begin{figure}[hbt]
\centering
    \subfloat[Throughput]{
    \label{chap4-fig:1ru1ue_mobile_iperf_th}
    \centering
    \setlength\fwidth{.7\columnwidth}
    \setlength\fheight{.2\columnwidth}
    \input{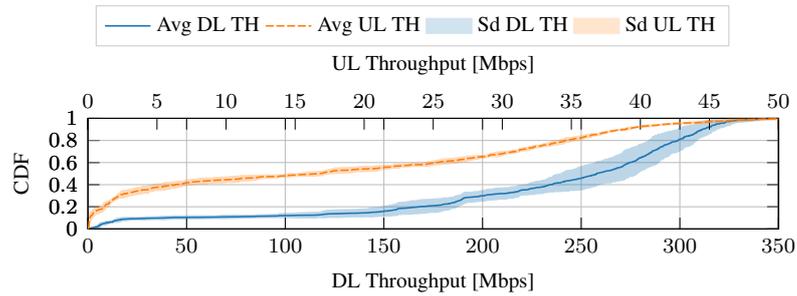}
    }

    \hfill
    
    \subfloat[\gls{mcs}]{
    \label{chap4-fig:1ru1ue_mobile_iperf_mcs}
    \centering
    \setlength\fwidth{.7\columnwidth}
    \setlength\fheight{.2\columnwidth}
    \input{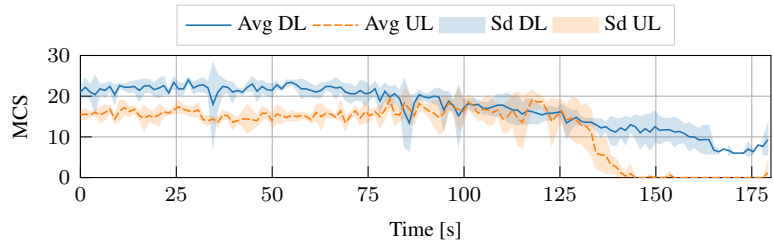}
    }

    \hfill
    
    \subfloat[\gls{rsrp}]{
    \label{chap4-fig:1ru1ue_mobile_iperf_rsrp}
    \centering
    \setlength\fwidth{.7\columnwidth}
    \setlength\fheight{.2\columnwidth}
    \input{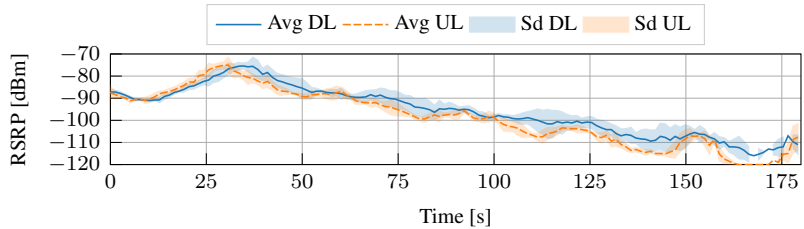}
    }
\caption{Performance profiling with one \acrshort{ru} and one mobile \acrshort{ue} in the iPerf use case: (a)~\acrshort{cdf} of \acrshort{dl} and \acrshort{ul} throughputs with averages (solid lines) and \acrshort{sd} (shaded areas); (b)~averages (solid lines) and \acrshort{sd} (shaded areas) of the \acrshort{dl} \acrshort{mcs} during \acrshort{dl} transmissions (blue) and of the \acrshort{ul} \acrshort{mcs} during \acrshort{ul} transmissions (orange); (c)~averages (solid lines) and \acrshort{sd} (shaded areas) of the \acrshort{rsrp} reported by the \acrshort{ue} during \acrshort{dl} (blue) and \acrshort{ul} (orange) data transmissions.}
\label{chap4-fig:1ru1ue_mobile_iperf}
\end{figure}

%
We assess the network performance by measuring throughput as the \gls{ue} follows the walking pattern around the laboratory space depicted by the dashed green line in Figure~\ref{chap4-fig:node-locations}. The entire walk from the start to the end point spans approximately $3$\:minutes at regular walking speed.
The mobile use case results are illustrated in Figure~\ref{chap4-fig:1ru1ue_mobile_iperf}. The application layer throughput is depicted by \gls{cdf} plots in Figure~\ref{chap4-fig:1ru1ue_mobile_iperf_th}, where solid lines represent the averaged curve for all runs, while the shaded areas around these lines illustrate the variation across different runs, indicating the range of values within one \gls{sd} above and below the mean.
The \gls{mcs} and \gls{rsrp} results at the \gls{mac} layer are shown in Figure~\ref{chap4-fig:1ru1ue_mobile_iperf_mcs} and Figure~\ref{chap4-fig:1ru1ue_mobile_iperf_rsrp}, respectively. Also here, the average values are depicted using solid lines, while the shaded areas indicate the \gls{sd}.

The throughput results of Figure~\ref{chap4-fig:1ru1ue_mobile_iperf_th} highlight notable variability in network quality influenced by mobility. Throughout the test, the \gls{ue} achieves peaks of up to $350$\:Mbps in \gls{dl} and $50$\:Mbps in \gls{ul}. However, significant fluctuations in performance are observed, particularly as the \gls{ue} moves further from the initial \gls{ru} position.
%
%
%
Figure~\ref{chap4-fig:1ru1ue_mobile_iperf_mcs} illustrates a significant drop in \gls{ul} \gls{mcs} values around the 100-second mark, where averages initially above 10 drop sharply, while \gls{dl} \gls{mcs} fluctuate more gradually until they fall below 10.
This sudden decline in \gls{ul} \gls{mcs} at this specific time is likely due to deteriorating signal conditions, as corroborated by the corresponding \gls{rsrp} trends in Figure~\ref{chap4-fig:1ru1ue_mobile_iperf_rsrp}. This is most probably due to increased distance from the base station or physical obstructions, leading to a necessary reduction in \gls{mcs} to maintain connectivity under compromised signal strength.


The \gls{mcs} results of Figure~\ref{chap4-fig:1ru1ue_mobile_iperf_mcs} show that both \gls{dl} and \gls{ul} \gls{mcs} values start relatively high but decrease as the \gls{ue} moves further from the \gls{ru}. The \gls{dl} \gls{mcs} exhibits more variability and sharper declines compared to the \gls{ul}, which maintains a more stable profile until the final part of the walk. This suggests that the uplink benefits from more aggressive modulation and coding strategies due to \gls{5g} adaptive power control mechanisms that mitigate the impact of increasing distance and obstacles more effectively.
Additionally, the \gls{rsrp} data, shown in Figure~\ref{chap4-fig:1ru1ue_mobile_iperf_rsrp}, indicates a gradual decline in signal strength as the \gls{ue} moves along its trajectory. \gls{rsrp} values for both \gls{dl} (blue) and \gls{ul} (orange) cases decrease over time, with the most significant drops observed after $100$ seconds. This reduction in signal quality corresponds with declines in throughput and \gls{mcs}, highlighting the strong dependency of these metrics on signal strength.
%


\subsection{Video Streaming Experiments}
\label{chap4-sec:exp-video}

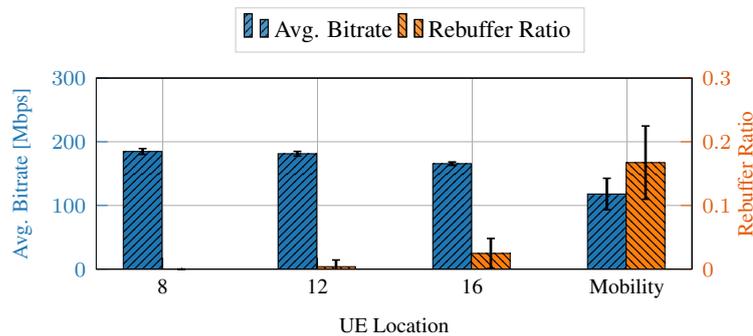
\begin{figure}[htb]
\centering
    \setlength\fwidth{.65\linewidth}
    \setlength\fheight{.18\linewidth}
\begin{tikzpicture}
\pgfplotsset{every tick label/.append style={font=\scriptsize}}

\definecolor{darkgray176}{RGB}{176,176,176}
\definecolor{darkorange25512714}{RGB}{255,127,14}
\definecolor{darkorange2309111}{RGB}{230,91,11}
\definecolor{lightgray204}{RGB}{204,204,204}
\definecolor{steelblue31119180}{RGB}{31,119,180}

\begin{axis}[
width=0.951\fwidth,
height=1.5\fheight,
at={(0\fwidth,0\fheight)},
legend cell align={left},
legend columns=2,
legend style={fill opacity=0.8, 
draw opacity=1, 
text opacity=1, 
draw=lightgray204, 
font=\footnotesize,
at={(0.82, 1.37)}},
x grid style={darkgray176},
xmajorticks=false,
xmin=-0.3, xmax=3.55,
xtick style={color=steelblue31119180},
xtick={0.13,1.13,2.13,3.13},
xticklabel style={rotate=45.0},
xticklabels={8,12,16,Mobility},
y grid style={darkgray176},
ylabel=\textcolor{steelblue31119180}{Avg. Bitrate [Mbps]},
ylabel style={font=\scriptsize},
xlabel style={font=\scriptsize},
ymin=0, ymax=300,
ytick pos=left,
ytick style={color=steelblue31119180},
yticklabel style={color=steelblue31119180},
xmajorgrids,
ymajorgrids
]

\addlegendimage{ybar,ybar legend,draw=black,fill=steelblue31119180,postaction={pattern=north east lines,pattern color=black}}
\addlegendentry{Avg. Bitrate}
\addlegendimage{ybar,ybar legend,draw=black,fill=darkorange25512714,postaction={pattern=north west lines,pattern color=black}}
\addlegendentry{Rebuffer Ratio}

\draw[draw=black,fill=steelblue31119180,postaction={pattern=north east lines,pattern color=black}] (axis cs:-0.125,0) rectangle (axis cs:0.125,184.684677810809);

\draw[draw=black,fill=steelblue31119180,postaction={pattern=north east lines,pattern color=black}] (axis cs:0.875,0) rectangle (axis cs:1.125,181.192263115416);

\draw[draw=black,fill=steelblue31119180,postaction={pattern=north east lines,pattern color=black}] (axis cs:1.875,0) rectangle (axis cs:2.125,165.600715680784);

\draw[draw=black,fill=steelblue31119180,postaction={pattern=north east lines,pattern color=black}] (axis cs:2.875,0) rectangle (axis cs:3.125,117.942643215823);

\path [draw=black, line width=1pt]
(axis cs:0,180.108878846041)
--(axis cs:0,189.260476775578);

\addplot [semithick, black, mark=-, mark size=1.5, mark options={solid}, only marks]
table {%
0 180.108878846041
};
\addplot [semithick, black, mark=-, mark size=1.5, mark options={solid}, only marks]
table {%
0 189.260476775578
};
\path [draw=black, line width=1pt]
(axis cs:1,177.525601291505)
--(axis cs:1,184.858924939326);

\addplot [semithick, black, mark=-, mark size=1.5, mark options={solid}, only marks]
table {%
1 177.525601291505
};
\addplot [semithick, black, mark=-, mark size=1.5, mark options={solid}, only marks]
table {%
1 184.858924939326
};
\path [draw=black, line width=1pt]
(axis cs:2,162.956109049625)
--(axis cs:2,168.245322311942);

\addplot [semithick, black, mark=-, mark size=1.5, mark options={solid}, only marks]
table {%
2 162.956109049625
};
\addplot [semithick, black, mark=-, mark size=1.5, mark options={solid}, only marks]
table {%
2 168.245322311942
};
\path [draw=black, line width=1pt]
(axis cs:3,93.2293861692557)
--(axis cs:3,142.65590026239);

\addplot [semithick, black, mark=-, mark size=1.5, mark options={solid}, only marks]
table {%
3 93.2293861692557
};
\addplot [semithick, black, mark=-, mark size=1.5, mark options={solid}, only marks]
table {%
3 142.65590026239
};
\end{axis}

\begin{axis}[
width=0.951\fwidth,
height=1.5\fheight,
at={(0\fwidth,0\fheight)},
axis y line*=right,
legend cell align={left},
legend style={
  fill opacity=0.8,
  draw opacity=1,
  text opacity=1,
  font=\footnotesize,
  draw=lightgray204
},
x grid style={darkgray176},
xmin=-0.3, xmax=3.55,
xtick pos=left,
xtick style={color=black},
xtick={0.13,1.13,2.13,3.13},
xticklabels={8,12,16,Mobility},
x label style={at={(axis description cs:0.5,-0.20)},anchor=north},
ylabel style={font=\scriptsize},
xlabel style={font=\scriptsize},
xlabel={UE Location},
y grid style={darkgray176},
ylabel=\textcolor{darkorange2309111}{Rebuffer Ratio},
ymin=0, ymax=0.3,
ytick pos=right,
legend columns=4,
ytick style={color=darkorange2309111},
yticklabel style={anchor=west,color=darkorange2309111},
ylabel shift=-5pt
]
\draw[draw=black,fill=darkorange25512714,postaction={pattern=north west lines,pattern color=black}] (axis cs:0.125,0) rectangle (axis cs:0.375,0);


\draw[draw=black,fill=darkorange25512714,postaction={pattern=north west lines,pattern color=black}] (axis cs:1.125,0) rectangle (axis cs:1.375,0.0037973392);
\draw[draw=black,fill=darkorange25512714,postaction={pattern=north west lines,pattern color=black}] (axis cs:2.125,0) rectangle (axis cs:2.375,0.0246761052);
\draw[draw=black,fill=darkorange25512714,postaction={pattern=north west lines,pattern color=black}] (axis cs:3.125,0) rectangle (axis cs:3.375,0.167254825);
\path [draw=black, line width=1pt]
(axis cs:0.25,0)
--(axis cs:0.25,0);

\addplot [semithick, black, mark=-, mark size=1.5, mark options={solid}, only marks, forget plot]
table {%
0.25 0
};
\addplot [semithick, black, mark=-, mark size=1.5, mark options={solid}, only marks, forget plot]
table {%
0.25 0
};
\path [draw=black, line width=1pt]
(axis cs:1.25,-0.00674576463461573)
--(axis cs:1.25,0.0143404430346157);

\addplot [semithick, black, mark=-, mark size=1.5, mark options={solid}, only marks, forget plot]
table {%
1.25 -0.00674576463461573
};
\addplot [semithick, black, mark=-, mark size=1.5, mark options={solid}, only marks, forget plot]
table {%
1.25 0.0143404430346157
};
\path [draw=black, line width=1pt]
(axis cs:2.25,0.00120152041884096)
--(axis cs:2.25,0.048150689981159);

\addplot [semithick, black, mark=-, mark size=1.5, mark options={solid}, only marks, forget plot]
table {%
2.25 0.00120152041884096
};
\addplot [semithick, black, mark=-, mark size=1.5, mark options={solid}, only marks, forget plot]
table {%
2.25 0.048150689981159
};
\path [draw=black, line width=1pt]
(axis cs:3.25,0.1097362637768)
--(axis cs:3.25,0.2247733862232);

\addplot [semithick, black, mark=-, mark size=1.5, mark options={solid}, only marks, forget plot]
table {%
3.25 0.1097362637768
};
\addplot [semithick, black, mark=-, mark size=1.5, mark options={solid}, only marks, forget plot]
table {%
3.25 0.2247733862232
};
\end{axis}

\end{tikzpicture}
    \setlength\abovecaptionskip{-0.1cm}
    \label{chap4-fig:static_video_1ru1ue}
\caption{Video streaming performance with one \acrshort{ue} and single \acrshort{ru} across both static (8—close, 12—mid, 16—far) and mobile use cases.}
\label{chap4-fig:video_static}
\end{figure}

%
We place the \gls{ue} at three static locations at different distances from the \gls{ru}: location~8 (close); 12 (mid); and 16 (far). We run each video session for three minutes, streaming five distinct profiles simultaneously at various resolutions as described in Section~\ref{chap4-sec:exp-setup-overview}. We then plot the mean bitrate over five runs, as well as the rebuffer ratio, in Figure~\ref{chap4-fig:video_static}.
As expected, the average bitrate decreases and the rebuffer ratio increases as further distances between \gls{ue} and \gls{ru} are considered, transitioning from close to far static locations. We observe that the \gls{ue} can achieve a steady mean bitrate of around $180$\:Mbps in all static cases. Note that, unlike test results achieved through iPerf backlogged traffic, the mean bitrate for video streaming is lower.
The video client fetches segments in an intermittent fashion (causing flows to be short), which depends on parameters, e.g., video buffer and segment size. Because of this, throughput sometimes does not increase to the fullest during that short period of time, and the client algorithm \gls{abr} downgrades the bitrate based on the estimate it gets. This is due to a slow \gls{mcs} selection loop in the \gls{oai} L2, which will be improved as part of our future work. However, this shows that our setup is capable of supporting up to 8K \gls{hdr} videos that require $150-300$\:Mbps bitrates according to YouTube guidelines~\cite{Youtube}.
During mobility, the average bitrate is $120$\:Mbps, and the rebuffer ratio increases to $15$\%. This is once again because the \gls{ue} moves away from the \gls{ru}, gradually entering low-coverage regions and eventually disconnecting.


\subsection{\blue{Peak Performance Experiments}}
\label{chap4-sec:exp-peak}

\blue{In this second set of experiments, we expand our evaluation to stress-test the system and attain peak performance results. To achieve these compared to previous tests, we leverage a \gls{gh} \gls{ran} server with a DDDDDDSUUU \gls{tdd} pattern, a 4x4~\gls{mimo} configuration, 4~layers \gls{dl}, 1~layer \gls{ul} and a $Q_{m}$) up to 256-\gls{qam}. We compare system output with a single and double commercial \gls{ota} \glspl{ue} connected to our network at a fixed location using Open5GS as \gls{cn} and iPerf to generate traffic, as well as with the Keysight RuSIM emulator device, emulating both \gls{ru} and up to 25 \glspl{ue}, using Keysight CoreSIM to emulate the \gls{cn}. The average throughput and $95$\% confidence intervals are plotted in Figure~\ref{chap4-fig:2ue_static_rusim}.
We observe that in \gls{ota} at a fixed location, the \glspl{ue} achieve steady throughput in all cases, as indicated by the small confidence interval values, with a peak of up to $1.05$\:Gbps in \gls{dl} and $100$\:Mbps in \gls{ul} for a single \gls{ue} (\textit{1-ota}). Furthermore, the combined throughput increases with the number of connected \glspl{ue} (\textit{2-ota}), reaching a maximum of $1.2$\:Gbps in \gls{dl}, while remaining close to $100$\:Mbps in \gls{ul}.}

\blue{By using the Keysight RuSIM emulator with the same configuration as \gls{ota}, performance improves to over $1.26$\:Gbps with a single \gls{ue} (\textit{1-sim}) and $1.42$\:Gbps with two \glspl{ue} in \gls{dl} (\textit{2-sim}), and close to $110$\:Mbps in \gls{ul} for both one and two \glspl{ue}. This performance increase can be attributed to the more controlled environment provided by RuSIM, which eliminates external interference and impairments.
In this case, an \textit{ExcellentRadioConditions} channel model---also used for \gls{bs} conformance testing as specified in the \gls{3gpp}
\begin{figure}[hbt]
\centering
    \setlength\fwidth{0.9\linewidth}
    \setlength\fheight{.25\linewidth}
    \input{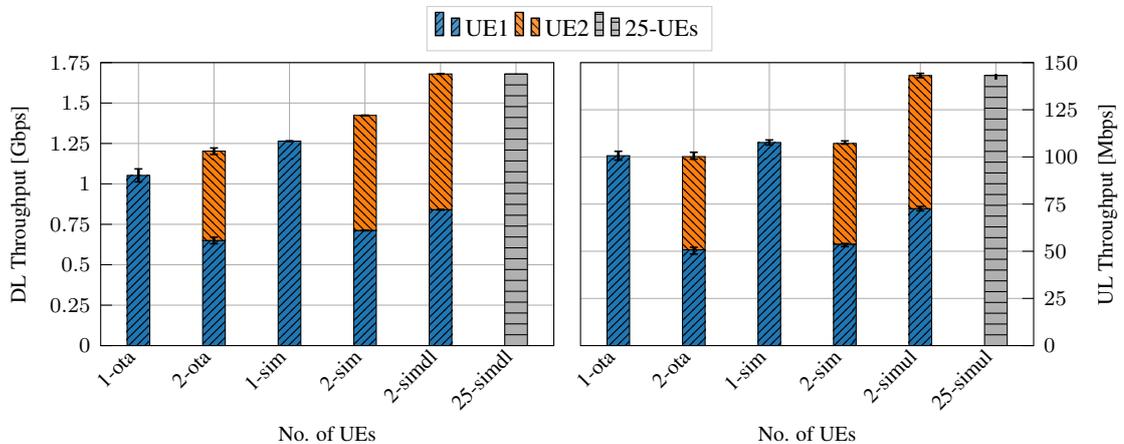}
\caption{\blue{Performance profiling to achieve peak network throughput, leveraging: one (\textit{1-ota}) and two (\textit{2-ota}) \acrshort{ota} \acrshortpl{ue} at a fixed location using iPerf and a DDDDDDSUUU \acrshort{tdd} pattern; one (\textit{1-sim}) and two (\textit{2-sim}) emulated \acrshortpl{ue} using Keysight RuSIM and CoreSIM with a DDDDDDSUUU \acrshort{tdd} pattern; and two (\textit{2-simdl, 2-simul}) and twenty-five (\textit{25-simdl, 25-simul}) emulated \acrshortpl{ue} with Keysight RuSIM and CoreSIM, a reduced number of guard symbols, a DDDDDDDSUU \acrshort{tdd} pattern for \acrshort{dl} cases, and a DDDSU \acrshort{tdd} pattern for \acrshort{ul} cases.}}
\label{chap4-fig:2ue_static_rusim}
\end{figure}
specifications~\cite{3gpptesting}---is enabled to simulate ideal radio conditions.
To achieve the current peak cell throughput, we leverage a DDDDDDDSUU \gls{tdd} pattern in \gls{dl} and a DDDSU pattern in \gls{ul}, utilizing a reduced number of guard symbols (only one) enabled by RuSIM during two separate experiment runs with two emulated \glspl{ue}. This approach results in an aggregate throughput of $1.68$\:Gbps in \gls{dl} (\textit{2-simdl}) and $143$\:Mbps in \gls{ul} (\textit{2-simul}).
Moreover, we stress-test the system by simultaneously connecting up to 25 emulated \glspl{ue} while exchanging traffic, achieving similar performance (\textit{25-simdl, 25-simul}). This demonstrates that the network can reliably sustain multiple \glspl{ue} and reaches its peak with two \glspl{ue}, while fairly distributing resources when more devices are connected.
These results highlight the maximum performance currently achievable by X5G, showcasing values comparable to those of production-level systems.}

\subsection{\blue{Long-running Experiments}}
\label{chap4-sec:exp-long}

\blue{To validate stability and reliability, we evaluate X5G through long-running experiments with a single \gls{ue} performing continuous operations. The cell configuration remains the same as in Section~\ref{chap4-sec:exp-peak}, utilizing the DDDDDDSUUU \gls{tdd} pattern, a 4x4~\gls{mimo} setup with 4~\gls{dl} layers and 1~\gls{ul} layer. The \gls{ue} is a Samsung S23 phone, which cycles randomly every $10$\:minutes between three different operations:
\begin{itemize}
    \item \gls{dl} test, a 1-minute \gls{udp} downlink iPerf data test targeting $50$\:Mbps.
    \item \gls{ul} test, a 1-minute \gls{udp} uplink iPerf data test targeting $10$\:Mbps.
    \item Disconnection, the \gls{ue} disconnects from the network, remains disconnected for the remaining $10$\:minutes, and then reconnects.
\end{itemize}
Figure~\ref{chap4-fig:ue-ops} shows the results of the long-running experiment, where the operations performed by the \gls{ue} every $10$\:minutes are represented with colored bars. The system can sustain indefinite uptime, as highlighted in the figure with over $180$\:hours of operation before the cell was manually shut down to vacate the spectrum for other planned experiments in the area.
Additionally, Figure~\ref{chap4-fig:ue-ops-gpu} presents some of the metrics available on the \gls{ran} server side,
\begin{figure}[htb]
    \centering
    \includegraphics[width=0.9\linewidth]{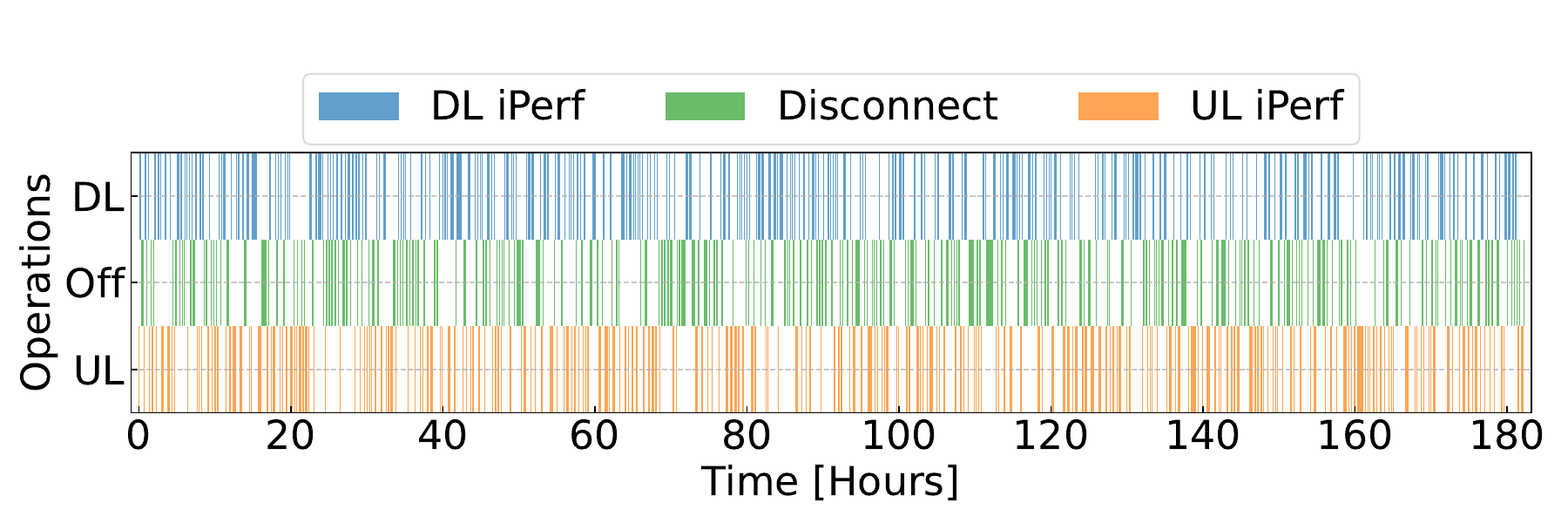}
    \caption{\blue{Long-running stability experiment involving one UE randomly cycling for over 180 hours among three operations, repeated every 10 minutes: (blue) DL iPerf for 1 minute; (orange) UL iPerf for 1 minute; and (green) disconnection from the network for the remainder of the 10-minute cycle window.}}
    \label{chap4-fig:ue-ops}
\end{figure}
\begin{figure}[htb]
    \centering
    \includegraphics[width=0.9\linewidth]{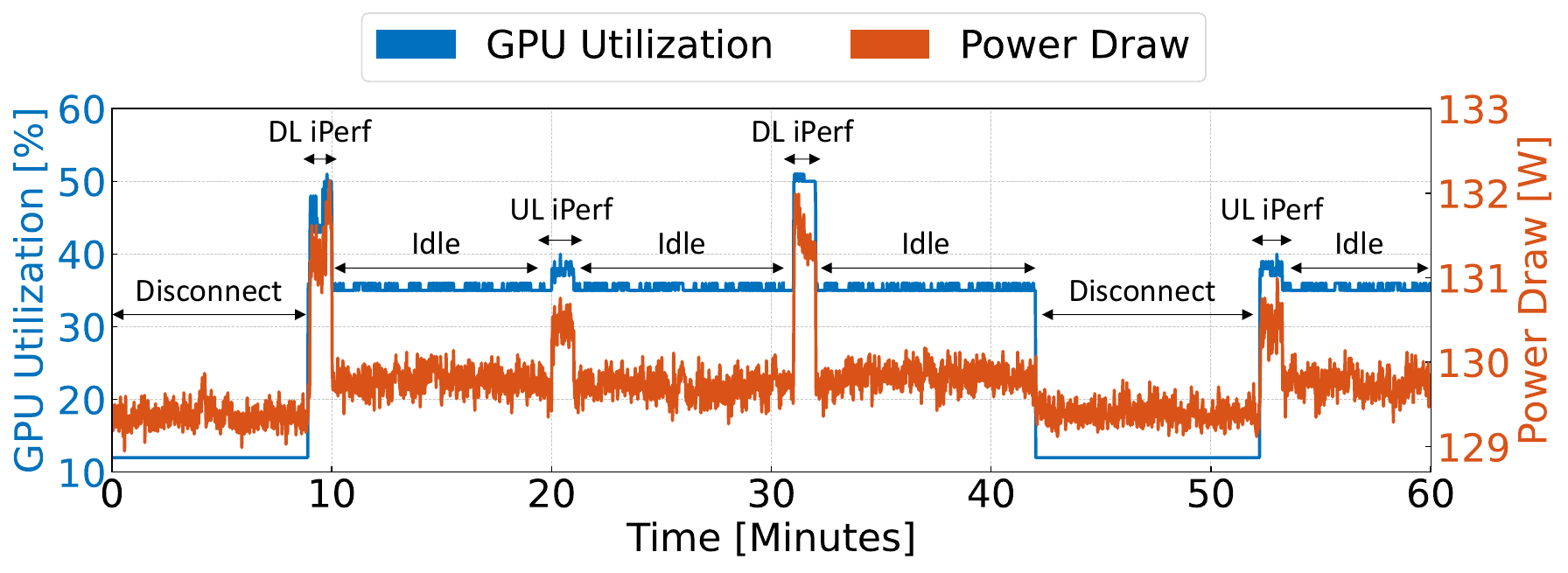}
    \caption{\blue{GPU utilization (blue) and power draw (orange) of the NVIDIA Grace Hopper server node during a one-hour window of the long-running stability experiment. The results show the behavior of the system when the \acrshort{ue} cycles through three operations: disconnecting for 10 minutes, performing a DL iPerf test for 1 minute, followed by 10 minutes of idling, and performing a UL iPerf test for 1 minute, followed by 10 minutes of idling.}}
    \label{chap4-fig:ue-ops-gpu}
\end{figure}
showing the resource utilization required to run the NVIDIA \gls{arc-ota} \gls{gnb} with a single cell on a GH200 \gls{gpu}. Specifically, \gls{gpu} utilization and power draw are depicted for a \gls{gh} server during a one-hour window of the previous long-running stability experiment. We can see how the utilization (in blue) drops to nearly 10\% when no \gls{ue} is connected, and rises to approximately 50\% during \gls{dl} data traffic, reflecting the system's computation demand. On the other hand, during idle periods and \gls{ul} communication, \gls{gpu} utilization remains stable between 35\% and 40\%, respectively. The power draw (in orange) follows a similar trend, ranging from $129$ to $132$~W. It is important to note that these results apply to a single cell, but the resource requirements for multiple cells do not scale linearly. Each \gls{gh} server can support up to 20 cells~\cite{fujitsu1} while maintaining a high-level of energy efficiency for \gls{ran} communications~\cite{kundu2024energy}.
Overall, these results highlight the high reliability of X5G in terms of both performance and stability, positioning it as a suitable candidate for \gls{p5g} deployments, as well as a valuable playground to develop, test, and evaluate novel \gls{ai}/\gls{ml} algorithms and solutions for the \gls{ran}.}

\section{GPU-Accelerated dApp Framework}
\label{chap4-sec:dapp-framework}

This section presents a GPU-accelerated framework for real-time \glspl{dapp} on an NVIDIA \gls{arc-ota} \gls{gnb}, such as X5G.
Section~\ref{chap4-sec:dapp-background} provides background on AI-native RANs and dApps, and identifies key design challenges. Section~\ref{chap4-sec:dapp-architecture} describes the framework architecture, including the real-time \gls{adl}, \gls{shm} design, and E3 interface integration. Section~\ref{chap4-sec:dapp-evaluation} evaluates the framework performance through \gls{e2e} latency benchmarks.

\subsection{dApp Background and Motivation}
\label{chap4-sec:dapp-background}

\subsubsection{AI-Native RAN Vision.}
Recent work in both standardization and industry is shifting towards an \gls{ai}-native \gls{ran} vision, where \glspl{gnb} are becoming programmable edge nodes that expose data and compute to \gls{ai} workloads. The O-RAN ALLIANCE promotes this transition through a disaggregated architecture with open interfaces and \glspl{ric} for data-driven control~\cite{polese2023understanding}, while initiatives such as the AI-RAN Alliance further refine how \gls{ai} workloads and services should be co-designed with the \gls{ran}~\cite{airan}.

\subsubsection{dApps: Real-Time Distributed Applications.}
\glspl{dapp} are lightweight applications, deployed directly on the \gls{gnb}, capable of accessing \gls{phy}/\gls{mac} telemetry and executing control actions at sub-$10$~ms timescales~\cite{doro2022dapps}. The architecture in~\cite{lacava2025dapps} defines a new E3 interface to allow real-time interactions between \gls{ran} (e.g., \gls{cu} and \glspl{du}) and \glspl{dapp}, and an \gls{e3ap} to expose structured user-plane messages and control primitives, and demonstrates feasibility with use cases such as spectrum sharing and positioning on an \gls{oai}-based \gls{gnb}. The O-RAN nGRG research report~\cite{ngrg-dapp-1} further analyzes dApp use cases and requirements.
\cite{neasamoni2025interforan} extends this line of work by embedding a GPU-accelerated dApp for \gls{ul} interference detection directly into the NVIDIA Aerial pipeline. 
In contrast, our work introduces a generic and comprehensive GPU-driven framework that exposes a shared-memory telemetry plane and structured E3 interface to support multiple data types and models, with dApps running outside the \gls{ran} process and decoupled from custom embedded kernels.

%


\subsubsection{NVIDIA Aerial Data Lake}
\label{subsec:arc-ota-datalake}
The X5G \gls{arc-ota} deployment described in Section\ref{chap4-sec:software} includes the NVIDIA \gls{adl} tool,  a data capture platform that records \gls{ota} signals and associated L2 metadata from \glspl{gnb} built on the Aerial stack, stores them in a database for offline post-processing~\cite{arc-ota}. \texttt{pyAerial} complements this by providing Python tools to query Data Lake, decode and process captured signals, and build datasets and \gls{ai} pipelines.
While \gls{adl} enables offline analysis, the dApp framework presented in this section extends it with a \gls{shm} data path to expose telemetry in real-time, as described in Section~\ref{chap4-sec:dapp-architecture}.

\subsubsection{Key Challenges and Requirements}
\label{subsec:challenges}

Enabling dApps on a GPU platform such as \gls{arc-ota} poses several key challenges, motivating the design of our \gls{gpu}-accelerated dApp framework presented in this section.

\noindent \textbf{C1: Real-time data access.} dApps require low-latency access to selected data (e.g., \gls{iq} samples, channel estimates) without the overhead and latency of extra operations and remote \glspl{ric}.

\noindent \textbf{C2: Co-location with loose coupling.} dApps should be co-located with the \gls{gnb} to meet strict timing constraints, and also remain loosely coupled so they can be deployed and updated without modifying the production \gls{ran}.

\noindent \textbf{C3: GPU-native \gls{ai} tooling.} The framework should be able to take advantage of the existing \gls{gpu} infrastructure used by the \gls{ran} and support various \gls{ai} toolchains, providing a simple path to develop and deploy new models.

\noindent \textbf{C4: Isolation and scalability.} Third-party dApps must be isolated from the \gls{ran}, with controlled access to shared-memory regions and compute resources, and the framework should support multiple concurrent dApps without degrading \gls{ran} performance.

\noindent \textbf{C5: Alignment with standardization and existing frameworks.} Interfaces and message formats should align with existing frameworks and build on E3 abstractions and serve as a reference for ongoing O-RAN, AI-RAN, and \gls{3gpp} standardization efforts on AI-native \glspl{ran} and \gls{isac} use cases.

\subsection{GPU-accelerated dApp Framework Design}
\label{chap4-sec:dapp-architecture}


This section describes our \gls{gpu}-accelerated dApp framework, including its main components and design choices.

\subsection{High-Level Architecture}



\begin{figure}[htb]
  \centering
  \includegraphics[width=\linewidth]{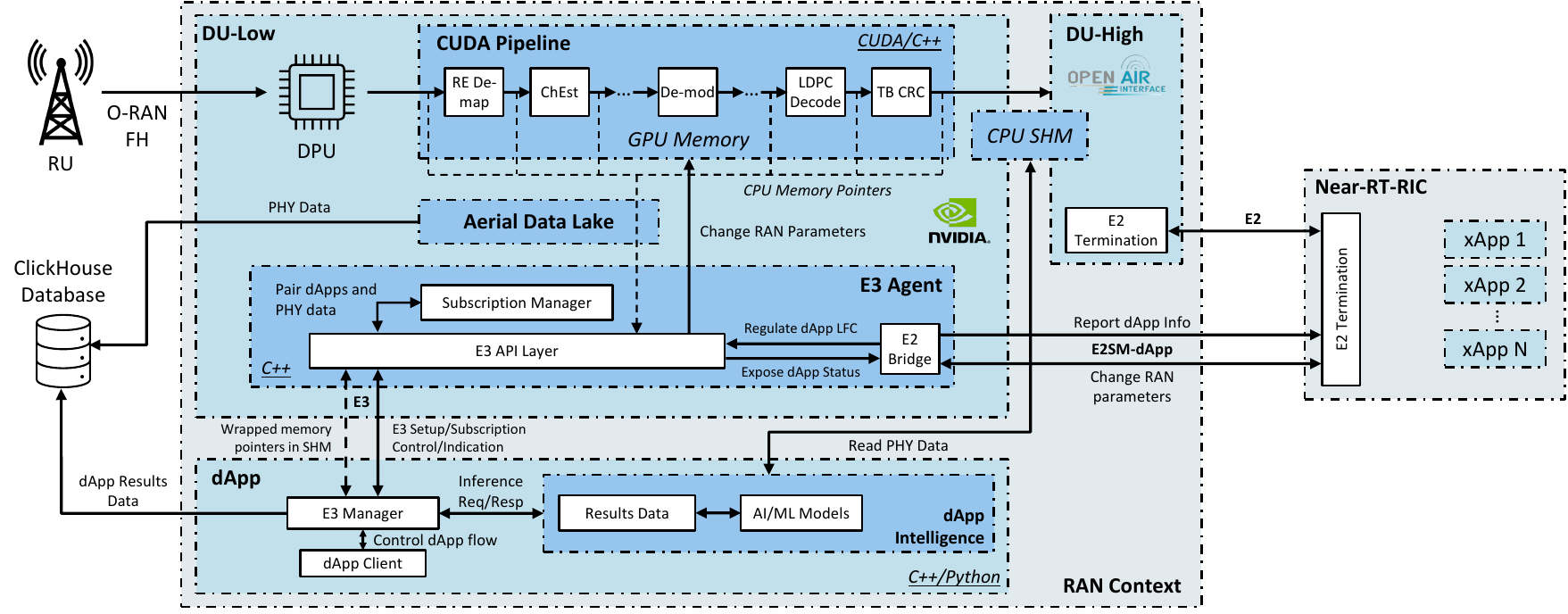}
  \caption{NVIDIA ARC-OTA dApp Integration Architecture.}
  \label{chap4-fig:high-level-diagram}
\end{figure}

Figure~\ref{chap4-fig:high-level-diagram} shows the architecture of an \gls{arc-ota} \gls{gnb}, extended to support our \gls{gpu}-accelerated dApp framework, with three main components: (i) \gls{du}-Low running the NVIDIA Aerial CUDA L1 pipeline, \gls{adl}, and an E3 Agent component; (ii) \gls{du}-High and above running \gls{oai} open-source stack; and (iii) one or more dApps.
The entire framework runs on the same physical host, requiring partitioning of host resources, such as CPU core pinning and \gls{gpu} sharing mechanisms like NVIDIA \gls{mps} or \gls{mig}, to avoid degradation of \gls{ran} performance, which remains the highest priority.

\subsubsection{Real-Time ADL and Shared Memory}
\label{subsec:rt-adl}



\begin{figure}[hbt]
  \centering
  \includegraphics[width=0.75\linewidth]{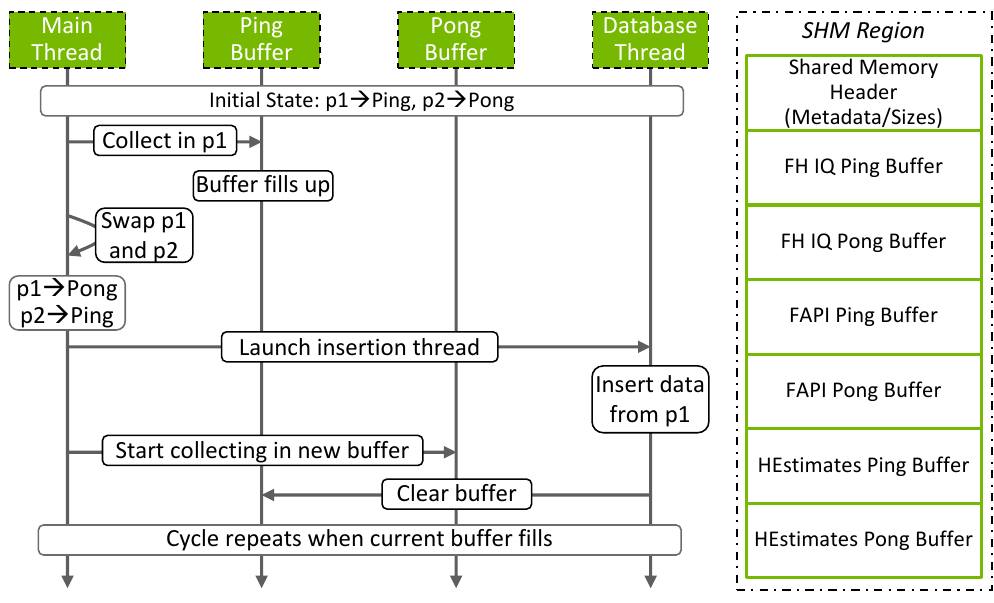}
  \caption{Aerial Data Lake ping-pong mechanism and shared memory structure.}
  \label{chap4-fig:adl-ping-pong}
\end{figure}

\textbf{Real-time Aerial Data Lake.}
To satisfy \textit{C1} (real-time data access), we extend \gls{adl}, discussed in Section~\ref{subsec:arc-ota-datalake}, to a real-time version which uses a double buffering mechanism to capture and manage incoming data efficiently. As shown in Figure~\ref{chap4-fig:adl-ping-pong}, the main thread uses two pointers ($p1$ and $p2$) that point to two buffers ($ping$ and $pong$) and alternates writes between them. When one buffer is full, the thread swaps the pointers, starts collecting into the other buffer, and triggers a database insertion thread to write the completed buffer to the ClickHouse DB. The database can be configured at start time to use in-memory RAM tables or SSD storage. Since insertion is slower than capture, this ping-pong pattern allows continuous data collection without interruptions, as long as the buffers are large enough.
%

\begin{figure}[hbt]
  \centering
  \includegraphics[width=0.7\linewidth]{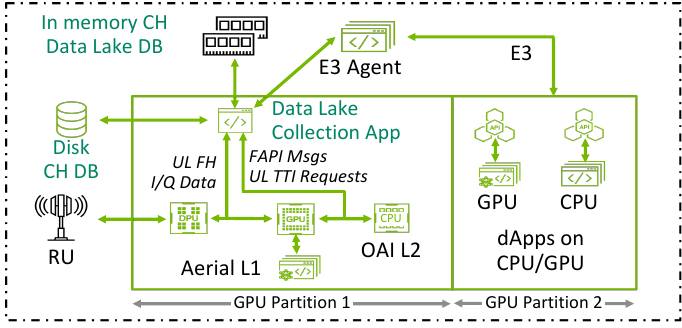}
  \caption{Data path integration between Aerial L1, Real-time ADL, shared-memory, and the E3 Agent. The steps (Op.~1–4) match the operations in Table~\ref{chap4-tab:datapath}.}
  \label{chap4-fig:adl-real-time}
\end{figure}

\textbf{E3 Agent Integration.}
We integrate an E3 Agent into the NVIDIA Aerial software, leveraging the same initial data path pipelines, triggering mechanisms (once per uplink \gls{tti}), and double-buffering abstraction of \gls{adl}.
We expose the ping-pong buffers as a POSIX shared-memory object in pinned host memory with the structure shown in Figure~\ref{chap4-fig:adl-ping-pong}, enabling direct access by external dApps.
As illustrated in Figure~\ref{chap4-fig:adl-real-time}, when \gls{adl} and/or the E3 Agent are enabled, the \gls{pusch} pipeline is instructed to copy (Op.~1) selected \gls{ul} data from device memory (GPU) into pinned host memory via an asynchronous CUDA \texttt{memcpy}. This allows the CPU to proceed with other work while the \gls{gpu} performs the transfers, minimizing the impact on the critical real-time L1 \gls{phy} path. Currently, we export \gls{iq} samples, \gls{hest}, \gls{mac} \glspl{pdu}, and a set of \gls{fapi} metadata, but additional data types can be added with minimal changes.

An atomic notification (Op.~2) from the \gls{pusch} processing then triggers the \gls{adl} routines, which write (Op.~3.1) the data from pinned host memory into the next slot of the appropriate \gls{shm} ping-pong buffer. Although these \texttt{memcpy} routines are fast, an additional copy is currently required to decouple data ownership between \texttt{cuPHY} and \gls{adl}. We plan to optimize this path by removing this extra copy in future versions, for example, by allowing the \gls{pusch} pipeline to copy directly into \gls{shm}.
If the E3 Agent is enabled, it is notified (Op.~3.2) and then packs and sends the corresponding pointers and metadata to subscribed dApps via E3 Indication messages (Op.~4). In particular, for small scalar values or metadata (e.g., slot index, cell ID), the indication carries the value directly. For larger data structures stored in \gls{shm}, the indication includes: (i) the data type (e.g., \gls{iq}, \gls{hest}); (ii) the buffer index (ping or pong); (iii) the offset within the buffer, expressed in units of \glspl{tti}; and (iv) optionally, the size of the written data. The dApp can then read the referenced memory while \gls{adl} and the E3 Agent prepare the next batch. Here, \gls{adl} and the E3 Agent act as read-write producers, while dApps act as read-only consumers.

\textbf{Shared Memory.}
This pointer-based access pattern provides zero-copy data sharing between \gls{ran} and dApps, enabled by a copy path that decouples from critical GPU signal processing pipelines. Placing the buffers in host \gls{shm} also ensures portability across different deployment configurations. While NVIDIA provides direct GPU-to-GPU mechanisms, such as NVLink for peer-to-peer access and GPUDirect RDMA for inter-node communication, these require specific hardware topologies and do not support memory sharing across \gls{mig} partitions, which present isolated memory spaces even within the same physical GPU. By routing the data through pinned host memory and exposing it via \gls{shm}, our design provides a general access path for dApps regardless of whether they execute on a different \gls{mig} partition, a separate GPU, or the CPU. This also allows the same data to be accessed safely by multiple concurrent dApps while still meeting the low-latency requirements of \gls{isac} and other sensitive applications.
While the current implementation focuses on a specific set of data types, the \gls{shm} layout is generic and can be extended with additional structures by defining new regions, up to a practical limit beyond which further additions may incur performance penalties. Additionally, buffer sizes are configurable at startup, allowing for a trade-off between the amount of accessible history and memory usage.
\gls{adl} and the E3 Agent share the same initial data paths and buffering mechanism, however, they can be enabled independently without changes to the \gls{ran}.

\subsubsection{E3 Agent and E3 Manager}
\label{subsec:e3}

\begin{figure}[htb]
  \centering
  \includegraphics[width=0.7\linewidth]{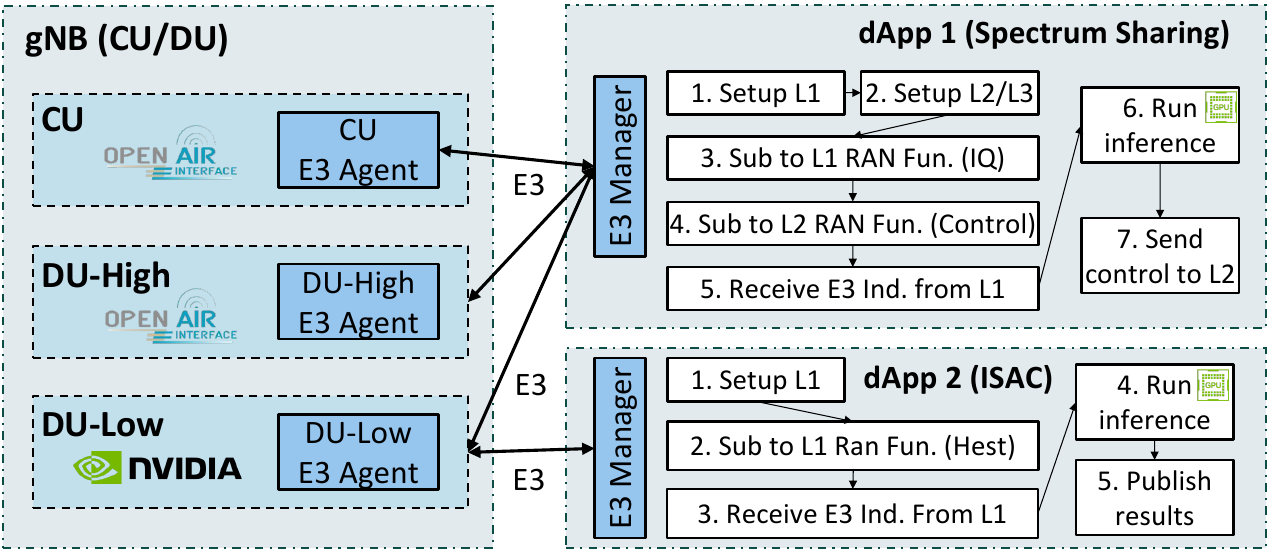}
  \caption{Support for multiple \acrshort{ran} nodes, E3 Agents, and dApps in the same gNB.}
  \label{chap4-fig:multi-e3}
\end{figure}

The E3 Agent and E3 Manager implement the communication over the recently proposed E3 interface between a \gls{ran} node and a dApp, respectively, following the \gls{e3ap} procedures described in~\cite{lacava2025dapps} of setup, subscription, indication, and control (\textit{C5: Standardization}). 
%
%
As shown in Figure~\ref{chap4-fig:multi-e3}, our framework supports multiple E3 Agents running within a \gls{gnb}, one for each \gls{ran} function (e.g., NVIDIA DU-Low, \gls{oai} DU-High).
%
An E3 Manager can independently register with multiple E3 Agents and create multiple subscriptions, for example, at different sampling rates or telemetry streams. This design enables system scalability (\textit{C4: Scalable}), allowing additional agents to be added without modifying existing dApps and for multiple dApps to subscribe to the same telemetry at different rates.
%
%
Figure~\ref{chap4-fig:multi-e3} shows two dApp examples: a spectrum sharing dApp that subscribes to L1 telemetry and sends control messages to L2 for scheduling decisions, and an \gls{isac} dApp like cuSense that only consumes channel estimates for inference without closing the control loop.

\subsubsection{dApp Reference Architecture}
\label{subsec:dapp-container}





\begin{figure}[hbt]
  \centering
  \includegraphics[width=0.7\linewidth]{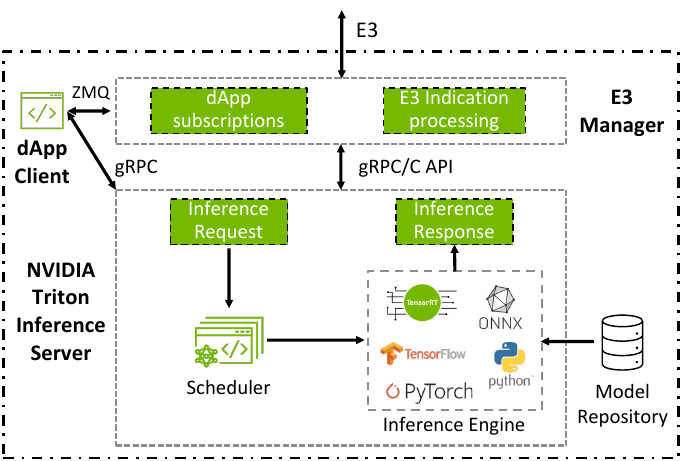}
  \caption{dApp container reference architecture with E3 Manager, NVIDIA Triton Inference Server, and dApp Client.}
  \label{chap4-fig:dapp-container}
\end{figure}

To address \textit{C3 (GPU-native \gls{ai} tooling)} and \textit{C4 (isolation and scalability)}, we design a reference dApp container architecture with three main components, as shown in Figure~\ref{chap4-fig:dapp-container}: (i) an E3 Manager, the required dApp orchestrator; (ii) an NVIDIA Triton Inference Server~\cite{triton}, our design choice for inference tasks; and (iii) a dApp client, an optional application controller. This design serves as a concrete blueprint for deploying a \gls{gpu}-accelerated dApp within a single container. 

\textbf{E3 Manager.}
The E3 Manager is a mandatory component that serves as the dApp orchestrator, interacting with the \gls{ran} nodes via the E3 interface. It manages all communication with the E3 Agent (e.g., E3 Setup and Subscription procedures) and with the dApp Client to implement the desired behavior, including application-specific \glspl{sm}. Additionally, it processes received E3 Indications by: (i) mapping the logical buffer information into the corresponding \gls{shm} locations; (ii) validating the presence and format of all inputs required by the selected \gls{ai} model; (iii) preparing and sending the inference request to Triton; and (iv) post-processing the inference results for possible control actions or external reporting.

\textbf{NVIDIA Triton Inference Server.}
In our reference design, each dApp container hosts an instance of Triton~\cite{triton}, an open-source inference serving software developed by NVIDIA and optimized for \gls{gpu}-accelerated inference. Developers can replace it with alternative inference engines or even lightweight \gls{cpu}-only scripts, while still using the same dApp architecture. It exposes a model repository with models packaged as standard Triton directories and supports different backends running on \gls{cpu} or \gls{gpu}, including Python, PyTorch (LibTorch), \gls{onnx} Runtime, and \gls{trt}. The E3 Manager and dApp client interact with Triton via its \gls{grpc}/\gls{http} or C \glspl{api} to handle model life cycles, consume memory pointers, schedule inference requests, and retrieve inference responses. Triton also allows combining different backends or running multiple models in parallel or in sequence, providing a flexible playground for benchmarking and experimentation.

\textbf{dApp client.}
The dApp client is an optional user-level service that controls the overall dApp flow. It implements the application-specific logic by: (i) querying and managing Triton models via its \gls{grpc} \glspl{api} (e.g., selecting the model and backend to use); (ii) configuring the E3 Manager with subscription options, including the target \gls{ran} node, a list of \gls{ran} functions, Triton model to run, polling interval, duration, and \gls{sm}; and (iii) performing additional post-processing of the results and triggering external events or control actions when needed. In some configurations, developers can choose to include all or part of the dApp client logic within the E3 Manager component.

This dApp design enables flexibility, as shown in Section~\ref{chap4-sec:dapp-evaluation}, and simplifies the modular deployment of different inference pipelines, as demonstrated by the cuSense \gls{isac} use case in Section~\ref{chap5-sec:cusense}. The source code will be released as open source, with documentation, guides, and inline code comments, providing developers with reference points for developing and deploying their own dApp use cases.

\subsection{Profiling dApp Framework Performance}
\label{chap4-sec:dapp-evaluation}


We now evaluate the performance of the proposed framework, integrated with NVIDIA \gls{arc-ota}, in terms of \gls{e2e} control-loop latency, overhead on the \gls{ran} pipeline, and model backend flexibility. All experiments share the same \gls{arc-ota} setup and traffic configuration, and employ a simple reference dApp.

\subsection{ARC-OTA Experimental Setup}
\label{subsec:setup}


We perform our integration and evaluation on a standard \gls{arc-ota} node as described in the documentation~\cite{arc-ota}. It consists of: (i) a single NVIDIA GH200 \gls{gh} server for the \gls{ran} workloads, featuring a 72-core Grace CPU, an NVIDIA H100 Tensor Core \gls{gpu}, two BlueField-3 \glspl{dpu}, and two ConnectX-7 \glspl{nic}; (ii) a VIAVI Qulsar QG-2 device as grandmaster synchronization clock; (iii) an NVIDIA Spectrum-2 SN3750-SX as \gls{fh} switch; (iv) a 4T4R Foxconn \gls{cbrs} \gls{fr1} \gls{ru} centered at $3.65$~GHz with $100$~MHz of bandwidth; and (v) a Samsung S23 acting as \gls{ue}.
The complete \gls{gnb} stack, comprising NVIDIA Aerial L1, \gls{oai} L2 and above, the dApp framework, and \gls{oai} \gls{cn}, runs within the GH200 node. The server used for the experiments presented here is not \gls{mig}-partitioned, but we have also validated the dApp framework on \gls{mig}-partitioned servers with other L2+ stacks like \gls{odc}. CPU cores are pinned separately for L1/L2 and dApp workloads, and the \gls{gpu} is shared using CUDA \gls{mps}.
%

%

\subsection{Framework Benchmarks}

To evaluate the performance characteristics of our framework, we conduct comprehensive benchmarking experiments measuring \gls{e2e} control-loop latency and the impact of different model backends.
These experiments use a simple dApp, \texttt{iq\_processor}, a \gls{phy}-layer telemetry processing application that is representative of use cases such as interference detection, spectrum monitoring, and \gls{isac}.

\textbf{Reference dApp Model: \texttt{iq\_processor}.}
Our benchmark model \texttt{iq\_processor} accepts as input a tensor of \gls{iq} samples with dimensions $[4, 14, 273, 12, 2]$, corresponding to $4$ antenna ports, $14$ \gls{ofdm} symbols, $273$ \glspl{prb}, $12$ subcarriers per \gls{prb}, and $2$ values (the $I$ and $Q$ components) in \gls{fp16} format. This corresponds to one full \gls{ul} slot of $100$~MHz bandwidth data with numerology~$\mu=1$.
The model computes the average power per \gls{prb} using
\begin{equation}
P_{\text{PRB}}(p) = \frac{1}{N_a N_s N_{sc}} \sum_{a=1}^{N_a} \sum_{s=1}^{N_s} \sum_{k=1}^{N_{sc}} \left( I^2_{a,s,p,k} + Q^2_{a,s,p,k} \right),
\end{equation}
where $P_{\text{PRB}}(p)$ is the average power for \gls{prb} $p \in \{1, ..., 273\}$, $N_a~=~4$ number of antennas, $N_s~=~14$ the number of \gls{ofdm} symbols, $N_{sc}~=~12$ the number of subcarriers per \gls{prb}, and $I_{a,s,p,k}$ and $Q_{a,s,p,k}$ are the \gls{iq} components. The output is a $273$-element \gls{fp32} vector containing the power measurements for each \gls{prb}.

\textbf{End-to-end Complete Control Loop.}
%
Table~\ref{chap4-tab:datapath} shows the complete control-loop data path, from the time an \gls{ul} slot completes processing on the \gls{gpu} pipeline, through the dApp routines, to the application of a control decision at the \gls{ran}. The communication between the E3 Agent and the E3 Manager is implemented with \gls{zmq}, a high-performance asynchronous messaging library for distributed systems that provides efficient request/reply and publish/subscribe socket patterns.
The framework currently contributes a fixed average overhead of approximately $495~\mu\mathrm{s}$ plus the model-dependent inference time $\delta$ (Operation~7). Operations~1–4 ($\approx 130~\mu\mathrm{s}$) represent the data collection phase involving the \mbox{\gls{gpu}$\rightarrow$\gls{cpu}$\rightarrow$\gls{shm}} transfers and atomic notifications.
Operations~5-8 correspond to the inference pipeline, measured as \emph{client latency} in our backend comparison analysis. This phase includes interactions with Triton, \gls{shm} access, and tensor preparation, and model inference time. For simplicity, we currently use gRPC as the communication protocol between the E3 Manager and Triton. This choice simplifies integration but dominates the overall framework overhead. We plan to replace gRPC with Triton C-based \glspl{api} to further reduce latency.
Finally, Operations~9-10 close the loop by sending an optional control message back to the E3 Agent and applying it. In this benchmark, the control logic is not implemented, so the additional processing beyond \gls{zmq} messaging is excluded.


\begin{table}[htb]
    \centering
    \caption{Data messaging path and cumulative latency for a single dApp end-to-end control-loop iteration.}
    \label{chap4-tab:datapath}
    \scriptsize
    \begin{tabular}{lccc}
        \toprule
        \textbf{Operation} & \textbf{Protocol} & \textbf{Overhead [$\mu\mathrm{s}$]} & \textbf{Total [$\mu\mathrm{s}$]} \\
        \midrule
        1. cuPHY copies data from GPU to CPU & memcpy & 35 & 35 \\
        2. cuBB notifies ADL data is ready & Atomic & 30 & 65 \\
        3. ADL copies data from CPU to SHM & memcpy & 50 & 115 \\
        4. E3 Agent sends pointers to E3 Manager & ZMQ & 15 & 130 \\
        5. E3 Manager sends request to Triton & gRPC & 150 & 280 \\
        6. Triton accesses and prepares input data & SHM & 50 & 330 \\
        7. AI Model performs inference & CUDA & $\delta$ & 330 + $\delta$ \\
        8. Triton sends results to E3 Manager & gRPC & 150 & 480 + $\delta$ \\
        9. E3 Manager sends control to E3 Agent & ZMQ & 15 & 495 + $\delta$ \\
        10. E3 Agent receives and applies control & API & $\sim$0 & 495 + $\delta$ \\
        \bottomrule
        \multicolumn{4}{l}{\parbox{0.85\linewidth}{\raggedright\small\textit{Note:} $\delta$ represents the model-dependent inference time, which varies with model complexity and backend as shown in Figure~\ref{chap4-fig:backend-results}.}} \\
    \end{tabular}
\end{table}

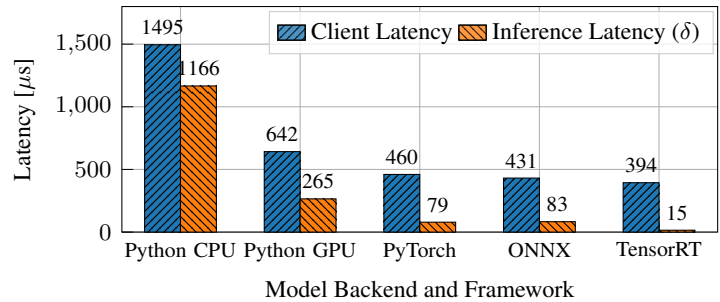
\begin{figure}[hbt]
    \centering
    \setlength\fwidth{.65\linewidth}
    \setlength\fheight{.3\linewidth}
    \begin{tikzpicture}
\pgfplotsset{every tick label/.append style={font=\footnotesize}}
\definecolor{darkgray176}{RGB}{176,176,176}
\definecolor{darkorange25512714}{RGB}{255,127,14}
\definecolor{lightgray204}{RGB}{204,204,204}
\definecolor{steelblue31119180}{RGB}{31,119,180}
\begin{axis}[
width=0.951\fwidth,
height=\fheight,
at={(0\fwidth,0\fheight)},
legend cell align={left},
legend columns=2,
legend style={fill opacity=0.8, 
    draw opacity=1, 
    text opacity=1, 
    draw=lightgray204, 
    font=\footnotesize,
    at={(0.62, 0.98)},
    anchor=north},
legend image post style={xscale=0.6},
x grid style={darkgray176},
xmin=-0.49, xmax=4.49,
xtick style={color=black},
xtick={0,1,2,3,4},
xticklabel style={font=\scriptsize, text width=2.5cm, align=center},
xticklabels={
    Python CPU,
    Python GPU,
    PyTorch,
    ONNX,
    TensorRT
},
xtick pos=bottom,
y grid style={darkgray176},
ylabel={Latency [$\mu$s]},
xlabel={Model Backend and Framework},
ylabel style={font=\footnotesize},
xlabel style={font=\footnotesize},
ymin=0, ymax=1800,
ytick pos=left,
ytick style={color=black},
xmajorgrids,
ymajorgrids,
bar width=0.35cm,
]
\addlegendimage{area legend,draw=black,fill=steelblue31119180,postaction={pattern=north east lines,pattern color=black}}
\addlegendentry{Client Latency}
\draw[draw=black,fill=steelblue31119180,postaction={pattern=north east lines,pattern color=black}] 
    (axis cs:-0.3,0) rectangle (axis cs:0,1495);
\draw[draw=black,fill=steelblue31119180,postaction={pattern=north east lines,pattern color=black}] 
    (axis cs:0.7,0) rectangle (axis cs:1,642);
\draw[draw=black,fill=steelblue31119180,postaction={pattern=north east lines,pattern color=black}] 
    (axis cs:1.7,0) rectangle (axis cs:2,460);
\draw[draw=black,fill=steelblue31119180,postaction={pattern=north east lines,pattern color=black}] 
    (axis cs:2.7,0) rectangle (axis cs:3,431);
\draw[draw=black,fill=steelblue31119180,postaction={pattern=north east lines,pattern color=black}] 
    (axis cs:3.7,0) rectangle (axis cs:4,394);
\addlegendimage{area legend,draw=black,fill=darkorange25512714,postaction={pattern=north west lines,pattern color=black}}
\addlegendentry{Inference Latency ($\delta$)}
\draw[draw=black,fill=darkorange25512714,postaction={pattern=north west lines,pattern color=black}] 
    (axis cs:0,0) rectangle (axis cs:0.3,1166);
\draw[draw=black,fill=darkorange25512714,postaction={pattern=north west lines,pattern color=black}] 
    (axis cs:1,0) rectangle (axis cs:1.3,265);
\draw[draw=black,fill=darkorange25512714,postaction={pattern=north west lines,pattern color=black}] 
    (axis cs:2,0) rectangle (axis cs:2.3,79);
\draw[draw=black,fill=darkorange25512714,postaction={pattern=north west lines,pattern color=black}] 
    (axis cs:3,0) rectangle (axis cs:3.3,83);
\draw[draw=black,fill=darkorange25512714,postaction={pattern=north west lines,pattern color=black}] 
    (axis cs:4,0) rectangle (axis cs:4.3,15);
\node[font=\scriptsize, above] at (axis cs:-0.15,1495) {1495};
\node[font=\scriptsize, above] at (axis cs:0.15,1166) {1166};
\node[font=\scriptsize, above] at (axis cs:0.85,642) {642};
\node[font=\scriptsize, above] at (axis cs:1.15,265) {265};
\node[font=\scriptsize, above] at (axis cs:1.85,460) {460};
\node[font=\scriptsize, above] at (axis cs:2.15,79) {79};
\node[font=\scriptsize, above] at (axis cs:2.85,431) {431};
\node[font=\scriptsize, above] at (axis cs:3.15,83) {83};
\node[font=\scriptsize, above] at (axis cs:3.85,394) {394};
\node[font=\scriptsize, above] at (axis cs:4.15,15) {15};
\end{axis}
\end{tikzpicture}
    \caption{Client latency (Operations~5-8 of Table~\ref{chap4-tab:datapath}) and inference latency ($\delta$) for the \texttt{iq\_processor} model across different backends.}
    \label{chap4-fig:backend-results}
\end{figure}

\textbf{Model Backend Performance Comparison.}
%
%
To assess the flexibility of the Triton-based architecture, we deploy the same \texttt{iq\_processor} model with five different backends: (i) Python/NumPy as CPU baseline; (ii) \gls{gpu}-accelerated Python using torch tensor operations; (iii) PyTorch/LibTorch with TorchScript-compiled execution; (iv) \gls{onnx} Runtime for cross-platform deployment; and (v) \gls{trt} for NVIDIA-optimized inference with \gls{fp16} precision.
We benchmark these models using Triton \textit{perf\_analyzer}, a tool that generates continuous inference requests and reports both the pure mode inference time ($\delta$) and the \gls{e2e} client latency (Operations~5-8 in Table~\ref{chap4-tab:datapath}). The results are shown in Figure~\ref{chap4-fig:backend-results}. The measurements highlight a substantial reduction in inference latency from more than $1$~ms for the CPU baseline to just $16~\mu\mathrm{s}$ with the optimized \gls{trt} backend, with all \gls{gpu}-based backends achieving sub-millisecond latency.
On the other hand, the difference between client and inference times remains roughly constant across backends (around $440$–$480~\mu\mathrm{s}$), reflecting the framework overhead due to gRPC-based communication and Triton internal request handling.
Combining these client latencies with the overheads of approximately $315~\mu\mathrm{s}$ (Operations~1-4 and 9-10) results in full \gls{e2e} control-loop latencies ranging from approximately $1.8$~ms for the CPU baseline down to around $710~\mu\mathrm{s}$ for \gls{trt} backend. While these values are obtained from our benchmarks and may vary slightly with different dApps and subscription configurations, they provide representative reference points.

Overall, these results validate the ability of our framework to meet real-time dApp requirements, with end-to-end control-loop latencies well below $10$~ms and, for GPU-accelerated backends, even below $1$~ms. At the same time, it provides a flexible and performant playground for developers to experiment, prototype, and deploy a variety of applications without any architectural change.

%

\section{Related Work}
\label{chap4-sec:related-work}

This section compares the features and capabilities of the X5G testbed within the context of similar programmable open RAN and \gls{5g}, highlighting its unique features and contributions beyond \gls{5g} research and experimentation. Surveys of testbeds for open and programmable wireless networks can also be found in~\cite{bonati2020open,polese2023understanding}.

The \gls{pawr}~\cite{pawr} offers a set of geographically and technically diverse testbeds designed to enhance specific wireless communication areas. These include POWDER, AERPAW, COSMOS, ARA, and Colosseum, each equipped with specialized technologies to address varied research needs.

The POWDER facility, located at the University of Utah in Salt Lake City, UT, supports a wide spectrum of research areas, including next-generation wireless networks and dynamic spectrum access~\cite{breen2020powder}. Its \gls{5g} stack is based primarily on a combination of open-source stacks, combined with \glspl{sdr} or \glspl{ru} but not accelerated at the physical layer, and on a commercial Mavenir system, which does not support access to the source code from the \gls{phy} to the core network, differently from the X5G stack. 

Similarly, AERPAW, deployed on the campus of North Carolina State University in Raleigh, NC, focuses on aerial and drone communications, diverging from our emphasis on private \gls{5g} network configurations~\cite{panicker2021aerpaw}. The AERPAW facility hosts an Ericsson \gls{5g} deployment with similar limitations with respect to stack programmability for research use cases.

The COSMOS project~\cite{chen2023open-access} leverages an array of programmable and software-defined radios, including USRP and Xilinx RFSoC boards, to facilitate mmWave communication experiments across a city-scale environment. The outdoor facilities of COSMOS are deployed in the Harlem area, in New York City, while its indoor wireless facilities are on the Rutgers campus in North Brunswick, NJ. Unlike X5G, COSMOS is designed for broad academic and industry use and is more focused on mmWave deployments enabling diverse external contributions to its development without specific emphasis on any single network architecture.

The ARA testbed \cite{zhang2021ara}, deployed across Iowa State University (ISU), in the city of Ames, and surrounding rural areas in central Iowa, serves as a large-scale platform for advanced wireless research tailored to rural settings. ARA includes diverse wireless platforms ranging from low-UHF massive \gls{mimo} to mmWave access, long-distance backhaul, free-space optical, and \gls{leo} satellite communications, utilizing both \gls{sdr} and programmable \gls{cots} platforms and leveraging open-source software like \gls{oai}, srsRAN, and SD-RAN~\cite{ARASDR}. However, unlike the X5G testbed,
ARA focuses primarily on rural connectivity without focusing on specialized hardware for \gls{phy} layer optimization or digital twin frameworks for \gls{rf} planning.

Colosseum is the world's largest Open RAN digital twin~\cite{villa2024dt,polese2024colosseum}. This testbed allows users to quickly instantiate softwarized cellular protocol stacks, e.g., the \gls{oai} one, on its 128~compute nodes. These nodes control 128~\glspl{sdr} that are used as \gls{rf} front-ends and are connected to a massive channel emulator, which enables experimentation in a variety of emulated \gls{rf} environments. However, the Colosseum servers are not equipped to offload lower-layer cellular operations on \glspl{gpu}, and the available \glspl{sdr} are USRP~X310 from NI, instead of commercial \glspl{ru}. 

The \gls{osc} is also involved in the creation of laboratory facilities~\cite{SoAOSC} that comply with O-RAN standards and support the testing and integration of O-RAN-compliant components. These testing facilities, distributed across multiple laboratories, foster a diverse ecosystem through their commitment to open standards and collaborative development. However, unlike X5G, they do not explicitly focus on the deployment complexities of private networks, nor do they provide any \gls{phy} layer acceleration technology or utilize a digital twin for \gls{rf} planning. Instead, they aim to promote multi-vendor interoperability within an open collaborative framework.


6G-SANDBOX~\cite{6GSANDBOX} is a versatile facility that includes four geographically displaced platforms in Europe, each equipped to support a variety of advanced wireless technologies and experimental setups. It uses a mix of commercial solutions (for example, Nokia microcells, Ericsson \gls{bbu}, and the Amarisoft stack) and open source solutions (for example, \gls{oai} and srsRAN) in diverse environments ranging from urban to rural settings. Unlike X5G, 6G-SANDBOX primarily facilitates wide-ranging 6G research through its extensive, multi-location infrastructure. Its predecessor, 5GENESIS~\cite{5Genesis}, featured a modular and flexible experimentation methodology, supporting both per-component and \gls{e2e} validation of \gls{5g} technologies and \gls{kpi} across five European locations. This testbed emphasizes a comprehensive approach to \gls{5g} performance assessment, integrating diverse technologies such as \gls{sdn}, \gls{nfv}, and network slicing to enable rigorous testing of vertical applications but not including O-RAN architectures.


The \gls{oaic} testbed~\cite{oaic}, developed at Virginia Tech, is an open-source \gls{5g} O-RAN-based platform designed to facilitate AI-based \gls{ran} management algorithms. It includes the \gls{oaic}-Control framework for designing AI-based \gls{ran} controllers and the \gls{oaic}-Testing framework for automated testing of these controllers. The \gls{oaic} testbed introduces a new real-time \gls{ric}, zApps, and a Z1 interface to support use cases requiring latency under $10$\:ms, integrated with the CORNET infrastructure for remote accessibility.


The CCI xG Testbed provides a comprehensive platform for advanced wireless research, particularly in the realm of \gls{5g} and beyond. It features a disaggregated architecture with multiple servers distributed across geographically disparate cloud sites, leveraging a combination of central and edge cloud infrastructures to optimize resource allocation and latency. The testbed includes several \gls{sdr}-based \gls{cbrs} Base Station Device (CBSD) integrated with an open-source \gls{sas} for dynamic spectrum sharing in the \gls{cbrs} band~\cite{CCI1}. Additionally, the testbed supports a full O-RAN stack using srsRAN and Open5GS and features both non-RT \gls{ric} and near-RT \gls{ric} for real-time and non-real-time radio resource management~\cite{CCI2,CCI3}.

The testbed in~\cite{NEC} provides a prototypical environment designed to experiment with vRAN deployments and evaluate resource allocation and orchestration algorithms. It focuses on the decoupling of radio software components from hardware to facilitate efficient and cost-effective \gls{ran} deployments. This testbed includes datasets that characterize computing usage, energy consumption, and application performance, which are made publicly available to foster further research. Unlike the X5G testbed, the O-RAN platform primarily addresses the flexibility and cost efficiency of virtualized RANs without incorporating specialized hardware for PHY layer tasks.

\blue{The disaggregated 5G testbed for live audio production~\cite{ReviewSoA1} emphasizes ultra-reliable low-latency communication for media applications. Its scope is narrower than X5G, which supports a broader range of experimental scenarios and computationally intensive network configurations.}

\blue{The data usage control framework~\cite{ReviewSoA2} addresses privacy challenges in hybrid private-public 5G networks. While it highlights the importance of secure orchestration and policy management, its focus on analytics differs from X5G capabilities in physical layer acceleration and network performance experimentation.}

\blue{Finally, the Microsoft enterprise-scale Open RAN testbed~\cite{bahl2023accelerating} highlights the potential of virtualized RAN functions on commodity servers, employing disaggregated architectures to demonstrate scalability and flexibility. By integrating Kubernetes for dynamic orchestration and using Intel FlexRAN with ACC100 accelerators for \gls{ldpc} look-aside offloading, this testbed achieves functional disaggregation of RAN workloads.}

Regarding real-time programmability, recent work extends O-RAN toward real-time and user-plane control, building on initial visions and implementations~\cite{doro2022dapps,ngrg-dapp-1,lacava2025dapps}. Follow-up work demonstrates dApps for CPU power management and interference detection on GPU-accelerated \gls{5g} \glspl{phy}~\cite{crespo2025energy,neasamoni2025interforan}. Janus takes a complementary approach, embedding eBPF-based “codelets” directly into vRAN functions to enable flexible telemetry and real-time control with strict safety guarantees~\cite{xenofonjanus}, while \cite{oaic} zApps target similar real-time functionality below the $10$~ms timescale at the \gls{ric} level. Compared to these systems, the dApp framework presented 
in Section~\ref{chap4-sec:dapp-framework} focuses on \gls{phy} telemetry on \gls{gpu} on a production-grade \gls{arc-ota} deployment and uses dApps to implement intelligent GPU-accelerated applications by creating complete \gls{e2e} pipelines for generic real-time data access and consumption, while keeping dApps outside the \gls{gnb} process and decoupled from vendor-specific \gls{phy} kernels.

\blue{While state-of-the-art software stacks such as srsRAN already offer similar performance in terms of core \gls{5g} functionalities, including handovers, X5G distinguishes itself through its integration of \gls{gpu} acceleration, enabling enhanced flexibility and computational power for future innovations. Unlike traditional platforms that rely on \gls{cpu}-based architectures, which achieve performance parity for standard \gls{ran} tasks, it leverages \glspl{gpu} not only for optimized \gls{phy} processing but also as a unified platform for \gls{ai}/\gls{ml} workloads. Indeed, the \gls{gpu} architecture of X5G supports the development and deployment of dApps~\cite{lacava2025dapps} that utilize \gls{ai}/\gls{ml} models for real-time network optimization. This capability aligns directly with the vision outlined by the AI-RAN Alliance~\cite{airan}, which emphasizes the integration of \gls{ai}-driven decision-making processes across three key development areas: (i) AI-for-RAN, (ii) AI-and-RAN, and (iii) AI-on-RAN, making our platform an ideal candidate for advancing these areas. Moreover, the modular design of X5G guarantees compatibility with both open-source and commercial cores, facilitating future experiments with advanced technologies like massive \gls{mimo}, mmWave, and beamforming that are currently under development.}

\section{Conclusions and Discussion}
\label{chap4-sec:conclusions}

This chapter introduced X5G, an open, programmable, and multi-vendor private 5G O-RAN testbed deployed at Northeastern University in Boston, MA.
We demonstrated the integration of NVIDIA Aerial, a \gls{phy} layer implementation on \glspl{gpu}, with higher layers based on \gls{oai}, resulting in the NVIDIA \gls{arc-ota} platform.
We provided an overview of the \gls{arc-ota} software and hardware implementations, designed for a multi-node deployment, including a Red Hat OpenShift cluster for the \gls{osc} \gls{ric} deployment, as well as examples of a \gls{kpm} xApp and a slicing xApp.
Additionally, we conducted a ray-tracing study using our \gls{dt} framework to determine the optimal placement of X5G \glspl{ru} in an indoor space.
Finally, we discussed platform performance with varying numbers of \gls{cots} and emulated \glspl{ue} and applications, such as iPerf and video streaming, as well as through long-running and stress-test experiments to evaluate its stability.

Beyond the core infrastructure, we designed, implemented, and evaluated a GPU-accelerated framework for real-time O-RAN dApps on NVIDIA \gls{arc-ota} that exposes \gls{phy}/\gls{mac} telemetry via shared memory and an E3-based interface. Coupled with NVIDIA Triton-based inference pipelines running on the same GPU as the gNB, the framework enables complete end-to-end control loops with less than $500~\mu\mathrm{s}$ overhead. This design addresses the key challenges of real-time data access, co-location with loose coupling, \gls{gpu}-native \gls{ai} tooling, isolation and scalability, and alignment with ongoing O-RAN and AI-RAN standardization efforts.

Similar to Colosseum explained in Chapter~\ref{chap:2}, X5G and the dApp framework address the key research challenges outlined in Chapter~\ref{chap:intro} for experimental wireless platforms. From a \emph{deployment} perspective, we designed and deployed a multi-vendor platform following disaggregated O-RAN architectures, achieving production-level stability, with the testbed operating continuously for extended periods. The platform provides \emph{realism} by validating an \gls{e2e}, 3GPP-compliant private 5G network with commercial \glspl{ue}, capturing hardware impairments and \gls{rf} propagation effects that emulation cannot fully reproduce. In terms of \emph{usability}, X5G provides reference design architectures and experimental pipelines for data collection and testing, including the dApp framework with its E3 interface and Triton-based inference integration. Finally, the platform supports a broad range of \emph{use cases}, including interference mitigation, network slicing, security, \gls{isac}, and a wide range of 5G/6G research applications---some of which are demonstrated in Chapter~\ref{chap:5}.
Together with the digital twin platforms presented in Chapters~\ref{chap:2} and~\ref{chap:3}, X5G takes the experimental idea journey one step further by providing a comprehensive methodology from controlled emulation to real-world validation.

\chapter{Intelligent RAN Control and Security on Physical Platforms}
\label{chap:5}

\section{Introduction}
\label{chap5-sec:intro}


Chapters~\ref{chap:2} and~\ref{chap:3} demonstrated how large-scale emulation and \gls{dt} platforms such as Colosseum, combined with \gls{cast}, can be used to prototype spectrum sharing, \gls{ai}-driven propagation modeling, and synthetic \gls{rf} data generation in a controlled environment, while Chapter~\ref{chap:4} introduced X5G as a production-grade private 5G O-RAN testbed that brings these ideas into the physical world. With X5G, we now have a stable, GPU-accelerated, end-to-end \gls{5g} platform with commercial \glspl{ue}, multi-vendor \glspl{ru}, and integrated O-RAN control, capable of supporting long-running experiments under realistic propagation and hardware conditions. This chapter leverages X5G and its dApp framework to explore how intelligent control and security mechanisms can be realized and validated \gls{ota} on a physical platform.

X5G provides an open and programmable platform that spans both the data and the control plane. On one hand, the GPU-accelerated dApp framework introduced in Section~\ref{chap4-sec:dapp-framework} allows \gls{ai}-native applications to run co-located with the \gls{gnb}, operating directly on \gls{phy}/\gls{mac} telemetry with sub-millisecond latency. On the other hand, the \gls{osc} near-real-time \gls{ric} deployed on X5G enables xApps to control \gls{ran} behavior at timescales of tens to hundreds of milliseconds. Additionally, the platform is also leveraged to study the security of open interfaces in realistic a deployment. Together, these capabilities enable X5G to expand the research possibilities from intelligent \gls{ran} control to security fields, where solutions can be evaluated under real \gls{rf} propagation, hardware constraints, and dynamic environmental conditions.

Within the four key research challenges introduced in Chapter~\ref{chap:intro}, this chapter primarily targets the \emph{use case} dimension on physical platforms, building on the \emph{deployment}, \emph{realism}, and \emph{usability} foundations established with Colosseum and X5G. It focuses on completed studies on \gls{isac}, interference detection, resource management, and security, while additional works, both completed and ongoing, address mobility, handover, automation, and energy saving, among other topics.

This chapter is organized as follows. Section~\ref{chap5-sec:cusense} presents cuSense, an uplink \gls{dmrs} \gls{csi}-based \gls{isac} dApp for real-time person detection using the GPU-accelerated framework. Section~\ref{chap5-sec:interforan} describes InterfO-RAN, a dApp for in-band uplink interference detection that leverages a custom implementation to monitor and identify anomalous transmissions. Section~\ref{chap5-sec:oranslice} turns to the control plane and discusses ORANSlice, an xApp-based platform for dynamic resource management and network slicing integrated with the OSC \gls{ric}. Section~\ref{chap5-sec:timesafe} focuses on security, introducing TIMESAFE and its \gls{ml}-based detection of timing attacks on the fronthaul. Finally, Section~\ref{chap5-sec:conclusions} summarizes the main findings, and concludes the chapter with discussions and lessons learned.

\section{cuSense: Uplink CSI-based ISAC dApp for Person Detection}
\label{chap5-sec:cusense}


Building upon the dApp framework presented in Section~\ref{chap4-sec:dapp-framework}, this section introduces cuSense, an \gls{isac} dApp for indoor person localization using uplink \gls{csi}. This case study validates the framework's ability to support intensive \gls{ai} workloads while showcasing the potential of \gls{isac} applications in \gls{5g} networks.

\begin{figure}[htb]
  \centering
  \includegraphics[width=0.95\linewidth]{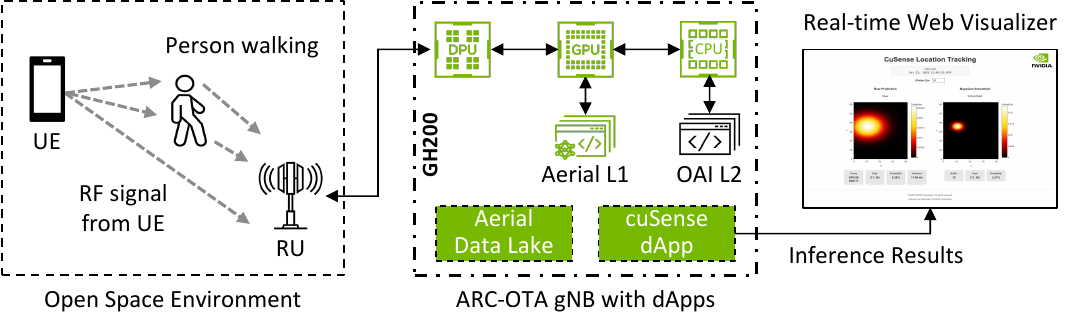}
  \caption{Overview of the cuSense UL DMRS-based ISAC dApp for person localization.}
  \label{chap5-fig:cusense-overview}
\end{figure}

\subsection{Overview and Goals}
\label{chap5-subsec:cusense-overview}


cuSense targets indoor location estimation in a single-cell \gls{cbrs} deployment using \gls{csi} derived from \gls{ul} \gls{pusch} transmissions. In particular, it leverages \gls{dmrs} signals transmitted by a \gls{ue} and exposed by the E3 Agent, in what is known as \gls{ul}-collaborative \gls{isac}~\cite{isacolab}. Following the reference dApp container architecture of Section~\ref{chap4-sec:dapp-architecture}, the cuSense dApp processes these \glspl{dmrs} through a lightweight signal-processing pipeline deployed in Triton, and runs a \gls{nn} model to capture the variations in the channel given by the moving object, as shown in Figure~\ref{chap5-fig:cusense-overview}. The goal is to estimate the 2D position $(x_t,y_t)$ of a person walking in the space through a probability map over the area of interest.
The system addresses the following key challenges in \gls{ul}-\gls{csi}-based sensing: (i) extracting real-time channel perturbations from static multipath; (ii) learning spatial mapping from high-dimensional \gls{csi} data; and (iii) achieving real-time inference.

Finally, cuSense is designed as a reference \gls{isac} dApp to showcase our framework's ability to support real-time, slot-level inference, and provide a reusable dataset and base pipelines for future \gls{isac} experiments. In the current prototype, we focus on single-user localization, but the same design can be extended without modifying the \gls{ran} to other tasks, such as 
occupancy monitoring and \gls{uav} or ground vehicles detection.

\begin{figure}[htb]
  \centering
  \includegraphics[width=\linewidth]{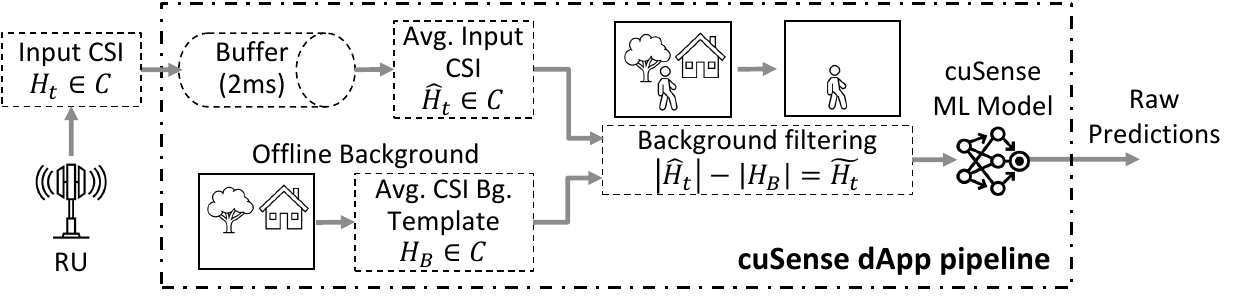}
  \caption{cuSense dApp processing pipeline overview.}
  \label{chap5-fig:cusense-pipeline}
\end{figure}

\subsection{Uplink CSI Processing Pipeline}
\label{chap5-subsec:cusense-pipeline}

cuSense processes raw \gls{ul} \gls{dmrs} \gls{csi} measurements to estimate spatial locations of targets within an indoor environment through a three-stage pipeline: (i) background environment characterization (computed offline); (ii) temporal noise reduction and feature normalization (online in the cuSense dApp pipeline); and (iii) \gls{nn}-based location estimation (online in Triton), as shown in Figure~\ref{chap5-fig:cusense-pipeline}.

\textbf{Stage 1: Background Environment Characterization.}
The objective of this first stage is to establish a baseline characterization of the wireless channel in the absence of dynamic targets. This background template captures an averaged \textit{snapshot} of multi-path reflections from static objects in a deployment location, such as walls, furniture, and other stationary objects, which can then be removed from live measurements to isolate target-induced perturbations.

Let $H_i[a,k,s] \in \mathbb{C}$ denote the complex \gls{csi} for measurement $i$ at antenna $a$, subcarrier $k$, and \gls{dmrs} symbol $s$, obtained after frequency-domain interpolation over all active subcarriers and before time-domain interpolation to all \gls{ofdm} symbols. To handle variable-length subcarrier allocations with zero-padding, we define a validity set for each $(a,k,s)$ position:
\begin{equation}
  \mathcal{V}_{a,k,s} = \{i : |H_i[a,k,s]| > \tau\}
\end{equation}
where $\tau = 10^{-10}$ is a tolerance power threshold, and $\mathcal{V}_{a,k,s}$ contains the indices of non-zero measurements. This ensures that statistics are computed exclusively from active subcarriers, avoiding contamination from zero-padded frequency bins that may vary across different channel allocations.

We compute a complex-valued background template by averaging over the $N_{a,k,s} = |\mathcal{V}_{a,k,s}|$ valid background measurements:
\begin{equation}
  H_{\mathrm{B}}[a,k,s] = \frac{1}{N_{a,k,s}}\sum_{i \in \mathcal{V}_{a,k,s}} H_i[a,k,s].
\end{equation}
The output of this stage is a complex background template $H_{\mathrm{B}}$ over active subcarriers, which is reused at inference time to remove static multi-path components.

\textbf{Stage 2: Temporal Noise Reduction and Feature Normalization.}
This stage applies temporal averaging to mitigate measurement noise, subtracts the static background to isolate dynamic components, and normalizes features to enable stable gradient-based optimization.
For each time $t$ in a target run, we construct a causal temporal window
\begin{equation}
  W(t) = \{ H_i \mid t - \Delta t \le t_i \le t \},
\end{equation}
containing all \gls{csi} samples whose timestamp $t_i$ falls within a window of length $\Delta t$ ending at $t$. Based on empirical hyper-parameter tuning, we set $\Delta t = 2$~ms, which provided the best trade-off between noise reduction and temporal resolution in our experiments.
We then compute the phase-coherent average element-wise in the complex domain for each $W(t)$:
\begin{equation}
  \hat{H}_{\mathrm{t}}[a,k,s]  = \frac{1}{|W(t)|} \sum_{i\in W(t)} H_i[a,k,s],
\end{equation}
excluding zero-valued samples on inactive subcarriers. Temporal averaging reduces noise variance by approximately a factor of $1/|W(t)|$ under an \gls{awgn} assumption, improving the effective \gls{snr} while preserving the slowly varying channel characteristics associated with target motion.

To isolate dynamic components from static environment reflections, we subtract the background template in the magnitude domain from the averaged measurements,
\begin{equation}
  \tilde{H}_{\mathrm{t}}[a,k,s] =
  \bigl| \hat{H}_{\mathrm{t}}[a,k,s] \bigr|
  - \bigl| H_{\mathrm{B}}[a,k,s] \bigr|,
\end{equation}
and then average across the $S$ \gls{dmrs} symbols in the \gls{ul} slot,
\begin{equation}
  \tilde{H}_{\mathrm{t}}^{\mathrm{D}}[a,k] =
  \frac{1}{S} \sum_{s=1}^{S} \tilde{H}_{\mathrm{t}}[a,k,s].
\end{equation}
This produces a real-valued tensor $\tilde{H}_{\mathrm{t}}^{\mathrm{D}} \in \mathbb{R}^{A \times K_{\mathrm{v}}}$, where $A$ is the total number of receive antennas and $K_{\mathrm{v}}$ the number of active (or valid) subcarriers. This step further reduces noise through intra-slot averaging under the assumption that the channel remains approximately constant within a single \gls{ofdm} slot.

Finally, we apply global $z$-score normalization,
\begin{equation}
  X_{\mathrm{t}}[a,k] =
  \frac{\tilde{H}_{\mathrm{t}}^{\mathrm{D}}[a,k] - \mu_{\mathrm{global}}}{\sigma_{\mathrm{global}} + \epsilon},
\end{equation}
where $\mu_{\mathrm{global}}$ and $\sigma_{\mathrm{global}}$ are computed over all available samples, antennas, and valid subcarriers, and are then applied identically to all splits, and $\epsilon = 10^{-8}$ is a small constant to prevent division by zero. This produces standardized input features $X_{\mathrm{t}}[a,k]$ for our \gls{nn} model with approximately zero mean and unit variance, preserving relative power relationships between samples, which facilitates stable training and inference.
In a deployed system, Stage~1 runs offline once per environment, while Stage~2 executes online inside the cuSense dApp, using the \gls{csi} data delivered by the E3 Agent via \gls{shm}.

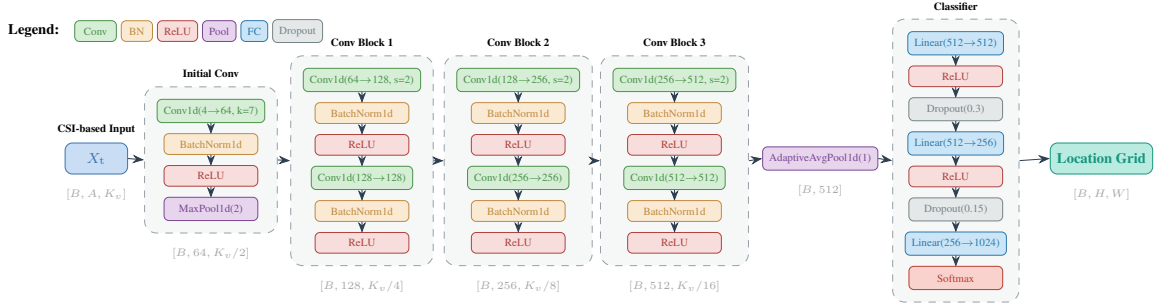
\begin{figure}[htb]
\centering
\resizebox{\textwidth}{!}{\begin{tikzpicture}[
    node distance=0.15cm,
    >=Stealth,
    scale=0.75, transform shape,
    input/.style={rectangle, rounded corners=3pt, minimum width=1.2cm, minimum height=0.6cm,
        draw=inputcolor!80!black, fill=inputcolor!30, font=\scriptsize\bfseries, text=inputcolor!80!black},
    conv/.style={rectangle, rounded corners=2pt, minimum width=1.8cm, minimum height=0.45cm,
        draw=convcolor!80!black, fill=convcolor!25, font=\tiny, text=convcolor!80!black},
    bn/.style={rectangle, rounded corners=2pt, minimum width=1.8cm, minimum height=0.35cm,
        draw=bncolor!80!black, fill=bncolor!25, font=\tiny, text=bncolor!80!black},
    activation/.style={rectangle, rounded corners=2pt, minimum width=1.8cm, minimum height=0.35cm,
        draw=relucolor!80!black, fill=relucolor!20, font=\tiny, text=relucolor!80!black},
    pool/.style={rectangle, rounded corners=2pt, minimum width=1.8cm, minimum height=0.45cm,
        draw=poolcolor!80!black, fill=poolcolor!25, font=\tiny, text=poolcolor!80!black},
    fc/.style={rectangle, rounded corners=2pt, minimum width=1.8cm, minimum height=0.45cm,
        draw=fccolor!80!black, fill=fccolor!25, font=\tiny, text=fccolor!80!black},
    dropout/.style={rectangle, rounded corners=2pt, minimum width=1.8cm, minimum height=0.35cm,
        draw=dropcolor!80!black, fill=dropcolor!25, font=\tiny, text=dropcolor!80!black},
    softmax/.style={rectangle, rounded corners=2pt, minimum width=1.8cm, minimum height=0.45cm,
        draw=softmaxcolor!80!black, fill=softmaxcolor!25, font=\tiny, text=softmaxcolor!80!black},
    output/.style={rectangle, rounded corners=3pt, minimum width=1.2cm, minimum height=0.6cm,
        draw=outputcolor!80!black, fill=outputcolor!30, font=\scriptsize\bfseries, text=outputcolor!80!black},
    resblock/.style={rectangle, rounded corners=5pt, draw=arrowcolor!50, dashed, fill=resblockcolor!50, inner sep=5pt},
    arrow/.style={->, very thin, arrowcolor},
    bigarrow/.style={->, thin, arrowcolor},
    dimtext/.style={font=\tiny\itshape, gray!70},
]

\def\centerY{0}

\node[input, label={[font=\tiny\bfseries, yshift=1pt]above:CSI-based Input}] (input) at (0, \centerY - 0.26) {\textbf{$X_{\mathrm{t}}$}};
\node[dimtext, below=0.15cm of input] {$[B, A, K_v]$};

\node[conv] (conv0) at (2.2, \centerY + 0.66) {Conv1d(4$\to$64, k=7)};
\node[bn, below=0.2cm of conv0] (bn0) {BatchNorm1d};
\node[activation, below=0.2cm of bn0] (relu0) {ReLU};
\node[pool, below=0.2cm of relu0] (pool0) {MaxPool1d(2)};
\begin{scope}[on background layer]
    \node[resblock, fit=(conv0)(bn0)(relu0)(pool0), label={[font=\tiny\bfseries, yshift=1pt]above:Initial Conv}] (initblock) {};
\end{scope}
\node[dimtext, below=0.15cm of initblock] {$[B, 64, K_v/2]$};

\node[conv] (conv1a) at (5.1, \centerY + 1.26) {Conv1d(64$\to$128, s=2)};
\node[bn, below=0.2cm of conv1a] (bn1a) {BatchNorm1d};
\node[activation, below=0.2cm of bn1a] (relu1a) {ReLU};
\node[conv, below=0.2cm of relu1a] (conv1b) {Conv1d(128$\to$128)};
\node[bn, below=0.2cm of conv1b] (bn1b) {BatchNorm1d};
\node[activation, below=0.2cm of bn1b] (relu1b) {ReLU};
\begin{scope}[on background layer]
    \node[resblock, fit=(conv1a)(bn1a)(relu1a)(conv1b)(bn1b)(relu1b), label={[font=\tiny\bfseries, yshift=1pt]above:Conv Block 1}] (block1) {};
\end{scope}
\node[dimtext, below=0.15cm of block1] {$[B, 128, K_v/4]$};

\node[conv] (conv2a) at (8.1, \centerY + 1.26) {Conv1d(128$\to$256, s=2)};
\node[bn, below=0.2cm of conv2a] (bn2a) {BatchNorm1d};
\node[activation, below=0.2cm of bn2a] (relu2a) {ReLU};
\node[conv, below=0.2cm of relu2a] (conv2b) {Conv1d(256$\to$256)};
\node[bn, below=0.2cm of conv2b] (bn2b) {BatchNorm1d};
\node[activation, below=0.2cm of bn2b] (relu2b) {ReLU};
\begin{scope}[on background layer]
    \node[resblock, fit=(conv2a)(bn2a)(relu2a)(conv2b)(bn2b)(relu2b), label={[font=\tiny\bfseries, yshift=1pt]above:Conv Block 2}] (block2) {};
\end{scope}
\node[dimtext, below=0.15cm of block2] {$[B, 256, K_v/8]$};

\node[conv] (conv3a) at (11.1, \centerY + 1.26) {Conv1d(256$\to$512, s=2)};
\node[bn, below=0.2cm of conv3a] (bn3a) {BatchNorm1d};
\node[activation, below=0.2cm of bn3a] (relu3a) {ReLU};
\node[conv, below=0.2cm of relu3a] (conv3b) {Conv1d(512$\to$512)};
\node[bn, below=0.2cm of conv3b] (bn3b) {BatchNorm1d};
\node[activation, below=0.2cm of bn3b] (relu3b) {ReLU};
\begin{scope}[on background layer]
    \node[resblock, fit=(conv3a)(bn3a)(relu3a)(conv3b)(bn3b)(relu3b), label={[font=\tiny\bfseries, yshift=1pt]above:Conv Block 3}] (block3) {};
\end{scope}
\node[dimtext, below=0.15cm of block3] {$[B, 512, K_v/16]$};

\node[pool] (gpool) at (13.9, \centerY - 0.26) {AdaptiveAvgPool1d(1)};
\node[dimtext, below=0.15cm of gpool] {$[B, 512]$};

\node[fc] (fc1) at (16.5, \centerY + 1.97) {Linear(512$\to$512)};
\node[activation, below=0.2cm of fc1] (relu4) {ReLU};
\node[dropout, below=0.2cm of relu4] (drop1) {Dropout(0.3)};
\node[fc, below=0.2cm of drop1] (fc2) {Linear(512$\to$256)};
\node[activation, below=0.2cm of fc2] (relu5) {ReLU};
\node[dropout, below=0.2cm of relu5] (drop2) {Dropout(0.15)};
\node[fc, below=0.2cm of drop2] (fc3) {Linear(256$\to$1024)};
\node[softmax, below=0.2cm of fc3] (softmax) {Softmax};
\begin{scope}[on background layer]
    \node[resblock, fit=(fc1)(relu4)(drop1)(fc2)(relu5)(drop2)(fc3)(softmax), label={[font=\tiny\bfseries, yshift=1pt]above:Classifier}] (classifier) {};
\end{scope}

\node[output] (output) at (19.3, \centerY - 0.26) {\textbf{Location Grid}};
\node[dimtext, below=0.15cm of output] {$[B, H, W]$};

\draw[bigarrow] (input.east) -- (initblock.west);
\draw[bigarrow] (initblock.east) -- (block1.west);
\draw[bigarrow] (block1.east) -- (block2.west);
\draw[bigarrow] (block2.east) -- (block3.west);
\draw[bigarrow] (block3.east) -- (gpool.west);
\draw[bigarrow] (gpool.east) -- (classifier.west);
\draw[bigarrow] (classifier.east) -- (output.west);

\draw[arrow] (conv0) -- (bn0);
\draw[arrow] (bn0) -- (relu0);
\draw[arrow] (relu0) -- (pool0);

\draw[arrow] (conv1a) -- (bn1a);
\draw[arrow] (bn1a) -- (relu1a);
\draw[arrow] (relu1a) -- (conv1b);
\draw[arrow] (conv1b) -- (bn1b);
\draw[arrow] (bn1b) -- (relu1b);

\draw[arrow] (conv2a) -- (bn2a);
\draw[arrow] (bn2a) -- (relu2a);
\draw[arrow] (relu2a) -- (conv2b);
\draw[arrow] (conv2b) -- (bn2b);
\draw[arrow] (bn2b) -- (relu2b);

\draw[arrow] (conv3a) -- (bn3a);
\draw[arrow] (bn3a) -- (relu3a);
\draw[arrow] (relu3a) -- (conv3b);
\draw[arrow] (conv3b) -- (bn3b);
\draw[arrow] (bn3b) -- (relu3b);

\draw[arrow] (fc1) -- (relu4);
\draw[arrow] (relu4) -- (drop1);
\draw[arrow] (drop1) -- (fc2);
\draw[arrow] (fc2) -- (relu5);
\draw[arrow] (relu5) -- (drop2);
\draw[arrow] (drop2) -- (fc3);
\draw[arrow] (fc3) -- (softmax);

\node[above=1.9cm of input, xshift=-1.2cm] (legendtitle) {\scriptsize\bfseries Legend:};
\node[conv, right=0.15cm of legendtitle, minimum width=0.8cm, minimum height=0.3cm] (legconv) {\tiny Conv};
\node[bn, right=0.08cm of legconv, minimum width=0.6cm, minimum height=0.3cm] (legbn) {\tiny BN};
\node[activation, right=0.08cm of legbn, minimum width=0.6cm, minimum height=0.3cm] (legrelu) {\tiny ReLU};
\node[pool, right=0.08cm of legrelu, minimum width=0.6cm, minimum height=0.3cm] (legpool) {\tiny Pool};
\node[fc, right=0.08cm of legpool, minimum width=0.5cm, minimum height=0.3cm] (legfc) {\tiny FC};
\node[dropout, right=0.08cm of legfc, minimum width=0.7cm, minimum height=0.3cm] (legdrop) {\tiny Dropout};

\end{tikzpicture}}
\caption{Proposed \acrshort{nn} architecture of cuSense dApp for CSI-based location estimation.}
\label{chap5-fig:cusense-architecture}
\end{figure}

\textbf{Stage 3: Neural Network Architecture and Training.}
In the last stage, we use a ResNet-inspired \gls{cnn} optimized for frequency-domain, multi-antenna \gls{csi} magnitude data to predict a 2D probability grid representing the likelihood of target presence at each location within the environment. In order to account for the relatively shallow architecture and to keep the parameter count low and allow fast inference, we consider only \textit{plain} blocks without residual connections for our architecture design (see \cite{He_2016_CVPR} for the original model).
The normalized features $X_{\mathrm{t}} \in \mathbb{R}^{A\times K_{\mathrm{v}}}$ are fed to our proposed architecture, as shown in Figure~\ref{chap5-fig:cusense-architecture}. The network consists of: (i) an initial 1D convolution and max-pooling layer; (ii) three sequential blocks with progressive channel expansion ($64\rightarrow128\rightarrow256\rightarrow512$); (iii) an adaptive global average pooling layer, to allow inputs of different shapes in case not all RBs are utilized in the \gls{ul} transmission; (iv) a three-layer \gls{mlp} stack that outputs an $H\times W$ (grid height and width) spatial probability map; and (v) a final softmax to ensure $\sum_{i,j} P_\mathrm{t}(i,j) = 1$ for each $(i,j)$ grid cell probability value $P_\mathrm{t}$.

Training uses the \gls{kl} divergence between the predicted distribution $P_{\text{X}}$ and a smoothed target distribution $P_{\text{Y}}$ as loss:
\begin{equation}
  \mathcal{L}_{\mathrm{KL}} =
  \sum_{i=0}^{H-1} \sum_{j=0}^{W-1}
  P_{\text{Y}}(i,j)
  \log \frac{P_{\text{Y}}(i,j)}{P_{\text{X}}(i,j)}.
\end{equation}
This formulation encourages learning the entire spatial distribution rather than only peak locations, improving robustness to measurement noise. The target distribution is obtained by convolving a one-hot probability grid\footnote{All grid probability values are 0, except for the target location with value equal to 1.} with a 2D Gaussian kernel $G_\sigma$ (label smoothing):
\begin{equation}
P_{\text{Y}} =
\frac{(G_\sigma * P_{\text{one-hot}})(i,j)}
     {\sum_{i,j} (G_\sigma * P_{\text{one-hot}})(i,j)},
\end{equation}
with $\sigma = 8.0$ in our experiments. Spatial smoothing encodes the intuition that nearby locations should receive similar probabilities, reflecting uncertainty in ground-truth annotations and measurement noise.
We train the model using Adam optimizer, mini-batches of size $B = 256$, and a learning rate $l_r = 10^{-3}$ for 100 epochs.
At inference time, cuSense runs Stage~2 pre-processing and the \gls{nn} inside the dApp container (via Triton), receiving \gls{csi} for each \gls{ul} slot $t$, and outputting a 2D probability map $P_{\mathrm{t}}(i,j)$ over the $H\times W$ grid.
From this distribution, we can derive a maximum-likelihood estimate 
$(\hat{i}_t,\hat{j}_t) = \arg\max_{i,j} P_t(i,j)$ to obtain the grid coordinates of highest target location probability, which can be then mapped to the relative world-coordinates.
%
In order to reduce output predictions errors, we average the output probability for the last 10 predictions and apply Kalman filtering before returning the final estimated tracked 2D location, as detailed in Section~\ref{chap5-subsubsec:cusense-results}.

\subsection{Experimental Evaluation}
\label{chap5-subsec:cusense-eval}

To assess the feasibility and performance of cuSense, we perform an extensive \gls{ota} data-collection campaign, develop labeling pipelines to obtain ground truth and build datasets, and train and test the cuSense \gls{nn} under realistic environmental conditions.

\subsubsection{Data Collection and Labeling Methods}

\begin{figure}[htb]
  \centering
  \includegraphics[width=0.8\linewidth]{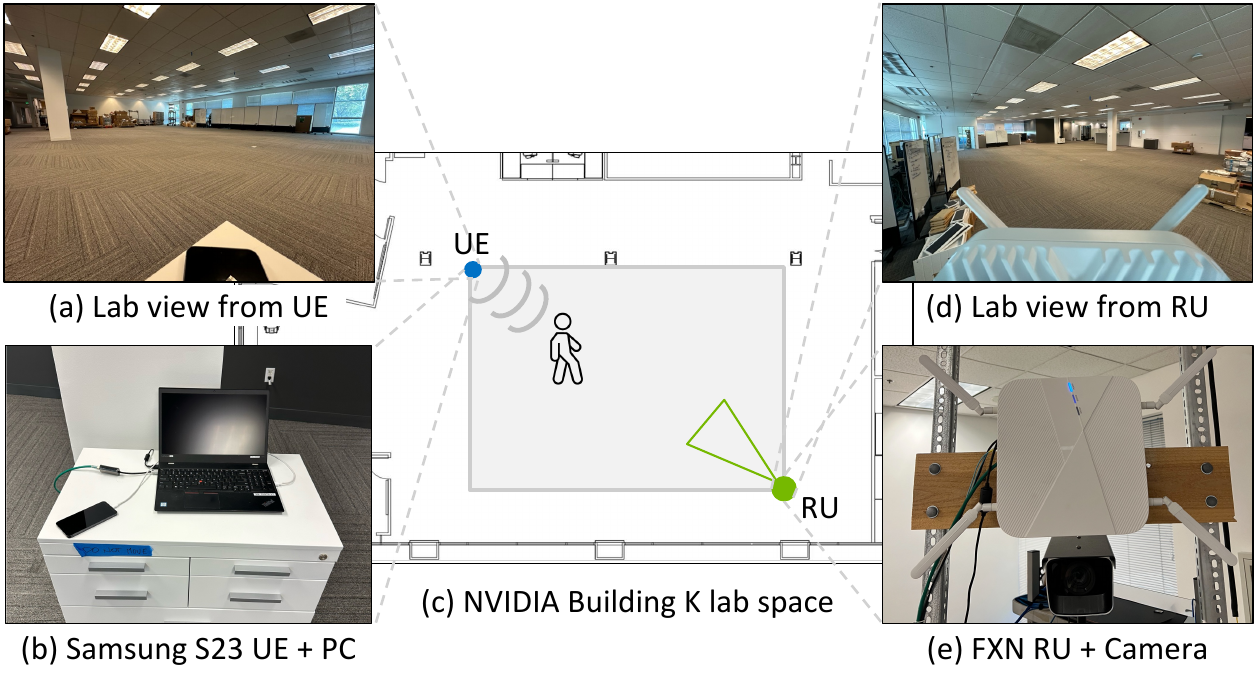}
  \caption{cuSense experimental environment.}
  \label{chap5-fig:cusense-setup}
\end{figure}

\textbf{Measurement Campaigns.}
\label{chap5-subsec:cusense-data}
All cuSense experiments are conducted on the same \gls{arc-ota} setup described in Chapter~\ref{chap:4} and illustrated in Figure~\ref{chap5-fig:cusense-setup}. The \gls{cbrs} Foxconn \gls{ru} is mounted on a custom stand~\cite{aerialsdk} facing an open indoor space in NVIDIA Lab. The Samsung \gls{ue}, together with a laptop used for remote access and traffic generation, is placed at the opposite end of a rectangular target area of approximately $6.78 \times 10.06$~m.
We collect multiple runs under two conditions: (i) \emph{background runs}, with no moving target present, to characterize the static channel; and (ii) \emph{target runs}, in which a person walks along various trajectories (e.g., lawnmower, spiral, random) at normal walking speed. Each run follows the same procedure: (i) start camera recording at $30$ or $60$~\gls{fps} (for target runs); (ii) start the \gls{arc-ota} \gls{gnb} with \gls{adl} enabled; (iii) connect the \gls{ue}; (iv) start uplink UDP traffic through \texttt{iperf} with a target rate of $100$~Mbps; (v) perform the walking trajectory within the target area (for target runs); and (vi) stop the run after approximately $2$--$3$ minutes.
Across all runs, we collect more than $400.000$ \gls{csi} records together with more than $30.000$ video frames, which we use in the next step for labeling and dataset construction.

\textbf{Temporal Video and Sensing Synchronization.}
To build the complete dataset for training and testing, we collect \gls{ul} \gls{csi} data (i.e., \gls{dmrs} \gls{hest}) together with a camera video recording the scene at a fixed frame rate, used to generate ground-truth 2D trajectories of the person walking. Synchronizing \gls{csi} data and video frames requires handling both time-reference differences and clock skew between devices.
\gls{csi} records from \gls{adl} use \gls{tai} timestamps, while video frames from the camera (an iPhone in our experiments, though any commercial \gls{5g} camera could be used) employ standard \gls{utc} UNIX timestamps. \gls{tai} is a continuous atomic time scale that differs from \gls{utc} by a constant offset $\Delta_{\text{leap}}$ ($37$~s at the time of our experiments), such that $t_{\text{TAI}} = t_{\text{UTC}} + \Delta_{\text{leap}}$ due to the leap seconds introduced since 1972 to account for Earth's rotation variations.
In addition to this systematic offset, we compensate for residual clock skew between the camera and the \gls{gnb} server with respect to a common \gls{ntp} server. The \gls{gnb} maintains microsecond-level timing accuracy through \gls{ptp} synchronization with the grandmaster clock, making its contribution to skew negligible, while the camera achieves millisecond-level precision, which we estimate during the measurement campaigns.
In a second step, we run an offline pipeline that parses the video file, extracts per-frame timestamps and images, and aligns them with the \gls{csi} after compensating for the $37$~s \gls{tai} offset and the measured server–camera clock skew. We assign each \gls{csi} record to the closest video frame within a threshold of half the frame period, generating tightly synchronized \gls{csi}–video pairs, where a single frame may correspond to multiple \gls{csi} records.

\textbf{Ground-Truth Extraction.}
To extract the ground-truth locations, we derive the 2D person coordinates from the camera stream using an offline computer-vision pipeline based on YOLOv8~\cite{yolov8}. For each video frame, we run a person detector and retain the bounding box with the highest confidence (our experiments are limited to a single walking subject). The box centroid $(u,v)$ in image coordinates is then projected onto the floor plane using a planar homography $H$ estimated from the manually annotated image corners of the rectangular area of interest in the environment. This generates a continuous trajectory $\{(x_t,y_t)\}$ in meters, sampled at the camera frame rate.
Each \gls{csi} record then inherits a physical position $(x_t,y_t)$, since it is associated with the closest video frame in time. For training, we discretize the floor plan into an $H \times W$ grid and quantize each position to the corresponding cell $(i,j)$, producing a one-hot label $P_{\text{one-hot}}(i,j)$ for that \gls{csi} sample. These one-hot labels are then converted into smoothed target distributions $P_{\text{target}}$ using the Gaussian label-smoothing procedure described in Stage~3.


\subsubsection{Performance Results}
\label{chap5-subsubsec:cusense-results}

We now evaluate cuSense on the dataset described in Section~\ref{chap5-subsec:cusense-data}, focusing on three aspects: (i) localization accuracy against the ground-truth; (ii) comparison with \gls{3gpp} sensing service categories; and (iii) real-time inference capability within our dApp framework.

\begin{figure}[htb]
  \centering
  \setlength\fwidth{\linewidth}
  \setlength\fheight{\linewidth}
  \begin{tikzpicture}
\definecolor{lightgray204}{RGB}{204,204,204}
\definecolor{avgcolor}{RGB}{230,159,0}
\definecolor{kalmancolor}{RGB}{86,180,233}
\definecolor{unseengray}{RGB}{220,220,220}

\node[font=\footnotesize, anchor=east] at (0.55,1.4) {Training runs};

\draw[draw=black,fill=white] (0.6,1.2) rectangle (2.1,1.6);

\draw[draw=black,fill=avgcolor,postaction={pattern=north west lines,pattern color=black}] (2.1,1.2) rectangle (2.65,1.6);
\node[font=\tiny\bfseries,white] at (2.375,1.4) {10\%};

\draw[draw=black,fill=white] (2.65,1.2) rectangle (3.85,1.6);

\draw[draw=black,fill=kalmancolor,postaction={pattern=north east lines,pattern color=black}] (3.85,1.2) rectangle (4.4,1.6);
\node[font=\tiny\bfseries,white] at (4.125,1.4) {10\%};

\draw[draw=black,fill=white] (4.4,1.2) rectangle (5.2,1.6);

\node[font=\footnotesize, anchor=east] at (0.55,0.7) {Unseen runs};

\draw[draw=black,fill=unseengray,postaction={pattern=vertical lines,pattern color=black}] (0.6,0.5) rectangle (5.2,0.9);

\draw[->] (0.6,0.2) -- (5.2,0.2);
\node[font=\footnotesize] at (2.9,0.0) {Samples};

\node[
    draw=lightgray204,
    fill=white,
    fill opacity=0.8,
    font=\footnotesize,
    inner sep=3pt,
    anchor=west,
] at (5.4,0.9) {
    \begin{tabular}{@{}cl@{}}
        \tikz{\draw[draw=black,fill=white] (0,0) rectangle (0.25,0.15);} & Train \\[2pt]
        \tikz{\draw[draw=black,fill=avgcolor,postaction={pattern=north west lines,pattern color=black}] (0,0) rectangle (0.25,0.15);} & Val \\[2pt]
        \tikz{\draw[draw=black,fill=kalmancolor,postaction={pattern=north east lines,pattern color=black}] (0,0) rectangle (0.25,0.15);} & Test \\[2pt]
        \tikz{\draw[draw=black,fill=unseengray,postaction={pattern=vertical lines,pattern color=black}] (0,0) rectangle (0.25,0.15);} & Unseen \\
    \end{tabular}
};

\end{tikzpicture}
  \caption{Temporal-random split strategy with example of contiguous validation/test blocks within training runs and fully unseen runs.}
  \label{chap5-fig:cusense-dataset}
\end{figure}
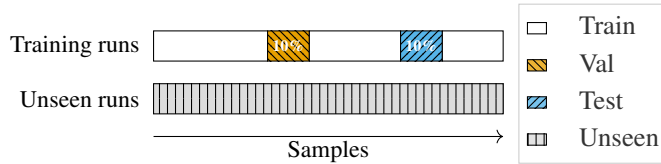

\textbf{Dataset and Evaluation Setup.}
The cuSense model is trained on \gls{csi} data from 5 runs with a moving target, for a total of $362{,}318$ slot samples each with 4 antennas and 3 \gls{dmrs} symbols. We use a temporal-random split strategy of 80\% training, 10\% validation, 10\% test as shown in the examples of Figure~\ref{chap5-fig:cusense-dataset}. Validation and test sets are two contiguous temporal blocks randomly positioned within each run's timeline, rather than randomly sampling individual points. This design choice is critical because consecutive \gls{csi} values may be highly correlated, leading to overfitting and poor generalization to unseen trajectories. 
The background template $H_{\text{B}}$ is computed from a single dedicated run of $54{,}400$ samples in a static environment with no moving target. We also evaluate cuSense generalization across two independent unseen runs (Unseen-Run~1 and Unseen-Run~2) collected separately from all training, testing, or validation data.
We measure localization performance using standard \gls{rmse}, median, and success rate within different thresholds over all unseen \gls{csi} records. Additionally, the inference pipeline applies a temporal arithmetic mean over 10 consecutive predictions from stage~3 processing output for noise reduction, followed by a Kalman filter tracking with process noise $Q = 10^{-5}$ and measurement noise base $R_{\text{base}} = 750$ to produce temporally coherent trajectory estimates.

\textbf{Localization Accuracy.}
Table~\ref{chap5-tab:cusense-accuracy} summarizes the localization performance on the unseen runs. cuSense achieves a mean localization error of $77.4$~cm with a median of $58.7$~cm, demonstrating sub-meter accuracy for the majority of predictions. The \gls{rmse} of $103$~cm reflects occasional outliers in challenging multipath conditions. Notably, performance is consistent across both test campaigns ($75.1$ and $79.7$~cm), indicating robust generalization to unseen measurement campaigns in the same environment.

\begin{table}[htbp]
    \centering
    \caption{cuSense localization accuracy on unseen runs.}
    \label{chap5-tab:cusense-accuracy}
    \small
    \begin{tabular}{lccc}
        \toprule
        \textbf{Metric} & \textbf{Unseen-Run 1} & \textbf{Unseen-Run 2} & \textbf{Average} \\
        \midrule
        Samples & 47,211 & 47,404 & - \\
        Mean Error [cm] & 75.1 & 79.7 & 77.4 \\
        Median Error [cm] & 54.9 & 62.6 & 58.7 \\
        Std. Dev. [cm] & 67.6 & 68.3 & 68.0 \\
        RMSE [cm] & 101.0 & 105.0 & 103.0 \\
        \bottomrule
    \end{tabular}
\end{table}

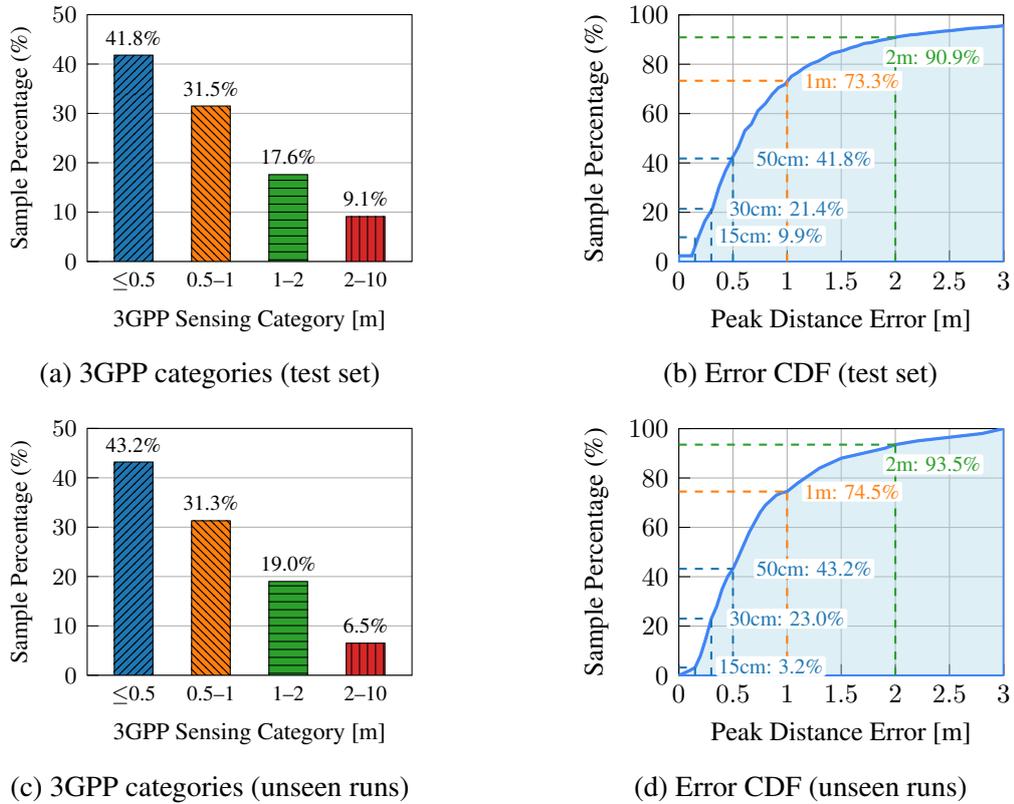
\begin{figure}[htb]
    \centering
\begin{subfigure}{0.49\linewidth}
    \centering
        \setlength\fwidth{0.8\linewidth}
        \setlength\fheight{0.65\linewidth}
\begin{tikzpicture}
\pgfplotsset{every tick label/.append style={font=\footnotesize}}
\definecolor{darkgray176}{RGB}{176,176,176}
\definecolor{lightgray204}{RGB}{204,204,204}
\definecolor{steelblue31119180}{RGB}{31,119,180}
\definecolor{darkorange25512714}{RGB}{255,127,14}
\definecolor{forestgreen4416044}{RGB}{44,160,44}
\definecolor{crimson2143940}{RGB}{214,39,40}

\begin{axis}[
width=0.98\fwidth,
height=\fheight,
at={(0\fwidth,0\fheight)},
x grid style={darkgray176},
xmin=-0.6, xmax=3.6,
xtick style={color=black},
xtick={0,1,2,3},
xticklabel style={font=\scriptsize},
xticklabels={$\leq$0.5, 0.5--1, 1--2, 2--10},
xtick pos=bottom,
y grid style={darkgray176},
ylabel={Sample Percentage (\%)},
xlabel={3GPP Sensing Category [m]},
ylabel style={font=\footnotesize},
xlabel style={font=\footnotesize},
ymin=0, ymax=50,
ytick={0,10,20,30,40,50},
ytick pos=left,
ytick style={color=black},
ymajorgrids,
bar width=0.5cm,
]

\draw[draw=black,fill=steelblue31119180,postaction={pattern=north east lines,pattern color=black}] 
    (axis cs:-0.25,0) rectangle (axis cs:0.25,41.8);
\node[font=\scriptsize, above] at (axis cs:0,41.8) {41.8\%};

\draw[draw=black,fill=darkorange25512714,postaction={pattern=north west lines,pattern color=black}] 
    (axis cs:0.75,0) rectangle (axis cs:1.25,31.5);
\node[font=\scriptsize, above] at (axis cs:1,31.5) {31.5\%};

\draw[draw=black,fill=forestgreen4416044,postaction={pattern=horizontal lines,pattern color=black}] 
    (axis cs:1.75,0) rectangle (axis cs:2.25,17.6);
\node[font=\scriptsize, above] at (axis cs:2,17.6) {17.6\%};

\draw[draw=black,fill=crimson2143940,postaction={pattern=vertical lines,pattern color=black}] 
    (axis cs:2.75,0) rectangle (axis cs:3.25,9.1);
\node[font=\scriptsize, above] at (axis cs:3,9.1) {9.1\%};

\end{axis}
\end{tikzpicture}
    \caption{\centering 3GPP categories (test set)}
    \label{chap5-fig:cusense-3gpp-train}
\end{subfigure}
    \hfill
\begin{subfigure}{0.49\linewidth}
    \centering
        \setlength\fwidth{0.8\linewidth}
        \setlength\fheight{0.65\linewidth}
\begin{tikzpicture}
\pgfplotsset{every tick label/.append style={font=\small}}
\definecolor{darkgray176}{RGB}{176,176,176}
\definecolor{cdfblue}{RGB}{66,133,244}
\definecolor{cdffill}{RGB}{173,216,230}
\definecolor{steelblue31119180}{RGB}{31,119,180}
\definecolor{darkorange25512714}{RGB}{255,127,14}
\definecolor{forestgreen4416044}{RGB}{44,160,44}
\definecolor{crimson2143940}{RGB}{214,39,40}

\begin{axis}[
width=0.98\fwidth,
height=\fheight,
at={(0\fwidth,0\fheight)},
x grid style={darkgray176},
y grid style={darkgray176},
xmin=0, xmax=3,
ymin=0, ymax=100,
xtick={0,0.5,1,1.5,2,2.5,3},
ytick={0,20,40,60,80,100},
xtick style={color=black},
ytick style={color=black},
xtick pos=bottom,
ytick pos=left,
xlabel={Peak Distance Error [m]},
ylabel={Sample Percentage (\%)},
ylabel style={font=\small},
xlabel style={font=\small},
xmajorgrids,
ymajorgrids,
]

\addplot[
    thick,
    color=cdfblue,
    fill=cdffill,
    fill opacity=0.4,
] coordinates {
    (0.00,2.4) (0.06,2.4) (0.12,2.4) (0.18,9.9) (0.24,16.4) (0.31,21.4) (0.37,30.1) (0.43,36.7) (0.49,41.8) (0.55,46.5) (0.61,53.1) (0.67,55.7) (0.73,61.1) (0.80,64.1) (0.86,67.7) (0.92,70.5) (0.98,71.8) (1.04,75.1) (1.10,76.7) (1.16,78.6) (1.22,80.1) (1.29,81.3) (1.35,82.8) (1.41,84.3) (1.47,85.0) (1.53,85.7) (1.59,86.7) (1.65,87.4) (1.71,88.4) (1.78,88.8) (1.84,89.5) (1.90,90.0) (1.96,90.5) (2.02,91.1) (2.08,91.5) (2.14,91.9) (2.20,92.1) (2.27,92.5) (2.33,92.8) (2.39,93.1) (2.45,93.4) (2.51,93.6) (2.57,93.9) (2.63,94.2) (2.69,94.5) (2.76,94.7) (2.82,94.9) (2.88,95.1) (2.94,95.3) (3.00,95.6)
} \closedcycle;

\addplot[very thick, color=cdfblue, mark=none] coordinates {
    (0.00,2.4) (0.06,2.4) (0.12,2.4) (0.18,9.9) (0.24,16.4) (0.31,21.4) (0.37,30.1) (0.43,36.7) (0.49,41.8) (0.55,46.5) (0.61,53.1) (0.67,55.7) (0.73,61.1) (0.80,64.1) (0.86,67.7) (0.92,70.5) (0.98,71.8) (1.04,75.1) (1.10,76.7) (1.16,78.6) (1.22,80.1) (1.29,81.3) (1.35,82.8) (1.41,84.3) (1.47,85.0) (1.53,85.7) (1.59,86.7) (1.65,87.4) (1.71,88.4) (1.78,88.8) (1.84,89.5) (1.90,90.0) (1.96,90.5) (2.02,91.1) (2.08,91.5) (2.14,91.9) (2.20,92.1) (2.27,92.5) (2.33,92.8) (2.39,93.1) (2.45,93.4) (2.51,93.6) (2.57,93.9) (2.63,94.2) (2.69,94.5) (2.76,94.7) (2.82,94.9) (2.88,95.1) (2.94,95.3) (3.00,95.6)
};


\draw[thick, dashed, color=forestgreen4416044] (axis cs:2.00,0) -- (axis cs:2.00,90.9);
\draw[thick, dashed, color=forestgreen4416044] (axis cs:0,90.9) -- (axis cs:2.00,90.9);
\node[fill=white, fill opacity=1, text opacity=1, font=\scriptsize, 
      rounded corners=1pt, inner sep=1pt] 
    at (axis cs:2.35,82.9) {\textcolor{forestgreen4416044}{2m: 90.9\%}};

\draw[thick, dashed, color=darkorange25512714] (axis cs:1.00,0) -- (axis cs:1.00,73.3);
\draw[thick, dashed, color=darkorange25512714] (axis cs:0,73.3) -- (axis cs:1.00,73.3);
\node[fill=white, fill opacity=1, text opacity=1, font=\scriptsize, 
      rounded corners=1pt, inner sep=1pt] 
    at (axis cs:1.6,73.3) {\textcolor{darkorange25512714}{1m: 73.3\%}};

\draw[thick, dashed, color=steelblue31119180] (axis cs:0.50,0) -- (axis cs:0.50,41.8);
\draw[thick, dashed, color=steelblue31119180] (axis cs:0,41.8) -- (axis cs:0.50,41.8);
\node[fill=white, fill opacity=1, text opacity=1, font=\scriptsize, 
      rounded corners=1pt, inner sep=1pt] 
    at (axis cs:1.25,41.8) {\textcolor{steelblue31119180}{50cm: 41.8\%}};

\draw[thick, dashed, color=steelblue31119180] (axis cs:0.30,0) -- (axis cs:0.30,21.4);
\draw[thick, dashed, color=steelblue31119180] (axis cs:0,21.4) -- (axis cs:0.30,21.4);
\node[fill=white, fill opacity=1, text opacity=1, font=\scriptsize, 
      rounded corners=1pt, inner sep=1pt] 
    at (axis cs:1.0,21.4) {\textcolor{steelblue31119180}{30cm: 21.4\%}};

\draw[thick, dashed, color=steelblue31119180] (axis cs:0.15,0) -- (axis cs:0.15,9.9);
\draw[thick, dashed, color=steelblue31119180] (axis cs:0,9.9) -- (axis cs:0.15,9.9);
\node[fill=white, fill opacity=1, text opacity=1, font=\scriptsize, 
      rounded corners=1pt, inner sep=1pt] 
    at (axis cs:0.85,10.4) {\textcolor{steelblue31119180}{15cm: 9.9\%}};
    
\end{axis}
\end{tikzpicture}
    \caption{\centering Error CDF (test set)}
    \label{chap5-fig:cusense-cdf-train}
\end{subfigure}
\\[0.2cm]
\begin{subfigure}{0.49\linewidth}
    \centering
        \setlength\fwidth{0.8\linewidth}
        \setlength\fheight{0.65\linewidth}
\begin{tikzpicture}
\pgfplotsset{every tick label/.append style={font=\footnotesize}}
\definecolor{darkgray176}{RGB}{176,176,176}
\definecolor{lightgray204}{RGB}{204,204,204}
\definecolor{steelblue31119180}{RGB}{31,119,180}
\definecolor{darkorange25512714}{RGB}{255,127,14}
\definecolor{forestgreen4416044}{RGB}{44,160,44}
\definecolor{crimson2143940}{RGB}{214,39,40}

\begin{axis}[
width=0.98\fwidth,
height=\fheight,
at={(0\fwidth,0\fheight)},
x grid style={darkgray176},
xmin=-0.6, xmax=3.6,
xtick style={color=black},
xtick={0,1,2,3},
xticklabel style={font=\scriptsize},
xticklabels={$\leq$0.5, 0.5--1, 1--2, 2--10},
xtick pos=bottom,
y grid style={darkgray176},
ylabel={Sample Percentage (\%)},
xlabel={3GPP Sensing Category [m]},
ylabel style={font=\footnotesize},
xlabel style={font=\footnotesize},
ymin=0, ymax=50,
ytick={0,10,20,30,40,50},
ytick pos=left,
ytick style={color=black},
ymajorgrids,
bar width=0.5cm,
]

\draw[draw=black,fill=steelblue31119180,postaction={pattern=north east lines,pattern color=black}] 
    (axis cs:-0.25,0) rectangle (axis cs:0.25,43.2);
\node[font=\scriptsize, above] at (axis cs:0,43.2) {43.2\%};

\draw[draw=black,fill=darkorange25512714,postaction={pattern=north west lines,pattern color=black}] 
    (axis cs:0.75,0) rectangle (axis cs:1.25,31.3);
\node[font=\scriptsize, above] at (axis cs:1,31.3) {31.3\%};

\draw[draw=black,fill=forestgreen4416044,postaction={pattern=horizontal lines,pattern color=black}] 
    (axis cs:1.75,0) rectangle (axis cs:2.25,19.0);
\node[font=\scriptsize, above] at (axis cs:2,19.0) {19.0\%};

\draw[draw=black,fill=crimson2143940,postaction={pattern=vertical lines,pattern color=black}] 
    (axis cs:2.75,0) rectangle (axis cs:3.25,6.5);
\node[font=\scriptsize, above] at (axis cs:3,6.5) {6.5\%};

\end{axis}
\end{tikzpicture}
    \caption{\centering 3GPP categories (unseen runs)}
    \label{chap5-fig:cusense-3gpp-test}
\end{subfigure}
    \hfill
\begin{subfigure}{0.49\linewidth}
    \centering
        \setlength\fwidth{0.8\linewidth}
        \setlength\fheight{0.65\linewidth}
\begin{tikzpicture}
\pgfplotsset{every tick label/.append style={font=\small}}
\definecolor{darkgray176}{RGB}{176,176,176}
\definecolor{cdfblue}{RGB}{66,133,244}
\definecolor{cdffill}{RGB}{173,216,230}
\definecolor{steelblue31119180}{RGB}{31,119,180}
\definecolor{darkorange25512714}{RGB}{255,127,14}
\definecolor{forestgreen4416044}{RGB}{44,160,44}
\definecolor{crimson2143940}{RGB}{214,39,40}

\begin{axis}[
width=0.98\fwidth,
height=\fheight,
at={(0\fwidth,0\fheight)},
x grid style={darkgray176},
y grid style={darkgray176},
xmin=0, xmax=3,
ymin=0, ymax=100,
xtick={0,0.5,1,1.5,2,2.5,3},
ytick={0,20,40,60,80,100},
xtick style={color=black},
ytick style={color=black},
xtick pos=bottom,
ytick pos=left,
xlabel={Peak Distance Error [m]},
ylabel={Sample Percentage (\%)},
ylabel style={font=\small},
xlabel style={font=\small},
xmajorgrids,
ymajorgrids,
]

\addplot[
    thick,
    color=cdfblue,
    fill=cdffill,
    fill opacity=0.4,
] coordinates {
    (0,0) (0.05,1) (0.10,2) (0.15,3.2) (0.20,8) (0.25,15)
    (0.30,23.0) (0.35,28) (0.40,35) (0.45,40) (0.50,43.2)
    (0.55,48) (0.60,53) (0.65,58) (0.70,62) (0.75,66)
    (0.80,69) (0.85,71) (0.90,73) (0.95,74) (1.00,74.5)
    (1.10,78) (1.20,81) (1.30,84) (1.40,86) (1.50,88)
    (1.60,89) (1.70,90) (1.80,91) (1.90,92) (2.00,93.5)
    (2.20,95) (2.40,96) (2.60,97) (2.80,98) (3.00,100)
} \closedcycle;

\addplot[very thick, color=cdfblue, mark=none] coordinates {
    (0,0) (0.05,1) (0.10,2) (0.15,3.2) (0.20,8) (0.25,15)
    (0.30,23.0) (0.35,28) (0.40,35) (0.45,40) (0.50,43.2)
    (0.55,48) (0.60,53) (0.65,58) (0.70,62) (0.75,66)
    (0.80,69) (0.85,71) (0.90,73) (0.95,74) (1.00,74.5)
    (1.10,78) (1.20,81) (1.30,84) (1.40,86) (1.50,88)
    (1.60,89) (1.70,90) (1.80,91) (1.90,92) (2.00,93.5)
    (2.20,95) (2.40,96) (2.60,97) (2.80,98) (3.00,100)
};


\draw[thick, dashed, color=forestgreen4416044] (axis cs:2.00,0) -- (axis cs:2.00,93.5);
\draw[thick, dashed, color=forestgreen4416044] (axis cs:0,93.5) -- (axis cs:2.00,93.5);
\node[fill=white, fill opacity=1, text opacity=1, font=\scriptsize, 
      rounded corners=1pt, inner sep=1pt] 
    at (axis cs:2.35,85.5) {\textcolor{forestgreen4416044}{2m: 93.5\%}};

\draw[thick, dashed, color=darkorange25512714] (axis cs:1.00,0) -- (axis cs:1.00,74.5);
\draw[thick, dashed, color=darkorange25512714] (axis cs:0,74.5) -- (axis cs:1.00,74.5);
\node[fill=white, fill opacity=1, text opacity=1, font=\scriptsize, 
      rounded corners=1pt, inner sep=1pt] 
    at (axis cs:1.6,74.5) {\textcolor{darkorange25512714}{1m: 74.5\%}};

\draw[thick, dashed, color=steelblue31119180] (axis cs:0.50,0) -- (axis cs:0.50,43.2);
\draw[thick, dashed, color=steelblue31119180] (axis cs:0,43.2) -- (axis cs:0.50,43.2);
\node[fill=white, fill opacity=1, text opacity=1, font=\scriptsize, 
      rounded corners=1pt, inner sep=1pt] 
    at (axis cs:1.25,43.2) {\textcolor{steelblue31119180}{50cm: 43.2\%}};

\draw[thick, dashed, color=steelblue31119180] (axis cs:0.30,0) -- (axis cs:0.30,23.0);
\draw[thick, dashed, color=steelblue31119180] (axis cs:0,23.0) -- (axis cs:0.30,23.0);
\node[fill=white, fill opacity=1, text opacity=1, font=\scriptsize, 
      rounded corners=1pt, inner sep=1pt] 
    at (axis cs:1.0,23.0) {\textcolor{steelblue31119180}{30cm: 23.0\%}};

\draw[thick, dashed, color=steelblue31119180] (axis cs:0.15,0) -- (axis cs:0.15,3.2);
\draw[thick, dashed, color=steelblue31119180] (axis cs:0,3.2) -- (axis cs:0.15,3.2);
\node[fill=white, fill opacity=1, text opacity=1, font=\scriptsize, 
      rounded corners=1pt, inner sep=1pt] 
    at (axis cs:0.85,3.7) {\textcolor{steelblue31119180}{15cm: 3.2\%}};
    
\end{axis}
\end{tikzpicture}
    \caption{\centering Error CDF (unseen runs)}
    \label{chap5-fig:cusense-cdf-test}
\end{subfigure}
    \caption{cuSense localization accuracy on test set (a, b) and unseen runs (c, d): 3GPP wireless sensing service category distributions and CDF of peak distance error.}
    \label{chap5-fig:cusense-accuracy}
\end{figure}

\textbf{3GPP Sensing Service Categories.}
We benchmark these results against the 3GPP Release 19 sensing accuracy categories defined for 5G wireless sensing applications~\cite{3gpp22137}. Figures~\ref{chap5-fig:cusense-3gpp-train} and~\ref{chap5-fig:cusense-cdf-train} show the cumulative accuracy at each category threshold, and the error \gls{cdf} for the test set (approximately $35k$ samples from the temporal-random split of training), while Figures~\ref{chap5-fig:cusense-3gpp-test} and~\ref{chap5-fig:cusense-cdf-test} present the same metrics for the two completely unseen runs (over $94$k samples).
On the test set, around 42\% of predictions achieve Category~4 accuracy ($\leq50$~cm), with cumulative sub-meter accuracy of 75\% (Categories~4 and 3). We note that performance on unseen runs remains consistent with 43\% of predictions falling within Category~4 and over 74\% within Category~3, despite these runs being entirely excluded from training. This indicates that the model generalizes well to new trajectories within the same environment rather than overfitting to training data. Over 93\% of predictions meet Category~2 requirements ($\leq2$~m) for the unseen runs, with the remaining predictions within Category~1 ($2$--$10$~m). These results demonstrate that cuSense meets \gls{3gpp} sensing requirements for indoor localization applications such as asset tracking, occupancy monitoring, and intelligent building automation.
%

\begin{figure}[!htb]
  \centering

\begin{tikzpicture}
    \begin{groupplot}[
        group style={
            group size=1 by 2,
            vertical sep=0.15cm,
            xlabels at=edge bottom,
        },
        width=\linewidth,
        height=3cm,
        xmin=0, xmax=2000,
        grid=major,
        grid style={gray!30},
        tick label style={font=\scriptsize},
        label style={font=\small},
    ]
    
    \nextgroupplot[
        ylabel={X Position},
        ymin=0, ymax=43,
        xticklabels={},
        legend columns=2,
        legend style={
            fill opacity=0.8,
            draw opacity=1,
            text opacity=1,
            draw=lightgray204,
            font=\footnotesize,
            at={(0.75, 1.65)},
            anchor=north east,
            column sep=5pt,
        },
    ]
    
    \addplot[rawcolor, opacity=0.6, line width=0.8pt] 
        table[x=sample, y=raw_x, col sep=comma] {fig/chap5/cusense/run12_kalman_ultra_smooth_trajectory_tikz.csv};
    \addlegendentry{Raw}
    
    \addplot[avgcolor, line width=1.2pt] 
        table[x=sample, y=avg_x, col sep=comma] {fig/chap5/cusense/run12_kalman_ultra_smooth_trajectory_tikz.csv};
    \addlegendentry{Bayesian Averaged}
    
    \addplot[kalmancolor, line width=1.5pt] 
        table[x=sample, y=kalman_x, col sep=comma] {fig/chap5/cusense/run12_kalman_ultra_smooth_trajectory_tikz.csv};
    \addlegendentry{Kalman Filtered}
    
    \addplot[gtcolor, line width=1.5pt, densely dashed] 
        table[x=sample, y=gt_x, col sep=comma] {fig/chap5/cusense/run12_kalman_ultra_smooth_trajectory_tikz.csv};
    \addlegendentry{Ground Truth}
    
    \nextgroupplot[
        ylabel={Y Position},
        ymin=0, ymax=64,
        xlabel={Sample Index},
    ]
    
    \addplot[rawcolor, opacity=0.6, line width=0.8pt] 
        table[x=sample, y=raw_y, col sep=comma] {fig/chap5/cusense/run12_kalman_ultra_smooth_trajectory_tikz.csv};
    
    \addplot[avgcolor, line width=1.2pt] 
        table[x=sample, y=avg_y, col sep=comma] {fig/chap5/cusense/run12_kalman_ultra_smooth_trajectory_tikz.csv};
    
    \addplot[kalmancolor, line width=1.5pt] 
        table[x=sample, y=kalman_y, col sep=comma] {fig/chap5/cusense/run12_kalman_ultra_smooth_trajectory_tikz.csv};
    
    \addplot[gtcolor, line width=1.5pt, densely dashed] 
        table[x=sample, y=gt_y, col sep=comma] {fig/chap5/cusense/run12_kalman_ultra_smooth_trajectory_tikz.csv};
    
    \end{groupplot}
\end{tikzpicture}
  \vspace{-10pt}
  \caption{Trajectory tracking comparison of an unseen run segment for the X and Y axis.}
  \label{chap5-fig:cusense-trajectory}
\end{figure}
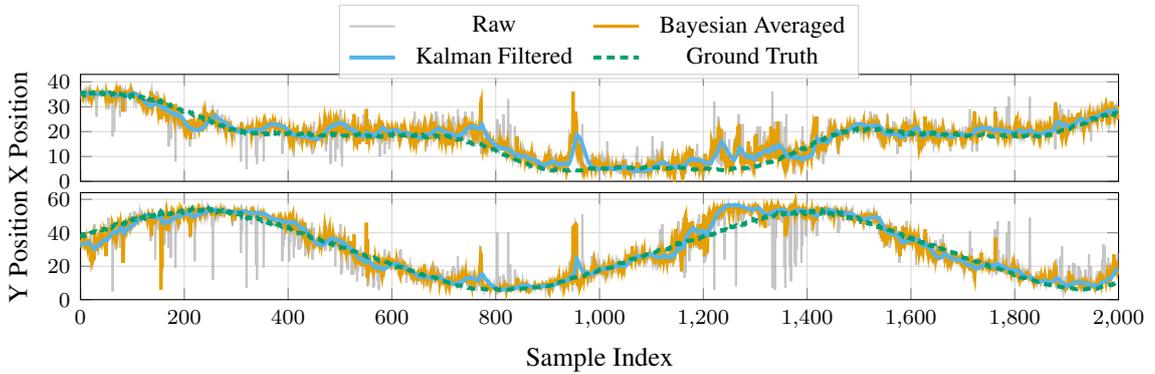

\textbf{Trajectory Tracking.}
Figure~\ref{chap5-fig:cusense-trajectory} visualizes the X and Y position estimates over time for a representative test segment (downsampled to 10 for clarity), comparing raw model predictions, temporally averaged predictions, Kalman-filtered output, and ground truth. The raw predictions exhibit significant frame-to-frame jitter from measurement noise.
Temporal averaging reduces this variability, while the Kalman filter produces smooth, physically plausible trajectories that closely track the ground truth. The multi-stage refinement provides cumulative error reduction of 30--40\% compared to raw predictions, with the Kalman filter successfully handling both slow movements and rapid direction changes.

\textbf{Real-Time dApp Inference Performance.}
Table~\ref{chap5-tab:cusense-latency} shows the computational performance of the cuSense dApp inference pipeline. The complete pipeline--including \gls{csi} preprocessing (background removal and temporal averaging), \gls{nn} inference via Triton, prediction averaging, and Kalman filtering--achieves a mean \gls{e2e} latency of $1.36$~ms per sample with PyTorch, and $0.65$~ms with \gls{trt} acceleration.
Combined with the dApp framework overhead of 495~$\mu$s (Table~\ref{chap4-tab:datapath}), the total cuSense control-loop latency is approximately $1.85$~ms (or $1.15$~ms with \gls{trt}), well within real-time requirements for tracking applications.

\begin{table}[htb]
\centering
\small
\caption{cuSense inference latency breakdown.}
\label{chap5-tab:cusense-latency}
\begin{tabular}{lc}
\toprule
\textbf{Operation} & \textbf{Overhead [ms]} \\
\midrule
CSI preprocessing & 0.3 \\
Neural network inference & 0.8 (PyTorch) - 0.1 (TRT) \\
Bayesian temporal averaging & 0.1 \\
Kalman filter tracking & 0.15 \\
Total inference Latency & 1.35 (PyTorch) - 0.65 (TRT) \\
\bottomrule
\end{tabular}
\end{table}

\subsection{Related Work}
\label{chap5-subsec:cusense-related}



\gls{isac} has been widely studied as a key enabler for beyond-\gls{5g} and \gls{6g} networks, with recent surveys reviewing system architectures, performance limits, and open challenges~\cite{isac2024lu,isac2022liu}. In the context of \gls{5g}, recent work has explored communication-centric \gls{isac} schemes for cooperative target localization and \gls{ul}-collaborative sensing using \gls{ofdm}/\gls{nr} signals~\cite{zhang2025isac,huang2025ul}, as well as passive radar sensing and \gls{csi}-based localization using \gls{nr} reference signals~\cite{dwivedi2024radar,ruan2022hiloc,bouknana2025oran}. Earlier Wi-Fi sensing systems~\cite{adib2014tracking,adib2013wifi,wang2019iccv} have demonstrated that commodity Wi-Fi signals and \gls{csi} can enable fine-grained localization and person perception. cuSense is complementary to this literature: rather than using Wi-Fi or standalone receivers, it implements \gls{csi}-based indoor localization as an uplink-collaborative \gls{isac} service directly on a \gls{3gpp}-compliant, GPU-accelerated \gls{5g} \gls{gnb}, exposing it as a programmable dApp within an O-RAN/AI-RAN framework without dedicated sensing hardware or changes to the \gls{ran} stack.

\subsection{Summary}
\label{chap5-subsec:cusense-summary}

This section presented cuSense, an uplink \gls{dmrs} \gls{csi}-based indoor localization dApp operating in real time on a production-grade \gls{5g} network without dedicated sensing hardware or \gls{ran} modifications. Building on the GPU-accelerated dApp framework introduced in Section~\ref{chap4-sec:dapp-framework}, cuSense processes channel estimates through a three-stage pipeline (background characterization, temporal noise reduction, and neural network inference) to achieve sub-meter localization accuracy with over 74\% of predictions within the \gls{3gpp} Category~3 threshold ($\leq 1$~m).

The experimental evaluation on X5G validates both the \emph{realism} and \emph{use-case} dimensions of this dissertation. On one hand, real \gls{ota} channels with commercial \glspl{ue} provide ground-truth validation that would be difficult to achieve in emulation. On the other hand, the complete data-collection, training, and inference pipelines demonstrate the feasibility of \gls{isac} deployment on physical platforms, particularly on a production-grade \gls{5g} \gls{ran} testbed.

The current \gls{ul}-collaborative bistatic architecture with a transmitting \gls{ue} offers a practical deployment model but limits spatial diversity and relies on environment-specific calibration. These constraints motivate future work on dynamic environments, multi-cell configurations, and sensing modes that reduce the need for explicit device collaboration. The dApp framework and cuSense pipelines are planned for open-source release to enable further research on \gls{gpu}-accelerated real-time applications for \gls{ai}-native \gls{ran}.

\section{InterfO-RAN: In-band Uplink Interference Detection}
\label{chap5-sec:interforan}


This section explores another real-time dApp use case focusing on interference detection: InterfO-RAN. InterfO-RAN is a custom GPU-accelerated dApp implementation embedded within the NVIDIA Aerial CUDA pipeline of X5G, which targets in-band \gls{ul} interference that degrades network performance in dense deployments.
This work further validates the flexibility of the X5G physical platform and its dApp architecture, demonstrating how the same infrastructure can support diverse \gls{ai}-driven applications operating at sub-millisecond timescales.

\subsection{Introduction}
\label{chap5-subsec:interforan-intro}

The evolution of wireless communication networks has been driven by the increasing demand for high-speed, low-latency, and ultra-reliable connectivity to support emerging applications such as autonomous systems, industrial automation, and immersive experiences. To meet these demands, 5G-and-beyond networks leverage massive densification to improve throughput and latency within the limited spectrum allocated to mobile communications~\cite{ultradense}. 
In ultra-dense networks, multiple neighboring small cells are configured to reuse the same frequency bands to increase frequency reuse~\cite{ultradense5g}. 
Directional millimeter wave systems further push this concept, with commercial deployments often leveraging the same $400$~MHz or $800$~MHz bands at $28$~GHz and $39$~GHz for all base stations~\cite{aarayanan2022comparative}.
Finally, private network deployments (e.g., for enterprise scenarios) use limited portions of shared spectrum, with all deployments constrained in the same $100$~MHz (e.g., Germany, Brazil, Netherlands) or $150$~MHz (e.g., U.S., with \acrshort{cbrs})~\cite{spec_availability}.

\begin{figure}[htb]
  \centering
   \includegraphics[width=0.8\linewidth]{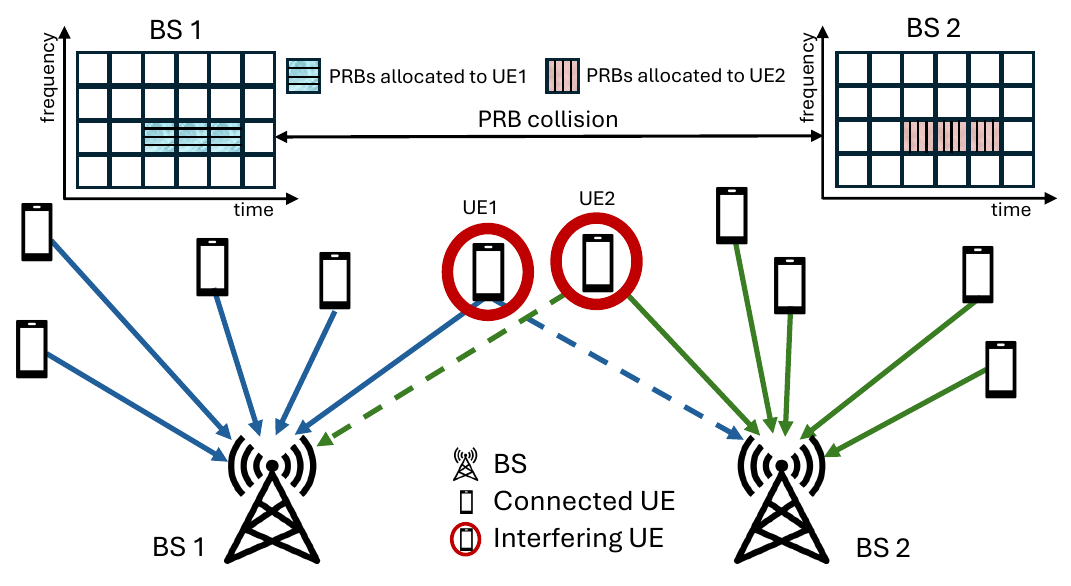}
  \caption{In-band \acrshort{ul} interference overview (dashed lines).}
  \label{chap5-fig:interforan-schematic}
\end{figure}

%
However, densification, sharing, and spectrum reuse also increase inter-cell interference. Although significant research has focused on coordination mechanisms to reduce interference in downlink~\cite{8938771,9405679,9411723,7536929},
in-band \gls{ul} interference remains a challenging problem, especially considering the unpredictability of user mobility and configurations and thus of the source of interference. This occurs when unwanted transmissions from \glspl{ue}, operating within the same frequency band but connected to different \glspl{gnb}, interfere with the \gls{ul} reception of a serving \gls{gnb}, as shown in Figure~\ref{chap5-fig:interforan-schematic}. Such an overlap can significantly degrade the \gls{ul} \gls{sinr} at the affected \gls{gnb}, potentially leading to unrecoverable packets and thus reduced network performance~\cite{7127550}.
This is particularly significant when \glspl{ue} are located at the cell edge, where overlapping regions between two neighboring cells force the devices to increase their transmit power to overcome the link budget constraints. 
In such cases, the unpredictability of potential in-band \gls{ul} interference is closely associated with \gls{ue} power control. Similarly, \glspl{ue} may transmit at excessive power levels to compensate for signal degradation when experiencing \gls{nlos} conditions with respect to their serving node. While this adjustment helps maintain reliable links, it can also lead to unintended interference in adjacent cells. 
Moreover, in scenarios involving directional \gls{ue} transmissions, as in millimeter wave links, beamforming introduces an additional source of unpredictability. It can impact adjacent cells through strong sidelobes or, under highly dynamic conditions, even affect them with the main lobe. These conditions are often unknown a priori (e.g., at scheduling time), and can vary on timescales faster than a subframe (i.e., sub-ms). This calls for real-time interference detection to enable a prompt reaction across the protocol stack.

The impact of in-band \gls{ul} interference extends beyond immediate signal degradation, and can disrupt protocol operations like scheduling and resource allocation at the \gls{gnb}. To execute these operations, the \gls{gnb} relies on quality indicators like \gls{rsrp} and \gls{rssi}. However, interference distorts the \gls{gnb}'s assessment of \gls{ue} channel conditions, introducing errors that propagate across multiple slots and affect both \gls{ul} and \gls{dl} transmissions. This disruption becomes particularly critical when interference affects control signaling (e.g., \gls{uci}) or \gls{dmrs} used for channel estimation. 
Because of the inherently lower power of \gls{ul} signals compared to \gls{dl} transmissions, even interference at low power levels can significantly impair communication quality. This forces the \gls{gnb} to allocate additional resources through retransmissions and more robust coding schemes, thereby reducing spectral efficiency and limiting overall user capacity. In scenarios with severe interference, connection drops can occur, underscoring the importance of developing and implementing robust interference detection and mitigation methods in practical deployments.

To address this challenge, this section presents InterfO-RAN, a real-time programmable solution that integrates a \gls{cnn} within the \gls{gnb} physical layer to process and analyze \gls{iq} samples. The solution achieves interference detection with accuracy exceeding $91$\% in less than $650$ $\mu$s, enabling practical implementation in production cellular networks.
Designed as a pluggable, programmable component, InterfO-RAN extends \gls{gpu} acceleration to \glspl{dapp} for real-time \gls{cnn}-based interference detection. It is integrated with the NVIDIA Aerial \gls{5g} \gls{nr} \gls{phy} layer, following the emerging O-RAN \gls{dapp} paradigm~\cite{lacava2025dapps} and aligning with AI-for-RAN use cases within the AI-RAN Alliance~\cite{airan}.
InterfO-RAN is implemented as a custom application embedded in the Aerial CUDA pipeline to access the necessary \glspl{kpm}, including \gls{iq} samples and channel quality data, and to allocate the computational resources required for proper execution. Comparable performance can also be achieved by deploying its intelligence within the dApp framework presented in Section~\ref{chap4-sec:dapp-framework}.

We design, develop, and test InterfO-RAN on the X5G testbed, leveraging this real-world setup to collect data with and without interference from a deployment spanning two different buildings. More than seven million \gls{nr} \gls{ul} slots, together with artificially generated data in MATLAB, are used to train and test eight configurations of the \gls{cnn}, employing \gls{tl} techniques as well. The selected solution is then evaluated \gls{ota} across various scenarios with different pairs of interfering \glspl{ru}. Our results demonstrate the robustness of InterfO-RAN in detecting interference, achieving over 91\% accuracy in less than $650~\mu$s, while imposing minimal strain on \gls{phy} operations.

The remainder of this section is organized as follows. Section~\ref{chap5-subsec:interforan-system} presents the system architecture overview. Section~\ref{chap5-subsec:interforan-phy} provides some background on the Aerial framework. Section~\ref{chap5-subsec:interforan-design} describes the system design and implementation of the InterfO-RAN dApp, including details on the \gls{cnn} and data pipelines. Section~\ref{chap5-subsec:interforan-framework} outlines the experimental setup and data collection operations, while Section~\ref{chap5-subsec:interforan-results} discusses the results. Related work is reviewed in Section~\ref{chap5-subsec:interforan-related}. Finally, Section~\ref{chap5-subsec:interforan-summary} summarizes the main findings and outlines directions for future research.

\subsection{System Architecture}
\label{chap5-subsec:interforan-system}

\begin{figure}[htb]
  \centering
    \includegraphics[width=0.8\linewidth]{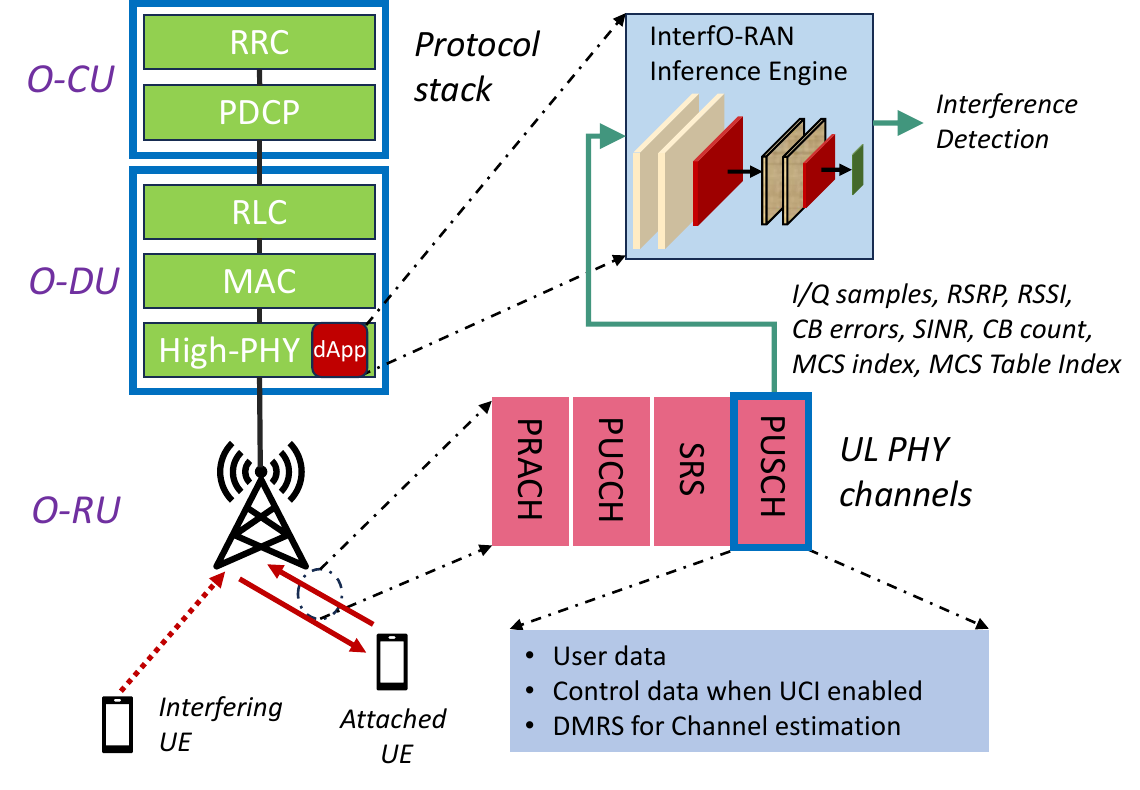}
  \caption{System Architecture.}
  \label{chap5-fig:interforan-system-archi}
\end{figure}

Figure~\ref{chap5-fig:interforan-system-archi} illustrates our system architecture, where InterfO-RAN is integrated as a programmable dApp into the high-\gls{phy} of a \gls{5g} \gls{nr} protocol stack. 
%
Specifically, we design the system to harness \gls{gpu} resources available as part of the NVIDIA Aerial \gls{gpu}-accelerated platform. 
Aerial already uses \gls{gpu} to offload computationally intensive and time-sensitive high-\gls{phy} layer workloads, contributing approximately $85$\% of the overall computational complexity and processing demands in \glspl{gnb}~\cite{kundu2023hardwareaccelerationopenradio}.
This architecture enables efficient resource sharing, allowing the dApp to perform inference with minimal computational overhead while preserving real-time processing capabilities. 
As explained in Section~\ref{chap5-subsec:interforan-design}, InterfO-RAN is implemented through customized functions utilizing a \gls{cnn} architecture, executed on a \gls{gpu} for real-time interference detection. The output of InterfO-RAN is a binary indicator that denotes the presence of interference. Once interference is detected, different strategies can be put in place for mitigation (e.g., coordinated resource allocation), which are left for future work.

The system processes input features comprising a combination of raw \gls{iq} samples within a slot carrying \gls{pusch} data, along with \gls{rssi}, \gls{rsrp}, the number of \gls{cb} errors, the total count of \glspl{cb}, \gls{sinr}, \gls{mcs} index, and \gls{mcs} table index. The \gls{iq} samples are extracted from the \gls{pusch} physical channel, without any processing such as equalization, demodulation, or decoding. InterfO-RAN does not interfere with ongoing \gls{ul}/\gls{dl} processes, let alone \gls{pusch} channel processes, while determining whether transmissions are affected by unwanted signals within the same frequency band.

Figure~\ref{chap5-fig:interforan-system-archi} further illustrates the \gls{ul} interaction between \gls{gnb} and \gls{ue}, providing a comprehensive view of the data flow and the \gls{pusch} channel engaged by InterfO-RAN. The \gls{pusch} carries data, control information, and \gls{dmrs} for channel estimation. The figure also highlights the other uplink channels, i.e., \gls{pucch}, \gls{srs}, and \gls{prach}.

In the next sections, we provide a preliminary overview of the GPU-accelerated framework for physical layer processing, and then detail how InterfO-RAN is designed and implemented in detail.

\subsection{GPU-Accelerated Physical Layer Processing}
\label{chap5-subsec:interforan-phy}

This section provides a detailed overview of the framework we leverage as the basis for the InterfO-RAN implementation, as well as notions on the physical layer \gls{pusch}.

\subsubsection{NVIDIA Aerial Physical Layer}
\label{chap5-subsubsec:interforan-aerial}

InterfO-RAN is embedded as a functional plugin in NVIDIA Aerial \gls{cubb}, a \gls{sdk} developed by NVIDIA that provides a \gls{5g} signal processing pipeline for \glspl{gpu}, implemented in \gls{cuda} and C++~\cite{arc-ota}. \gls{cubb} operates with slot-level granularity, with each slot lasting $500$~$\mu$s, aligned with a $30$~kHz subcarrier spacing. InterfO-RAN complements \gls{cubb} with real-time, high-performance inference.
NVIDIA Aerial uses a \gls{gpu} for inline acceleration of resource-intensive tasks such as channel estimation, \gls{ldpc} encoding/decoding, and channel equalization, among others.

From an implementation standpoint, \gls{cubb} combines several software components, which we extend to design and implement InterfO-RAN. The \gls{cuphy} controller serves as the primary coordinator, initializing \gls{gpu} resources and creating the initial context for \gls{ru} connections. During \gls{ota} \gls{ul}/\gls{dl} transmissions, the L2 adapter within the \gls{cuphy} controller translates control plane messages from the \gls{scf} \gls{fapi} interface into slot commands for \gls{phy}-layer data traffic. These commands are processed by the \gls{cuphy} controller, converted into tasks, and assigned to specific \gls{ul} and \gls{dl} channel worker threads. Mapped to appropriate \gls{cpu} cores, these threads delegate computational tasks to the \gls{gpu}. Each thread represents a physical channel implemented as a pipeline, with operations executed on \gls{cuda} kernels. \gls{cuda} kernels are functions executed in parallel on a \gls{gpu}. The pipeline is supported by \glspl{api} provided by the \gls{cuphy} library, invoked by the \gls{cuphy} driver to manage creation, configuration, and execution. 

\subsubsection{PUSCH Operations}
\label{subsec:pusch-cubb}

InterfO-RAN focuses on the \gls{ul} receive path, particularly the \gls{pusch}, to detect interference. \gls{ul} processing starts with slot configuration from the L2 layer and ends with \gls{ul} signal processing. Among \gls{ul} channels (\gls{pusch}, \gls{pucch}, \gls{prach}, \gls{srs}), InterfO-RAN analyzes the \gls{pusch}, the primary \gls{ul} channel for data transmission, to detect interference from other \glspl{ue} sharing the same radio resources. 

\gls{pusch} transmissions are dynamically scheduled by the \gls{gnb}, which allocates frequency and time resources based on real-time network conditions. The resource allocation is signaled to the \gls{ue} through \gls{dci} messages, specifying parameters such as \gls{mcs}, resource block allocation, and transmission power~\cite{dahlman2020}.
The \gls{pusch} channel in \gls{cubb} is implemented as a pipeline, for operations such as \gls{re} demapping, channel estimation, channel equalization, de-rate matching, de-layer mapping, de-scrambling, \gls{ldpc} decoding, and \gls{crc} checks for both \glspl{cb} and \glspl{tb}. Additionally, it includes processing for \gls{uci} on \gls{pusch} and transform precoding, if enabled, executed through individual \gls{cuda} kernels. Specialized kernels also manage \gls{ta} and \gls{cfo}.

\begin{figure}[hbt]
  \centering
    \includegraphics[width=.8\linewidth]{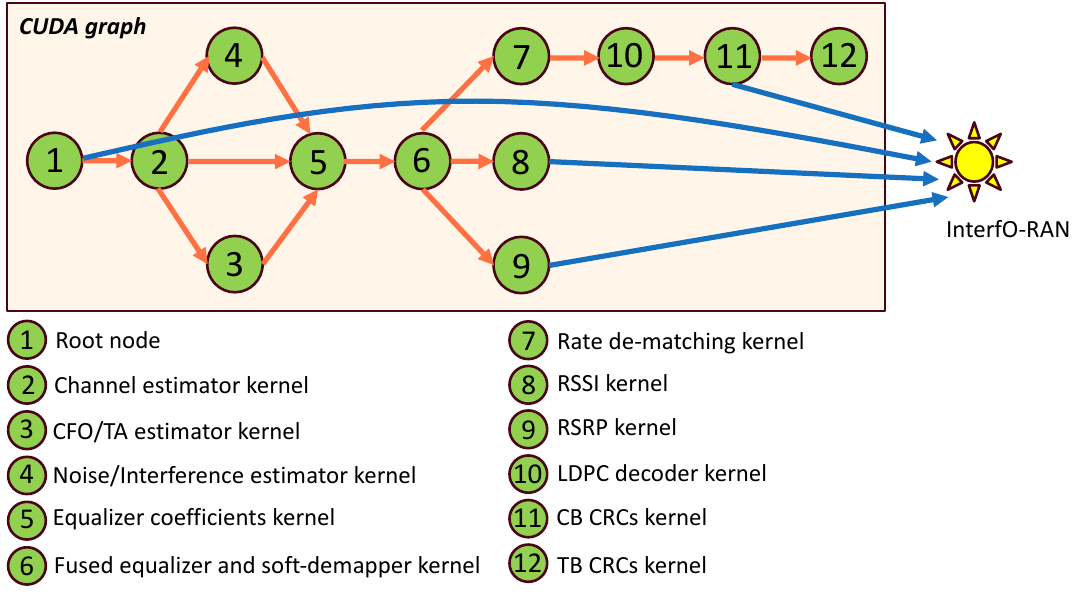}
  \caption{\justifying InterfO-RAN as a post-processing component following the PUSCH CUDA graph, where each node represents a CUDA kernel.}
  \label{chap5-fig:interforan-cuda-graph}
\end{figure}

This pipeline is further optimized and structured using an advanced feature known as \gls{cuda} graph, which represents an acyclic graph where nodes correspond to kernels, and edges define their inter-dependencies, as illustrated in Figure~\ref{chap5-fig:interforan-cuda-graph}. This graph-based architecture effectively delineates the intricate workflow of kernel operations, enabling the management of dependencies within the graph itself. By facilitating the concurrent launch of all kernels as a unified entity, the \gls{gpu} autonomously oversees the execution process, thereby reducing the necessity for continuous \gls{cpu} intervention. This method significantly enhances temporal efficiency by minimizing launch overhead, which is the latency associated with initiating \gls{cuda} kernels.

\subsection{InterfO-RAN Design and Implementation}
\label{chap5-subsec:interforan-design}

We design and implement InterfO-RAN on top of the NVIDIA Aerial \gls{sdk}, leveraging the spare capacity of the \gls{gpu} to execute the \gls{cnn}-based dApp.
As discussed in detail in the results analysis in Section~\ref{chap5-subsec:interforan-results}, the typical execution of the \gls{phy} layer in \gls{cubb} utilizes at most $50$\% of the \gls{gpu} resources, leaving substantial computational capacity available for additional tasks. This excess capacity enables InterfO-RAN to take advantage of the remaining \gls{gpu} resources for interference detection.
In doing this, however, it becomes necessary to design the dApp to efficiently access the available resources and to avoid interfering with real-time processing constraints of the \gls{gnb} physical layer. 

\subsubsection{dApp Design and PHY Integration}
\label{chap5-sec:interforan-dappdesign}

\textbf{Design and Implementation Challenges.} To design and integrate InterfO-RAN within the 5G NR GPU-accelerated PHY, we addressed the following challenges. First, the dApp needs to access information available across different processing steps of the \gls{pusch} pipeline (i.e., \gls{iq} samples, \gls{rsrp}, \gls{rssi}, the sum of \gls{cb} \gls{crc} errors, count of \glspl{cb} derived from \gls{cb} \glspl{crc}, \gls{sinr}, \gls{mcs} Index and \gls{mcs} table index). This can amount to 183,484 bytes for each \gls{ul} transport block, thus requiring an efficient mechanism to expose such information. Second, the \gls{cnn} needs to return results in less than a millisecond (i.e., a 5G NR subframe) to make sure that the output is relevant to take further decisions across the stack. Therefore, the GPU-based implementation needs to be efficient in running inference with the provided input. In addition, there may be a need to support different models or configurations for the \gls{ai} processing, thus managing the lifecycle of the overall model. 
For these reasons, we select \gls{ort} as the provider, making it possible to efficiently deploy trained models on the GPU-based dApp. At the same time, \gls{ort} requires a significant loading and configuration time the first time it is executed. In general, as part of our design, it is important to avoid disrupting operations of the rest of the physical layer, thus guaranteeing that the timing of any operation within the dApp does not affect physical layer processing. 
Finally, for development purposes, the InterfO-RAN implementation needs to enable automated data collection and labeling, to streamline the gathering of samples for the training of \gls{ai} models.

\begin{figure}[htb]
  \centering
  \includegraphics[width=\linewidth]{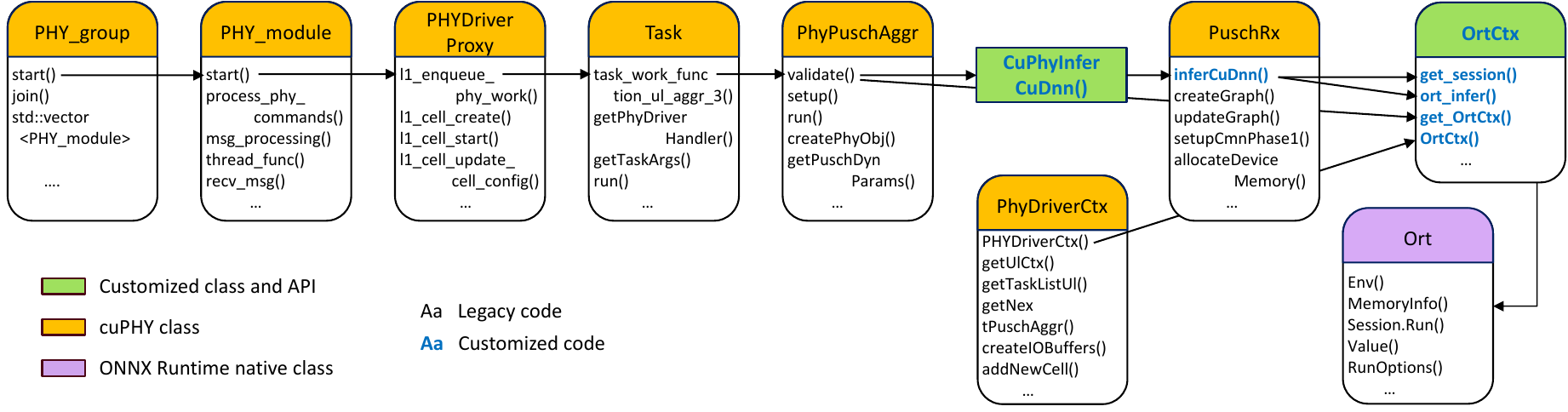}
  \caption{The workflow diagram for the OrtCtx class for performing inference.}
  \label{chap5-fig:interforan-uml}
\end{figure}

\textbf{InterfO-RAN System Structure.}
As the core component of InterfO-RAN, the inference module functions as a post-processing unit following the execution of the \gls{pusch} pipeline, ensuring that ongoing \gls{ul}/\gls{dl} pipeline processes remain unaffected, as illustrated in Figure~\ref{chap5-fig:interforan-cuda-graph}. This functionality is efficiently implemented through the function \texttt{CuPhyInferCuDnn()}---a customized \gls{api} hosted by \gls{cuphy} and managed by the \gls{cuphy} driver, as indicated in Figure~\ref{chap5-fig:interforan-uml}. 
This \gls{api} further employs the \texttt{Session.Run()} method from the \gls{ort} framework to execute the actual inference process. To prevent InterfO-RAN from interfering with the tightly time-coupled operations of the \gls{pusch} or other channels, the \gls{cuphy} driver delegates the processing to a newly instantiated and independent \gls{cpu} thread. 
This thread is assigned to a dedicated \gls{cpu} core---isolated from other operations---to improve stability. 

\textbf{ORT Integration and Tuning.}
\gls{ort} is built to execute models in the \gls{onnx} format, which is an open standard for representing \gls{ml} models. \gls{ai} operations using \gls{ort} can be executed on either the \gls{cpu}, utilizing the \gls{cpu} execution provider, or on the \gls{gpu}, utilizing \gls{cuda} or \gls{trt} execution providers. Our implementation supports all three options, with processing latency shown in Table~\ref{chap5-tab:interforan-ort-times}, for inference using standalone \gls{ort} (i.e., independently of the dApp implementation) and the model discussed in Section~\ref{chap5-subsubsec:interforan-cnn}. \gls{cpu}-based inference takes approximately $30.354$~ms for a 24-core Intel Xeon Gold CPU, while \gls{gpu}-based inference using \gls{trt} and \gls{cuda} takes $0.524$~ms and $3.675$~ms using an NVIDIA A100, respectively. 

\begin{table}[htb]
    \centering
    \caption{Inference time between different execution providers for standalone ORT.}
    \label{chap5-tab:interforan-ort-times}
    \scriptsize
    \begin{tabular}{lccc}
        \toprule
        \textbf{ORT Execution Provider} & CPU & CUDA & TensorRT \\
        \midrule
        \textbf{Inference Time [ms]} & 30.354 & 3.675 & 0.524 \\
        \bottomrule
    \end{tabular}
\end{table}

Although both \gls{cuda} and \gls{trt} utilize the \gls{gpu} for inference tasks, \gls{trt} is specifically optimized for high performance. NVIDIA \gls{trt} is an \gls{sdk} for deep learning model optimization, featuring an inference optimization engine and runtime environment. As evidenced in Table~\ref{chap5-tab:interforan-ort-times}, it achieves lower execution times than both \gls{cpu} and \gls{cuda} providers. The framework parses the \gls{onnx} model, applies optimizations such as layer fusion, and selects efficient \gls{cuda} kernels by leveraging libraries like \gls{cudnn} and \gls{cublas}. Consequently, \gls{trt} is used exclusively in our experiments to maximize performance. 

Based on our measurements, however, the \gls{gpu}-based \gls{ort} mode comes with a significant delay during initial execution---$1.25$\:s and $9.65$\:s with \gls{cuda} and \gls{trt}, respectively---for the model described in Section~\ref{chap5-subsubsec:interforan-cnn}. 
These findings identify the root cause of the timing issue as the long \gls{gpu} warm-up phase.
This is because \gls{ort} requires several seconds to initialize and allocate \gls{gpu} resources essential for inference. In particular, during the first call to \texttt{Session.Run()}, \gls{ort} performs several critical setup steps:
(i) Memory allocation for the model tensors;
(ii) Graph optimization and compilation to ensure efficient runtime execution;
(iii) Execution provider setup, such as configuring \gls{cuda} for \gls{gpu} inference; and
(iv) Caching mechanisms for future executions.
This warm-up latency conflicts with the strict timing constraints of the \gls{ul} slot in \gls{5g} \gls{nr} systems, making it a challenge for real-time execution. However, once initialized, these setup steps are cached, allowing subsequent invocations to bypass the preparatory steps and run significantly faster.

\textbf{PHY Layer and dApp Setup.} Therefore, we design the InterfO-RAN dApp so that it performs an initial warm-up inference before the base station becomes operational. 
Particularly, during the initialization of \gls{cubb}, a preliminary inference step is performed on dummy data encompassing all input features. Then a pointer to the \gls{ort} run session (already initialized) is passed to InterfO-RAN. 
To support this, InterfO-RAN is encapsulated within a customized \gls{ort} context class, \texttt{OrtCtx}, which is instantiated within the \gls{cuphy} driver context class, \texttt{PhyDriverCtx}, as illustrated in Figure~\ref{chap5-fig:interforan-uml}.
This manages instances of transmit and receive processing pipelines for \gls{ul} and \gls{dl} slots, as well as individual classes corresponding to each \gls{phy} channel, among others. Additionally, it handles the allocation of \gls{gpu} resources and manages the \gls{gpu} footprint for all \gls{ul} and \gls{dl} executions. \texttt{PhyDriverCtx} is instantiated as the whole physical layer instance starts, making it possible to execute the warm-up phase for \gls{ort}. 

As part of this process, the \texttt{OrtCtx} class constructor initializes and configures the dynamically allocated instance of an \gls{ort} inference session as a smart pointer, using the native \gls{ort} \gls{api} and leveraging its comprehensive classes and methods, as shown in Figure~\ref{chap5-fig:interforan-uml}.  
The process starts with creating an \texttt{Ort::Env()} object for logging and runtime management, followed by configuring session options for optimization, including graph processing, threading, memory allocation, and \gls{gpu}-specific settings for \gls{cuda} and \gls{trt}.
The dummy input data for warm-up inference is allocated in \gls{cpu} memory and transferred to the \gls{gpu} for execution, encapsulated as \texttt{Ort::Value()} tensors. \texttt{Session.Run()} processes the data through the pre-trained \gls{onnx} model and generates output tensors containing class probabilities, selecting the highest probability as the final prediction. 

\textbf{Runtime Inference.} Upon initialization, the \texttt{OrtCtx} object is passed to the \gls{pusch} channel via the physical \gls{pusch} aggregate class, \texttt{PhyPuschAggr}, and invoked during the validation phase as a callable function, \texttt{get\_OrtCtx()}. This validation phase is a post-processing step always executed after the \gls{pusch} pipeline. Inference in the \gls{pusch} pipeline is orchestrated by \texttt{inferCuDnn()}, a member of \texttt{PuschRx} class, and invoked by \texttt{CuPhyInferCuDnn()} \gls{api}, as illustrated in Figure~\ref{chap5-fig:interforan-uml}. It retrieves the inference session from the \texttt{OrtCtx} object via \texttt{get\_session()}, reusing the session configuration and runtime environment. We encapsulate input features into \gls{onnx}-compatible tensors using \texttt{ort\_infer()}, using \gls{gpu} memory for \gls{cuda}/\gls{trt} or \gls{cpu} memory otherwise. When performing inference on the \gls{gpu}, the input features are transferred directly within \gls{gpu} memory without overhead. Tensors follow predefined shapes and types, with input/output names built dynamically for runtime flexibility.

\textbf{dApp-based Automated Data Collection.} We also designed the InterfO-RAN dApp to perform automated data collection to create datasets for offline model training. This leverages a \texttt{CuPhyInfer\-CuDnn()} \gls{api} to stream data into an \gls{hdf5} file at a frequency of one every 10~\gls{ul} slots. Different from what happens for online inference, the \gls{api} asynchronously transfers features to be stored from \gls{gpu} to \gls{cpu} and writes to disk thereafter. 
This avoids many \gls{gpu} operations, including
data conversion to tensors and subsequent \gls{gpu} memory deallocation. 
Besides, to avoid interference with the tightly coupled operations of \gls{pusch} or other channels, the \gls{cuphy} driver delegates data logging to a newly instantiated \gls{cpu} thread operating in detached mode, ensuring uninterrupted processing.

\subsubsection{AI-Based Interference Detection}
\label{chap5-subsubsec:interforan-cnn}

InterfO-RAN leverages a \gls{cnn} for real-time interference classification, using a data-driven methodology that allows the network to identify complex patterns in incoming \gls{ul} data. 

\textbf{Input Features.}
As discussed above, the input features include \gls{iq} samples for the \gls{pusch}, as well as additional \glspl{kpm} representing \gls{rssi}, \gls{rsrp}, \gls{sinr}, \gls{mcs} index, \gls{mcs} table index, total count of \gls{cb} errors, and number of \glspl{cb} derived from \gls{cb}-\glspl{crc}. As part of the dApp design, one challenge was ensuring the compatibility of heterogeneous numerical formats, such as \textit{float16} (for \gls{iq}), \textit{float32} (for \gls{rssi}, \gls{rsrp}, \gls{sinr}), and \textit{int32} (for \gls{mcs} and \gls{cb} inputs), across frameworks such as \gls{tf} (offline training) and \gls{cubb}.

Unlike traditional approaches that rely on heatmaps of \gls{iq} samples for interference detection~\cite{8885870,9600524}, InterfO-RAN processes raw \gls{iq} data directly and incorporates additional features to improve classification accuracy.
\gls{iq} samples are extracted prior to the \gls{mmse}-\gls{irc} equalizer, which is designed to mitigate channel distortion and interference. This pre-equalization extraction provides an unaltered view of interference effects, allowing for more accurate analysis. 
In \gls{cubb}, the \gls{pusch} channel \gls{iq} samples are stored as 32-bit words (with two 16 bits floats for each component), in blocks of $14\times273\times12$ contiguous words, where $14$ is the number of symbols per slot, $273$ is the maximum number of \glspl{prb} with $12$ subcarriers each.  
\gls{ort} converts the \gls{iq} samples into a matrix with dimensions $14, 3276, 2$, to align with \gls{tf}'s expected format for inference, interpreting the third dimension as the real and imaginary components of the \glspl{iq}.

Additionally, specific input features are transformed to ensure data compatibility. In the native \gls{cb}-\gls{crc} \gls{cuda} kernel, the \gls{cb} errors for each \gls{pusch} channel are output as an array, where each non-zero value represents a corrupted \gls{cb} within the \gls{tb}, resulting in a dynamic size. To meet the fixed-size input requirements of the learning model, two additional variables are added to the \gls{gpu}: one to determine the number of corrupted \glspl{cb} in a \gls{tb} and another to find the total number of \glspl{cb}. These operations are seamlessly integrated into the kernel 11 in Figure~\ref{chap5-fig:interforan-cuda-graph}.

\begin{figure}[htb]
  \centering
  \includegraphics[width=.9\linewidth]{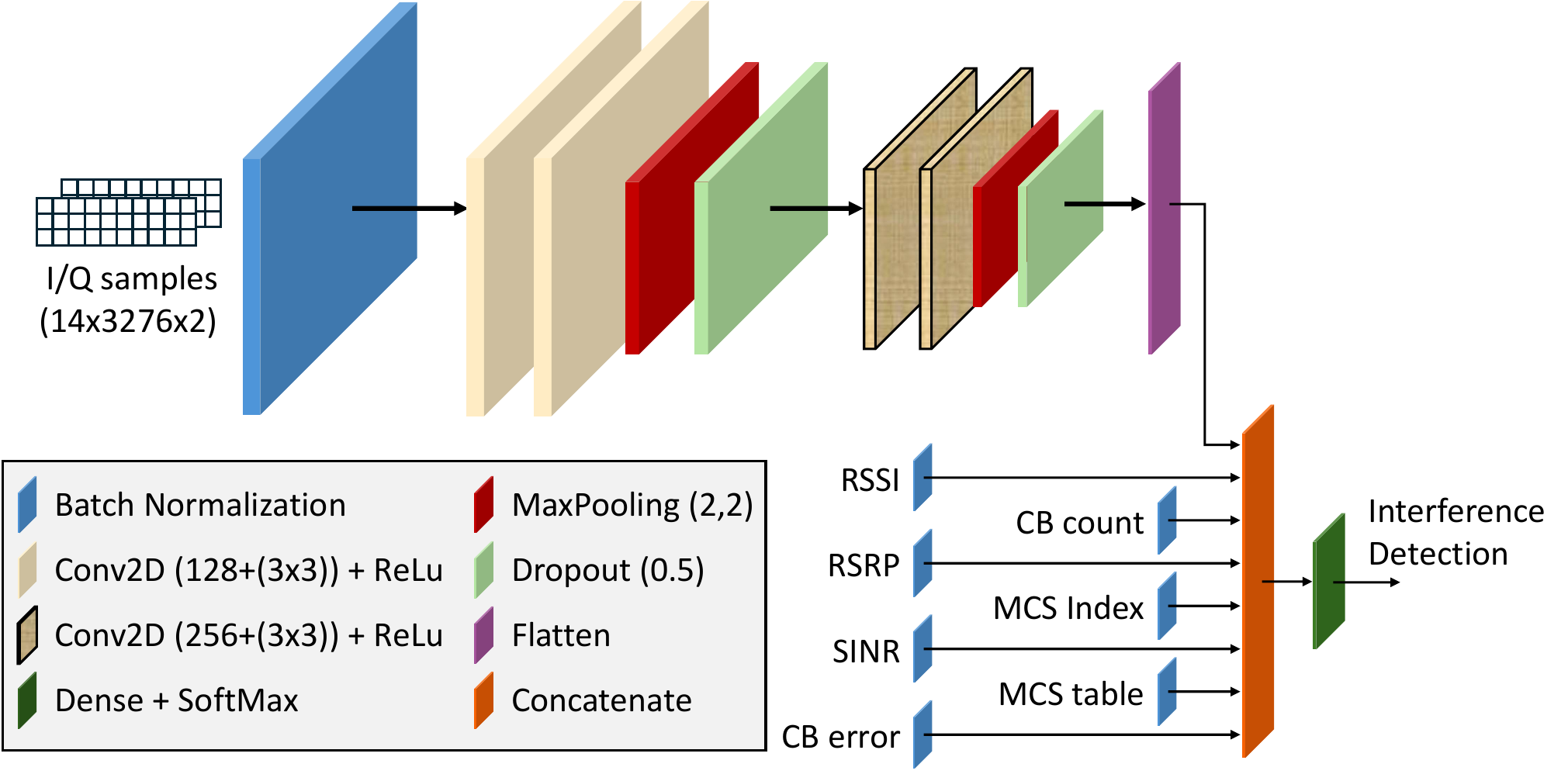}
  \caption{\justifying InterfO-RAN's CNN architecture showing inputs (I/Q and scalar features), layer specifications, and the binary interference detection output.}
  \label{chap5-fig:interforan-cnn}
\end{figure}

\textbf{CNN Architecture.}
We choose CNNs for their ability to extract complex patterns from OFDM-based IQ samples and their consistent outperformance of alternatives like LSTMs and ResNets. Figure~\ref{chap5-fig:interforan-cnn} illustrates the selected architecture. The network features seven layers. Two convolutional blocks perform feature extraction on \gls{pusch} \gls{iq} samples, each containing two Conv2D layers (128 and 256 filters, $3\times3$ kernel, ReLU activation) and one $2\times2$ MaxPooling layer. Then, the extracted features are flattened and concatenated with the additional normalized structured inputs (i.e., \gls{rssi}, \gls{rsrp}, \gls{sinr}, \gls{mcs} index, \gls{mcs} table index, total count of \gls{cb} errors, and number of \glspl{cb}). This final representation is processed through a fully connected Dense layer with softmax activation, which performs the final interference detection.
The starting point of our proposed \gls{cnn} is the VGG16 model, a widely recognized architecture in image processing. We systematically reduced its depth and identified a two-block variant that provided the best trade-off between computational efficiency and classification performance.

To enhance generalization and mitigate overfitting, we integrate dropout layers and an L2 regularizer. Dropout layers randomly deactivate neurons during training, encouraging the model to learn underlying patterns rather than memorizing the training data, while the L2 regularizer penalizes large weights, promoting simpler, more generalized models. Additionally, to address class imbalances (e.g., in case of data collections with different number of samples for different radios, or interference conditions), we use a weighted loss function instead of a standard one, which tends to favor the majority class. By assigning a greater weight to the minority class, the loss function ensures that its contributions have a stronger influence on the overall loss computation.

\textbf{Transfer Learning.}
We employ \gls{tl} using heterogeneous datasets from diverse \gls{rf} environments, detailed in Section~\ref{chap5-subsubsec:interforan-data}, to enhance model generalization capabilities. \gls{tl} is a machine learning technique wherein knowledge acquired from a source domain is leveraged to improve performance on a target domain through model adaptation. In our implementation, we partially fine-tune the foundational \gls{cnn} model (Figure~\ref{chap5-fig:interforan-cnn}), trained on data from multiple deployment locations (see Figure~\ref{chap5-fig:interforan-map}), and then tune to specific deployments by freezing the first block and fine-tuning the remaining layers with site-specific data. Thus, we allow the model to learn from a broader range of conditions and fine-tune it for a specific deployment, retaining knowledge of diverse scenarios by preserving foundational feature representations and enhancing performance.

\subsection{Data and Evaluation Framework}
\label{chap5-subsec:interforan-framework}

This section presents the setup for the training, testing, and evaluation of InterfO-RAN, from the generation of synthetic data with simulations to experimental \gls{ota} data collection and evaluation. We also discuss preprocessing and offline training procedures.
Real-world data collection, experiments, and validation of InterfO-RAN are all performed using the X5G platform.

\subsubsection{Empirical OTA Data Collection, Automated Labeling, and Preprocessing}
\label{chap5-subsubsec:interforan-data}

\begin{figure}[htbp]
  \centering
  \includegraphics[width=.95\linewidth]{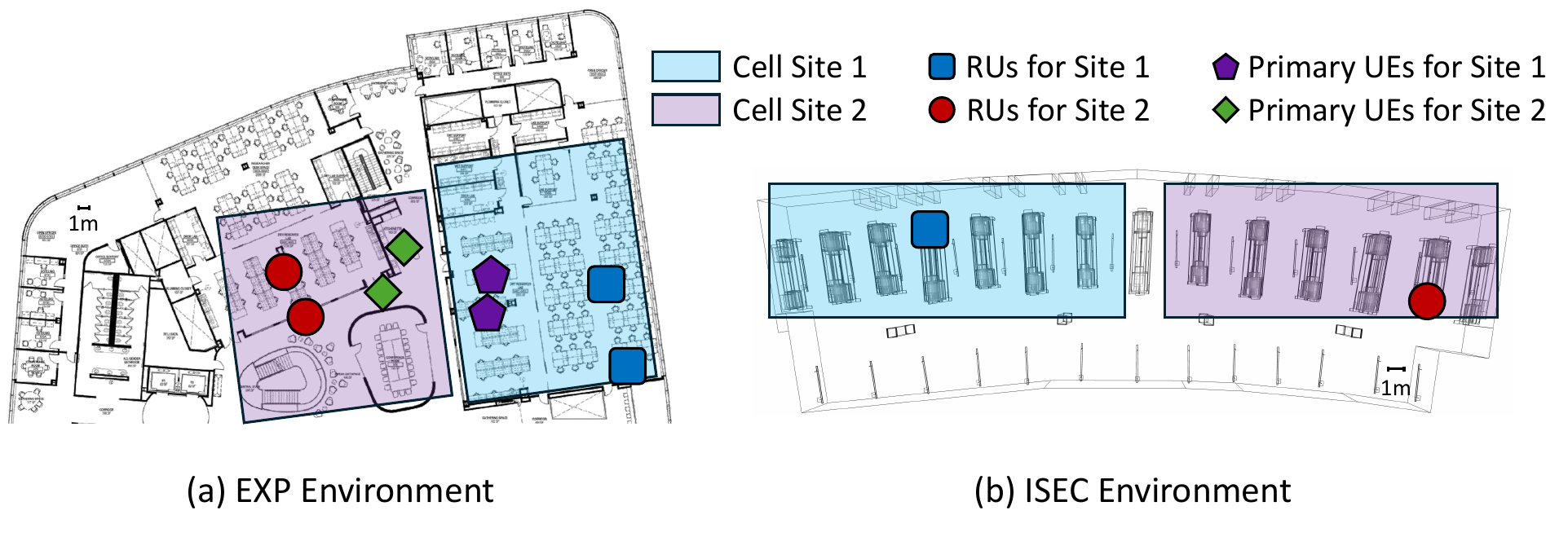}
  \caption{Indoor experiment layout maps of Northeastern University EXP (a) and ISEC (b) buildings, showing: the designated regions for the two cell sites, where the corresponding UEs can be located; RU locations for each respective cell site; and primary UE locations, where most of the data is collected.}
  \label{chap5-fig:interforan-map}
\end{figure}

All experiments involving interference are conducted in a controlled indoor environment within the Northeastern University EXP building in Boston, MA, as depicted in the indoor experiment map shown in Figure~\ref{chap5-fig:interforan-map}a. The layout includes: two cell sites (shown in light blue and pink regions), where the corresponding \glspl{ue} can be located; the \gls{ru} locations for each cell site, with only one active at a time per site (represented by blue and red icons); and the two primary \gls{ue} locations for the two cell sites, where most of the data is collected (violet and green icons).
Each experiment involves both cell sites and one \gls{ue} per cell at a time, tested under \gls{los} and \gls{nlos} channel conditions.
Similarly, we collected additional data in a second indoor environment within the Northeastern University ISEC building in Boston, MA, (Figure~\ref{chap5-fig:interforan-map}b).

During each data collection session, the \gls{ue} connects to its \gls{ru} and generates \gls{ul} traffic using iPerf.
We categorize the \glspl{gnb} traffic levels into two distinct classes---High Traffic and No Traffic---based on the \gls{tb} size, to improve interference identification accuracy. In the No Traffic scenario, the \gls{ue} remains connected to the \gls{gnb} while exchanging minimal control information necessary for maintaining the connection stability. In this case, we observe that the \gls{tb} size is typically around $185$~bytes and does not exceed $1056$~bytes, with the number of \glspl{cb} limited to one. 
The \glspl{cb} from a \gls{tb} are segmented and \gls{ldpc}-encoded using base graphs defined by the 3GPP for error correction. \gls{ldpc} base graph 2 is used for smaller \glspl{cb}, with each \gls{cb} sized at $480$~bytes, while base graph 1 is used for larger ones, with each \gls{cb} sized at $1056$~bytes. 
In contrast, High Traffic \gls{ul} transmissions produce sufficiently strong signals that can interfere with neighboring cells. In this scenario, the \gls{tb} size typically exceeds $1056$~bytes, and the number of \glspl{cb} is greater than one.

\begin{figure}[htb]
    \centering
    \begin{subfigure}{0.7\linewidth} 
        \centering
    \setlength\fwidth{\linewidth}
    \setlength\fheight{.29\linewidth}
    \input{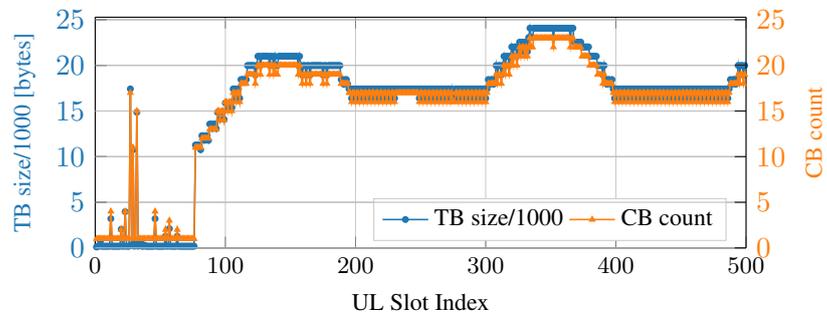}
        \caption{\centering TB size in bytes (scaled down by 1000) versus CB count}
        \label{chap5-fig:interforan-cb-tbsize}
    \end{subfigure}
    \hfill
        \begin{subfigure}{0.7\linewidth} 
        \centering
    \setlength\fwidth{\linewidth}
    \setlength\fheight{.29\linewidth}
    \input{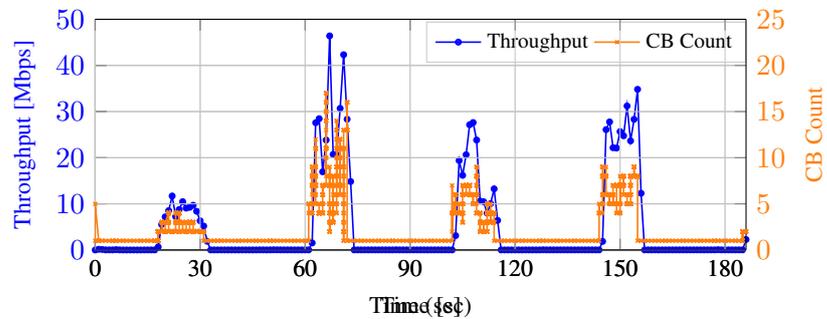}
        \caption{\centering UL throughput versus CB count}
        \label{chap5-fig:interforan-cb-ulthru}
    \end{subfigure}
    \caption{UL throughput, TB size, and CB count across different scenarios, demonstrating their close correlation.}
    \label{chap5-fig:interforan-combined-plots}
\end{figure}

As depicted in Figure~\ref{chap5-fig:interforan-combined-plots}, there is a strong correlation between the \gls{tb} size and the number of \glspl{cb} (Figure~\ref{chap5-fig:interforan-cb-tbsize}), as well as between the \gls{ul} throughput and the \gls{cb} count (Figure~\ref{chap5-fig:interforan-cb-ulthru}), reinforcing the decision to use the \gls{cb} count as the primary criterion for traffic classification.
Building upon this, we create a systematic labeling methodology based on the \gls{ul} traffic levels to classify the \gls{ul} transmissions as either affected by interference ('INTERF') or unaffected ('CLEAN').
The labeling rules, shown in Table~\ref{chap5-tab:interforan-lookup}, are defined as follows:
\begin{itemize}
    \item No Traffic/High Traffic: if one \gls{ue} transmits with 'High Traffic' while the other has 'No Traffic', the 'High Traffic' time windows on one side are used to label the other side as being affected by interference, designated as 'INTERF', while the transmitting side is labeled as ‘CLEAN’.
    \item High Traffic on both sides: if both \glspl{ue} transmit simultaneously to different \glspl{ru} with 'High Traffic', the corresponding samples are labeled as 'INTERF'.
    \item No Traffic on both sides: if both \glspl{ue} are transmitting with 'No Traffic', the samples are labeled as 'CLEAN', i.e., no interference.
    \item Single-active Transmission: if only one \gls{ue} transmits while the other remains in airplane mode (i.e., not connected to the base station), the samples are labeled as ‘CLEAN’.
\end{itemize}
A threshold of 1 is set for the \gls{cb} count in an \gls{ul} slot to identify high-traffic windows on one \gls{gnb}, which are subsequently used to mark instances of interference on the other \gls{gnb}.
Notably, the exact thresholding mechanism is utilized to evaluate inference performance in \gls{ota} data scenarios by identifying high-traffic time windows, which are then used to assign ground truth labels of ‘CLEAN’ or ‘INTERF’ to each input example, thereby enabling the effective monitoring of spectrum sensing capabilities.

\begin{table}[!htb]
    \centering
    \caption{Lookup Table for gNB UL Traffic and Labels}
    \label{chap5-tab:interforan-lookup}
    \footnotesize
    \begin{tabular}{p{0.2\linewidth} p{0.2\linewidth} p{0.2\linewidth} p{0.2\linewidth}}  
        \toprule
       \textbf{gNB1 Traffic} & \textbf{gNB2 Traffic} & \textbf{gNB1 UL Label} & \textbf{gNB2 UL Label} \\
        \midrule
        High Traffic & No Traffic & CLEAN & INTERF \\
        No Traffic & High Traffic & INTERF & CLEAN \\
        High Traffic & High Traffic & INTERF & INTERF \\
        No Traffic & No Traffic & CLEAN & CLEAN \\
        No UE & No/High Traffic & NA & CLEAN \\
        No/High Traffic & No UE & CLEAN & NA \\
        \bottomrule
    \end{tabular}
\end{table}

To efficiently manage \gls{gpu} memory during offline training---particularly given the relatively large input feature consisting of 3D \gls{iq} samples of size $14\times3276\times2$---we use \gls{tf}’s Sequence class to implement a custom data generator. The data generator dynamically loads small batches into \gls{gpu} memory, rather than loading the entire dataset at once, thereby optimizing memory usage and avoiding excessive memory strain.

\subsubsection{Synthetic Data}
\label{chap5-subsubsec:interforan-synthetic}
We also perform an extensive data collection campaign using the NVIDIA Aerial simulator nr\_sim, which complements the \gls{cubb} framework by extending the 5G MATLAB Toolbox. Through this, we collect data and evaluate the model in a larger variety of configurations compared to those supported in our experimental setup. The parameters that we sweep include \gls{snr}, \gls{sir}, \gls{mcs} index, channel type, delay profile, interference channel delay profile, numerology, and carrier frequency. In addition, we simulate up to 5 interfering \glspl{ue}, so that the classification can be extended to include additional \glspl{ue}.

To simulate interference, an in-band interference signal is artificially generated by transmitting \gls{ofdm} signals from each antenna independently through a \gls{tdl} channel characterized by high delay spread and Doppler shift. These conditions emulate a dynamic and realistic environment affected by multipath propagation and high mobility, thus creating a robust testbed for interference mitigation. The processed signal is then multiplied by a modified all-zero matrix with randomly distributed contiguous ones, introducing localized and burst interference in the \glspl{re} of the \gls{ofdm} slot. This process ensures random patterns for contiguous ones across antenna streams while maintaining the average power of the interference signal. The resulting combination of randomized interfering signals, thermal noise, and received \gls{ofdm} signal at the \gls{gnb} closely mimics real-world interference effects.

\subsection{Performance Evaluation and Experimental Results}
\label{chap5-subsec:interforan-results}

This section presents experimental results to benchmark the performance of InterfO-RAN across various model configurations, using the simulation setup described in Section~\ref{chap5-subsubsec:interforan-synthetic} as well as over \gls{ota} tests.

\begin{figure}[htb]
    \centering
    \setlength\fwidth{.7\linewidth}
    \setlength\fheight{.2\linewidth}
\begin{tikzpicture}

\definecolor{black38}{RGB}{38,38,38}
\definecolor{gray176}{RGB}{176,176,176}

\definecolor{color0}{RGB}{230,240,255}    
\definecolor{color1}{RGB}{198,219,239}    
\definecolor{color2}{RGB}{158,202,225}    
\definecolor{color3}{RGB}{107,174,214}    
\definecolor{color4}{RGB}{66,146,198}     
\definecolor{color5}{RGB}{33,113,181}     
\definecolor{color6}{RGB}{8,81,156}       
\definecolor{color7}{RGB}{0,55,120}       

\begin{axis}[
    width=.937\fwidth,
    height=1.5\fheight,
    tick align=inside,
    tick pos=left,
    x grid style={gray176},
    xlabel={Predicted label},
    xlabel style={font=\footnotesize},
    xmin=0, xmax=6,
    xtick style={color=black},
    xtick={0.5,1.5,2.5,3.5,4.5,5.5},
    xticklabels={0,1,2,3,4,5},
    y dir=reverse,
    y grid style={gray176},
    ylabel={True label},
    ylabel style={font=\footnotesize},
    ymin=0, ymax=6,
    ytick style={color=black},
    ytick={0.5,1.5,2.5,3.5,4.5,5.5},
    yticklabel style={rotate=90.0},
    yticklabels={0,1,2,3,4,5},
    xticklabel style={font=\footnotesize},
    yticklabel style={font=\footnotesize},
    colormap={bluegrad}{
        rgb255(0pt)=(230,240,255);
        rgb255(1pt)=(230,240,255);
        rgb255(3pt)=(198,219,239);
        rgb255(10pt)=(158,202,225);
        rgb255(30pt)=(107,174,214);
        rgb255(60pt)=(66,146,198);
        rgb255(90pt)=(33,113,181);
        rgb255(95pt)=(8,81,156);
        rgb255(100pt)=(0,55,120)
    },
    colorbar,
    colorbar style={
        title={Accuracy (\%)},
        title style={font=\footnotesize, rotate=90, yshift=-10mm, xshift=-12.5mm},
        ylabel style={font=\footnotesize},
        yticklabel style={font=\footnotesize},
        ytick={0,20,40,60,80,100},
        yticklabels={0,20,40,60,80,100},
        width=2mm,
        yticklabel pos=right
    },
    point meta min=0,
    point meta max=100
]


\fill[color7] (0,0) rectangle (1,1); 
\fill[color1] (1,0) rectangle (2,1); 
\fill[color0] (2,0) rectangle (3,1); 
\fill[color0] (3,0) rectangle (4,1); 
\fill[color0] (4,0) rectangle (5,1); 
\fill[color0] (5,0) rectangle (6,1); 

\fill[color2] (0,1) rectangle (1,2); 
\fill[color6] (1,1) rectangle (2,2); 
\fill[color1] (2,1) rectangle (3,2); 
\fill[color0] (3,1) rectangle (4,2); 
\fill[color0] (4,1) rectangle (5,2); 
\fill[color0] (5,1) rectangle (6,2); 

\fill[color0] (0,2) rectangle (1,3); 
\fill[color0] (1,2) rectangle (2,3); 
\fill[color7] (2,2) rectangle (3,3); 
\fill[color0] (3,2) rectangle (4,3); 
\fill[color0] (4,2) rectangle (5,3); 
\fill[color0] (5,2) rectangle (6,3); 

\fill[color0] (0,3) rectangle (1,4); 
\fill[color0] (1,3) rectangle (2,4); 
\fill[color0] (2,3) rectangle (3,4); 
\fill[color7] (3,3) rectangle (4,4); 
\fill[color0] (4,3) rectangle (5,4); 
\fill[color0] (5,3) rectangle (6,4); 

\fill[color0] (0,4) rectangle (1,5); 
\fill[color0] (1,4) rectangle (2,5); 
\fill[color0] (2,4) rectangle (3,5); 
\fill[color0] (3,4) rectangle (4,5); 
\fill[color7] (4,4) rectangle (5,5); 
\fill[color0] (5,4) rectangle (6,5); 

\fill[color0] (0,5) rectangle (1,6); 
\fill[color0] (1,5) rectangle (2,6); 
\fill[color0] (2,5) rectangle (3,6); 
\fill[color0] (3,5) rectangle (4,6); 
\fill[color0] (4,5) rectangle (5,6); 
\fill[color7] (5,5) rectangle (6,6); 

\draw (axis cs:0.5,0.5) node[scale=1.2, text=white, rotate=0.0, align=center]{\footnotesize 96.93};
\draw (axis cs:1.5,0.5) node[scale=1.2, text=black38, rotate=0.0, align=center]{\footnotesize 2.66};
\draw (axis cs:2.5,0.5) node[scale=1.2, text=black38, rotate=0.0, align=center]{\footnotesize 0.41};
\draw (axis cs:3.5,0.5) node[scale=1.2, text=black38, rotate=0.0, align=center]{\footnotesize 0.00};
\draw (axis cs:4.5,0.5) node[scale=1.2, text=black38, rotate=0.0, align=center]{\footnotesize 0.00};
\draw (axis cs:5.5,0.5) node[scale=1.2, text=black38, rotate=0.0, align=center]{\footnotesize 0.00};

\draw (axis cs:0.5,1.5) node[scale=1.2, text=black38, rotate=0.0, align=center]{\footnotesize 4.02};
\draw (axis cs:1.5,1.5) node[scale=1.2, text=white, rotate=0.0, align=center]{\footnotesize 93.87};
\draw (axis cs:2.5,1.5) node[scale=1.2, text=black38, rotate=0.0, align=center]{\footnotesize 2.11};
\draw (axis cs:3.5,1.5) node[scale=1.2, text=black38, rotate=0.0, align=center]{\footnotesize 0.00};
\draw (axis cs:4.5,1.5) node[scale=1.2, text=black38, rotate=0.0, align=center]{\footnotesize 0.00};
\draw (axis cs:5.5,1.5) node[scale=1.2, text=black38, rotate=0.0, align=center]{\footnotesize 0.00};

\draw (axis cs:0.5,2.5) node[scale=1.2, text=black38, rotate=0.0, align=center]{\footnotesize 0.00};
\draw (axis cs:1.5,2.5) node[scale=1.2, text=black38, rotate=0.0, align=center]{\footnotesize 0.00};
\draw (axis cs:2.5,2.5) node[scale=1.2, text=white, rotate=0.0, align=center]{\footnotesize 99.88};
\draw (axis cs:3.5,2.5) node[scale=1.2, text=black38, rotate=0.0, align=center]{\footnotesize 0.12};
\draw (axis cs:4.5,2.5) node[scale=1.2, text=black38, rotate=0.0, align=center]{\footnotesize 0.00};
\draw (axis cs:5.5,2.5) node[scale=1.2, text=black38, rotate=0.0, align=center]{\footnotesize 0.00};

\draw (axis cs:0.5,3.5) node[scale=1.2, text=black38, rotate=0.0, align=center]{\footnotesize 0.00};
\draw (axis cs:1.5,3.5) node[scale=1.2, text=black38, rotate=0.0, align=center]{\footnotesize 0.00};
\draw (axis cs:2.5,3.5) node[scale=1.2, text=black38, rotate=0.0, align=center]{\footnotesize 0.00};
\draw (axis cs:3.5,3.5) node[scale=1.2, text=white, rotate=0.0, align=center]{\footnotesize 100.00};
\draw (axis cs:4.5,3.5) node[scale=1.2, text=black38, rotate=0.0, align=center]{\footnotesize 0.00};
\draw (axis cs:5.5,3.5) node[scale=1.2, text=black38, rotate=0.0, align=center]{\footnotesize 0.00};

\draw (axis cs:0.5,4.5) node[scale=1.2, text=black38, rotate=0.0, align=center]{\footnotesize 0.00};
\draw (axis cs:1.5,4.5) node[scale=1.2, text=black38, rotate=0.0, align=center]{\footnotesize 0.00};
\draw (axis cs:2.5,4.5) node[scale=1.2, text=black38, rotate=0.0, align=center]{\footnotesize 0.00};
\draw (axis cs:3.5,4.5) node[scale=1.2, text=black38, rotate=0.0, align=center]{\footnotesize 0.00};
\draw (axis cs:4.5,4.5) node[scale=1.2, text=white, rotate=0.0, align=center]{\footnotesize 100.00};
\draw (axis cs:5.5,4.5) node[scale=1.2, text=black38, rotate=0.0, align=center]{\footnotesize 0.00};

\draw (axis cs:0.5,5.5) node[scale=1.2, text=black38, rotate=0.0, align=center]{\footnotesize 0.00};
\draw (axis cs:1.5,5.5) node[scale=1.2, text=black38, rotate=0.0, align=center]{\footnotesize 0.00};
\draw (axis cs:2.5,5.5) node[scale=1.2, text=black38, rotate=0.0, align=center]{\footnotesize 0.00};
\draw (axis cs:3.5,5.5) node[scale=1.2, text=black38, rotate=0.0, align=center]{\footnotesize 0.00};
\draw (axis cs:4.5,5.5) node[scale=1.2, text=black38, rotate=0.0, align=center]{\footnotesize 0.00};
\draw (axis cs:5.5,5.5) node[scale=1.2, text=white, rotate=0.0, align=center]{\footnotesize 100.00};

\textbf{}\end{axis}

\end{tikzpicture}
    \caption{Confusion matrix of the model tested on synthetic data with 13800 samples per class.}
    \label{chap5-fig:interforan-conf-matrix-nrsim}
\end{figure}
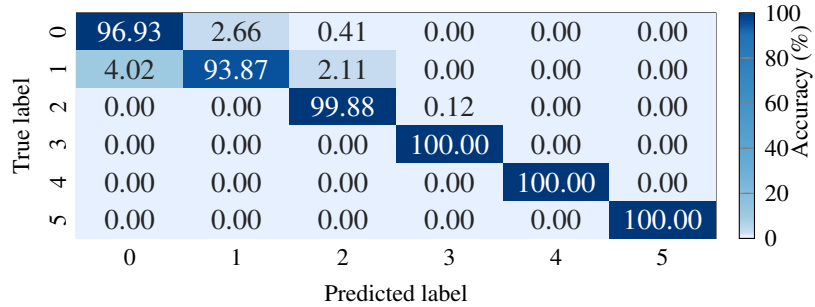

We test the model, shown in Figure~\ref{chap5-fig:interforan-cnn}, on synthetic data with scenarios involving 0 to 5 interferers, totaling $13800$~samples per class, to ensure consistent performance across different parameters described in Section~\ref{chap5-subsubsec:interforan-synthetic} before \gls{ota} deployment in InterfO-RAN. The resulting confusion matrix in Figure~\ref{chap5-fig:interforan-conf-matrix-nrsim} indicates that the model achieves high accuracy (above 96\%) for scenarios with no interferer and 2 or more interferers. For scenarios with a single interferer, accuracy remains solid at $93.87$\%, with the slight drop likely due to the lower impact of a single interferer on the signal. Thus, the model becomes a strong candidate for \gls{ota} deployment, along with several other models with minor modifications explained in Section~\ref{chap5-subsubsec:interforan-ota}.

\subsubsection{OTA Evaluation}
\label{chap5-subsubsec:interforan-ota}

To select the best generalized model suitable for diverse \gls{rf} environments, we conduct a series of \gls{ota} experiments evaluating eight configurations of the underlying \gls{cnn} architecture shown in Figure~\ref{chap5-fig:interforan-cnn}.
We represent each model as \{[$\alpha$,$\beta$],$\gamma$\}, where $\alpha$ and $\beta$ denote the number of filters in the convolutional layers of the first and second blocks, and $\gamma$ represents the batch size.
We select two configurations for [$\alpha$, $\beta$]: [64, 128] and [128, 256], inspired by VGG16's first blocks for a lightweight model, with $\gamma$ ranging over 16, 32, 64, and 128.

We carry out experiments at various \gls{ru} locations and \gls{ue} positions across the map, as shown in Figure~\ref{chap5-fig:interforan-map}.
For fair comparability, we evaluate different models using the same dataset with data logging. The data collected from one side is labeled with ground truth based on the traffic level on the other side, as detailed in Table~\ref{chap5-tab:interforan-lookup}.
We then compute metrics such as accuracy, measuring the ratio of correctly predicted samples to the total number of samples, specificity, evaluating the identification of 'CLEAN' examples, and recall, assessing the detection of 'INTERF' examples.

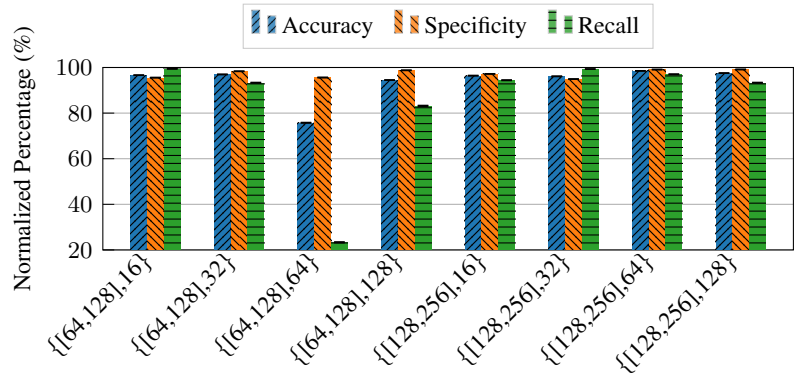
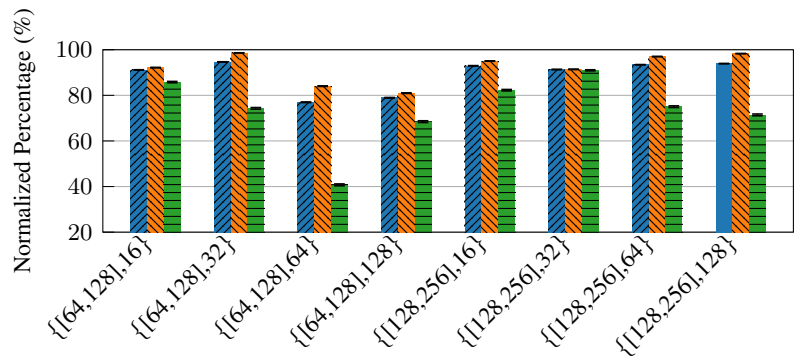
\begin{figure}[t]
    \centering
    \begin{subfigure}{0.7\linewidth} 
            \centering
        \setlength\fwidth{\linewidth}
        \setlength\fheight{.15\linewidth}
\begin{tikzpicture}

\definecolor{gray176}{RGB}{176,176,176}
\definecolor{gray31119180}{RGB}{31,119,180}
\definecolor{green4416044}{RGB}{44,160,44}
\definecolor{orange25512714}{RGB}{255,127,14}
\definecolor{white204}{RGB}{204,204,204}

\definecolor{gray176}{RGB}{176,176,176}
\definecolor{blue31119180}{RGB}{31,119,180}
\definecolor{green4416044}{RGB}{44,160,44}
\definecolor{orange25512714}{RGB}{255,127,14}
\definecolor{white204}{RGB}{204,204,204}

\begin{axis}[
width=\fwidth,
height=2.5\fheight,
legend cell align={left},
legend style={
  at={(.5,1.35)},
  anchor=north,
  draw=white204,
  fill opacity=0.8,
  draw opacity=1,
  font=\footnotesize,
    legend columns=-1, 
  /tikz/every even column/.append style={column sep=0.2cm}, 
  text opacity=1
},
tick label style={font=\footnotesize},
ylabel style={font=\footnotesize},
tick align=inside,
tick pos=left,
x grid style={gray176},
xmin=-0.63, xmax=7.63,
xtick style={color=black},
xticklabel style={rotate=45, anchor=east},
xtick={0,1,2,3,4,5,6,7},
xticklabels={
  {\{[64,128],16\}},
  {\{[64,128],32\}},
    {\{[64,128],64\}},
  {\{[64,128],128\}},
  {\{[128,256],16\}},
  {\{[128,256],32\}},
  {\{[128,256],64\}},
  {\{[128,256],128\}}
},
y grid style={gray176},
ylabel={Normalized Percentage (\%)},
ymajorgrids,
ymin=20, ymax=100,
ytick style={color=black}
]


\draw[draw=none,fill=blue31119180,postaction={pattern=north east lines,pattern color=black}] (axis cs:-0.3,0) rectangle (axis cs:-0.1,96.6244530285244);
\addlegendimage{ybar,ybar legend,draw=none,fill=blue31119180,postaction={pattern=north east lines,pattern color=black}}
\addlegendentry{Accuracy}

\draw[draw=none,fill=blue31119180,postaction={pattern=north east lines,pattern color=black}] (axis cs:0.7,0) rectangle (axis cs:0.9,96.9601209353509);
\draw[draw=none,fill=blue31119180,postaction={pattern=north east lines,pattern color=black}] (axis cs:1.7,0) rectangle (axis cs:1.9,75.7722838723673);
\draw[draw=none,fill=blue31119180,postaction={pattern=north east lines,pattern color=black}] (axis cs:2.7,0) rectangle (axis cs:2.9,94.4872958001403);
\draw[draw=none,fill=blue31119180,postaction={pattern=north east lines,pattern color=black}] (axis cs:3.7,0) rectangle (axis cs:3.9,96.4372427553466);
\draw[draw=none,fill=blue31119180,postaction={pattern=north east lines,pattern color=black}] (axis cs:4.7,0) rectangle (axis cs:4.9,96.1228197428093);
\draw[draw=none,fill=blue31119180,postaction={pattern=north east lines,pattern color=black}] (axis cs:5.7,0) rectangle (axis cs:5.9,98.5354213812887);
\draw[draw=none,fill=blue31119180,postaction={pattern=north east lines,pattern color=black}] (axis cs:6.7,0) rectangle (axis cs:6.9,97.5444610276115);

\draw[draw=none,fill=orange25512714,postaction={pattern=north west lines,pattern color=black}] (axis cs:-0.1,0) rectangle (axis cs:0.1,95.5637739609773);
\addlegendimage{ybar,ybar legend,draw=none,fill=orange25512714,postaction={pattern=north west lines,pattern color=black}}
\addlegendentry{Specificity}

\draw[draw=none,fill=orange25512714,postaction={pattern=north west lines,pattern color=black}] (axis cs:0.9,0) rectangle (axis cs:1.1,98.3358507606783);
\draw[draw=none,fill=orange25512714,postaction={pattern=north west lines,pattern color=black}] (axis cs:1.9,0) rectangle (axis cs:2.1,95.6087830567541);
\draw[draw=none,fill=orange25512714,postaction={pattern=north west lines,pattern color=black}] (axis cs:2.9,0) rectangle (axis cs:3.1,98.7623600492066);
\draw[draw=none,fill=orange25512714,postaction={pattern=north west lines,pattern color=black}] (axis cs:3.9,0) rectangle (axis cs:4.1,97.2005169809558);
\draw[draw=none,fill=orange25512714,postaction={pattern=north west lines,pattern color=black}] (axis cs:4.9,0) rectangle (axis cs:5.1,94.9191128496039);
\draw[draw=none,fill=orange25512714,postaction={pattern=north west lines,pattern color=black}] (axis cs:5.9,0) rectangle (axis cs:6.1,99.1063392453947);
\draw[draw=none,fill=orange25512714,postaction={pattern=north west lines,pattern color=black}] (axis cs:6.9,0) rectangle (axis cs:7.1,99.1401298681076);

\draw[draw=none,fill=green4416044,postaction={pattern=horizontal lines,pattern color=black}] (axis cs:0.1,0) rectangle (axis cs:0.3,99.4285973051858);
\addlegendimage{ybar,ybar legend,draw=none,fill=green4416044,postaction={pattern=horizontal lines,pattern color=black}}
\addlegendentry{Recall}

\draw[draw=none,fill=green4416044,postaction={pattern=horizontal lines,pattern color=black}] (axis cs:1.1,0) rectangle (axis cs:1.3,93.3230689429462);
\draw[draw=none,fill=green4416044,postaction={pattern=horizontal lines,pattern color=black}] (axis cs:2.1,0) rectangle (axis cs:2.3,23.329943888914);
\draw[draw=none,fill=green4416044,postaction={pattern=horizontal lines,pattern color=black}] (axis cs:3.1,0) rectangle (axis cs:3.3,83.1851995175188);
\draw[draw=none,fill=green4416044,postaction={pattern=horizontal lines,pattern color=black}] (axis cs:4.1,0) rectangle (axis cs:4.3,94.4193552371033);
\draw[draw=none,fill=green4416044,postaction={pattern=horizontal lines,pattern color=black}] (axis cs:5.1,0) rectangle (axis cs:5.3,99.305095281811);
\draw[draw=none,fill=green4416044,postaction={pattern=horizontal lines,pattern color=black}] (axis cs:6.1,0) rectangle (axis cs:6.3,97.0260712771344);
\draw[draw=none,fill=green4416044,postaction={pattern=horizontal lines,pattern color=black}] (axis cs:7.1,0) rectangle (axis cs:7.3,93.3259506568249);
\path [draw=black, semithick]
(axis cs:-0.2,96.5866260101472)
--(axis cs:-0.2,96.6622800469017);

\path [draw=black, semithick]
(axis cs:0.8,96.9241501231867)
--(axis cs:0.8,96.9960917475151);

\path [draw=black, semithick]
(axis cs:1.8,75.6829109134345)
--(axis cs:1.8,75.8616568313);

\path [draw=black, semithick]
(axis cs:2.8,94.4395556928677)
--(axis cs:2.8,94.5350359074129);

\path [draw=black, semithick]
(axis cs:3.8,96.3984247792036)
--(axis cs:3.8,96.4760607314895);

\path [draw=black, semithick]
(axis cs:4.8,96.0824009492529)
--(axis cs:4.8,96.1632385363657);

\path [draw=black, semithick]
(axis cs:5.8,98.5101831638592)
--(axis cs:5.8,98.5606595987182);

\path [draw=black, semithick]
(axis cs:6.8,97.5120117602985)
--(axis cs:6.8,97.5769102949245);

\addplot [semithick, black, mark=-, mark size=2, mark options={solid}, only marks]
table {%
-0.2 96.5866260101472
0.8 96.9241501231867
1.8 75.6829109134345
2.8 94.4395556928677
3.8 96.3984247792036
4.8 96.0824009492529
5.8 98.5101831638592
6.8 97.5120117602985
};
\addplot [semithick, black, mark=-, mark size=2, mark options={solid}, only marks]
table {%
-0.2 96.6622800469017
0.8 96.9960917475151
1.8 75.8616568313
2.8 94.5350359074129
3.8 96.4760607314895
4.8 96.1632385363657
5.8 98.5606595987182
6.8 97.5769102949245
};
\path [draw=black, semithick]
(axis cs:0,95.5131426678483)
--(axis cs:0,95.6144052541063);

\path [draw=black, semithick]
(axis cs:1,98.3042730617697)
--(axis cs:1,98.3674284595869);

\path [draw=black, semithick]
(axis cs:2,95.5583958441253)
--(axis cs:2,95.659170269383);

\path [draw=black, semithick]
(axis cs:3,98.7350267142959)
--(axis cs:3,98.7896933841173);

\path [draw=black, semithick]
(axis cs:4,97.1598887886572)
--(axis cs:4,97.2411451732544);

\path [draw=black, semithick]
(axis cs:5,94.865132722317)
--(axis cs:5,94.9730929768908);

\path [draw=black, semithick]
(axis cs:6,99.0830264496901)
--(axis cs:6,99.1296520410993);

\path [draw=black, semithick]
(axis cs:7,99.1172523325935)
--(axis cs:7,99.1630074036216);

\addplot [semithick, black, mark=-, mark size=2, mark options={solid}, only marks]
table {%
0 95.5131426678483
1 98.3042730617697
2 95.5583958441253
3 98.7350267142959
4 97.1598887886572
5 94.865132722317
6 99.0830264496901
7 99.1172523325935
};
\addplot [semithick, black, mark=-, mark size=2, mark options={solid}, only marks]
table {%
0 95.6144052541063
1 98.3674284595869
2 95.659170269383
3 98.7896933841173
4 97.2411451732544
5 94.9730929768908
6 99.1296520410993
7 99.1630074036216
};
\path [draw=black, semithick]
(axis cs:0.2,99.3978312320774)
--(axis cs:0.2,99.4593633782943);

\path [draw=black, semithick]
(axis cs:1.2,93.2231146679983)
--(axis cs:1.2,93.4230232178941);

\path [draw=black, semithick]
(axis cs:2.2,23.1621787482034)
--(axis cs:2.2,23.4977090296247);

\path [draw=black, semithick]
(axis cs:3.2,83.0359467528689)
--(axis cs:3.2,83.3344522821686);

\path [draw=black, semithick]
(axis cs:4.2,94.3273662735299)
--(axis cs:4.2,94.5113442006767);

\path [draw=black, semithick]
(axis cs:5.2,99.2712717387075)
--(axis cs:5.2,99.3389188249146);

\path [draw=black, semithick]
(axis cs:6.2,96.957772701783)
--(axis cs:6.2,97.0943698524859);

\path [draw=black, semithick]
(axis cs:7.2,93.2260162269535)
--(axis cs:7.2,93.4258850866963);

\addplot [semithick, black, mark=-, mark size=2, mark options={solid}, only marks]
table {%
0.2 99.3978312320774
1.2 93.2231146679983
2.2 23.1621787482034
3.2 83.0359467528689
4.2 94.3273662735299
5.2 99.2712717387075
6.2 96.957772701783
7.2 93.2260162269535
};
\addplot [semithick, black, mark=-, mark size=2, mark options={solid}, only marks]
table {%
0.2 99.4593633782943
1.2 93.4230232178941
2.2 23.4977090296247
3.2 83.3344522821686
4.2 94.5113442006767
5.2 99.3389188249146
6.2 97.0943698524859
7.2 93.4258850866963
};
\end{axis}

\end{tikzpicture} 
        \caption{\centering Performance in a familiar RF environment}
        \label{chap5-fig:interforan-barplot-perf1}
    \end{subfigure}
    \hfill
        \begin{subfigure}{0.7\linewidth} 
        \centering
        \setlength\fwidth{\linewidth}
        \setlength\fheight{.15\linewidth}
\begin{tikzpicture}

\definecolor{gray176}{RGB}{176,176,176}
\definecolor{gray31119180}{RGB}{31,119,180}
\definecolor{green4416044}{RGB}{44,160,44}
\definecolor{orange25512714}{RGB}{255,127,14}
\definecolor{white204}{RGB}{204,204,204}

\definecolor{gray176}{RGB}{176,176,176}
\definecolor{blue31119180}{RGB}{31,119,180}
\definecolor{green4416044}{RGB}{44,160,44}
\definecolor{orange25512714}{RGB}{255,127,14}
\definecolor{white204}{RGB}{204,204,204}

\begin{axis}[
width=\fwidth,
height=2.5\fheight,
tick label style={font=\footnotesize},
ylabel style={font=\footnotesize},
tick align=inside,
tick pos=left,
x grid style={gray176},
xmin=-0.63, xmax=7.63,
xtick style={color=black},
xticklabel style={rotate=45, anchor=east},
xtick={0,1,2,3,4,5,6,7},
xticklabels={
  {\{[64,128],16\}},
  {\{[64,128],32\}},
    {\{[64,128],64\}},
  {\{[64,128],128\}},
  {\{[128,256],16\}},
  {\{[128,256],32\}},
  {\{[128,256],64\}},
  {\{[128,256],128\}}
},
y grid style={gray176},
ylabel={Normalized Percentage (\%)},
ymajorgrids,
ymin=20, ymax=100,
ytick style={color=black}
]
\draw[draw=none,fill=blue31119180,postaction={pattern=north east lines,pattern color=black}] (axis cs:-0.3,0) rectangle (axis cs:-0.1,91.1331611478894);

\draw[draw=none,fill=blue31119180,postaction={pattern=north east lines,pattern color=black}] (axis cs:0.7,0) rectangle (axis cs:0.9,94.6514769424988);
\draw[draw=none,fill=blue31119180,postaction={pattern=north east lines,pattern color=black}] (axis cs:1.7,0) rectangle (axis cs:1.9,76.9659786364011);
\draw[draw=none,fill=blue31119180,postaction={pattern=north east lines,pattern color=black}] (axis cs:2.7,0) rectangle (axis cs:2.9,78.9346254356729);
\draw[draw=none,fill=blue31119180,postaction={pattern=north east lines,pattern color=black}] (axis cs:3.7,0) rectangle (axis cs:3.9,92.9575746856904);
\draw[draw=none,fill=blue31119180,postaction={pattern=north east lines,pattern color=black}] (axis cs:4.7,0) rectangle (axis cs:4.9,91.3325872641725);
\draw[draw=none,fill=blue31119180,postaction={pattern=north east lines,pattern color=black}] (axis cs:5.7,0) rectangle (axis cs:5.9,93.42351215615);
\draw[draw=none,fill=gray31119180] (axis cs:6.7,0) rectangle (axis cs:6.9,93.9184182520636);

\draw[draw=none,fill=orange25512714,postaction={pattern=north west lines,pattern color=black}] (axis cs:-0.1,0) rectangle (axis cs:0.1,92.1704307775276);

\draw[draw=none,fill=orange25512714,postaction={pattern=north west lines,pattern color=black}] (axis cs:0.9,0) rectangle (axis cs:1.1,98.6226196462537);
\draw[draw=none,fill=orange25512714,postaction={pattern=north west lines,pattern color=black}] (axis cs:1.9,0) rectangle (axis cs:2.1,84.027124918924);
\draw[draw=none,fill=orange25512714,postaction={pattern=north west lines,pattern color=black}] (axis cs:2.9,0) rectangle (axis cs:3.1,80.976482505812);
\draw[draw=none,fill=orange25512714,postaction={pattern=north west lines,pattern color=black}] (axis cs:3.9,0) rectangle (axis cs:4.1,95.0439814308723);
\draw[draw=none,fill=orange25512714,postaction={pattern=north west lines,pattern color=black}] (axis cs:4.9,0) rectangle (axis cs:5.1,91.4033975382059);
\draw[draw=none,fill=orange25512714,postaction={pattern=north west lines,pattern color=black}] (axis cs:5.9,0) rectangle (axis cs:6.1,97.0105744914989);
\draw[draw=none,fill=orange25512714,postaction={pattern=north west lines,pattern color=black}] (axis cs:6.9,0) rectangle (axis cs:7.1,98.3026884423942);
\draw[draw=none,fill=green4416044,postaction={pattern=horizontal lines,pattern color=black}] (axis cs:0.1,0) rectangle (axis cs:0.3,85.8130711869475);

\draw[draw=none,fill=green4416044,postaction={pattern=horizontal lines,pattern color=black}] (axis cs:1.1,0) rectangle (axis cs:1.3,74.283738576261);
\draw[draw=none,fill=green4416044,postaction={pattern=horizontal lines,pattern color=black}] (axis cs:2.1,0) rectangle (axis cs:2.3,40.7498084362794);
\draw[draw=none,fill=green4416044,postaction={pattern=horizontal lines,pattern color=black}] (axis cs:3.1,0) rectangle (axis cs:3.3,68.4620703833143);
\draw[draw=none,fill=green4416044,postaction={pattern=horizontal lines,pattern color=black}] (axis cs:4.1,0) rectangle (axis cs:4.3,82.2565271833592);
\draw[draw=none,fill=green4416044,postaction={pattern=horizontal lines,pattern color=black}] (axis cs:5.1,0) rectangle (axis cs:5.3,90.9694058721289);
\draw[draw=none,fill=green4416044,postaction={pattern=horizontal lines,pattern color=black}] (axis cs:6.1,0) rectangle (axis cs:6.3,75.0256975722803);
\draw[draw=none,fill=green4416044,postaction={pattern=horizontal lines,pattern color=black}] (axis cs:7.1,0) rectangle (axis cs:7.3,71.4317752817388);
\path [draw=black, semithick]
(axis cs:-0.2,91.0353876589776)
--(axis cs:-0.2,91.2309346368012);

\path [draw=black, semithick]
(axis cs:0.8,94.5739454644055)
--(axis cs:0.8,94.7290084205921);

\path [draw=black, semithick]
(axis cs:1.8,76.8215564490273)
--(axis cs:1.8,77.1104008237748);

\path [draw=black, semithick]
(axis cs:2.8,78.7947244535381)
--(axis cs:2.8,79.0745264178077);

\path [draw=black, semithick]
(axis cs:3.8,92.86950088183)
--(axis cs:3.8,93.0456484895508);

\path [draw=black, semithick]
(axis cs:4.8,91.2358065398065)
--(axis cs:4.8,91.4293679885385);

\path [draw=black, semithick]
(axis cs:5.8,93.3381674230466)
--(axis cs:5.8,93.5088568892535);

\path [draw=black, semithick]
(axis cs:6.8,93.836106320805)
--(axis cs:6.8,94.0007301833223);

\addplot [semithick, black, mark=-, mark size=2, mark options={solid}, only marks]
table {%
-0.2 91.0353876589776
0.8 94.5739454644055
1.8 76.8215564490273
2.8 78.7947244535381
3.8 92.86950088183
4.8 91.2358065398065
5.8 93.3381674230466
6.8 93.836106320805
};
\addplot [semithick, black, mark=-, mark size=2, mark options={solid}, only marks]
table {%
-0.2 91.2309346368012
0.8 94.7290084205921
1.8 77.1104008237748
2.8 79.0745264178077
3.8 93.0456484895508
4.8 91.4293679885385
5.8 93.5088568892535
6.8 94.0007301833223
};
\path [draw=black, semithick]
(axis cs:0,92.0693330279699)
--(axis cs:0,92.2715285270853);

\path [draw=black, semithick]
(axis cs:1,98.5783282341417)
--(axis cs:1,98.6669110583657);

\path [draw=black, semithick]
(axis cs:2,83.8895825351456)
--(axis cs:2,84.1646673027025);

\path [draw=black, semithick]
(axis cs:3,80.8292060449198)
--(axis cs:3,81.1237589667041);

\path [draw=black, semithick]
(axis cs:4,94.9621486948013)
--(axis cs:4,95.1258141669433);

\path [draw=black, semithick]
(axis cs:5,91.2979417017762)
--(axis cs:5,91.5088533746357);

\path [draw=black, semithick]
(axis cs:6,96.946199848389)
--(axis cs:6,97.0749491346087);

\path [draw=black, semithick]
(axis cs:7,98.2536806224567)
--(axis cs:7,98.3516962623317);

\addplot [semithick, black, mark=-, mark size=2, mark options={solid}, only marks]
table {%
0 92.0693330279699
1 98.5783282341417
2 83.8895825351456
3 80.8292060449198
4 94.9621486948013
5 91.2979417017762
6 96.946199848389
7 98.2536806224567
};
\addplot [semithick, black, mark=-, mark size=2, mark options={solid}, only marks]
table {%
0 92.2715285270853
1 98.6669110583657
2 84.1646673027025
3 81.1237589667041
4 95.1258141669433
5 91.5088533746357
6 97.0749491346087
7 98.3516962623317
};
\path [draw=black, semithick]
(axis cs:0.2,85.5148593061729)
--(axis cs:0.2,86.1112830677221);

\path [draw=black, semithick]
(axis cs:1.2,73.9116704777929)
--(axis cs:1.2,74.6558066747292);

\path [draw=black, semithick]
(axis cs:2.2,40.3341446863203)
--(axis cs:2.2,41.1654721862385);

\path [draw=black, semithick]
(axis cs:3.2,68.0670397162607)
--(axis cs:3.2,68.857101050368);

\path [draw=black, semithick]
(axis cs:4.2,81.9305114547347)
--(axis cs:4.2,82.5825429119837);

\path [draw=black, semithick]
(axis cs:5.2,90.7235997665339)
--(axis cs:5.2,91.2152119777239);

\path [draw=black, semithick]
(axis cs:6.2,74.6571392000012)
--(axis cs:6.2,75.3942559445593);

\path [draw=black, semithick]
(axis cs:7.2,71.0474842089785)
--(axis cs:7.2,71.8160663544992);

\addplot [semithick, black, mark=-, mark size=2, mark options={solid}, only marks]
table {%
0.2 85.5148593061729
1.2 73.9116704777929
2.2 40.3341446863203
3.2 68.0670397162607
4.2 81.9305114547347
5.2 90.7235997665339
6.2 74.6571392000012
7.2 71.0474842089785
};
\addplot [semithick, black, mark=-, mark size=2, mark options={solid}, only marks]
table {%
0.2 86.1112830677221
1.2 74.6558066747292
2.2 41.1654721862385
3.2 68.857101050368
4.2 82.5825429119837
5.2 91.2152119777239
6.2 75.3942559445593
7.2 71.8160663544992
};
\end{axis}

\end{tikzpicture} 
        \caption{\centering Performance in an unseen RF environment}
        \label{chap5-fig:interforan-barplot-perf2}
    \end{subfigure}
\caption{Model performance (accuracy, specificity, and recall) for Transfer Learning: (a) familiar RF environment and (b) unseen RF environment to evaluate generalization.}
    \label{chap5-fig:interforan-perf-comparison}
\end{figure}

Figure~\ref{chap5-fig:interforan-perf-comparison} compares accuracy, specificity, and recall for \gls{tl} models, described in Section~\ref{chap5-subsubsec:interforan-tl}, tested in a familiar \gls{rf} environment (Figure~\ref{chap5-fig:interforan-barplot-perf1}), and an unseen \gls{rf} environment (Figure~\ref{chap5-fig:interforan-barplot-perf2}), to evaluate generalization capabilities. We select the \{[128,256], 32\} model as the best-performing configuration, based on its strong results in familiar settings (96.12\% accuracy, 94.92\% specificity, and 99.31\% recall) and its superior generalization in the unseen \gls{rf} environment (91.33\% accuracy, 90.97\% specificity, and 91.40\% recall), outperforming the other model configurations.

\subsubsection{Impact of Transfer Learning}
\label{chap5-subsubsec:interforan-tl}
We evaluate InterfO-RAN performance across various \gls{rf} configurations to assess the impact of the \gls{tl} method on the model.
We use ISEC dataset (Figure~\ref{chap5-fig:interforan-map}b), dominated by High Traffic instances on both sides, as the foundational model for \gls{tl}, then fine-tuned on the EXP dataset (Figure~\ref{chap5-fig:interforan-map}a), detailed in Section~\ref{chap5-subsubsec:interforan-cnn}, which contains sufficient data for all traffic instances, as described in Table~\ref{chap5-tab:interforan-lookup}. Additionally, we train a baseline model using only the EXP dataset, and compare both performances to assess the benefits of \gls{tl}.

\begin{figure}[htb]
    \centering
        \begin{subfigure}{0.3\linewidth}
        \centering
        \setlength\fwidth{\linewidth}
        \setlength\fheight{.6\linewidth}
\begin{tikzpicture}

\definecolor{black38}{RGB}{38,38,38}
\definecolor{gray176}{RGB}{176,176,176}
\definecolor{myred}{RGB}{230,240,255}  
\definecolor{myblue}{RGB}{0,55,120} 

\definecolor{color0}{RGB}{230,240,255}    
\definecolor{color1}{RGB}{198,219,239}    
\definecolor{color2}{RGB}{158,202,225}    
\definecolor{color3}{RGB}{107,174,214}    
\definecolor{color4}{RGB}{66,146,198}     
\definecolor{color5}{RGB}{33,113,181}     
\definecolor{color6}{RGB}{8,81,156}       
\definecolor{color7}{RGB}{0,55,120}       

\begin{axis}[
    width=0.951\fwidth,
    height=1.5\fheight,
    tick align=inside,
    tick pos=left,
    x grid style={gray176},
    xlabel={Predicted label},
    xlabel style={font=\footnotesize},
    xmin=0, xmax=2,
    xtick style={color=black},
    xtick={0.5,1.5},
    xticklabels={INTERF,CLEAN},
    y dir=reverse,
    y grid style={gray176},
    ylabel={True label},
    ylabel style={font=\footnotesize},
    ymin=0, ymax=2,
    ytick style={color=black},
    ytick={0.5,1.5},
    yticklabel style={rotate=90.0},
    yticklabels={INTERF,CLEAN},
    xticklabel style={font=\scriptsize},
    yticklabel style={font=\scriptsize}
]

\fill[color6] (0,0) rectangle (1,1); 
\fill[color1] (1,0) rectangle (2,1); 
\fill[color0] (0,1) rectangle (1,2); 
\fill[color7] (1,1) rectangle (2,2); 


\draw (axis cs:0.5,0.5) node[
    scale=1.1,
    text=white,  
    rotate=0.0,
    align=center
]{\footnotesize 88.73\\
\scriptsize(215527)};

\draw (axis cs:1.5,0.5) node[
    scale=1.1,
    text=black38,  
    rotate=0.0,
    align=center
]{\footnotesize 11.27\\
\scriptsize(27384)};

\draw (axis cs:0.5,1.5) node[
    scale=1.1,
    text=black38,  
    rotate=0.0,
    align=center
]{\footnotesize 2.97\\
\scriptsize(19051)};

\draw (axis cs:1.5,1.5) node[
    scale=1.1,
    text=white,
    rotate=0.0,
    align=center
]{\footnotesize 97.03\\
\scriptsize(623139)};

\end{axis}

\end{tikzpicture} 
        \caption{\centering No TL in familiar RF environment}  
        \label{chap5-fig:interforan-conf-matrix-exp}
    \end{subfigure}
    \hfill
    \begin{subfigure}{0.3\linewidth}
        \centering
        \setlength\fwidth{\linewidth}
         \setlength\fheight{.6\linewidth}
\begin{tikzpicture}
\definecolor{black38}{RGB}{38,38,38}
\definecolor{gray176}{RGB}{176,176,176}
\definecolor{myred}{RGB}{230,240,255}  
\definecolor{myblue}{RGB}{0,55,120} 
\definecolor{color0}{RGB}{230,240,255}    
\definecolor{color1}{RGB}{198,219,239}    
\definecolor{color2}{RGB}{158,202,225}    
\definecolor{color3}{RGB}{107,174,214}    
\definecolor{color4}{RGB}{66,146,198}     
\definecolor{color5}{RGB}{33,113,181}     
\definecolor{color6}{RGB}{8,81,156}       
\definecolor{color7}{RGB}{0,55,120}       
\begin{axis}[
    width=0.951\fwidth,
    height=1.5\fheight,
    tick align=inside,
    tick pos=left,
    x grid style={gray176},
    xlabel={Predicted label},
    xlabel style={font=\footnotesize},
    xmin=0, xmax=2,
    xtick style={color=black},
    xtick={0.5,1.5},
    xticklabels={INTERF,CLEAN},
    y dir=reverse,
    y grid style={gray176},
    ylabel={True label},
    ylabel style={font=\footnotesize},
    ymin=0, ymax=2,
    ytick style={color=black},
    ytick={0.5,1.5},
    yticklabel style={rotate=90.0},
    yticklabels={INTERF,CLEAN},
    xticklabel style={font=\scriptsize},
    yticklabel style={font=\scriptsize}
]
\fill[color7] (0,0) rectangle (1,1); 
\fill[color0] (1,0) rectangle (2,1); 
\fill[color1] (0,1) rectangle (1,2); 
\fill[color7] (1,1) rectangle (2,2); 
\draw (axis cs:0.5,0.5) node[
    scale=1.1,
    text=white,  
    rotate=0.0,
    align=center
]{\footnotesize 99.31\\
\scriptsize (241223)};
\draw (axis cs:1.5,0.5) node[
    scale=1.1,
    text=black38,  
    rotate=0.0,
    align=center
]{\footnotesize 0.69\\
\scriptsize (1688)};
\draw (axis cs:0.5,1.5) node[
    scale=1.1,
    text=black38,  
    rotate=0.0,
    align=center
]{\footnotesize 5.08\\
\scriptsize (32629)};
\draw (axis cs:1.5,1.5) node[
    scale=1.1,
    text=white,
    rotate=0.0,
    align=center
]{\footnotesize 94.92\\
\scriptsize (609562)};
\end{axis}
\end{tikzpicture} 
        \caption{\centering TL in familiar RF environment}  
        \label{chap5-fig:interforan-conf-matrix-transf1}
    \end{subfigure}
    \hfill
        \begin{subfigure}{0.3\linewidth}
        \centering
        \setlength\fwidth{\linewidth}
         \setlength\fheight{.6\linewidth}
\begin{tikzpicture}
\definecolor{black38}{RGB}{38,38,38}
\definecolor{gray176}{RGB}{176,176,176}
\definecolor{myred}{RGB}{230,240,255}  
\definecolor{myblue}{RGB}{0,55,120} 
\definecolor{color0}{RGB}{230,240,255}    
\definecolor{color1}{RGB}{198,219,239}    
\definecolor{color2}{RGB}{158,202,225}    
\definecolor{color3}{RGB}{107,174,214}    
\definecolor{color4}{RGB}{66,146,198}     
\definecolor{color5}{RGB}{33,113,181}     
\definecolor{color6}{RGB}{8,81,156}       
\definecolor{color7}{RGB}{0,55,120}       
\begin{axis}[
    width=0.951\fwidth,
    height=1.5\fheight,
    tick align=inside,
    tick pos=left,
    x grid style={gray176},
    xlabel={Predicted label},
    xlabel style={font=\footnotesize},
    xmin=0, xmax=2,
    xtick style={color=black},
    xtick={0.5,1.5},
    xticklabels={INTERF,CLEAN},
    y dir=reverse,
    y grid style={gray176},
    ylabel={True label},
    ylabel style={font=\footnotesize},
    ymin=0, ymax=2,
    ytick style={color=black},
    ytick={0.5,1.5},
    yticklabel style={rotate=90.0},
    yticklabels={INTERF,CLEAN},
    xticklabel style={font=\scriptsize},
    yticklabel style={font=\scriptsize}
]
\fill[color7] (0,0) rectangle (1,1); 
\fill[color1] (1,0) rectangle (2,1); 
\fill[color1] (0,1) rectangle (1,2); 
\fill[color7] (1,1) rectangle (2,2); 
\draw (axis cs:0.5,0.5) node[
    scale=1.1,
    text=white,  
    rotate=0.0,
    align=center
]{\footnotesize 90.97\\
\scriptsize (48675)};
\draw (axis cs:1.5,0.5) node[
    scale=1.1,
    text=black38,  
    rotate=0.0,
    align=center
]{\footnotesize 9.03\\
\scriptsize (4832)};
\draw (axis cs:0.5,1.5) node[
    scale=1.1,
    text=black38,  
    rotate=0.0,
    align=center
]{\footnotesize 8.60\\
\scriptsize (23592)};
\draw (axis cs:1.5,1.5) node[
    scale=1.1,
    text=white,
    rotate=0.0,
    align=center
]{\footnotesize 91.40\\
\scriptsize (250842)};
\end{axis}
\end{tikzpicture} 
        \caption{\centering TL in unseen RF environment}  
        \label{chap5-fig:interforan-conf-matrix-transf2}
    \end{subfigure}
    \caption{Classification accuracy (percentage, sample count) with and without TL across different RF environments: (a) no TL in a familiar environment, (b) TL tested in a familiar environment, and (c) TL tested in an unseen environment, using the same color bar as Figure~\ref{chap5-fig:interforan-conf-matrix-nrsim}.}
    \label{chap5-fig:interforan-conf-matrix-comparison}
\end{figure}
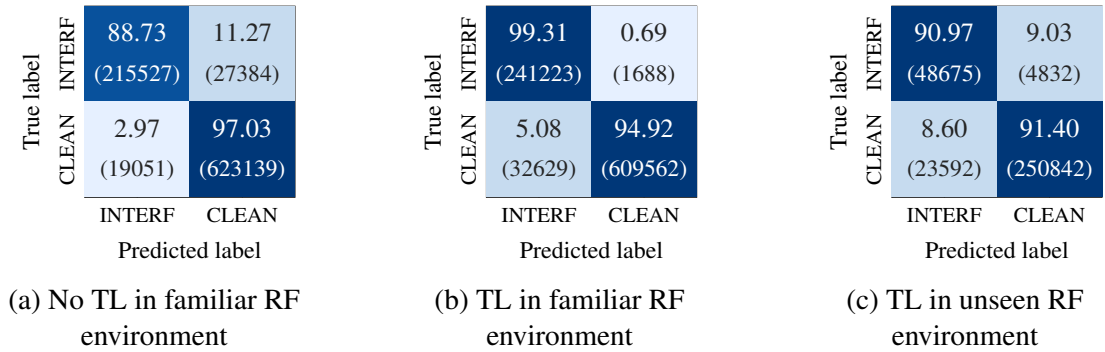

Figure~\ref{chap5-fig:interforan-conf-matrix-exp} shows the confusion matrix of the baseline model without \gls{tl}, tested in a familiar \gls{rf} environment using the same pair of \glspl{ru} locations as in the training. In contrast, Figure~\ref{chap5-fig:interforan-conf-matrix-transf1} and Figure~\ref{chap5-fig:interforan-conf-matrix-transf2} present the performance of the \gls{tl} model, evaluated both in the same familiar \gls{rf} setting as the non-\gls{tl} model, and in a new set of \gls{ru} locations, i.e., an unseen environment.
The \gls{tl} model achieves overall better performance in detecting interference when tested in both a familiar (+10.52\%) and unseen (+2.24\%) \gls{rf}. However, it is outperformed by the non-\gls{tl} model in No Traffic scenarios, with performance drops of -2.11\% in the familiar and -5.63\% in the unseen one, likely due to the influence of the foundational model.

\subsubsection{Timing and Power Benchmarking}
\label{chap5-subsubsec:interforan-benchmarking}
We assess InterfO-RAN's robustness by analyzing its inference time, GPU utilization, and power consumption, both with and without it.

\begin{figure}[htb]
    \centering
\begin{subfigure}{0.49\linewidth}
    \centering
        \setlength\fwidth{.9\linewidth}
        \setlength\fheight{.4\linewidth}
    \input{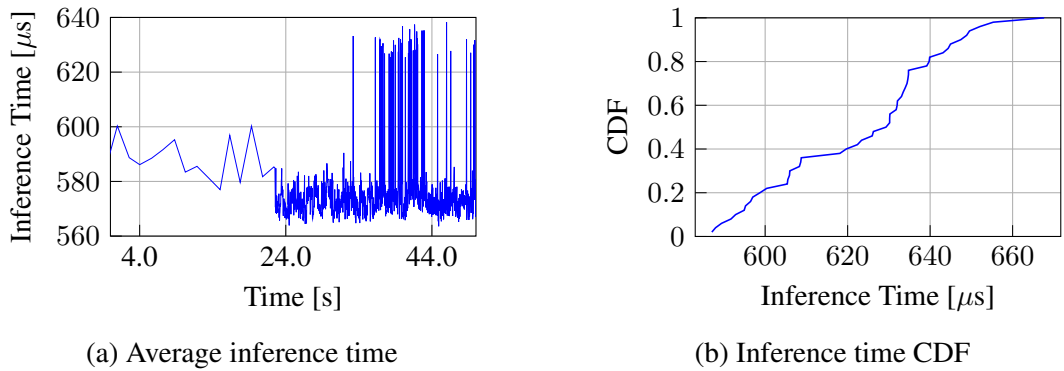} 
    \caption{\centering Average inference time}
    \label{chap5-fig:interforan-inf-time}
\end{subfigure}
    \hfill
\begin{subfigure}{0.49\linewidth}
    \centering
        \setlength\fwidth{.9\linewidth}
        \setlength\fheight{.4\linewidth}
\begin{tikzpicture}

\definecolor{gray176}{RGB}{176,176,176}
\definecolor{white204}{RGB}{204,204,204}

\begin{axis}[
width=0.951\fwidth,
height=1.5\fheight,
legend cell align={left},
legend style={
  fill opacity=0.8,
  draw opacity=1,
  text opacity=1,
  at={(0.03,0.97)},
  anchor=north west,
  draw=white204
},
tick align=inside,
tick pos=left,
x grid style={gray176},
xlabel={Inference Time [$\mu$s]},
xmajorgrids,
xmin=583.04525, xmax=671.67775,
xtick style={color=black},
y grid style={gray176},
ylabel={CDF},
ymajorgrids,
ymin=0.00, ymax=1.0,
ytick style={color=black}
]
\addplot [semithick, blue]
table {%
587.074 0.02
587.979 0.04
589.421 0.06
591.721 0.08
592.835 0.1
594.919 0.12
595.244 0.14
596.461 0.16
596.997 0.18
598.733 0.2
600.378 0.22
605.362 0.24
605.497 0.26
605.904 0.28
605.978 0.3
608.3 0.32
608.624 0.34
608.761 0.36
618.123 0.38
619.66 0.4
622.375 0.42
623.376 0.44
625.962 0.46
626.247 0.48
629.299 0.5
630.186 0.52
630.186 0.54
630.244 0.56
631.777 0.58
632.001 0.6
632.081 0.62
633.051 0.64
633.465 0.66
633.962 0.68
634.401 0.7
634.603 0.72
634.746 0.74
634.764 0.76
639.182 0.78
639.759 0.8
639.898 0.82
643.119 0.84
644.369 0.86
644.943 0.88
647.437 0.9
648.853 0.92
649.553 0.94
652.063 0.96
655.3 0.98
667.649 1
};
\end{axis}

\end{tikzpicture} 
    \caption{\centering Inference time CDF}
    \label{chap5-fig:interforan-cdf-time}
\end{subfigure}
    \caption{\justifying Inference time from first iteration: (a) temporal variation with a 5-instance moving average with No (before $24$~s) and High Traffic (after $24$~s), (b) inference time CDF.}
    \label{chap5-fig:interforan-inference-analysis}
\end{figure}

\textbf{Analysis of Inference Time.}
\label{subsec:inferencetime}
Figure~\ref{chap5-fig:interforan-inf-time} shows the temporal variation of the inference time using a 5-instance moving average between No (before $24$~s) and High Traffic (after $24$~s), while Figure~\ref{chap5-fig:interforan-cdf-time} presents the corresponding \gls{cdf}.
We observe fluctuations during high-traffic scenarios, when InterfO-RAN performs more frequent inferences, likely due to the increased number of operations and contention for \gls{gpu} resources. However, the system maintains overall stability, effectively managing workload distribution.
Additionally, Table~\ref{chap5-tab:interforan-ort-times-model} shows the average inference time for different model configurations, varying with filter counts. We notice a $220$~$\mu$s improvement in the smaller configuration due to the reduced computational complexity.


\begin{table}[htb]
    \centering
    \caption{Average inference times across different models.}
    \label{chap5-tab:interforan-ort-times-model}
    \begin{tabular}{lcc}
        \toprule
        \textbf{Model Configuration} & \{[64,128],any\} & \{[128,256],any\} \\
        \midrule
        \textbf{Inference Time [$\mu$s]} & 401.8 & 621.6  \\
        \bottomrule
    \end{tabular}
\end{table}

\begin{figure}[t]
    \centering
    \begin{subfigure}{0.8\linewidth} 
        \centering
                \includegraphics[width=\linewidth]{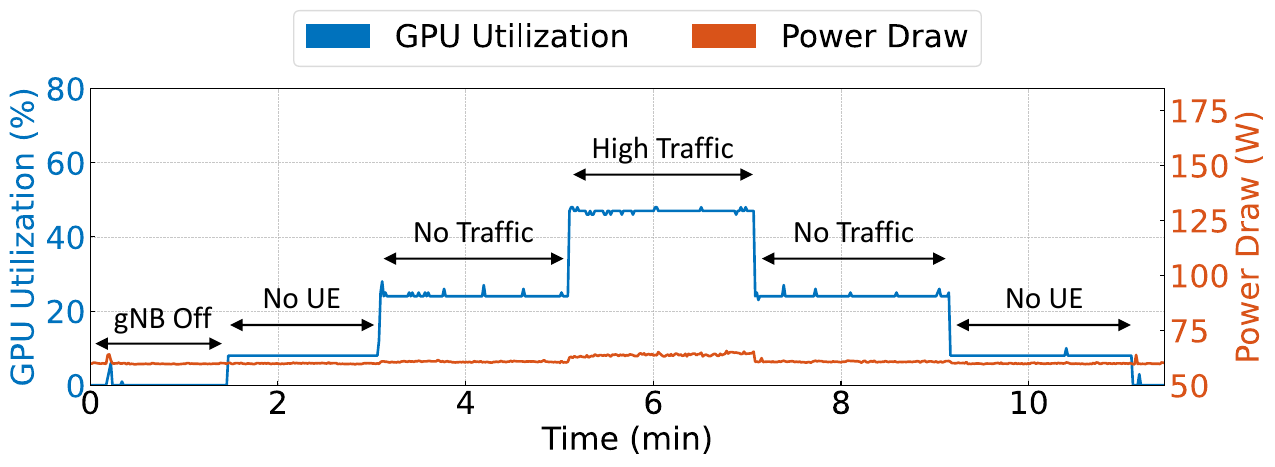}
  \caption{\centering Without InterfO-RAN dApp}
  \label{chap5-fig:interforan-gpu-nodapp}
    \end{subfigure}
    \hfill
        \begin{subfigure}{0.8\linewidth} 
        \centering
\includegraphics[width=\linewidth]
        {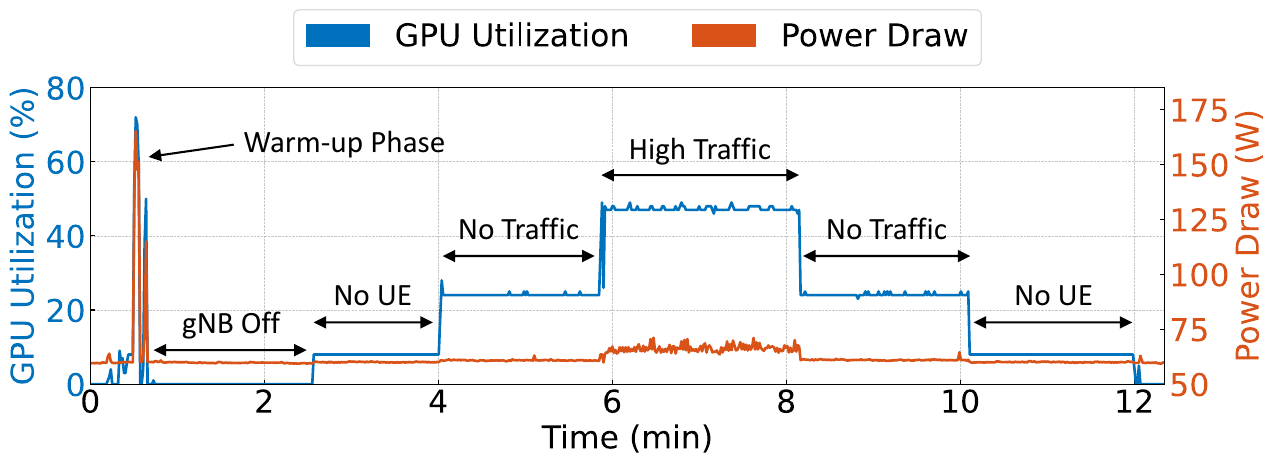}
  \caption{\centering With InterfO-RAN dApp}
  \label{chap5-fig:interforan-gpu-dapp}
    \end{subfigure}
    \caption{GPU utilization (blue) and power draw (orange) comparison with and without InterfO-RAN dApp.}
    \label{chap5-fig:interforan-gpu-comparison}
\end{figure}

\textbf{Analysis of GPU Utilization and Power.}
Figure~\ref{chap5-fig:interforan-gpu-nodapp} and Figure~\ref{chap5-fig:interforan-gpu-dapp} compare A100 \gls{gpu} utilization and power draw with and without InterfO-RAN in the same end-to-end scenario. During the warm-up phase (see Section~\ref{chap5-subsec:interforan-design}), \gls{gpu} usage spikes to 21.2\% (vs. 3.3\% without InterfO-RAN), and the power rises to $82.73$~W (vs. $61.79$~W). In other phases, power remains similar, except in High Traffic conditions, where it increases from $63.83$~W to $66.13$~W and a higher standard deviation (1.77 vs. 0.67) when using InterfO-RAN.
We conclude that InterfO-RAN effectively balances the workload without straining \gls{ran} operations, while leveraging \gls{trt} optimizations.

\label{subsec:gpu_usage_power}

\subsection{Related Work}
\label{chap5-subsec:interforan-related}

Limited research has been conducted on \gls{ul} interference in \gls{5g} \gls{nr} within real-time timescales and tested on real-world testbed environments.
Existing literature predominantly relies on simulations to predict the presence of interference. For instance, \cite{9860900} and \cite{9475442} detect intermodulation interference in \gls{5g} \gls{nr} using linear regression and \glspl{cnn}, respectively. Similarly, \cite{8885870} and \cite{9600524} detect \gls{lte} \gls{ul} interference using a novel approach that preprocesses time-domain signals into spectral waterfall representations and addresses the problem using an image classification \gls{cnn}. Additionally, authors in \cite{8104767} identify interference among IEEE 802.11b/g, IEEE 802.15.4, and IEEE 802.15.1 using a deep \gls{cnn}.

In the context of spectrum monitoring, \cite{8325299} develops a \gls{cnn} for wireless signal identification, while \cite{9120462} applies logistic regression to detect co-channel interference between \gls{lte} and WiFi users, though both approaches still lack validation beyond simulations. The authors of \cite{uvaydov2021deepsense} propose a framework based on \glspl{cnn} to sense and classify wideband spectrum portions, supported by real-world data collection and testing.
Several studies are conducted to characterize interference signals in various scenarios, such as \cite{s21020597}, which addresses downlink interference in massive \gls{mimo} \gls{5g} macro-cells using geometric channel models, and \cite{testolina2024modeling}, which focuses on modeling interference at higher frequencies for \gls{6g} networks.
Furthermore, \cite{Hattab_2019} and \cite{9860578} discuss interference mitigation mechanisms. The former focuses on angular-based exclusion zones and spatial power control in mmWave frequency ranges, while the latter employs supervised learning-based \gls{iw} selection methods.
InterfO-RAN differs from these approaches by implementing end-to-end interference detection on a \gls{gpu}-based \gls{ran} node, tested in real time on a production-ready testbed, processing raw \gls{iq} samples directly within the physical-layer pipeline.

%
%
%

\subsection{Summary}
\label{chap5-subsec:interforan-summary}



This section presented InterfO-RAN, a \gls{gpu}-accelerated O-RAN dApp that detects in-band \gls{ul} interference with over 91\% accuracy at sub-millisecond speeds (650~$\mu$s). This solution leverages a \gls{cnn} to analyze \gls{iq} samples directly within the \gls{gnb} physical layer, demonstrating seamless integration with both NVIDIA Aerial's 5G NR stack and \gls{oai} on the X5G platform. 

From the perspective of the research dimensions introduced in Chapter~\ref{chap:intro}, InterfO-RAN validates the \emph{use case} one by demonstrating interference detection on the X5G physical platform. Additionally, \emph{realism} is shown through extensive \gls{ota} data collection and testing across multiple indoor environments under realistic interference conditions, while \emph{usability} is demonstrated through the data-collection and labeling pipelines that enable rapid model development and deployment.

Future work on InterfO-RAN includes exploring scalability to higher \gls{ue} densities and mobility scenarios, repurposing it for \gls{aoa} estimation/5G positioning, implementing interference mitigation via resource allocation tuning to enhance performance, and developing online training capabilities allowing the \gls{cnn} to adapt continuously to specific \gls{rf} environments and scenarios.

\section{ORANSlice: Dynamic Resource Management}
\label{chap5-sec:oranslice}


The previous sections demonstrated real-time dApp use cases operating at sub-millisecond timescales for \gls{isac} and interference detection. This section explores a complementary capability of the X5G platform: near-real-time control via the O-RAN near-real-time \gls{ric}. ORANSlice is an open-source network slicing framework that extends \gls{oai} to support \gls{3gpp}-compliant \gls{ran} slicing with xApp-based control. In this work, we integrate ORANSlice with the \gls{osc} near-real-time \gls{ric} deployed on X5G and validate slice-aware resource management with commercial \glspl{ue} and realistic traffic patterns, showcasing a complete, standards-compliant slicing solution that integrates with the broader O-RAN ecosystem and operates on a production-ready \gls{ota} testbed.

\subsection{Introduction}

Network slicing has been identified as a key technology to deliver bespoke services and superior performance in \gls{5g} networks. 
Specifically, \gls{ran} slicing makes it possible to dynamically allocate a certain amount of \gls{ran} resources, e.g., \glspl{prb}, to each slice based on their \gls{qos}, current network conditions, and traffic load. 

The importance of network slicing is emphasized by the O-RAN ALLIANCE, which has defined it as a critical use case and technology in the context of Open \gls{ran} systems~\cite{oran-wg1-slicing-architecture}. The O-RAN architecture foresees network slicing in the \gls{ran}, controlled through an xApp deployed on the Near-RT \gls{ric}. The latter is connected to the \gls{ran} through the E2 interface, whose functionalities can be specified through \glspl{sm}. For slicing, the \gls{e2sm} Cell Configuration and Control (E2SM-CCC)~\cite{oran-wg3-ccc} allows for near-real-time adaptation of slicing parameters via xApps~\cite{oran-wg3-ricarch}. 

Despite the potential of network slicing, which has generated tremendous momentum and technological advancements, practical deployment of \gls{ran} slicing in commercial \gls{5g} networks remains largely unrealized. Numerous studies and research have demonstrated the benefits of \gls{ran} slicing, exploring various aspects, including the use of optimization~\cite{9411723} and \gls{ai}-based solutions~\cite{zhang2022federated}. 
However, most existing \gls{ran} slicing implementations are confined to research environments and bench setups.

\textbf{Contributions.}
With ORANSlice, we advance the state of the art by developing and implementing network slicing models compliant with both \gls{3gpp} and O-RAN specifications. First, we extend the \gls{oai} 5G protocol stack to support \gls{ran} slicing, redesigning the original proportional-fair scheduler into a two-tier radio resource scheduler that operates at both slice and \gls{ue} level. We also develop multi-\gls{pdu} support for the \gls{oai} softwarized 5G \gls{ue} (nrUE), enabling multiple concurrent slices on the same device. Second, we implement an E2SM-CCC-based service model and \gls{ran} slicing xApp for closed-loop control via the Near-RT \gls{ric}. Third, we conduct extensive testing and validation on Arena~\cite{bertizzolo2020comnet} and X5G~\cite{villa2024x5g}, demonstrating that ORANSlice can enforce and control slicing policies with both \gls{cots} 5G modules and softwarized \glspl{ue}, and across different radio devices and Near-RT \glspl{ric}. By releasing ORANSlice as open source, we provide the community with a \gls{3gpp}- and O-RAN-compliant \gls{ran} slicing framework that can serve as a reference design for future research and experimentation.
\footnote{\href{https://github.com/wineslab/ORANSlice}{https://github.com/wineslab/ORANSlice}}

\textbf{Organization.} The remainder of this section is organized as follows. The overview of network slicing is given in Section~\ref{chap5-sec:oranslice-openran_framework}. The details of the developed software components in ORANSlice are elaborated in Section~\ref{chap5-sec:oranslice-main_func}. The testbeds and experiment results are presented in Section~\ref{chap5-sec:oranslice-use_cases}. Finally, Section~\ref{chap5-sec:oranslice-relatedwork} discusses related work, and we conclude in Section~\ref{chap5-sec:oranslice-conclusion}.

\subsection{Network Slicing Background}
\label{chap5-sec:oranslice-openran_framework}


\begin{figure}[htb]
    \centering
    \includegraphics[width=.8\linewidth]{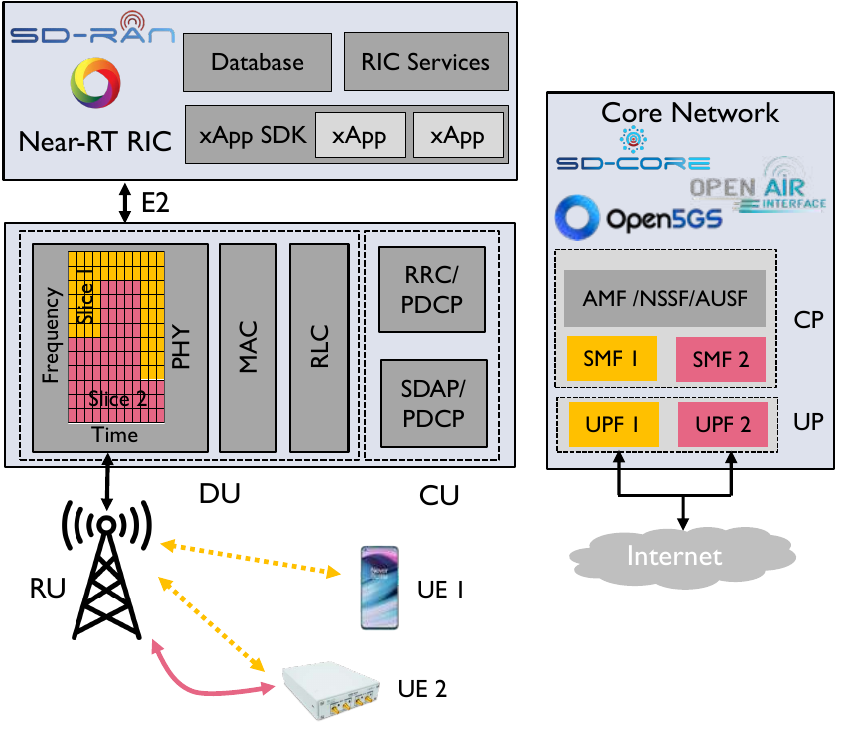}
    \caption{End-to-end network slicing in O-RAN.}
    \label{chap5-fig:oranslice-e2e_netslicing}
\end{figure}

A network slice is an end-to-end logical network spanning both \gls{ran} and \gls{cn}. It can be dynamically created and configured to provide bespoke services to serve a diverse set of applications and use cases.
In a cellular network, each network slice is uniquely identified by a \gls{snssai}, which consists of a \gls{sst} and a \gls{sd}.  
\gls{sst} is an 8-bit mandatory field identifying the slice type. 
\gls{sd} is a 24-bit optional field that differentiates among slices with the same \gls{sst}. 
It is worth noting that the same \gls{ue} can be subscribed to up to 8 slices, which benefits applications with varying network requirements.


\textbf{RAN Slicing.}~Radio resources are organized in a time and frequency grid (Figure~\ref{chap5-fig:oranslice-e2e_netslicing}). Each grid element, referred to as a \gls{prb}, is used to schedule \gls{ue} transmissions and to broadcast control messages, among other uses.  
Network slicing in the \gls{ran} (i.e., \gls{ran} slicing) refers to the problem of allocating (and dedicating) such radio resources (i.e., \glspl{prb}) to different slices according to certain slicing policies. 


\textbf{Core Network Slicing.}
The \gls{5g} \gls{cn} enables secure and reliable connectivity between the \gls{ue} and the Internet. As shown in Figure~\ref{chap5-fig:oranslice-e2e_netslicing}, the \gls{5g} \gls{cn} is decoupled into \gls{up} and \gls{cp}.
To enable network slicing in the \gls{cn} (i.e., core slicing), dedicated \glspl{nf} in both \gls{cp} and \gls{up} need to be created for each slice. For example, a dedicated \gls{smf} and \gls{upf} pair can be created for each slice (Figure~\ref{chap5-fig:oranslice-e2e_netslicing}). 
The other \glspl{nf} in the \gls{cn}, such as the \gls{amf}, can be shared instead.

\textbf{Multi-slice Support.}
\gls{3gpp} specifies a set of procedures to link \gls{ue} traffic to a certain network slice. Specifically, during the \gls{pdu} session establishment phase, the \gls{ue} can specify the \gls{snssai} of the target network slice.
Upon establishment, the \gls{pdu} session is assigned an IP address, allowing applications in the \gls{ue} to access the network service of the network slice by binding to that IP. 
This procedure enables multiple \gls{pdu} sessions on the same \gls{ue}, and each \gls{pdu} can be bound to a dedicated network slice tailored to the use case or application executed by the \gls{ue}.  

\textbf{RAN Slicing in O-RAN.}
The O-RAN WG3 has identified \gls{ran} slicing as a key use case in the context of the Near-RT \gls{ric} specifications~\cite{oran-wg3-ucr}. It has also released the O-RAN service model E2SM-CCC~\cite{oran-wg3-ccc} which details the structure and procedures necessary to enable \gls{ran} slicing in O-RAN following the 3GPP specifications described above. This is performed via xApps executing at the Near-RT \gls{ric}, where limited radio resources are managed in near-real-time to meet \gls{qos} requirements and to accommodate highly varying \gls{ran} load and conditions.

\subsection{ORANSlice Implementation}
\label{chap5-sec:oranslice-main_func}

This section details the design and implementation of the missing blocks required to support the \gls{ran} slicing use cases based on the O-RAN specifications.
From a software point of view, the open-source ecosystem already offers all architectural blocks necessary to instantiate and operate a 5G network.
For example, disaggregated 5G base stations can be instantiated via \gls{oai}~\cite{oai5g} and srsRAN~\cite{srsRAN}, which also offer O-RAN integration and functionalities. A 5G core network can be instantiated via \gls{oai}, Open5GS \cite{open5gs}, or SD-Core~\cite{onf}, all of which support core slicing. Similarly, the \gls{osc} and Aether offer open-source implementations of Near-RT \glspl{ric}. OpenRAN Gym, an open-source project for collaborative research in the O-RAN ecosystem, provides components to connect across \gls{ran} and \glspl{ric}~\cite{bonati2022openrangym-pawr}. 


What has been missing is an open-source, 3GPP- and O-RAN-compliant implementation of \gls{ran} slicing functionalities and the support for multi-slice applications at the same \gls{ue}. 


\subsubsection{RAN Slicing Enabled Protocol Stacks}

The protocol stack of ORANSlice is based on \gls{oai}~\cite{oai5g}. 
To enable \gls{ran} slicing, ORANSlice advances and extends the functionalities of the \gls{mac} layer, including slice information of \gls{pdu} sessions of each \gls{ue} and a re-designed two-tier radio resource scheduler.  
Moreover, we extend the \gls{oai} nrUE to support the instantiation and management of multiple \gls{pdu} sessions for different slices.


\textbf{Slice-Aware MAC.} 
In \gls{oai}, the \gls{mac} layer implements a \emph{proportional-fair} scheduling algorithm to allocate radio resources between different \glspl{ue}.
However, to realize \gls{ran} slicing, the \gls{mac} scheduler needs to be aware of \gls{pdu}-slice associations so as to properly allocate resources among users.

There are two basic types of \gls{ran} slicing schemes. The first one implements \textit{slice-isolation}, where a fixed amount of resources is exclusively dedicated to each slice to guarantee resource availability~\cite{bonati2021scope}. This method completely prevents resource sharing, resulting in low resource utilization efficiency because unused \glspl{prb} are not reallocated to other slices. However, it offers a low-complexity implementation.
The second one is the \textit{slice-aware} scheme, in which resources can be exclusively allocated to each slice or shared among slices via a priority-based access mechanism. The slice-aware scheme increases resource utilization efficiency by allowing \glspl{prb} allocated to a slice but unused to be distributed across other slices on demand.

\begin{figure}[tb]
    \centering
    \includegraphics[width=.7\linewidth]{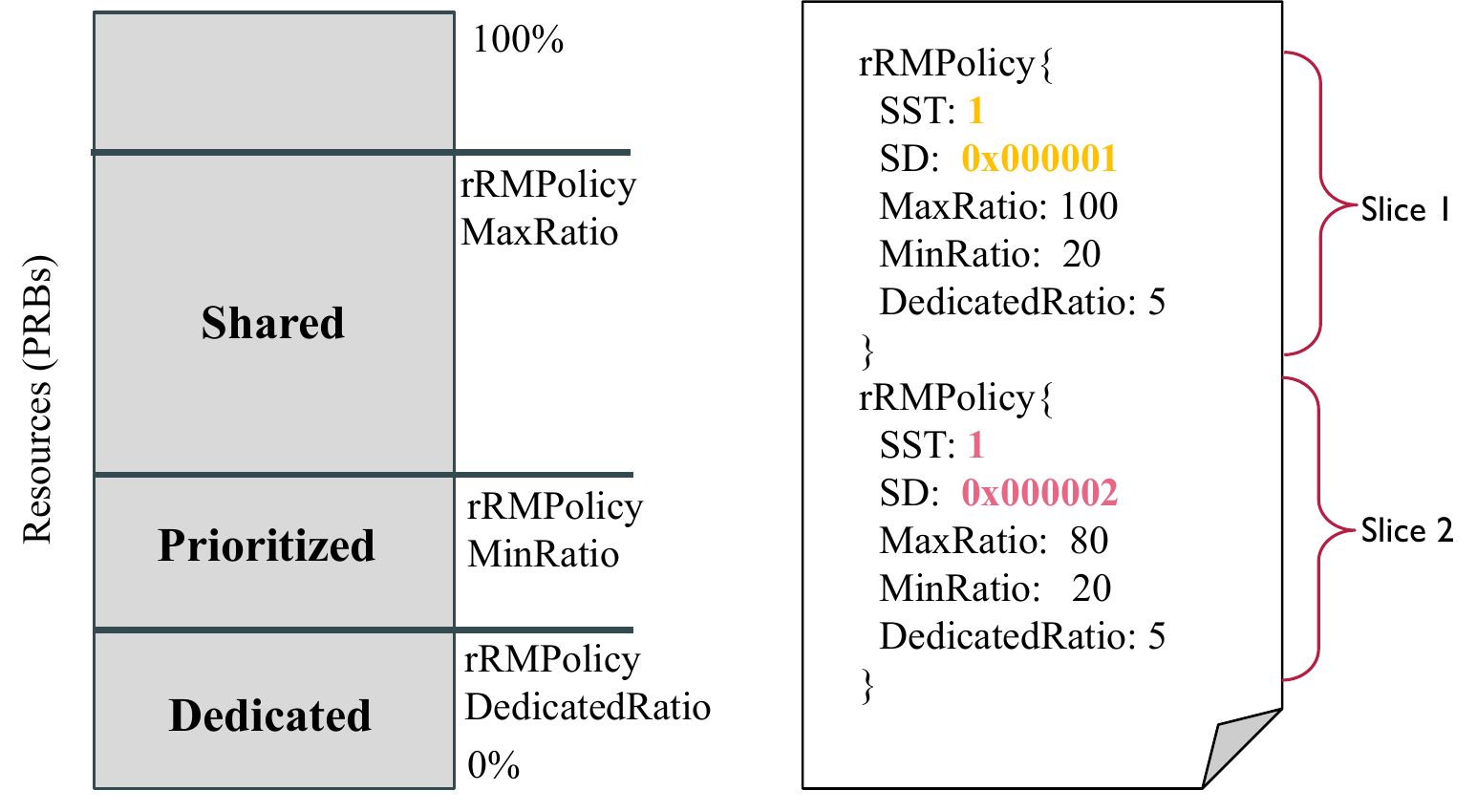}
    \caption{Illustration of RRM Policy Ratio Message}
    \label{chap5-fig:oranslice-rrmPolicy}
\end{figure}

Since the second approach is more efficient, it has been selected by the 3GPP~\cite{3gpp.28.541} and the O-RAN ALLIANCE~\cite{oran-wg1-slicing-architecture} as a reference implementation for \gls{ran} slicing, and for this reason, we consider it in ORANSlice. 
In the context of \gls{ran} slicing, the 3GPP introduces the concept of \gls{rrm} \gls{ran} slicing policy~\cite{3gpp.28.541}, as illustrated in Figure~\ref{chap5-fig:oranslice-rrmPolicy}. The \texttt{rRMPolicyDedicatedRatio} represents the dedicated percentage of \glspl{prb} allocated to the network slice and is exclusively allocated to the slice even if the slice has no active traffic demand. 
The 3GPP also defines the concept of prioritized \gls{prb} access via two parameters: \texttt{rRMPolicyMinRatio} and \texttt{rRMPolicyMaxRatio}.
The former represents the minimum percentage of \glspl{prb} that would be prioritized for allocation to the network slice.
The latter represents the maximum percentage of \glspl{prb} that can be allocated to the slice. At a high level, the percentage of \glspl{prb} that falls within the prioritized range (i.e., defined as the difference between the maximum ratio and the dedicated ratio) is guaranteed to be allocated to the slice, but only if the slice has active users requesting \glspl{prb}. 

ORANSlice extends the \emph{proportional-fair} scheduler of \gls{oai} to integrate the slice-aware scheme described above. As we will describe later, this is achieved by implementing a two-tier resource allocation mechanism that considers: (i) inter-slice resource allocation and sharing according to \gls{ran} slicing policies, which will be elaborated in Section~\ref{chap5-sec:oranslice-e2sm_rrmpolicy}; and (ii) intra-slice resource allocation to schedule transmissions for all \glspl{ue} that belong to the same slices.\footnote{We would like to mention that support for the dedicated ratio defined by the \gls{rrm} policy in Figure~\ref{chap5-fig:oranslice-rrmPolicy} will be added in the future.}

\textbf{UE Slicing Support in OAI nrUE.} 
The other important contribution of ORANSlice is the introduction of support for multiple slices on the same \gls{ue}. 
Prior to ORANSlice, \gls{oai} softwarized \gls{5g} \gls{ue}, i.e., \gls{oai} nrUE, supported only one \gls{pdu} session per \gls{ue}, limiting each \gls{ue} to a single active slice at any time. To enable multi-slice support, we have extended nrUE functionalities to: (i) enable the coexistence of multiple \gls{pdu} sessions on each \gls{ue}; (ii) instantiate/delete new \glspl{pdu} on-demand; and (iii) assign each \gls{pdu} to a slice. 
We already contributed these functionalities to \gls{oai} nrUE repository, and they are available to the community as open-source components. It is worth mentioning that currently available 5G smartphones do not support these features, and we hope that this will 
help the community to investigate and explore \gls{ran} slicing topics that go beyond single \gls{ue}-slice associations.

\subsubsection{Implementing the E2SM-CCC}
\label{chap5-sec:oranslice-e2sm_rrmpolicy}

To enable the O-RAN \gls{ran} slicing use case, we also developed the corresponding \gls{e2sm} to realize the O-RAN E2SM-CCC service model. We developed two versions of the E2SM-CCC. One has been integrated with the Linux Foundation's Aether SD-RAN Near-RT \gls{ric}~\cite{onf} and already made public,\footnote{\href{https://github.com/onosproject/onos-e2-sm/pull/392}{https://github.com/onosproject/onos-e2-sm/pull/392}}
while a simplified version has been integrated with the \gls{osc} Near-RT \gls{ric}. 

\noindent\textbf{E2SM-CCC Service Model.}
The \gls{e2sm} describes how a specific \gls{ran} function (the \textit{service}) within an E2 node interacts with the Near-RT \gls{ric} and its xApps. 
It consists of an \gls{e2ap} and a data schema accepted by both the \gls{ran} function and Near-RT \gls{ric}.
The flexible, programmable E2SM-CCC is critical to the autonomous, intelligent \gls{ran} control loop in O-RAN.

To support \gls{ran} slicing reconfiguration via xApps, we have implemented a simplified version of the E2SM-CCC service model integrated with the \gls{osc} Near-RT \gls{ric} to support 3GPP-compliant \gls{ran} slicing reconfiguration via xApps. 

Only the \textit{O-RRMPolicyRatio} configuration of E2SM-CCC is supported in the implemented service model.
Other configurations defined in E2SM-CCC are omitted as they are beyond the scope of this work.
Protobuf is used to send control messages that include the \gls{rrm} policy necessary to update \gls{ran} slicing strategies enforced by the \gls{ran}. 

\subsubsection{RAN Slicing xApp and Data-driven Control Automation}

We developed a \gls{ran} Slicing xApp to compute \gls{rrm} policies necessary to update \gls{ran} slicing strategies. 
Specifically, the \gls{ran} Slicing xApp periodically reads the \gls{ran} \glspl{kpi} from an InfluxDB database, processes them, and generates the slicing policies that are sent to the gNB through the E2 interface via a \gls{ric} Control message. 
Control messages from the xApp are serialized into Protobuf objects and sent via the \glspl{sm} described above. 
The data from the gNB is periodically sent to a dedicated \gls{osc} \gls{kpm} xApp, which reads \glspl{kpm} received over the E2 termination via a \gls{ric} Indication message. The \gls{kpm} xApp, which is also running on the Near-RT \gls{ric} and subscribes to the gNB reporting through a \gls{ric} Subscription message, inserts the received \gls{ran} \glspl{kpm} into the InfluxDB database that is leveraged by the \gls{ran} Slicing xApp. Similar to the \gls{ric} Control messages, the payload of the \gls{ric} Indication messages is serialized into Protobuf objects.
\subsection{Testbeds and Experiment Results}
\label{chap5-sec:oranslice-use_cases}

In this section, we demonstrate the effectiveness and portability of ORANSlice by deploying an end-to-end O-RAN cellular network on two physical testbeds: Arena, the \gls{sdr}-based \gls{ota} platform described in Section~\ref{chap2-sec:arena}~\cite{bertizzolo2020comnet}, and X5G, described in Chapter~\ref{chap:4}~\cite{villa2025tmc}. 
We demonstrate the functionality of ORANSlice by performing two experiments.
In the first experiment (Section~\ref{chap5-sec:oranslice-ran_slicing_ctrl}), we show an end-to-end network slicing deployment with \gls{kpm} and \gls{ran} slicing xApp running in the \gls{osc} Near-RT \gls{ric}. The goal of the first experiment is to demonstrate the \gls{ran} slicing functionalities enabled by ORANSlice.
In the second experiment (Section~\ref{chap5-sec:oranslice-qos}), we show a simple, yet illustrative, example of how to guarantee access to a minimum amount of radio resources via tailored \gls{ran} slicing policies in ORANSlice. In this latter experiment, we also demonstrate the multi-slice support offered by ORANSlice. 
In addition to the \gls{ota} transmission experiments on two testbeds, we also replicate the two experiments in \gls{oai} RFSim mode to show the ability of ORANSlice working without physical \gls{ue} or radio devices.

\noindent \textbf{5G UE.}
We test both \gls{cots} and softwarized \glspl{ue}. 
The \gls{cots} Sierra Wireless EM9191 \gls{5g} module is selected for flexible experiments, as it supports multiple \gls{pdu} sessions and customized \gls{snssai} for each \gls{pdu} session. 
We also tested ORANSlice with the \gls{oai} nrUE, a softwarized \gls{ue} that we have extended to enable multi-\gls{pdu} sessions support and thus \gls{ue} multi-slicing.

\noindent \textbf{5G Core Network.}
To validate slicing operations, we tested ORANSlice with \gls{oai} \gls{cn}, Open5GS, and SD-Core. These are modular \glspl{cn} and support network slicing natively. 
With consistent slicing configuration at \gls{ue}, \gls{gnb}, and the core network, ORANSlice can achieve an end-to-end \gls{5g} network slicing.

\noindent \textbf{Near-RT \gls{ric}.}
The Near-RT \gls{ric} and the xApps of Figure~\ref{chap5-fig:oranslice-e2e_netslicing} are deployed via Red Hat OpenShift. Specifically, we deployed the ``E'' release of the \gls{osc} Near-RT \gls{ric} together with the \gls{osc} \gls{kpm} xApp and the \gls{ran} Slicing xApp. 



\subsubsection{Testing RAN Slicing Control}
\label{chap5-sec:oranslice-ran_slicing_ctrl}



In the Near-RT \gls{ric}, the deployed \gls{kpm} xApp and \gls{ran} Slicing xApp work together to enable \gls{ran} slicing control. Every $0.5$ seconds, the \gls{kpm} xApp acquires \glspl{kpi} for all connected \glspl{ue} from the E2 interface. This includes user information (e.g., \gls{rnti} and \gls{snssai}) and \glspl{kpm} such as \gls{bler}, \gls{mcs}, and throughput. This data is stored in an InfluxDB database in the form of time series data. 

The \gls{ran} Slicing xApp reads the \glspl{kpm} from the database and calculates the average downlink throughput for each slice for the previous $5$ seconds. Then, it identifies the slice with the lowest and highest reported throughput and sets their \texttt{rRMPolicyMaxRatio} to $90\%$ and $10\%$, respectively. 
The goal of this experiment is to demonstrate the correctness of the \gls{rrm} policy update. 
How to use ORANSlice to satisfy a target \gls{qos} will be shown in Section~\ref{chap5-sec:oranslice-qos}.

\begin{figure}[htb!]
    \centering
    \includegraphics[width=.7\linewidth]{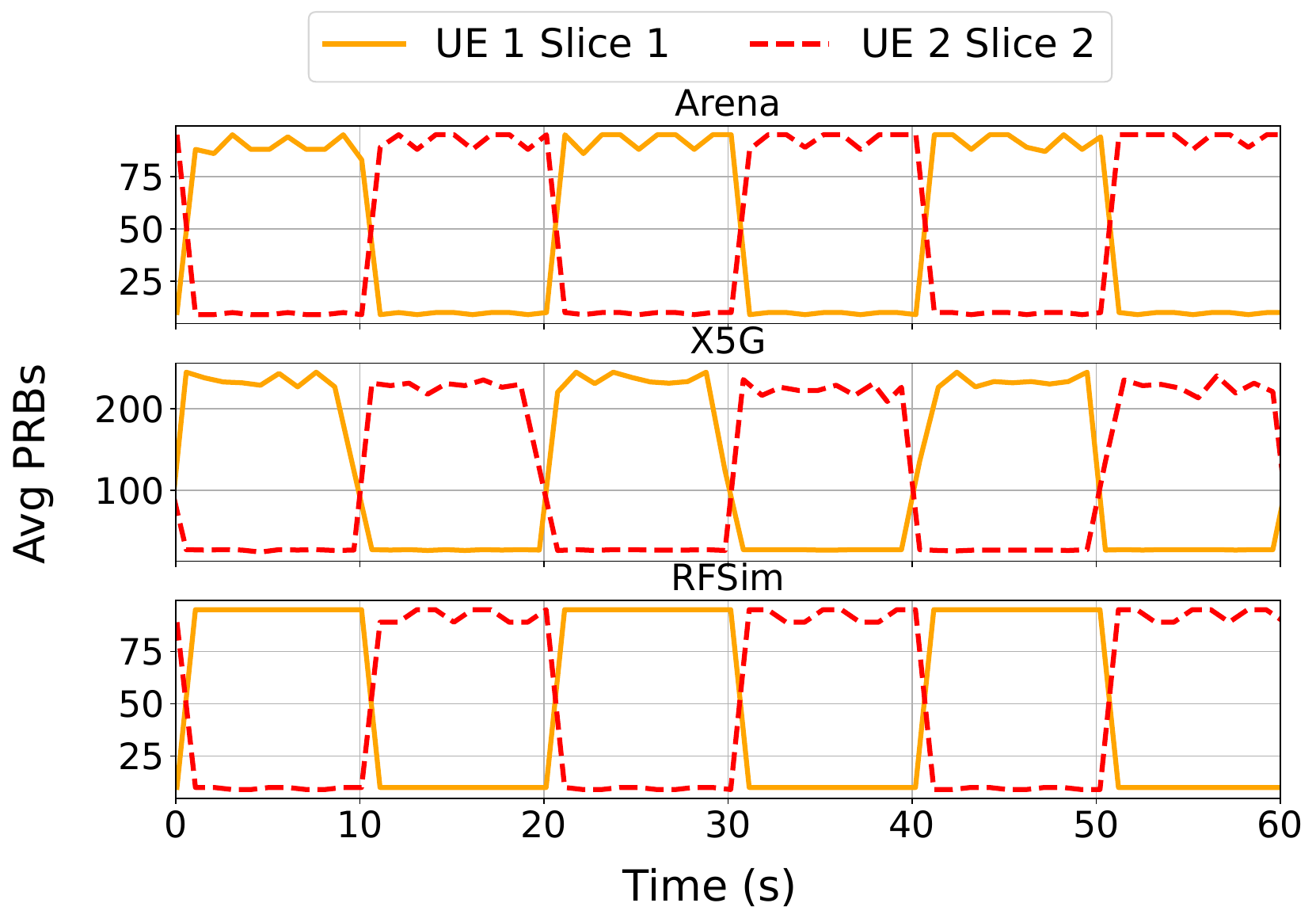}
    \caption{Average number of DL \acrshortpl{prb} allocated to each slice in the first experiment.}
    \label{chap5-fig:oranslice-dl_prb_slicing_ctrl}
\end{figure}

\begin{figure}[htb!]
    \centering
    \includegraphics[width=.7\linewidth]{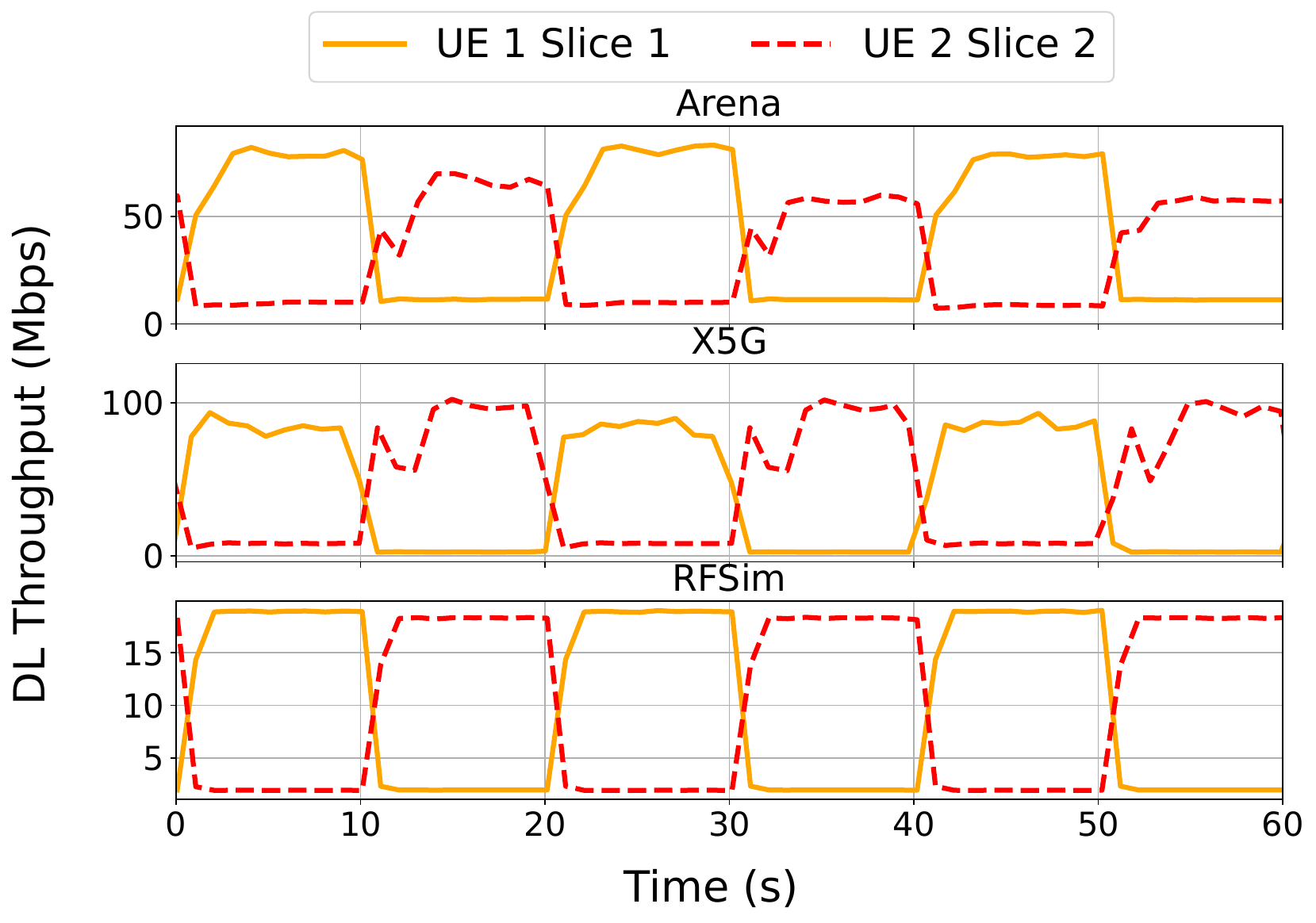}
    \caption{DL Throughput for each slice in the first experiment.}
    \label{chap5-fig:oranslice-dl_thruput_slicing_ctrl}
\end{figure}

We consider two slices and two \glspl{ue}, each associated with one slice. The control logic computes a new \gls{rrm} policy every $10$ seconds. 
As shown in Figure~\ref{chap5-fig:oranslice-dl_prb_slicing_ctrl}, the averaged \glspl{prb} per frame for \gls{ue} 1 and \gls{ue} 2 changes every $10$ seconds. For the Arena testbed (which can allocate up to 106 \glspl{prb}), the \glspl{prb} of each \gls{ue} varies between $10$ and $90$. For the X5G testbed, the \glspl{prb} of each \gls{ue} range from $27$ to $230$ (the gNB in this testbed can allocate up to 273 \glspl{prb}). RFSim has the same 5G numerology as Arena, and the average number of allocated \glspl{prb} follows the same pattern.
In all cases, the number of \glspl{prb} allocated to each slice is approximately equal to $10\%$ and $90\%$ of the number of available \glspl{prb} in each testbed, which proves that the \gls{rrm} policy is applied correctly.
In Figure~\ref{chap5-fig:oranslice-dl_thruput_slicing_ctrl}, we also show the impact that the slicing policy has on the downlink throughput of each slice, which depends on the amount of \glspl{prb} available to the slice.
We see the downlink throughput for the two \glspl{ue} varies following the number of \glspl{prb} in Figure~\ref{chap5-fig:oranslice-dl_prb_slicing_ctrl}. The difference between the throughput values of \glspl{ue} is related to the differences in the hardware and RF channel peculiar to each testbed.


\subsubsection{Multi-Slice with Guaranteed PRBs}
\label{chap5-sec:oranslice-qos}

In this experiment, we show how ORANSlice handles multi-slice applications and guarantees minimum \gls{prb} allocation by fine-tuning \texttt{rRMPolicyMinRatio} in the \gls{rrm} policy. 
We consider two slices and 2 \glspl{ue}, and \gls{ue} 1 activates two \glspl{pdu} each with a different \gls{snssai}. This means that \gls{ue} 1 has active \glspl{pdu} on both slices.  
We set $\texttt{rRMPolicyMinRatio} = 0$ and $\texttt{rRMPolicyMaxRatio} = 100$ for all slices, i.e., no minimum \gls{prb} guarantee.


\begin{figure}[htb]
    \centering
    \includegraphics[width=.7\linewidth]{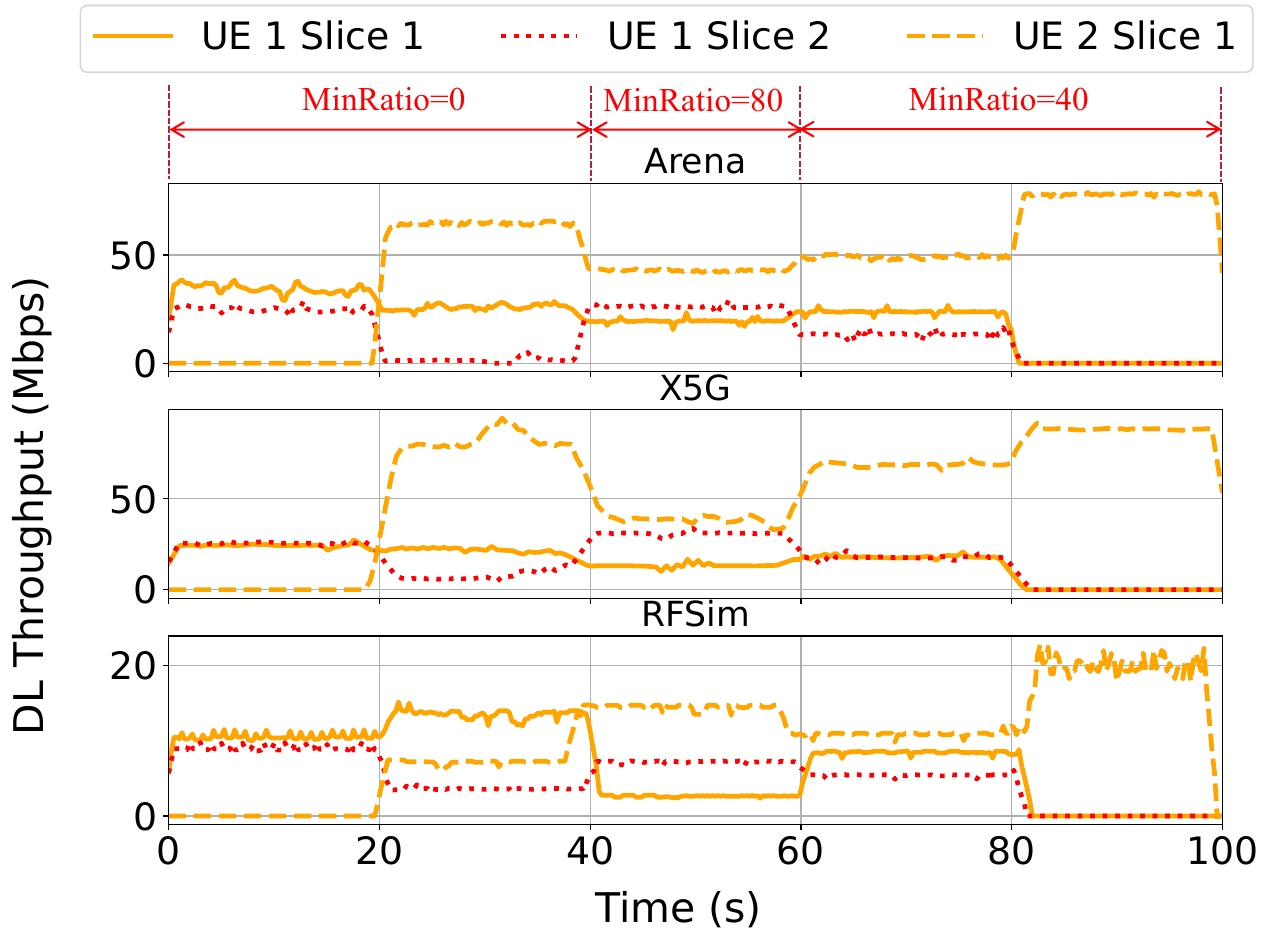}
    \caption{DL throughput for each slice and \acrshort{ue} in the second experiment.}
    \label{chap5-fig:oranslice-slices_qos}
\end{figure}

The experiment evolution is illustrated in Figure~\ref{chap5-fig:oranslice-slices_qos}. 
In step~1 ($0\sim20$s), \gls{ue} 1 establishes two \gls{pdu} sessions, with one \gls{pdu} session associated with each slice. For each \gls{pdu}, \gls{ue}~1 generates \gls{tcp} downlink data via \texttt{iPerf3}, while
\gls{ue} 2 is inactive.
We observe that the throughput achieved by the two \glspl{pdu} of \gls{ue} 1 is comparable, thanks to the proportional-fair scheduler.
In step~2 ($20\sim40$s), \gls{ue} 2 establishes a \gls{pdu} session associated to slice $1$, and starts an \texttt{iPerf3} \gls{tcp} downlink data transmission. The throughput of \gls{ue} 1's \gls{pdu} associated with slice $2$ decreases to about 1~Mbps, due to the resource competition caused by the new establishment of \gls{ue} 2's PDU on slice $1$. 
Moreover, since $\texttt{rRMPolicyMinRatio} = 0$ for all slices, the gNB does not attempt to improve the throughput of \gls{ue} 1's PDU associated with slice $2$ as the \gls{rrm} policy does not specify any guaranteed \gls{prb} provisioning for slice $2$.
In step~3 ($40\sim60$s), the $\texttt{rRMPolicyMinRatio}$ of slice $2$ is set to $80$, which makes the throughput of \gls{ue} 1's \gls{pdu} associated to slice $2$ increase thanks to the minimum \glspl{prb} guarantee.
Similarly, in step~4 ($60\sim80$s), the $\texttt{rRMPolicyMinRatio}$ of slice $2$ is decreased from $80$ to $40$ and the throughput of \gls{ue} 1 on slice $2$ decreases.
In step~5 (80 to 100 seconds), \gls{ue} 1 stops the \texttt{iPerf3} transmission at the beginning of this step, while \gls{ue} 2 stops it at the end.
This experiment demonstrates how fine-tuning the $\texttt{rRMPolicyMinRatio}$ parameter for each slice can guarantee a minimum \gls{prb} level for individual slices.

\subsection{Related Work}
\label{chap5-sec:oranslice-relatedwork}

\gls{ran} software stacks, such as \gls{oai}~\cite{oai5g} and srsRAN~\cite{srsRAN}, have been instrumental in advancing the field by welcoming contributions from the open-source community, for example, to integrate and release network slicing technologies for both 5G~\cite{FlexSlice2023} and 4G~\cite{bonati2021scope} systems.
Examples of this are provided by SCOPE~\cite{bonati2021scope} and FlexSlice~\cite{FlexSlice2023}. The former is a network slicing framework based on srsRAN 4G, which, however, uses custom \glspl{sm} to communicate with the xApps.
%
The latter is a \gls{ran} slicing framework with a recursive radio resource scheduler based on \gls{oai} 5G. However, only a single slice can be associated with each \gls{ue}, and it lacks integration with the O-RAN E2SM.

Efforts similar to ours have been documented in the literature. 
ProSlice in~\cite{ProSlice2022} developed a customized E2SM and xApp to support \gls{ran} slicing. However, ProSlice does not follow the \gls{3gpp} \gls{ran} slicing model, and the E2SM is not O-RAN-compliant. Besides, ProSlice is not open-source.
RadioSaber~\cite{RadioSaber2023nsdi} is a channel-aware \gls{ran} slicing framework implemented in a \gls{ran} simulator.
Finally, Zipper~\cite{Zipper2024nsdi} is a \gls{ran} slicing framework based on closed-source commercial protocol stacks.
The closest works to ours are~\cite{ProSlice2022} and~\cite{FlexSlice2023}. 
Compared to these works, ORANSlice advances the state-of-the-art by introducing support for multi-\gls{pdu} and thus multi-slice management on the same \gls{ue}.
This enhancement enables more efficient, flexible handling of different types of traffic, catering to the diverse needs of modern applications where multiple classes of traffic with different \gls{qos} requirements coexist on the same device and must be satisfied simultaneously.

\subsection{Conclusions}
\label{chap5-sec:oranslice-conclusion}


This section presented ORANSlice, an open-source \gls{5g} framework for network slicing in O-RAN. ORANSlice extends \gls{oai} to deliver \gls{3gpp}-compliant \gls{ran} slicing and support for multi-slice applications, integrating an E2 service model based on E2SM-CCC and a \gls{ran} Slicing xApp working with the \gls{osc} Near-RT \gls{ric}. The \gls{ran} slicing functionalities were demonstrated on multiple \gls{ota} O-RAN testbeds, including X5G and Arena, and on the \gls{oai} RF simulator for different use cases.

The release of ORANSlice as open-source software aims to bridge the gap between research and practice, providing the community with a valuable resource for further exploration and development of \gls{ran} slicing technologies in the O-RAN ecosystem.

\section{TIMESAFE: Security of Open Interfaces}
\label{chap5-sec:timesafe}


The previous use cases focused on leveraging the X5G physical platform for \gls{ai}-driven applications and intelligent control, such as sensing, interference detection, and resource management. This section explores a different capability, security assessment, through the development of TIMESAFE.
TIMESAFE investigates vulnerabilities in the O-RAN \gls{fh} synchronization plane, demonstrating how attacks against \gls{ptp} can cause catastrophic failures in production \gls{5g} networks. This work highlights the unique value of physical platforms like X5G for security research, enabling the validation of attack impact on real hardware with commercial \glspl{ru} and \glspl{ue}, and providing insights that cannot be obtained through simulation alone.

\subsection{Introduction}
\label{chap5-sec:timesafe-intro}

\begin{figure}[tb]
    \centering
    \includegraphics[width=.8\linewidth]{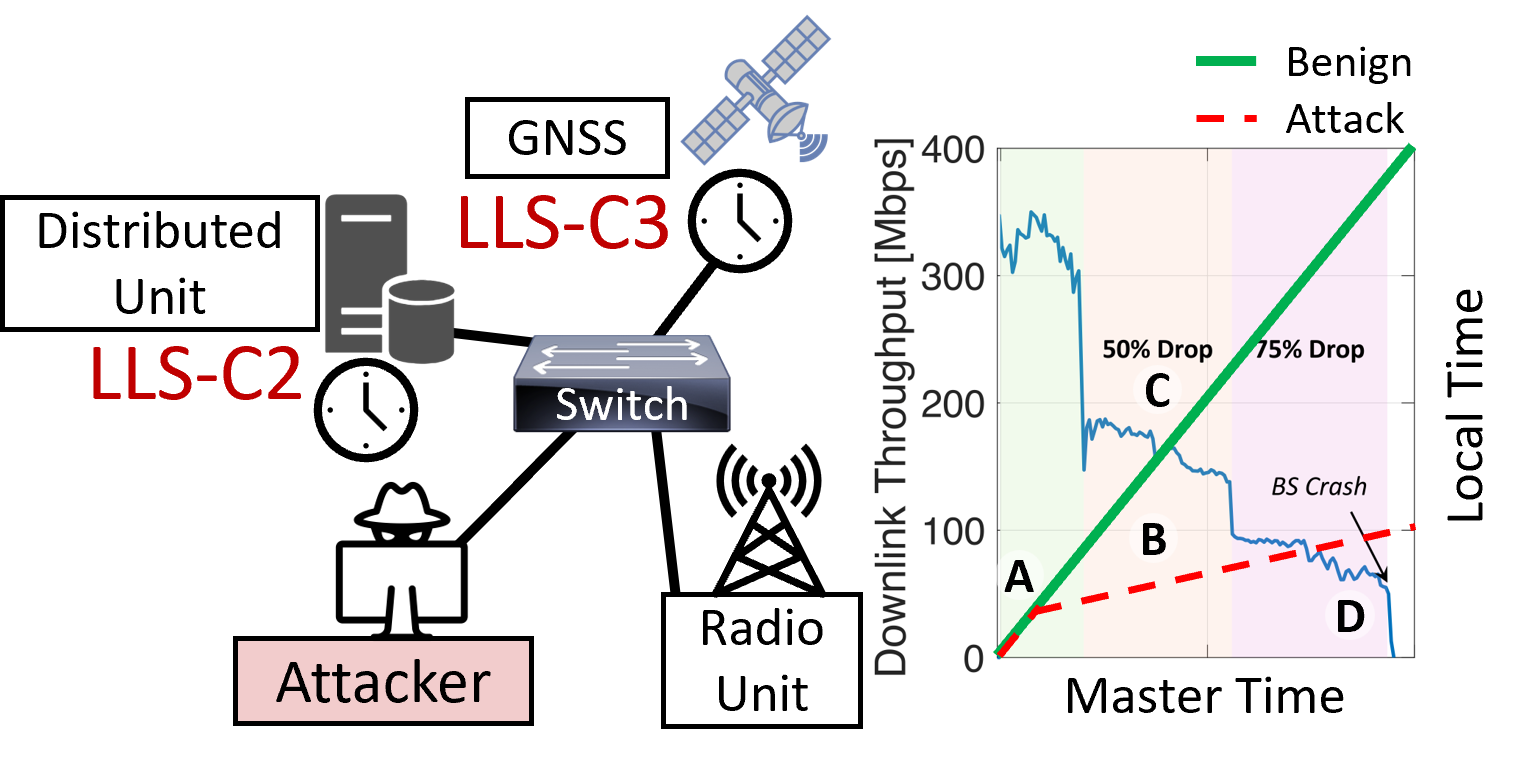}
    \caption{The Open Fronthaul employs PTP to synchronize the master clock with distributed base station components over a switched network. In this scenario, an attacker compromises a device on the network and, at time \textbf{A}, initiates a Spoofing Attack. This leads to gradual synchronization drift (\textbf{B}), first degrading performance (\textbf{C}) and then causing the base station to crash at time \textbf{D}.}
    \label{chap5-fig:timesafe-overview}
\end{figure}

Recent advancements in \gls{5g} and beyond have focused on disaggregating the \gls{gnb} and core networks, leading to the evolution of more open and modular systems. As a result, the \gls{fh} has emerged as a critical component, enabling the separation of radio elements from baseband processing.
Disaggregated FH deployments were initially based on the \gls{cpri}. 
While \gls{cpri} facilitated high-speed, centralized processing of radio signals, it lacked built-in security measures due to its design for single-vendor, co-located deployments~\cite{wong2022security}. The subsequent \gls{ecpri} introduced some security improvements, but significant gaps remain, particularly in synchronization~\cite{ecpri}. The \gls{ecpri}-based O-RAN ALLIANCE's open \gls{fh} now connects disaggregated components from multiple vendors, often using switched Ethernet topologies~\cite{ORA2023,etsiFh}. While Ethernet's flexibility supports diverse traffic types, it also exposes the \gls{fh} to threats, as demonstrated by potential attacks on interconnected devices~\cite{dik2021transport, abdalla2022toward, hung2024security, atalay2023securing, 10.1145/3643833.3656118}. These attacks can disrupt synchronization and cause outages. For example, as shown in Figure~\ref{chap5-fig:timesafe-overview}, an attacker that gains access to the \gls{fh} network can carry out spoofing attacks that cause synchronization drift, resulting in a reduction in throughput and eventually a complete crash of the base station.
Although industry, government, and academia are working to address these security gaps~\cite{wong2022security, ericcson, OranWG11-secreqspec, DoD, abdalla2022toward, groen2024securing}, no comprehensive standards for securing the \gls{fh} have yet been established.

The open \gls{fh} relies on precise timing through the \gls{sp} to perform critical operations like OFDM synchronization, carrier aggregation, and handovers, which require time accuracy within tight margins~\cite{municio2023ran, s_plane_survey}. This synchronization depends on the \gls{ptp}, which, like the \gls{fh}, was initially developed without any security mechanisms~\cite{shi2023ms}. While recent updates to the \gls{ptp} standard include additional security features~\cite[Annex P]{ptp_standard}, these have not been widely implemented in \gls{oran} environments, leaving vulnerabilities that attackers could exploit.

Security is critical for the success and widespread adoption of any system, and \gls{oran} has been criticized for lacking robust security mechanisms. In response, the O-RAN ALLIANCE established WG11~\cite{OranWG11-secreqspec} to define a security framework, including zero-trust architectures~\cite{zta}, procedures, and defense mechanisms essential for \gls{oran}'s success~\cite{groen2024securing}. While WG11 acknowledges that attacks on the \gls{sp} can have a high impact, they consider the likelihood low due to the perceived sophistication required by an attacker~\cite{OranWG11-threatmodel}. However, a significant gap remains in analyzing security risks and proposing solutions for the open \gls{fh} \gls{sp} in production-grade \gls{oran} environments.

This section presents TIMESAFE (\underline{T}iming \underline{I}nterruption \underline{M}onitoring and \underline{S}ecurity \underline{A}ssessment for \underline{F}ronthaul \underline{E}nvironments), a framework for assessing the impact and likelihood of attacks against \gls{ptp} in the \gls{fh} of O-RAN and \gls{5g} networks. The key contributions relevant to this dissertation include: (i) the first experimental analysis of timing attack impact on a production-ready private cellular network, following O-RAN ALLIANCE testing procedures~\cite{OranWG11-sectestspec}, revealing that such attacks can cause catastrophic \gls{gnb} outages; and (ii) demonstration that timing attacks on \gls{ptp} in the \gls{fh} \gls{sp} are straightforward and require minimal sophistication, countering the assumption that such attacks are highly complex. The broader TIMESAFE framework also includes a transformer-based \gls{ml} detection mechanism achieving over 97.5\% accuracy. However, the focus in this dissertation is on the experimental security assessment enabled by the X5G platform~\cite{groen2024timesafe}.



The remainder of this section is organized as follows. Section~\ref{chap5-sec:timesafe-background} provides background on \gls{fh} synchronization and \gls{ptp}. Section~\ref{chap5-sec:timesafe-threat-model} describes the threat model and attack methods. Section~\ref{chap5-sec:timesafe-setup} presents the experimental setup on X5G, and Section~\ref{chap5-sec:timesafe-results} demonstrates the production network impact. Section~\ref{chap5-subsec:timesafe-ml} describes an ML-based attack detector. Finally, Section~\ref{chap5-sec:timesafe-related-work} surveys related work and Section~\ref{chap5-sec:timesafe-conclusion} concludes.

\subsection{Background}
\label{chap5-sec:timesafe-background}

\subsubsection{Fronthaul Synchronization Plane}

\begin{figure}[tb]
    \centering
    \includegraphics[width=.6\linewidth]{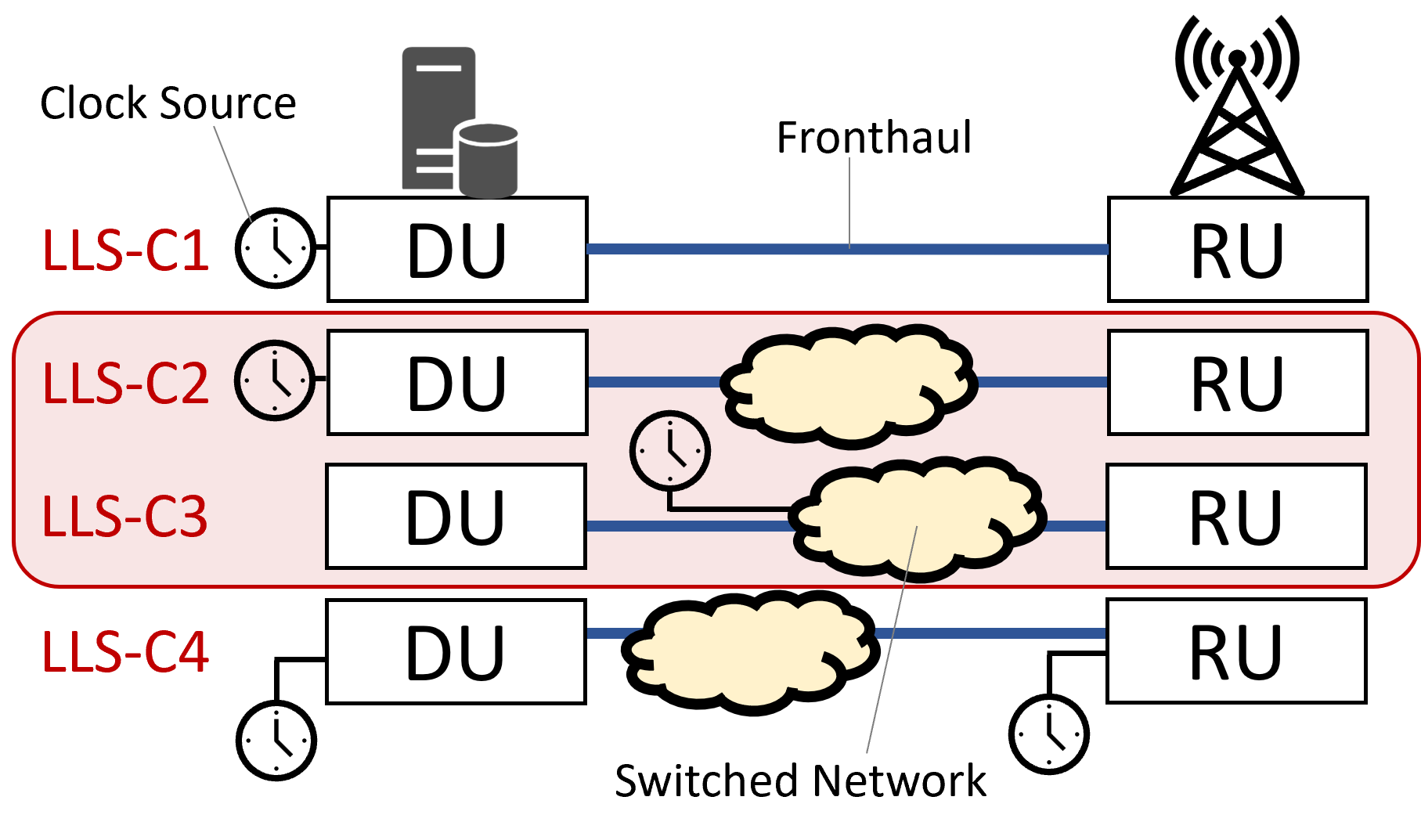}
    \caption{The Fronthaul connects the Distributed Unit (DU) to the Radio Unit (RU). There are four Synchronization Plane topologies based on the location of the clock source and switched network. LLS-C2 and LLS-C3 present the greatest risk to \acrshort{ptp} because the switched network is used to transport the timing messages. 
    }
    \label{chap5-fig:timesafe-sync_overview}
\end{figure}

The \gls{fh} \gls{sp} standards define four clock models and synchronization topologies based on the Lower Layer Split Control Plane (LLS-C): LLS-C1, LLS-C2, LLS-C3, and LLS-C4~\cite{ORA2023}, as illustrated in Figure~\ref{chap5-fig:timesafe-sync_overview}. LLS-C1, where the \gls{du} and \gls{ru} are directly connected without a switched network, poses the least security risk due to the absence of intermediate attack vectors. LLS-C4 provides timing directly to the \gls{du} and \gls{ru} using local GNSS receivers, reducing \gls{ptp} attack opportunities, except for potential GNSS jamming.

In contrast, LLS-C2 and LLS-C3 are more vulnerable due to their reliance on the switched network for \gls{ptp} message transport. In these scenarios, every device in the switched network represents a potential attack vector. In LLS-C2, the \gls{du} is part of the synchronization chain to the \gls{ru} with one or more Ethernet switches in between. LLS-C3 differs in that the \gls{du} is not in the synchronization chain; instead, timing is distributed from the \gls{prtc} to the \gls{du} and \gls{ru}(s) through the network. The increased complexity and multiple entities involved in securing the network infrastructure expand the risk surface, making LLS-C2 and LLS-C3 the most susceptible to attacks~\cite{groen2023implementing,OranWG11-threatmodel}.

\subsubsection{Precision Time Protocol Overview}
\label{chap5-sec:timesafe-ptp-overview}


The \gls{ptp} standard (IEEE 1588~\cite{ptp_standard}) ensures precise network clock synchronization in the nanosecond range. It achieves this through hardware time stamping, which minimizes delays from the networking stack~\cite{rezabek2022ptp}. The process involves electing a master clock that synchronizes all slave clocks in the network through a series of exchanges and messages.






The initial step involves leader election through the \gls{bmca}. Each clock periodically broadcasts \texttt{Announce} messages containing attributes such as: \textit{Priority1}, a configurable priority setting used as the first comparison feature; \textit{ClockClass}, indicating the clock's traceability and suitability as a time source; \textit{ClockAccuracy}, reflecting the clock's precision; and \textit{ClockIdentity}, a unique identifier used to resolve ties. 
Each clock evaluates the \textit{Announce} messages based on the available attributes, selecting the clock with the highest quality as the master. This process is dynamic, allowing for continuous re-evaluation and re-selection if higher-quality clocks are introduced or current attributes change.

\begin{figure}[tb]
    \centering
    \includegraphics[width=\linewidth]{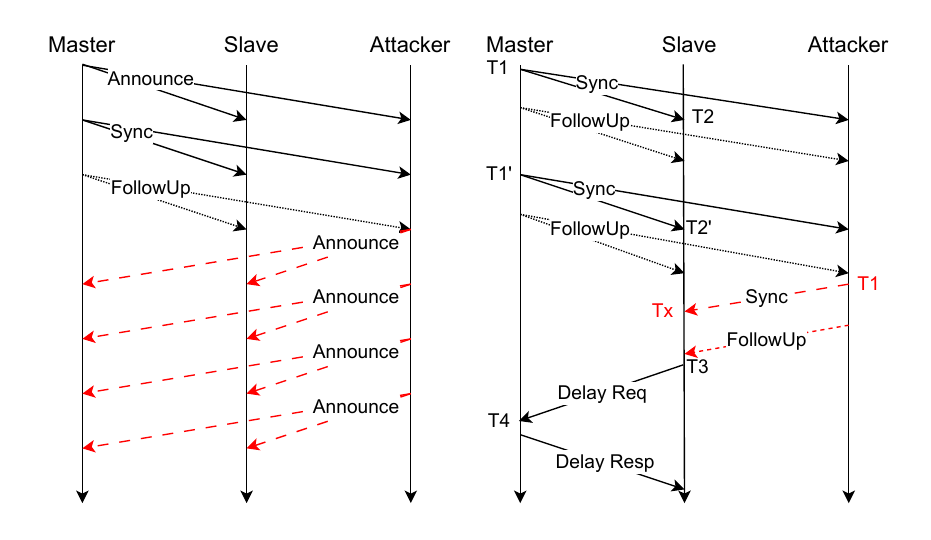}
    \caption{
    The black (solid/dotted) lines represent normal traffic, while the red (dashed) lines on the left side illustrate the flow of messages during a Spoofing Attack, where an attacker manipulates the \acrshort{bmca} to become the master. The red (dashed) on the right side illustrates the flow of messages during a Replay Attack,  showing the re-transmission of Sync and FollowUp messages.}
    \label{chap5-fig:timesafe-attacks}
\end{figure}

After election, the master clock begins synchronization with slave clocks through three critical aspects, illustrated in Figure~\ref{chap5-fig:timesafe-attacks}. \textit{Frequency} refers to the rate at which a clock oscillates; the master sends \texttt{Sync} messages and takes a timestamp ($T_1$), with TwoStep clocks sending a separate \texttt{FollowUp} message containing this timestamp. The slave records the receive timestamp ($T_2$), and by comparing successive \texttt{Sync} messages, calculates the drift to adjust its oscillation rate. \textit{Delay} refers to the propagation time through the network; the slave sends a \texttt{DelayReq} message and records timestamp ($T_3$), while the master records the receive timestamp ($T_4$) and returns it in a \texttt{DelayResp} message, allowing the slave to compute the one-way delay as $\textit{Delay} = \frac{(T_4 - T_1) - (T_3 - T_2)}{2}$. \textit{Phase} refers to the relative alignment between clocks; after calculating delay, slaves compute the offset as $\textit{Offset} = (T_2 - T_1) - \textit{Delay}$ to minimize phase difference with the master. This message exchange process is repeated at defined intervals to maintain synchronization.

\subsection{Threat model}
\label{chap5-sec:timesafe-threat-model}

The threat model of this work assumes an attacker has gained access to a device within the \gls{fh} network. The complexity of the O-RAN architecture, with its many interconnected components, heightens security risks through several vectors. \textit{Increased remote access}: O-RAN's multi-vendor architecture involves diverse suppliers and service providers with varying levels of access control, increasing the risk of misconfigured authentication or compromised credentials~\cite{ibm2023, verizon}. \textit{Increased physical access}: the distributed nature of \gls{gnb} edge servers, often located in accessible public spaces, makes on-site physical access an increasingly viable entry point, where adversaries can insert rogue devices to intercept or replay traffic~\cite{xing2024criticality}. \textit{Supply chain vulnerabilities}: software-based network functions are susceptible to supply chain attacks, with recent studies uncovering high-risk dependencies, misconfigurations, and weak security practices across multiple O-RAN components~\cite{10.1145/3643833.3656118, hung2024security, atalay2023securing}.

Once an attacker gains access to the switched network, they can observe, replay, or inject \gls{ptp} broadcast packets. Within the TIMESAFE framework, we focus on two specific attacks targeting \gls{ptp} synchronization and chosen for their direct impact on \gls{ptp} synchronization mechanisms: \textit{Spoofing Attack} and \textit{Replay Attack}.


\subsubsection{Spoofing Attack}

The TIMESAFE framework spoofing attack targets \gls{ptp} nodes by sending counterfeit \textit{Announce} messages to manipulate the \gls{bmca} leader election and assume the role of the master clock. Initially, the attacker monitors benign traffic to gather critical information about the protocol configuration and node attributes. With this data, the attacker crafts \textit{Announce} packets designed to have superior attributes crucial for master clock selection. The \gls{bmca} relies on up to nine features to determine the master clock, starting with \textit{Priority1} (Section~\ref{chap5-sec:timesafe-ptp-overview}). The attacker manipulates this value by setting it to 1. Similarly, the attacker sets the values for \textit{ClockClass} 
to 1. The \textit{ClockIdentity} field is created by inserting priority values between the manufacturer ID and device ID portions of the MAC address. The attacker inserts \texttt{ffff} 
ensuring it has higher priority values than the legitimate Master. Other values are copied from the legitimate master.

These crafted messages are sent periodically to maintain the attacker's role as master. Apart from \textit{Announce} messages, no synchronization packets 
are sent, and \textit{DelayReq} messages from slaves remain unanswered. By disrupting the synchronization process, all clocks operate independently, leading to a gradual drift. Consequently, the attacker introduces synchronization errors among all nodes in the network, causing a complete outage for the \gls{gnb}, as we will show in Section~\ref{chap5-sec:timesafe-results} and illustrated in Figure~\ref{chap5-fig:timesafe-overview}.

\subsubsection{Replay Attack}
The Replay Attack involves sniffing time synchronization messages (\textit{Sync} and \textit{FollowUp}) from the Master clock, storing, and retransmitting them after a delay. When the Slave clock receives the retransmitted messages, it uses the outdated timestamp \( T_{1} \) and a new timestamp \( T_{x} \) to compute drift, delay, and offset (see Figure~\ref{chap5-fig:timesafe-attacks}). The formulas in Section~\ref{chap5-sec:timesafe-ptp-overview} become: $\textit{Drift} = \frac{(T_{2}' - T_{x}) - (T_{1}' - T_{1})}{T_{1}' - T_{1}}$, $\textit{Delay} = \frac{(T_{4} - T_{1}) - (T_{3} - T_{x})}{2}$, and $\textit{Offset} = (T_{x} - T_{1}) - \textit{Delay}$. The actual difference between when $T_1$ is generated and received in the malicious \gls{ptp} message significantly exceeds the calculated difference, causing offset to increase by orders of magnitude and severely disrupting network synchronization.



\subsection{Experimental Setup}
\label{chap5-sec:timesafe-setup}

The experimental analysis is performed on the X5G platform described in Section~\ref{chap:4}. The platform comprises multiple NVIDIA \gls{arc} nodes with dedicated \gls{cn} and \gls{fh} infrastructure. NVIDIA \gls{arc} combines \gls{oai} for the higher layers of the protocol stack with NVIDIA Aerial, a physical layer implementation accelerated on \gls{gpu} (NVIDIA A100) with a 7.2 split implementation using NVIDIA Mellanox smart \glspl{nic}. The \gls{gpu} is coupled with the programmable \gls{nic} (Mellanox ConnectX-6 Dx) via \gls{rdma}, bypassing the CPU for direct packet transfer.
The \gls{fh} infrastructure uses a Dell S5248F-ON switch and a Qulsar QG-2 as the grandmaster clock, distributing \gls{ptp} and SyncE synchronization to the \gls{du} and \gls{ru} with an O-RAN LLS-C3 configuration. For \gls{ptp} synchronization, \textit{ptp4l} is used, an IEEE 1588 compliant implementation for Linux. The \gls{ru} is a Foxconn 4T4R unit operating in the $3.7$--$3.8$~GHz band, with \gls{cots} \gls{5g} smartphones from OnePlus as \glspl{ue}.
Based on the threat model, one of the \glspl{du} is assumed to be the compromised machine from which the attacker can send malicious \gls{ptp} packets. The experimental validation focuses on the spoofing attack, as proper switch port configuration in the production environment prevented the replay attack from reaching other nodes.

\subsection{Impact of the Attacks}
\label{chap5-sec:timesafe-results}

The spoofing attack is tested in the production-ready private \gls{5g} network to assess their impact following the recently published guidelines in the O-RAN ALLIANCE testing specifications~\cite{OranWG11-sectestspec}. According to NIST's National Vulnerability Database~\cite{NVD}, impact levels are categorized as none, low, or high. Low impact attacks cause performance degradation or intermittent availability without fully denying service. High impact attacks result in complete outages, denying service to all \glspl{ue}.

\subsubsection{Switch Configuration Effects}

In our initial tests, we found that switch configurations can provide protection against attacks. Malicious packets are not always received by other nodes, as verified through Wireshark captures on both the malicious machine and other nodes. This protection is tied to the switch port's \gls{ptp} role settings. On the Dell S5248F-ON switch, ports set to master only send \gls{ptp} packets, slave mode only receives, and dynamic mode allows both. Additionally, switch configurations can block duplicate MAC addresses. While a properly configured switch prevented our replay attack, assuming correct configurations is risky, especially when \gls{du} and \gls{ru} vendors lack control over network settings, and the complexity and diversity of equipment and configurations in the \gls{fh} path exacerbate this risk~\cite{verizon, ibm2023, rapid7}. The spoofing attack, which does not rely on MAC address duplication, remained effective under various switch configurations.

\subsubsection{Attack Results}

\begin{figure}[htb]
    \centering
    \includegraphics[width=.7\linewidth]{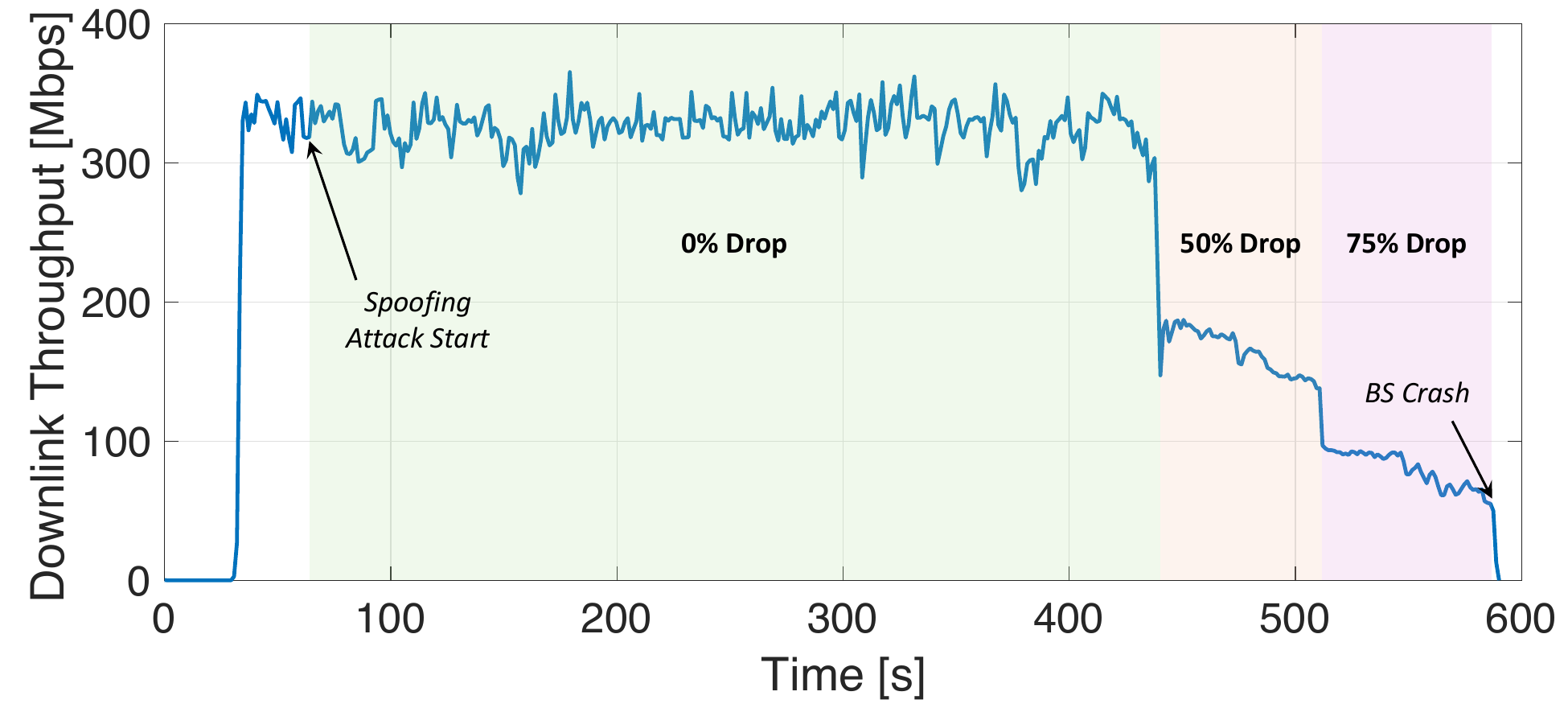}
    \caption{
    Downlink throughput to a \gls{ue} during a spoofing attack with the switch port set to \gls{ptp} master. 
    Approximately 60 seconds into the experiment, the spoofing attack is initiated. Around second 440 there is a 50\% drop in throughput, progressing to a 75\% drop by second 510, and ultimately causing the base station to crash at around second 580. 
    }
    \label{chap5-fig:timesafe-SpoofingX5_first_experiment}
\end{figure}

In our first successful attack, we configure the port of our malicious machine with a \gls{ptp} master role. We start our private 5G network, connect the \gls{ue}, and begin sending around 300 Mbps downlink iPerf traffic. After approximately 60 seconds, we launch the spoofing attack, as shown in Figure~\ref{chap5-fig:timesafe-SpoofingX5_first_experiment}. Monitoring the \gls{ptp} service logs at the \gls{du} reveals that the attack causes the \gls{du} to lose synchronization with the grandmaster clock, leading to a drift in network synchronization as the \gls{du} uses its internal clock. This drift degrades \gls{ue} performance, eventually causing disconnection from the network. About 380 seconds after the beginning of the attack, there is a sudden 50\% drop in throughput. From this point, the throughput continues to drop roughly linearly as the clock drift increases. Finally, the \gls{du} loses synchronization with the \gls{ru}, the throughput drops to 0, and the \gls{ue} disconnects. When this attack stops, the \gls{du} resynchronizes with the grandmaster clock, and the network returns to a healthy state.


\begin{figure}[htb]
    \centering
    \includegraphics[width=.7\linewidth]{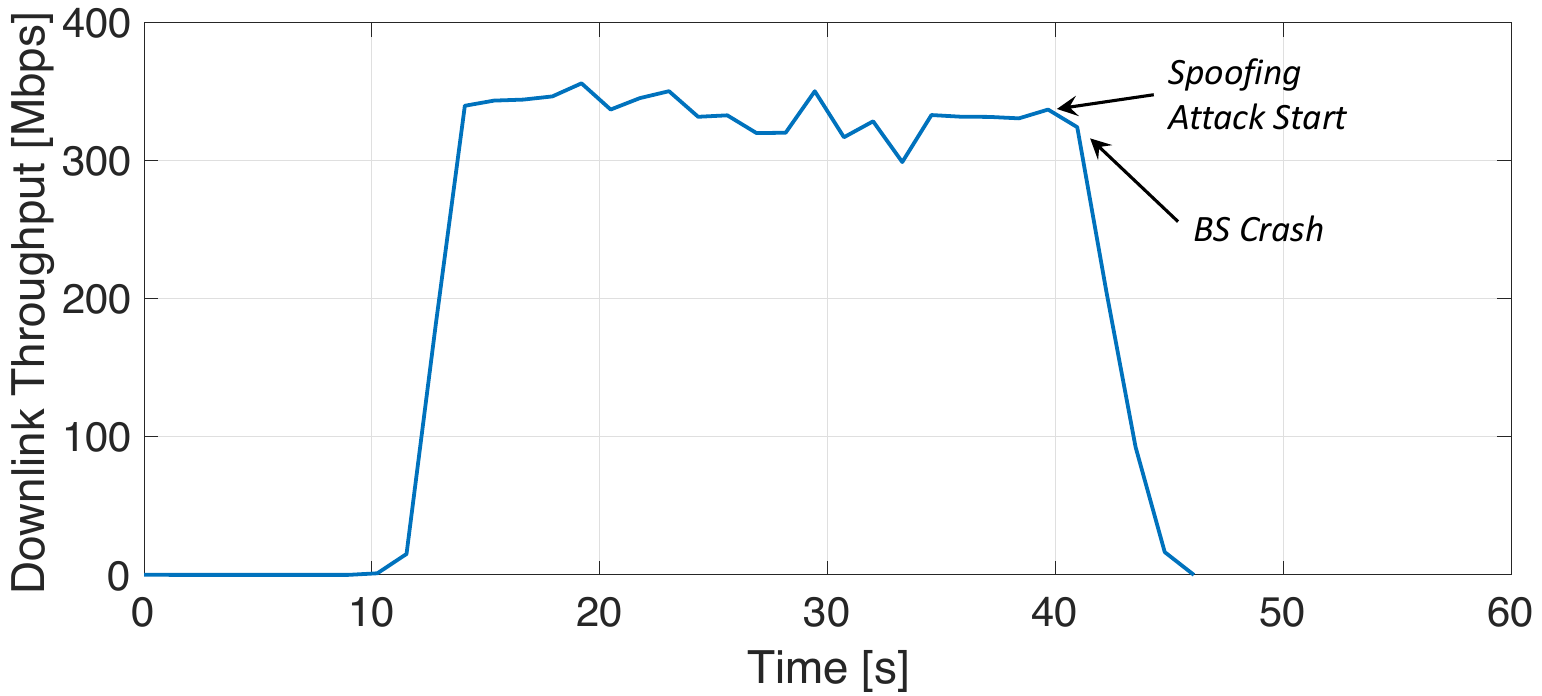}
    \caption{
    Downlink throughput to a \gls{ue} during a spoofing attack with switch ports set to \gls{ptp} dynamic role. Throughput remains stable at around 350 Mbps initially. About 40 seconds in, the spoofing attack begins, causing the 5G cell to drop and the \gls{ue} to lose connectivity. }
    \label{chap5-fig:timesafe-SpoofingX5}
\end{figure}

Next, we configure all ports as \gls{ptp} dynamic so that malicious packets can be received by the other nodes and repeat the experiment. This time, the malicious \gls{ptp} packets reach all the machines in the network. Approximately 2 seconds after the attack begins, the \gls{ru} crashes, stopping its operations, causing the 5G cell to drop and the \gls{ue} to lose connection, as shown in Figure~\ref{chap5-fig:timesafe-SpoofingX5}. While the \gls{du} was able to recover on its own after the first attack, the \gls{ru} required a manual reboot to regain functionality after the second attack. This behavior might be device specific, in our case the Foxconn \gls{ru}. Regardless of any potential vendor-specific variations, these attacks proved to be very powerful, causing significant disruption and necessitating manual intervention to restore full network operations. These tests demonstrate the impact to the \gls{oran} \gls{gnb} is \textit{high}. Failing to secure the \gls{sp} against spoofing attacks results in a complete outage of the \gls{gnb}.

\subsection{ML-Based Attack Detection}
\label{chap5-subsec:timesafe-ml}

Given the demonstrated impact of synchronization attacks, a detection mechanism is essential. Traditional rule-based solutions require extensive expert knowledge and fixed thresholds that struggle to adapt to evolving threats. In contrast, \gls{ml} offers a more adaptive approach: models can learn complex patterns in \gls{ptp} traffic to distinguish between benign fluctuations and actual attacks, without requiring manually specified rules for each attack type.

\subsubsection{Detection Pipeline}

The TIMESAFE detection pipeline consists of three parallel components deployed on the \gls{du}. The \textit{packet acquisition} stage captures Ethernet frames, filters for \gls{ptp} packets, and extracts six features from each packet: Ethernet source and destination addresses (to identify flows and packet direction), packet length (to detect malformed packets), \gls{ptp} sequence ID and message type (to understand the nature and order of messages), and inter-arrival time (a key indicator since O-RAN specifies fixed intervals for announce and sync messages). The \textit{pre-processing} stage maps observed MAC addresses to integers, enabling the model to learn traffic patterns rather than specific addresses. The \textit{decision-making} stage uses a sliding window approach to classify each sequence as benign or malicious. With a stride of two, at most two malicious packets are sent before the model detects an attack.

\subsubsection{Model Architecture}

\begin{figure}[tb]
    \centering
    \includegraphics[width=.8\linewidth]{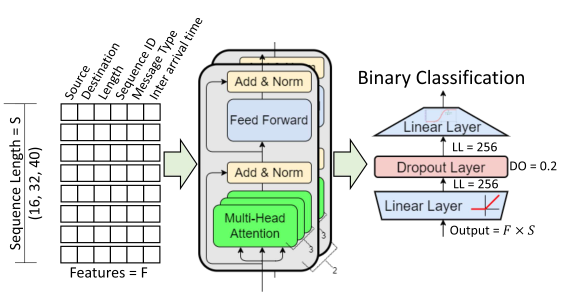}
    \caption{The final TIMESAFE transformer model uses 2 transformer layers, a fully connected linear layer with ReLU activation function, a dropout layer, and a second linear layer with a sigmoid activation function. The output is a single binary classification for the entire sequence.}
    \label{chap5-fig:timesafe-transformer}
\end{figure}

Several \gls{ml} architectures were evaluated, including Random Forest, LSTM, CNN, and Transformer models. The task of classifying packet sequences as malicious or benign parallels sentiment analysis and image classification, where context and sequential features are key. CNNs excel at extracting spatial features from packet patterns, while Transformers are superior at capturing contextual dependencies in sequential data. Both significantly outperformed other approaches.

The selected Transformer model (Figure~\ref{chap5-fig:timesafe-transformer}) uses only the encoder portion of the full transformer architecture. It processes input sequences of fixed length $S=32$ packets with $F=6$ features. The transformer encoder outputs a feature matrix fed into a fully connected linear layer with ReLU activation, followed by a dropout layer (probability 0.2) and a final linear layer with sigmoid activation for binary classification. The model uses 2 attention heads and comprises 115,609 parameters (463~KB)---small enough for deployment on production \gls{du} hardware.

\subsubsection{Training and Evaluation}

Training data was generated by capturing \gls{ptp} traffic during attack and normal operation periods using the \gls{dt} environment described fully in~\cite{groen2024timesafe}. The data was split into equal-sized chunks (1000 messages each) without overlaps, with 80\% for training, 10\% for validation, and 10\% for testing. Training employed weighted binary cross-entropy loss to address class imbalance, the Adam optimizer with initial learning rate of 0.001, and a StepLR scheduler reducing the learning rate by 0.1 every 10 epochs. Early stopping monitored validation loss with patience of 20 epochs.

Given that the impact of false negatives (missing an attack) is catastrophic while false positives have lower impact, both accuracy and recall were considered in model selection. The Transformer achieved the best balance with accuracy exceeding 97.5\% and low false negative rate on offline evaluation.

\subsubsection{Production Detection Results}

\begin{figure}[htb]
    \centering
    \includegraphics[width=.4\linewidth]{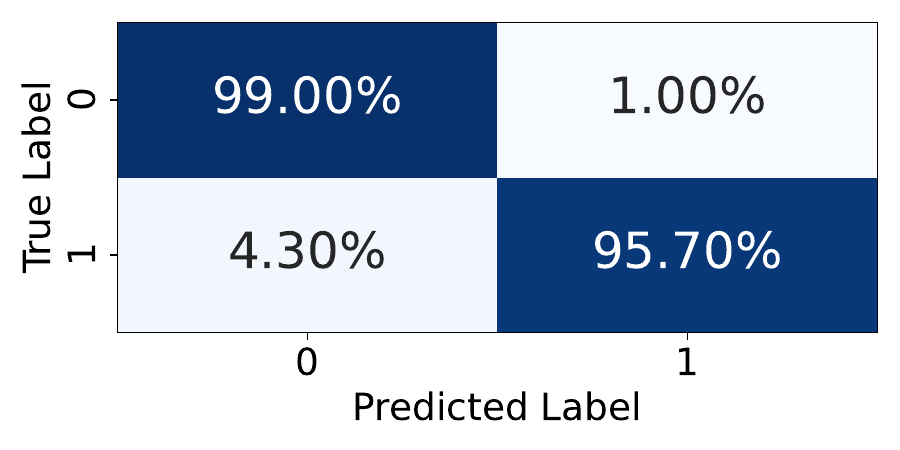}
    \caption{Confusion matrix for production environment attack detection. The Transformer model achieves over 99\% accuracy in detecting \acrshort{ptp} attacks on the X5G platform.}
    \label{chap5-fig:timesafe-detection}
\end{figure}

The detection pipeline was evaluated using packet traces (\texttt{.pcap} files) captured during successful attacks on the X5G production network. As shown in Figure~\ref{chap5-fig:timesafe-detection}, the Transformer model achieves accuracy exceeding 99\%, demonstrating strong ability to distinguish between malicious and benign traffic in a real-world setting. This result validates that \gls{ml}-based monitoring provides a cost-effective approach to detecting synchronization attacks without the latency and complexity overhead of cryptographic solutions.

\subsection{Related Work}
\label{chap5-sec:timesafe-related-work}

There is a growing awareness of the need to secure \gls{ptp}. 
Several attacks are demonstrated in a data center context by DeCusatis \emph{et al.}~\cite{decusatis2019impact}. Itkin and Wool developed 
a detailed analysis of threats against \gls{ptp} and proposed using an efficient elliptic-curve public-key signature for Prong A~\cite{itkin2017security}. Shereen \emph{et al.}~\cite{shereen2019next} 
experimentally evaluated implementing Prong A, finding that software-based authentication adds at least an additional 70$\mu$s of delay. Similarly, Rezabek \emph{et al.}~\cite{rezabek2022ptp} evaluated software-based authentication and concluded that there is visible degradation of clock synchronization for each hop in the network with standard deviations between $118$ and $571$~ns.

There are several works that investigate security threats and propose related defense mechanisms for \gls{ptp} in smart grids. 
Moussa \emph{et al.}~\cite{moussa2016detection} proposed adding a new type of \gls{ptp} clock and modifying the \gls{ptp} slave functionality. Then in subsequent work, Moussa \emph{et al.}~ 
proposed adding further message types to aid in detection and mitigation~\cite{moussa2019extension}, and adding feedback from both slaves and master clocks to a reference entity~\cite{moussa2018securing}.
%
A method targeting Prong C, where clock and network path redundancy are added, is proposed by Shi \emph{et al.}~\cite{shi2023ms}. Likewise, the approach by Finkenzeller \emph{et al.}~\cite{finkenzeller2024ptpsec} uses redundant, or cyclic paths, to detect and mitigate time delay attacks.
Maamary \emph{et al.} 
give an overview of several threats to the \gls{fh} \gls{sp} and discuss possible countermeasures~\cite{s_plane_survey}, though they do not implement either attacks or countermeasures. \gls{fh} security considerations are further described and MACsec is proposed as a solution by Dik \emph{et al.}~\cite{dik2021transport,dik2023100,dik2023macsec,dik2023open}. While these works are the most similar to ours, they primarily address Prong B and could be implemented in parallel to our work. 
%
%
In contrast to the above works, TIMESAFE demonstrates successful attacks against \gls{ptp} used for the \gls{sp} causing the gNB to fail and demonstrating how O-RAN-connected devices can be leveraged as attack vectors. Additionally, we propose a \gls{ml}-based method to monitor \gls{ptp} in accordance with Prong D, successfully discriminating between benign and malicious traffic with over 97.5\% accuracy.

\subsection{Conclusion}
\label{chap5-sec:timesafe-conclusion}




This section presented TIMESAFE, a security assessment framework that demonstrates the vulnerability of O-RAN \gls{fh} synchronization to \gls{ptp} attacks. Experimental validation on the X5G platform showed that spoofing attacks can cause production-ready \gls{5g} base stations to fail within 2 seconds, with the \gls{ru} requiring manual reboot to restore operations. While proper switch configuration prevented replay attacks in the production environment, the spoofing attack remained effective, showing that network configuration alone cannot guarantee security. To counter these threats, TIMESAFE employs an \gls{ml}-based detection pipeline that achieves over 99\% accuracy in identifying malicious \gls{ptp} traffic on the production network.

Addressing synchronization security involves balancing security costs against performance implications. While robust measures like cryptographic authentication introduce latency and computational overhead---particularly challenging for FPGA-based \glspl{ru}---TIMESAFE's monitoring approach offers a cost-efficient alternative. By enabling real-time detection, higher-cost security measures can be applied only when necessary, optimizing both security and performance. Future work should investigate additional attack vectors such as GNSS jamming (affecting LLS-C4), packet removal, and selective delay manipulation, as well as multi-class classification to distinguish between different attack types.

This work highlights a unique capability of physical, production-grade testbeds for security research. Unlike simulation-based studies, attacks validated on real hardware with commercial \glspl{ru} and \glspl{ue} provide evidence of vulnerability effects. The results show the importance of securing the \gls{sp} in O-RAN deployments through proactive security measures and demonstrate how platforms like X5G enable security research that would be difficult on emulated or simulated systems.

\section{Summary and Discussion}
\label{chap5-sec:conclusions}



This chapter demonstrated the \emph{use case} dimension of physical platforms, moving one step further in the experimental journey from \glspl{dt} and emulation environments to real-world \gls{ota} experimentation. Building on the X5G platform architecture and dApp framework presented in Chapter~\ref{chap:4}, we showcased four representative use cases highlighting the unique capabilities of production-grade systems and demonstrating how an open and programmable \gls{5g} \gls{ran} can support advanced \gls{ai}-native functionalities under realistic propagation and hardware conditions for advancing \gls{5g} and \gls{6g} research.

First, cuSense demonstrated \gls{isac} capabilities by leveraging \gls{csi} from existing \gls{5g} uplink signals for real-time person localization, achieving sub-millisecond inference latency while operating as a sensing overlay on the \gls{gpu}-accelerated \gls{ran}. Second, InterfO-RAN showcased real-time interference detection using a CNN-based classifier on IQ samples, achieving over 91\% accuracy with latency under $650~\mu\mathrm{s}$. Third, ORANSlice provided an open-source, \gls{3gpp}-compliant \gls{ran} slicing platform with E2SM-CCC integration and Near-RT \gls{ric} xApp control, validated on X5G with commercial hardware. Finally, TIMESAFE exposed \gls{fh} synchronization vulnerabilities, demonstrating that \gls{ptp} spoofing attacks can crash production base stations within $2$~seconds, while an \gls{ml}-based detector achieves over 99\% accuracy.

These results demonstrate the complementary role of physical platforms compared to \glspl{dt} and emulation environments discussed in Chapters~\ref{chap:2} and~\ref{chap:3}. While emulation provides controlled, repeatable conditions for initial design and large-scale data generation, physical testbeds such as X5G are necessary to capture hardware impairments, synchronization issues, environmental dynamics, and implementation details for a proper full end-to-end behavior. 

From the perspective of the four key research challenges introduced in Chapter~\ref{chap:intro}, this chapter mainly addressed the \emph{use case} dimension on physical platforms, building on the \emph{deployment}, \emph{realism}, and \emph{usability} foundations established in Chapter~\ref{chap:4}. The design, development, deployment, and validation of the presented applications (cuSense, InterfO-RAN, ORANSlice, and TIMESAFE) on X5G confirm that the software and hardware infrastructure enable the execution of \gls{ai}-native workloads and security experiments through the use of commercial \glspl{ue}, multi-vendor \glspl{ru}, and realistic traffic patterns. Additionally, the open-source components and pipelines developed and released by these works contribute reusable tools that can be leveraged by the whole research community.

Finally, the four case studies presented in this chapter represent only a subset of the research that physical platforms like X5G can enable. Several additional studies are ongoing or already published, including: mobility and handover optimization, e.g., via traffic-aware F1 handover strategies using data-driven controllers; \gls{ai}-based channel estimation, such as \cite{ford2025sim}, which combines \gls{dt}-based training with \gls{ota} validation; network automation and zero-touch provisioning~\cite{maxenti2025autoran}; energy-efficient \gls{ran} operation; multi-vendor integration testing~\cite{gemmi2024open6g}; agentic \gls{ai} for scheduling and autonomous control~\cite{elkael2025allstar,elkael2025agentran}.

\chapter{Conclusions}
\label{chap:conclusion}




This dissertation has explored the design, development, deployment, and evaluation of next-generation cellular networks through open and programmable platforms, moving from digital twins and emulation environments to production-grade physical \gls{ota} testbeds.
Starting from the observation that modern \glspl{ran} are becoming more complex, are being realized as software running on general-purpose and accelerated hardware, and that the Open RAN paradigm with its open interfaces, such as those defined by O-RAN, is making the network programmable and observable at multiple timescales, this work has focused on two complementary goals: building realistic platforms to enable the development and testing of novel ideas in a safe and reproducible environment; and using those platforms to realize intelligent applications for monitoring, control, sensing, and security.

On the digital side, we demonstrated the use of Colosseum, the world's largest wireless network emulator with hardware-in-the-loop, as a digital twin for experimental wireless research at scale. Through the development of the Channel Emulation Scenario generator and Sounder Toolchain (CaST), we enabled the automated creation of emulated environments that faithfully replicate real-world deployments. CaST combines 3D modeling, ray-tracing-based channel generation, and channel sounding validation to bridge physical and digital domains. On top of this toolchain, we realized end-to-end emulation pipelines for spectrum sharing, \gls{ai}-driven propagation modeling, and synthetic \gls{rf} data generation, demonstrating how digital platforms can serve both as safe experimentation environments and as data factories for \gls{ai}-native \gls{ran} solutions.

On the physical side, we designed, deployed, and evaluated X5G, an open, programmable, and multi-vendor private 5G O-RAN testbed. X5G integrates \gls{gpu}-accelerated physical layer processing through NVIDIA Aerial with open-source higher layers from OpenAirInterface, connected via O-RAN fronthaul to commercial and programmable \glspl{ru} and multiple \glspl{cn}, and connects to a Near-RT \gls{ric} running xApps on an OpenShift cluster. Beyond the core infrastructure, we developed a \gls{gpu}-accelerated dApp framework that exposes real-time \gls{phy}/\gls{mac} telemetry through shared memory and E3-based interfaces and couples it with Triton-based inference pipelines on the same accelerator as the \gls{gnb}, enabling \gls{ai}-native distributed applications to close control loops at sub-millisecond timescales.

We leveraged these platforms through a set of representative intelligent \gls{ran} applications. In emulation, Colosseum and CaST have been used to study spectrum sharing between cellular and incumbent systems, \gls{ai}-assisted radio map generation, and generative models for synthetic \gls{rf} datasets. On X5G, we realized cuSense for uplink-\gls{csi}-based \gls{isac}, InterfO-RAN for real-time uplink interference detection, ORANSlice for dynamic resource management and network slicing via xApps, and TIMESAFE for security assessment of fronthaul synchronization.
Colosseum and X5G play complementary roles. Colosseum offers a controlled, reproducible environment where solutions can be safely prototyped and validated at scale, while X5G provides a production-grade \gls{ota} network with commercial devices and real-world \gls{rf} conditions. Together, they form a pipeline from emulation to real deployment, allowing researchers to refine ideas while gradually increasing experimental fidelity without risking commercial infrastructure.

Throughout this dissertation, we can now fully revisit the four key research dimensions from Chapter~\ref{chap:intro} that guided our work.
In terms of \emph{deployment}, we developed CaST for automating \gls{dt} scenario creation and validation on Colosseum and we designed and deployed a multi-vendor O-RAN-compliant platform with \gls{gpu}-acceleration, namely X5G.
For \emph{realism}, we proved that Colosseum digital twins can closely reproduce real setups, and we validated X5G as an end-to-end, 3GPP-compliant private 5G network with commercial smartphones.
Regarding \emph{usability}, we released open-source tools, \gls{dt} scenarios, reference design architectures, experimental pipelines, and comprehensive documentation that lower the barrier for researchers to develop and test their own solutions.
Finally, the \emph{use case} dimension has been demonstrated through an extensive set of applications across both platforms, showcasing the capabilities and how these systems can help address key challenges and advance research.

Beyond these specific contributions, this work highlights how modern wireless networks are evolving into more open, programmable systems that require realistic and controlled infrastructures to test bold ideas and collect data at scale. By combining digital twins on Colosseum with a private 5G cell like X5G, we obtain an end-to-end pipeline from emulation to over-the-air validation. The resulting tools, platforms, and use cases form a reusable blueprint where a new idea can follow the experimental journey from concept through simulation and emulation on Colosseum, to \gls{ota} validation on X5G, and finally to deployment as an \gls{ai}-native dApp or xApp, lowering the barrier to designing and testing the next generation of intelligent, open, and secure wireless networks.

\bibliographystyle{IEEEtran}

\bibliography{bib/thesis,bib/chap1,bib/chap2,bib/chap3/chap3,bib/chap3/chap3-twinning,bib/chap3/chap3-airmap,bib/chap3/chap3-gentwin,bib/chap4,bib/chap5/chap5-cusense,bib/chap5/chap5-interforan,bib/chap5/chap5-oranslice,bib/chap5/chap5-timesafe}

\appendix

\cleardoublepage

\begin{acknowledgements}

\markboth{ACKNOWLEDGMENTS}{ACKNOWLEDGMENTS}

The Ph.D. journey is a path that everyone who decides to take it experiences in their own way. I have met many other students over these years, and most of the time they would tell me that they could not wait to graduate and move on. In my case, I truly enjoyed the journey, and even when it was not always easy, all the days spent in the lab--and all the effort to complete that last experiment late at night--were surely worth it in the end.

But none of this would have been possible without my advisor, Tommaso Melodia. He believed in me and welcomed me into this journey. He is the first person I would like to thank, and I will always be grateful for his immense guidance and support.
I am also grateful to my Ph.D. committee members, Stefano Basagni and Josep Jornet, for their valuable insights and discussions over all these years.
I owe special thanks to Michele Polese and Leonardo Bonati, who patiently guided me from the very beginning and from whom I learned so much. I could not have done this without them.
In addition, I would like to recognize my co-authors, who provided valuable discussions, advice, and collaboration across all these projects, including: Salvatore D'Oro, Pedram Johari, Dimitrios Koutsonikolas, Hai Cheng, Neagin Neasamoni Santhi, Chris Dick, Mauro Belgiovine, Florian Kaltenberger, Ali Saeizadeh, Joshua Groen, Miead Tehrani-Moayyed, Nicholas Hedberg, Stefano Maxenti, Daniel Uvaydov, Simone Di Valerio, Osman Basaran, Russell Ford, Hao Chen, Imran Khan, Kurt Yoder, Paresh Patel, Ventz Petkov, Michael Seltser, Abhimanyu Sheshashayee, Clifton Robinson, Matteo Bordin, Pietro Brach del Prever, Gabriele Gemmi, Maria Tsampazi, Andrea Lacava, Ruben Soares da Silva, Anupa Kelkar, Abhimanyu Gosain, Kaushik Chowdhury, J. Gordon Beattie Jr., Ian Wong, Eric Anderson, Rajeev Gangula, Sakthivel Velumani, Claudio Fiandrino, Falko Dressler, Utku Demir, Eduardo Baena, Subhramoy Mohanti, Elisa Bertino, Francesca Cuomo, Chiara Petrioli, Robert Schmidt, Sagar Arora, Mikel Irazabal, Irfan Ghauri.

I also extend my acknowledgment to all other colleagues at the Wireless Networks and Embedded Systems (WiNES) Lab and, more broadly, at the Institute for the Wireless Internet of Things. Since I started in ISEC and now that we have moved to EXP, the group has grown a lot, and many people have passed through--some for just a few months, others for longer, and some who decided to stay--and they made the environment the dynamic and fun place that I enjoyed going to every day. So, if not already mentioned, I would like to thank: Ravis, Maxime, Paolo, MJ, Angelo, Eugenio, Mattia, Sergi, Ergest, Chris, Niloofar, Reena, Evgeniia, Fiona, Noah, Neko, Sandra, Bill, Kuntheary, Rosy, Ed, Deniz, Amani, Emrecan, Kerem, Luca, Kengo, Haruki, Nicoló, Filippo, Noemi, Nicole, Riccardo, Alessio, Manuel, Rawlings, Luca, Cleverson, Pierluigi, Tommaso, Matteo, Francesca, Raoul, Ivan, Leonardo, Antonio, Simone, Pietro, Luis, Rafael, Armando, Andrea, Hithesh, Madhukar, Reshma, Tamerlan, Dat, Francesco, Antonio, Daniele, Alessandro, Lorenzo, Raffaele, Francesco, Sofia, Alessandro, Aditya, Alberto, Moinak, Stella, Vitaly, Amin, Duschia, Samar, Xavi, Phuc, Andrew, Fabio, Sara.

Boston has been my home for the past few years, and I have had the pleasure of meeting many good friends who made this time truly special. In particular, I would like to express my gratitude to, among others: Francesca, Maria, Gabriel, Farah, Sasha, Aldo, Marianna, Matteo, Federica, Davide, Valentina, Sandro, Francesca, Elisa, Farnoush, Justin, Luca, Pietro, Aurelio, Yvonne, Samuel, Teresa, Walter, Alessio, Nicolas, Zamora, Tommaso, Niccolò, Mariasilvia, Bernard, Lorenzo, Gaia, Kayland, Aruna, Benedetta, Amedeo, Sen, Filippo, Raffaela, Leonardo, Chiara, Maria, Matteo, Marco, Bianca, Michela, Simone, Andrea, Alessandro, Cristina, Pasquale, Helena, Francesco.

Moreover, I also had the chance to travel around, doing internships outside of Boston and experiencing different environments, and I always felt lucky to meet great people and make meaningful connections. I sincerely appreciated all the conversations and experiences we shared, and so I would like to thank, among others: Junho, Ana, Gavin, Humza, Nelson, Jing, JF, Fredrik, Kuntal, Lopa, Chason, Mingjun, Reinhard, Jacob, Caitriona, Ragav, Xiaowen, Jiayu, Andrew, Yirong, Longfei, Vitali, Debashri, Pranav, Sebastian, Carlo, Brighton, TK, Josh, Francesco, Chieh, Bimo, Neel, Manu.
%

In addition, I want to express my appreciation to my old friends from overseas who, even across this long distance, were able to maintain a great connection with me, including: Giancarlo, Francesco, Carlo, Francesco, Lorenzo, Willy, Giulio, Serena, Manuel, Mirko, Roberto, Marco, Marcello, Silvia, Giovanni, Giulia, Nacho, Erica, Federico, Simone.

Lastly, but certainly not least, I want to acknowledge all of my family for their constant support. Most of all, I would like to thank my parents, Mirella and Franco, to whom I was, am, and always will be immensely grateful for everything they give me every day from so far away, and who continue to make me the person I am today.


\end{acknowledgements}

\printindex

\end{document}